\documentclass[12pt,a4paper]{article}

\usepackage[dvipsnames]{xcolor}
 \definecolor{newred}{rgb}{0.43, 0, 0}

\usepackage[colorlinks=true, linkcolor=newred, citecolor=newred]{hyperref}

\usepackage{comment}
\usepackage[T1]{fontenc}
\usepackage{color}
\usepackage{adjustbox}
\usepackage{amssymb,bbold,MnSymbol}
\usepackage{epsf}
\usepackage{epsfig}
\usepackage{relsize}
\usepackage{longtable}
\usepackage{enumerate}
\usepackage{slashed}
\usepackage{latexsym,mathrsfs,dsfont}
\usepackage[normalem]{ulem} 
\usepackage[compress]{cite}
\usepackage{graphicx}
\usepackage{makecell}
\usepackage{cancel}
\usepackage{url}
\usepackage{booktabs}
\usepackage{float}
\usepackage{multirow}
\usepackage{changepage}
\usepackage[hypcap]{caption, subcaption}
\usepackage{stackengine}

\def\CC{{C\nolinebreak[4]\hspace{-.05em}\raisebox{.4ex}{\tiny\bf ++}}}

\usepackage{tikz}
\usetikzlibrary{decorations.pathmorphing} 
\usetikzlibrary{fit, matrix}                                    
\usetikzlibrary{backgrounds}    
\usetikzlibrary{shapes,arrows,fit,calc,positioning}

\usepackage[titles]{tocloft}

 \usepackage{amsmath}

\usepackage{tabularx,colortbl}

\usepackage{fancyhdr}
\advance\headheight by 3.0truept       

\allowdisplaybreaks[1]

\definecolor{red}{cmyk}{0,1,1,0.4}
\definecolor{darkgreen}{rgb}{0.0,0.6,0.0}
\definecolor{cDarkGrey}{RGB}{91,91,91}
\definecolor{cGrey}{RGB}{245,243,238}
\definecolor{cBlue}{RGB}{0,110,191}
\definecolor{cLightBlue}{RGB}{214,237,252}
\definecolor{cRed}{RGB}{196,0,100}
\definecolor{cLightRed}{RGB}{254,222,237}
\definecolor{cGreen}{RGB}{0,166,80}
\definecolor{cLightGreen}{RGB}{254,222,237}
\definecolor{cOrange}{RGB}{221,74,44}
\definecolor{cLightOrange}{RGB}{255,215,210}
\definecolor{cPurple}{RGB}{93,35,125}
\definecolor{cLightPurple}{RGB}{241,230,252}
\definecolor{cYellow}{RGB}{252,191,10}
\definecolor{cISSRBlue}{RGB}{0,111,174}
\definecolor{cISSRGrey}{RGB}{167,169,172}

\newcommand{\be}{\begin{equation}}
\newcommand{\ee}{\end{equation}}

\newcommand{\nn}{\nonumber}

\newcommand{\mev}{\, {\rm MeV}}

\newcounter{TODO}

\DeclareMathOperator{\im}{Im}

\newcommand{\ord}[1]{\mathcal{O}\left(#1 \right)}

\newcommand{\hc}{\mathrm{h.c.}}

\newcommand{\DF}{\Delta\,F}

\newcommand{\GeV}{\,\text{GeV}}

\newcommand{\TeV}{\,\text{TeV}}

\newcommand{\vcb}{|V_{cb}|}
\newcommand{\vtd}{|V_{td}|}
\newcommand{\vub}{|V_{ub}|}
\newcommand{\vts}{|V_{ts}|}

\newcommand{\yuk}{{Y}}

\def\klpll{K_L\to\pi^0\ell^+\ell^-}

\newcommand{\epe}{\varepsilon'/\varepsilon}

\def\kpn{K^+\rightarrow\pi^+\nu\bar\nu}
\def\klpn{K_{L}\rightarrow\pi^0\nu\bar\nu}

\newcommand{\KKbar}{K^0-\bar{K}^0}
\newcommand{\DDbar}{D^0-\bar{D}^0}
\newcommand{\BBbar}{B_{s,d}-\bar{B}_{s,d}}

\newcommand{\alS}{\alpha_s}

\newcommand{\muLow}{{\mu_\text{had}}}
\newcommand{\muEW}{{\mu_\text{ew}}}
\newcommand{\muNP}{{\Lambda}}

\newcommand{\nc}{\newcommand}
\nc{\eps}{\varepsilon}
\nc{\vp}{H}
\nc{\tvp}{\widetilde{H}}
\nc{\D}{\mbox{$\not\!\!D$}}
\nc{\Db}{\mbox{${\raisebox{2mm}{\boldmath ${}^\leftarrow$}\hspace{-4mm} D}$}}
\nc{\Dfb}{\mbox{$\raisebox{2mm}{\boldmath ${}^\leftrightarrow$}\hspace{-4mm} D$}}
\nc{\vpj}{\mbox{${H^\dag i\,\raisebox{2mm}{\boldmath ${}^\leftrightarrow$}\hspace{-4mm} D_\mu\, H}$}}
\nc{\vpjt}{\mbox{${H^\dag i\,\raisebox{2mm}{\boldmath ${}^\leftrightarrow$}\hspace{-4mm} D_\mu^{\,I}\,H}$}}

\def\wt{\widetilde}
\newcommand{\wc}[3][{}]{{C_{\! \underset{ #3}{ #2}}^{{#1}}} }

\newcommand{\wcs}[3][{}]{{C_{\! \underset{ #3}{ #2}}^{{#1}}} }

\newcommand{\ops}[3][{}]{{Q_{\! \underset{ #3}{ #2}}^{{#1}}} }
\newcommand{\opsup}[3][{}]{{\tilde{Q}_{\! \underset{ #3}{ #2}}^{{#1}}} }
\newcommand{\dotwc}[3][{}]{{\dot{C}_{\! \underset{ #3}{ #2}}^{{#1}}} }

\newcommand{\wcup}[3][{}]{{\tilde{C}_{\! \underset{ #3}{ #2}}^{{\tiny #1}}} }

\newcommand{\opup}[3][{}]{{\tilde{Q}_{\! \underset{ #3}{ #2}}^{{\tiny #1}}} }

\newcommand{\wcL}[3][{}]{{\mathcal{C}_{\! \underset{#3}{ #2}}^{{\tiny #1}}} }

\newcommand{\opL}[3][{}]{{\mathcal{O}_{\! \underset{#3}{ #2}}^{{\tiny #1}}} }

\newcommand{\Op}[2][{}]{\mathcal{O}_{#2}^{#1}}

\tikzstyle{input}=[circle,
                                    thick,
                                    minimum size=1cm,
                                    draw=red!30,
                                    fill=red!20]

\definecolor{Grayd}{gray}{0.8}
\definecolor{Gray}{gray}{0.95}

\newcommand{\Wc}[2][{}]{{C}_{#2}^{#1}}

\newcommand{\op}[3]{\mathcal{O}^{#2,#3}_{#1}}

\newcommand{\WCL}[2][{}]{C_{#2}^{#1}}

\newcommand{\OpL}[2][{}]{Q_{#2}^{#1}}

\newcommand{\e}{\mathrm{e}}

\def\as{\alpha_s}

\newcommand{\bea}{\begin{eqnarray}}
\newcommand{\eea}{\end{eqnarray}}

\usepackage{tocloft}

\makeatother%

\numberwithin{equation}{section}

\begin{document}


\vspace{-14mm}
\begin{flushright}
  AJB-25-1\\
  CERN-TH-2025-129\\
  LA-UR-24-24665
\end{flushright}

\medskip

\begin{center}
{\Large\bf\boldmath{SMEFT ATLAS:\\ The Landscape Beyond the Standard Model}
}
\\[1.2cm]
{\bf
  Jason~Aebischer$^{a}$,
    Andrzej~J.~Buras$^{b,c}$,
  Jacky Kumar$^{d}$
  }\\[0.5cm]

{\small
$^a$Theoretical Physics Department, CERN, 1211 Geneva, Switzerland \\[0.2cm]
$^b$TUM Institute for Advanced Study,
  Lichtenbergstr. 2a, D-85747 Garching, Germany \\[0.2cm]
  $^c$ Physik Department, TUM School of Natural Sciences, TU M\"unchen,\\ James-Franck-Stra{\ss}e, D-85748 Garching, Germany\\[0.2cm]
$^d$ Theoretical Division T-2, Los Alamos National Laboratory, Los Alamos, NM 87545, USA \\[0.2cm]
}
\end{center}

\vskip 1.0cm

\begin{abstract}
\noindent
The Standard Model Effective Field Theory (SMEFT) based on the unbroken gauge group $\text{SU(3)}_C\otimes\text{SU(2)}_L\otimes\text{U(1)}_Y$\, and containing only particles of the Standard Model (SM) has developed in the last decade to a mature field. It is the framework to be used in the energy gap from  scales sufficiently higher than the electroweak scale up to the lowest energy scale at which new particles show up. We summarize the present status of this theory with a particular emphasize on its role in the indirect search for new physics (NP). While flavour physics of both quarks and leptons is the main topic of our review, we also discuss electric dipole moments, anomalous magnetic moments $(g-2)_{\mu,e}$, $Z$-pole observables, Higgs observables and high-$p_T$ scattering processes within the SMEFT. We group the observables into ten classes and list for each class the most relevant operators and the corresponding renormalization group equations (RGEs). We exhibit the correlations between different classes implied both by the operator mixing and the $\text{SU(2)}_L$ gauge symmetry. Our main goal is to provide an insight into the complicated operator structure of this framework which hopefully will facilitate the identification of valid ultraviolet completions behind  possible anomalies observed in future data. Numerous colourful charts, and 85 tables, while representing rather complicated RG evolution from the NP scale down to the electroweak scale, beautify the involved SMEFT landscape. Over 1000 references to the literature underline the importance and the popularity of this field. We discuss both top-down and bottom-up approaches as well as their interplay. This allows us eventually to present an atlas of different landscapes beyond the SM that includes heavy gauge bosons and scalars, vector-like quarks and leptons and leptoquarks. 
\end{abstract}

\setcounter{page}{0}
\thispagestyle{empty}
\newpage

\setcounter{tocdepth}{3}

\setlength{\cftbeforesecskip}{0.21cm}
\tableofcontents
\clearpage   

\part{SMEFT at Large}
\section{Introduction}\label{intro}

One of the main frontiers in elementary particle physics is the search for new particles and new forces beyond those present in the Standard Model (SM) of particle physics. As the direct searches at the Large Hadron Collider (LHC) at CERN, even thirteen years after the Higgs discovery, did not provide any definitive hint on what these new particles and forces could be, the indirect searches for new physics (NP) through very rare processes caused by virtual exchanges of heavy particles gained in importance. They allow in fact to see footprints of new particles and forces acting at much shorter distance scales than it is possible to explore  at the LHC and presently planned high energy colliders. While the LHC can at best explore through Heisenberg's uncertainty principle distance scales as short as $3\cdot 10^{-20}$m, corresponding roughly to energies in the ballpark of $7\,\TeV$,\footnote{In a proton-proton collider even the production of particles with masses as large as $7\,\TeV$ will be a great challenge so that this estimate is optimistic.} the indirect search with the help of suitably chosen processes can offer us information about scales as short as a Zeptometre ($10^{-21}$m) which cannot be probed even by the planned $100\TeV$ proton-proton collider at CERN. Even shorter scales can be explored in this manner \cite{Buras:2015nta}.

In fact rare processes like $K_L\to \mu^+\mu^-$, known since the late 1960s, implied the existence of the charm quark prior to its discovery in 1974 as only then its branching ratio could be suppressed in the SM with the help of the Glashow-Iliopoulos-Maiani (GIM)  mechanism~\cite{Glashow:1970gm}, to agree with experiment. Moreover, it was possible to predict successfully its mass with the help of the $K_L-K_S$ mass difference $\Delta M_K$ in the $K^0-\bar K^0$ mixing prior to its discovery~\cite{Gaillard:1974hs}. Similarly the size of $B_d^0-\bar B_d^0$ mixing, discovered in the late 1980s, implied a heavy top quark that has been confirmed only in 1995. It is then natural to expect that this indirect search for NP will also be successful at much shorter distance scales.

In this context, rare weak decays of mesons played already for decades a prominent role besides the transitions between particles and antiparticles in which flavours of quarks are changed. In particular $\kpn$, $\klpn$, $K_{L,S}\to\mu^+\mu^-$, $B_s^0\to\ell^+\ell^-$, $B_d^0\to\ell^+\ell^-$, $B\to K(K^*)\ell^+\ell^-$ and $B_d^0\to K(K^*)\nu\bar\nu$ decays, but also $B^0_s-\bar B^0_s$, $B^0_d-\bar B^0_d$, $K^0-\bar K^0$, $D^0-\bar D^0$ mixings and CP-violation in $K\to\pi\pi$, $B_d\to\pi K$ decays among others  provide important constraints on NP. Most of these transitions are very strongly loop-suppressed within the SM due to the GIM mechanism and also due to the small elements $V_{cb}$, $V_{ub}$, $V_{td}$ and $V_{ts}$ of the CKM matrix~\cite{Cabibbo:1963yz,Kobayashi:1973fv}. The predicted
branching ratios for some of them are as low as $10^{-11}$. But as the GIM mechanism is generally violated by NP contributions these branching ratios could in fact be much larger.

Of particular interest are also flavour changing decays of leptons like $\mu\to e \gamma$, $\tau\to e\gamma$, $\tau \to\mu\gamma$ as well as three-body decays like $\mu^-\to e^-e^+e^-$, $\tau^-\to e^-e^+e^-$ among others. These decays are forbidden in the SM and therefore observing them would be a clear indication for NP. Indeed, even if neutrino masses are taken into account, these decays are suppressed within the SM by many orders of magnitude relative to the sensitivity of the experiments planned in the coming decades.
A similar comment applies to decays like $K_{L,S}\to\mu\, e$, $K_L\to\pi^0\mu \,e$, $B_{d,s}\to\mu\, e$, $B_d\to K\mu \,e$ and alike, in which both quark and lepton flavours are changed.

Furthermore, flavour conserving observables like electric dipole moments (EDMs) of particles and atoms are of great interest because SM predictions for them are by several orders of magnitude below the corresponding present experimental upper bounds. They are particularly interesting because while being flavour conserving, they are CP-violating. Finally, anomalous magnetic moments $(g-2)_{\mu,e}$ play these days an important role due to a possible deviation of the data from SM predictions.

There is a vast literature on this broad field. In particular detailed reviews, books and lectures can be found in \cite{Buchalla:1995vs,Buras:1998raa,Branco:1999fs,Bigi:2000yz,Zupan:2019uoi,Silvestrini:2019sey,Buras:2020xsm,Bigi:2021hxw,Artuso:2022ijh,Buras:2011we,Grossman:2023wrq,Altmannshofer:2024ykf,Isidori:2025iyu}.

The first step in this indirect strategy is to search for deviations in branching ratio measurements of the decays in question from SM predictions and similar for the $K_L-K_S$  mass difference $\Delta M_K$, and analogous mass differences  $\Delta M_s$ and $\Delta M_d$  in $B^0_s-\bar{B}^0_s$ and $B^0_d-\bar B^0_d$ mixings, respectively. But while these processes are governed by quark interactions at the fundamental level, the decaying objects are mesons, the bound states of quarks and antiquarks. In particular in the case of non-leptonic transitions like $B^0_s-\bar B^0_s$, $B_d^0-\bar B_d^0$, $K^0-\bar K^0$, $D^0-\bar D^0$ mixings and CP-violation in $K\to\pi\pi$ and  $B\to K\pi$ decays, QCD plays an important role. It enters at short distance scales, where, due to the asymptotic freedom in QCD, perturbative calculations can be performed, and at long-distance scales where non-perturbative methods are required. QCD has also an impact on semi-leptonic decays like $\kpn$, $\klpn$, $B\to K(K^*)\nu\bar\nu$, $B\to K(K^*)\ell^+\ell^-$ and even  on leptonic ones like $K_{L,S}\to\mu^+\mu^-$, $B_s^0\to\ell^+\ell^-$ and $B_d^0\to\ell^+\ell^-$. In order to be able to identify the departures of various experimental results from the SM predictions that would signal NP at work, the latter predictions must be accurate, and this means the effects of QCD have to be brought under control.

But this is not the whole story. To make predictions for rare processes in the SM one has to determine the four parameters of the unitary CKM matrix
\be
V_{us}\,, \qquad V_{cb}\,, \qquad V_{ub}\,, \qquad \gamma\,,
\ee
with $\gamma$ being the sole phase in this matrix. It allows to describe CP-violating effects observed in a number of experiments. We will return to the present status of these parameters and the status of the short-distance QCD calculations and the one of non-perturbative calculations of hadronic parameters. But assuming that the QCD effects are under control and the masses of all SM particles are known, the SM has only four free parameters in the quark sector, the ones listed above, when neutrino masses are set to zero and one has determined the three gauge parameters $\alpha_s$, $\alpha_2$ and $\alpha_1$ in collider processes.

With four parameters to be determined from experiments and having hundreds of measured observables at our disposal, not only these parameters but also relations between various branching ratios, free of CKM dependence, can be derived~\cite{Buras:2003td,Buras:2021nns,Buras:2022wpw,Buras:2022qip,Aebischer:2023mbz,Buras:2024per}. Such relations can rather quickly test the SM and subsequently assuming the dominance of SM contributions in $\varepsilon_K$ and mass differences like  $\Delta M_s$ and $\Delta M_d$ provide SM predictions for various branching ratios. While these tests of the SM often require some novel ideas with the goal to reduce hadronic uncertainties and the use of computer codes, the phenomenological SM analyses are generally transparent because the number of free parameters that have to be determined is very small.

The situation changes dramatically when some inconsistencies in the SM description of the experimental data are found, signalling thereby the presence of new interactions and new particles, in short NP. If these departures from SM expectations are accompanied by new particles discovered in a high-energy collider, the number of NP models will be restricted to the ones which, while accommodating these new particles, simultaneously describe the deviations of the data from the SM predictions, in particular those for rare decays of mesons and leptons listed above. This would facilitate the identification of the so-called ultraviolet (UV) completion of the SM.

But what if the LHC will not discover any new particles and the SM will fail to describe several different processes? In fact  already now there are several profound anomalies in rare processes which signal that NP could be at work and the goal of theorists is to find out somehow what this NP could be. While it is not excluded that some of these anomalies will be dwarfed or will disappear as soon as data improves, the fact that the SM cannot answer a number of well known questions makes us believe that some of these anomalies will remain and other anomalies will be found in the coming years. For a recent review of present anomalies see \cite{Crivellin:2025txc}.

There are basically two classes of strategies to find out what the data is telling us about new forces and new particles possibly responsible for these anomalies.

\begin{center}
{\bf Top-Down Approach }
\end{center}

One constructs a new model, the UV completion of the SM, which is supposed to describe the anomalies in question. It is an extension of the SM  that contains new particles and new forces. The latter are described by new gauge symmetries that are spontaneously broken directly or in steps down to the SM gauge group at some scale $\Lambda$ at which the lightest new particles have been integrated out. Below $\Lambda$ only SM particles appear and the gauge symmetry is the unbroken gauge group of the SM,
\be\label{SMGAUGE}
\text{SU(3)}_C\otimes\text{SU(2)}_L\otimes\text{U(1)}_Y\,,
\ee
provided $\Lambda$ is significantly larger than the electroweak (EW) scale $\muEW$.

The Effective Field Theory (EFT) with the unbroken SM gauge group  is usually called the Standard Model Effective Field Theory (SMEFT) because at low-energies, it should reduce to the SM, provided no undiscovered weakly coupled {\em light} particles exist, like axions or sterile neutrinos.\footnote{For recent reviews on the SMEFT see for instance \cite{Brivio:2017vri,Isidori:2023pyp}.}
Yet, as we will see below, the presence of new operators, in particular dimension six ones, whose Wilson coefficients (WCs) can be calculated in a given model as functions of its parameters, can introduce very significant modifications of SM predictions for flavour observables. Also the values of SM parameters can be modified in this manner. This is also the case for the WCs of the involved operators which are generated in the SM and, being modified, contain information about NP beyond the SM. Moreover there is a multitude of operators that are strongly suppressed in the SM but can be important in the SMEFT. Therefore the name SMEFT is to some extent  a misnomer but as the basic gauge symmetry is the SM one, it has been accepted by the particle physics community.

The structure of the interactions in the SMEFT, even if governed by the SM gauge group  depends on the UV completion considered. This is important, because otherwise there would be no chance to find out anything about the dynamics above the scale $\Lambda$. In particular the number of free parameters will depend on the fundamental theory or model considered. These parameters can be conveniently defined at the scale $\Lambda$. Their number depends on the theory considered and it will not change when the renormalization group (RG) evolution down to the electroweak scale is performed. Only their values will change as well as the values of the WCs in the effective theory describing the physics between the scale $\Lambda$ and the electroweak scale. This physics, beyond the one of the SM, is described by an effective Lagrangian
\begin{equation}\label{BASICNP}
\mathcal L^{(6)} =   \sum_{k} C_k^{(6)} Q_k^{(6)}\,,
\,
\end{equation}
where for the time being we consider only the  dimension-6 operators $Q_k^{(6)}$ and their {\em dimensionfull} WCs $C_k^{(6)}$ with dimension $\text{GeV}^{-2}$. Absorbing the usual factor $1/\Lambda^2$ into  $C_k^{(6)}$ simplifies the matching of the SMEFT with the Weak Effective Theory (WET) for which the WCs will also be kept {\em dimensionfull.}

The number of operators and their Dirac structure depends generally on the model considered. However, to be prepared for all possible models and to develop a general framework one can classify operators in full generality by imposing only the SM gauge symmetry in (\ref{SMGAUGE}). This has been done in~\cite{Buchmuller:1985jz,Grzadkowski:2010es}. The second of these papers removed certain redundant operators present in the first one and we will use the results of~\cite{Grzadkowski:2010es} here. The corresponding RG analysis at leading order (LO) of all these operators has been presented in~\cite{Jenkins:2013zja,Jenkins:2013wua,Alonso:2013hga}. The results of this analysis will play  a very important role in our review and will be presented in detail in ten classes of observables in Part II of our review.

It turns out that for three generations of fermions, there  are 2499 independent operators $ Q_k^{(6)}$ (59 irreducible flavour representations) that do not violate baryon and lepton number. This means that the RG analysis involves in full generality a $2499\times 2499$ anomalous dimension matrix with the evolution governed by the Higgs self-coupling $\lambda$~\cite{Jenkins:2013zja}, {the} Yukawa couplings~\cite{Jenkins:2013wua} and the SM gauge interactions~\cite{Alonso:2013hga}. While several of the entries in this matrix have been known before, even beyond LO,  these papers provided a large number of new entries so that RG effects can be investigated, albeit only at LO, in full generality. We will summarize the present status of next-to-leading order (NLO) corrections later.

The WCs $C_k^{(6)}$ contain in full generality 1350 CP-even and 1149 CP-odd parameters. They contain information about NP at and above the scale  $\Lambda$. It should also be emphasized that in addition to the 2499 operators in question there are 273 $\Delta B=1$ dimension six operators (4 irreducible flavour representations) and their Hermitian conjugates, which are the leading operators that permit proton decay in the SMEFT~\cite{Buchmuller:1985jz,Grzadkowski:2010es,Alonso:2014zka,Weinberg:1979sa,Wilczek:1979hc,Abbott:1980zj}.

The SMEFT RGEs allow calculating the WCs of the SMEFT at the electroweak scale. Below the electroweak scale the interactions are governed not by the full gauge symmetry of the SM but by
\be\label{SMWET}
\text{SU(3)}_C\otimes\text{U(1)}_{\text{QED}}\,,
\ee
which is very familiar to us. However, in the presence of NP there are new operators generated already at the NP scale and/or at the electroweak scale through RG running in the SMEFT and through the matching onto the low-energy theory described by this reduced symmetry. Consequently, the starting point for the RG evolution from the electroweak scale down to the low energy scale differs from the one encountered within the SM. Not only the initial conditions for SM operators can be modified by NP but also new WCs might be present. The resulting effective theory below the electroweak scale, although as the SM based on the gauge symmetry of QCD~$\times$~QED, differs then from the SM and is called the WET. It should be stressed that this theory does not involve $W$, $Z$ gauge bosons, the Higgs boson and the top quark as dynamical degrees of freedom. Moreover, there are no elementary scalars in this theory.

The counting of all possible gauge-invariant operators of the WET has been accomplished in~\cite{Jenkins:2017jig}. There are 70 
hermitian $D=5$ and 3631 hermitian $D=6$ operators that conserve baryon and lepton numbers, as well as $\Delta B=\pm \Delta L=\pm 1$, $\Delta L=\pm 2$, and $\Delta L=\pm 4$ operators. Among the 3631 operators in question 1933 are CP-even and 1698 CP-odd. 

This counting shows clearly that in order for a model to be predictive the number of free parameters has to be reduced by much. One of the goals of our review is to discuss various possibilities with a reduced number of parameters. In particular flavour symmetries play here an important role. However, this general classification of operators and the results for RG effects turn out to be very useful in concrete models so that we will have a closer look in full generality at them as well. In Section~\ref{sec:smeft-lag}, the SMEFT Lagrangian is discussed in detail. In particular the operators are grouped in different classes. Further, in Section~\ref{sec:smeft-weak-basis}, various choices for weak bases of the SMEFT are described. What this means will be clear from the formulae presented there but the choice of a weak basis has to be made prior to any calculations. Concrete examples will be presented in Part II of our review.

\begin{center}
{\bf Bottom-Up Approach}
\end{center}

In this approach, one first tries to describe various departures of the low energy data from SM expectations by modifying the WCs of SM operators and/or adding new operators. Sometimes only a small number of new operators and related WCs is sufficient to describe the data. Having their values at the electroweak scale one can then as a first step construct simplified NP models which could imply the presence of these new operators and the modifications of SM WCs. Subsequently one can try to develop a more sophisticated model like in the top-down approach from which these WCs would result.

Thus in practice one tries to borrow the lessons from both approaches because while by definition in the top-down approach one knows better the fundamental new theory at short distance scales, in the bottom-up approach one has close contact with the data which the fundamental theory is supposed to describe.

This seems at first sight to be rather straightforward but the contrary is the case. In fact, knowing the WCs at low energy scales, it is rather non-trivial to find their values at the NP scale $\Lambda$ which would be the optimal strategy for the construction of  UV completions that are capable to describe the experimental data both at very low energies but also those explored directly at the LHC.

The main reasons are as follows
\begin{itemize}
\item
The large number of operators both in the SMEFT and WET.
\item
The mixing between the operators not only through QCD interactions but in particular through Yukawa interactions that in the case of the top quark can have a significant impact on the flavour pattern of contributing operators. This means that the pattern of the WCs and their number observed at low energy can differ significantly from the one at the NP scale $\Lambda$.
\item
Modification of Higgs, electroweak and also Yukawa couplings that are caused by the presence of dim-6 operators and also affected by RG evolutions of both gauge and Yukawa couplings.
\item
Modifications of the CKM parameters by dim-6 operators and RG evolutions. Hence, the corresponding SMEFT parameters at $\Lambda$ have to generate after RG evolutions CKM parameters measured in low energy experiments, which is certainly non-trivial.
\end{itemize}

This short description of both strategies demonstrates clearly that the expedition to very short distance scales, beyond the ones explored by the LHC, with the help of quantum fluctuations is a true mammoth project. Therefore, it requires a systematic and solid preparation.

In this spirit, we postpone  the outline of our review to Section~\ref{sec:GV} in which the {\em Grand View} of our SMEFT expedition to short distance scales will be presented. The main goal of this expedition is to provide the technology of the SMEFT and the insight in its structure. In the future when more observables will be measured with higher precision and possibly new anomalies will be found, by consulting our review one will be able to select efficiently the class of models which offer the best description of the low-energy data. But first we have to recall the basic notion of effective field theories as viewed presently.

\section{Modern Effective Field Theories}\label{sec:efts}

An effective field theory is defined with respect to a full theory (known or unknown), that contains more degrees of freedom than the respective EFT. The masses of the additional particles in the full theory are at least of the order of the matching scale, below which the full theory is replaced by the effective one. In that sense an EFT is the low-energy limit of an underlying full theory. It is based on the fact that the low-energy behaviour described by the full theory is to a good approximation independent of its heavy degrees of freedom. For instance, heavy particles cannot be produced directly at low enough energies and therefore only enter the calculations indirectly, via quantum corrections. The formal ground for effective field theories is given by the Appelquist-Carazzone decoupling theorem~\cite{Appelquist:1974tg}, which states that heavy fields decouple at low energies except for their contribution to renormalization effects. An effective field theory is therefore valid up to the matching scale, at which these heavy degrees of freedom become important again. Hence, in order to determine the EFT from the full theory, the heavy particles of the full theory are integrated out at the matching scale. In the path integral formalism this corresponds to actually integrating the generating functional over the heavy field configurations of the theory. In the canonical quantization picture, the {\it matching} (procedure) is performed by expanding the amplitudes in small external momenta and masses:\footnote{A recent discussion on EFT matching using analyticity and unitarity can be found in \cite{DeAngelis:2023bmd}. A numerical method to perform on-shell matching calculations was presented in \cite{Chala:2024llp} and on-shell matching using the unitary cut method has been discussed in \cite{Dong:2025wvf}. Furthermore, one-loop matching computations using covariant diagrams in the covariant derivative expansion are discussed in \cite{Zhang:2016pja}.} The expanded amplitudes are computed at the matching scale, both in the EFT and the full theory. The comparison of the two results fixes the parameters of the EFT, which are typically called {\it Wilson coefficients} and denoted here by WCs.

As a consequence of this general procedure, the basis for any phenomenology of weak decays of hadrons and leptons, discussed in this review, is the {\it Operator Product Expansion} (OPE)~\cite{Wilson:1969zs,Zimmermann:1972tv}. It allows to write down the effective weak Lagrangians for various transitions within the SM as follows\footnote{The minus sign is introduced in order to be consistent with the effective Hamiltonians in the SM literature \cite{Buchalla:1995vs,Buras:2020xsm}.}
\be\label{SMH}
{\cal{L}}^\text{SM}_\text{eff}= -\mathcal{N}\sum_i C_i^\text{SM} \mathcal{O}_i^\text{SM},
\ee
where the overall factor $\mathcal{N}$, to be specified later, is usually introduced to make the WCs $C_i^\text{SM}$ of the SM local operators $\mathcal{O}_i^\text{SM}$ {\em dimensionless}. Despite the fact that in the WET and the SMEFT we will work with {\em dimensionfull WCs} for reasons stated after \eqref{BASICNP}, for the SM we will use the standard SM conventions to avoid changing numerous formulae present in the literature.

The SM Lagrangian is obtained by integrating out $W^\pm$ and $Z$ gauge bosons as well as the Higgs boson and the top quark. Subsequently, performing RG evolution down to low energy scales $(\muLow)$ relevant for weak decays of mesons one obtains the values of $C_i^\text{SM}(\muLow)$. There is a reach literature on the details of this procedure and some aspects will be discussed in the context of our review. For the time being we refer to~\cite{Buchalla:1995vs,Buras:2020xsm}. There, one finds the lists of operators contributing in the SM to leptonic, semileptonic and non-leptonic decays of mesons and also those contributing to meson mixings $K^0-\bar K^0$, $D^0-\bar D^0$, $B_d^0-\bar B^0_d$ and $B_s^0-\bar B^0_s$.

Denoting the hadronic scale by $\muLow$ and the electroweak scale by $\muEW$, the general expression for $ C^\text{SM}_i(\muLow) $ is given by:
\begin{equation}\label{CV}
\vec C^\text{SM}(\muLow) = \hat U_\text{SM}(\muLow,\, \muEW) \vec C^\text{SM}(\muEW)~,
\end{equation}
where $ \vec C^\text{SM}$ is a column vector built out of $C_i^\text{SM}$. $\vec C^\text{SM}(\muEW)$ are the initial conditions for the RG evolution down to the low-energy scale $\muLow$. They depend on short-distance physics at high energy scales. In particular they depend on $m_t$ and the masses of the $W$ and $Z$.

The RG evolution matrix for SM operators is known including one-loop and two-loop QCD contributions and in several cases also NNLO contributions. Including LO and NLO contributions it is given as follows
\begin{align}
\label{eq:USM}
\hat U_\text{SM}(\muLow,\, \muEW) &
= \left[\hat 1 + \hat J_\text{SM}\frac{\alS(\muLow)}{4\pi} \right]
\hat U_\text{SM}^{(0)}(\muLow,\, \muEW)
\left[\hat 1 - \hat J_\text{SM} \frac{\alS(\muEW)}{4\pi} \right]\,,
\end{align}
where $\hat U_\text{SM}^{(0)}$ is the renormalization scheme (RS) independent LO evolution matrix. On the other hand, $ \hat J_\text{SM}$ stems from the RS-dependent two-loop Anomalous Dimension Matrices (ADMs), which makes them sensitive to the renormalization scheme considered. This scheme dependence is cancelled by the one of $\vec C^\text{SM}(\muEW)$ and by the one of the hadronic matrix elements at $\muLow$. Explicit general expressions for $\hat U_\text{SM}^{(0)}$ and $ \hat J_\text{SM}$ in terms of the coefficients of the one-loop and two-loop perturbative expansions for the ADM $\hat\gamma$ and the QCD $\beta$-function can be found including their derivations in Chapter 5 of~\cite{Buras:2020xsm}. 

The status of the NLO and NNLO QCD and NLO electroweak corrections within the SM has been recently updated in~\cite{Buras:2011we}. Basically, for all relevant decays, these corrections are by now known. 

Beyond the SM the effective weak Lagrangian in \eqref{SMH} is generalized as follows
\be\label{BSMH}
{\cal L}_\text{eff}= -\mathcal{N}\sum_i C_i \mathcal{O}_i^\text{SM} -\mathcal{N}^\text{NP}\sum_j C_j^\text{NP} 
\mathcal{O}_j^\text{NP}\,,\qquad C_i=C_i^\text{SM}+\Delta_i^\text{NP},
\ee
where the overall factor $\mathcal{N}^\text{NP}=1$ underlines that WCs $C_j^\text{NP}$ of the NP local operators $\mathcal{O}_j^\text{NP}$, as opposed to {\em dimensionless} $C_i$, are {\em dimensionfull} for reasons stated after \eqref{BASICNP}.

Here
\begin{itemize}
\item
$\mathcal{O}_i^\text{SM}$ are again local operators present in the SM and $\mathcal{O}_j^\text{NP}$ are new local operators having typically new Lorentz (Dirac) structures, in particular scalar-scalar and tensor-tensor ones.
\item
$C_i$ and $C_j^\text{NP}$ are the WCs of these operators. NP effects modify not only the WCs of the SM operators but also generate
new operators with non-vanishing $C_j^\text{NP}$.
\end{itemize}

In the Top-Down approach, the contributing operators and their WCs at the NP scale $\Lambda$ are found by integrating out all new particles, performing the RG evolution within the SMEFT down to $\muEW$ and subsequently within the WET down to $\mu_{\muLow}$. These RG evolutions are much more involved than the ones within the SM, primarily because of the large number of operators. We will discuss them in some detail in the context of our review.

After this introductory description of the SMEFT and the WET we are ready to present the Grand View of our SMEFT expedition. We hope that it will give the reader a general idea of our strategy for presenting this rather advanced topic. Furthermore, this section should facilitate following more technical parts without losing the goal we are attempting to reach: the development of strategies for the identification of NP at very short distance scales.

\section{Grand View of the SMEFT Expedition}\label{sec:GV}
\subsection{Top-Down Technology}
As our goal is to find eventually a UV completion or at least a few favourite UV completions responsible for possible anomalies observed now and in the future, it is strategically useful to discuss first the top-down approach. Let us then assume that we have a NP model motivated not only by the observed anomalies but also by open questions which the SM cannot answer. Let us then formulate the steps one has to take to eventually be able to make predictions for various observables within a given NP model that could be compared with the experimental data. These steps will be first formulated in general terms. Subsequently, they will be executed in the following sections. In order not to get lost in complicated formulae, we will often refer to the literature.

\paragraph{Step 1: Renormalization Group Evolution in the NP Model}
If the NP model contains an entire spectrum of new heavy particles with vastly different masses, this step consists of several steps and corresponding effective theories. It includes performing RG evolution to lower energies and consecutively integrating out heavy particles, i.e. at every threshold the matching between the two effective theories has to be computed. Eventually one arrives at a scale $\Lambda_\text{NP}$ at which all new particles have been integrated out. This means that below this scale only SM particles are present as dynamical degrees of freedom so that all effects of new particles are present in the WCs that multiply operators, which exclusively contain SM fields. The last appearance of new particles is in the next step, which takes place at the scale $\Lambda_\text{NP}$ to be denoted often simply by $\Lambda$ in what follows.

We will discuss this first step in detail only in Part III of our review, where we will consider a number of simplified NP scenarios. The technology involving RG evolution and the matching between different effective theories will be developed in the following steps.

\paragraph{ Step 2: Matching of a NP Model onto the SMEFT}
At the scale $\Lambda_\text{NP}$ the NP is matched onto the SMEFT. To this end we have to choose a useful basis of operators. We will first introduce the operators in Sec.~\ref{sec:smeft-lag} in the so-called {\em Warsaw-Basis} without specifying their flavour indices. Subsequently in Sec.~\ref{sec:smeft-weak-basis} we will take care of these indices which will result in the discussion of two common bases used in the literature: {\em the Warsaw-Down} and {\em the Warsaw-Up} basis. This will exhibit the complicated structure of the SMEFT mentioned at the beginning of our review.

After having specified the basis the matching will result in a set of non-vanishing SMEFT WCs at $\mu=\Lambda_{\text{NP}}$ in the chosen basis. In the matching procedure new particles are present in tree- and loop diagrams so that the WCs depend on their masses and in particular on their couplings to SM particles. But the couplings between purely new physics particles also affect the WCs in question. It should be emphasized that among the steps listed here, this is the last one which is model-dependent. This dependence enters precisely through the generated operators at this scale and through the values of their WCs. The set of operators with non-vanishing WCs at $\Lambda_\text{NP}$ and also the values of their WCs depend on the model considered. Once this step has been completed the remaining steps are model-independent in the sense that only SM fields and their interactions are involved. In particular in the SMEFT below $\mu=\Lambda_{\text{NP}}$ even the loop diagrams do not involve any new particles because they all have been integrated out and are not dynamical degrees of freedom anymore. But their footprints can still be seen in tree-level factors like for instance $1/M^2_{Z^\prime}$ in $Z^\prime$ models. We will be more explicit about it in Part III of our review.

In this regard the matching procedure is very important, since in this step the remnants of NP become apparent for the last time, in terms of the generated operators and their WCs that usually depend on new parameters of a given NP model. If the model cannot fix the latter parameters by itself, one has to determine them by comparing various observables with the experimental data in the last
step of this procedure. This is in fact always the case when using simplified models but also often in more advanced models.

\paragraph{Step 3: Renormalization Group Evolution in the SMEFT}
Starting then with the operators and their WCs at $\mu=\Lambda_{\text{NP}}$ we perform the RG evolution in the SMEFT, which modifies the WCs of the operators we started with and generates new ones. This step involves all SM particles including the top quark, $W^\pm$, $Z$ and Higgs. Therefore, in all RG beta functions one has to set the number of quark flavours to $f=6$ and similar for leptons. In this step the most important interactions result from QCD and the top Yukawa interactions, with the latter having a significant impact on the mixing of operators with different flavour structure. Eventually one arrives at the scale $\mu=\muEW$ at which the top quark, $W^\pm$, $Z$ and Higgs are integrated out.

Yet, this step is more involved than the description given above. The reason is that in the RG evolution not only WCs of the operators are modified, but one also has to take care of the modifications of the parameters involved, in particular of the CKM parameters. That is the values of the latter parameters at $\mu=\Lambda_{\text{NP}}$, that are needed for this evolution, have to result after the RG evolution to low scales in the ones extracted with the help of the low-energy measurements. This important issue and the issues related to other parameters are discussed first in Secs.~\ref{sec:SM-smeft-corrections} and \ref{sec:ren-gro-run} and later in Part II of our review.

\paragraph{Step 4: Matching of the SMEFT onto the WET}
At the electroweak scale $\muEW$ the SMEFT is matched onto the WET. This results in non-vanishing WCs of the WET at $\mu=\muEW$. In this context the change of operator basis used in the SMEFT has to be made to obtain the so-called JMS basis, which is useful for discussing the physics below $\mu=\muEW$. Here the issue of the relative signs of SMEFT WCs and WET WCs has to be considered. Another important issue is the involvement of the evanescent operators, which have to be taken into account in the calculations. This is the case not only in the process of two-loop computations of anomalous dimensions but also when performing changes in operator bases at the NLO level. This step is the last one at which the top quark, $W^\pm$ and $Z$ gauge bosons and the Higgs are present as dynamical degrees of freedom. They are all integrated out at this step and their footprints are only seen in various one-loop functions
or in the factors like $1/M^2_{Z}$ or $1/M^2_{W}$. These issues are discussed in Sec.~\ref{sec:smeft-beyond-leading}.

\paragraph{Step 5: Renormalization group evolution in the WET}
In the absence of the top quark as a dynamical degree of freedom, this evolution is dominated by QCD and at some level by QED effects. Since these interactions are both flavour conserving, neither of them have a direct impact on the flavour structure generated at $\mu=\muEW$, but indirectly they can change it by enhancing or suppressing WCs of operators with different Dirac and flavour structures. Bottom Yukawa effects could have in principle some impact, but for all practical purposes they can be neglected at present. Despite the fact that only QCD and QED interactions together with the SM fermions are present, this step is very different from the usual RG evolution within the SM. This is due to new operators that can have an impact on the evolution of SM operators in addition to their own evolution, including the mixing between these new operators. In performing this step one encounters three effective theories characterized by the number of flavours $f=5,4,3$. At every threshold between two effective theories the matching between the two effective theories in question has to be performed. Moreover, the RG beta functions depend on $f$. Typically, the matching between the $f=5$ and $f=4$ theory is done around the $m_b$ scale, while the one between the $f=4$ and $f=3$ theory around the $m_c$ scale.

\paragraph{Step 6: Calculation of the observables}
Finally, after having evaluated the WCs at scales relevant for the observables of interest, the latter have to be computed. This is often a difficult step, in particular in the case of non-leptonic decays of mesons for which hadronic matrix elements of four-quark operators have to be evaluated by non-perturbative methods. But also in semi-leptonic decays and leptonic decays of mesons non-perturbative QCD effects enter through formfactors and weak decay constants of mesons and their good knowledge is required. While a detailed presentation of these issues is beyond the scope of our review, we will list the relevant references where this topic is reviewed.

\subsection{Bottom-Up Technology}

As far as RG technology is concerned, it is the same as in the top-down approach, except that one tries to extract from the data the WCs of new operators and the modifications of SM WCs. Often only the modification of a single WC or a small number of WCs is considered and global fits allow to give some hints for possible NP responsible for given anomalies. Once some selection of possible models is found one can proceed as in the top-down approach. This interplay of these two approaches will be illustrated in Sec.~\ref{Searching}.

\subsection{The Map of the Expedition}
The map of our expedition is represented by the table of contents. Our expedition starts in Sec.~\ref{sec:smeft-lag} with the introduction of the SMEFT and the WET and in particular with the main actors of both theories, the operators. In addition to Dim-6 operators, we briefly discuss the SMEFT beyond the Dim-6 level.

The choice of useful bases for these operators is discussed in Sec.~\ref{sec:smeft-weak-basis}. Further, in Sec.~\ref{sec:SM-smeft-corrections} we discuss in detail the corrections from dimension-six operators to the dimension-four parameters in the SM, such as the Higgs parameters, electroweak parameters, Yukawa couplings and the CKM parameters after electroweak symmetry breaking.

Having this preparation we are ready to discuss general aspects of the renormalization group evolution both in the SMEFT and the WET. Sec.~\ref{sec:ren-gro-run} covers the LO analysis, for which only one-loop ADMs and tree-level matchings between a UV completion and the SMEFT and also between the SMEFT and the WET are necessary. The discussion is extended to the NLO case in Sec.~\ref{sec:smeft-beyond-leading} for which two-loop ADMs and one-loop matchings in question are necessary. Here the issues like renormalization scheme dependences and their cancellation as well the evanescent operators are discussed.

The general SMEFT at dimension six contains 2499 parameters and a general global fit in this framework, in particular in the bottom-up approach is not feasible. In order to reduce the number of parameters of the general SMEFT in a number of papers various flavour symmetries have been imposed, reducing significantly the number of operators so that global fits become feasible. We discuss these strategies in Sec.~\ref{Fsym}.

Already at this stage it is evident that to perform any phenomenological applications of the SMEFT and also to get an insight in its structure powerful computer tools are required. We review the existing tools in Sec.~\ref{SMEFTtools}.

In Sec.~\ref{sec:obs} we collect SMEFT analyses of different observables that have been performed until now. This section is supposed to give an overview of the different developments in the SMEFT.
 
Having the technology developed in Part I at hand we are in the position to move to Part II of our review with the goal to construct the SMEFT-Atlas.

In Sec.~\ref{classification} we first group observables, to be considered by us later on, into classes. Subsequently, in Tables \ref{tab:cor-bosonic}-\ref{tab:cor-fermionic} we list the classes of observables affected by a given single WC $C_j^{\text{NP}}(\Lambda)$.

Next in Sec.~\ref{sec:10} we construct Tabs.~\ref{tab:mixing-bosonic-color}-\ref{tab:mixing-fermionic-color4} which allow one to find out quickly which operators are generated at the EW scale through RG running from a given operator having non-vanishing $C_i(\Lambda)$ at the NP scale $\Lambda$. In doing so we introduce criteria which allow to exhibit in colours the ranking in the strength of the mixings so that one can see quickly which mixings are most important. Additional insight concerning correlations between sets of operators through RG operator mixing is provided by Tabs.~\ref{tab:mixing-bosonic-fermionic-color2new}-\ref{tab:impact CF}. These tables should give some insight into the material in PART II.

Next Sec.~\ref{sec:11} can be considered as a guide to our SMEFT ATLAS that we present in Secs.~\ref{class1}-\ref{class10}. As evident from the title of our review and from the next paragraph they constitute the central part of our review.

Since in the Tabs.~\ref{tab:mixing-bosonic-color}-\ref{tab:impact CF} flavour indices are suppressed, they give only an overall view on the effects of RG evolution. In order to study explicit flavour dependence we discuss in Secs.~\ref{class1}-\ref{class10} the classes of processes listed in Sec.~\ref{classification} one by one. In addition to listing the most relevant operators in each case, we present the corresponding charts which exhibit the flavour structure of the RG evolution of operators that play the leading role in various processes discussed in PART II.

Having this technology and these results at hand allows us in PART III to move from the operator picture to the particle picture of the NP. First in Sec.~\ref{sec:5} we concisely describe a number of popular NP scenarios within the top-down approach. This includes models with heavy gauge bosons and scalars as well as vector-like quark, vector-like lepton and leptoquark models.

Sec.~\ref{sec:6} deals with the bottom-up approach. Here we discuss in some detail the RG evolution from hadronic scales up to the scale $\Lambda$, which is opposite to the one in the top-down approach.

Finally, in Sec.~\ref{Searching} the top-down and bottom-up approaches are combined with the goal to develop strategies for the identification of NP that are more efficient than strategies formulated separately in one of these approaches.

In Sec.~\ref{sec:7} a brief summary of our review is made and an outlook for the future of the SMEFT and of the WET is given. In five appendices we provide details on the JMS and SMEFT bases, comment on the extensions of the SMEFT and WET, give additional insight into the SMEFT-WET matching and collect formulae that give some insight into the structure of the RG charts mentioned above.

\section{SMEFT Lagrangian}\label{sec:smeft-lag}

\begin{table}[tbp]
\centering
\begin{tabular}{clclcl}
\toprule
fermion  & Rep. & gauge boson  & Rep.  & scalar  & Rep.
\\
\midrule
$q$ & $\left(3,2\right)_{\frac 16}$ & $G$ & $\left(8,1\right)_{0}$ & $H$ & $\left(1,2\right)_{\frac 12}$ \\
$\ell$ & $\left(1,2\right)_{-\frac 12}$ & $W$ & $\left(1,3\right)_{0}$ &  &  \\
$u$ & $\left(3,1\right)_{\frac 23}$ & $B$ & $\left(1,1\right)_{0}$ &  &  \\
$d$ & $\left(3,1\right)_{-\frac 13}$ &  &  &  &  \\
$e$ & $\left(1,1\right)_{-1}$ &  &  &  &  \\
\bottomrule
\end{tabular}
\caption{SM particle content and their representations under the SM gauge group \eqref{SMGAUGE}. The gauge fields are in the adjoint representation.}
\label{tab:SM_fields}
\end{table}

The building blocks of the SMEFT are the SM fields given in Tab.~\ref{tab:SM_fields}. The resulting Lagrangian is obtained by constructing all non-redundant Poincar\'e and gauge invariant structures under the SM gauge group \eqref{SMGAUGE}. In the following we discuss the most relevant parts of $\mathcal{L}_{\text{SMEFT}}$, focusing mainly on the dimension six part.

\subsection{Dim-4 Lagrangian}
We begin by reporting the four-dimensional interactions which form the leading term of the SMEFT Lagrangian. They can be split into the following five parts:

\begin{equation}\label{eq:Ldim4}
\mathcal{L}^{(4)}_{\text{SMEFT}} = \mathcal{L}_{\text{G,kin}}+\mathcal{L}_{\text{Higgs}}+\mathcal{L}_{\text{f,kin}}+\mathcal{L}_{\text{Y}}+\mathcal{L}_{\theta}\,.
\end{equation}

The kinetic terms for the $\text{SU(3)}_C$, $\text{SU(2)}_L$ and $\text{U(1)}_Y$ gauge fields are given by
\begin{equation}
\mathcal{L}_{\text{G,kin}} = -\frac{1}{4}G_{\mu\nu}^AG^{A\,\mu\nu}-\frac{1}{4}W_{\mu\nu}^IW^{I\,\mu\nu}-\frac{1}{4}B_{\mu\nu}B^{\mu\nu}\,,
\end{equation}

with the definitions of the field-strength tensors
\begin{align}
G_{\mu\nu}^A &= \partial_\mu G_\nu^A - \partial_\nu G_\mu^A-g_s f^{ABC}G_\mu^BG_\nu^C\,, \\
W_{\mu\nu}^I &= \partial_\mu W_\nu^I - \partial_\nu W_\mu^I-g \epsilon^{IJK}W_\mu^JW_\nu^K\,, \\
B_{\mu\nu} &= \partial_\mu B_\nu - \partial_\nu B_\mu\,.
\end{align}
Here, the $\text{SU(3)}_C$ and $\text{SU(2)}_L$ structure constants are denoted by $f^{ABC}$ and $\epsilon^{IJK}$, respectively. The Higgs Lagrangian is given by

\begin{equation}
\mathcal{L}_{\text{Higgs}} = (D_\mu H)^\dag (D^\mu H)+\mu^2 (H^\dag H)-\frac{1}{2}\lambda (H^\dag H)^2\,,
\end{equation}

\noindent
where $\mu$ and $\lambda$ denote the mass parameter and self-interaction of the Higgs. The kinetic terms for the fermions are given by:

\begin{equation}
\mathcal{L}_{\text{f,kin}} = i(\overline{\ell}\slashed{D} \ell+\overline{e}\slashed{D} e+\overline{q}\slashed{D}q+\overline{u}\slashed{D}u+\overline{d}\slashed{D}d)\,,
\end{equation}
\noindent
where the covariant derivative for a fermion $f$ takes the general form:

\begin{equation}
D_\mu f = (\partial_\mu+ig_s T^A G^A_\mu+ig t^I W^I_\mu+ig' Y_f B_\mu)f \,,
\end{equation}
where $T^A=\frac{1}{2}\lambda^A$ and $t^I=\frac{1}{2}\sigma^I$, with the Gell-Mann and Pauli matrices $\lambda^A$ and $\sigma^I$, respectively.
\noindent
The Yukawa Lagrangian is given by:
\begin{equation}\label{eq:Lyuk}
-{\cal L}_{\yuk} = \hat Y_u (\bar q\, \widetilde H \,u)+\hat Y_d (\bar q\, 
H\,d)+\hat Y_e (\bar \ell\, H \,e)+\text{h.c.}\,,
\end{equation}
where $\hat Y_{u,d,e}$ are the Yukawa matrices and the conjugated Higgs field is ${\widetilde H_j} =\varepsilon_{jk}(H^k)^*$.
Finally, the theta terms are given by

\begin{equation}\label{eq:Lyuk1}
{\cal L}_{\theta} = \frac{\theta_s g_s^2}{32\pi^2}G^A_{\mu\nu}{\tilde G}^{A,\mu\nu}+\frac{\theta_g g^2}{32\pi^2}W^I_{\mu\nu}{\tilde W}^{I,\mu\nu}+\frac{\theta_{g'} g'^2}{32\pi^2}B_{\mu\nu}{\tilde B}^{\mu\nu}\,,
\end{equation}
with the dual field strength tensors $\tilde X_{\mu\nu}=\frac{1}{2}\epsilon_{\mu\nu\rho\sigma}X^{\rho\sigma}$. These terms are however of topological origin and we will not discuss them further in the following.

Furthermore, we do not include the gauge-fixing Lagrangian in \eqref{eq:Ldim4}, which is given for instance in \cite{Helset:2018fgq}.\footnote{For a general discussion on $R_\xi$-gauges in EFTs see \cite{Misiak:2018gvl}.}

The equations of motion (EOMs) following from the four-dimensional Lagrangian in \eqref{eq:Ldim4} are given for the fermions by 
\begin{equation}
i\slashed{D}\ell = \hat Y_e e H\, \quad i\slashed{D}e = \hat Y_e^\dag H^\dag \ell\,, \quad i\slashed{D}q = \hat Y_u u \widetilde{H}+\hat Y_d dH\,,\quad i\slashed{D}u = \hat Y_u^\dag \widetilde{H}^\dag q\,,\quad i\slashed{D}d = \hat Y_d^\dag H^\dag q\,,
\end{equation}

and for the bosons by

\begin{align}
(D^\rho G_{\rho\mu})^A &= g_s\left(\overline q \gamma_\mu T^A q+\overline u \gamma_\mu T^A u+\overline d \gamma_\mu T^A d\right)\,, \\
(D^\rho W_{\rho\mu})^I &= g\left(\overline q \gamma_\mu t^I q+\overline \ell \gamma_\mu t^I \ell +\frac{1}{2} H^\dag i\overleftrightarrow D^I_\mu H\right)\,, \\
\partial^\rho B_{\rho\mu} &= g'\left( \sum_f Y_f \overline f \gamma_\mu f + Y_H H^\dag i\overleftrightarrow D_\mu H\right)\,, \\
(D^\mu D_\mu H)^j &= -\overline e \hat Y_e^\dag \ell^j+\epsilon_{jk}\overline q^k\hat Y_u u-\overline d \hat Y_d^\dag q^j+\mu^2 H^j-\lambda (H^\dag H)H^j\,.
\end{align}
where
\begin{align}
\overleftrightarrow D_\mu &= D_\mu-\overleftarrow{D}_\mu\,,\\
\overleftrightarrow D^I_\mu & = \tau^ID_\mu-\overleftarrow{D}_\mu\tau^I\,.
\end{align}

\subsection{Basis Construction}

To find a complete and non-redundant operator basis for a given EFT typically involves the usage of many different identities like Fierz identities, EOMs and integration by parts (IBP), that relate large sets of operators to each other. A systematic way to construct a non-redundant operator basis is the Hilbert series approach \cite{Henning:2017fpj}, which was used to count the number of non-redundant SMEFT operators up to mass dimension 12 \cite{Henning:2015alf}. Nowadays, several sophisticated codes are available, which allow to perform this Hilbert series counting in an automated way, such as the Mathematica package \texttt{Sym2Int} \cite{Fonseca:2017lem} and the Python package \texttt{BasisGen} \cite{Criado:2019ugp}. The Hilbert series approach has however the disadvantage that it requires the use of a specific operator basis, namely the one with the minimal number of derivatives. An automated tool which does not rely on this assumption is the Mathematica package \texttt{DEFT} \cite{Gripaios:2018zrz}, which allows to find the SMEFT operator basis up to arbitrary mass dimensions. Yet another code that allows to generate a general SMEFT basis is \texttt{ABC4EFT} \cite{Li:2022tec}. It is based on the on-shell amplitude basis \cite{Shadmi:2018xan,Ma:2019gtx,Durieux:2019siw,AccettulliHuber:2021uoa,Balkin:2021dko,Durieux:2019eor,Durieux:2020gip,Dong:2021vxo}, and uses the algorithm proposed in \cite{Li:2020gnx,Li:2020xlh,Li:2020zfq}.

Constructing a complete and non-redundant basis for dimension five and six operators in the SMEFT was however achieved without any computer tools and dates back several decades: The dimension-five Weinberg operator, first discussed in \cite{Weinberg:1979sa} marks the starting point of this long journey. Next, the construction of a complete set of dimension-six operators was made in the year 1985 by Buchm\"uller and Wyler \cite{Buchmuller:1985jz}. However, it was found later by the Warsaw group that some of the operators are redundant in their list. Therefore, the first non-redundant SMEFT basis carries the name {\em Warsaw basis} \cite{Grzadkowski:2010es}. The Warsaw operator basis, which was constructed by eliminating as many covariant derivatives as possible, satisfies the criterion of having the maximal number of Potential-Tree-Generated (PTG) operators. This criterion was formulated in \cite{Arzt:1994gp,Einhorn:2013kja} and states that PTG operators should be given preference, if equivalent to Loop-Generated (LG) operators, since the former may have larger coefficients.

\subsection{Warsaw Basis at Dim-6 Level}
\label{sub:warsaw-basis}

The SMEFT effective Lagrangian ($\mathcal{L}^{\rm eff}=-\mathcal{H}^{\rm eff}$) at dim-6 is given by 
\be \label{Lag-WCxf}
\mathcal{L}_{\rm SMEFT}^{\rm eff} = \left[\sum_{\Op[\dagger]{a} =\Op[]{a}} \Wc[(6)]{a} \Op[(6)]{a}
+ \sum_{\Op[\dagger]{a} \ne \Op[]{a}}  \Wc[(6)]{a} \Op[(6)]{a} +(\Wc[(6)]{a})^* (\Op[(6) ]{a})^\dagger \right] \,,
 \ee
 where we have suppressed flavour indices.
We will use a non-redundant SMEFT bases (the so-called Warsaw-down or Warsaw-up {basis}) following {\tt WCxf} conventions\cite{Aebischer:2017ugx}. Moreover, in this review, we will primarily focus on the SMEFT operators up to mass dimension six (i.e. $d\le 6$).

The Warsaw basis has a total of 59 baryon number conserving operators (up to flavour indices). They are grouped into eight classes, which carry the following names:
\be
f^4, \quad f^2 H^2 D, \quad f^2 X H, \quad f^2 H^3, \quad X^3, \quad H^6, \quad H^4 D^2, \quad X^2 H^2.
\ee
 Here,
$X_{\mu\nu}= G_{\mu \nu}, W_{\mu\nu}, B_{\mu \nu} $ or their duals, $H$ is the SM Higgs-doublet and $f= \Psi = q, \ell$ or $f= \psi=u, d, e$ are the LH $\text{SU(2)}_L$ doublet and RH singlet fermionic fields, respectively. An alternative SMEFT basis is the SILH basis, which is further discussed in \cite{Giudice:2007fh,Contino:2013kra}, and a mapping between the Warsaw basis and on-shell amplitudes was recently discussed in \cite{Aoude:2019tzn}. The complete list of SMEFT operators can be found in Tabs.~\ref{tab:no4fermsmeft} and \ref{tab:4fermsmeft}. The Feynman rules generated by these operators in the general $R_\xi$-gauge can be found in \cite{Dedes:2017zog}. The corresponding Mathematica package \texttt{SmeftFR} v3~\cite{Dedes:2023zws} allows to derive the Feynman rules for dimension-5, -6, and all bosonic dimension-8 operators. The Feynman rules in the back-ground field method are discussed in \cite{Corbett:2020bqv}.\footnote{The Ward identities in the SMEFT were studied at the tree-level in \cite{Corbett:2019cwl} and at the one-loop level in \cite{Corbett:2020ymv}. Furthermore, massive Ward identities and their violation in the SMEFT for EW amplitudes were studied in \cite{Li:2025ikn}.}

For those readers who see these tables for the first time their appearance could cause a shock. Therefore, in the following, we briefly recap the different classes of operators in the Warsaw basis and the role they play after EW symmetry breaking. In the course of the study of numerous RGEs in Part II of our review it is useful to look up their description given below.

\boldmath
\paragraph{Four-fermion operators, $f^4$ :}
\unboldmath
The four-fermion operators play a crucial role in the SMEFT as they directly govern the weak decays of hadrons and leptons. They are constructed out of the SM fields $\Psi = q, \ell,$ or $\psi=u, d, e$ in the unbroken phase. Therefore, upon expanding in component form, these operators possibly match onto more than one WET four-fermion operator. We will see this explicitly at various places in our review. Depending on the chiralities, there are five different types of baryon number conserving four-fermion operators in the SMEFT, namely, $(\overline LL)(\overline LL)$, $(\overline RR)(\overline RR)$, $(\overline LL)(\overline RR)$, $(\overline LR)(\overline RL)$, $(\overline LR)(\overline LR)$.

The left-left four-quark, four-lepton, and semileptonic operators can be represented by
\bea
\Op[(1)]{\Psi_1 \Psi_2} &=& (\overline \Psi_1 \gamma_\mu \Psi_1 ) (\bar \Psi_2 \gamma^\mu \Psi_2)\,, \quad
\Op[(3)]{\Psi_1 \Psi_2} = (\overline \Psi_1 \gamma_\mu \tau^I \Psi_1 ) (\bar \Psi_2 \gamma^\mu \tau^I \Psi_2)\,,
\eea
with $\Psi_1, \Psi_2 = \ell$ or $q$ and $\tau^I$ being the Pauli matrices. The pure leptonic triplet operator $\Op[(3)]{\ell \ell}$ can be eliminated using $\text{SU(2)}$ matrix relations, so there is no such operator in the Warsaw basis. The right-right operators are given by
\be
\Op[]{\psi_1 \psi_2} = (\overline \psi_1 \gamma_\mu \psi_1 ) (\overline \psi_2 \gamma^\mu \psi_2)\,,
\ee
with $\psi_1, \psi_2= e, u, d$. As compared to the $(\overline LL)(\overline LL)$ sector, in this case we have both colour singlet and octet four-quark operators
\be
\Op[(1)]{ud} = (\overline u \gamma_\mu u ) (\overline d \gamma^\mu d)\,, \quad  \Op[(8)]{ud} = (\overline u \gamma_\mu T^A u ) (\overline d \gamma^\mu T^A d),
\ee
where $T^A$ are the SU(3) generators.

Coming to the left-right operators, we have four-lepton and semileptonic operators
\be
\Op[]{\ell \psi} = (\overline \ell \gamma_\mu \ell ) (\overline \psi \gamma^\mu \psi)\,, \quad \Op[]{qe} = (\overline q \gamma_\mu q ) (\overline e \gamma^\mu e)\,,
\ee
with $\psi= e, u, d$. Similarly, left-right four-quark operators are
\be
\Op[(1)]{q \psi} = (\overline q \gamma_\mu q) (\overline \psi \gamma^\mu \psi)\,, \quad \Op[(8)]{q \psi} = (\overline q \gamma_\mu T^A q) (\overline \psi \gamma^\mu T^A \psi)\,,
\ee
with $\psi= u, d$. Finally, the scalar and tensor four-fermion operators are given as
\bea
\Op[(1)]{quqd} &=& (\bar q^j u) \epsilon_{jk} (\bar q^k d)\,, \quad \Op[(8)]{quqd} = (\bar q^j T^A u) \epsilon_{jk} (\bar q^k T^A d)\,, \\
\Op[(1)]{\ell e qu} &=& (\bar \ell^j e) \epsilon_{jk} (\bar q^k u)\,, \quad \Op[(3)]{\ell e qu} = (\bar \ell^j \sigma_{\mu\nu} e) \epsilon_{jk} (\bar q^k \sigma^{\mu\nu} u)\,, \\
&& \quad \quad \quad \Op[]{\ell e dq} = (\bar \ell^j e) (\bar d q^j).
\eea

\paragraph{\boldmath $f^2 H^2 D$ operators:}
Next, we have $f^2 H^2 D$ class of operators which have the form
\begin{eqnarray}
\Op[(1)]{H \Psi} &=& (H^\dagger i \overleftrightarrow D_\mu H)
 (\bar \Psi \gamma^\mu \Psi)\,, \,\ \,\
\Op[(3)]{H \Psi} = (H^\dagger i \overleftrightarrow D_\mu^I H)
 (\bar \Psi \gamma^\mu \tau^I \Psi)\,, \\
&& \qquad\quad\quad\Op[]{H \psi} = (H^\dagger i \overleftrightarrow D_\mu H)
 (\bar \psi \gamma^\mu \psi)\,.
\end{eqnarray}

Here, the fields $\Psi = \ell, q$ and $\psi= e,u,d$. After electroweak symmetry breaking these operators give rise to the SMEFT corrections of gauge boson couplings to the fermions. Furthermore, we note that all operators containing $(H^\dagger i \overleftrightarrow D_\mu H)$ break custodial symmetry \cite{Kribs:2020jgn}. Additionally, we have another operator in this class
\begin{equation}\label{eq:Hud}
\Op[]{H ud} = i(\widetilde H^\dagger D_\mu H) (\bar u \gamma^\mu d)\,,
\end{equation}
which generates $W$-couplings to the RH quarks.

\boldmath
\paragraph{\boldmath $f^2 X H$ operators:}
\unboldmath
The $f^2 X H$ operators correspond to the dipole operators in the WET after EW breaking. In the case of quarks they have the form
\begin{equation}
\Op[]{\psi G} =  (\bar q \sigma^{\mu \nu} T^A \psi) H G^A_{\mu\nu}\,,  \quad
\Op[]{\psi W} =  (\bar q \sigma^{\mu \nu} \tau^I \psi) H W^I_{\mu\nu}\,, \quad
\Op[]{\psi B} =  (\bar q \sigma^{\mu \nu} \psi) H B_{\mu\nu}\,,
\end{equation}
with $\psi =u,d$ and where $H$ has to be replaced by the conjugate Higgs-doublet $\widetilde H$ for the up-type fermions. In the case of leptons we have
\begin{equation}
\Op[]{e W} =  (\bar \ell \sigma^{\mu \nu} \tau^I e) H W^I_{\mu\nu}\,,  \quad
\Op[]{e B} =  (\bar \ell \sigma^{\mu \nu}  e) H B_{\mu\nu}\,.
\end{equation}

\boldmath
\paragraph{\boldmath $f^2 H^3$ operators:}
\unboldmath
The operators of $f^2 H^3$ give dimension-six corrections to the SM Yukawa couplings and fermion masses after the EW symmetry breaking. They are defined as
\begin{eqnarray}
\Op[]{eH} &=& (H^\dagger H) (\bar \ell e H) \,, \,\ \,\
\Op[]{uH} = (H^\dagger H) (\bar q u \widetilde H) \,, \,\ \,\
\Op[]{dH} = (H^\dagger H) (\bar q d H).
\end{eqnarray}

The remaining classes of the Warsaw basis are purely bosonic.
\paragraph{\boldmath $X^3$ operators}
This class contains triple-gauge interactions:

\begin{equation}
 \Op[]{X} = f^{ijk}X^{i\,\, \nu}_\mu X^{j\,\, \rho}_\nu X^{k\,\, \mu}_\rho\,,\quad \Op[]{\tilde X} = f^{ijk}{\tilde X}^{i\,\, \nu}_\mu X^{j\,\, \rho}_\nu X^{k\,\, \mu}_\rho\,,
\end{equation}
with $X_{\mu\nu} = G_{\mu\nu}, W_{\mu\nu}$, the dual field strength $\tilde{X}_{\mu\nu}=\tilde{G}_{\mu\nu}, \tilde{W}_{\mu\nu}$ and where $f^{ijk}$ denote the structure constants of the respective gauge theories.

\paragraph{\boldmath $H^6$ $\&$ $H^4 D^2$ operators}
The purely Higgs operators read:

\begin{equation}\label{eq:Hops}
\Op[]{H} = (H^\dag H)^3\,,\quad \Op[]{H\square} = (H^\dag H)\square (H^\dag H)\,,\quad \Op[]{HD} = (H^\dag D_\mu H)^*(H^\dag D^\mu H)\,,
\end{equation}
where the last operator again breaks custodial symmetry \cite{Kribs:2020jgn}. The operators in \eqref{eq:Hops} contribute to the Higgs potential and the gauge boson masses \cite{Dedes:2017zog, Dekens:2019ept}.

\paragraph{\boldmath $X^2 H^2$ operators}
Finally, the operators made out of Higgs fields and field strength tensors are given by

\begin{equation}
\Op[]{HX} = (H^\dag H) X_{\mu\nu}X^{\mu\nu}\,,\quad \Op[]{H\tilde X} = (H^\dag H) {\tilde X}_{\mu\nu}X^{\mu\nu}\,,
\end{equation}
where $X_{\mu\nu} = G_{\mu\nu}, W_{\mu\nu},B_{\mu\nu}$. Furthermore, this sector contains the mixed operators

\begin{equation}
\Op[]{HWB} = (H^\dag\tau^I H) W^I_{\mu\nu}B^{\mu\nu}\,,\quad \Op[]{H\tilde WB} = (H^\dag\tau^I H) {\tilde W}^I_{\mu\nu}B^{\mu\nu}\,.
\end{equation}

After EW symmetry breaking, the operators in this sector redefine the gauge couplings and gauge fields, as well as the theta terms of the theory \cite{Dekens:2019ept}.

\subsection{\boldmath SMEFT Beyond Dim-6 Level}

As stated earlier, we will restrict ourselves to dimension-six operators, even though it is worth outlining the theoretical advancements made beyond this order. One can imagine several motivations to go beyond dimension six in the power series of the WET and SMEFT:

Firstly, the increasing precision at the various experiments such as the LHC possibly calls for going beyond dimension six \cite{Passarino:2019yjx}. For example, Drell-Yan (DY) processes can be sensitive to the dimension-eight SMEFT effects\cite{Alioli:2020kez, Alioli:2022fng} and they can even play a dominant role in $pp\to hjj$ \cite{Assi:2024zap}. Furthermore, neutral triple gauge couplings (nTGCs) are absent for dimension-6 operators in the SMEFT and first arise at the dim-8 level \cite{Degrande:2013kka,Ellis:2024omd}. The importance of dimension-eight operators in electroweak precision observables is outlined in \cite{Corbett:2021eux,Adhikary:2025gdh}. On the other hand, low-energy observables such as the neutron decay as well as EDMs are sensitive to the dimension-eight operators in the SMEFT~\cite{Alioli:2022fng}. Moreover, dim-8 effects also play a role in $\mu \to e$ transitions \cite{Ardu:2021koz}. Another important reason to go beyond dimension-six in the SMEFT is the fact that certain features can be first seen only at the dimension-eight level. For instance, some WET operators obtain the first non-zero matching contribution from the SMEFT only at the dimension-eight level\cite{Burgess:2021ylu}. But there is yet another reason to include dim-8 operators, which one should keep in mind when performing calculations of branching ratios. While at the level of the decay amplitudes the impact of dim-6 operators is weighted by $1/\Lambda^2$, at the level of branching ratios this is only the case for the interference of the SM and NP contributions. In the absence of this interference, that is often the case for scalar currents, NP contributions of dim-6 operators to branching ratios are weighted by $1/\Lambda^4$ like the interferences between the SM and dim-8 contributions. Finally, based on helicity selection rules the interference of dimension-six operators with the SM might vanish, making dim-8 terms the leading contributions to a given process \cite{Azatov:2016sqh}.

For the following discussion we note, that for odd powers of mass dimension the corresponding operators in the Lagrangian are either baryon number violating, lepton number violating, or both \cite{Kobach:2016ami,Helset:2019eyc}. This is reflected in the following formula, that relates the violation of baryon number $\Delta\mathcal{B}$, the violation of lepton number $\Delta L$ and the mass dimension of the operator $d$:

\begin{equation}\label{eq:DBDL}
\frac{\Delta\mathcal{B}-\Delta L}{2} = d \,\,\text{mod}\,\,2 \,.
\end{equation}

As for the Warsaw basis, we will in our review mainly focus on $\Delta\mathcal{B}=\Delta L=0$, that is $\mathcal{B}$ and $L$ conserving operators. Given inputs of $\mathcal{B}$ and $L$ based on \eqref{eq:DBDL} a lower bound on the mass dimension of SMEFT operators was found in \cite{Heeck:2025btc}.

\paragraph{SMEFT basis: Dimension 7}
By now the full SMEFT operator basis up to dimension-seven is known \cite{Lehman:2014jma, Liao:2016hru}. There are only 20 new operators, but all of them violate either baryon number, lepton number, or both, as is implied by \eqref{eq:DBDL}.

\paragraph{SMEFT basis: Dimension 8}
A first discussion on dimension 8 SMEFT operators using Hilbert series techniques can be found in \cite{Lehman:2015coa} and the full bases are given in \cite{Li:2020gnx} and \cite{Murphy:2020rsh}. At the dimension-eight level, there are 13 new classes of operators (baryon and lepton number conserving) \cite{Murphy:2020rsh}. The classes that consist of four-fermion fields are ${f^4 H^2, f^4 X, f^4 HD, f^4 D^2}$. The operator classes involving two-fermions are $f^2 X^2 H$, $f^2 X H^3$, $f^2 H^2D^3$, $f^2 H^5$, $f^2 H^4 D$, $f^2 X^2 D$, $f^2 X H^2 D$, $f^2 XH D^2$, $f^2 H^3 D^2$. In addition, there are several other purely bosonic classes. A basis for anomalous quartic gauge couplings (aQGCs) was proposed in \cite{Durieux:2024zrg}. Finally, for the Universal SMEFT, for which NP dominantly couples to gauge bosons, a full operator basis up to dimension eight was derived in \cite{Corbett:2024yoy}.

\paragraph{SMEFT basis: Dimension 9}
The complete operator basis at mass dimension nine is given in \cite{Li:2020xlh}, and was found using amplitude-operator correspondence. Another derivation using Hilbert Series techniques is given in \cite{Liao:2020jmn}. Again, all operators are either baryon- or lepton number violating, or both.

\paragraph{SMEFT basis: Dimension 10,11,12}
The complete operator basis up to mass dimension 12 is provided in digital form in \cite{Harlander:2023psl}. The operators were derived using the code \texttt{AutoEFT}, which is based on the algorithm described in \cite{Li:2020gnx,Li:2020xlh,Li:2022tec}. Besides the standard SMEFT operators also the ones for the SMEFT augmented with gravitons, the so-called GRSMEFT, are derived up to mass dimension 12 \cite{Harlander:2023psl}.

\subsection{\boldmath SMEFT Green's bases}
In this section we mention the developments regarding Green's bases in the SMEFT. As opposed to a physical basis a Green's basis is related to off-shell one-particle irreducible Green's functions. Consequently, the resulting operators form a complete basis, with the caveat that they are redundant under field redefinitions. The Green's basis is especially useful when performing off-shell matching onto the SMEFT or WET.

\paragraph{SMEFT Green's basis: Dimension 6}
The complete Green's basis at dimension six is given in \cite{Gherardi:2020det}. It was derived as a byproduct of the complete one-loop matching of $(\overline{3},1)_{1/3}$ and $(\overline{3},3)_{1/3}$ scalar leptoquark models onto the SMEFT. Furthermore , the reduction formulae of the Green's basis to the Warsaw basis are given in \cite{Gherardi:2020det}.

\paragraph{SMEFT Green's basis: Dimension 7}
A Green's basis for dim-7 operators together with a novel physical basis at the dimension seven level can be found in \cite{Zhang:2023kvw}. Also the reduction relations to the physical basis are provided.

\paragraph{SMEFT Green's basis: Dimension 8}
The bosonic Green's basis at dim-8 was first found in \cite{Chala:2021cgt}. The complete Green's basis was then derived in \cite{Ren:2022tvi}, using the off-shell amplitude formalism.

\subsection{\boldmath WET Beyond Dim-6 Level}
The WET basis, the so-called JMS-basis at dim-6, is presented in Appendix~\ref{app:jmsbasis}. Furthermore, an evanescent operator basis for the WET in the 't Hooft-Veltman (HV) scheme is given in \cite{Naterop:2023dek}. For completeness, we mention here also the higher-dimensional operators of the WET.

\paragraph{WET basis: Dimension 7}
The complete WET basis at dimension-seven was presented in \cite{Liao:2020zyx}. The operators can be grouped into the classes $f^2 X^2$, $f^2 X \widetilde X$ and $f^4 D$. Also the complete tree-level matching between the SMEFT and WET at dimensions seven is provided in \cite{Liao:2020zyx}.

\paragraph{WET basis: Dimension 8}
The complete set of dim-8 WET operators has been classified in \cite{Murphy:2020cly}. At the dimension-eight level, one finds the baryon number conserving classes $X^4$, $f^2 X^2 D$, $f^4 X$, $f^4 D^2$. The tree-level matching between dim-8 and dim-6 SMEFT operators and dim-6 WET operators has been recently computed in Ref.~\cite{Burgess:2021ylu}. The matching among operators with different mass dimensions occurs when the Higgs field acquires its vacuum expectation value (VEV), thereby reducing the dynamical degrees of freedom of the initial operator.

\paragraph{WET basis: Dimension 9}
The complete dim-9 SMEFT Lagrangian has been derived in \cite{Li:2020tsi}, using the massive spinor helicity formalism. At dimension nine one encounters the following operators classes: $f^2X^2D^2, f^2X^3$ containing two fermion fields, $f^4D^3, f^4XD$ involving four fermions as well as six-fermion operators $f^6$.

\section{Choice of Weak Bases}
\label{sec:smeft-weak-basis}
\noindent
In a UV completion, having the complete Lagrangian including the Yukawa matrices along with the Wilson coefficients at the NP scale the flavour basis can be fixed. That is a concrete UV completion determines how Yukawa and other interactions are oriented in flavour space.

On the other hand in an effective theory like the flavoured SMEFT, the choice of the flavour basis in the space of three fermion generations for each of the quark and lepton fields is rather subtle. Since all fermions are massless in the $\text{SU(2)}_L\times \text{U(1)}_Y$-symmetric phase, there is no a priori preferred basis, such as the mass basis below the electroweak scale. Yet, having the set of linearly independent SMEFT operators at hand, we have to specify the weak-eigenstate basis in which we plan to perform calculations including the RG evolution above the electroweak scale. In what follows we will describe briefly how this issue is addressed in the literature. More details can be found in \cite{Aebischer:2017ugx} and also in Section~\ref{SMEFTtools}.

\noindent
The gauge interactions in the SM are invariant under the U(3)$^5$ flavour symmetry\footnote{To our knowledge this has been pointed out first in \cite{Gerard:1982mm}.}
\begin{align} 
q_L & \rightarrow \hat U_q\, q_L \,, &
u_R & \rightarrow \hat U_{u_R}\, u_R \,, &
d_R & \rightarrow \hat U_{d_R}\, d_R \,,
\label{eq:RotationSMquarks1}
\\
\ell_L & \rightarrow \hat U_\ell\, \ell_L \,, &
e_R & \rightarrow \hat U_e\, e_R \,,
\label{eq:RotationSMleptons1}
\end{align}
where $\hat U_{q}, \hat U_{u_R}, \hat U_{d_R}, \hat U_\ell$ and $\hat U_e$ are unitary $3\times 3$ matrices. This is the consequence of gauge coupling universality for all left- and right-handed fermions. However, the SM fermions acquire their masses through Higgs interactions after EW symmetry breaking. The Yukawa sector in~\eqref{eq:Lyuk} breaks this universality and consequently the U(3)$^5$ symmetry explicitly, simply because the Yukawa couplings to fermions are not subject to further symmetry constraints. A consequence of this is the known mass spectrum of quarks and leptons \cite{Gerard:1982mm}. The preferred basis for calculations is the mass-eigenstate basis in which the Yukawa and consequently mass matrices are diagonalized as explicitly given by
\begin{align}
\label{equ:LYukRdiag1}
\hat U_{d_L}^\dagger {\hat Y_d} \hat U_{d_R} & = \hat{Y}_d^D \,, &
\hat U_{u_L}^\dagger {\hat Y_u} \hat U_{u_R} & = \hat{Y}_u^D \,, &
\hat U_\ell^\dagger {\hat Y_e} \hat U_e & = \hat{Y}_e^D \,,
\end{align}
with ${\hat{Y}^D}_{i}$ being diagonal. Here $\hat U_{u_L}$ and $\hat U_{d_L}$ rotate the $\text{SU(2)}_L$ components of $q_L$ individually contrary to $\hat U_q$ in \eqref{eq:RotationSMquarks1}.

The CKM matrix is then given by
\be\label{eq:CKM}
\hat V = \hat U_{u_L}^\dag \hat U_{d_L}\,.
\ee

Because of the freedom \eqref{eq:RotationSMquarks1}-\eqref{eq:RotationSMleptons1}, different choices of bases for the SMEFT WCs are possible above the weak scale. This is in contrast to the SM where only CKM and PMNS matrices appear in the mass basis and hence in the physical quantities. On the other hand, in the SMEFT the unitary matrices can explicitly appear in the expressions for observables. Strictly speaking, different choices of the weak bases within the SMEFT correspond to different NP models and consequently to different NP effects in observables. This is at first sight surprising. How can the outcome of a given SMEFT analysis depend on the chosen basis? We will elaborate on this important issue below. But first let us present two convenient choices that are popular in the literature.

\subsection{Warsaw-Down and Up Bases}
A particular choice of basis is the {\em down-basis} that can be realized by setting $\hat U_q = \hat U_{d_L}$, and performing $\text{U(3)}^5$ rotations in the SMEFT Lagrangian.\footnote{The down-basis was first discussed in \cite{Aebischer:2015fzz}.} This leads to diagonal down-type Yukawas but the SMEFT WCs have to be redefined, which now become functions of the original WCs and $\hat U_{d_L}$. {After these steps} the overall form of the Lagrangian remains unchanged. Interestingly, this procedure allows to absorb the unknown parameters like $\hat U_{d_L}$ in the WCs, so the theory can now be expressed entirely in terms of the redefined WCs and the known dim-4 SM parameters. 

Subsequently, at the EW scale, for the purpose of matching onto to the WET the $q^i$ fields take a specific form:
\begin{equation}
q^i= \quad
\begin{pmatrix}
{ V^*_{ji} u_L^j} \\
d_L^i
\end{pmatrix}
\,, \qquad (\text{\bf down-basis})
\end{equation}
where $V_{ji}$ are the elements of the CKM matrix. Summation over $j$ is understood. The above form is a result of rotating the weak basis quarks into mass basis states $u_L^j$ and $d_L^j$ at the EW scale. 

Another popular basis choice is the {\em up-basis} with $\hat U_q= \hat U_{u_L}$, and hence diagonal up-type Yukawa matrices. For the left-handed quark-fields one finds
\begin{equation}
         q^i= \quad
\begin{pmatrix}
u_L^i  \\
 V_{ij} d_L^j
\end{pmatrix}
\,.\qquad (\text{\bf up-basis})
\end{equation}
Changing between these two bases is achieved by rotating the corresponding parameters by CKM factors. 

As an example we consider the WCs $\wcup[(1)]{qq}{}$ and $\wcup[(8)]{qu}{}$ with four flavour indices, which transform according to
\be
\begin{aligned}
\wcup[(1)]{qq}{ijkl} & = 
{V_{ip}^{} V^*_{jr} V_{ks}^{} V^*_{lt} \, \wc[(1)]{qq}{prst}} \,, 
\\
\wcup[(8)]{qu}{ijkl} & = 
{V_{ip}^{} V^*_{jr} \, \wc[(1)]{qu}{prkl}} \,.
\end{aligned}
\ee
The tilded and un-tilded WCs correspond to the up-basis and the down-basis, respectively. Repeated indices are implicitly summed over. The choice of basis might have certain advantages in phenomenological analyses. For instance, if we are interested in FCNCs in the down (up)-sector, it is more convenient to work in the down (up)-basis.

In the context of solving RGEs, it is useful to have the explicit expressions for the Yukawa matrices in the two bases. For the {\bf down-basis} we have 

\be
\begin{aligned}
{[\hat Y_u]}_{zy} &= y_t V^*_{3z} \delta_{y3}\,, \quad [\hat Y_u^\dagger]_{zy} =  y_t  \delta_{z3}V_{3y}\,, \\
{[\hat Y_d]}_{zy} &=  [\hat Y_d^\dagger]_{zy} = y_b \delta_{z3} \delta_{y3}  \,, \\
{[\hat Y_e]}_{zy} &=  [\hat Y_e^\dagger]_{zy} = y_\tau \delta_{z3} \delta_{y3}\,.
\end{aligned}
\ee

where we have ignored the lighter Yukawas. Then to an excellent approximation we have 
  \be
\begin{aligned}
{[\hat Y^\dagger_u \hat Y_u]}_{zy} &=y^2_t \delta_{z3}\delta_{y3}\,, \quad 
[\hat Y_u \hat Y^\dagger_u]_{zy}=y^2_t {V^*_{3z}V_{3y}}\,, \\
[\hat Y^\dagger_d \hat Y_d]_{zy} &= [\hat Y_d \hat Y^\dagger_d]_{zy}=y^2_b \delta_{z3}\delta_{y3}\,, \\
[\hat Y^\dagger_e \hat Y_e]_{zy} &= [\hat Y_e \hat Y^\dagger_e]_{zy}=y^2_\tau \delta_{z3}\delta_{y3}\,, \\
[\hat Y_u^\dagger \hat Y_d]_{zy}  &= [\hat Y_d^\dagger \hat Y_u]_{zy}     =y_t y_b \delta_{z3} \delta_{y3}\,, \\
[\hat Y_u \hat Y_d^\dagger]_{zy}  & =y_t y_b V^*_{3z} \delta_{y3}\,, \quad [\hat Y_d \hat Y_u^\dagger]_{zy}   =y_t y_b \delta_{z3} V_{3y}\,, 
\end{aligned}
\ee
\be
\begin{aligned}
& {[\hat Y_u \hat Y_u^\dagger \hat Y_u]_{zy}}  = {y_t^3 V^*_{3z} \delta_{y3}\,, \quad [\hat Y_u^\dagger \hat Y_u \hat Y_u^\dagger]_{zy}  = y_t^3 \delta_{z3}V_{3y} }\,, \\
& {[\hat Y_u \hat Y_u^\dagger \hat Y_d]_{zy}}  = {y_by_t^2V_{33} V^*_{3z} \delta_{y3}\,, \quad [\hat Y_d \hat Y_d^\dagger \hat Y_u]_{zy}  = y_b^2y_tV^*_{33} \delta_{z3}\delta_{y3} }\,, \\
{[\hat Y_d \hat Y_d^\dagger \hat Y_d]_{zy}}  & = {[\hat Y_d^\dagger \hat Y_d \hat Y_d^\dagger]_{zy}= y_b^3 \delta_{z3} \delta_{y3}\,, \quad [\hat Y_e \hat Y_e^\dagger \hat Y_e]_{zy}   = [\hat Y_e^\dagger \hat Y_e \hat Y_e^\dagger]_{zy}= y_\tau^3 \delta_{z3} \delta_{y3}}\,.
\end{aligned}
  \ee

Likewise in the {\bf  up-basis} 

\be
\begin{aligned}
{[\tilde Y_u]}_{zy} &=  {[\tilde Y_u^\dagger]}_{zy} = y_t  \delta_{z3} \delta_{y3}\,, \\
{[\tilde Y_d]}_{zy} &=  y_b V_{z3} \delta_{y3} \,, \quad  {[\tilde Y_d^\dagger]}_{zy} =  y_b \delta_{z3}  V^*_{y3}  \,, \\
{[\tilde Y_e]}_{zy} &=  [\tilde Y_e^\dagger]_{zy} = y_\tau \delta_{z3} \delta_{y3}\,.
\end{aligned}
\ee
Then to an excellent approximation we have 
  \be
\begin{aligned}
{[\tilde Y^\dagger_u \tilde Y_u]}_{zy} &=[\tilde Y_u \tilde Y^\dagger_u]_{zy}=y^2_t \delta_{z3}\delta_{y3}\,,  \\
[\tilde Y^\dagger_d \tilde Y_d]_{zy} &=y^2_b \delta_{z3}\delta_{y3}\,,\quad
[\tilde Y_d \tilde Y^\dagger_d]_{zy}={y^2_b V_{z3}V^*_{y3}} \,,
\\
[\tilde Y^\dagger_e \tilde Y_e]_{zy} &= [\tilde Y_e \tilde Y^\dagger_e]_{zy}=y^2_\tau \delta_{z3}\delta_{y3}\,, \\
[\tilde Y_u^\dagger \tilde Y_d]_{zy}  &=  [\tilde Y_d^\dagger \tilde Y_u]_{zy}= y_t y_b \delta_{z3} \delta_{y3}\,, \\
[\tilde Y_u \tilde Y_d^\dagger]_{zy}  & =y_t y_b  \delta_{z3}V^*_{y3}\,, \quad [\tilde Y_d \tilde Y_u^\dagger]_{zy}   =y_t y_b V_{z3}\delta_{y3}\,, 
\end{aligned}
\ee
\be
\begin{aligned}
& {[\tilde Y_d \tilde Y_d^\dagger \tilde Y_d]_{zy}}  = {y_b^3 V_{z3}\delta_{y3}\,, \quad [\tilde Y_d^\dagger \tilde Y_d \tilde Y_d^\dagger]_{zy}  = y_b^3 \delta_{z3}V^*_{y3} }\,, \\
& {[\tilde Y_u \tilde Y_u^\dagger \tilde Y_d]_{zy}}  = {y_by_t^2V_{33} \delta_{z3} \delta_{y3}\,, \quad [\tilde Y_d \tilde Y_d^\dagger \tilde Y_u]_{zy}  = y_b^2y_tV^*_{33} V_{z3}\delta_{y3} }\,, \\
{[\tilde Y_u \tilde Y_u^\dagger \tilde Y_u]_{zy}}  & = {[\tilde Y_u^\dagger \tilde Y_u \tilde Y_u^\dagger]_{zy}= y_t^3 \delta_{z3} \delta_{y3}\,, \quad [\tilde Y_e \tilde Y_e^\dagger \tilde Y_e]_{zy}   = [\tilde Y_e^\dagger \tilde Y_e \tilde Y_e^\dagger]_{zy}= y_\tau^3 \delta_{z3} \delta_{y3}}\,.
\end{aligned}
\ee

We have set $\mathrm{U}_\mathrm{PMNS} = \mathbb{1}$. It is instructive to have them also in the numerical form. In the down- and up-basis respectively, we find:
\be
\begin{aligned}
\hat  Y_u &\approx
\begin{pmatrix}
0 &0 & 8.6 \cdot 10^{-3}  \\
0 &0 & 4.0 \cdot 10^{-2} \\
0 &0 &  1.0\\
\end{pmatrix} ,
\quad
\hat Y_d \approx
\begin{pmatrix}
0  &0 &0 \\
0 & 0 &0 \\
0 & 0& 2.7\cdot 10^{-2} \\
\end{pmatrix},
\quad
\hat Y_e \approx
\begin{pmatrix}
0  &0 &0 \\
0 & 0 &0 \\
0 & 0& 1.0\cdot 10^{-2} \\
\end{pmatrix}, \\
\tilde  Y_u & \approx
\begin{pmatrix}
0 &0 & 0  \\
0 &0 & 0\\
0 &0 & 1.0\\
\end{pmatrix} ,
\quad
\tilde Y_d \approx
\begin{pmatrix}
0  &0 & 9.9 \cdot 10^{-5} \\
0 & 0 & 1.1 \cdot 10^{-3} \\
0 & 0& 2.7 \cdot 10^{-2} \\
\end{pmatrix},
\quad
\tilde Y_e \approx
\begin{pmatrix}
0  &0 &0 \\
0 & 0 &0 \\
0 & 0& 1.0 \cdot 10^{-2} \\
\end{pmatrix}.
\end{aligned}
\ee

In the SM, due to universal gauge couplings and unitary rotation matrices, FCNCs are absent at tree-level. Consequently, flavour changes appear only in the charged currents parameterized by CKM and PMNS matrices. It should be stressed that in the SM it is equivalent to either rotate the down-quarks from flavour to mass eigenstates together with the assumption that flavour and mass eigenstates in the up-quark sector are equal, or vice versa. This is due to the fact, that the interactions in the mass-eigenstate basis remain unchanged. The same applies to the lepton sector.

Another SMEFT basis is the Warsaw-mass basis \cite{Dedes:2017zog, Aebischer:2016xmn} or the mixed Higgs-Warsaw up basis \cite{LHCHiggsCrossSectionWorkingGroup:2016ypw}. The complete list of non-redundant operators in the flavour space for various bases was first defined in \cite{Aebischer:2017ugx} and can be found in \url{https://wcxf.github.io}.

Let us then return to the issue of the dependence of the outcome of a SMEFT analysis on the chosen weak basis of operators to convince ourself that a particular choice corresponds to a particular class of UV completions. A simple $Z^\prime$ model is very instructive in this context.

\subsection{NP example: $Z^\prime$}

Let us discuss a NP example, involving the bases mentioned above, by assuming NP contributions with non-universal but generation-diagonal gauge couplings, that break the U(3)$^5$ flavour symmetry explicitly. In order to see the consequences of this breakdown let us consider a $Z^\prime$ model and choose the up-basis, i.e. $\hat U_{u_L} = \mathbb{1}$ and $\hat U_{u_R} = \mathbb{1}$. As a consequence the Yukawa matrix or equivalently the mass matrix for up-quarks is diagonal and the same holds for interactions of up-quarks with the $Z^\prime$. There is no flavour violation in the up-quark sector mediated by the $Z^\prime$, up to contributions from matching and back-rotation in the SMEFT, with the latter effect discussed in detail in Sec.~\ref{sub:back-rot}. But with the basis choice $\hat U_{u_L} = \mathbb{1}$ it follows from \eqref{eq:CKM} that $\hat U_{d_L} = \hat V$. Therefore, performing the usual rotations in the down sector from flavour- to mass-eigenstate basis we find FCNC transitions in the down-quark sector. More explicitly, the $Z^\prime$ couplings to left- and right-handed down-quarks are given by
\begin{align}
\Delta^{ij}_L(Z^{\prime}) &
= g_{Z^\prime} \big[\hat V^\dagger\, \hat Z^{d}_L\, \hat V \big]_{ij}\,, &
\Delta^{ij}_R(Z^{\prime}) &
= g_{Z^\prime} \big[(\hat U_{d_R})^\dagger\, \hat Z^{d}_R\, \hat U_{d_R} \big]_{ij} \,.
\end{align}
Here $(i,j=d,s,b)$ denote the different down-type flavours, $g_{Z^\prime}$ is the $\text{U(1)}^\prime$ gauge coupling and the diagonal matrices $\hat Z^{d}_{L,R}$ denote the $\text{U(1)}^\prime$ charges of left- and right-handed down-quarks. Since the diagonal matrices $\hat Z^{d}_{L,R}$ are not proportional to a unit matrix, both couplings $\Delta^{ij}_L(Z^{\prime})$ and $\Delta^{ij}_R(Z^{\prime})$ are flavour violating.

However, $\hat U_{u_L} = \mathbb{1}$ and $\hat U_{u_R} = \mathbb{1}$ is an assumption which specifies a particular model, actually a class of models that could still differ by other parameters than the ones present in rotation matrices. It assumes that in the basis in which Yukawa matrices for up-quarks are diagonal, also the interactions of the $Z^\prime$ with up-quarks are flavour diagonal. In other words $\hat Y_u$ and $Z^\prime$ interactions for the up-quarks are aligned. But we could as well have chosen $\hat U_{d_L} = \mathbb{1}$ and $\hat U_{d_R} = \mathbb{1}$ which would result in FCNCs mediated by the $Z^\prime$ in the up-quark sector and no FCNCs in the down-quark sector, again up to contributions from matching and back-rotation.

These simple examples show that in the absence of a U(3)$^5$ flavour symmetry in the gauge sector we have more freedom and the physics depends on how the Yukawa and other interaction matrices are oriented in flavour space. This also explains why the bounds on various coefficients found in \cite{Silvestrini:2018dos} for the down-basis and up-basis differ from each other.

These findings underline the importance of the construction of UV completions in which also a flavour theory is specified, so that the orientation between Yukawa matrices and the matrices describing the interactions are known. Interesting model constructions in this direction can be found in \cite{Bordone:2017bld,Gherardi:2019zil,Davighi:2023evx,Davighi:2022bqf,Davighi:2023iks,Allwicher:2023shc,Davighi:2023xqn,Marzocca:2024hua,Moreno-Sanchez:2025bzz,Giarnetti:2025idu,Banks:2025baf,Isidori:2025rci}. 

Our presentation was rather brief. Much more details on rotations in flavour space with physical implications can be found in \cite{Alonso:2014csa,Aebischer:2015fzz,Feruglio:2017rjo,Buttazzo:2017ixm,Aebischer:2017ugx,Kumar:2018kmr,Aebischer:2018iyb,Silvestrini:2018dos,Ciuchini:2019usw,Aebischer:2019mlg,Gherardi:2019zil}.

\subsection{Generic Basis}
One has to be careful while working with the up and down basis. Especially the scenarios in which a subset of operators are assumed at the NP scale are unlikely to be realized in an ultraviolet completion. Moreover, the subset of operators chosen at $\Lambda$ is not stable under renormalization group running \cite{Aebischer:2018bkb, Aebischer:2020lsx}. Recently, in \cite{Datta:2025csr} the \emph{generic basis} was proposed which does not make any a priori assumptions about the alignment in the flavour space and interpolates between the up and down bases.\footnote{Again the analyses in \cite{Silvestrini:2018dos,Bordone:2017bld,Gherardi:2019zil,Davighi:2023evx,Davighi:2022bqf,Davighi:2023iks,Allwicher:2023shc,Davighi:2023xqn,Marzocca:2024hua} are useful in this context.} When NP is integrated out, it can be directly mapped with the generic basis. Subsequently, to compute low energy observables one can map the generic basis operators onto up or down basis in SMEFT at scale $\Lambda$ and eventually to mass basis in WET at $\muEW$. In this review we work in the down basis, however we do not make any assumptions about subset of operators at the NP scale. Therefore, the full set of operators in the down basis can span the flavour space.

\subsection{Non-Redundant Flavour Basis}\label{NONRED}

The Warsaw basis described in Sec.~\ref{sub:warsaw-basis} is redundant with respect to flavour indices. This is due to the fact, that all possible combinations of flavour indices in fermionic operators are included in the Lagrangian. Consequently, operators with a flavour index symmetry appear multiple times in the basis. The anomalous dimension matrices \cite{Alonso:2013hga,Jenkins:2013zja,Jenkins:2013wua} as well as the matching conditions at the EW scale \cite{Jenkins:2017dyc} have been computed in such a redundant Warsaw basis only. The first non-redundant flavour basis for the SMEFT was proposed in \cite{Aebischer:2017ugx}.

\subsubsection{Flavour Redundant Operators}
The redundant operators arise due to symmetric flavour structures in the operator definitions. A typical example are four-fermion operators with identical fermion currents \cite{Aebischer:2018iyb, Aebischer:2017ugx, Aebischer:2018bkb}. An example of such an operator is
\begin{equation}
\ops[]{dd}{ijkl} =(\bar d_i \gamma_{\mu}d_j)(\bar d_k\gamma^{\mu}d_l)\,.
\end{equation}
Since, $\ops[]{dd}{ijkl} = \ops[]{dd}{klij}$, the set of non-redundant operators in this case would be\footnote{Up to hermitian conjugation and considering only the first two generations.}
\be \label{eq:nonred-dd}
\ops[]{dd}{1111}\,,\,  \ops[]{dd}{2222}\,, \,  \ops[]{dd}{1122}\,, \,  \ops[]{dd}{1112}\,, \, \ops[]{dd}{1222}\,, \,
\ops[]{dd}{1212}\,, \, \ops[]{dd}{1221}\,,
\ee
and the following set of operators are considered to be redundant
\be
\ops[]{dd}{2211}\,,\,\  \ops[]{dd}{1211}\,, \,\  \ops[]{dd}{2212}\,, \,\  \ops[]{dd}{2112}.
\ee
For a complete list of non-redundant operators we refer to {the} Wilson Coefficient Exchange Format (\texttt{WCxf}) \cite{Aebischer:2017ugx} for the Warsaw-down basis.\footnote{\texttt{WCxf} webpage: \url{https://wcxf.github.io/assets/pdf/SMEFT.Warsaw.pdf}} 

For the computation of physical processes it is typically more convenient to choose a minimal basis, in which all operators are independent of each other. Such a choice avoids unwanted symmetry factors in the Lagrangian. To make this clearer, consider the example of the Lagrangian written in a redundant basis featuring the operator $O_{dd}$
\begin{align}\label{eq:redLag}
\mathcal{L}^{\text{SMEFT}}_\text{red}\supset& \, \wc[]{dd}{1122}  \ops[]{dd}{1122} + \wc[]{dd}{2211}  \ops[]{dd}{2211} \\\notag
=& \,\wc[]{dd}{1122}  {\ops[]{dd}{1122}} + \wc[]{dd}{2211}  \ops[]{dd}{1122} \\\notag
=& \, (  \wc[]{dd}{1122} + \wc[]{dd}{2211} ) \ops[]{dd}{1122} \\\notag
=& \,2 \wc[]{dd}{1122}  \ops[]{dd}{1122} 
= \,2 \wc[]{dd}{2211}  \ops[]{dd}{2211}\,,
\end{align}
whereas in a non-redundant basis only one flavour combination is taken into account:
\begin{equation}\label{nonred1}
\mathcal{L}^{\text{SMEFT}}_{\text{non-red}} \supset \wc[]{dd}{1122}  \ops[]{dd}{1122}\,,
\end{equation}
and the redundant contribution is not part of the Lagrangian. Factors like in the last line of \eqref{eq:redLag} have to be taken into account when using results from the literature, like the ADMs and matching computations of the SMEFT, as will be discussed in the following subsection.

But this is not the whole story. Due to Fierz identities one also has the following relations 
\be\label{nonred2}
\ops[]{ee}{ijkl} = \ops[]{ee}{ilkj}\,,
\ee
that are consistent with the $\text{SU(2)}_L$ gauge symmetry. Again, in a non-redundant basis only one of these two flavour combinations should be kept.

In this context we should emphasize that in the lists of RGEs in \cite{Alonso:2013hga,Jenkins:2013zja,Jenkins:2013wua} the rules in \eqref{nonred1} and \eqref{nonred2} have not been used. In particular in the case of operators with field combinations $qq$, $\ell\ell$, $dd$, $uu$, $ee$ both operators with indices $prst$ and $stpr$ or $psrt$ and $rtps$ are kept. In Part II of our review, when presenting the lists of RGEs relevant for a given class of observables we will remove one of such contributions from the RGEs in \cite{Alonso:2013hga,Jenkins:2013zja,Jenkins:2013wua}.

\subsubsection{Redundant Parts in ADMs and WCs}
The fact that ADMs and matching conditions for the SMEFT are computed in a redundant basis creates complications once a non-redundant basis is adopted. For instance symmetry factors that result when redundant operators are eliminated can enter the ADMs of the non-redundant WCs. For example the beta function of the Wilson coefficient $\wc[]{dd}{prst}$ in a redundant SMEFT basis contains terms of the form \cite{Alonso:2013hga}:
\be \label{eq:red-adm}
\dotwc[]{dd}{prst} =\frac{2}{3}g_1^2N_cy_d^2( \wc[]{dd}{prww} + \wc[]{dd}{wwpr} )\delta_{st}+\ldots\,\,\,.
\ee
The operators that have a flavour index symmetry, like e.g. $prst=iijj,\, i\neq j$, get the same contribution from $\wc[]{dd}{iiww}$ and $\wc[]{dd}{wwii}$. However, in a non-redundant basis, only one of these contributions is present, since the other operator is not included in the basis.

Therefore, one has to be careful when using SMEFT ADMs for a non-redundant basis: Either one should discard the parts of the ADM corresponding to the redundant operators or the redundant operators have to be divided by their corresponding symmetry factors.\footnote{In \texttt{wilson}~\cite{Aebischer:2018bkb} the proper division (multiplication) by the corresponding symmetry factors before (after) running is implemented for the complete SMEFT and WET ADM.} The same holds for the matching and running at the EW scale and below.

As an additional complication, the operator basis has to be chosen before reducing the non-redundant basis to a minimal one. The reason is that a change of basis might reintroduce redundant operators. To illustrate this point let us consider the operator $O^{(1),prst}_{qq}$ in the Warsaw up-basis (denoted with a tilde), together with the basis change to the down-basis (denoted without a tilde). For the index combination $prst=1122$ {one finds \cite{Aebischer:2015fzz}}:
\begin{equation}
\opsup[(1)]{qq}{1122} =V_{ui}\,V^*_{uj}\,V_{ck}\,V^*_{cl}\, \ops[(1)]{qq}{ijkl} \,.
\end{equation}
Hence, the operator $\opsup[(1)]{qq}{1122}$ depends for instance on the operator $\ops[(1)]{qq}{2211}$ as well as on its redundant counterpart $\ops[(1)]{qq}{1122}$.

The issues above appear to be very complicated, but the good news is that they are automatically taken into account in computer tools such as \texttt{wilson}~\cite{Aebischer:2018bkb}, so the user does not have to worry about them.

\section{Effective Parameters}
\label{sec:SM-smeft-corrections}

Within the SMEFT, the dimension-four parameters in the SM, such as the Higgs parameters, electroweak parameters, Yukawa couplings and the CKM parameters receive corrections from dimension-six operators, after electroweak symmetry breaking. As a result, all the SM parameters need to be redefined in order to incorporate SMEFT corrections. We will use the tilde $(\widetilde p)$ and non-tilde $(p)$ notation for the parameters in the SMEFT and the SM, respectively. Furthermore, the fields have to be redefined, in order to guarantee canonical normalization of the corresponding propagators. We refrain from presenting the explicit expressions for the corresponding renormalization constants and refer to \cite{Dedes:2017zog,Dekens:2019ept} instead, where these have been worked out in detail.

\subsection{Higgs, Electroweak and Yukawa Couplings}
\label{sub:HiggsEWYukawa}

The SMEFT operators $\Op[]{H D} = (H^\dagger D^\mu H)^* (H^\dagger D_\mu H) $ and $\Op[]{H \Box} = (H^\dagger H) \Box (H^\dagger H)$ contribute to the Higgs kinetic term after electroweak symmetry breaking. Similarly, $\Op[]{H} = (H^\dag H)^3$ gives corrections to the Higgs potential. As a result, after electroweak symmetry breaking, the Higgs-doublet needs to be redefined as
\begin{equation}
H = \frac{1}{ \sqrt{2}}
\begin{pmatrix}
0\\
(1+ c_H) h + \tilde v
\end{pmatrix} \,,
\end{equation}
where $c_H = (\Wc[]{H\Box} - \frac{1}{4} \Wc[]{HD}){\tilde{v}^2}$ and $\tilde v$ is the effective VEV of the Higgs.\footnote{A study where the VEV takes value in the complex projective plane $\mathbb{CP}^2$ rather than in $\mathbb{C}^2$, is provided in~\cite{Manton:2024eli}.} It is given by\footnote{We stress that {in our convention $\lambda$ is twice as large compared to} \cite{Jenkins:2013zja,Jenkins:2013wua,Alonso:2013hga}.}
\be \label{eq:eff-vev}
\tilde v = \sqrt{\frac{2\mu^2}{\lambda}} + \frac{3\mu^3\Wc[]{H}}{\sqrt{2} \lambda^{5/2}} ~=~{v + \frac{3 v^3 \Wc[]{H}}{4\lambda}.}
\ee
The effective Higgs mass in the SMEFT is then given by 
\be \label{eq:eff-Hmass}
\tilde m_H^2 = 2 \mu^2\left(1+2c_H-3\frac{\Wc[]{H}\mu^2}{\lambda^2}\right)\,.
\ee

The Yukawa matrices on the other hand receive corrections from the operators of class $f^2 H^3$. As a result the effective Yukawas and fermion mass matrices in the SMEFT can be defined as
\begin{eqnarray} \label{eq:eff-yukawas}
\widetilde Y_\psi  & = &  \frac{1}{\sqrt{2}} \left(Y_\psi (1+c_H)  - \frac{3}{2}
{\tilde v^2 \Wc[]{\psi H}}   \right)\,,  \quad
\widetilde M_\psi   =   \frac{\tilde v}{\sqrt{2}} \left(Y_\psi  - \frac{1}{2} {\tilde v^2 \Wc[]{\psi H}}   \right)\,,
\end{eqnarray}
with $\psi = u,d,e$ for the up-type, down-type quarks and leptons, respectively.

The gauge couplings and masses in the electroweak sector are redefined as
\bea \label{eq:eff-gauge}
\tilde g_I &=& g_I (1+ \Wc[]{HX} \tilde v^2)\,, \quad\quad \widetilde M_W^2 =  \frac{1}{4} \tilde g_2^2 \tilde v^2 \,, \\
\widetilde M_Z^2 &= & \frac{\tilde v^2}{4} (\tilde g_1^2 + \tilde g_2^2) (1+ \frac{1}{2} \Wc[]{HD} \tilde v^2) + {\frac{\tilde v^4}{2}\tilde g_1\tilde g_2\Wc[]{HWB} }\,,
\eea
where, $\Op[]{HX} = (H^\dagger H) X_{\mu\nu}  X^{\mu \nu}$. Since not all EW parameters are independent, one has to make a choice concerning the used input values. Typical input schemes are $\{\hat M_W,\hat M_Z,\hat G_F,\hat M_h, \hat \alpha_s \}$ \cite{Brivio:2017bnu,Brivio:2017btx} as well as $\{\hat \alpha_{ew},\hat M_Z,\hat G_F,\hat M_h, \hat \alpha_s\}$
\cite{Alonso:2013hga,Falkowski:2014tna,Brivio:2017vri}.

Clearly, only a few SMEFT operators entering the above relations can affect the dimension-four couplings directly. However, through the operator mixing effects, other operators can also affect dimension-four couplings indirectly. We stress that, in order to study the impact of BSM physics on any observable it is crucial to include these SMEFT corrections to the SM parameters, as the NP effects could enter via such an indirect route. Furthermore, at this stage, it is worth mentioning that, all of the above discussion applies only at the electroweak scale. However, in order to solve the SMEFT RGEs, the non-tilde parameters such as the gauge and Yukawa couplings along with the other SMEFT WCs have to be known at the input scale $\Lambda$, because these parameters are part of the SMEFT RGEs. We will come back to this issue in Section~\ref{sec:ren-gro-run}.

\subsection{Cabibbo-Kobayashi-Maskawa Matrix}

Last but not least we discuss the CKM parameters. These are extracted from the low-energy measurements of $B$ and $K$ meson observables, which in principle can be well described within the framework of the WET. However, one can easily connect the WET WCs to the corresponding SMEFT ones by matching the two theories at the EW scale. Following, this approach, GFFAV \cite{Descotes-Genon:2018foz} introduced a procedure for the treatment of the CKM parameters in the SMEFT. Their idea is analogous to the extraction of the Fermi-constant $G_F$ from the muon decay process $\mu \to e \nu_\mu \bar \nu_\e$, which is contaminated by the dimension-six SMEFT coefficients. But, one can absorb the SMEFT corrections to the muon-decay in the effective Fermi-constant $\widetilde G_F$, which is a function of $G_F$ and the SMEFT WCs entering the muon-decay. As a result, all other observables can be expressed in terms of $\widetilde G_F$, and the SMEFT WCs affecting the muon-decay indirectly also enter into these observables.

In the GFFAV scheme the CKM matrix is parametrized in terms of four Wolfenstein parameters $\lambda$, $A$, $\rho$, and $\eta$ \cite{Wolfenstein:1983yz} but in contrast to the latter paper, following \cite{Buras:1994ec}, it includes higher order corrections in $\lambda$ so that the CKM unitarity is satisfied including $\lambda^5$ terms.

In order to determine $\tilde\lambda$, $\tilde A$, $\tilde\rho$, and $\tilde\eta$ GFFAV choose as four input observables 
\be\label{4Q}
\Gamma(K\to \mu \nu_\mu)/\Gamma(\pi \to \mu \nu_\mu)\,, \quad  \Gamma(B\to \tau \nu_\tau)\,, \quad \Delta M_d\,, \quad \Delta M_s\,.
\ee
The respective relations between the tilde and non-tilde CKM elements entering these observables are given by \cite{Descotes-Genon:2018foz} ($q=d,s$)
\bea
\frac{|\widetilde V_{us}|^2}{|\widetilde V_{ud}|^2} &=& \frac{|V_{us}|^2}{|V_{ud}|^2} (1+ \Delta_{K/\pi})\,, \\
|\widetilde V_{ub} |^2 &=& | V_{ub} |^2  (1+ \Delta_{B\tau 2})\,, \\
|\widetilde V_{tb} \widetilde V_{tq}|^2 &=& | V_{tb} V_{tq}|^2   (	1+ \Delta_{\Delta M_q})\,,
\eea
where $\Delta_{K/\pi}$, $\Delta_{B \tau 2}$, and $\Delta_{\Delta M_q}$ are functions of the SMEFT WCs. One finds then
\bea
\tilde \lambda  &=& \lambda + \delta \lambda =  0.22537\pm 0.00046 \,, \quad
\tilde A =  A + \delta A   = 0.828 \pm 0.021  \,, \\
\tilde \rho &=& \rho + \delta \rho = 0.194 \pm 0.024 \,, \qquad \quad
\tilde \eta = {\eta} + \delta \eta  = 0.391 \pm 0.048.
\eea
Here, the shifts $\delta \lambda$, $\delta A$, $\delta \rho$ and $\delta \eta$ are functions of the NP parameters $\Delta_{K/\pi}$, $\Delta_{B \tau 2}$, $\Delta_{ \Delta M_q}$. The explicit expressions for them can be found in \cite{Descotes-Genon:2018foz}.

Correspondingly, one finds the following numerical constraints on the tilde CKM elements
\bea
|\tilde V_{us}/\tilde V_{ud}| &=&  0.23131\pm 0.00050 \,, \quad
|\tilde V_{ub}| = 0.00425 \pm 0.00049 \,, \\
|\tilde V_{tb}\tilde V_{td}| &=& 0.00851 \pm 0.00025 \,, \quad
|\tilde V_{tb}\tilde V_{ts}| = 0.0414 \pm 0.0010\, .
\eea

Any other observable which depends on the CKM parameters can be expressed in terms of $\tilde \lambda$, $\tilde A$, $\tilde \rho$, $\tilde \eta$ and $\delta \lambda$, $\delta A$, $\delta \rho$, $\delta \eta$. Hence, the NP operators entering these shifts get distributed over a wide range of observables. A similar treatment of the CKM in the SMEFT can be found in \cite{Aebischer:2018iyb}, which has been implemented in the Python package {\tt smelli}.

\subsection{Strong coupling constant}

As for the EW gauge couplings in \ref{eq:eff-gauge}, the strong coupling constant is shifted within the SMEFT:
\begin{equation}
\tilde g_s =g_s(1+\Wc[]{HG} \tilde v^2)\,.
\end{equation}
Furthermore, a similar discussion like for the CKM matrix has been presented recently for the strong coupling constant $\alpha_s$ in \cite{Trott:2023jrw}. It has been found, that shifts from higher-dimensional operators to this parameter can be neglected and that the lattice-averaged value \cite{FlavourLatticeAveragingGroup:2019iem} can be used in global SMEFT analyses.

\section{SMEFT Renormalization Group Evolution}
\label{sec:ren-gro-run}
\subsection{Preliminaries}
The important point stressed already in the past and in particular emphasized in \cite{Alonso:2013hga} is that RG evolution between the NP scale $\Lambda$ and the electroweak scale involves operator mixing and the pattern of deviations from SM expectations observed at the electroweak scale and low-energy scales can differ significantly from the pattern of WCs at the NP scale $\Lambda$. Here we just quote one example known already for decades in the context of QCD RG evolution. Tiny effects of right-handed currents originating at a very high scale $\Lambda$, when accompanied by left-handed currents, also generated by NP or present in the SM, can in $\Delta F=2$ observables cause very large effects in flavour observables. This is due to the presence of left-right operators with enhanced WCs through RG evolution to hadronic scales and chirally enhanced hadronic matrix elements, in particular in the $K$-meson system. 

However, the point made in \cite{Alonso:2013hga} and in subsequent papers \cite{Feruglio:2016gvd,Bobeth:2016llm,Bobeth:2017xry,Feruglio:2017rjo,Gonzalez-Alonso:2017iyc,Buttazzo:2017ixm,Kumar:2018kmr,Aebischer:2018iyb,Silvestrini:2018dos} is not related to QCD but in particular to top Yukawa couplings. Even in the absence of left-handed currents generated by NP, the presence of right-handed currents in the presence of Yukawa couplings can generate in the process of RG evolution left-right operators at the electroweak scale, because left-handed currents are already present in the SM. Moreover, this effect is further enhanced through QCD RG evolution to low energy scales as stated above. This underlines the importance of RG studies in the search for NP far beyond the LHC scales. As stressed in particular in \cite{Silvestrini:2018dos} the bounds from $\Delta F=2$ processes can affect in turn $\Delta F=1$ and $\Delta F=0$ processes. In this section we will recall the basic equations governing the RG evolution which will play a crucial role in the rest of the review.

\subsection{Renormalization Group Evolution}
EFTs are renormalizable order by order in the power series expansion. However, also the loop order has to be taken into account, in order to correctly determine the power counting of the EFT, as was pointed out in \cite{Buchalla:2022vjp}.\footnote{A related discussion on the concept of minimal coupling can be found in \cite{Jenkins:2013fya}.} In the process of renormalization, the parameters of the theory such as those of dim-4 as well as the WCs become scale-dependent quantities. This is known as the renormalization group (RG) running and can be taken into account using RG equations (RGEs)
\be \label{eq:rge}
\frac{\text{d} C_i(\mu)}{\text{d} \ln \mu} = {(\hat\gamma_C)}_{ij} C_j(\mu)\,,
\ee
where {$\hat\gamma_C$} is the ADM for the WCs which can be related to ADMs for the operators $\hat\gamma$ as $\hat\gamma_C = \hat\gamma^T$. In the process of RG running different EFT operators can mix with each other, which means in general that $\hat\gamma$ contains off-diagonal terms. In terms of renormalization, mixing of an operator $A$ into $B$ means, that renormalizing the operator $A$ requires a counter-term proportional to operator $B$. Note that the mixing of $A$ with $B$ does not guarantee operator mixing of $B$ with $A$. In other words, $\hat\gamma_C$ in \eqref{eq:rge} is often an asymmetric matrix. 

For the case of SMEFT, the full-fledged one-loop renormalization is known at the dimension-six level \cite{Jenkins:2013zja, Jenkins:2013wua, Alonso:2013hga}.\footnote{Visit \url{http://einstein.ucsd.edu/} for erratum.} Here, by complete, we mean the full dependence of $\hat\gamma_C^{\rm SMEFT}$ on the strong, weak, Yukawa couplings as well as the quartic Higgs coupling 
has been kept in these references. That is
\be
\hat\gamma_C^{\rm SMEFT}  = \frac{g_i^2}{16\pi^2} \hat \gamma_C^{(i)} + \frac{y_t^2}{16\pi^2} \hat \gamma_C^{(t)} 
+  \frac{\lambda^2}{16\pi^2} \hat \gamma_C^{(\lambda)} +\dots\,,
\ee
where $g_i$ for $i=1,2,3$ are the $\text{U(1)}_Y$, $\text{SU(2)}_L$ and $\text{SU(3)}_C$ gauge couplings, respectively. The $y_t$ is top-Yukawa and $\lambda$ the quartic coupling of the Higgs-doublet. The dots represent the terms depending on the other Yukawas and the ones which depend on more than one type of coupling. Generally, the exact dependence of the ADM on the coupling constants can be determined using naive dimensional analysis (NDA) \cite{Jenkins:2013sda}.

\begin{figure}[htb]%
\centering
\includegraphics[width=0.7\textwidth]{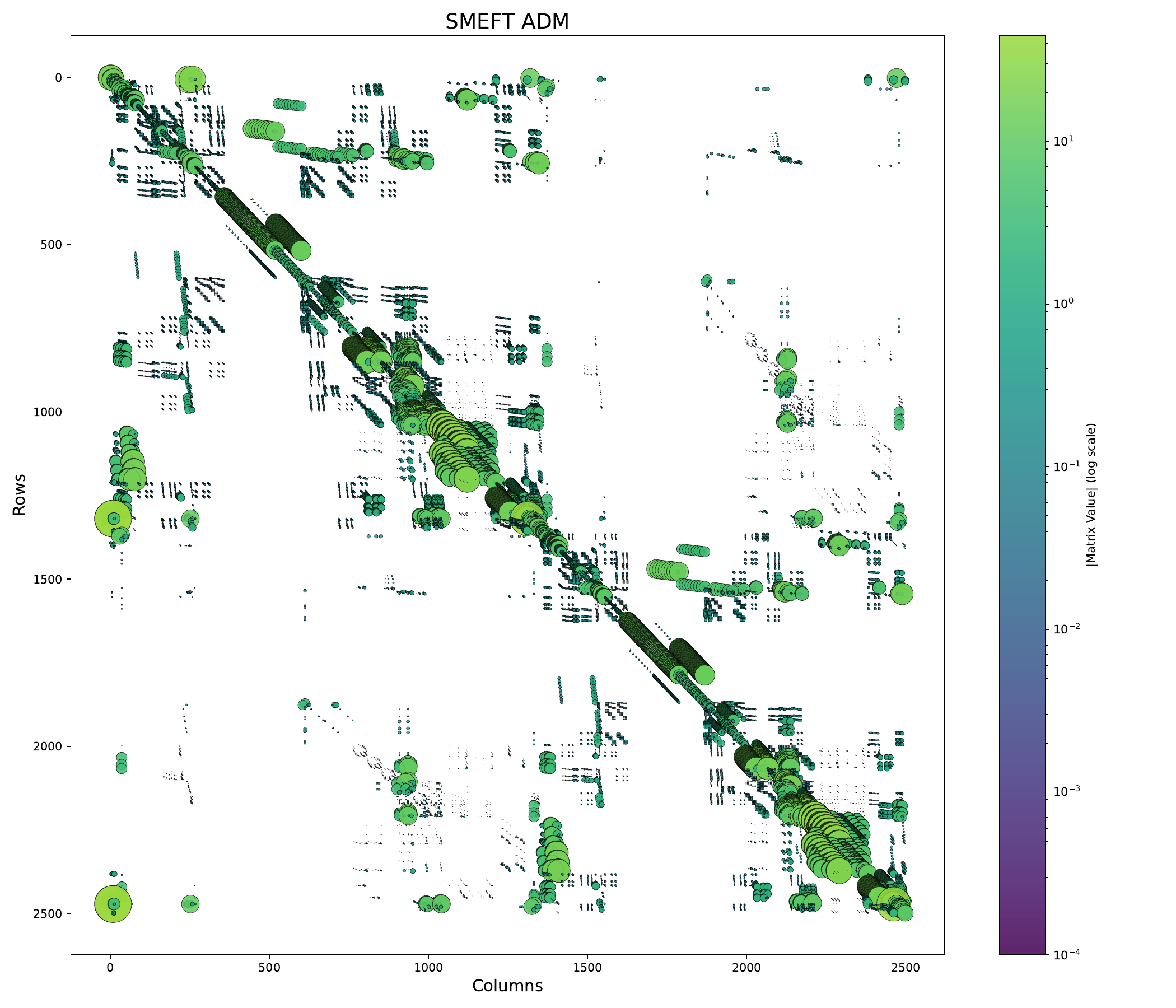}
\caption{ Here we show the non-zero entries of the ADM.}%
\label{fig:ADMplot}%
\end{figure}

The ADM $\hat\gamma_C^{\rm SMEFT}$ plays a crucial role in the physical implications of the SMEFT and we will look at it in some detail in Part II of our review. But already at this stage it is useful to have its grand picture to realize how involved it is. To this end we show in Fig.~\ref{fig:ADMplot} an artistic modern view of this matrix to indicate that it has many oases in a big desert. Some insight into these oases and the desert between them will be gained first in Section~\ref{A-F} and in more detail in Part II of our review.

This is in contrast to the WET, where only $\alpha_{\rm em}$ and $\alpha_s$ are present and the WET ADM $\hat\gamma_C^{\rm WET}$ depends on these two couplings
\be
{\hat\gamma_C^{\rm WET}}  = \frac{e^2}{16\pi^2} \hat \gamma_C^{\rm (QED)} + \frac{g_s^2}{16\pi^2} \hat \gamma_C^{\rm (QCD)} .
\ee
The complete list of WET ADMs at the one-loop level is given in \cite{Jenkins:2017dyc}. Partial results both in the SMEFT and the WET were known before from the SM and concrete beyond SM calculations in specific models as reviewed in \cite{Buchalla:1995vs,Buras:2011we}.

The analytic solution of the RG equations in \ref{eq:rge}, which form a coupled system of differential equations, is given as
\be
{\vec C(\mu) = \exp \left [ \int_{\text{ln} (\Lambda)}^{\text{ln} (\mu)} {\hat\gamma_C(\tilde \mu)} ~  
\text{d}~ \text{ln} (\tilde \mu)   \right ]\vec C(\Lambda)\,.}
\ee
The RGEs can however also be solved numerically, using for instance the codes {\tt Dsixtools} \cite{Celis:2017hod} or {\tt wilson} \cite{Aebischer:2018bkb}. Alternatively, analytical approximations such as the LO approximation can be used, for which the solution can be written as 
\be \label{eq:LO}
{\vec C(\mu) = \exp\left[- \hat \gamma_C \ln\left(\frac{\Lambda}{\mu}\right)\right] \vec C(\Lambda)\,,}
\ee
with the one-loop anomalous dimension $\hat\gamma_C$. This solution sums logarithms to all orders of perturbation theory. The first leading-log approximation (LLA) is obtained by expanding the exponential in \eqref{eq:LO} and keeping only the leading logarithm
\be \label{eq:LLA}
{\vec C(\mu) = \left [\mathbb{1} - \hat\gamma_C \ln\left (\frac{\Lambda}{\mu}  \right )  
    \right ] \vec C(\Lambda)\,.}
\ee
One of the important features of SMEFT running is that even the operators from different classes as described in Sec.~\ref{sub:warsaw-basis} can mix among each other, thereby rendering the operator mixing patterns very complicated. This often obscures the low-energy predictions of a UV completion. Secondly, the operator mixing resulting from Yukawa interactions in the SMEFT can cause mixing among different flavours of a single operator, as well as mixing among different operator types. Last but not least, the RG running of dim-4 interactions such as Yukawa couplings, render the choice of flavour basis scale dependent. This leads to back-rotation effects, which are discussed in Sec.~\ref{sub:back-rot}.

\subsection{RGEs of SM Parameters}
\label{sec:smrges}

The RGEs for the SM parameters in terms of the $\beta$-functions read
\be \label{eq:smbeta}
16\pi^2 \frac{\text{d} c_i (\mu) }{\text{d} \ln (\mu)} = \beta_i\,,
\ee
where the $c_i$ denote the SM couplings. The $\beta$-functions are given by\footnote{{The full expressions can be found in App. B of \cite{Celis:2017hod}.}}
\bea
\beta_{g_1} &=& \frac{41}{6} g_1^3+\dots\,, \quad \beta_{g_2} = -\frac{19}{6} g_2^3+\dots\,, \quad
\beta_{g_3} = -7 g_3^2+\dots\,, \\
\beta_{Y_u}  &=& \frac{3}{2} \left (Y_u Y_u^\dagger  - Y_d Y_d^\dagger   \right ) Y_u+\dots\,, 
\quad  \beta_{Y_d}  = \frac{3}{2} \left (Y_d Y_d^\dagger  - Y_u Y_u^\dagger   \right ) Y_d+\dots\,,
\label{eq:YuYdrun} \\
\beta_{Y_e}  &=& \frac{3}{2} Y_e Y_e^\dagger  Y_e  + \dots\,, \qquad {\beta_{\mu^2} = 6\mu^2\lambda+\dots\,,}\qquad {\beta_{\lambda} = 12\lambda^2+\dots\,,}\label{eq:beta_Ye}
\eea
where we have shown only one-loop contributions and set $f=6$. In order to solve the SMEFT RGEs \eqref{eq:rge} the SM RGEs have to be solved simultaneously. Observe that the running of $Y_d(Y_u)$ involves not only $Y_d(Y_u)$ but also $Y_u(Y_d)$. This makes the choice of flavour basis in the SMEFT scale-dependent. Next, we will discuss this issue in detail.

{\boldmath
\subsection{The SM Parameters at the Scale $\Lambda$}
}
\noindent
The ADMs constituting the RGEs \eqref{eq:rge} are functions of the SM parameters such as gauge couplings, Yukawa couplings and Higgs potential parameters, which have to be simultaneously evolved with the SMEFT WCs. Typically, the initial conditions of the WCs are given at the scale $\Lambda$. Hence, in order to solve the set of coupled differential equations, the SM parameters are also needed at $\Lambda$. The latter are however extracted from low-energy observables, and hence need to be evolved up to the high scale. This evolution requires however the knowledge of the SMEFT WCs at the low scale, because the SM beta-functions also depend on SMEFT WCs. This makes the evolution of these parameters a convoluted task. Therefore, certain assumptions are required to get around this issue. A practicable solution to this problem was proposed in \cite{Aebischer:2018bkb}, which is outlined in the following. Specifically, the SM parameters required for the RG evolution are
\bea
\text{Gauge Couplings}  &:&  g_1(\Lambda),\,\  g_2(\Lambda), \,\  g_3(\Lambda) \,, \nn  \\
\text{Yukawa Matrices} &:&  Y_u(\Lambda),\,\   Y_d(\Lambda), \,\  Y_e(\Lambda) \,,  \\
\text{Higgs Potential}  &: & \mu(\Lambda), \,\  \lambda(\Lambda).  \nn
\eea
Note that, the Lagrangian parameters (non-tilde) enter the ADMs, but effective SM parameters (tilde) are measured from low-energy experiments. Setting the effective parameters equal to the $\overline{\text{MS}}$ parameters (see Table 1 of \cite{Aebischer:2018bkb}) at the EW scale we can get the Lagrangian parameters by inverting \eqref{eq:eff-gauge} and \eqref{eq:eff-yukawas}
\be \label{eq:lag-gauge}
g_I (\muEW)=   \tilde g_I (\muEW) (1- C_{HX} (\muEW)  {\tilde v}^2)\,, \quad \forall \, I\,,
\ee 
assuming $C_{HX} (\muEW) = C_{HX}(\Lambda),\, X= B, W, G$. Likewise, the non-tilde Yukawa matrices can be obtained from
\begin{align} \label{eq:lag-yukawa}
\widetilde M_u(\muEW) &=  \hat V^\dagger~ \text{diag}(m_u(\muEW),m_c(\muEW),m_t(\muEW)) \,, \nn \\
\widetilde M_d(\muEW) &=  \text{diag}(m_d(\muEW),m_s(\muEW),m_b(\muEW)) \,, \\
\widetilde M_e(\muEW) &=  \text{diag}(m_e(\muEW),m_\mu(\muEW),m_\tau(\muEW)) \nn. 
\end{align}
Here, we have used the down-basis and $m_i$ are the $\overline{\text{MS}}$ running quark masses at the EW scale. The $\hat V$ is the CKM matrix which is extracted from low-energy experiments as discussed in the previous section. Then using \eqref{eq:eff-yukawas}, the non-tilde Yukawa matrices are given by 
\be\label{eq:Ya}
{Y_\psi(\muEW) = \frac{\sqrt{2} \widetilde M _\psi(\muEW)}{\tilde v} +  \frac{1}{2}\tilde v^2 C_{\psi H}(\muEW)\,, \quad \psi= u,d,e.}
\ee
Finally, the Higgs sector parameters $\lambda (\muEW)$ and $\mu^2(\muEW)$ can be expressed in terms of the effective Higgs mass and the VEV by inverting \eqref{eq:eff-vev} and \eqref{eq:eff-Hmass}, respectively.

Now, since the WCs entering in \eqref{eq:lag-gauge}-\eqref{eq:Ya} are not known a priori, it is assumed that $C(\muEW)=C(\Lambda)$ for the purpose of obtaining the non-tilde parameters. Further, it is assumed that the beta functions do not depend on the SMEFT WCs. With these two assumptions, we can evolve the weak scale non-tilde parameters up to the scale $\Lambda$ using the RGEs \eqref{eq:smbeta}-{\eqref{eq:beta_Ye}}. Once we have the SM parameters at $\Lambda$, we can simultaneously evolve them with the WCs down to the weak scale. Finally, we need to make sure that the weak scale SM parameters agree with the measured values. For that purpose, the evolution can be iteratively performed between $\muEW$ and $\Lambda$ until this is achieved.

Considering for instance the evolution of the down-basis Yukawa matrices from $\Lambda=3$ TeV down to the EW scale one finds, using the singular value decomposition, the down-type rotation matrices defined in \eqref{equ:LYukRdiag1} \cite{Aebischer:2020lsx}
\begin{align}\label{eq:UdRnum}
\hat U_{d_R}=&\left [
\begin{array}{ccc}
-0.93\, +0.37 i & 1.6\cdot 10^{-6}\, + 2.5 \cdot 10^{-8}\, i
& -7.2 \cdot 10^{-7} \\
-1.1\cdot 10^{-6}\,+ 1.1\cdot 10^{-6}\, i & -0.93\,+ 0.37 i &
6.2\cdot 10^{-5}\, -2.6\cdot 10^{-5}\, i \\
5.2\cdot 10^{-7}\, -5 \cdot 10^{-7} \,i& -6.3 \cdot 10^{-5}\, + 2.4 \cdot 10^{-5}\,i &
-0.93\, +0.37 i \\
\end{array}
\right ] \,,\\\label{eq:UdLnum}
\hat U_{d_L}=&\left [
\begin{array}{ccc}
-0.93\, +0.37 i & 1.6 \cdot 10^{-5}\, + 2.5 \cdot 10^{-7}\, i & -3.8 \cdot 10^{-4} \\
-1.2\cdot 10^{-5}\, +1.1\cdot 10^{-5}\, i & -0.93\, +0.37 i & 1.6\cdot 10^{-3}\, -6.7\cdot 10^{-4}\,i  \\
2.7\cdot 10^{-4}\, -2.6 \cdot 10^{-4}\,i & -1.6\cdot 10^{-3}\, + 6.1 \cdot 10^{-4}\,i& -0.93\, +0.37\,i \\
\end{array}
\right ] \,.
\end{align}
The rotation matrix $\hat U_{u_L}$ is fixed through the relation $\hat V = \hat U_{u_L}^\dag \hat U_{d_L}$ in (\ref{eq:CKM}) and $\hat U_{e_L} = \hat U_{e_R} = \mathbb{1}$, assuming there are no right-handed neutrinos.

\subsection{Back-Rotation}
\label{sub:back-rot}

Below the EW scale, left- and right-handed quark fields can be transformed separately, using different unitary transformations \eqref{equ:LYukRdiag1}. However, for solving the SMEFT RGEs the Yukawa matrices have to be specified above the EW scale. Above $\muEW$ only five unitary transformations can be performed, corresponding to the five fermion representations of the full SM gauge group. We denote them in the following way:
\begin{equation}\label{eq:rot}
q'=\hat U^{\text{SMEFT}}_{q} q\,,\quad u'=\hat U^{\text{SMEFT}}_{u_R} u\,,\quad d'
=\hat U^{\text{SMEFT}}_{d_R} d\,,\quad \ell'=\hat U^{\text{SMEFT}}_{\ell} \ell\,,\quad e'=\hat U^{\text{SMEFT}}_{e} e\,.\quad
\end{equation}
Having only five rotation matrices at hand allows to diagonalize just two of the three Yukawa matrices above the EW scale. 

If we are interested in flavour observables in the down-sector, then we adopt the Warsaw-down basis defined in \cite{Aebischer:2017ugx}\footnote{See also \cite{Dedes:2017zog} for a generalization to the full Warsaw basis.}. In this basis, the down-type and lepton Yukawa matrices are diagonal, whereas the up-type Yukawa matrix is rotated by the CKM matrix $\hat V$. At the NP scale $\Lambda$ one has
\begin{equation}
\hat Y_d(\Lambda)={\rm diag}(y_d,y_s,y_b)\,,\quad \hat Y_u(\Lambda)=\hat V^\dag{\rm diag}(y_u,y_c,y_t)\,,\quad \hat Y_e(\Lambda)={\rm diag}(y_e,y_\mu,y_\tau)\,.
\end{equation}
However, this simple form of the Yukawa matrices holds only at a single scale, $\Lambda$ in this case, and is broken once RG evolution is considered. This is because running effects generate off-diagonal entries in the Yukawa matrices and therefore the theory parameters are not given in the Warsaw-down basis anymore. The generation of the off-diagonal entries can be understood from the first lines of the Yukawa $\beta$-functions in \eqref{eq:YuYdrun}.

Considering $\hat Y_d$, the leading term of its $\beta$-function \eqref{eq:YuYdrun} involves the up-quark Yukawa matrix $\hat Y_u$, which is non-diagonal in the down-basis. Indeed, at the EW scale $\muEW$ in the first LL approximation \eqref{eq:LLA} and keeping only the dominant $y_t$-contribution one finds:

\begin{equation}\label{eq:YdEW}
\hat Y_d(\muEW) = \hat Y_d(\Lambda)  -\delta \hat Y_d\frac{3 y_t^2}{32\pi^2}
\ln \left (\frac{\muEW}{\Lambda} \right )  + ...\,,
\end{equation}
where
\begin{align}
\delta \hat Y_d =
&\left(
\begin{array}{ccc}
y_d \lambda^{dd}_t &
y_s \lambda^{ds}_t  &
y_b \lambda^{db}_t  \\
y_d \lambda^{sd}_t &
y_s \lambda^{ss}_t &
y_b \lambda^{sb}_t \\
y_d \lambda^{bd}_t  &
y_s \lambda^{bs}_t &
y_b \lambda^{bb}_t  \\
\end{array}
\right)\,,\quad \lambda_t^{ij} = V_{ti}^* V_{tj}\,.
\end{align}
Note that for simplicity we have not shown the SMEFT contribution to the RG running of $\hat Y_d$. As shown in \eqref{eq:YdEW}, the down-type Yukawa matrix is off-diagonal at the EW scale. However, in order to examine physical processes, a basis change to the mass basis has to be performed \eqref{equ:LYukRdiag1}.

This concludes the evolution of the down-type Yukawa matrix from the NP scale down to $\muEW$. In short, $\hat Y_d$ started by construction from a diagonal form at the high scale $\Lambda$, became off-diagonal at the EW scale through RGE effects, and is finally diagonalized at the EW scale. In \cite{Aebischer:2020lsx} the latter diagonalization was coined to be {\it back-rotation} to the down-basis.
 
It should be noted that the effect of the down-type Yukawa matrix on the evolution of the up-Yukawa matrix in the up-basis is much smaller because of small Yukawa couplings of down-quarks. This implies that the running of WCs in the up-basis is much simpler than in the down-basis as stressed for example in Sec.~\ref{class1} in Part II of our review.

\subsection{RG Evolution in SMEFT}

\begin{figure}[tb]
\begin{center}
 \includegraphics[clip, trim=0.1cm 9cm -1.0cm 10cm,width=1.1\textwidth]{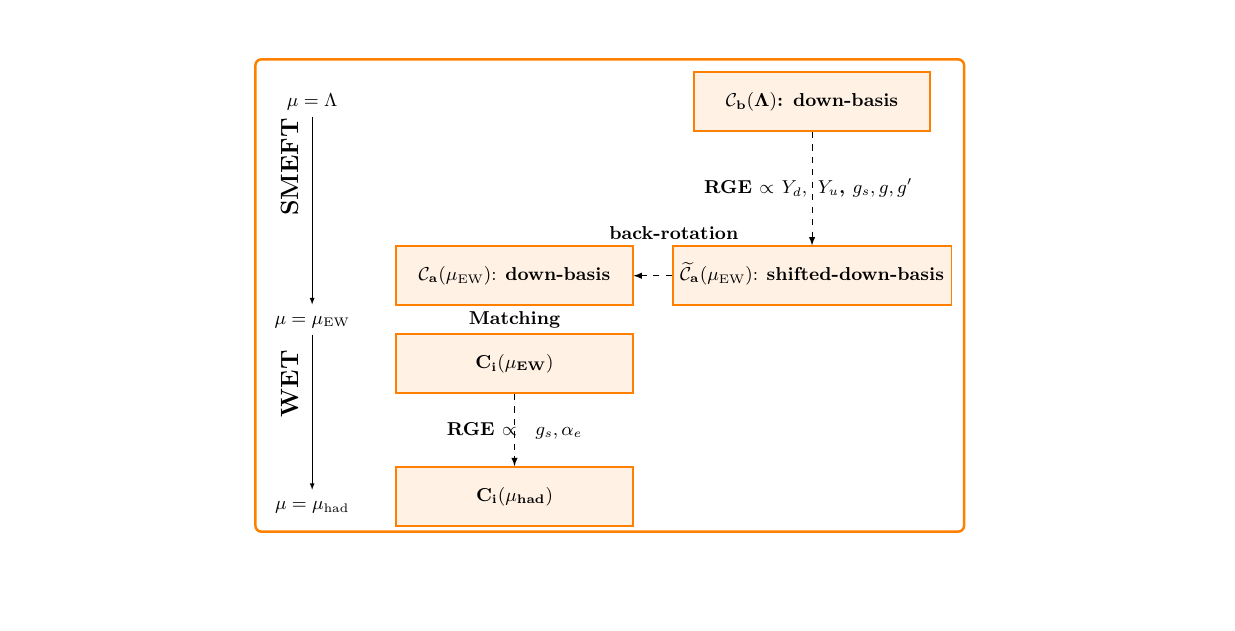}
 \vspace{-1.4cm}
\caption{\small The RG running of the down-basis SMEFT Wilson coefficients from the new physics scale $\Lambda$ to the EW scale 
$\muEW$ is shown. Down-type Yukawa running generates a tilde-basis ($\widetilde{\cal{C}}_a$), which has to be rotated back to the down-basis ($\mathcal{C}_a$) at the EW scale. Subsequently, the Wilson coefficients are matched onto the WET and further evolved down to lower scales {($\muLow$)} to estimate flavour observables.}
\label{fig:runrot}
\end{center}
\end{figure}

We are now in a position to discuss the complete procedure of the RG evolution of the SMEFT WCs together with the Yukawa matrices. The evolution of the Wilson coefficients from $\Lambda$ down to the EW scale proceeds in two steps which are shown in Fig.~\ref{fig:runrot} and described in the following:

{\bf Step 1:} The Wilson coefficients are evolved from the high scale $\Lambda$ down to the EW scale $\muEW$ using the full SMEFT RGEs. As an example, considering WCs with two flavour indices such as $\wc[(1)]{Hq}{ij}$ in the first LLA \eqref{eq:LLA}, one finds their evolution to be
\begin{equation}\label{eq:wcEWLL}
\big[{\widetilde{{C}}}_a(\muEW)\big]_{i j } = \big[{ C}_a(\Lambda)\big]_{i j}
+\frac{(\beta_{ab})^{ijkl}}{16\pi^2}\ln{\left(\frac{\muEW}{\Lambda}\right)}
\big[{ C}_b(\Lambda)\big]_{kl}\,,
\end{equation}
where $a,b$ label different Wilson coefficients, $i,j,k,l$ are flavour indices and $\beta$ denotes the $\beta$-function of the corresponding Wilson coefficient. A similar expression exists for four-fermi operators with four flavour indices. The tilde ($\sim$) on the left-hand side of \eqref{eq:wcEWLL} denotes the fact that Wilson coefficients at the EW scale are not in the down-basis anymore, but in a shifted-down basis which we will call the {\it tilde-basis} due to simultaneous evolution of the $Y_d$ according to \eqref{eq:YdEW}. As explained in the previous subsection, this is due to the off-diagonal Yukawa elements generated through the running from $\Lambda$ to $\muEW$. If we are interested in flavour observables in the down-sector, the next step would consist of changing from the tilde-basis $\widetilde{{C}}_i$ back to the down-basis ${C}_i$:

{\bf Step 2:} At the EW scale, the Wilson coefficients $\widetilde{{ C}}_a(\muEW)$ are rotated back to the down-basis using
\begin{equation}\label{eq:backrot}
\big[{ C}_a(\muEW)\big]_{ij} = \hat U^\dag_{ik} 
\big[\widetilde{{ C}}_a(\muEW)\big]_{kl} \hat U_{lj}\,,
\end{equation}
where $i,j,k,l$ are flavour indices and $\hat U_{ij}$ denote the rotation matrices in \eqref{equ:LYukRdiag1}. This {\it back-rotation} to the down-basis is key for the study of down-type flavour observables, since it transforms the involved fields into mass eigenstates. 

It is important to note that for the SMEFT running due to the gauge couplings or the running below the EW scale in WET no such back-rotation is necessary, since off-diagonal Yukawa elements can not be generated through QCD or QED interactions. We refer to the original \cite{Aebischer:2020lsx} for the impact of the back-rotation in the context of flavour processes within the SMEFT framework.

This completes the discussion of RG evolution within the SMEFT in the LO approximation. However, in order to increase the precision and in particular to match the WCs at $\muLow$ the hadronic matrix elements, calculated usually by Lattice QCD (LQCD), one has to go beyond this approximation. Indeed, LQCD calculations are done in a renormalization scheme that differs from the $\overline{\text{MS}}$ used for the calculations of the WCs. As renormalization scheme dependence enters first at the NLO level, in order to combine LQCD results for hadronic matrix elements with the WCs calculated in the $\overline{\text{MS}}$ scheme one has to go beyond the leading order which we will do in the next section.

In the following subsections we will summarize important references concerning the matching and running in the SMEFT and WET.
   
\subsection{RG Evolution at dim-5}

The one-loop RG evolution of the Weinberg operator was computed in \cite{Chankowski:1993tx,Babu:1993qv,Antusch:2001ck} and the complete two-loop renormalization was calculated recently in \cite{Ibarra:2024tpt}. Double-insertions of the Weinberg operator, mixing into dimension-six operators are discussed in \cite{Davidson:2018zuo}. In the WET at dim-5, two-loop anomalous dimensions were recently calculated in \cite{Naterop:2025nqv}. 

\subsection{RG Evolution and Matching in the SMEFT and WET at dim-6}\label{RGMA}
It will be evident from the next section that in order to increase the precision of the theoretical calculations one has to go beyond the LO of the RG improved perturbation theory and include in addition to one-loop ADMs and tree-level matching of the SMEFT onto WET also two-loop ADMs both in the SMEFT and the WET. At this NLO level also the one-loop matching between the SMEFT and the WET has to be included. It allows to cancel various unphysical renormalization scheme and renormalization scale dependences in physical amplitudes. Here we summarize what is presently known in this context.

\subsubsection{Matching of UV onto SMEFT at $\Lambda$}
In \cite{deBlas:2017xtg} the tree-level dictionary between general UV completions containing heavy scalars, fermions and vector bosons was presented. It has been generalized recently to the one-loop level for heavy scalars and fermions \cite{Guedes:2023azv,Guedes:2024vuf}. More about it can be found in Sections~\ref{SMEFTtools} and \ref{sec:5}. In particular in Table~\ref{TAB3} we list the papers in which the matchings in various NP scenarios on the SMEFT have been performed. Moreover in Sec.~\ref{sec:5} examples of explicit expressions for the WCs at $\Lambda$ in models with heavy gauge bosons, scalars, vector-like quarks and leptoquarks are presented.

\subsubsection{RG SMEFT Running from $\Lambda$ down to $\muEW$}
\begin{itemize}
\item
The one-loop ADMs relevant for the RG in the SMEFT at dimension six are known \cite{Jenkins:2013zja,Jenkins:2013wua,Alonso:2013hga}. They include the effects of the gauge couplings \cite{Alonso:2013hga}, of the Yukawa couplings \cite{Jenkins:2013wua} and of the $\lambda$-dependent terms \cite{Jenkins:2013zja}. The results for bosonic operators have been reproduced using super-heat-kernel expansion in \cite{Buchalla:2019wsc}.\footnote{Recently, the complete one-loop RGEs for general bosonic theories were derived in \cite{Fonseca:2025zjb,Misiak:2025xzq,Aebischer:2025zxg} and higher-order running of scalar theories and their connection to conformal field theories was studied in \cite{Henriksson:2025hwi}.} Moreover, the one-loop renormalization of the SMEFT using amplitude methods is illustrated in \cite{Baratella:2020lzz}. In this framework selection rules for flavour invariant blocks within the dim-6 SMEFT RGE were derived in \cite{Machado:2022ozb}. 
\item  
The two-loop QCD ADMs relevant for RG evolutions of $\Delta F=2$ transitions in SMEFT are also known \cite{Aebischer:2022anv} and the ones for $\Delta F=1$ transitions should be known soon. In this context master formulae for $\Delta F=2$ amplitudes both in the SMEFT and the WET have been presented in \cite{Aebischer:2020dsw} and illustrated with tree-level exchanges of heavy gauge bosons $(Z^\prime,~G^\prime)$ and corresponding heavy scalars.\footnote{A NLO RG analysis for purely scalar leptoquarks has been performed in \cite{Banik:2023ogi}.}
\item
On-shell methods for the computation of the one-loop and two-loop ADMs in the SMEFT have been developed in \cite{EliasMiro:2020tdv,Bern:2020ikv,Baratella:2022nog,Machado:2022ozb,Baratella:2020dvw}. They allow a good insight into the flavour structure of the ADMs. In particular in \cite{Bern:2020ikv} parts of the one-loop SMEFT RGEs as well as several zeros in the two-loop SMEFT ADM have been computed using these methods. The results were obtained for general EFTs and the authors show how to find a convenient renormalization scheme such that parts in the two-loop ADM vanish.
\item
In \cite{EliasMiro:2020tdv} several two-loop mixing contributions among various operators have been computed, again using on-shell methods. Results are given in general terms for the mixing of $n$-point into $(n-1)$-point functions for $n=4,5,6$.  
\item
The two-loop contributions of the dipole and triple gauge operators to the gluon and quark field renormalization constants were recently computed in \cite{Duhr:2025zqw}.
\item
The four-top quark operator contributions to the two-loop renormalization constants of the fermion and gluon fields together with a discussion on scheme dependence can be found in \cite{DiNoi:2025arz}.
\item
The two-loop top-quark Yukawa contributions to effective Higgs-gluon couplings were recently computed in \cite{DiNoi:2024ajj}.
\item 
The two-loop Yukawa-induced running of Higgs-gluon operators was presented in \cite{DiNoi:2025tka}.
\item 
Recently, the complete two-loop running of baryon number violating dim-6 operators was presented in \cite{Banik:2025wpi}.
\item 
The two-loop and even part of the three-loop running of the gluonic Weinberg operator was performed in \cite{deVries:2019nsu}.
\item
In the context of electric dipole moments the authors of \cite{Panico:2018hal} have computed parts of the two-loop RGEs for EW dipole operators. The results include the EW mixing of semi-leptonic, four-lepton as well as Yukawa-type operators into the EW dipoles.
\item
The scalar sector of the dim-6 SMEFT operators was renormalized at the two-loop level in \cite{Jenkins:2023rtg,Jenkins:2023bls}.
\item
The complete bosonic sector in the SMEFT was renormalized for the first time, using functional methods \cite{Born:2024mgz}.\footnote{A recent guide on functional methods can be found in \cite{Fuentes-Martin:2024agf}.}
\item
Symmetry-restoring finite counter terms in the HV scheme were presented in \cite{Fuentes-Martin:2025meq}, using functional methods.
\item
The two-loop renormaliztion of the quark and gluon fields was performed in \cite{Duhr:2025yor}.
\end{itemize}

\subsubsection{Matching of SMEFT onto WET at $\muEW$}
The matching between the SMEFT and WET is known by now both at tree-level \cite{Jenkins:2017jig} and one-loop level \cite{Dekens:2019ept}. Previous partial results can be found, for example, in \cite{Aebischer:2015fzz,Bobeth:2016llm,Bobeth:2017xry,Hurth:2019ula, Endo:2018gdn,Grzadkowski:2008mf}.

The one- and two-loop matching between the SMEFT and WET, focusing on the contributions of third-generation four-quark operators have been presented recently in \cite{Haisch:2024wnw}. They are relevant for the study of electroweak precision measurements and flavour physics observables.

\subsubsection{RG WET Running from $\muEW$ down to $\muLow$}
The one-loop ADMs relevant for the RG in the WET \cite{Jenkins:2017dyc,Aebischer:2017gaw} are known. A discussion on flavour and CP invariant subsets of the RGEs for the four-fermion sector can be found in \cite{Renner:2025cmd}. The two-loop QCD ADMs relevant for RG evolutions for both $\Delta F=2$ and $\Delta F=1$ non-leptonic transitions in WET are also known \cite{Buras:2000if,Aebischer:2021raf,Aebischer:2020dsw}. The complete two-loop renormalization of the four-fermion WET operators was recently presented for the first time in \cite{Aebischer:2025hsx}. The RGEs in the WET for dim-6 baryon-number-violating operators at two loops have been presented in the HV scheme in \cite{Naterop:2025lzc} and for dim-6 operators that conserve baryon number in \cite{Naterop:2025cwg}.

\subsection{RG Evolution at dim-7}
The one-loop ADM for the six dim-7 operators satisfying $(-\Delta L = \Delta B =\pm 1)$ was computed in \cite{Liao:2016hru}. The renormalization of the complete set of dim-7 operators for a flavour-specific basis was presented in \cite{Liao:2019tep}. The $\mathcal{O}(\Lambda^{-3})$ running contributions to dim-5 and dim-7 operators was derived in \cite{Zhang:2023kvw}. Finally, the full mixing among dim-7 operators was presented for the first time in \cite{Zhang:2023ndw,Zhang:2024clp}, completing the running of dim-7 operators.

\subsection{RG Evolution at dim-8}
Besides general aspects of one-loop RGEs in the on-shell method, the renormalization of several dim-8 bosonic operators is discussed in \cite{Jiang:2020mhe}. Single insertions of bosonic operators were considered in \cite{DasBakshi:2022mwk}, and double-insertions of dim-6 operators contributing to the running of bosonic dim-8 operators were computed in \cite{Chala:2021pll}. The contributions from lepton flavour violating operators to the bosonic dim-8 renormalization were computed in \cite{DasBakshi:2023htx}. Finally, the evolution of dim-8 fermion operators is discussed in \cite{Bakshi:2024wzz,Boughezal:2024zqa}.

\section{SMEFT Beyond Leading Order}
\label{sec:smeft-beyond-leading}
\subsection{Preliminaries}
In this section, we will elaborate on several unphysical dependences that are present in the intermediate steps of any renormalization group analysis that have to be removed to obtain physically meaningful predictions. These are
\begin{itemize}
\item
The dependence on the renormalization scheme that enters first at the NLO level through two-loop ADMs.
\item
The renormalization scheme dependence present in the one-loop matching of a particular effective theory to another one, in our case in the matching of the SMEFT to the WET.\footnote{The one-loop scheme change between the Breitenlohner Maison t'Hooft Veltman (BMHV) scheme and the NDR scheme in the SMEFT was recently computed in \cite{DiNoi:2025uan}.}
\item
The dependence on the scale at which a given theory, being either a full theory or an effective theory, like the SMEFT, is matched to another effective theory like the WET.
\item
The treatment of evanescent operators that appear at the NLO level.
\item
The related issues in basis transformations like for instance from the SMEFT basis to the JMS basis.
\end{itemize}

In the context of the SM all these topics except for the last one are discussed in detail in Chapter 5 of \cite{Buras:2020xsm}. They all are relevant for precise predictions within the SM. As far as the matching of the SMEFT onto the WET is concerned, the inclusion of the matching effects for specific processes was already emphasized in \cite{Aebischer:2015fzz} (for $\Delta F=1 $ processes) and \cite{ Bobeth:2017xry} (for $\Delta F=2$ processes). More recently the matching conditions have been extended to the full set of operators at one-loop level \cite{Dekens:2019ept}.

It should also be stressed that matching effects are important because certain SMEFT operators can be mapped onto the WET only through one-loop matching, not via 1-loop ADMs and tree-level matching. In this context we recommend the analyses of EDMs in \cite{Kley:2021yhn,Kumar:2024yuu}.

Here we summarize briefly these topics, concentrating on the issues closely related to our review. We will first discuss the topic of renormalization scheme and renormalization scale dependence. Subsequently, the issues of evanescent operators and basis transformations will be addressed.

\subsection{Renormalization Scheme Dependence}\label{RSDEP}
It is well known that two-loop anomalous dimensions of fields and parameters (including the WCs) depend on the RS. There is a correlation between RS dependencies of the WCs and matrix elements of the corresponding operators. In dimensional
regularization, the $\overline{\rm MS}$ scheme definition includes specifying the necessary evanescent operators, as well as choosing particular Dirac structures in the physical operator basis~\cite{Buras:2020xsm}. The final physical amplitudes in the WET, with the operators given in the JMS basis read
\begin{align}\label{FullA}
\mathcal{A} &
= \langle \vec{\OpL{}}_\text{JMS}(\muLow)\rangle^T \,
\vec{C}_\text{JMS}(\muLow)\,,
\end{align}
and are RS-independent. In particular, they do not depend on the definitions of evanescent operators. Such dependencies are canceled between
$\langle \vec{\OpL{}}_\text{JMS} (\muLow) \rangle$ and $\vec{C}_\text{JMS}(\muLow)$. The latter WCs can be expressed in terms of the SMEFT ones at the NP scale $\muNP$ as follows: 
\begin{align}
\label{eq:fullEvolX}
\vec{C}_\text{JMS}(\muLow) &
= {\hat U_\text{JMS}(\muLow, \muEW) \; \hat K(\muEW) \;
\hat U_\text{SMEFT}(\muEW,\muNP) \;
\vec{\mathcal{C}}_\text{SMEFT}(\muNP)} \,.
\end{align}
Here, the matrix $\hat K$ summarizes matching relations between the WET in the JMS basis and the SMEFT in the Warsaw basis. It will be given for various WCs in PART II. It has been calculated including one-loop contributions in \cite{Dekens:2019ept}. Explicitly
\begin{align}
\label{eq:SMEFTWET}
\vec{\mathcal{C}}_\text{JMS}(\muEW) &
= \hat K(\muEW) \; \vec{\mathcal{C}}_\text{SMEFT}(\muEW)\,.
\end{align} 
Various contributions to \eqref{eq:fullEvolX} must be evaluated using the same RS to guarantee proper cancellation of the RS dependences. To describe it in more detail, let us factorize out the NLO contributions to the QCD evolution matrices\footnote{For simplicity, only the RG running due to QCD interactions is discussed here.}
\begin{align}
\label{eq:UBMU}
\hat U_\text{JMS}(\muLow,\, \muEW) &
= \left[\mathbb{1} + \hat J_\text{JMS}\frac{\alS(\muLow)}{4\pi} \right]
\hat U_\text{JMS}^{(0)}(\muLow,\, \muEW)
\left[\mathbb{1} - \hat J_\text{JMS} \frac{\alS(\muEW)}{4\pi} \right]\,,
\\
\label{eq:USMEFT}
\hat U_\text{SMEFT}(\muEW,\, \muNP) &
= \left[\mathbb{1} + \hat J_\text{SMEFT} \frac{\alS(\muEW)}{4\pi} \right]
\hat U_\text{SMEFT}^{(0)}(\muEW,\, \muNP)
\left[\mathbb{1} - \hat J_\text{SMEFT} \frac{\alS(\muNP)}{4\pi} \right],
\end{align}
where $\hat U_i^{(0)}$ are the RS-independent LO evolution matrices. On the other hand, $\hat J_i$ stem from the RS-dependent two-loop ADMs, which makes them sensitive to the evanescent operator definitions. Explicit general expressions for $\hat U_i^{(0)}$ and $\hat J_i$ can be found in \cite{Buras:2020xsm}. 

Perturbative expansions of $\hat K(\muEW)$ and $\vec{\mathcal{C}}_\text{SMEFT}(\muNP)$ take the form
\begin{align}\label{SMEFTWETM}
\hat K(\muEW) &
= \hat K^{(0)} +\frac{\alS(\muEW)}{4\pi} \hat K^{(1)}\,,
&
\vec{\mathcal{C}}_\text{SMEFT}(\muNP) &
= \vec{\mathcal{C}}^{(0)}_\text{SMEFT} +
\frac{\alS(\muNP)}{4\pi} \vec{\mathcal{C}}_\text{SMEFT}^{(1)}\,.
\end{align}

Having all the above expressions at hand, we can easily trace out cancellations of the RS dependencies:
\begin{itemize}
\item
The RS dependence of $\vec{\mathcal{C}}_\text{SMEFT}^{(1)}$ is canceled by the one in $\hat J_\text{SMEFT}$ entering the last factor
on the RHS of \eqref{eq:USMEFT}.
\item
The one in $\hat J_{\rm JMS}$ entering the first factor on the RHS of \eqref{eq:UBMU} is canceled by $\vec{B}$ in 
\be\label{MAT5}
\langle\vec{\mathcal{O}}_\text{JMS}(\muLow)\rangle=\langle \vec{\mathcal{O}}\rangle_\text{tree}\left[\mathbb{1}+
\frac{\as(\muLow)}{4\pi}\vec{B}\right]\,,
\ee
where we did not expose the scale uncertainty in $\muLow$ which will be done in \eqref{MAT5}.

\item
Finally, the remaining RS dependencies, including those related to the evanescent operator definitions cancel in the product
\begin{align}
\left[\mathbb{1} - \hat J_{\rm JMS} \frac{\alS(\muEW)}{4\pi} \right]
\left[\hat K^{(0)} +\frac{\alS(\muEW)}{4\pi}\hat K^{(1)} \right]
\left[\mathbb{1} + \hat J_{\rm SMEFT}\frac{\alS(\muEW)}{4\pi} \right]\,.
\end{align}
\end{itemize}
To obtain RS-independent results, one has to ensure that evaluating the matching matrix $\hat K$ is consistent with the scheme dependencies in $\hat J_\text{JMS}$ and $\hat J_\text{SMEFT}$. As far as $\hat K$ in \cite{Dekens:2019ept} is concerned, it uses precisely the same RS as ours and it is recommended to use it in future calculations of $\hat J_\text{SMEFT}$. Further details on RS dependencies can be found in \cite{Buras:2000if, Chetyrkin:1997gb,Gorbahn:2004my}.

\subsection{Renormalization Scale Dependence}\label{SCALEDEP}
Having all these expressions at hand let us discuss the issue of the scale dependence. There are three scales which are not precisely
specified. These are
\be
\muLow=\ord{m_b},\qquad \muEW=\ord{M_W}, \qquad \Lambda=\ord{M_{Z^\prime}},
 \ee
where, as an example we have chosen $B$ decays in the context of a $Z^\prime$ model. We will now demonstrate that the final amplitude as given in (\ref{FullA}) and (\ref{eq:fullEvolX}) does not depend on these scales provided $\alpha_s$ NLO corrections at these three scales both in the RG evolution and various matchings are included.
 
Keeping only the first leading logarithms in the RG evolution and restricting the discussion to a single WC one finds using (5.191) of \cite{Buras:2020xsm} 
\be\label{UNEW1}
 U_\text{JMS}(\muLow,\, \muEW)=\left (1+
\frac{\as(\muLow)}{4\pi} \frac{\gamma_\text{JMS}^{(0)}}{2}
\ln\left(\frac{\muLow^2}{m_b^2}\right)\right)
\left[\frac{\as(M_W)}{\as(m_b)}\right]^{P_1}
\left(1+ \frac{\as(\muEW)}{4\pi} \frac{\gamma_\text{JMS}^{(0)}}{2}
\ln\left(\frac{M_W^2}{\muEW^2}\right)\right)\,,
\ee
and
\be\label{UNEW2}
U_\text{SMEFT}(\muEW,\, \Lambda)=\left (1+
\frac{\as(\muEW)}{4\pi} \frac{\gamma_\text{SMEFT}^{(0)}}{2}
\ln\left(\frac{\muEW^2}{M_W^2}\right)\right)
\left[\frac{\as(M_{Z^\prime})}{\as(M_W)}\right]^{P_2} 
\left(1+ \frac{\as(\Lambda)}{4\pi} \frac{\gamma_\text{SMEFT}^{(0)}}{2}
\ln\left(\frac{M^2_{Z^\prime}}{\Lambda^2}\right)\right)\,,
\ee
where 
\be
P_1=\frac{\gamma_\text{JMS}^{(0)}}{2\beta^0_\text{JMS}}\,,\qquad
P_2=\frac{\gamma_\text{SMEFT}^{(0)}}{2\beta^0_\text{SMEFT}}\,.
\ee

Moreover, we have
\be\label{cp}
{\mathcal{C}}_\text{SMEFT}(\muNP) 
= \mathcal{C}^{(0)}_\text{SMEFT} +\frac{\as(\Lambda)}{4\pi}\left ( \frac{\gamma_\text{SMEFT}^{(0)}}{2}\ln\left(\frac{\Lambda^2}{M^2_{Z^\prime}}\right)+
{\mathcal{C}}_\text{SMEFT}^{(1)}\right)\,,
\ee

\be\label{MAT5}
\langle\mathcal{O}_\text{JMS}(\muLow)\rangle=\langle \mathcal{O}\rangle_\text{tree}\left[1+
\frac{\as(\muLow)}{4\pi}\left( \frac{\gamma_\text{JMS}^{(0)}}{2}
\ln\left(\frac{m_b^2}{\muLow^2}\right)+{B} \right)\right]\,,
\ee
with ${\mathcal{C}}_\text{SMEFT}^{(1)}$ and $B$ relevant for the cancellation of the renormalization scheme dependence as already discussed above but playing no role in the cancellation of scale dependence as evident from the discussion below.

Inserting these expressions into \eqref{FullA} and \eqref{eq:fullEvolX} we confirm right away the removal of the $\muLow$ and $\Lambda$ dependences. As $P_1$ may differ from $P_2$ the final removal of the dependence on $\muEW$ is provided by the matching in \eqref{eq:SMEFTWET}, provided $\alpha_s$ corrections are included as already done in \eqref{SMEFTWETM} in the context of the cancellation of the RS dependences.

In App.~\ref{app:WETmatching} we present two examples of one-loop matchings of the SMEFT onto the WET which demonstrate the cancellation of the $\muEW$ dependence in explicit terms.

\subsection{Evanescent Operators}

The notion of evanescent operators becomes important when performing higher-order loop computations in dimensional regularization. Typical examples include the one-loop matching of one EFT onto another, or the two-loop running of WCs in an EFT. In such computations, when using dimensional regularization, four-dimensional identities involving Dirac structures, such as Fierz identities or the Chisholm identity, cannot be used directly, but have to be generalized to the $D$-dimensional case. Such generalizations are not unique, since the only condition for them to satisfy is to reproduce the four-dimensional case in the limit $D\to 4$. This condition is satisfied by any function that vanishes in $D=4$ dimensions. Therefore, the choice of how to reduce Dirac structures in general $D$ dimensions determines the renormalization scheme \cite{Herrlich:1994kh}. 

In general terms, once a Dirac structure $\Gamma$ is reduced in $D$ dimensions, the corresponding evanescent operator $E_\Gamma$ is defined by the difference between the $D$-dimensional Dirac structure and its four-dimensional version:

\begin{equation}
E_\Gamma=\Gamma_D-\Gamma_4\,.
\end{equation}

Let us illustrate this point by considering the example of a four-fermion operator containing three gamma matrices in each current:

\begin{equation}
O^{3\gamma}=(\overline f_1 \gamma_\mu\gamma_\nu\gamma_\rho P_L f_2)(\overline f_3 \gamma^\mu\gamma^\nu\gamma^\rho P_L f_4)\,.
\end{equation}

To reduce this operator to the standard vector operator 

\begin{equation}
O^{V,LL} = (\overline f_1 \gamma_\mu P_L f_2)(\overline f_3 \gamma^\mu P_L f_4)\,,
\end{equation}
one can use the Chisolm identity in $D=4$ dimensions

\begin{equation}\label{eq:Chisholm}
\gamma_\mu\gamma_\nu\gamma_\rho = \gamma_\mu g_{\nu\rho}-\gamma_\nu g_{\mu\rho}+\gamma_\rho g_{\mu\nu}+i\varepsilon_{\mu\nu\rho\alpha}\gamma^{\alpha}\gamma_5\,,
\end{equation}
on both currents to obtain the following relation between the two operators

\begin{equation}\label{4D}
O^{3\gamma}=16\,O^{V,LL}\,.
\end{equation}

In $D=4-2\varepsilon$ dimensions the Chisholm identity in~\eqref{eq:Chisholm} is not valid anymore and the reduction has to be performed differently. One example to reduce such structures is the so-called Greek method \cite{Tracas:1982gp, Buras:1989xd}, which we illustrate in the following.

For simplicity we consider only  the tensor product for which, inspired by the $D=4$ reduction \eqref{4D}, we make the following Ansatz for the reduction in $D$ dimensions:
\begin{equation}
\gamma_\mu\gamma_\nu\gamma_\rho P_L \otimes \gamma^\mu\gamma^\nu\gamma^\rho P_L = A \gamma_\mu P_L \otimes \gamma^\mu P_L\,,
\end{equation}
where the coefficient $A$ needs to be determined. In the Greek method, this is achieved by collapsing the tensor products on both sides of the equation by simply replacing the tensor product symbol with another gamma matrix, $\gamma_\tau$:\footnote{Dirac objects that reduce to scalar and tensor operators in $D=4$ dimensions are reduced in the Greek method using a more general Ansatz that involves scalar and tensor currents on the RHS. The two constants are then fixed by replacing the tensor product symbol by both $\mathbb{1}$ and $\gamma_{\tau_1}\gamma_{\tau_2}$ {and solving the resulting set of linear equations}.}

\begin{equation}
\gamma_\mu\gamma_\nu\gamma_\rho P_L \gamma_\tau \gamma^\mu\gamma^\nu\gamma^\rho P_L = A \gamma_\mu P_L \gamma_\tau \gamma^\mu P_L\,,
\end{equation}
Using now standard Dirac Algebra in $D$ dimensions one finds:
\begin{equation}
-(2-D)(D^2-10D+8)\gamma_\tau P_L = (2-D)A \gamma_\tau P_L\,,
\end{equation}
and therefore $A=-D^2+10D-8$. Expanding in $\varepsilon$ one therefore finds first:
\begin{equation}
O^{3\gamma}=(16-4\varepsilon)O^{V,LL}\,,
\end{equation}
which has the correct limit for $\varepsilon\to 0$, given in \eqref{4D}. Despite of this as pointed out in \cite{Buras:1989xd}, this formula is incorrect. Generally one has to add evanescent operators on the r.h.s of this equation so that the correct equation reads
\begin{equation}\label{4ED}
O^{3\gamma}=(16-4\varepsilon)O^{V,LL}+E^{3\gamma}\,,
\end{equation}
with the additional operator called the evanescent operator. It is given by the difference 
\begin{equation}
E^{3\gamma}=O^{3\gamma}-(16-4\varepsilon)\,O^{V,LL}.
\end{equation}
As \eqref{4ED} has to reduce to \eqref{4D} in $D=4$, the operator $E^{3\gamma}$ vanishes for $\varepsilon\to 0$.

\subsection{Basis transformation at tree-level and beyond}

Several bases are used in the literature that we will encounter on our route. A prominent example is the JMS basis \cite{Jenkins:2017jig} in the WET, which is very useful for matching the WET to the SMEFT. Another one is the so-called BMU \cite{Buras:2000if} basis that is useful in particular for the calculation of NLO QCD corrections to non-leptonic meson decays. Comparing results in these different bases is straightforward but tedious even at the tree level. Basis transformations typically involve Fierz identities as well as completeness (sometimes called "colour-Fierz") relations for the gauge group generators. Other important aspects to consider are field redefinitions \cite{Criado:2018sdb,Cohen:2022uuw,Cohen:2023ekv,Manohar:2024xbh,Cohen:2024fak,Criado:2024mpx} and EOMs \cite{Arzt:1993gz}. At the one-loop level basis transformations become more complicated, since also the schemes in the two different bases have to be transformed into each other. For that reason also evanescent operators need to be considered in the basis transformation since they define the used scheme. In \cite{Chetyrkin:1997gb} this procedure was explained in general terms, including the contributions from evanescent operators (EVs).

There is a rich literature on EVs and we will comment here only on selected papers.
\begin{itemize}
\item
In \cite{Gorbahn:2004my} the role of EVs in the context of NNLO QCD corrections to $\Delta F=1$ decays were highlighted.
\item
In \cite{Aebischer:2022tvz} simple rules for evanescent operators in one-loop transformations have been presented.
\item
Various aspects of evanescent operators, related to one-loop Fierz transformations in the process of one-loop matching computations are discussed in detail in \cite{Aebischer:2022aze} for four-fermion operators and in \cite{Aebischer:2022rxf} for dipole operators. The scheme factorization and simultaneous change of basis and scheme in the context of one-loop Fierz identities was discussed in \cite{Aebischer:2023djt}.
\item In a more recent work the authors of \cite{Aebischer:2024xnf} propose a renormalization scheme that is free of EVs. Several standard results were reproduced using this simple approach.
\item In \cite{Fuentes-Martin:2022vvu} the authors define an EV operator basis for the SMEFT, which can be used in matching calculations from NP models onto the SMEFT.
\end{itemize}

Numerical basis transformations can for instance be performed using the packages \texttt{WCxF} \cite{Aebischer:2017ugx} and \texttt{Rosetta} \cite{Falkowski:2015wza}. Because of the very large number of WCs such codes are indispensable when performing global phenomenological analyses.

\section{Flavour Symmetries in the SMEFT}\label{Fsym}

\subsection{Preliminaries}
The general SMEFT at dim-6 contains 2499 baryon number conserving parameters and a global fit in this framework in a bottom-up approach is not feasible. In usual phenomenological analyses of a given set of observables it is often assumed that only one or a few WCs are non-vanishing at the hadronic scale. The choice of them is governed by the observed deviations from SM expectations. However, such an approach can only be considered as a first approximation because important correlations due to RG effects are not taken into account and this is also the case for correlations between large classes of observables.

In order to reduce the number of parameters of the general SMEFT in a number of papers various flavour symmetries have been imposed, so that global fits become feasible. In this section we list the most interesting flavour symmetries in the context of the SMEFT that have been analyzed so far. Indeed they provide an effective organizing principle for the vast parameter space of the SMEFT.

While flavour symmetries reduce the number of parameters of the general SMEFT, it should be emphasized that the main motivation for them is the hope to explain the origin of a number of flavour puzzles related to the pattern of fermion masses and the puzzling structure of the CKM and PMNS matrices. A very recent compact review of these efforts can be found in \cite{Altmannshofer:2024jyv,Altmannshofer:2025rxc}.

\boldmath
\subsection{$\text{U(3)}^5$ }
\unboldmath
This flavour symmetry group is the maximal flavour symmetry group allowed by the SM gauge symmetry. The underlying concept is MFV formulated for quarks in \cite{DAmbrosio:2002vsn} and for leptons in \cite{Cirigliano:2005ck}. It assumes that the Yukawa couplings are the only sources of flavour violation and CP violation. See also \cite{Kagan:2009bn} and \cite{Feldmann:2006jk}. Other profound discussions of various aspects of MFV can be found in \cite{Paradisi:2008qh,Mercolli:2009ns,Feldmann:2009dc,Paradisi:2009ey}. Excellent compact formulations of MFV in the context of effective field theories have been given by Isidori \cite{Isidori:2010gz} and Nir \cite{Nir:2007xn}. We also recommend the reviews in \cite{Hurth:2008jc,Isidori:2012ts}, where phenomenological aspects of MFV are summarized. See also Section 15 of \cite{Buras:2020xsm}.

The SMEFT with MFV has been analyzed recently in \cite{Hurth:2019ula,Aoude:2020dwv,Faroughy:2020ina,Bruggisser:2021duo,Bruggisser:2022rhb,Greljo:2022cah,Bartocci:2023nvp,Allwicher:2023shc,Fajfer:2023gie,Grunwald:2023nli,Grunwald:2024yuq,Bartocci:2024fmm}. In particular a detailed counting of the operators at different orders in the symmetry breaking of $\text{U(3)}^5$ can be found in Table 1 of \cite{Faroughy:2020ina}. In the limit of exact symmetry one finds 41 CP-even and 6 CP-odd operators, a dramatic reduction of the number of operators relative to the general case. Including the symmetry breaking terms to first order in Yukawa couplings increases the number of operators to 52 and 17, respectively. Also, in this case a global fit analysis is feasible. For a very recent analysis of flavour invariants in the case of the $\text{U(3)}^5$ symmetry including also dim-8 operators see \cite{Sun:2025axx}.

Detailed phenomenological analyses of a multitude of observables, assuming the SMEFT with $\text{U(3)}^5$ or $\text{U(3)}^3$ symmetry at the NP scale have been performed including RG evolution down to hadronic scales in \cite{Aoude:2020dwv,Bartocci:2023nvp,Fajfer:2023gie}. In particular in \cite{Fajfer:2023gie}, concentrating on quark dipole transitions, a detailed investigation of RG-induced correlations between different flavour-violating processes and EDMs was presented. This includes dipole contributions to observables in non-leptonic and radiative $B$, $D$ and $K$ decays as well as the neutron and electron EDMs. Next the SMEFT analysis in \cite{Bruggisser:2021duo}, based exclusively on effective four-quark and two-quark couplings shows that the combination of top and bottom observables allows to pin down possible sources of flavour symmetry breaking from UV physics. In a follow-up analysis in \cite{Bruggisser:2022rhb} that included flavour, top-quark, electroweak and dijet observables the flavour structure of the four-quark couplings has been resolved without leaving blind directions in the parameter space.

In \cite{Greljo:2023adz} in the spirit of the tree-level dictionary of \cite{deBlas:2017xtg} for the general SMEFT, a classification of possible irreducible representations under $\text{U(3)}^5$ of new heavy spin-$0$, $1/2$ and $1$ fields that match onto dimension-6 operators at the tree-level has been presented. For a general perturbative NP model, the resulting flavour-symmetric interactions turn out to be very restrictive. In most cases they predict a single Hermitian operator with a definite sign. The authors call these {\em leading directions in the SMEFT}. In \cite{Greljo:2023bdy} this analysis is extended to include RG running from the NP scale to the electroweak scale. The authors find that the leading directions, corresponding to a single-mediator dominance, provide an important indirect probe of new interactions.

In all the above analysis spurions play an important role. Recently most general EFTs resulting from spurion analyses have been constructed in \cite{Grinstein:2024iyf}, and the construction was illustrated for the case of MFV. Finally, a spurion matching from the WET onto the Chiral Lagrangian was recently performed in \cite{Song:2025snz}.

\boldmath
\subsection{$\text{U(2)}^5$}
\unboldmath
$\text{U(2)}^5$ flavour symmetry is another popular symmetry in which the two light generations transform as doublets under $\text{U(2)}^5$ while the third generation is a singlet \cite{Barbieri:2011ci,Barbieri:2012uh,Isidori:2012ts}. A detailed counting of the operators at different order in the symmetry breaking of $\text{U(2)}^5$ can be found in Table 6 of \cite{Faroughy:2020ina}. In the limit of exact symmetry one finds 124 CP-even and 23 CP-odd operators, still a dramatic reduction of the number of operators relative to the general case. Including the symmetry breaking terms to first order in Yukawa couplings increases the number of operators to 182 and 81, respectively. This becomes a challenge for a global fit analysis without further assumptions. The recent analysis in \cite{Allwicher:2023shc} including RG effects and assuming that the new degrees of freedom couple mostly to the third generation demonstrates that still definitive results can be obtained. This is in particular the case when present electroweak, flavour and collider bounds are taken into account. It is found that all present bounds are consistent with an effective NP scale as low as $1.5\TeV$. It is also shown that a future circular $e^+e^-$ collider program such as FCC-ee would push most of these bounds by an order of magnitude. In this paper a list of several models with $\text{U(2)}^5$ flavour symmetry can be found. For a very recent analysis of flavour invariants in the case of the $\text{U(2)}^5$ symmetry including also dim-8 operators see \cite{Sun:2025axx}.

We also recommend the analysis in \cite{Fajfer:2023gie}, mentioned already above, in which a detailed investigation of RG-induced correlations between different flavour-violating processes and EDMs, this time in the $\text{U(2)}^3$ framework, has been performed.

\boldmath
\subsection{Other Favour Symmetries}
\unboldmath
In addition to the symmetries just discussed the authors of \cite{Greljo:2022cah} building on the work of\cite{Faroughy:2020ina} defined several well-motivated flavour symmetries and symmetry-breaking patterns that can serve as competing hypotheses about the UV dynamics beyond the SM, not far above the TeV scale. They consider in total four different symmetry structures in the quark sector and seven in the charged lepton sector so that one deals in total with 28 flavour scenarios with the number of operators listed in Table 1 of that paper. The set of flavour-breaking spurions is taken to be the minimal one needed to reproduce the observed charged fermion masses and mixings. The Mathematica package \texttt{SMEFTflavor} is provided to automatically construct the SMEFT operators for a given flavour group. For other related discussions see \cite{Smolkovic:2019jow,Bordone:2019uzc,Antusch:2023shi,Capdevila:2024gki,Grinstein:2023njq,Greljo:2023bix,Greljo:2024zrj}.

Selection rules for charged lepton flavour violating (cLFV) processes from residual flavour groups have been investigated in \cite{Calibbi:2025fzi}. The allowed flavour structures of operators in the SMEFT lead to distinctive and observable patterns of cLFV processes. A very nice summary of these rules is presented in Fig.~1 of that paper.

Finally, the exploration of flavour symmetries in the broad scenario of a strong interacting light Higgs has been performed in \cite{Glioti:2024hye}.

\boldmath
\subsection{Constrained Minimal Flavour Violation (CMFV)}
\unboldmath
Finally, we describe briefly the so-called Constrained Minimal Flavour Violation (CMFV) proposed in \cite{Buras:2000dm}. This is possibly the simplest class of extensions of the SM. It is defined pragmatically as follows \cite{Buras:2000dm}:
\begin{itemize}
\item
The only source of flavour and CP violation is the CKM matrix. This implies that the only CP-violating phase is the KM phase. Moreover, it is assumed that CP-violating flavour blind phases, discussed in \cite{Kagan:2009bn}, are absent.
\item
The only relevant operators in the effective Hamiltonian {\it below} the electroweak scale are the ones present within the SM.
\end{itemize}

Detailed expositions of phenomenological consequences of this NP scenario have been given already long time ago in \cite{Buras:2003jf,Blanke:2006ig,Buras:2012ts,Blanke:2016bhf} and were recently summarized in Chapter 15 of \cite{Buras:2020xsm}.

Here we will collect the most important properties of this class of models. Their simplest formulation can be made with the help of the master one-loop functions resulting from calculations of penguin and box diagrams that within the SM depend only on the top quark mass \cite{Inami:1980fz,Buchalla:1990qz}. Beyond the SM they are given as follows \cite{Buras:2003jf}
\be\label{masterf1}
S(\omega),~X(\omega),~Y(\omega),~Z(\omega),~E(\omega),~D'(\omega),~E'(\omega),
\ee
\noindent
where the variable $\omega$ collects the parameters of a given model. Two properties of these functions should be emphasized
\begin{itemize}
\item
They are {\it real} valued as the only complex phases reside in CKM factors that multiply these functions in effective Hamiltonians and flavour observables.
\item
They are flavour universal, as the full flavour dependence comes from the CKM factors.
\end{itemize}

These properties imply relations between various observables, not only within a given meson system but also between observables in different meson systems. Such relations are valid for the whole class of CMFV models and most importantly do not depend on any NP parameters. These relations are in fact the same as in the SM and as such do not allow the distinction between various CMFV models, which is only possible by means of observables that explicitly depend on the functions in \eqref{masterf1}. On the other hand violation of any of these relations by experimental data would automatically signal new sources of flavour and CP violation beyond the CMFV framework and would be a problem for all models of this class.

Yet, one should stress that MFV as formulated in \cite{DAmbrosio:2002vsn} is in contrast to CMFV, based on a renormalization-group-invariant symmetry argument, which can easily be extended above the electroweak scale where new degrees of freedom, such as extra Higgs doublets or SUSY partners of the SM fields are included. It can also be extended to strongly coupled gauge theories, although in this case the expansion in powers of the Yukawa spurions is not necessarily a rapidly convergent series. In this case, a resummation of the terms involving the top-quark Yukawa coupling needs to be performed~\cite{Feldmann:2008ja}.

This model-independent structure does not hold in CMFV in which one assumes that the effective FCNC operators playing a significant role within the SM are the only relevant ones also beyond the SM. This condition is realized only in weakly coupled theories at the TeV scale with only one light Higgs doublet. In the Minimal Supersymmetric Standard Model (MSSM), where two Higgs doublets are present, it still works for small $\tan\beta$ but does not work for large $\tan\beta$ as then new scalar operators become important. A very nice summary of this situation can be found in \cite{Isidori:2010gz} and further details are presented in \cite{D'Ambrosio:2002ex}.

Therefore, CMFV can only be formulated at the electroweak scale, and within the SMEFT the NP at $\Lambda$ must be such that at the electroweak scale after the RG evolution the important properties of CMFV listed above must be satisfied. The CMFV framework is much simpler than MFV, but shares some properties with the latter, like no new sources of flavour violation beyond the CKM matrix or equivalently Yukawa couplings. It is a useful starting point to investigate correlations between various observables, not only within a given meson system but in particular between different meson systems. More details on CMFV and in particular the list of various correlations between various observables can be found in Chapter 15 of \cite{Buras:2020xsm}.

\section{SMEFT Tools}\label{SMEFTtools}

In this section we give a brief overview over the plethora of computer tools that are available to perform various tasks in the SMEFT and the WET. The current tools on the market range from basis generation and automated matching computations over RG evolution and fitting tools for WCs. For further recent reviews we refer to \cite{Aebischer:2023irs,Proceedings:2019rnh,Dawson:2022ewj,Belvedere:2024nzh}. 

\subsection{Basis generation}

Deriving a complete and non-redundant basis is a straightforward but tedious task. It involves the use of different identities and symmetries such as:
\begin{itemize}
\item Poincar\'e invariance
\item gauge invariance
\item completeness relations for gauge and Dirac structures
\item integration by parts
\item field redefinitions
\end{itemize}
Typically, in a first step all possible gauge- and Poincar\'e invariant structures up to a given mass dimension are written down. In the next step this set is reduced according to certain rules, by considering the previously mentioned identities. In the case of the dimension six Warsaw basis for the SMEFT these rules correspond to reducing the number of covariant derivatives in the basis as much as possible. Apart from specifying the rules to eliminate certain operators in favour of others, the process of generating an overcomplete basis can be easily automated. In this respect, there are several different tools that allow to perform one or several tasks relevant for basis generation.

The Mathematica package \texttt{Sym2Int} \cite{Fonseca:2017lem} allows to count the (non-redundant) number of operators of a given operator basis. It is based on the Mathematica package \texttt{GroupMath} \cite{Fonseca:2020vke}, with which one can perform manipulations involving semi-simple Lie algebras as well as permutation groups.\footnote{Another useful tool to perform group theory manipulations is the Mathematica package LieART \cite{Feger:2019tvk}.} With the use of \texttt{Sym2Int} it was shown, that the number of given independent operators in the dim-8 Murphy basis \cite{Murphy:2020rsh} is too large. Instead of the quoted 1030 real structures there should only be 1019 operators. 

Also the Python package \texttt{BasisGen} \cite{Criado:2019ugp} allows to count the number of independent operators in a basis for arbitrary gauge groups, field representations, operator dimensions and flavour indices. It is based on the Hilbert series approach \cite{Lehman:2015via,Henning:2015daa,Lehman:2015coa,Henning:2015alf,Henning:2017fpj} and can be used for further operations such as tensor product decomposition and calculating weights of group representations.

The Python-based computer program \texttt{DEFT} \cite{Gripaios:2018zrz} is specifically tailored to the SMEFT, and allows to find the SMEFT operator basis up to arbitrary mass dimensions, neglecting flavour indices. Several manipulations can be performed using \texttt{DEFT}, such as checking whether a set of input operators forms a basis, completing the set if possible as well as changing it to another basis.

Another code that generates general operator bases is the Mathematica package \texttt{ABC4EFT} \cite{Li:2022tec}. Different types of bases can be constructed with \texttt{ABC4EFT}: Besides the fully independent basis the code allows to construct bases up to flavour indices as well as operator sets containing gauge-variant operators. The operator sets are constructed using Young Tableau tensors as well as several novel algorithms, which are further explained in \cite{Li:2022tec}.

A very recent code is the Python package \texttt{AutoEFT} \cite{Harlander:2023ozs,Schaaf:2023mpw}. It allows to construct on-shell bases for EFTs by taking into account EOMs, IBPs, Fierz identities, algebraic identities as well as redundancies due to repeated fields. It is based on the algorithm developed in \cite{Li:2020gnx,Li:2020xlh} and allows to construct bases for EFTs with gauge groups that are direct products of $\text{SU(N)}$ and $\text{U(N)}$ factors.

Furthermore, in the context of flavour symmetries the Mathematica package \texttt{SMEFTflavor} can be used to construct and count the number of operators for a given spurion expansion \cite{Greljo:2022cah}. It allows to study different flavour symmetries that can be imposed on the SMEFT Lagrangian up to dimension six.

\subsection{Numerical basis transformations}
Besides analytic basis transformations which can be performed for instance by \texttt{DEFT}, a basis transformation can also be performed for numerical values of the WCs. Several codes allow to do this numerical transformation for predefined operator bases in the SMEFT and the WET. 

\texttt{Rosetta} \cite{Falkowski:2015wza} was the first code which was able to perform such transformations. It allows to change from the Warsaw to the SILH basis. Furthermore, other bases can be added to the code in order to change them into the predefined bases.

Another Python package that allows changing between various operator bases in the SMEFT and WET is \texttt{WCxf} \cite{Aebischer:2017ugx}.\footnote{\texttt{WCxf} is integrated in the Python package \texttt{wilson} \cite{Aebischer:2018bkb}.} More than 15 bases are predefined in \texttt{WCxf} and customized bases can be defined by the user.\footnote{The bases and their definitions can be found in \href{https://wcxf.github.io/bases.html}{WCxf bases}.} Also the matching between the SMEFT and WET basis can be performed using \texttt{WCxf}, since the complete tree-level and one-loop matching is implemented as well.

\subsection{Matching tools}
The matching procedure is an essential part of each EFT calculation. It is conceptually straightforward but is error-prone when done by hand. Therefore, having a tool at hand that automates the process is of great advantage. 

Automated tree-level matching can be performed for instance with the Python package \texttt{MatchingTools} \cite{Criado:2017khh}. The matching conditions are obtained using equations of motion to eliminate heavy fields in the full Lagrangian.

Recently, the Mathematica package \texttt{mosca} \cite{LopezMiras:2025gar} was developed, which allows to perform automated on-shell tree-level matching. Besides basis transformations the reduction from a Green's basis to a physical basis can be performed automatically using this tool.

Furthermore, there are several one-loop matching codes that we will list in the following:

The matching code \texttt{CoDEx}~\cite{Bakshi:2018ics} is a Mathematica package that performs the NLO matching procedure from a full theory onto the SMEFT automatically using functional methods. It is based on the Universal One-Loop Effective Action (UOLEA) approach \cite{Henning:2014wua,Drozd:2015rsp,Fuentes-Martin:2016uol,Ellis:2017jns} that was developed in the past few years. Progress is made in including also dim-8 contributions to the matching including heavy scalars \cite{Banerjee:2023iiv,Banerjee:2022thk} and heavy fermions \cite{Chakrabortty:2023yke}.

Another advanced matching tool is the Mathematica package \texttt{Matchete}\cite{Fuentes-Martin:2022jrf}. It allows to compute automatic tree-level matching as well as one-loop matching for heavy scalars and fermions onto the SMEFT and WET. \texttt{Matchete} is based on functional methods and uses the package \texttt{SuperTracer}~\cite{Fuentes-Martin:2020udw} in order to compute the necessary supertraces for the matching. Current effort is put in implementing the effects of evanescent operators as discussed in \cite{Fuentes-Martin:2022vvu}.

Supertraces can also be calculated using the Mathematica package \texttt{STrEAM}, for which a detailed manual can be found in \cite{Cohen:2020qvb}. It can be used for matching calculations of generic UV models onto arbitrary EFTs using functional methods~\cite{Cohen:2020fcu}. The underlying algorithm is based on the covariant derivative expansion.

A further matching tool on the market is \texttt{Matchmakereft} \cite{Carmona:2021xtq}. The Mathematica code allows to perform the matching procedure at NLO onto the SMEFT. It makes use of several preexisting packages such as \texttt{QGRAF} \cite{NOGUEIRA1993279} for Feynman diagram generation and \texttt{FeynRules} \cite{Alloul:2013bka} to specify the interactions that are used to characterize the full theory. Furthermore it is based on different computer languages such as Mathematica for pattern matching, \texttt{FORM} for the NLO computation and Python for interfacing the different components. There are several extensions of \texttt{Matchmakereft} that allow to interface it with other codes, as for example the Mathematica package \texttt{MATCH2FIT} \cite{terHoeve:2023pvs}, with which \texttt{Matchmakereft} output can be used in \texttt{SMEFiT} \cite{Hartland:2019bjb}. A similar extension will soon be available for the Python package \texttt{smelli}~\cite{Aebischer:2018iyb}.

In \cite{Guedes:2023azv,Guedes:2024vuf} the matching code \texttt{SOLD} has been developed that includes an arbitrary number of heavy scalars and fermions. It consists of an EFT to UV SMEFT one-loop dim-6 dictionary, that generalizes the tree-level dictionary of \cite{deBlas:2017xtg}. More about it can be found in Sec.~\ref{sec:5}.

Finally, a computational tool that is of interest when performing one-loop matching calculations is \texttt{GoSam} \cite{Braun:2025afl}. It allows to input UFO files for various EFTs and to perform general one-loop calculations in different renormalization schemes.

\subsection{Running tools}

A Mathematica package to perform automated running of SMEFT and WET WCs is \texttt{DsixTools} \cite{Celis:2017hod,Fuentes-Martin:2020zaz}, in which the full one-loop RGE evolution in the SMEFT \cite{Alonso:2013hga,Jenkins:2013zja,Jenkins:2013wua} and the WET \cite{Jenkins:2017dyc} is implemented. It supports \texttt{WCxf} \cite{Aebischer:2017ugx} in- and output and gives analytic expressions for the EFT parameters in both theories. Furthermore, the full tree-level \cite{Jenkins:2017jig} and one-loop \cite{Dekens:2019ept} matching from the SMEFT onto the WET is implemented in \texttt{DsixTools}, which allows for a full LO evolution from scales above the EW scale down to scales below it.

Automated RG evolution can also be performed using the Python package \texttt{wilson}~\cite{Aebischer:2018bkb}, which allows to perform the complete SMEFT and WET RG evolution at the one-loop level. Since recently the complete NLO matching from the SMEFT onto the WET is also implemented in \texttt{wilson}. Furthermore, the package contains the basis change module \texttt{WCxF} \cite{Aebischer:2017ugx} and therefore allows to change between different bases before or after the running/matching procedure. The recent release of {\tt wilson} is operational for the SMEFT with sterile neutrinos \cite{Aebischer:2024csk}.

A third code that allows to run the SMEFT parameters to different energy scales is \texttt{RGESolver}~\cite{DiNoi:2022ejg}. It is a \CC~ library, that allows for LO running of baryon and lepton number conserving SMEFT operators.

The Python package \texttt{D7RGESolver} \cite{Liao:2025qwp} allows to solve RGEs in an automated way for dim-7 operators. It is the first implementation that allows to solve RGEs for SMEFT operators other than dim-6 and includes dim-5 and dim-7 effects in the running.

\subsection{Observable calculators and fitting tools}
There are several tools available that collect theory predictions for a large number of observables, together with the corresponding experimental values.

The Python package \texttt{flavio}~\cite{Straub:2018kue} contains over 200 observables, ranging from flavour physics over Higgs and neutrino physics to low-energy observables like electric and magnetic dipole moments. The different observables are implemented as functions of WCs evaluated at a certain scale. In order to run the WCs to the scale of the process as well as performing the matching and changing between different bases \texttt{flavio} makes use of the packages \texttt{wilson} and \texttt{WCxf}. A fitting tool that is based on \texttt{flavio} is the SMEFT likelihood calculator \texttt{smelli}~\cite{Aebischer:2018iyb}. It allows to compute the likelihood function in the space of SMEFT WCs based on the observables implemented in \texttt{flavio}.

A \CC~code to perform fits and predictions of observables in flavour physics is the package \texttt{EOS}~\cite{EOSAuthors:2021xpv}. It has a Python front-end and also supports the \texttt{WCxf} format for WCs. One feature of \texttt{EOS} is that it allows for Bayesian inference for the theory parameters. 

\texttt{kolya} \cite{Fael:2024fkt} is a rather novel Python tool, which allows to compute inclusive semileptonic $B$ decays. It includes all available QCD perturbative corrections as well as $1/m_b^5$ power corrections and is interfaced with \texttt{CRunDec} \cite{Chetyrkin:2000yt,Schmidt:2012az,Herren:2017osy}.

A rather new tool that allows for parameter fits in the SMEFT is the Mathematica package \texttt{HighPT}~\cite{Allwicher:2022mcg,Allwicher:2022gkm}. It is mainly focused on high-$p_T$ observables but there are current efforts to also include low-energy observables, as well as the running of WCs together with the matching onto the WET.

Another SMEFT fitting tool, that focuses mainly on top physics observables is the package \texttt{SMEFiT}~\cite{Hartland:2019bjb,Giani:2023gfq,Celada:2024mcf}. In this tool the likelihood function in the space of SMEFT WCs is obtained using Monte Carlo methods, including NLO QCD corrections to most of the considered observables.

\texttt{HEPfit}~\cite{DeBlas:2019ehy} is another \CC~code that allows to perform likelihood calculations for SMEFT WCs, by comparing experimental predictions to a large data base of experimental observables. The fitting procedure is performed via Bayesian Markov Chain Monte Carlo analyses.

Recently, the code \texttt{SIMUnet} was implemented, which is a machine-learning based framework that allows for simultaneous fits of WCs and Parton Distribution Functions (PDFs) \cite{Costantini:2024xae}. It also includes different fit metrics, like quality metrics, uncertainties as well as PDF and SMEFT correlations. 

 Another machine learning approach to constrain SMEFT WCs was studied in \cite{Krause:2025qnl}. The BITNET architecture, typically used for Large Language Models, was used to study WCs via regression.

 A fitting tool that allows for model creation and Bayesian fitting of SMEFT parameters to differential cross-section data is \texttt{dEFT} \cite{Keaveney:2021dfa}. It is based on Python and allows the study of linear and quadratic dim-6 SMEFT effects in cross-section fits.
 
 Finally, we mention the code \texttt{FlavBit}~\cite{Workgroup:2017myk} which is a GAMBIT \cite{GAMBIT:2017yxo} module that allows to compute flavour observables as well as likelihood functions in and beyond the SM.

\subsection{Monte-Carlo tools}

The \texttt{SMEFTsim}~\cite{Brivio:2017btx} package is a set of \texttt{FeynRules} implementations that allows to study the full SMEFT as well as the $\text{U(3)}^5$-SMEFT and MFV-SMEFT theories. Different theoretical predictions can be made, using the parameter schemes \{$\alpha_e, M_Z, G_F$\} or \{$M_W, M_Z, G_F$\}. 

Another tool for Monte-Carlo (MC) simulations, that is frequently used in the literature is the code \texttt{SMEFT@NLO}~\cite{Degrande:2020evl}, which is a \texttt{FeynRules} implementation of the SMEFT Warsaw basis. The implemented operators obey a $\text{U(2)}_q\times \text{U(2)}_u \times \text{U(2)}_d$ flavour symmetry and the evanescent operators are dealt with using Greek projections. Using the MC generator MadGraph5\_aMC@NLO 5 \cite{Alwall:2014hca}, \texttt{SMEFT@NLO} allows to compute all implemented observables at the linear and quadratic order in the WCs.

The package \texttt{SmeftFR} v3 allows to derive the Feynman rules for dimension-5, -6, and all bosonic dimension-8 operators \cite{Dedes:2023zws}. Different output formats can be chosen for further use in established codes such as \texttt{FeynRules} or \texttt{FeynArts} \cite{Hahn:2000kx}. The main application of \texttt{SmeftFR} is however to use it in combination with Monte-Carlo generators like MadGraph5\_aMC@NLO 5, Sherpa \cite{Gleisberg:2008ta} and several further codes.

\subsection{Clustering tools}
The Python package \texttt{ClusterKing}~\cite{Aebischer:2019zoe} allows to cluster the parameter space spanned by WCs of the SMEFT and WET. The user can choose between different clustering algorithms, customized metrics and various visualization methods. 

The R package \texttt{Pandemonium}~\cite{Laa:2021dlg} allows to perform hierarchical clustering. It features several visualization routines and an interactive user interface. 

Both \texttt{ClusterKing} and \texttt{Pandemonium} are not limited to the SMEFT or WET, but can be used to cluster the parameter space of arbitrary physical systems.

\section{A Guide to existing SMEFT Analyses}\label{sec:obs}

During the last ten years numerous analyses of a multitude of observables have been performed within the SMEFT.\footnote{{See \cite{Bechtle:2022tck} for a philosophical discussion on the change of focus in particle physics towards model-independent analysis like the SMEFT.}}
This section is supposed to give an overview of the different developments in the SMEFT. The main goal of these analyses was to constraint the most relevant WCs for a given observable or set of observables. We list these observables and related operators in Tabs.~\ref{tab:LOSMEFTobsflavor}-\ref{tab:fitSMEFTobscoll}. There is no space to review this rich literature and we invite the reader to have a look at these analyses in order to get an impression on the complexity of such analyses.

Right at the beginning the following point should be made. At first sight one might expect that arbitrary values of the WCs in the SMEFT are allowed, and that they are bound only by the experimental data, but this is false. Fundamental principles of QFT such as unitarity and analyticity constrain the space of WCs, in particular allowing only their specific signs dependently on whether the NP particles are scalars or vectors as stressed in particular in \cite{Adams:2006sv,Remmen:2019cyz,Remmen:2020vts,Remmen:2020uze,Remmen:2022orj}. We will be more explicit about these different analyses below.

\subsection{Tree-level}
In this subsection we collect analyses that were performed for SMEFT WCs at the tree-level. The following tables include observables for quark and lepton flavour violation (LFV) (Tab.~\ref{tab:LOSMEFTobsflavor}), low energy observables (Tab.~\ref{tab:LOSMEFTobslowen}), Higgs physics (Tab.~\ref{tab:LOSMEFTobsHiggs}), EWP and top processes (Tab.~\ref{tab:LOSMEFTobsEW}) and quark-quark as well as quark-lepton interactions (Tab.~\ref{tab:LOSMEFTobsqqll}). Miscellaneous tree-level observables are collected in Tab.~\ref{tab:LOSMEFTobsmisc}. 
Furthermore, we briefly mention here methods other than the study of particular observables to set bounds on SMEFT WCs: Constraints can for instance be obtained by deriving spin sum rules, based on analyticity and unitarity arguments \cite{Remmen:2020uze,Remmen:2022orj}. Unitarity bounds for the Warsaw and SILH basis were studied in \cite{Cao:2024vfc}, and unitarity bounds on EFTs from the LHC were studied in detail in \cite{Cohen:2021gdw}. Recently a detailed analysis of the implications of these constraints on top quark decays has been presented in \cite{Altmannshofer:2023bfk,Altmannshofer:2025lun}.

Another possibility to constrain WCs is to study positivity bounds (PBs) \cite{Adams:2006sv}.\footnote{One-loop PBs for scalar theories were studied in \cite{Ye:2024rzr}.} Such bounds arise from analyticity, unitarity, crossing symmetry, locality and Lorentz invariance, and imply fixed signs of WC combinations. They apply however only at the order $\mathcal{O}(1/\Lambda^4)$, i.e. for double insertions of dim-6 operators, or for dim-8 operators. PBs from vector boson scattering (VBS) for dim-8 operators were studied in \cite{Zhang:2018shp}. Bounds for quartic gauge boson couplings were derived in \cite{Bi:2019phv}. Elastic positivity and extremal positivity are constructed for dim-8 operators in \cite{Yamashita:2020gtt}. Furthermore, PBs can be used to distinguish between the HEFT and the SMEFT \cite{Remmen:2024hry}. Besides PBs also bounds on the magnitude of dim-8 CP-odd operators can be obtained from two-to-two bosonic scattering amplitudes \cite{Remmen:2019cyz}. Constraints from flavour physics on four-fermion dim-8 operators were studied in \cite{Remmen:2020vts}. The gluonic dim-6 and dim-8 operators were discussed in \cite{Ghosh:2022qqq} and PBs of dim-8 Higgs operators were studied in \cite{Chen:2023bhu}. Furthermore, the convex geometry perspective of the SMEFT parameter space was described in \cite{Zhang:2020jyn}. It allows to draw conclusions on possible UV completions, simply by looking at the geometric properties of the parameter space that describes scattering amplitudes. This parameter space is typically given by a complex cone. Such cones can then be expressed in terms of the $J$-basis, which allows for a systematic derivation of PBs \cite{Yang:2023ncf}.\footnote{The $J$-basis is given by the linearly independent eigenstates of the squared Pauli-Lubanski vector.} Recently, an $a$-theorem for EFTs was derived in \cite{Liao:2025npz}, which can be used to determine whether renormalization effects preserve given PBs for SMEFT WCs. Furthermore, PBs were also used to tackle the inverse problem, i.e. finding the corresponding UV completions for a given set of measured SMEFT WCs \cite{Zhang:2021eeo}. Finally, PBs can also be used to determine non-trivial zeros and signs in the dim-8 ADM of the SMEFT \cite{Chala:2023xjy}.

Other strong constraints on WCs can arise from CP-violation. In this context in \cite{Bonnefoy:2021tbt}, similar to the Jarlskog invariant in the SM \cite{Jarlskog:1985ht,Jarlskog:1985cw} all CP-invariants in the SMEFT up to $\mathcal{O}(1/\Lambda^2)$ have been derived. The authors find 699 invariants. Furthermore, CP-invariant bases in the SMEFT and $\nu$SMEFT have been studied in \cite{Darvishi:2023ckq,Darvishi:2024cwe}, using the Ring-Diagrams formalism. CP violation and flavour invariants in the leptonic sector have been analyzed in \cite{Wang:2021wdq,Yu:2021cco,Yu:2022ttm}. For the corresponding analyses within the SM including the RG running of the invariants see in particular \cite{Jenkins:2009dy,Feldmann:2015nia}.

Possible first-order EW phase transitions in the SMEFT, together with the resulting constraints on WCs are studied in \cite{Camargo-Molina:2024sde,Chala:2025xlk}, and the high-temperature limit of the SMEFT is discussed in \cite{Chala:2025aiz}. {Higher-order corrections and their impact on strong phase transitions in the SMEFT were studied in \cite{Chala:2025oul}.}
\begin{table}[H]
\centering
\resizebox{\columnwidth}{!}{%
\begin{tabular}{lccc}
\toprule
process  & operators & comments & ref
\\
\midrule
$\varepsilon'/\varepsilon$ & $f^2XH$, $f^2H^2D$, $f^4$,    & $\cancel{\text{\text{CP}}}$, $f=q,u,d$ & \cite{Aebischer:2018csl} \\
$\varepsilon'/\varepsilon$, $\Delta M_K$, $K\to\pi\bar\nu\nu$ & $f^2H^2D$, $f^4$  & $Z'$ model & \cite{Aebischer:2020mkv} \\
$\Delta F =2$ & $f^4$, dim-8 & - & \cite{Liao:2024xel} \\
$\overline{B}\to X_c\ell\nu$  & $f^2H^2D$, $f^4$  & {$\ell = \mu,e$} & \cite{Carvunis:2025vab} \\
$B\to K\nu\bar\nu$  & $f^2H^2D$, $f^4$  & - & \cite{Marzocca:2024hua} \\
$K\to \pi\nu\nu$  & dim-5, dim-7, dim-9  & {LNV} & \cite{Deppisch:2020oyx} \\
$\Lambda_b\to \Lambda_c(\to\Lambda \pi)\tau \overline{\nu}_\tau$ &  $f^2H^2D$, $f^4$  & - & \cite{Karmakar:2023rdt} \\
$\mu^+\mu^-\to tc$ &  $f^2H^2D$, $f^4$  & - & \cite{Bhattacharya:2023beo} \\
$\gamma q\to t\gamma$ &  $f^2XH$ & FCC-$\mu p$ & \cite{Alici:2024eez} \\
flavour, EWPO  & $f^2H^2D$, $f^4$  & - & \cite{Kumar:2021yod} \\
LFV $B$ decays  & $f^2H^3$, $f^2XH$, $f^2H^2D$, $f^4$  & - & \cite{Ali:2023kua} \\
$\Delta \mathcal{B}=2$ &  dim-9  & RGE & \cite{ThomasArun:2025rgx} \\
\hline \\
cLFV &  {$f^4$} & -- & \cite{Chattopadhyay:2025air} \\
LFV &  $f^2H^3$, $f^2XH$, $f^2H^2D$, $f^4$ & Muon collider & \cite{Asadi:2025dii} \\
$\ell\to \ell'\gamma$,$\ell\to \ell(')\ell\ell$,$Z\to \ell'\ell$, $a_\ell$, $d_\ell$ &  $f^2H^3$, $f^2XH$, $f^2H^2D$, $f^4$ & $f=\ell,e$ or SL & \cite{Crivellin:2013hpa} \\
$\ell\to \ell'\gamma$, $a_\ell$, $d_\ell$ &  $f^2H^3$, $f^2XH$, $f^2H^2D$, $f^4$ & \cancel{CP} & \cite{Ardu:2025awk} \\
$\mu\to e\gamma$ &  $f^2H^3$, $f^2XH$, $f^2H^2D$, $f^4$ & $f=\ell,e$ or SL & \cite{Pruna:2014asa} \\
$\mu\to e\gamma$ \& $\mu\to\tau\to e$ & $f^2H^3$, $f^2XH$, $f^2H^2D$, $f^4$, dim-8  & - & \cite{Ardu:2022pzk} \\
$\mu\to 5e$, $\mu\to 7e$ etc. & $f^2XH$, $f^4$, dim-8, dim-10  & light mediators & \cite{Greljo:2025ljr} \\
$\ell\to \ell'\gamma$, $\ell'\to \ell\ell\ell$, $Z\to \ell\ell'$ & $f^2H^3$, $f^2XH$, $f^2H^2D$, $f^4$  & $f=\ell,e$ or SL & \cite{Pruna:2015jhf} \\
$\mu\to e$, $\tau\to 3\ell$ & dim-5, $f^2H^3$, $f^2XH$, $f^2H^2D$, $f^4$  & - & \cite{Ardu:2024bua} \\
$\mu^+ e^-\to t q$ & $f^4$  & {$\mu$TRISTAN} & \cite{Sarkar:2025bgo} \\
$a_\mu$, $\mu\to e$, $\tau\to \mu \gamma$ &  $f^2H^3$, $f^2XH$, $f^4$  & flavour align. & \cite{Isidori:2021gqe} \\
$Z\to \ell_i \ell_j$ &  $f^2H^3$, $f^2XH$, $f^2H^2D$, $f^4$  & FCC-ee, CEPC & \cite{Calibbi:2021pyh} \\
$e^+e^-\to \tau\mu$ &  $f^2XH$, $f^2H^2D$, $f^4$  & X & \cite{Altmannshofer:2023tsa,Altmannshofer:2025nbp} \\
$e^+e^-\to \tau\ell,e\mu$ &  $f^2XH$, $f^2H^2D$, $f^4$  & optimized obs. & \cite{Jahedi:2024kvi} \\
$M_{\nu}$ & dim-5,6,7,  $\Delta L \neq 0$  & $Y_e,Y_d= 0$ & \cite{Chala:2021juk} \\
$P_{\bar{\nu}_i\to \bar{\nu}_i}$, $\nu_i\leftrightarrow \nu_j$  & $f^2H^2D$, $f^4$  & - & \cite{Falkowski:2019xoe} \\
neutrino oscill./scattering &  $H^4D^2$, $f^2H^3$, $X^2H^2$, $f^2H^2D$, $f^4$  & - & \cite{Du:2021rdg} \\
neutrino oscillations &  $H^4D^2$, $f^2HD$, $f^4$  & - & \cite{Du:2020dwr} \\
neutrino oscillations &  $f^4$  & {quasi-elastic det.} & \cite{Kopp:2025ffx} \\
$0\nu\beta\beta$ &  dim-5, dim-7  & LFV & \cite{Graf:2025cfk} \\
$M_1^-\to M_2^+\ell_1^-\ell_2^-$ &  dim-9  & large $N_c$ & \cite{Donini:2025cuy} \\
$\Delta L=2$ &  dim-7  & - & \cite{Fridell:2023rtr} \\
$\Delta L=2$ &  $f^4$  & - & \cite{Heeck:2024uiz} \\
$\Delta L=2$ &  dim-5, dim-7  & muon collider & \cite{Bhattacharya:2025xwv} \\
$\Delta L=3$ &  dim-7, dim-10  & - & \cite{Heeck:2025jfs} \\
\bottomrule
\end{tabular}
}
\caption{LO analyses involving quark- and lepton-flavour transitions in the SMEFT. X=FCC-ee, CEPC, ILC, CLIC.}
\label{tab:LOSMEFTobsflavor}
\end{table}

\begin{table}[H]
\centering
\begin{tabular}{lccc}
\toprule
process  & operators & comments & ref
\\
\midrule
$a_{\mu,e}$, $d_{\mu,e}$ & all except $H^6$  & 1L matching & \cite{Aebischer:2021uvt} \\
$(g-2)_\ell$ \& $h\to\ell^+\ell^-$ &  $f^2H^3$, $f^2XH$, $f^4$  & $y_t$ running & \cite{Fajfer:2021cxa} \\
EDMs, CP-odd Higgs obs. & $X^3$, $X^2H^2$  & $\cancel{\text{\text{CP}}}$ & \cite{Cirigliano:2019vfc} \\
EDMs &  $f^2XH$, $f^2H^2D$, $f^4$  & {$\cancel{\text{\text{CP}}}$} & \cite{Unal:2023wct} \\
FC $q$-dipoles  &  $f^2H^3$, $f^2XH$, $f^2H^2D$, $f^4$  & $\cancel{\text{\text{CP}}}$, MFV, $\text{U(2)}^3$ & \cite{Fajfer:2023gie} \\
up-type dipoles  &  $f^2XH$, $f^2H^2D$  &  MFV & \cite{Bonnefoy:2024gca} \\
atomic parity violation &  $f^4$  & $\cancel{\text{\text{P}}}$, LQs & \cite{Bischer:2021jqn} \\
\bottomrule
\end{tabular}
\caption{LO analyses of low energy observables like electric dipole moments (EDMs) in the SMEFT.}
\label{tab:LOSMEFTobslowen}
\end{table}

\begin{table}[H]
\centering
\begin{tabular}{lccc}
\toprule
process  & operators & comments & ref
\\
\midrule
$pp\to t\overline{t}h$ &   $f^2H^3$, $f^2H^2D$, dim-8  & VL top & \cite{Dawson:2021xei} \\
$pp\to thj$ &  all except $H^6$  & $\text{U(2)}^2\times \text{U(3)}^3$ & \cite{Guchait:2022ktz} \\
$pp\to hW^+$ &  $H^6$, $H^4D^2$, $X^2H^2$, $f^2H^2D$, dim-8  & $(\text{dim}$-6$)^2$ & \cite{Hays:2018zze} \\
$pp\to \overline{t}tZ(ll)h(bb)$ &  $H^4D^2$, $X^2H^2$, $f^2H^2D$  & {FCC-hh} & \cite{Banerjee:2025dsh} \\
$h\to ZZ^*$ &   $X^2H^2$, $f^2H^2D$  & entanglement & \cite{Subba:2024aut} \\
$q\overline{q}\to q\overline{q} h\gamma$  & $X^2H^2$  & MC & \cite{Biekotter:2020flu} \\
$qq\to hjj$ &   $f^2H^2D$  & ML & \cite{Araz:2020zyh} \\
$gg\to t\overline th$ &  $f^2H^3$, $X^2H^2$, $f^2XH$  & $p_T$ & \cite{Battaglia:2021nys} \\
$WW\to hh$ &  $H^4D^2$, dim-8  & HEFT & \cite{Domenech:2022uud} \\
$Wh,Zh$ prod. &   $X^2H^2$  & ML & \cite{Freitas:2019hbk} \\
$Wh\to \ell \nu b \bar{b}$  &   $X^2H^2$  & {MC} & \cite{Silva:2025hzo} \\
$\mu^+\mu^-\to \nu_\mu\overline{\nu}_\mu t\overline{t}h$ &  $f^2H^3$  & MC & \cite{Barger:2023wbg} \\
$W_LW_L\to 2h,3h,4h$ &  $H^4D^2$, dim-8  & MC & \cite{Delgado:2023ynh} \\
$\Gamma_h$ &  all except $f^2XH$  & - & \cite{Brivio:2019myy} \\
$h\to \gamma\gamma,Z\gamma,ZZ^*$,$Z\to f\overline{f}$ &  $H^4D^2$, $f^2HD$, $f^4$, dim-8  & geo & \cite{Hays:2020scx} \\
$h\to gg,\gamma\gamma,Z\gamma$ &  $H^4D^2$, $X^2H^2$, dim-8  & UV models & \cite{Grojean:2024tcw} \\
\bottomrule
\end{tabular}
\caption{LO analyses of Higgs physics in the SMEFT.}
\label{tab:LOSMEFTobsHiggs}
\end{table}

Finally, a guiding principle to constrain and correlate WCs are the symmetry arguments. For instance, correlations resulting from $\text{SU(2)}_L$-invariance is the key feature of SMEFT. A numerous studies have exploited this symmetry to relate different processes. A non-exhaustive list of such references include \cite{Bhattacharya:2014wla,Alonso:2014csa,Bause:2020auq,Karmakar:2024gla,Cirigliano:2017tqn,Aebischer:2018iyb,Kumar:2018kmr,Buras:2014fpa}. At tree-level all such $SU(2)_L$ relations directly follow from the matching conditions between SMEFT and WET \cite{Jenkins:2017jig,Hamoudou:2022tdn }.

A rigorous discussion on accidental symmetries in the SMEFT and in general EFTs is provided in \cite{Grinstein:2024jqt}. {Lately also goofy symmetries have been discussed in the literature \cite{deBoer:2025jhc,Trautner:2025prm}, which transform the kinetic term in a non-trivial way and which leave parts of the renormalization group invariant.}

\begin{table}[H]
\centering
\resizebox{\columnwidth}{!}{%
\begin{tabular}{lccc}
\toprule
process  & operators & comments & ref
\\
\midrule
$pp\to ZZ,Z\gamma$  &  dim-8  & positivity & \cite{Bellazzini:2018paj} \\
$pp\to WZ,W\gamma$ &  all except $H^6$  & $\cancel{\text{\text{CP}}}$, $U(1)^{14,13}$ & \cite{Degrande:2021zpv} \\
$pp\to WW$ &  $X^3$, $X^2H^2$  & - & \cite{Degrande:2012wf} \\
$pp\to WW,WZ$ &  $X^3$, $H^4D^2$, $X^2H^2$, $f^2H^2D$  & aTGC & \cite{Falkowski:2016cxu} \\
$pp\to WW,WZ$ &  $X^3$  & aTGC & \cite{ElFaham:2024uop} \\
$pp\to W^\pm Z,W^+W^-$ &  $X^3$, $X^2H^2$  & jet substructure & \cite{Aoude:2019cmc} \\
$pp\to VVV$ &  $X^3$, $H^4D^2$, $X^2H^2$, $f^2H^2D$, $f^4$  & flavour universality & \cite{Celada:2024cxw} \\
$pp\to (W \to q\overline{q}) (Z \to \ell\overline{\ell})$ &  $X^3$  & ML & \cite{Chatterjee:2024pbp} \\
$pp\to W^\pm (\ell^\pm \nu)\gamma$ &  $X^3$, $X^2H^2$, $f^2H^2D$, $f^4$, dim-8  & amp. & \cite{Martin:2023tvi} \\

$q\overline q\to WW,WZ$  & $X^3$, dim-8  & $V_{\text{CKM}}$ diag. & \cite{Degrande:2023iob} \\
$f\overline f\to ZZ,\gamma Z$  & dim-8  & nTGC & \cite{Degrande:2013kka} \\
$VV'\to VV'$, $pp\to VV'(jj)$ &   $X^3$, $X^2H^2$, $f^2H^2D$, $f^2H^2D$  & MC, $\text{SU(3)}^5$ & \cite{Ethier:2021ydt} \\
$Z(+$jet) prod. &  $f^2XH$  & Lam-Tung rel. & \cite{Gauld:2024glt} \\
$Z\to 4\ell$ &  $H^4D^2$, $X^2H^2$, $f^2H^2D$, $f^4$  & flavor univ. & \cite{Boughezal:2020klp} \\
$Z\to c\bar{c}$ &  $f^4$  & {FCC-ee/CEPC} & \cite{DiCanto:2025fpk} \\
$M_W,\Gamma_W$ &  $H^4D^2$, $X^2H^2$, $f^2H^2D$, $f^4$  & - & \cite{Bjorn:2016zlr} \\
oblique parameters &  $X^3$, $H^4D^2$, $X^2H^2$, $f^2H^2D$, $f^4$  & spin sum rules & \cite{Remmen:2022orj} \\
oblique parameters &  $X^3$, $H^6$, $H^4D^2$, $f^2H^3$, $X^2H^2$, $f^2H^2D$, $f^4$  & universal EFTs & \cite{Wells:2015uba,Wells:2015cre} \\
$2\to 2$ with $t$ and $W/Z$  & all except $f^4$  & $\text{U(2)}^2\times \text{U(3)}^3$ & \cite{Maltoni:2019aot} \\
$pp\to t\overline{t}Z, tZj$ &  $f^2H^3$, $f^2XH$, $f^2H^2D$  & MC, ML & \cite{Barman:2022vjd} \\
$pp\to t\overline{t} t\overline{t}$  & all except $H^6$  & MC, $\text{U(2)}^2\times \text{U(3)}^3$ & \cite{Aoude:2022deh} \\
$e^+e^-\to t\overline{t} $  & $f^2XH$, $f^2H^2D$, $f^4$ & entanglement & \cite{Maltoni:2024csn}
\\
$e^+e^-\to WW$  & $X^3$, $f^2H^2D$ & aTGC & \cite{Jahedi:2024wnw}
\\
$e^+e^-\to ZZZ^*,ZZ\gamma^*$  & dim-8 & aTGC & \cite{Ellis:2025jgt}
\\
$e^+e^-\to Z\gamma^*$  & dim-8 & aTGC & \cite{Ellis:2025ghl}
\\
$e^+e^-\to Zh $  & $X^2H^2$ & ILC & \cite{Bhattacharya:2025jhs}
\\
\bottomrule
\end{tabular}
}
\caption{LO analyses involving EW gauge bosons and top quarks in the SMEFT.}
\label{tab:LOSMEFTobsEW}
\end{table}

\begin{table}[H]
\centering
\begin{tabular}{lccc}
\toprule
process  & operators & comments & ref
\\
\midrule
$\overline{f}f\to \overline{f_1'}f_2'\overline{f_3'}f_4'$ &  $X^3$, $H^4D^2$, $X^2H^2$, $f^2H^2D$, $f^4$  & $m_W^2/2\to 0$ & \cite{Helset:2017mlf} \\
$pp\to jj$ &  $X^3$, $f^4$  & MC & \cite{Alte:2017pme} \\
$pp\to jj$ &  $X^3$  & MC & \cite{Goldouzian:2020wdq} \\
$p\overline{p}\to jj$ &  $X^3$, $f^4$  & MC & \cite{Keilmann:2019cbp} \\
$pp\to \ell^-\ell^+,\ell^+\nu,\ell^-\overline\nu$  &  dim-8  & $U(3)^5$ & \cite{Kim:2022amu} \\
$pp,p\overline{p}\to \ell^+\ell^-$ &  $f^4$  & SL & \cite{Alte:2018xgc} \\
$pp\to \ell^+\ell^-$ &  $X^2H^2$, $f^2H^2D$, $f^4$  & MC & \cite{Horne:2020pot} \\
DIS \& DY &  $f^4$  & flavour univ. & \cite{Boughezal:2020uwq} \\
DY &  $f^4$  & binning & \cite{Grossi:2024tou} \\
DY &  $f^2XH$, $f^4$  & MET \& jet & \cite{Hiller:2025zov} \\
$\ell^+\ell^-\to f\overline{f}$ &  $X^2H^2$, $f^2HD$, $f^4$  & $\text{U(3)}^5$ & \cite{Berthier:2015oma} \\
$e^+e^-\to W^+W^-\to f_1\overline{f}_2f_3\overline{f}_4$ &  $X^3$, $H^4D^2$, $X^2H^2$, $f^2H^2D$, $f^4$  & $\text{U(3)}^3$ & \cite{Berthier:2016tkq} \\
$e^+e^-\to Z\gamma(\ell\bar\ell\gamma)$ &  dim-8  & aTGC & \cite{Liu:2024tcz} \\
$e^+e^-\to 2\ell 2j \slashed{E}$ &  $X^3$, $X^2H^2$, $f^2H^2D$,  & aTGC & \cite{Subba:2025pos} \\
$\mu^+\mu^-\to \gamma \nu \bar\nu$  & dim-8 & aTGC & \cite{Xie:2025izk}\\
\bottomrule
\end{tabular}
\caption{LO analyses involving quark-quark and quark-lepton interactions, including deep inelastic scattering (DIS) and Drell-Yan (DY) processes in the SMEFT.}
\label{tab:LOSMEFTobsqqll}
\end{table}

\begin{table}[H]
\centering
\resizebox{\columnwidth}{!}{%
\begin{tabular}{lccc}
\toprule
process  & operators & comments & ref
\\
\midrule
DM &   $X^3$, $H^6$, $H^4D^2$, $X^2H^2$, $f^2H^2D$, $f^4$  & {MFV} & \cite{Biondini:2025gpg} \\
DM &   $f^2H^2D$  & {Tera-Z} & \cite{Olgoso:2025jot} \\
grav. waves &   $H^6$, $H^4D^2$, $f^2H^3$, $f^2H^2D$, dim-8  & - & \cite{Hashino:2022ghd} \\
grav. waves &   $X^3$, $H^6$, $H^4D^2$, $f^2H^3$, $X^2H^2$, $f^2XH$, $f^2H^2D$, $f^4$  & RGEs & \cite{Hashino:2025nku} \\
EW phase tr., grav. waves &   $H^6$, $H^4D^2$  & - & \cite{Banerjee:2024qiu} \\
Single Transv. Spin Asym.  & $f^2XH$  & - & \cite{Wen:2023xxc} \\
$ggg$  & $X^3$  & MC, see also \cite{Krauss:2016ely} & \cite{Hirschi:2018etq} \\
$ggg$  &  $X^3$  & interference & \cite{Degrande:2020tno} \\
$p\to M\ell$, $n\to M\ell$  &  $f^4$, dim-7,8,9  & - & \cite{Gargalionis:2024nij} \\
\bottomrule
\end{tabular}
}
\caption{LO analyses of miscellaneous processes in the SMEFT.}
\label{tab:LOSMEFTobsmisc}
\end{table}

\subsection{NLO}
In this subsection we collect NLO SMEFT analyses, which are grouped into different types of observables. They include flavour transitions (Tab.~\ref{tab:NLOSMEFTobsflavor}), low energy observables (Tab.~\ref{tab:NLOSMEFTobslowen}), Higgs physics (Tab.~\ref{tab:NLOSMEFTobsHiggs}) as well as EWP and top processes (Tab.~\ref{tab:NLOSMEFTobsEW}). NLO analyses involving neutrinos are collected in Tab.~\ref{tab:NLOSMEFTobsneut}. As NLO analyses we consider here articles that go beyond tree-level matching and one-loop running analyses. This entails for instance including one-loop matrix elements like in several MC simulations. 

At the one-loop level gauge anomaly cancellations were discussed in \cite{Cata:2020crs} and more recently in \cite{Beis:2025zzd}. It was shown in \cite{Cohen:2023gap,Cohen:2023hmq}, using the Covariant Derivative Expansion, that anomaly cancellation does not lead to any new constraints on the SMEFT WCs. Constraints from the Veltman condition in order to solve the hierarchy problem are studied in \cite{Biswas:2020abl}.

\begin{table}[H]
\centering
\begin{tabular}{lccc}
\toprule
process  & operators & comments & ref
\\
\midrule
$\Delta F =2$ & $f^2XH$, $f^2H^2D$, $f^4$ & $f=q,u,d$ & \cite{Aebischer:2020dsw} \\
$\Delta F =2$ & $H^4D^2$, $f^2H^2D$, $f^4$ & 1L matching & \cite{Endo:2018gdn} \\
$b\to s\ell\ell$  & $f^2H^2D$, $f^4$  & 1L matching &  \cite{Endo:2020kie} \\
$\mu\to e\gamma,3e$, $\mu N\to eN$  & $f^2H^3$, $f^2XH$, $f^2H^2D$, $f^4$  & OS methods &  \cite{EliasMiro:2021jgu} \\
\bottomrule
\end{tabular}
\caption{NLO flavour analyses in the SMEFT.}
\label{tab:NLOSMEFTobsflavor}
\end{table}

\begin{table}[H]
\centering
\begin{tabular}{lccc}
\toprule
process  & operators & comments & ref
\\
\midrule
$a_\tau$, $d_\tau$ & $f^2XH$ & MC, $pp\to\tau\tau$  & \cite{Haisch:2023upo} \\
$d_\ell,d_n$ & all except $H^6$ & -  & \cite{Kley:2021yhn} \\
$d_e$ & $X^3$, $f^2H^3$, $X^2H^2$, $f^2XH$, $f^4$, dim-8 & -  & \cite{Ardu:2025rqy} \\
$\beta$ decay & $H^4D^2$, $f^2H^2D$, $f^4$ & LFU  & \cite{Dawid:2024wmp} \\
nucleon EDMs & $f^2H^2D$, $f^4$ & 1L matching  & \cite{Endo:2019mxw} \\
\bottomrule
\end{tabular}
\caption{NLO analyses of low energy observables like electric dipole moments (EDMs) in the
SMEFT.}
\label{tab:NLOSMEFTobslowen}
\end{table}

\begin{table}[H]
\centering
\resizebox{\columnwidth}{!}{%
\begin{tabular}{lccc}
\toprule
process  & operators & comments & ref
\\
\midrule
$pp\to hj$  &  $f^2H^3$, $X^2H^2$, $f^2XH$ & MC & \cite{Grazzini:2018eyk} \\
$pp\to ht\overline{t}$  & $f^2H^3$, $X^2H^2$, $f^2XH$, $f^4$  & MC &  \cite{Maltoni:2016yxb} \\
$pp\to ht\overline{t}$  & all except $f^2H^2D$  & $f=t$ &  \cite{DiNoi:2023onw} \\
$pp\to thj$  &   $H^4D^2$, $f^2H^3$, $f^4$ & MC & \cite{Bhattacharya:2022kje} \\
$pp\to thj, tZj$ & all except $H^6$ &  MC & \cite{Degrande:2018fog} \\
$pp \to th, t\gamma,  tZ$ & $f^2H^3$, $f^2XH$, $f^2H^2D$   & MC, $f=q,t$ & \cite{Degrande:2014tta} \\
$pp\to W/Z+h$, DY, VBF  &  $f^2H^3$, $X^2H^2$, $f^2XH$, $f^2H^2D$, $f^4$ & MC & \cite{Alioli:2018ljm} \\
$pp\to Zh,t\overline{t}$  & all except $X^3$, $H^6$  & ML &  \cite{GomezAmbrosio:2022mpm} \\
$pp\to Zh\to \ell^+\ell^-h$  & $H^4D^2$, $X^2H^2$, $f^2H^2D$, $f^4$  & NNLO+PS &  \cite{Gauld:2023gtb} \\
$pp\to Zh\to \ell^+\ell^-b\overline{b}$ & $X^3$, $H^4D^2$, $X^2H^2$, $f^2H^4$, $f^2XH$   &  MFV, $V_{\text{CKM}}=\mathbb{1}$ & \cite{Haisch:2022nwz} \\
$pp\to W^+h,hjj$ & $f^2H^2D$, $X^2H^2$ & SILH  & \cite{Degrande:2016dqg} \\
$pp\to Zh/Wh$ & $X^3$, $X^2H^2$ & \cancel{CP}  & \cite{Rossia:2024rfo} \\
$pp\to Wh$ &  $f^2XH$, $f^2H^2D$ & -  & \cite{Bonetti:2025hnb} \\
$pp\to W^{\pm}h,Zh,W^{+}W^{-},W^{\pm}Z$  & $X^3$, $H^4D^2$, $X^2H^2$, $f^2H^2D$  & MC &  \cite{Baglio:2020oqu} \\
$pp\to VV^{(')}V^{('')}$  & $X^3$, $H^4D^2$, $X^2H^2$  & MC &  \cite{Bellan:2023efn} \\
$pp\to V(\ell\ell)h(bb)$  & $H^4D^2$, $f^2H^2D$, $f^2H^3$, $X^2H^2$ & MFV &  \cite{Banerjee:2019twi} \\
$pp\to t\overline{t}W/Z/h+j$  & $f^2H^2D$, $f^2H^3$, $f^2XH$ & MC & \cite{Goldouzian:2020ekx} \\
$e^+e^-\to ZH$  & $H^6$, $H^4D^2$, $f^2H^3$, $f^2XH$, $f^2H^2D$, $f^4$ & $V_{\text{CKM}}=\mathbb{1}$ & \cite{Asteriadis:2024xts} \\
$gg\to h$ & $f^2H^3$, $X^2H^2$, $f^2XH$ & $f=q^3,t$  & \cite{Deutschmann:2017qum} \\
$gg\to h$ & all sectors & double ins.  & \cite{Asteriadis:2022ras} \\
$gg\to h$, $h\to gg,\gamma\gamma,ff$  &  all except $X^3$ & geo & \cite{Corbett:2021cil,Martin:2023fad} \\
$gg\to h\to \gamma\gamma$  &  $f^2H^3$, $X^2H^2$, $f^2XH$, $f^4$ & {$f^4$ contributions} & \cite{Haisch:2025vqj} \\
$gg\to h$ &  $X^3$, $X^2H^2$, $f^2XH$ & 2L ADM & \cite{Haisch:2025lvd} \\
$gg\to hh$  &  $H^6$, $H^4D^2$, $f^2H^3$, $X^2H^2$, & MC & \cite{Heinrich:2022idm} \\
$gg\to hh$  & all except $X^3$, $f^2H^2D$  & MC &  \cite{Heinrich:2023rsd} \\
$gg\to h$  & $f^2H^3$, $X^2H^2$, $f^2XH$, $f^2H^2D$, $f^4$  & dif. schemes &  \cite{DiNoi:2023ygk} \\
$gg\to Zh$, $pp\to t\bar tZ(\gamma)(\mu^+\mu^-)$ & $f^2XH$, $f^2H^2D$ & --  & \cite{BessidskaiaBylund:2016jvp} \\
$gg\to hh,Zh,ZZ,WW$ & all except $X^3$, $f^4$ & MC  & \cite{Rossia:2023hen} \\
$gg\to hh,ZZ,WW,\gamma Z,\gamma\gamma$ &  $f^2H^3$, $X^2H^2$, $f^2XH$ & \cancel{CP}  & \cite{Thomas:2024dwd} \\
$h$ prod./decay & $H^4D^2$, $f^2H^3$, $X^2H^2$, $f^2XH$, $f^2H^2D$ & top loop  & \cite{Vryonidou:2018eyv} \\
$h$ prod./decay & $f^4$ & $f=t_{L,R}, b_L$  & \cite{Alasfar:2022zyr} \\
$h\to b\overline{b}, \tau\overline{\tau}$ & all sectors   & $m_t\to \infty$ for $\tau\overline{\tau}$ & \cite{Gauld:2015lmb,Gauld:2016kuu,Cullen:2019nnr} \\
$h\to c\overline{c}, \tau\overline{\tau}, \mu\overline{\mu}$ & all sectors   & MFV & \cite{Cullen:2020zof} \\
$h\to 4\ell$  &  $H^4D^2$, $f^2H^3$, $X^2H^2$, $f^2XH$, $f^2H^2D$ & MC & \cite{Banerjee:2020vtm} \\
$h\to \overline{\ell}\ell Z$, $Z\to \overline{\ell}\ell$  &  $H^4D^2$, $X^2H^2$, $f^2H^2D$, $f^4$ & EW & \cite{Dawson:2024pft} \\
$h\to \overline{\ell}\ell\gamma$  &  $f^2H^3$, $X^2H^2$, $f^2H^2D$, dim-8 & geo & \cite{Corbett:2021iob} \\
$h \to \gamma\gamma$ & all sectors & see also \cite{Manohar:2006gz,Grojean:2013kd,Ghezzi:2015vva,Vryonidou:2018eyv,Hartmann:2015aia,Hartmann:2015oia,Dawson:2018liq} & \cite{Dedes:2018seb} \\
$h\to Z\gamma$ & all except $H^6$ &  see also \cite{Ghezzi:2015vva,Cirigliano:2016nyn,Vryonidou:2018eyv} & \cite{Dedes:2019bew,Dawson:2018pyl} \\
$h\to ZZ^*$ & all sectors & -  & \cite{Dawson:2018pyl} \\
$h\to WW^*$ & all except $H^6$ & $C_i = C_i \mathbb{1}$  & \cite{Dawson:2018liq} \\
$Vh(\to b\overline{b})$ & $f^2H^2D$ & MC  & \cite{Bishara:2022vsc} \\
$h\to (Z/W)f\bar f,(Z)gg,\gamma\gamma,\gamma Z$ & all except $X^3$ & {$U(3)^5$}  & \cite{Bellafronte:2025jbk} \\
\bottomrule
\end{tabular}
}
\caption{NLO Higgs physics analyses in the SMEFT.}
\label{tab:NLOSMEFTobsHiggs}
\end{table}

\begin{table}[H]
\centering
\begin{tabular}{lccc}
\toprule
process  & operators & comments & ref
\\
\midrule
$pp\to WZ\to (\ell^+\ell^-)(\ell'\overline{\nu}')$  & $X^3$, $H^4D^2$, $X^2H^2$, $f^2H^2D$  & MC &  \cite{Baglio:2019uty} \\
$pp\to Zjj,WZ,W\gamma$  & $X^3$ & MC &  \cite{Degrande:2024bmd} \\
$pp\to W^\pm Z$  & $X^3$, $X^2H^2$ & {$C_i \in \mathbb{R}$} &  \cite{Haisch:2025jqr} \\
$pp\to Z^*,\gamma^*\to \ell^+\ell^-$  &  all except $H^6$, $f^2H^3$ & -  & \cite{Dawson:2021ofa} \\
$pp\to W^+W^-$ & $X^3$, $H^4D^2$, $X^2H^2$, $f^2H^2D$, $f^4$ &  - & \cite{Baglio:2017bfe} \\
$pp\to \ell^+\ell^-$  & $f^2XH$, $f^2H^2D$, $f^4$, dim-8  & $\mathcal{O}(1/\Lambda^4)$ &  \cite{Boughezal:2021tih} \\
$f\bar f\to l^+l^-$ & $X^3$, $H^4D^2$, $X^2H^2$, $f^2H^2D$, $f^4$ & -  & \cite{Dawson:2018dxp} \\
EWPOs & all sectors & general flav. & \cite{Biekotter:2025nln} \\
$Z$, $W$ pole obs.  & all except $H^6$ & $V_{\text{CKM}}=\mathbb{1}$  & \cite{Dawson:2019clf,Dawson:2022bxd} \\
$\Gamma_Z, m_Z$  &  all except $X^3$ & $\text{U(3)}^5$ &  \cite{Hartmann:2016pil} \\
$pp\to t\bar tj$ & $f^2XH$, $f^4$ & MC, $A_E(\theta_J)$  & \cite{Basan:2020btr} \\
$pp,\mu\mu\to t\bar t$, DY & $f^4$ &  EW & \cite{ElFaham:2024egs} \\
$pp\to tWZ$ & $X^3$, $H^4D^2$, $X^2H^2$, $f^2XH$ & $\text{U(2)}^2\times \text{U(3)}^3$  & \cite{Faham:2021zet,Keaveney:2021dfa} \\
$t\overline{t}$ prod.  &  $f^2XH$ & MC, NNLO & \cite{Kidonakis:2023htm} \\
$pp\to t\overline{t}$  &  $X^3$, $X^2H^2$, $f^2XH$, $f^4$  & entanglement &  \cite{Aoude:2022imd} \\
$t$-prod. & $f^2XH$, $f^2H^2D$, $f^4$ & MC, 3rd gen.  & \cite{Zhang:2016omx} \\
$t\to Wb$ & $f^2XH$, $f^2H^2D$, $f^4$ & - & \cite{Boughezal:2019xpp} \\
$t$ decays & $f^2H^3$, $f^2XH$, $f^2H^2D$, $f^4$ & $f=q,u$  & \cite{Zhang:2014rja} \\
$t$ spin & $f^2XH$, $f^2H^2D$, $f^4$ & MC, $\text{U(2)}^3$  & \cite{Severi:2022qjy} \\
$t$ FCNC & $f^2H^3$, $f^2XH$, $f^2H^2D$, $f^4$ &  MC, up-basis & \cite{Durieux:2014xla} \\
\bottomrule
\end{tabular}
\caption{NLO analyses involving EW gauge bosons and top quarks in the SMEFT.}
\label{tab:NLOSMEFTobsEW}
\end{table}

\begin{table}[H]
\centering
\begin{tabular}{lccc}
\toprule
process  & operators & comments & ref
\\
\midrule
$pp\to \nu \overline{\nu}+X$  & $f^2XH$, $f^2H^2D$, $f^4$  & MC &  \cite{Hiller:2024vtr} \\
\bottomrule
\end{tabular}
\caption{NLO analyses involving neutrinos in the SMEFT.}
\label{tab:NLOSMEFTobsneut}
\end{table}

\subsection{Global fits of SMEFT WCs}

In this subsection we collect references to global SMEFT fits and describe them briefly. They involve flavour physics in Tab.~\ref{tab:fitSMEFTobsflavor}, top physics in Tab.~\ref{tab:fitSMEFTobstop}, EW precision and Higgs physics in Tab.~\ref{tab:fitSMEFTobsEWPO} and collider physics in Tab.~\ref{tab:fitSMEFTobscoll}. Furthermore, we mention several analyses that discuss important aspects of SMEFT fits. For instance, a process specific methodology to systematically assign theoretical uncertainties in the SMEFT within LO global fits has been proposed in \cite{Trott:2021vqa}. A general framework to perform simultaneous fits to pdfs and WCs is presented in \cite{Iranipour:2022iak}. In \cite{vanBeek:2019evb} the method of Bayesian reweighting, which allows to incorporate constraints from novel measurements into MC analyses is discussed. {A novel strategy to find regions in parameter space where the SMEFT is more favourable than the SM and which are well suited for discovery is discussed in \cite{Hirsch:2025qya}.} Differences between the profile likelihood method and Baysian marginalization have been studied in the case of Higgs physics in \cite{Brivio:2022hrb}. The efficient evaluation of profile likelihoods using machine learning (ML) techniques is discussed in \cite{Heimel:2024drk}. The importance of RGE effects in global fits was studied in \cite{Dawson:2020oco}, assuming different UV completions of the SMEFT. EFT uncertainties in LHC searches are discussed in detail in \cite{Chang:2025ohh}. Finally, in \cite{Camponovo:2022wwn} randomly generated SMEFT WCs were treated as pseudo-data in a bottom-up approach, in order to find deviations from the SM around the scale of NP physics $\Lambda$.

\begin{table}[H]
\centering
\begin{tabular}{lccc}
\toprule
process  & operators & comments & ref
\\
\midrule
$b\to u\ell\nu$  & $f^2H^2D$, $f^4$  & real WCs & \cite{Greljo:2023bab} \\
$b\to c\ell\nu$  & $f^2H^2D$, $f^4$  & - & \cite{Jung:2018lfu} \\
$b\to c \tau\nu$  &  $f^2H^2D$, dim-8 & no $f^4$ & \cite{Burgess:2021ylu} \\
$b\to q\ell\ell$ \& DY &  $f^4$ & - & \cite{Greljo:2022jac} \\
$b\to s\ell\ell$ \& $B\to \pi K$ &  $f^2H^2D$, $f^4$ & - & \cite{Datta:2025csr} \\ 
$b\to s\mu^{\pm} \tau^{\mp}$  &   $f^4$ & real WCs & \cite{Panda:2024ygr} \\
$B\to K^{*0}\tau^+\tau^-$  &  $f^4$ & angular distr. & \cite{Karmakar:2024dml} \\
$B\to K\nu \overline{\nu}$  &  $f^4$ & 3rd gen. NP & \cite{Chen:2024jlj} \\
$B\to K\nu \overline{\nu}$  &  $f^2H^2D$, $f^4$ & MFV & \cite{Hou:2024vyw} \\
$B_s\to \mu^+\mu^-\gamma$ &  $f^2H^2D$, $f^4$ & $b\to s\mu^+\mu^-, b\to s\gamma$ & \cite{Guadagnoli:2023ddc} \\
$B\to a_1(1260)\ell^- \overline{\nu}_\ell$  &  $f^2H^2D$, $f^4$ & down-basis & \cite{Mohapatra:2024knf} \\
$\Lambda_b\to (\Lambda^*\mu^+\mu^-,\Lambda \nu\bar\nu)$  & $f^2H^2D$, $f^4$   & also $b\to s( \ell^+\ell^-,\nu\bar\nu)$ & \cite{Das:2023kch} \\
$\Lambda_c^-\to \Lambda(\to p\pi)\mu^-\bar\nu_\mu$  & $f^2H^2D$, $f^4$   & $\nu_R$ & \cite{Boora:2025odj} \\
\hline
$\tau\to \ell$, $\mu\to e$  & $f^2H^3$, $f^2XH$, $f^2H^2D$, $f^4$ & 1. gen. quarks & \cite{Fernandez-Martinez:2024bxg} \\
collider, flavour \& low-energy & $f^4$ & {muon coll.} & \cite{Glioti:2025zpn} \\
\bottomrule
\end{tabular}
\caption{Global fit analyses involving quark- and lepton-flavour transitions in the SMEFT.}
\label{tab:fitSMEFTobsflavor}
\end{table}

Furthermore, there were many recent developments in the context of top physics: The benefits from including high momentum transfer final states in top EFT-fits was studied in \cite{Englert:2016aei}. A note that proposes common SMEFT standards to interpret top-quark measurements at the LHC was presented in \cite{Aguilar-Saavedra:2018ksv}. Distorting effects from four-fermion operators on resonance-shape measurements in top pair production are discussed in \cite{Egle:2025buk}. The impact of one-loop SMEFT RGEs on $\overline{t}t$ production at the LHC and other collider observables were studied in \cite{Aoude:2022aro}.

Global analysis of the electroweak data can be found in \cite{deBlas:2021wap} and global SMEFT fits at future colliders in \cite{deBlas:2022ofj}. Other studies involving future colliders, like FCC-ee can be found in \cite{Celada:2024oax, Allwicher:2024sso,terHoeve:2025gey,terHoeve:2025zmp,Allanach:2025wfi,Chala:2025utt}. In particular in~\cite{terHoeve:2025gey} the role of RGE effects in global fits at LEP, LHC and future colliders has been stressed. See also \cite{Bartocci:2024fmm}. On the other hand as shown in \cite{Allwicher:2024sso} nearly every new particle which matches at tree-level to dim-6 operators onto the SMEFT affects EWPOs, either through tree-level matching or through RG running. Moreover, the effect of momentum-dependent particle widths and propagators on gauge and Higgs bosons and the top quark were studied in \cite{Englert:2025onf}. Finally, comprehensive discussions on Higgs physics prospects at future colliders are provided in \cite{deBlas:2019rxi,Blondel:2024mry}.

Another interesting avenue to study SMEFT effects are kinematic tails of non-resonant process at the LHC and future colliders, for which energy-enhanced operators play a key role. In this context a power counting based on the number of external legs was formalized in \cite{Assi:2025zmp}, which allows to identify the energy-enhanced contribution of higher-dimensional operators to such processes.

Using an effective field theory approach, coherent neutrino scattering on nuclei, in the setup pertinent to the COHERENT experiment has been investigated in \cite{Breso-Pla:2023tnz}. This work shows that COHERENT data should be included in electroweak precision studies. See also \cite{Coloma:2024ict}, where the impact of neutrino oscillations and of coherent elastic neutrino-nucleus scattering on global SMEFT fits has been investigated. Furthermore, it was demonstrated in \cite{Beltran:2025ilg} that for certain operators the LHC will be able to probe NP with heavy neutral leptons at scales as high as 12 TeV.

\begin{table}[H]
\centering
\begin{tabular}{lccc}
\toprule
process  & operators & comments & ref
\\
\midrule
$t$ &  $X^3$, $X^2H^2$, $f^2XH$, $f^2H^2D$, $f^4$  & MFV & \cite{Buckley:2015lku} \\
$t$ &  $X^3$, $X^2H^2$, $f^2XH$, $f^2H^2D$, $f^4$  & MC & \cite{Buckley:2015nca} \\
$t$ &  $f^2XH$, $f^2H^2D$, $f^4$  & various unc. correlations & \cite{Bissmann:2019qcd} \\
$t$ &  $X^3$, $f^2XH$, $f^2H^2D$, $f^4$  & LHC run II, $\text{U(2)}^3$ & \cite{Brivio:2019ius} \\
$t$  &   $f^2XH$, $f^2H^2D$, $f^4$ & $\text{U(2)}$ & \cite{Elmer:2023wtr} \\
$t$ prod. &  $X^3$, $f^2XH$, $f^2H^2D$, $f^4$  & $\text{U(2)}^3$ & \cite{Hartland:2019bjb} \\
$t$ prod.  &   $f^2XH$, $f^2H^2D$, $f^4$ & pdfs, ML & \cite{Kassabov:2023hbm} \\
$pp\to t\overline t (j)$ &  $f^2XH$, $f^4$ & WCs \& PDFs, ML & \cite{Gao:2022srd} \\
$t$ \& EWPO &  $f^4$  & {$\gamma_5$} & \cite{DiNoi:2025uhu} \\
$t$, $b$, $Z$ &  $f^2XH$, $f^2H^2D$, $f^4$  & up-basis & \cite{Bissmann:2020mfi} \\
$t$, $b$, $Z$, DY &  $f^2XH$, $f^2H^2D$, $f^4$  & MFV & \cite{Grunwald:2023nli} \\
$t$, $h$, $WW$ \& EWPO &  all except $H^6$, $f^2XH$  & MFV & \cite{Ellis:2020unq} \\
$t$, $h$ \& diboson &  all except $H^6$  & MFV & \cite{Ethier:2021bye} \\
$h$, $t$, low-energy  &   all, except $X^3$, $H^6$ & ALP & \cite{Biekotter:2023mpd} \\
$t$, flavour \& EDMs &  all except $H^6$, $f^2HD$  & - & \cite{Cirigliano:2016nyn} \\
$t$, flavour \& EDMs  &  $f^2XH$, $f^2H^2D$, $f^4$ & $f=q,t$ & \cite{Garosi:2023yxg} \\
$t$, flavour, EWPO \& dijet  &  $f^2XH$, $f^4$ & MFV & \cite{Bruggisser:2022rhb} \\
$t$, $h$, DY, EWPO \& flavor &  all except $H^6$  & {$U(3)^5$, $U(2)^5$} & \cite{deBlas:2025xhe} \\
\bottomrule
\end{tabular}
\caption{Global fit analyses involving top quarks in the SMEFT.}
\label{tab:fitSMEFTobstop}
\end{table}

The effects of the one-loop renormalization group running/mixing of the Wilson coefficients in the SMEFT on the predictions for Higgs production at the LHC has been investigated in \cite{Maltoni:2024dpn}. The impact of these RG effects on the constraints that can be obtained on the relevant WCs through current and future measurements at the LHC has been investigated.

In \cite{Gisbert:2024sjw} indirect constraints on top quark operators that violate baryon number by one unit above the TeV scale were studied. One finds that these constraints are typically many orders of magnitude more stringent than the recently derived direct bounds from collider experiments. 

A global analysis of DY production of charged leptons and dineutrinos, the latter in missing energy plus jet events, in proton-proton collisions within the SMEFT was performed very recently in \cite{Hiller:2025hpf}, paying in particular attention to flavour violating effects. DY processes were also analyzed recently in the Universal SMEFT including dimension-eight operators \cite{Corbett:2025oqk}.

A very systematic recent analysis that includes the processes from Classes 1, 2, 4, 8 and 9 has been performed in \cite{Kala:2025srq}. It contains many references to previous analyses of anomalous $Wtb$ couplings obtained separately from different processes. Performing a global SMEFT analysis of a multitude of observables that constrained the WCs of the relevant operators allowed eventually to predict the branching ratios for top-FCNC processes such as $t\to u(c)\gamma$, $t\to u(c)Z$ and $t\to u(c)H$.

The implications of dimension-six operators on vector boson scattering in the $pp\to ZZjj$ channel have been investigated in \cite{Gomez-Ambrosio:2018pnl}. It was emphasized that operators of dimension-six should not be neglected in favour of those of dimension-eight.

An implementation to compute the likelihood of EWPOs for the LHC in the SMEFT is provided in \cite{Mildner:2024wbl}. The SMEFT predictions include NLO effects and are provided for five different input schemes. Systematic uncertainties in unbinned LHC data, focusing on SMEFT WCs constraints were discussed in \cite{Schofbeck:2024zjo}. 

\begin{table}[H]
\centering
\resizebox{\columnwidth}{!}{%
\begin{tabular}{lccc}
\toprule
process  & operators & comments & ref
\\
\midrule
EWPO  & $X^3$, $X^2H^2$, $f^2H^2D$  & red. basis & \cite{Falkowski:2014tna} \\
EWPO &   all sectors & - & \cite{Efrati:2015eaa} \\
EWPO  &  $H^4D^2$ $X^2H^2$, $f^4$ & NLO, $\text{U(3)}^3$ & \cite{Bellafronte:2023amz} \\
EWPO \& $\Delta_{\text{CKM}}$ &  $H^4D^2$, $f^2H^2D$, $f^4$ & 1L matching & \cite{ThomasArun:2023wbd} \\
EWPO \& $M_W$ &  $X^2H^2$, $f^2H^2D$ & - & \cite{Fan:2022yly} \\
EWPO \& Triple Gauge   &  $X^3$, $H^6$, $X^2H^2$, dim-8 & MC & \cite{Corbett:2023qtg} \\
EWPO \& $h$  &  $X^3$, $H^6$, $H^4D^2$ $X^2H^2$, $f^2H^2D$, $f^4$ & RGE & \cite{Elias-Miro:2013eta} \\
EWPO \& $h$  & $H^6$, $H^4D^2$ $f^2H^2D$, $f^4$ & portal & \cite{Ahmed:2024hpg} \\
EWPO \& $h$ \& flavour  & $X^3$, $H^4D^2$, $f^2H^3$, $X^2H^2$, $f^2H^2D$    & no flavor symm. & \cite{Falkowski:2019hvp} \\
EWPO, $h$ \& $M_W$ &  all sectors  & $S\text{U(3)}^5$ symmetry & \cite{Bagnaschi:2022whn} \\
EWPO, $h$ \& diboson   &  all except $H^6$ & $\text{U(3)}^5$ & \cite{Ellis:2018gqa} \\
EWPO, low-en. \& DY  &  $H^4D^2$, $X^2H^2$, $f^2H^2D$, $f^4$ & $\text{U(3)}^5$ & \cite{Cirigliano:2023nol} \\
EWPO, flav. \& DY  &  $f^4$, dim-8 & $Z', U_1$ & \cite{Allwicher:2024mzw} \\
$pp\to WW,WZ$  &   $X^3$, $H^4D^2$, $f^4$ & aTGC & \cite{Grojean:2018dqj} \\
$gg\to WW$  &   dim-8 & -- & \cite{Gillies:2024mqp} \\
$pp\to Zh$  &  $H^4D^2$, $X^2H^2$, $f^2XH$, $f^2H^2D$ & ML & \cite{Bhattacharya:2024sxl} \\
$h$ prod.  & $X^2H^2$  & $\cancel{CP}$ & \cite{Bernlochner:2018opw} \\
$h$-prod./decay  &   $H^6$, $H^4D^2$, $f^2H^3$, $X^2H^2$, $f^4$ & regularized lin. regr. & \cite{Murphy:2017omb} \\
$h$-prod./decay  &   all, except $H^6$ & ATLAS & \cite{ATLAS:2022xyx} \\
$h$ \& $WW$ & $X^3$, $H^4D^2$, $f^2H^3$, $f^2H^2D$ & aTGC & \cite{Falkowski:2015jaa}\\
$h$ pair prod. & $H^6$, $H^4D^2$, $f^2H^3$, $X^2H^2$, $f^2XH$, $f^2H^2D$, $f^4$ & HL-LHC, FCC-ee & \cite{terHoeve:2025yup}\\
collider, flav., EWPO \& $h$ &  $X^3$, $H^6$, $H^4D^2$ $X^2H^2$, $f^2H^2D$, $f^4$ & FCC-ee & \cite{Maura:2025rcv} \\
EWPO \& flav. &  $f^2H^2D$, $f^4$ & FCC-ee & \cite{Allwicher:2025bub} \\
EWPO  & $X^3$, $H^6$, $H^4D^2$, $X^2H^2$, $f^2H^2D$, $f^4$ & FCC-ee & \cite{Maura:2024zxz} \\
$Z\to b\bar b$ \& oblique parameters  & $f^2H^2D$ & {FCC-ee, MSSM} & \cite{Greljo:2025ggc} \\
EWPO \& $h$ \& EDMs \& LFV & $f^2H^3$, $X^2H^2$, $f^2XH$, $f^4$ & {\cancel{CP}} & \cite{Kosnik:2025srw}\\
\bottomrule
\end{tabular}
}
\caption{Global fit analyses involving EW precision observables (EWPO) and Higgs physics in the SMEFT.}
\label{tab:fitSMEFTobsEWPO}
\end{table}

\begin{table}[H]
\centering
\begin{tabular}{lccc}
\toprule
process  & operators & comments & ref
\\
\midrule
$A_{FB}$  & $X^2H^2$, $f^2H^2D$, $f^4$,   & $Z$ pole & \cite{Breso-Pla:2021qoe} \\
$A_{FB}$ for DY   &  $f^4$ & MFV & \cite{Boughezal:2023nhe} \\
DIS  &  $H^4D^2$, $X^2H^2$, $f^2H^2D$, $f^4$ & LHeC, FCC-eh, EIC & \cite{Bissolotti:2023vdw} \\
DIS  &  $f^4$ & $\cancel{P}$, MC & \cite{Boughezal:2022pmb} \\
collider  &  $f^4$ & PDFs & \cite{Greljo:2021kvv} \\
$pp\to \ell^+\ell^-$ &  $f^4$, dim-8 & (HL-)LHC & \cite{Boughezal:2022nof} \\
neutral DY  &  $f^2H^2D$ & MC & \cite{Abdolmaleki:2023jvw} \\
DY  &  $H^4D^2$, $f^2H^2D$, $f^4$, dim-8 & $\mathcal{O}(1/\Lambda^4)$ effects &   \cite{Corbett:2024evt} \\
$e^--$scat. \& DY &  $f^4$, dim-8 & $\cancel{P}$ & \cite{Boughezal:2021kla} \\
collider, flav. \& low-en.  &  $X^2H^2$, $f^2XH$, $f^2H^2D$ & NLO, MC & \cite{Alioli:2017ces} \\
collider, flav. \& low-en.  &  $H^4D^2$, $X^2H^2$, $f^2H^2D$, $f^4$ & - & \cite{Falkowski:2017pss} \\
collider \& flav. &  $f^4$ & diquarks & \cite{Englert:2024nlj} \\
collider \& flav. &  $f^4$ & FCC & \cite{Greljo:2024ytg} \\
\bottomrule
\end{tabular}
\caption{Global fit analyses involving collider constraints, including Drell-Yan (DY) and deep inelastic scattering (DIS) processes in the SMEFT.}
\label{tab:fitSMEFTobscoll}
\end{table}

It has been demonstrated in \cite{Mantani:2025bqu} that the flavour blind SMEFT can be constrained by low energy observables including flavour transitions. High-$p_T$ LHC constraints on SMEFT operators affecting rare kaon and hyperon decays were analysed in \cite{Roy:2024avj}. An analysis of $b\to c\ell\nu$ baryonic decay modes in the SMEFT approach was performed in \cite{Panda:2024oam}. This includes the branching ratio, the forward-backward asymmetry parameter, the lepton non-universal observable and the longitudinal polarization fraction of the $b$-baryonic decay channels. A global SMEFT survey of lepton flavour universal and non-universal NP in $b\to s \ell^+\ell^-$ transitions was presented in \cite{Ali:2025xkw}.

Finally, we mention several explicit examples of constraints for purely bosonic operators that result from fitting procedures: The WCs $\Wc[]{H\widetilde B}, \Wc[]{H\widetilde W},\Wc[]{H\widetilde WB}$ are strongly constrained from CP-violating observables \cite{Cirigliano:2019vfc}, whereas $\Wc[]{H}$ is rather weakly constrained \cite{DiVita:2017vrr,Chala:2018ari,Henning:2018kys}.

\newpage
\part{SMEFT Atlas}
\section{Classification of Processes}\label{classification}

In the presentation of the operator landscape it is useful to group various processes into classes that describe different particle transitions such as particle-antiparticle mixings, semileptonic, non-leptonic and leptonic decays of mesons, charged lepton decays, electric dipole moments, $(g-2)_{\mu,e}$, charged current processes and additional specific classes to be listed below. Each of them is dominated by a well-defined set of operators that will be exhibited during the presentation of our atlas. We find that, although the tree-level WET operators are unique to different classes, this is not the case within SMEFT, even at the tree-level. Moreover the RG evolution is not closed within classes. Both of these features lead to correlations between different classes. The structure of RGEs will be visualized by the corresponding SMEFT charts. 

As following the presentation of the classes in question and of the related RGEs and charts will be rather demanding, this section and the following two sections can be considered as the base camp of our expedition so that the readers will get a general idea of what to expect in the subsequent ten sections. In this context some additional information provided in these three sections should facilitate the following of the rather technical material presented in this part of our review.

\subsection{Observable Classes}\label{sub:obs-classes}
We consider the following 10 classes of the most important observables in the particle physics phenomenology, that can be $\Delta F=0,1$ or 2. Here $F$ corresponds to the quark or lepton flavour. 

{\bf Class 1: Meson-Antimeson Mixing (Sec.~\ref{class1})}

\begin{center} 
{\bf 1A: Down type Mesons}
\end{center}
\be \label{Class1}
K^0-\bar K^0\,, \qquad  B_d^0-\bar B_d^0\,, \qquad  B_s^0-\bar B_s^0\,.
\ee

\begin{center}
{\bf 1B: Up type Meson}
\end{center}
\be
D^0-\bar D^0\,.
\ee         

{\bf Class 2: FCNC Decays of Mesons (Sec.~\ref{class2})}

\begin{center}
{\bf 2A: Semileptonic Decays}
\end{center}
\be\label{Class2x}
B\to K (K^*)\ell^+\ell^-\,, \quad B\to K(K^*)\nu\bar\nu\,,\quad  B\to X_{s,d} \ell^+\ell^-\,,\quad  B\to X_{s,d}\nu\bar\nu \,, \nn
\ee
\be\label{Class2y} 
\kpn\,, \quad \klpn\,, \quad  K_L\to \pi^0 \mu^+\mu^-\,, \quad  K_L\to \pi^0 e^+e^-\,,
 \ee
\be\label{Class2z} 
B\to \pi\nu \bar \nu\,, \quad B\to \varrho \nu \bar \nu\,, \quad {D\to K \,\ell \,\nu_\ell\,, \quad D\to \pi \,e \,\nu_e\,.}\nn
\ee

\begin{center}
{\bf 2B: Leptonic Decays}
\end{center}
\be\label{Class2b}
B_{s,d}\to\mu^+\mu^-\,,\quad K_{L,S}\to\mu^+\mu^-\,,\quad D\to\mu^+\mu^-\,.
\ee

\begin{center}
{\bf 2C: Radiative Decays}
\end{center}
\be\label{Class2c}
B\to X_s\gamma\,, \quad  B\to X_d\gamma\,, \quad B\to K^*\gamma\,, \quad B\to \varrho\gamma\,, {\quad D\to \rho\gamma}\,.
\ee

{\bf Class 3: Non-Leptonic Decays of Mesons (Sec.~\ref{class3})}

The most studied decays in this class are
\be\label{Class3KB}
K_L\to \pi\pi\,, \quad \epe\,, \quad B\to\pi\pi\,,\quad B\to K\pi\,,\quad B\to KK\,.
\ee
\be\label{Class3D}
D\to \pi\pi\,,\quad D\to\pi K\,,\quad D\to KK\,.
\ee

{\bf Class 4: {Electroweak Precision Observables} (Sec.~\ref{class4})}

\be \label{Class4}
W, Z \text{-Pole~Observables}\,.
\ee

{\bf Class 5:  Leptonic LFV Decays (Sec.~\ref{class5})}

\begin{center}
{\bf 5A: $\Delta F=1$ Decays}
\end{center}
\be \label{Class5A}
\mu^-\to e^-e^+e^-\,, \quad \tau^-\to\mu^-\mu^+\mu^-\,, \quad\tau^-\to e^-e^+e^-\,,\quad
\tau^-\to\mu^-e^+e^-\,, \quad \tau^-\to e^-\mu^+\mu^-\,,
\ee
with the last two receiving also $\Delta F=2$ contributions.
\begin{center} 
{\bf 5B: $\Delta F=2$ Decays}
\end{center}
\be
\tau^-\to e^-\mu^+e^-\,, \qquad \tau^-\to\mu^- e^+\mu^-\,.
\ee

\begin{center} 
{\bf 5C: Radiative Decays}
\end{center}
\be
{{\mu\to e \gamma\,, \qquad \tau\to e\gamma\,, \qquad \tau\to \mu\gamma}\,.}
\ee

\begin{center}
{\bf 5D: LFV $Z$ Decays}
\end{center}
\be
Z \to f_i  \bar f_j.
\ee

{\bf Class 6: Semileptonic LFV Decays (Sec.~\ref{class6})}

\begin{center}
{\bf 6A: Quark Flavour Conserving Decays}
\end{center}
\be
{\tau^-}\to \pi\ell^-\,,\qquad {\tau^-}\to \rho\ell^-\,,\qquad {\tau^-}\to\phi \ell^-\,, \quad \mu\to e\,~\text{conversion}\,.
\ee

\begin{center}
{\bf 6B: Quark Flavour Violating Decays}
\end{center}
\be
K_{L,S}\to\mu e\,, \qquad K_{L,S}\to\pi^0\mu e \,, \nn
\ee
\be
B_{d,s} \rightarrow \mu e\,, \qquad B_{d,s} \rightarrow \tau e\,, \qquad B_{d,s} \rightarrow \tau \mu\,,
\ee
\be
B_d\to K^{(*)}\tau^\pm\mu^\mp\,, \qquad B_d\to K^{(*)}\mu^\pm e^\mp\,, \nn
\ee
\be
D^0\to \mu^\pm e^\mp\,, \qquad D^0\to \pi^0\mu^\pm e^\mp\,. \nn
\ee

{\bf Class 7: Electric and Magnetic Dipole Moments (Sec.~\ref{class7})}

\be
{\rm {7A:}~ Electric~ Dipole~ Moments}\,, \qquad {\rm 7B:}~ (g-2)_{e,\mu}\,.
\ee

{\bf Class 8: Charged Current Processes (Sec.~\ref{class8})}

\be
\bar B\to D \ell\bar\nu_l\,, \quad \bar B\to D^* \ell\bar\nu_l\,, \quad B_c\to J/\psi\tau\bar\nu\,,\quad   B_c^+\to \tau^+\nu_\tau\,, \quad  B^+\to \tau^+\nu_\tau\,,
\ee
\be
\pi^- \to e^- \bar \nu\,,\quad {\rm Beta-decay~ and ~neutron~ decay,}    \quad D^+\to e^+\nu\,. \nn
\ee

{\bf Class 9: Higgs Observables (Sec.~\ref{class9})}

\be
\text{Higgs Signal Strengths}
\ee

{\bf Class 10: High $p_T$ Scattering Processes (Sec.~\ref{class10})}

\be
\begin{aligned}
&
{\rm 10A:} ~\ell^+ \ell^- \to \ell^+ \ell^- \,, \qquad 
{\rm 10B:} ~\ell^+ \ell^- \to q \bar q'  \,, \qquad  \\
&
{\rm 10C:} ~pp \to \ell^+ \ell^-\,, \quad  p p \to \ell \bar \nu  \,, \qquad 
{\rm 10D:} ~ pp \to q \bar q' .
\end{aligned}
\ee

In principle one could also consider top quark decays, but there are very many SMEFT operators that have to be considered and we refer to a very comprehensive presentation of them in \cite{Aguilar-Saavedra:2018ksv} for details. Of particular interest are constraints for flavour-violating top decays from analyticity and unitarity that link flavour-conserving and flavour-violating four-fermion interactions. They have been recently analyzed in detail in \cite{Altmannshofer:2023bfk,Altmannshofer:2025lun}. Many useful references can be found in these papers. See also references in Sec.~\ref{sec:obs}.

Another important issue that we will not cover here in detail is the study of NP effects on the $Wtb$ couplings. This involves not only top decays but in particular loop induced processes. It turns out that the operators contributing to such couplings also appear in several processes discussed above. As a result, they provide indirect constraints on $Wtb$ couplings. The list of relevant tree-level and one-loop operators affecting the $Wtb$ vertex will be covered in Class 4, where $W$ boson couplings to all fermions are discussed in the context of $W$- and $Z$-pole measurements.

\subsection{Global View of Correlations between Different Classes}
Several of the processes listed above are correlated within a given NP model through various flavour and gauge symmetries present in this model. However independently of such symmetries specific to a given model, additional correlations between observables in a given class and also correlations between observables in different classes are generated through the unbroken $\text{SU(2)}_L$ symmetry of the SMEFT and operator mixing in the process of RG evolution and matching. This has often an important impact on phenomenology. Here we just mention two examples investigated in detail in the literature.

First, the explanation of the so-called $B$-physics anomalies in semileptonic $B$ decays (Class 2) with leptoquarks implies new effects in LFV decays (Class 5), thereby putting significant bounds on the coefficients of semileptonic operators as stressed in particular in \cite{Feruglio:2016gvd, Feruglio:2017rjo}.

Similar, the attempt to obtain significant enhancement of the ratio $\epe$ (Class 3) within leptoquark models \cite{Bobeth:2017ecx} is very much restricted by the decays of Class 2.

Generally such correlations are within processes in a given class and also between different classes. Moreover these correlations are mutual. For instance semileptonic operators (Class 2) can have through operator mixing an impact on WCs of non-leptonic operators (Classes 1 and 3). Consequently, the bounds on the latter WCs imply bounds on the semileptonic operators.

\begin{table}[H]
\centering
\setlength{\tabcolsep}{1.5pt}
\begin{center}
\begin{tabular}{l@{\ }|c|c|c|}
\toprule
\multicolumn{3}{|c}{\,\,\,\qquad\qquad\qquad\qquad Bosonic Operators} & \\ \hline
\rowcolor{Grayd}
Class & $C_i$   &  Tree-Level Matching & First LL RG Running  \\
\midrule
\multirow{1}{*}{\rotatebox{-270}{$H^6$ $\&$ $H^4 D$ }}  
&$\wc[]{H}{}$    &  9   &   --   \\

&$\wc[]{H \Box}{}$   & 9  & 2, 3, 4, 7A, 8  \\

&$\wc[]{H D}{}$  & 4,  8,  9  & 2, 3, 7A   \\
&&&\\
\hline

\rowcolor{Gray}
\multirow{6}{*}{\rotatebox{-270}{$X^3$}}  &$\wc[]{G}{}$    & -- & --  \\

&$\wc[]{\widetilde G}{}$    & 7A  &  --  \\

&$\wc[]{W}{}$    & -- &  4, 9   \\

\rowcolor{Gray}
&$\wc[]{\widetilde W}{}$  &  -- & 9 \\
\hline

\rowcolor{Gray}
\multirow{11}{*}{\rotatebox{-270}{$X^2 H^2$ }}  

&$\wc[]{HG}{}$  & 9 &  --  \\

\rowcolor{Gray}
&$\wc[]{H \widetilde G}{}$ & 9   & --  \\

\rowcolor{Gray}
&$\wc[]{HW}{}$  & 9 &  4, 5C, 7A   \\

&$\wc[]{H \widetilde W}{}$  & 9   & 5C, 7A  \\

&$\wc[]{HB}{}$  &9 & 4, 5C, 7A   \\

\rowcolor{Gray}
&$\wc[]{H \widetilde B}{}$  &9 & 5C, 7A    \\

\rowcolor{Gray}
&$\wc[]{H W B}{}$  & 4, 8,  9  & 5C, 7A    \\

\rowcolor{Gray}
&$\wc[]{H \widetilde W B}{}$  &9 & 5C, 7A   \\

\hline
\bottomrule
\end{tabular}
\caption{\small The impact of purely bosonic operators on the classes of observables.}
\label{tab:cor-bosonic}
\end{center}
\end{table}

Of interest are also correlations between various processes implied by gauge anomaly cancellation in NP models with extended gauge groups. In the context of simple $Z^\prime$ models such correlations within the quark sector, the lepton sector and most interestingly between quark flavour and lepton flavour violating processes have been analyzed in \cite{Aebischer:2019blw,Colangelo:2025nbd}. Since such correlations are model dependent we will not discuss them here.

In an effective theory the correlations between various processes are generally caused at different stages of the analysis.
\begin{itemize}
\item
At tree-level, when a given WC enters two or more observables and varying its value changes the predictions of these observables in a correlated manner.
\item
When RG evolution is performed from the NP scale $\Lambda$ down to $\muEW$ new operators enter the game. They can modify the correlations present already at tree-level but could also lead to new correlations between observables that were absent at tree-level. Being suppressed by gauge couplings they are generally smaller than at tree-level except of course for those which were absent at tree level. But in the case of large ADMs or top-Yukawa contributions they can still have a significant impact on the phenomenology.
\item
Also the matching between a UV completion and the SMEFT as well as the matching between the SMEFT and the WET can have an impact on the correlations between various observables. It can be significant but in contrast to RG effects above the EW scale, it is not enhanced by large logarithms of the type $\ln(\Lambda/\muEW)$. Yet, its inclusion is required in particular at the NLO level of a RG-improved analysis to guarantee cancellation of the unphysical scale and renormalization scheme dependencies, as discussed in Sec.~\ref{sec:smeft-beyond-leading}.
\item Next, correlations between different observables are induced by $\text{SU(2)}_L$ symmetry.
\item
Finally, as stressed already above, the correlations in question can enter through $1/\Lambda^2$ effects in the SM parameters, which is discussed in Sec.~\ref{sec:SM-smeft-corrections}.
\end{itemize}

\begin{table}[H]
\centering
\setlength{\tabcolsep}{1.5pt}
\begin{center}
\begin{adjustbox}{width=0.8\textwidth}
\begin{tabular}{l@{\ }|c|c|c|}
\toprule
\multicolumn{3}{c}{\qquad\qquad\qquad\qquad Fermionic-Bosonic Operators} & \\ \hline
\rowcolor{Grayd}
Class & $C_i$   &   Tree-Level Matching & First LL RG Running  \\     \midrule
\rowcolor{Gray}  \multirow{13}{*}{\rotatebox{-270}{$f^2 H^2 D$ }}  
&$\wc[(1)]{H\ell}{}$    &4, 5A, 5D, 6A, 10A, 10B, 10C  & 2,  9  \\
\rowcolor{Gray}
& $\wc[(3)]{H\ell}{}$ &4, 5A, 5D, 6A, 8, 9, 10A, 10B, 10C & 2  \\ 
\rowcolor{Gray}
&  $\wc[(1)]{Hq}{}$   &2A, 2B, 3, 4, 10B, 10C, 10D  & 1A, 1B, 8, 9  \\
& $\wc[(3)]{Hq}{}$  & 2A, 2B, 3, 4, 8, 10B, 10C, 10D  & 1A, 1B,  9
\\ 
&$\wc[]{Hd}{}$   & 2A, 2B, 3, 4, 10B, 10C, 10D & 1A, 7A, 8, 9   \\ 
&$\wc[]{Hu}{}$   &   2A, 2B, 4, 10B, 10C, 10D &  1B, 2, 3, 8, 9 \\ 
\rowcolor{Gray}
&$\wc[]{Hud}{}$  & 3, 7A,  8, 10B, 10C  & 2, 4, 9  \\ 
\rowcolor{Gray}
& $\wc[]{He}{}$  & 4, 5A, 5D, 6A, 10A, 10B, 10C & 2, 8,  9    \\ 
\hline
 
\multirow{4}{*}{\rotatebox{-270}{$f^2 H^3$ }}  
&$\wc[]{eH}{}$    &  5B, 9  &   --  \\ 

&$\wc[]{uH}{}$    & 9  & 2    \\ 

&$\wc[]{dH}{}$     &9  & 2    \\ 
\hline
\rowcolor{Gray}  \multirow{11}{*}{\rotatebox{-270}{$f^2  XH$ }}  
&$\wc[]{eW}{}$  &  5C, 7A, 7B, 10A, 10B, 10C  &   4, 9    \\ 
\rowcolor{Gray}
&$\wc[]{eB}{}$   &  5C, 7A, 7B, 10A, 10B, 10C  & 4, 9    \\ 
\rowcolor{Gray}
&$\wc[]{uG}{}$   &  7A,  10D &  2, 3   \\ 

&$\wc[]{dG}{}$   & 2C, 7A, 10D &   3   \\ 

&$\wc[]{dW}{}$  & 2C, 7A, 10B, 10C, 10D & 3, 4, 9    \\ 

\rowcolor{Gray}
&$\wc[]{uW}{}$     &  7A, 10B, 10C, 10D &  2, 3, 4, 9     \\ 

\rowcolor{Gray}
&$\wc[]{uB}{}$      &  7A, 10B, 10C, 10D & 2, 3, 4, 9     \\ 

\rowcolor{Gray}
&$\wc[]{dB}{}$    & 2C, 7A, 10B, 10C, 10D & 3, 4, 9   \\ 
\hline
\bottomrule
\end{tabular}
\end{adjustbox}
\caption{\small The impact of the fermionic-bosonic operators on the classes of observables.}
\label{tab:cor-bosonic-fermionic}
\end{center}
\end{table}

In Tables \ref{tab:cor-bosonic}, \ref{tab:cor-bosonic-fermionic} and \ref{tab:cor-fermionic} we provide a global view on the correlations in question. In the first column operators are listed and in the rows the classes of observables to which this operator contributes are shown. We distinguish between tree-level matching at the EW scale and RG mixing in the first LL approximation. We leave the one-loop matchings and the impact of NP on the SM parameters aside.

{\small
\begin{table}[H]
\centering
\setlength{\tabcolsep}{1.8pt}
\begin{center}
\begin{adjustbox}{width=0.7\textwidth}
\begin{tabular}{l@{\ }|c|c|c|}
\toprule
\multicolumn{3}{c}{\qquad\qquad\qquad\qquad Four-fermion Operators} & \\ \hline
\rowcolor{Grayd}
Class & $C_i$       &   Tree-Level Matching & First LL RG Running   \\
\midrule
\rowcolor{Gray}  \multirow{7}{*}{\rotatebox{-270}{$(\overline LL) (\overline LL)$}} 
&  $\wc[]{\ell \ell}{}$  & 4, 5A, 5B, 8, 9, 10A  & 5D, 6A   \\

& $\wc[(1)]{qq}{}$ & 1A, 1B, 3, 10D  & 2, 4, 8    \\ 

&  $\wc[(3)]{qq}{}$  & 1A, 1B, 3, 10D  & 2, 4, 8    \\

& $\wc[(1)]{\ell q}{}$  & 2A, 2B, 5A, 6A, 6B, 10B, 10C & 3, 4, 5D, 8   \\

\rowcolor{Gray}
& $\wc[(3)]{ \ell q}{}$  &2A, 2B, 5A, 6A,  6B, 8, 10B, 10C  &  3, 4, 5D,  8  \\ 

\hline

\rowcolor{Gray}  \multirow{10}{*}{\rotatebox{-270}{$(\overline RR) (\overline RR)$}}  &
$\wc[]{ee}{}$   & 5A, 5B, 10A  & 4, 5D, 6A, 8    \\ 

\rowcolor{Gray}
&$\wc[]{uu}{}$   & 1B, 3, 10D  &  4  \\ 

&$\wc[]{dd}{}$   & 1A, 3, 10D &  2, 4  \\ 

&$\wc[]{eu}{}$   &  2B, 5A, 6A, 6B, 10B, 10C  & 2, 3, 4, 5D, 8   \\

&$\wc[]{ed}{}$   & 2A, 2B, 5A, 6A, 6B, 10B, 10C & 3, 4, 5D, 8     \\

&$\wc[(1)]{ud}{}$  & 3, 10D  & 1A, 1B, 2, 4, 7A, 8  \\

\rowcolor{Gray}
&$\wc[(8)]{ud}{}$  & 3, 10D  & 1A, 1B, 7A, 8    \\

\hline

\rowcolor{Gray}  \multirow{11}{*}{\rotatebox{-270}{$(\overline LL) (\overline RR)$}}  
&$\wc[]{\ell e}{}$   & 5A, 5B, 7A, {7B,} 8, 10A  &  4, 5D, 6A, 6B, 8, 9   \\

\rowcolor{Gray}
&$\wc[]{\ell u}{}$   & 2A, 2B, 5A, 6A, 6B, 10B, 10C  & 2, 4, 5D, 8    \\

&$\wc[]{\ell d}{}$   & 2A, 2B, 5A, 6A, 6B, 10B, 10C  & 3, 4, 5D,  8 \\

&$\wc[]{qe}{}$    & 2A, 5A, 6A, 6B, 10B, 10C & 3, 4, 5D,  8   \\

&$\wc[(1)]{qu}{}$   & 1B, 3, 7A, 10D  & 1A, 2, 4, 9   \\

&$\wc[(8)]{qu}{}$  & 1B, 3, 7A, 10D  & 1A, 9    \\

\rowcolor{Gray}
&$\wc[(1)]{qd}{}$  & 1A, 3, 7A, 10D  & 1B, 2, 4, 9    \\

\rowcolor{Gray}
&$\wc[(8)]{qd}{}$   & 1A, 3, 7A, 10D & 1B, 9    \\

\hline

\multirow{10}{*}{\rotatebox{-270}{$(\overline LR) (\overline RL)$ $\&$  $(\overline LR) (\overline LR)$}}  
&&&\\

&$\wc[]{\ell e dq}{}$  & 2B,\, 6A,\, 6B,\, {7A,\,7B,\,} 8, {10B,} {10C}  & 5A, 5B, 9     \\

&$\wc[(1)]{quqd}{}$  & {3}, 7A, 10D  &  1A, 2, 9    \\ 

&$\wc[(8)]{quqd}{}$  & {3}, 7A, 10D &  1A, 2, 9    \\

& $\wc[(1)]{\ell e qu}{}$  &6A, 6B, 7A, {7B,} 8, {10B,} {10C}   & 5A, 5B, 9    \\

&$\wc[(3)]{\ell e qu}{}$   & 5C, 6A, 6B, 7A,{7B,} 8, {10B,} {10C} & --     \\

&&&\\
\hline
\bottomrule
\end{tabular}
\end{adjustbox}
\caption{\small The impact of the four-fermi operators on the classes of observables. }
\label{tab:cor-fermionic}
\end{center}
\end{table}
}

The important goal of Part II is to provide an insight into the dynamics behind the Tables~\ref{tab:cor-bosonic}, \ref{tab:cor-bosonic-fermionic} and \ref{tab:cor-fermionic} by listing the operators relevant for each class and subsequently to investigate various correlations between observables within a given class and also between observables of different classes.

However, before discussing these classes, related operators as well as the SMEFT charts, it will be useful to discuss the pattern of operator mixing in the SMEFT in general.

\section{Operator Mixing in the SMEFT}\label{sec:10}

As discussed in Sec.~\ref{sec:ren-gro-run}, the operator mixing pattern in the SMEFT is in general very complex, especially due to the presence of the large top quark Yukawa coupling above the weak scale. This leads to an intricate mixing among the SMEFT operators having distinct flavour structures. In addition, the operator mixing also takes place due to electroweak and strong interactions which are flavour diagonal by construction. Formally, operator-mixing means that insertions of a given single operator into one-loop diagrams mediated by the SM fields give rise to new operator structures. In the $\overline{\rm MS}$ scheme, the ADMs are given by the poles ($\propto 1/\varepsilon$) of the corresponding one-loop integrals. Furthermore, back-rotation effects introduce additional flavour mixing. This makes predictions of low-energy implications resulting from UV physics a tedious task. In the present section, we will try to shed some light on the operator mixing between different SMEFT classes. In particular, we want to exhibit how complicated this mixing can be. 

In the context of the presentation of different classes the bottom-up running turns out to be useful in the search for operators of a given class generated through this running. The top-down running on the other hand is crucial for the derivation of low energy implications for given UV scenarios, motivated by the discovery of new particles or anomalies in the experimental data.

\subsection{Bottom-Up Running}\label{Bottom-Up}
In this subsection, we discuss the bottom-up running. In Sec.~\ref{sec:GV} we have outlined a six-step procedure on how one should proceed in a top-down approach, knowing a specific UV completion, to obtain predictions for observables. However, in the search for operators at $\Lambda$ that are relevant for a given class of processes this is not the procedure to follow. This is evident from Tabs.~\ref{tab:cor-bosonic}-\ref{tab:cor-fermionic} in Sec.~\ref{classification}. A given operator at $\Lambda$ affects several classes simultaneously. Therefore, to identify the most important operators at $\Lambda$ for a given class one has to use a bottom-up approach that consists of the following steps.

\paragraph{Step 1: Determination of the WET Operators at $\muEW$}

To find them one has to inspect known formulae for processes in a given class.

\paragraph{Step 2: Determination of the SMEFT Operators at $\muEW$}

This is simply found through matching of the SMEFT operators onto the WET operators of a given class.

\paragraph{Step 3: Determination of the SMEFT Operators at $\Lambda$}
This step is completed by RG running from $\muEW$ to $\Lambda$, which is in the opposite direction than in the top-down procedure discussed in Sec.~\ref{sec:GV}. We will discuss this RG running in detail in Sec.~\ref{sec:6}. 
This will allow us to identify \emph{in principle}  relevant operators for each class. However, 
not all such operators will have significant impact on the phenomenology.
The quantitative relevance of different SMEFT operators at $\Lambda$ can be determined using $\rho$ and $\eta$ parameters defined in Sec.~\ref{GImpact}. Further, in realistic phenomenological analysis, 
 one should  also account for back-rotation effect, which will not be separately discussed for each class.
However, these effects can be included numerically using the {\tt wilson} package.

\paragraph{Step 4: Determination of Correlations to Other Classes}
Having the list of operators at $\Lambda$ corresponding to a given class the correlations between {\em different classes} induced purely by RG running from $\Lambda$ to $\muEW$ at the 1-loop level can be identified. This includes correlations between tree-generated processes in a given class and loop-induced processes from other classes. Finally, we can also have purely tree-level correlations due to $\text{SU(2)}_L$ and CKM rotations. 
In the presentation of some classes, examples of correlations are explicitly given. However, 
using the unique list of SMEFT operators, at scales $\muEW$ and $\Lambda$,  identified in this review for each class, it would be straightforward  
to identify additional possible correlations arising from RGE evolution or simply from $\text{SU(2)}_L$ symmetry. 

This four-step procedure will govern the presentation of different classes in this part of our review. Next, we list the criteria used to find for a given class relevant SMEFT operators at $\Lambda$ and the constructions of the corresponding charts.

\vspace*{0.5cm}
{\bf Selection Criterion-I}
\vspace*{0.3cm}

In a first step we use a rather loose criterion to decide which terms are kept in the calculation. It is defined as follows:
\be \label{eq:criteriaI}
\boxed{ 
\textrm{{\bf Criterion-I}: All terms depending on}~ y_b,\,\ y_t, \,\ g_1, \,\  g_2, \,\ {g_s}, \,\  \lambda^n\,,
}
\ee
for $n=1,..,6$. Following this criterion, we keep all terms in the tree-level matching and 1-loop running which depend upon the parameters $y_b, y_t, g_1, g_2, g_s$ and $\lambda^n$, where $\lambda$ is the Wolfenstein parameter. The light quark and leptonic Yukawas are neglected. These criteria alone leave still many contributions that are at least at present phenomenologically irrelevant. Therefore, we introduce an additional criterion that reduces the number of contributions significantly, while simultaneously keeping all relevant ones. Note however that in certain observable classes the lighter quark and leptonic Yukawas might be relevant. Therefore for some classes, for instance for EDMs and LFV observables, we will have to relax Criterion-I.

\vspace*{0.5cm}
{\bf Selection Criterion-II}
\vspace*{0.3cm}

It is useful to organize the SMEFT matching and running effects according to powers of a single parameter, that we choose to be the Wolfenstein parameter $\lambda$. The tree-level matching conditions and anomalous dimensions involve the parameter $\lambda$ due to the misalignment of the Yukawas in flavour space. Further, various other factors involved in the evolution matrices can be expressed in powers of $\lambda$, because
\be 
\begin{aligned}
\lambda &= 0.225\,, \,\, \lambda^2 = 0.05\,, \,\, \lambda^3 = 0.01\,, \,\, \lambda^4 =3\cdot10^{-3}\,,  \,\,
\lambda^5 = 6 \cdot 10^{-4}\,, \,\, \lambda^6 = 10^{-4}\,,\\
 y_t & \approx 1\,, \quad  y_b \approx 0.02\,, \quad  {1\over 16\pi^2} \approx 6\cdot 10^{-3}\,, 
\quad g_1^2\,  \approx 0.13\,, \quad  \, g_2^2\, \approx\,0.43\,,\\
g_s^2 & \approx 1.4\,,\quad log \approx 2-4\,,
\end{aligned}
\ee
where {\em log} stands for $\ln(\Lambda/\muEW)$. This implies 
\be \label{eq:inlambda}
{{log}\over 16\pi^2} \approx 2\lambda^3\,, \quad y_b \approx \lambda^2/2\,, \quad  g_1^2\,  \approx \lambda/2\,, \quad 
  \, g_2^2\, \approx 2\lambda \,.
\ee
The WET operators receive two kinds of effects in the SMEFT: i) from tree-level matching, ii) from 1-loop RG running, or both. The typical size of WET operators generated by the SMEFT is given by 
\be \label{eq:rge-lambda}
\begin{aligned}
\textrm{CKM-Mixing} &\approx c_0 \lambda^x \,,  \\
\textrm{Gauge-Mixing} &\approx c_1 {{\lambda^x}  g_i^2 {log} \over 16\pi^2}\,,  \\
\textrm{Yukawa-Mixing}& \approx c_2 {\lambda^{x}  y_i^2 {log}  \over 16\pi^2}\,,  \\
\textrm{Yukawa-Gauge}& \approx c_3 {\lambda^{x} g_i y_i {log}  \over 16\pi^2}\,.
\end{aligned}
\ee
Here, $c_i$ are fixed by the tree-level matching conditions and the numerical factors multiplying the coupling constants in the ADMs.\footnote{We do not consider 1-loop matching in this Atlas, because to be consistent one would need to include the two-loop RG running.} The first term in \eqref{eq:rge-lambda} comes from the tree-level matching conditions representing the contributions from trivial (independent of $\lambda$, for $x=0$) and non-trivial (for $x\ne 0$)  kinds of SMEFT operators. The non-trivial flavour operators bear $\lambda^x$ suppression present in the CKM-mixing.

The remaining three terms originate from one-loop RG evolution. In these terms, powers of $\lambda$ come either purely from the one-loop ADM or from a combination of tree-level matching and one-loop ADM. The latter results from operator mixing between UV SMEFT operators and non-trivial flavour operators in the tree-level matching at the EW scale. 

Using \eqref{eq:inlambda} one can quantify the size of WET operators solely in powers of $\lambda$. In Criterion II, we keep all the terms satisfying
\be \label{eq:criteriaII}
\boxed{
\textrm{\bf Criterion-II}: ~C_{\rm WET}^{\rm mass~ basis}(\muEW) \gtrapprox  {\lambda^n}  C_{\rm SMEFT}^{\rm weak~ basis}(\Lambda).
}
\ee
Here $n$ can be different for different classes. Unless specified otherwise the weak basis for the SMEFT is chosen to be the down-basis. One can use Criterion-II to find a list of the most important operators at $\Lambda$ for each class. This $\lambda$ criterion can also be used to asses the relevance of different SMEFT operators at $\Lambda$ in terms of the $\rho,\eta$ parameters defined in Sec.~\ref{GImpact}.

\subsection{Top-Down Running}
In the top-down running, we will turn on a single WC $C_i(\Lambda)$ at the NP scale and look at the WCs $C_j(\muEW)$ which are generated at the electroweak scale. For simplicity, we will first suppress the flavour indices. To deal with operators with specific flavour indices, we recommend using computer programs like {\tt wilson} and {\tt DsixTools} that are described in Sec.~\ref{SMEFTtools}. In the following sections, after defining various effective Hamiltonians relevant for phenomenology, we will see transparently how this mixing affects the observables.

In Tabs.~\ref{tab:mixing-bosonic-color}-\ref{tab:mixing-fermionic-color4}, we show the operator-mixing for a given single NP operator belonging to purely Bosonic, Bosonic-Fermionic and Four-fermion SMEFT classes, to all other possible operators of different classes.

To construct these tables the full numerical solution of the RGEs at the one-loop level has been used. That is all leading logarithms have been summed up to all orders of perturbation theory. Further, the ADMs due to Yukawas, gauge as well as quartic Higgs interactions are taken into account. Given the many operators involved, we have shown in these tables only the strongest connections given by the elements of evolution matrix 
\be
C_j(\muEW)=  U_{ji}(\muEW, \Lambda) C_i(\Lambda) . 
\ee
To this end scanning over all possible flavour combinations of the involved WCs has been performed, for instance:
\be
\begin{aligned}
\wc[]{G}{}(\Lambda) & \rightarrow \wc[(8)]{quqd}{ijkl}(\muEW), \quad \forall \,\ \textit{i,j,k,l} \in 1-3\,, \\
\wc[]{ll}{abcd}(\Lambda) & \rightarrow \wc[]{le}{ijkl}(\muEW), \quad \forall \,\ \textit{i,j,k,l,a,b,c,d} \in 1-3.
\end{aligned}
\ee
Out of all the evolution matrix elements, only the one with the largest absolute value is shown, highlighting the mountains in the ranking of operator mixing. The colour coding is used to indicate the size of the elements of the evolution matrix $U(\muEW, \Lambda)$. Assuming $\lambda\simeq 0.225$ the colours are chosen as follows:
\be
\begin{aligned}
\textit{\rm Red}: \,\,\,  \mathcal{O}(\lambda),   \quad
\textit{\rm Orange}: \,\,\, \mathcal{O}(\lambda^2), \quad
\textit{\rm Green}: \,\,\, \mathcal{O}(\lambda^3), \quad
\textit{\rm Cyan}: \,\,\,  \mathcal{O}(\lambda^4).
\end{aligned}
\ee
{
Here, $\mathcal{O}(\lambda^n)$ is defined by 
\be
\lambda^{n} \le | U_{ji}(\muEW, \Lambda) | \less \lambda^{n-1}. 
\ee
Specifically for $\lambda =0.225$, we get the range for $U_{ji}(\muEW, \Lambda)$ 
\be
 \mathcal{O}(\lambda):  \,\,\, \ge 0.225   \,, \quad
 \mathcal{O}(\lambda^2):~ { [}0.05, 0.225)\,, \quad
 \mathcal{O}(\lambda^3): ~{ [}0.01, 0.05)\,, \quad
 \mathcal{O}(\lambda^4):~  {[}0.003, 0.01).
\ee
 }
In order to increase transparency we did not exhibit mixings of $\mathcal{O}(\lambda^5) \sim 0.0006$ which are generally negligible. Studying these tables one notices that all SMEFT operators exhibit self-mixing. It is instructive, following the terminology introduced in \cite{Buras:2018gto}, to describe which types of operator mixing in the SMEFT are represented by Tabs.~\ref{tab:mixing-bosonic-color}-\ref{tab:mixing-fermionic-color4} beyond the first LLA:
\begin{itemize}
\item
In the latter approximation each row in the tables considered separately would describe only a family of operators with the {\em parent operator} in each family represented by its WCs $C_i(\Lambda)$ and the generated {\em child operators}. There would be no mixing in this approximation between different families and between children because only one coefficient is non-vanishing at $\Lambda$ and only the first LLA in \eqref{eq:LLA} is taken into account. 
\item
Remaining still within a given family but expanding the exponential in \eqref{eq:LO} and including $\log^2$- and higher-order contributions as done in these tables introduces mixing between child operators and also results in the generation of grandchildren operators which could not be generated directly from the parent operator in the first LLA at the $\Lambda$ scale.
\item
The latter step can be seen by considering other families in which the child operators of the first family are parent operators.
\item
In this manner also mixing between different families takes place.
\end{itemize}

\begin{table}[H]
\centering
\setlength{\tabcolsep}{1.5pt}
\begin{center}
\begin{tabular}{l@{\ }|c|ccccccccccccccccc{c}}
\multicolumn{12}{c}{Bosonic Operators} & \\
\toprule
\multicolumn{12}{c}{ \colorbox{RedOrange}{$\lambda$}, \colorbox{BurntOrange}{$\lambda^2$},
\colorbox{Green}{$\lambda^3$}, \colorbox{SkyBlue}{$\lambda^4$}} & \\
\midrule
\rowcolor{Grayd} Class & $C_i(\Lambda)$   &     \multicolumn{3}{l}{}  &$C_j(\muEW) $ & &\multicolumn{7}{l}{} \\
\midrule
\multirow{23}{*}{\rotatebox{-270}{$X^3$,  $H^6$, $H^4 D^2$, $X^2 H^2$ }}

&$\wcs[]{G}{}$
&\colorbox{RedOrange}{$\wcs[]{G}{}$}
&\colorbox{BurntOrange}{$\wcs[]{uG}{}$}
&\colorbox{BurntOrange}{$\wcs[]{uH}{}$}
&\colorbox{SkyBlue}{$\wcs[]{HG}{}$}
&\colorbox{SkyBlue}{$\wcs[]{H}{}$}
&\colorbox{SkyBlue}{$\wcs[]{uW}{}$}
&\colorbox{SkyBlue}{$\wcs[]{uB}{}$}
&\colorbox{SkyBlue}{$\wcs[]{dG}{}$}
&&&&\\ 
&$\wcs[]{\widetilde G }{}$
&\colorbox{RedOrange}{$\wcs[]{\widetilde G }{}$}
&\colorbox{BurntOrange}{$\wcs[]{uG}{}$}
&\colorbox{BurntOrange}{$\wcs[]{uH}{}$}
&\colorbox{SkyBlue}{$\wcs[]{H\widetilde G}{}$}
&\colorbox{SkyBlue}{$\wcs[]{uW}{}$}
&\colorbox{SkyBlue}{$\wcs[]{uB}{}$}
&\colorbox{SkyBlue}{$\wcs[]{dG}{}$}
&&&&&\\ 
&$\wcs[]{W}{}$
&\colorbox{RedOrange}{$\wcs[]{W}{}$}
&\colorbox{BurntOrange}{$\wcs[]{HW}{}$}
&\colorbox{SkyBlue}{$\wcs[]{H W B}{}$}
&\colorbox{SkyBlue}{$\wcs[]{uH}{}$}
&&&&&&&&\\ 
&$\wcs[]{\widetilde W }{}$
&\colorbox{RedOrange}{$\wcs[]{\widetilde W }{}$}
&\colorbox{BurntOrange}{$\wcs[]{H\widetilde W}{}$}
&\colorbox{SkyBlue}{$\wcs[]{H  \widetilde W B}{}$}
&\colorbox{SkyBlue}{$\wcs[]{uH}{}$}
&&&&&&&&\\ 
&$\wcs[]{H}{}$
&\colorbox{RedOrange}{$\wcs[]{H}{}$}
&&&&&&&&&&&\\ 
&$\wcs[]{H\Box}{}$
&\colorbox{RedOrange}{$\wcs[]{H\Box}{}$}
&\colorbox{BurntOrange}{$\wcs[]{uH}{}$}
&\colorbox{Green}{$\wcs[]{H}{}$}
&\colorbox{Green}{$\wcs[]{Hu}{ }$}
&\colorbox{SkyBlue}{$\wcs[]{HD}{}$}
&\colorbox{SkyBlue}{$\wcs[(1)]{Hq}{}$}
&\colorbox{SkyBlue}{$\wcs[(3)]{Hq}{}$}
&&&&&\\ 
&$\wcs[]{HD}{}$
&\colorbox{RedOrange}{$\wcs[]{HD}{}$}
&\colorbox{Green}{$\wcs[]{Hu}{ }$}
&\colorbox{SkyBlue}{$\wcs[]{uH}{}$}
&\colorbox{SkyBlue}{$\wcs[(1)]{Hq}{}$}
&\colorbox{SkyBlue}{$\wcs[]{H}{}$}
&\colorbox{SkyBlue}{$\wcs[]{H\Box}{}$}
&&&&&&\\ 
&$\wcs[]{HG}{}$
&\colorbox{RedOrange}{$\wcs[]{HG}{}$}
&\colorbox{RedOrange}{$\wcs[]{uH}{}$}
&\colorbox{BurntOrange}{$\wcs[]{H}{}$}
&\colorbox{BurntOrange}{$\wcs[]{uG}{}$}
&\colorbox{Green}{$\wcs[]{dH}{}$}
&&&&&&&\\ 
&$\wcs[]{HB}{}$
&\colorbox{RedOrange}{$\wcs[]{HB}{}$}
&\colorbox{Green}{$\wcs[]{uH}{}$}
&\colorbox{SkyBlue}{$\wcs[]{uB}{}$}
&\colorbox{SkyBlue}{$\wcs[]{H W B}{}$}
&&&&&&&&\\ 
&$\wcs[]{HW}{}$
&\colorbox{RedOrange}{$\wcs[]{HW}{}$}
&\colorbox{BurntOrange}{$\wcs[]{uH}{}$}
&\colorbox{Green}{$\wcs[]{uW}{}$}
&\colorbox{SkyBlue}{$\wcs[]{H W B}{}$}
&\colorbox{SkyBlue}{$\wcs[]{H}{}$}
&&&&&&&\\ 
&$\wcs[]{H W B}{}$
&\colorbox{RedOrange}{$\wcs[]{H W B}{}$}
&\colorbox{Green}{$\wcs[]{uB}{}$}
&\colorbox{Green}{$\wcs[]{HB}{}$}
&\colorbox{SkyBlue}{$\wcs[]{uW}{}$}
&\colorbox{SkyBlue}{$\wcs[]{HW}{}$}
&\colorbox{SkyBlue}{$\wcs[]{uH}{}$}
&&&&&&\\ 
&$\wcs[]{H\widetilde G}{}$
&\colorbox{RedOrange}{$\wcs[]{H\widetilde G}{}$}
&\colorbox{RedOrange}{$\wcs[]{uH}{}$}
&\colorbox{BurntOrange}{$\wcs[]{uG}{}$}
&\colorbox{Green}{$\wcs[]{dH}{}$}
&&&&&&&&\\ 
&$\wcs[]{H\widetilde B}{}$
&\colorbox{RedOrange}{$\wcs[]{H\widetilde B}{}$}
&\colorbox{Green}{$\wcs[]{uH}{}$}
&\colorbox{SkyBlue}{$\wcs[]{uB}{}$}
&\colorbox{SkyBlue}{$\wcs[]{H  \widetilde W B}{}$}
&&&&&&&&\\ 
&$\wcs[]{H\widetilde W}{}$
&\colorbox{RedOrange}{$\wcs[]{H\widetilde W}{}$}
&\colorbox{BurntOrange}{$\wcs[]{uH}{}$}
&\colorbox{Green}{$\wcs[]{uW}{}$}
&\colorbox{SkyBlue}{$\wcs[]{H  \widetilde W B}{}$}
&&&&&&&&\\ 
&$\wcs[]{H  \widetilde W B}{}$
&\colorbox{RedOrange}{$\wcs[]{H  \widetilde W B}{}$}
&\colorbox{Green}{$\wcs[]{uB}{}$}
&\colorbox{Green}{$\wcs[]{H\widetilde B}{}$}
&\colorbox{SkyBlue}{$\wcs[]{uW}{}$}
&\colorbox{SkyBlue}{$\wcs[]{H\widetilde W}{}$}
&\colorbox{SkyBlue}{$\wcs[]{uH}{}$}
&&&&&&\\ 
\bottomrule
\end{tabular}
\caption{\small {The pattern of operator mixing for the purely Bosonic operators in the SMEFT. The operator mixing due to all gauge interactions, Higgs quartic, as well as the Yukawa interactions are taken into account. Further, the operator mixing shown here corresponds to the full numerical solution which encompasses the parent-child, parent-grandchild etc, i.e. multi-step operator mixing.
The $\mathcal{O}(\lambda^n)$ indicates $\lambda^{n} \le {|C_j(\muEW)/C_i(\Lambda)|}  \less \lambda^{n-1}$}.}
\label{tab:mixing-bosonic-color}
\end{center}
\end{table}

This way of visualizing the mixing allows an insight into the Tabs.~\ref{tab:mixing-bosonic-color}-\ref{tab:mixing-fermionic-color4} that consider at the NP scale bosonic, bosonic-fermionic and four-fermionic operators. Simultaneously it shows that this mixing is rather involved. Further insight into this mixing can be gained by inspecting these tables together with the charts (see Sec.~\ref{SMEFTcharts}) that use first LLA and various RGEs, to be given in part II of this review, which allow in no time to find out which operators are mixed in this approximation. Additional insight will be given through the anatomy of operator mixing performed in Sec.~\ref{A-F}.

{\boldmath \subsubsection{Bosonic Operators}}
The operators corresponding to the SMEFT classes $X^3$, $H^6$, $H^4D^2$, and $X^2 H^2$ are bosonic in nature. However, as seen in Tab.~\ref{tab:mixing-bosonic-color} they can mix into operators in other classes, in particular into bosonic-fermionic ones that are collected in Tabs.~\ref{tab:mixing-bosonic-fermionic-color1} and \ref{tab:mixing-bosonic-fermionic-color2}. We observe 
the following.
\begin{itemize}
\item
The $X^3$ operators apart from self-mixing mix with the $f^2 XH$, $f^2 H^3$, $H^6$ and $X^2 H^2$ operators. The strongest connections of these operators turn out to be with $f^2 XH$, $X^2 H^2$ and $f^2H^3$ types operators.
\item
$H^6$ exhibit only self-mixing. $H^4D^2$ operators, apart from the self-mixing, can mix with the bosonic-fermionic operators $f^2 H^3$, $f^2 H^2 D$ and $H^6$. As discussed in Sec.~\ref{sub:HiggsEWYukawa}, the $H^4D^2$ and $f^2 H^3$ classes give rise to dim-6 corrections to the Yukawa couplings. Therefore, the $H^6$ operator can indirectly contribute to these couplings through mixing. 

Interestingly, operators from these sectors do exhibit mixing with four-fermion operators only at $\lambda^6$ level.
\item
Finally, the $X^2H^2$ operators can generate $f^2 H^3$, $f^2 X H$, and $H^6$ operators. Here too the mixing with four-fermions is absent at the shown levels.
\end{itemize}
We note that all bosonic operators exhibit a strong self-mixing. From the above discussion it is evident that in order to take all these effects into account efficient computer codes are necessary.
{\boldmath \subsubsection{Bosonic-Fermionic Operators}}
Next we discuss the mixing of Bosonic-Fermionic operators with other SMEFT classes. Such operators can be divided into the three SMEFT classes $f^2 H^3$, $f^2 XH$ and $f^2 H^2 D$. From Tabs.~\ref{tab:mixing-bosonic-fermionic-color1} and \ref{tab:mixing-bosonic-fermionic-color2} we learn the following.
\begin{itemize}
\item
The $f^2 H^3$ operators exhibit only self-mixing, except for $\wc[]{uH}{}$ which mixes into $H^6$. Note that at this level they do not have an impact on $f^4$ and  $X^2 H^2$ operators. 
\item
The $f^2 X H$ operators, apart from mixing among themselves, can mix with bosonic operators $X^2 H^2$ and $H^6$ in Tab.~\ref{tab:mixing-bosonic-color}, operators in class $f^2 H^3$ and various four-fermion operators listed in Tabs.~\ref{tab:mixing-fermionic-color1}-\ref{tab:mixing-fermionic-color4}.
\item
The $f^2 H^2 D$ operators can mix with $H^4 D^2$, $f^2 H^3$, $f^4$, $H^6$, as well as within $f^2 H^2 D$ class operators. In addition, all operators in this class exhibit a strong self-mixing.
\end{itemize}

{\boldmath \subsubsection{Four-fermion Operators}}
The Four-fermion operators can be divided into five types of SMEFT classes based upon the chiralities of the fermions. Their mixing is displayed in Tabs.~\ref{tab:mixing-fermionic-color1}-\ref{tab:mixing-fermionic-color4}. We observe the following patterns of this mixing.
\begin{itemize}
\item
The $(\overline LL) (\overline LL)$ class can mix with $f^2 H^2 D$ and $H^4 D^2$, $H^6$, $f^2 H^3$ operators as well as different kinds of other four-fermion operators.
\item
The $(\overline RR) (\overline RR)$ class can only mix with the $f^2 H^2 D $, $f^2 H^3$ and $H^4 D^2$ classes, in addition to the mixing with various four-fermion operators which is exhibited by all operators in this class. 
\item
The $(\overline LL) (\overline RR)$ operators mix with $f^2 H^2 D$, $H^4 D^2$, $H^6$ and $f^2 H^3$ operators in addition to various four-fermionic ones.
\item
Finally, the remaining two classes $(\overline LR) (\overline RL)$ and $(\overline LR) (\overline LR)$ also show a diverse mixing pattern. These operators can mix with $f^2 H^3$, $f^2 XH$, as well as several four-fermion classes.
\end{itemize}

\begin{table}[tbp]
\centering
\setlength{\tabcolsep}{2pt}
\begin{center}
\begin{adjustbox}{width=\textwidth}
\begin{tabular}{l@{\ }|c|ccccccccccccccccc{c}}
\toprule
\multicolumn{13}{c}{Fermionic-Bosonic Operators-I} & \\ 
\multicolumn{13}{c}{ \colorbox{RedOrange}{$\lambda$}, \colorbox{BurntOrange}{$\lambda^2$},
\colorbox{Green}{$\lambda^3$}, \colorbox{SkyBlue}{$\lambda^4$}}  & \\ \hline
\rowcolor{Grayd}
Class & $C_i(\Lambda)$   &    \multicolumn{4}{l}{} &$C_j(\muEW) $&  \multicolumn{6}{l}{} \\
\midrule 
\multirow{18}{*}{ \rotatebox{-270}{$f^2 H^3 $, $f^2 XH$}} 
&$\wcs[]{uH}{}$
&\colorbox{RedOrange}{$\wcs[]{uH}{}$}
&\colorbox{RedOrange}{$\wcs[]{H}{}$}
&&&&&&&&&&\\ 
&$\wcs[]{dH}{}$
&\colorbox{RedOrange}{$\wcs[]{dH}{}$}
&&&&&&&&&&&\\ 
&$\wcs[]{eH}{}$
&\colorbox{RedOrange}{$\wcs[]{eH}{}$}
&&&&&&&&&&&\\ 
&$\wcs[]{eW}{}$
&\colorbox{RedOrange}{$\wcs[]{eW}{}$}
&\colorbox{Green}{$\wcs[]{eH}{}$}
&\colorbox{Green}{$\wcs[(3)]{lequ}{}$}
&\colorbox{SkyBlue}{$\wcs[]{eB}{}$}
&\colorbox{SkyBlue}{$\wcs[(1)]{lequ}{}$}
&&&&&&&\\ 
&$\wcs[]{eB}{}$
&\colorbox{RedOrange}{$\wcs[]{eB}{}$}
&\colorbox{SkyBlue}{$\wcs[(3)]{lequ}{}$}
&&&&&&&&&&\\ 
&$\wcs[]{uG}{}$
&\colorbox{RedOrange}{$\wcs[]{uG}{}$}
&\colorbox{RedOrange}{$\wcs[]{uH}{}$}
&\colorbox{BurntOrange}{$\wcs[]{HG}{}$}
&\colorbox{BurntOrange}{$\wcs[]{H}{}$}
&\colorbox{Green}{$\wcs[]{uW}{}$}
&\colorbox{Green}{$\wcs[]{uB}{}$}
&\colorbox{SkyBlue}{$\wcs[(8)]{quqd}{}$}
&\colorbox{SkyBlue}{$\wcs[(1)]{quqd}{}$}
&\colorbox{SkyBlue}{$\wcs[]{H W B}{}$}
&&&\\ 
&$\wcs[]{uW}{}$
&\colorbox{RedOrange}{$\wcs[]{uW}{}$}
&\colorbox{BurntOrange}{$\wcs[]{uG}{}$}
&\colorbox{BurntOrange}{$\wcs[]{HW}{}$}
&\colorbox{BurntOrange}{$\wcs[]{H W B}{}$}
&\colorbox{BurntOrange}{$\wcs[]{uH}{}$}
&\colorbox{SkyBlue}{$\wcs[]{HG}{}$}
&\colorbox{SkyBlue}{$\wcs[]{uB}{}$}
&\colorbox{SkyBlue}{$\wcs[(8)]{quqd}{}$}
&\colorbox{SkyBlue}{$\wcs[]{H}{}$}
&&&\\ 
&$\wcs[]{uB}{}$
&\colorbox{RedOrange}{$\wcs[]{uB}{}$}
&\colorbox{BurntOrange}{$\wcs[]{H W B}{}$}
&\colorbox{BurntOrange}{$\wcs[]{HB}{}$}
&\colorbox{Green}{$\wcs[]{uH}{}$}
&\colorbox{Green}{$\wcs[]{uG}{}$}
&\colorbox{SkyBlue}{$\wcs[]{H}{}$}
&&&&&&\\ 
&$\wcs[]{dG}{}$
&\colorbox{RedOrange}{$\wcs[]{dG}{}$}
&\colorbox{RedOrange}{$\wcs[(8)]{quqd}{}$}
&\colorbox{RedOrange}{$\wcs[(1)]{quqd}{}$}
&\colorbox{Green}{$\wcs[]{dW}{}$}
&\colorbox{SkyBlue}{$\wcs[]{dH}{}$}
&\colorbox{SkyBlue}{$\wcs[]{dB}{}$}
&&&&&&\\ 
&$\wcs[]{dW}{}$
&\colorbox{RedOrange}{$\wcs[]{dW}{}$}
&\colorbox{RedOrange}{$\wcs[(8)]{quqd}{}$}
&\colorbox{BurntOrange}{$\wcs[]{dH}{}$}
&\colorbox{BurntOrange}{$\wcs[]{dG}{}$}
&\colorbox{BurntOrange}{$\wcs[(1)]{quqd}{}$}
&\colorbox{Green}{$\wcs[]{dB}{}$}
&&&&&&\\ 
&$\wcs[]{dB}{}$
&\colorbox{RedOrange}{$\wcs[]{dB}{}$}
&\colorbox{BurntOrange}{$\wcs[(8)]{quqd}{}$}
&\colorbox{Green}{$\wcs[(1)]{quqd}{}$}
&\colorbox{SkyBlue}{$\wcs[]{dH}{}$}
&\colorbox{SkyBlue}{$\wcs[]{dG}{}$}
&\colorbox{SkyBlue}{$\wcs[]{dW}{}$}
&&&&&&\\ 

\bottomrule
\end{tabular}
\end{adjustbox}
\caption{\small The pattern of operator mixing for the Fermionic-Bosonic operators $f^2H^3$ and $f^2 X H$ in SMEFT. Other details are the same as for Tab.~\ref{tab:mixing-bosonic-color}.}
\label{tab:mixing-bosonic-fermionic-color1}
\end{center}
\end{table}
\begin{table}[tbp]
\centering
\setlength{\tabcolsep}{2pt}
\begin{center}
\begin{adjustbox}{width=\textwidth}
\begin{tabular}{l@{\ }|c|ccccccccccccccccc{c}}
\toprule
\multicolumn{13}{c}{Fermionic-Bosonic Operators-II} & \\ 
\multicolumn{13}{c}{ \colorbox{RedOrange}{$\lambda$}, \colorbox{BurntOrange}{$\lambda^2$},
\colorbox{Green}{$\lambda^3$}, \colorbox{SkyBlue}{$\lambda^4$}}  & \\ \hline
\rowcolor{Grayd}
Class & $C_i(\Lambda)$   &    \multicolumn{4}{l}{} &$C_j(\muEW) $&  &\multicolumn{6}{l}{} \\
\midrule
 \multirow{14}{*}{\rotatebox{-270}{$f^2 H^2 D$ }} 
&$\wcs[(1)]{Hl}{}$
&\colorbox{RedOrange}{$\wcs[(1)]{Hl}{}$}
&\colorbox{Green}{$\wcs[]{lu}{}$}
&\colorbox{Green}{$\wcs[(1)]{lq}{}$}
&\colorbox{SkyBlue}{$\wcs[]{HD}{}$}
&\colorbox{SkyBlue}{$\wcs[]{He}{ }$}
&
&&&&&&\\ 
&$\wcs[(3)]{Hl}{}$
&\colorbox{RedOrange}{$\wcs[(3)]{Hl}{}$}
&\colorbox{Green}{$\wcs[]{H\Box}{}$}
&\colorbox{Green}{$\wcs[(3)]{lq}{}$}
&\colorbox{SkyBlue}{$\wcs[]{uH}{}$}
&\colorbox{SkyBlue}{$\wcs[]{ll}{}$}
&\colorbox{SkyBlue}{$\wcs[]{H}{}$}
&\colorbox{SkyBlue}{$\wcs[(3)]{Hq}{}$}
&&&&&\\ 
&$\wcs[]{He}{ }$
&\colorbox{RedOrange}{$\wcs[]{He}{ }$}
&\colorbox{Green}{$\wcs[]{eu}{}$}
&\colorbox{Green}{$\wcs[]{qe}{}$}
&\colorbox{SkyBlue}{$\wcs[]{HD}{}$}
&\colorbox{SkyBlue}{$\wcs[]{ee}{}$}
&
&&&&&&\\ 
&$\wcs[(1)]{Hq}{}$
&\colorbox{RedOrange}{$\wcs[(1)]{Hq}{}$}
&\colorbox{RedOrange}{$\wcs[]{HD}{}$}
&\colorbox{BurntOrange}{$\wcs[]{H\Box}{}$}
&\colorbox{Green}{$\wcs[(3)]{Hq}{}$}
&\colorbox{Green}{$\wcs[(1)]{qu}{}$}
&\colorbox{Green}{$\wcs[]{uH}{}$}
&\colorbox{Green}{$\wcs[]{Hu}{ }$}
&\colorbox{Green}{$\wcs[(1)]{qq}{}$}
&\colorbox{SkyBlue}{$\wcs[(8)]{qu}{}$}
&\colorbox{SkyBlue}{$\wcs[]{H}{}$}
&\colorbox{SkyBlue}{$\wcs[]{He}{ }$}
&\\ 
&$\wcs[(3)]{Hq}{}$
&\colorbox{RedOrange}{$\wcs[(3)]{Hq}{}$}
&\colorbox{RedOrange}{$\wcs[]{H\Box}{}$}
&\colorbox{BurntOrange}{$\wcs[(1)]{Hq}{}$}
&\colorbox{BurntOrange}{$\wcs[]{uH}{}$}
&\colorbox{BurntOrange}{$\wcs[]{H}{}$}
&\colorbox{Green}{$\wcs[]{HD}{}$}
&\colorbox{Green}{$\wcs[(3)]{Hl}{}$}
&\colorbox{Green}{$\wcs[(3)]{qq}{}$}
&\colorbox{SkyBlue}{$\wcs[(1)]{qq}{}$}
&\colorbox{SkyBlue}{$\wcs[]{Hu}{ }$}
&\colorbox{SkyBlue}{$\wcs[(3)]{lq}{}$}
&\colorbox{SkyBlue}{$\wcs[(1)]{qu}{}$}
&\\ 
&$\wcs[]{Hu}{ }$
&\colorbox{RedOrange}{$\wcs[]{Hu}{ }$}
&\colorbox{RedOrange}{$\wcs[]{HD}{}$}
&\colorbox{BurntOrange}{$\wcs[]{H\Box}{}$}
&\colorbox{Green}{$\wcs[]{uu}{}$}
&\colorbox{Green}{$\wcs[(1)]{qu}{}$}
&\colorbox{Green}{$\wcs[]{uH}{}$}
&\colorbox{SkyBlue}{$\wcs[(1)]{Hq}{}$}
&\colorbox{SkyBlue}{$\wcs[]{He}{ }$}
&\colorbox{SkyBlue}{$\wcs[(8)]{qu}{}$}
&\colorbox{SkyBlue}{$\wcs[(1)]{Hl}{}$}
&
&\\ 
&$\wcs[]{Hd}{ }$
&\colorbox{RedOrange}{$\wcs[]{Hd}{ }$}
&\colorbox{Green}{$\wcs[(1)]{ud}{}$}
&\colorbox{Green}{$\wcs[(1)]{qd}{}$}
&\colorbox{SkyBlue}{$\wcs[]{HD}{}$}
&\colorbox{SkyBlue}{$\wcs[(8)]{ud}{}$}
&\colorbox{SkyBlue}{$\wcs[(8)]{qd}{}$}
&\colorbox{SkyBlue}{$\wcs[]{He}{ }$}
&
&&&&\\ 
&$\wcs[]{Hud}{}$
&\colorbox{RedOrange}{$\wcs[]{Hud}{}$}
&\colorbox{Green}{$\wcs[]{dH}{}$}
&\colorbox{SkyBlue}{$\wcs[]{HD}{}$}
&\colorbox{SkyBlue}{$\wcs[]{H\Box}{}$}
&\colorbox{SkyBlue}{$\wcs[]{uH}{}$}
&&&&&&&\\ 

\bottomrule
\end{tabular}
\end{adjustbox}
\caption{\small The pattern of operator mixing for the Fermionic-Bosonic operators $f^2 H^2D$ in SMEFT. Other details are the same as for Tab.~\ref{tab:mixing-bosonic-color}.}
\label{tab:mixing-bosonic-fermionic-color2}
\end{center}
\end{table}
The presentation of the operator mixing pattern presented above is very useful for the top-down approach as well as for the bottom-up one. In the top-down approach, a given UV completion will generate a very specific set of SMEFT operators at the NP scale $\Lambda$, in which the operators could be related to each other via model couplings and/or by symmetries. These relations among the WCs are characteristic of that model. Given these initial conditions and using the above tables, one can immediately find the SMEFT operators which can be present at the EW scale. Indirectly this takes care of the one-loop RG effects of UV completions via a SMEFT route. Further, matching at the EW scale gives rise to finite corrections that need to be considered depending on the desired precision.

Finally, we remark that the phenomenological implication of operator mixing discussed above could affect the low energy predictions of new operators introduced at the NP scale in a very important way. Several groups have investigated this issue and a summary of various such studies is presented in Sec.~\ref{sec:obs}. We believe that there is still a lot to explore in this direction.

\begin{table}[tbp]
\centering
\setlength{\tabcolsep}{2pt}
\begin{center}
\begin{adjustbox}{width=\textwidth}
\begin{tabular}{l@{\ }|c|ccccccccccccccccccc{c}}
\toprule
\multicolumn{13}{c}{Four-fermion Operators-I}  \\ 
\multicolumn{13}{c}{ \colorbox{RedOrange}{$\lambda$}, \colorbox{BurntOrange}{$\lambda^2$},
\colorbox{Green}{$\lambda^3$}, \colorbox{SkyBlue}{$\lambda^4$}}   \\ \hline
\rowcolor{Grayd}
Class & $C_i(\Lambda)$   &     \multicolumn{4}{l}{} &$C_j(\muEW)$ & &\multicolumn{6}{l}{} \\
\midrule
\multirow{12}{*}{\rotatebox{-270}{ $(\overline LL) (\overline LL)$ }}

&$\wcs[]{ll}{}$
&\colorbox{RedOrange}{$\wcs[]{ll}{}$}
&\colorbox{Green}{$\wcs[]{le}{}$}
&\colorbox{SkyBlue}{$\wcs[]{lu}{}$}
&\colorbox{SkyBlue}{$\wcs[(3)]{Hl}{}$}
&\colorbox{SkyBlue}{$\wcs[(3)]{lq}{}$}
&\colorbox{SkyBlue}{$\wcs[(1)]{Hl}{}$}
&\colorbox{SkyBlue}{$\wcs[]{ld}{}$}
&&&&&\\ 
&$\wcs[(1)]{qq}{}$
&\colorbox{RedOrange}{$\wcs[(1)]{qq}{}$}
&\colorbox{RedOrange}{$\wcs[(1)]{Hq}{}$}
&\colorbox{BurntOrange}{$\wcs[(3)]{qq}{}$}
&\colorbox{BurntOrange}{$\wcs[(1)]{qu}{}$}
&\colorbox{BurntOrange}{$\wcs[(8)]{qu}{}$}
&\colorbox{BurntOrange}{$\wcs[(8)]{qd}{}$}
&\colorbox{BurntOrange}{$\wcs[]{HD}{}$}
&\colorbox{Green}{$\wcs[]{H\Box}{}$}
&\colorbox{Green}{$\wcs[(3)]{Hq}{}$}
&\colorbox{SkyBlue}{$\wcs[]{qe}{}$}
&\colorbox{SkyBlue}{$\wcs[]{Hu}{ }$}
&\colorbox{SkyBlue}{$\wcs[]{uH}{}$}
\\ &
&\colorbox{SkyBlue}{$\wcs[(8)]{quqd}{}$}
&\colorbox{SkyBlue}{$\wcs[(1)]{lq}{}$}
&\colorbox{SkyBlue}{$\wcs[(3)]{lq}{}$}
&\colorbox{SkyBlue}{$\wcs[(8)]{ud}{}$}
&\colorbox{SkyBlue}{$\wcs[(1)]{quqd}{}$}
\\ 
&$\wcs[(3)]{qq}{}$
&\colorbox{RedOrange}{$\wcs[(3)]{qq}{}$}
&\colorbox{RedOrange}{$\wcs[(1)]{qq}{}$}
&\colorbox{RedOrange}{$\wcs[(8)]{qu}{}$}
&\colorbox{RedOrange}{$\wcs[(8)]{qd}{}$}
&\colorbox{BurntOrange}{$\wcs[(3)]{Hq}{}$}
&\colorbox{Green}{$\wcs[(1)]{Hq}{}$}
&\colorbox{Green}{$\wcs[(3)]{lq}{}$}
&\colorbox{Green}{$\wcs[]{H\Box}{}$}
&\colorbox{Green}{$\wcs[(8)]{quqd}{}$}
&\colorbox{Green}{$\wcs[(1)]{qu}{}$}
&\colorbox{Green}{$\wcs[]{HD}{}$}
&\colorbox{SkyBlue}{$\wcs[]{uH}{}$}
\\ &
&\colorbox{SkyBlue}{$\wcs[(8)]{ud}{}$}
&\colorbox{SkyBlue}{$\wcs[(1)]{qd}{}$}
&\colorbox{SkyBlue}{$\wcs[(1)]{quqd}{}$}
&\colorbox{SkyBlue}{$\wcs[]{H}{}$}
&\colorbox{SkyBlue}{$\wcs[]{dd}{}$}
&\colorbox{SkyBlue}{$\wcs[]{uu}{}$}
&\colorbox{SkyBlue}{$\wcs[(3)]{Hl}{}$}
&\colorbox{SkyBlue}{$\wcs[(1)]{lq}{}$}
&\colorbox{SkyBlue}{$\wcs[]{qe}{}$}
\\ 
&$\wcs[(1)]{lq}{}$
&\colorbox{RedOrange}{$\wcs[(1)]{lq}{}$}
&\colorbox{BurntOrange}{$\wcs[(1)]{Hl}{}$}
&\colorbox{Green}{$\wcs[(3)]{lq}{}$}
&\colorbox{Green}{$\wcs[]{lu}{}$}
&\colorbox{SkyBlue}{$\wcs[]{le}{}$}
&\colorbox{SkyBlue}{$\wcs[]{qe}{}$}
&\colorbox{SkyBlue}{$\wcs[(1)]{qu}{}$}
&&&&&\\ 
&$\wcs[(3)]{lq}{}$
&\colorbox{RedOrange}{$\wcs[(3)]{lq}{}$}
&\colorbox{BurntOrange}{$\wcs[(3)]{Hl}{}$}
&\colorbox{BurntOrange}{$\wcs[(1)]{lq}{}$}
&\colorbox{Green}{$\wcs[]{ll}{}$}
&\colorbox{SkyBlue}{$\wcs[(3)]{qq}{}$}
&\colorbox{SkyBlue}{$\wcs[(3)]{Hq}{}$}
&\colorbox{SkyBlue}{$\wcs[(1)]{Hl}{}$}
&&&&&\\ 

\bottomrule
\end{tabular}
\end{adjustbox}
\caption{\small The pattern of operator mixing for the $(\overline LL) (\overline LL)$ four-fermion $\Delta B=\Delta L=0$ operators in the SMEFT. Other details are the same as for Tab.~\ref{tab:mixing-bosonic-color}.}
\label{tab:mixing-fermionic-color1}
\end{center}
\end{table}

\begin{table}[tbp]
\centering
\setlength{\tabcolsep}{2pt}
\begin{center}
\begin{adjustbox}{width=\textwidth}
\begin{tabular}{l@{\ }|c|ccccccccccccccccccc{c}}
\toprule
\multicolumn{13}{c}{Four-fermion Operators-II} & \\ 
\multicolumn{13}{c}{ \colorbox{RedOrange}{$\lambda$}, \colorbox{BurntOrange}{$\lambda^2$},
\colorbox{Green}{$\lambda^3$}, \colorbox{SkyBlue}{$\lambda^4$}}  & \\ \hline
\rowcolor{Grayd}
Class & $C_i(\Lambda)$   &     \multicolumn{4}{l}{} &$C_j(\muEW)$ & &\multicolumn{6}{l}{} \\
\midrule

\multirow{12}{*}{\rotatebox{-270}{ $(\overline RR) (\overline RR)$ }}
&$\wcs[]{ee}{}$
&\colorbox{RedOrange}{$\wcs[]{ee}{}$}
&\colorbox{SkyBlue}{$\wcs[]{eu}{}$}
&\colorbox{SkyBlue}{$\wcs[]{le}{}$}
&\colorbox{SkyBlue}{$\wcs[]{He}{ }$}
&\colorbox{SkyBlue}{$\wcs[]{ed}{}$}
&\colorbox{SkyBlue}{$\wcs[]{qe}{}$}
&&&&&&\\ 
&$\wcs[]{uu}{}$
&\colorbox{RedOrange}{$\wcs[]{uu}{}$}
&\colorbox{RedOrange}{$\wcs[]{Hu}{ }$}
&\colorbox{BurntOrange}{$\wcs[(8)]{qu}{}$}
&\colorbox{BurntOrange}{$\wcs[(8)]{ud}{}$}
&\colorbox{BurntOrange}{$\wcs[]{HD}{}$}
&\colorbox{Green}{$\wcs[(1)]{qu}{}$}
&\colorbox{Green}{$\wcs[]{eu}{}$}
&\colorbox{Green}{$\wcs[]{H\Box}{}$}
&\colorbox{SkyBlue}{$\wcs[]{lu}{}$}
&\colorbox{SkyBlue}{$\wcs[(1)]{ud}{}$}
&\colorbox{SkyBlue}{$\wcs[]{uH}{}$}
&\colorbox{SkyBlue}{$\wcs[(1)]{Hq}{}$}
&\\ 
&$\wcs[]{dd}{}$
&\colorbox{RedOrange}{$\wcs[]{dd}{}$}
&\colorbox{BurntOrange}{$\wcs[(8)]{qd}{}$}
&\colorbox{BurntOrange}{$\wcs[(8)]{ud}{}$}
&\colorbox{SkyBlue}{$\wcs[]{ed}{}$}
&\colorbox{SkyBlue}{$\wcs[(1)]{ud}{}$}
&\colorbox{SkyBlue}{$\wcs[]{Hd}{ }$}
&\colorbox{SkyBlue}{$\wcs[]{ld}{}$}
&&&&&\\ 
&$\wcs[]{eu}{}$
&\colorbox{RedOrange}{$\wcs[]{eu}{}$}
&\colorbox{BurntOrange}{$\wcs[]{He}{ }$}
&\colorbox{Green}{$\wcs[]{ee}{}$}
&\colorbox{Green}{$\wcs[]{qe}{}$}
&\colorbox{SkyBlue}{$\wcs[]{le}{}$}
&\colorbox{SkyBlue}{$\wcs[]{uu}{}$}
&\colorbox{SkyBlue}{$\wcs[]{ed}{}$}
&&&&&\\ 
&$\wcs[]{ed}{}$
&\colorbox{RedOrange}{$\wcs[]{ed}{}$}
&\colorbox{SkyBlue}{$\wcs[]{ee}{}$}
&\colorbox{SkyBlue}{$\wcs[(1)]{ud}{}$}
&\colorbox{SkyBlue}{$\wcs[]{eu}{}$}
&&&&&&&&\\ 
&$\wcs[(1)]{ud}{}$
&\colorbox{RedOrange}{$\wcs[(1)]{ud}{}$}
&\colorbox{RedOrange}{$\wcs[(8)]{ud}{}$}
&\colorbox{BurntOrange}{$\wcs[]{Hd}{ }$}
&\colorbox{Green}{$\wcs[(1)]{qd}{}$}
&\colorbox{SkyBlue}{$\wcs[]{ed}{}$}
&\colorbox{SkyBlue}{$\wcs[]{eu}{}$}
&\colorbox{SkyBlue}{$\wcs[(8)]{qu}{}$}
&\colorbox{SkyBlue}{$\wcs[]{ld}{}$}
&\colorbox{SkyBlue}{$\wcs[]{dd}{}$}
&\colorbox{SkyBlue}{$\wcs[]{uu}{}$}
&\colorbox{SkyBlue}{$\wcs[(8)]{qd}{}$}
&\colorbox{SkyBlue}{$\wcs[(8)]{quqd}{}$}
&\\ 
&$\wcs[(8)]{ud}{}$
&\colorbox{RedOrange}{$\wcs[(8)]{ud}{}$}
&\colorbox{BurntOrange}{$\wcs[(1)]{ud}{}$}
&\colorbox{Green}{$\wcs[(8)]{qu}{}$}
&\colorbox{Green}{$\wcs[(8)]{qd}{}$}
&\colorbox{SkyBlue}{$\wcs[]{dd}{}$}
&\colorbox{SkyBlue}{$\wcs[]{uu}{}$}
&\colorbox{SkyBlue}{$\wcs[]{Hd}{ }$}
&&&&&\\ 

\bottomrule
\end{tabular}
\end{adjustbox}
\caption{\small The pattern of operator mixing for the $(\overline RR) (\overline RR)$ four-fermion $\Delta B=\Delta L=0$ operators in the SMEFT. Other details are the same as for Tab.~\ref{tab:mixing-bosonic-color}.}
\label{tab:mixing-fermionic-color2}
\end{center}
\end{table}

\begin{table}[tbp]
\centering
\setlength{\tabcolsep}{2pt}
\begin{center}
\begin{adjustbox}{width=\textwidth}
\begin{tabular}{l@{\ }|c|ccccccccccccccccccc{c}}
\toprule
\multicolumn{13}{c}{Four-fermion Operators-III} & \\ 
\multicolumn{13}{c}{ \colorbox{RedOrange}{$\lambda$}, \colorbox{BurntOrange}{$\lambda^2$},
\colorbox{Green}{$\lambda^3$}, \colorbox{SkyBlue}{$\lambda^4$}}  & \\ \hline
\rowcolor{Grayd}
Class & $C_i(\Lambda)$   &     \multicolumn{4}{l}{} &$C_j(\muEW)$ & &\multicolumn{6}{l}{} \\
        \midrule

\multirow{17}{*}{\rotatebox{-270}{ $(\overline LL) (\overline RR)$ }}

&$\wcs[]{le}{}$
&\colorbox{RedOrange}{$\wcs[]{le}{}$}
&\colorbox{SkyBlue}{$\wcs[]{ee}{}$}
&\colorbox{SkyBlue}{$\wcs[]{lu}{}$}
&\colorbox{SkyBlue}{$\wcs[]{eu}{}$}
&&&&&&&&\\ 
&$\wcs[]{lu}{}$
&\colorbox{RedOrange}{$\wcs[]{lu}{}$}
&\colorbox{BurntOrange}{$\wcs[(1)]{Hl}{}$}
&\colorbox{Green}{$\wcs[(1)]{lq}{}$}
&\colorbox{SkyBlue}{$\wcs[]{le}{}$}
&\colorbox{SkyBlue}{$\wcs[]{ll}{}$}
&\colorbox{SkyBlue}{$\wcs[]{eu}{}$}
&\colorbox{SkyBlue}{$\wcs[]{uu}{}$}
&\colorbox{SkyBlue}{$\wcs[]{ld}{}$}
&&&&\\ 
&$\wcs[]{ld}{}$
&\colorbox{RedOrange}{$\wcs[]{ld}{}$}
&\colorbox{SkyBlue}{$\wcs[]{le}{}$}
&\colorbox{SkyBlue}{$\wcs[]{ed}{}$}
&\colorbox{SkyBlue}{$\wcs[(1)]{ud}{}$}
&\colorbox{SkyBlue}{$\wcs[]{lu}{}$}
&&&&&&&\\ 
&$\wcs[]{qe}{}$
&\colorbox{RedOrange}{$\wcs[]{qe}{}$}
&\colorbox{BurntOrange}{$\wcs[]{He}{ }$}
&\colorbox{Green}{$\wcs[]{eu}{}$}
&\colorbox{SkyBlue}{$\wcs[]{ee}{}$}
&\colorbox{SkyBlue}{$\wcs[(1)]{qu}{}$}
&&&&&&&\\ 
&$\wcs[(1)]{qu}{}$
&\colorbox{RedOrange}{$\wcs[(1)]{qu}{}$}
&\colorbox{RedOrange}{$\wcs[(8)]{qu}{}$}
&\colorbox{BurntOrange}{$\wcs[]{uH}{}$}
&\colorbox{BurntOrange}{$\wcs[]{Hu}{ }$}
&\colorbox{BurntOrange}{$\wcs[(1)]{Hq}{}$}
&\colorbox{Green}{$\wcs[]{HD}{}$}
&\colorbox{Green}{$\wcs[]{uu}{}$}
&\colorbox{Green}{$\wcs[(1)]{qq}{}$}
&\colorbox{Green}{$\wcs[]{H}{}$}
&\colorbox{SkyBlue}{$\wcs[]{H\Box}{}$}
&\colorbox{SkyBlue}{$\wcs[]{qe}{}$}
&\colorbox{SkyBlue}{$\wcs[(8)]{ud}{}$}
\\ &
&\colorbox{SkyBlue}{$\wcs[]{eu}{}$}
&\colorbox{SkyBlue}{$\wcs[(1)]{lq}{}$}
&\colorbox{SkyBlue}{$\wcs[(8)]{qd}{}$}
&\colorbox{SkyBlue}{$\wcs[(3)]{Hq}{}$}
&\colorbox{SkyBlue}{$\wcs[(1)]{qd}{}$}
&\colorbox{SkyBlue}{$\wcs[(8)]{quqd}{}$}
\\ 
&$\wcs[(1)]{qd}{}$
&\colorbox{RedOrange}{$\wcs[(1)]{qd}{}$}
&\colorbox{RedOrange}{$\wcs[(8)]{qd}{}$}
&\colorbox{BurntOrange}{$\wcs[]{Hd}{ }$}
&\colorbox{Green}{$\wcs[(1)]{ud}{}$}
&\colorbox{SkyBlue}{$\wcs[(8)]{ud}{}$}
&\colorbox{SkyBlue}{$\wcs[]{dd}{}$}
&\colorbox{SkyBlue}{$\wcs[]{qe}{}$}
&\colorbox{SkyBlue}{$\wcs[]{ed}{}$}
&\colorbox{SkyBlue}{$\wcs[(8)]{qu}{}$}
&\colorbox{SkyBlue}{$\wcs[(1)]{qu}{}$}
&\colorbox{SkyBlue}{$\wcs[(8)]{quqd}{}$}
&\\ 
&$\wcs[(8)]{qu}{}$
&\colorbox{RedOrange}{$\wcs[(8)]{qu}{}$}
&\colorbox{BurntOrange}{$\wcs[]{uH}{}$}
&\colorbox{BurntOrange}{$\wcs[(1)]{qu}{}$}
&\colorbox{Green}{$\wcs[(8)]{ud}{}$}
&\colorbox{Green}{$\wcs[(8)]{qd}{}$}
&\colorbox{Green}{$\wcs[]{H}{}$}
&\colorbox{Green}{$\wcs[]{uu}{}$}
&\colorbox{SkyBlue}{$\wcs[(3)]{qq}{}$}
&\colorbox{SkyBlue}{$\wcs[(1)]{qq}{}$}
&\colorbox{SkyBlue}{$\wcs[(1)]{Hq}{}$}
&\colorbox{SkyBlue}{$\wcs[]{Hu}{ }$}
&\\ 
&$\wcs[(8)]{qd}{}$
&\colorbox{RedOrange}{$\wcs[(8)]{qd}{}$}
&\colorbox{BurntOrange}{$\wcs[(1)]{qd}{}$}
&\colorbox{Green}{$\wcs[(8)]{ud}{}$}
&\colorbox{Green}{$\wcs[(8)]{qu}{}$}
&\colorbox{Green}{$\wcs[]{dd}{}$}
&\colorbox{SkyBlue}{$\wcs[(3)]{qq}{}$}
&\colorbox{SkyBlue}{$\wcs[(1)]{qq}{}$}
&\colorbox{SkyBlue}{$\wcs[]{Hd}{ }$}
&&&&\\ 
\bottomrule
\end{tabular}
\end{adjustbox}
\caption{\small The pattern of operator mixing for the $(\overline LL) (\overline RR)$ four-fermion $\Delta B=\Delta L=0$ operators in the SMEFT. Other details are the same as for Tab.~\ref{tab:mixing-bosonic-color}.}
\label{tab:mixing-fermionic-color3}
\end{center}
\end{table}

\begin{table}[tbp]
\centering
\setlength{\tabcolsep}{2pt}
\begin{center}
\begin{adjustbox}{width=0.9\textwidth}
\begin{tabular}{l@{\ }|c|ccccccccccccccccccc{c}}
\toprule
\multicolumn{13}{c}{Four-fermion Operators-IV} & \\ 
\multicolumn{13}{c}{ \colorbox{RedOrange}{$\lambda$}, \colorbox{BurntOrange}{$\lambda^2$},
\colorbox{Green}{$\lambda^3$}, \colorbox{SkyBlue}{$\lambda^4$}}  & \\ \hline
\rowcolor{Grayd}
Class & $C_i(\Lambda)$   &     \multicolumn{2}{l}{} &$C_j(\muEW)$  &\multicolumn{8}{l}{} \\
\midrule
\multirow{10}{*}{\rotatebox{-270}{ $(\overline LR) (\overline RL), (\overline LR) (\overline LR)$ }}

&$\wcs[]{ledq}{}$
&\colorbox{RedOrange}{$\wcs[]{ledq}{}$}
&&&&&&&&&&&\\ 
&$\wcs[(1)]{quqd}{}$
&\colorbox{RedOrange}{$\wcs[(1)]{quqd}{}$}
&\colorbox{RedOrange}{$\wcs[(8)]{quqd}{}$}
&\colorbox{RedOrange}{$\wcs[]{dH}{}$}
&\colorbox{Green}{$\wcs[]{dG}{}$}
&\colorbox{SkyBlue}{$\wcs[(8)]{qd}{}$}
&\colorbox{SkyBlue}{$\wcs[]{dW}{}$}
&\colorbox{SkyBlue}{$\wcs[(8)]{qu}{}$}
&\colorbox{SkyBlue}{$\wcs[]{dB}{}$}
&&&&\\ 
&$\wcs[(8)]{quqd}{}$
&\colorbox{RedOrange}{$\wcs[(8)]{quqd}{}$}
&\colorbox{BurntOrange}{$\wcs[(1)]{quqd}{}$}
&\colorbox{Green}{$\wcs[]{dH}{}$}
&\colorbox{SkyBlue}{$\wcs[]{dG}{}$}
&\colorbox{SkyBlue}{$\wcs[]{dW}{}$}
&\colorbox{SkyBlue}{$\wcs[]{dB}{}$}
&&&&&&\\ 
&$\wcs[(1)]{lequ}{}$
&\colorbox{RedOrange}{$\wcs[(1)]{lequ}{}$}
&\colorbox{BurntOrange}{$\wcs[]{eH}{}$}
&\colorbox{SkyBlue}{$\wcs[(3)]{lequ}{}$}
&&&&&&&&&\\ 
&$\wcs[(3)]{lequ}{}$
&\colorbox{RedOrange}{$\wcs[(3)]{lequ}{}$}
&\colorbox{RedOrange}{$\wcs[(1)]{lequ}{}$}
&\colorbox{BurntOrange}{$\wcs[]{eW}{}$}
&\colorbox{BurntOrange}{$\wcs[]{eB}{}$}
&\colorbox{Green}{$\wcs[]{eH}{}$}
&\colorbox{SkyBlue}{$\wcs[]{eu}{}$}
&\colorbox{SkyBlue}{$\wcs[]{qe}{}$}
&\colorbox{SkyBlue}{$\wcs[]{lu}{}$}
&&&&\\ 

\bottomrule
\end{tabular}
\end{adjustbox}
\caption{\small The pattern of operator mixing for the $(\overline LR) (\overline RL)$ and $(\overline LR) (\overline LR)$ four-fermion $\Delta B=\Delta L=0$ operators in the SMEFT. Other details are the same as for  Tab.~\ref{tab:mixing-bosonic-color}.}
\label{tab:mixing-fermionic-color4}
\end{center}
\end{table}

\subsection{Active and Passive Operators}
Inspecting Tabs.~\ref{tab:mixing-bosonic-color}-\ref{tab:mixing-fermionic-color4} we observe that certain operators are very active in the RG evolution contributing at $\mathcal{O}(\lambda^2)$-$\mathcal{O}(\lambda^4)$ and occasionally at the $\mathcal{O}(\lambda)$ level. Concentrating on operators involving fermions these are in particular the following sets of operators.
First
\be 
\ops[(1)]{Hq}{}\,,\quad \ops[(3)]{Hq}{}\,,\quad \ops[(1)]{lq}{}\,,\quad
\ops[(3)]{lq}{}\,,\quad \ops[(1)]{qq}{}\,,\quad \ops[(3)]{qq}{}\,,
\ee
that are important for semi-leptonic and non-leptonic decays. Next the operators
\be
\ops[(1,8)]{qu}{}\,,\quad  \ops[(1,8)]{qd}{}\,,\quad  \ops[]{uu}{}\,,\quad 
\ops[]{dd}{}\,,\quad  \ops[(1,8)]{ud}{}\,,\quad \ops[]{Hu}{}\,,\quad \ops[]{Hd}{}\,,
\ee
that are important for non-leptonic decays. Finally
\be
\ops[]{uG}{}\,,\quad \ops[]{uW}{}\,,\quad \ops[]{uB}{}\,, \quad \ops[]{dG}{}\,,\quad \ops[]{dW}{}\,,\quad \ops[]{dB}{}\,,\quad
\ops[(1)]{quqd}{}\,,\quad \ops[(8)]{quqd}{}\,,
\quad \ops[(1)]{lequ}{}\,,\quad\ops[(3)]{lequ}{}\,,
\ee
that play an important role for EDMs and magnetic moments.

On the other hand there are passive operators that do not have any impact on the remaining operators if $\mathcal{O}(\lambda^5)$ effects are neglected. These are
\be
\ops[]{H}{}\,, \qquad \ops[]{dH}{}\,,\qquad \ops[]{eH}{}\,,\qquad \ops[]{ledq}{}\,.
\ee
In addition there are operators that impact other operators mainly at the $\mathcal{O}(\lambda^4)$. These are
\be
{\ops[]{eB}{}\,,\qquad}\ops[]{ll}{}\,,\qquad  \ops[]{ee}{}\,,\qquad \ops[]{ed}{}\,,\qquad \ops[]{le}{}\,, \qquad \ops[]{ld}{}\,.
\ee
This means that these two sets of operators play a subleading role in the RG evolution in the SMEFT although they might be important in certain decays to which they contribute already at tree-level.

\subsection{Another View on Operator Mixing}\label{A-F}

\begin{table}[H]
\centering
\setlength{\tabcolsep}{3pt}
\begin{center}
\begin{adjustbox}{width=0.7\textwidth}
\begin{tabular}{l@{\ }|c|ccccccccccccccccc}
\toprule
\multicolumn{13}{c}{Operator-mixing for the SMEFT operators of {\bf set A}} \\ 
\multicolumn{13}{c}{ \colorbox{RedOrange}{$\lambda$}, \colorbox{BurntOrange}{$\lambda^2$},
\colorbox{Green}{$\lambda^3$}, \colorbox{SkyBlue}{$\lambda^4$}}   \\ \hline
\rowcolor{Grayd}
 &    &    \multicolumn{4}{l}{} & &  &\multicolumn{6}{l}{} \\
\hline
&$\wcs[]{ll}{}$
&\colorbox{RedOrange}{$\wcs[]{ll}{}$}
&\colorbox{SkyBlue}{$\wcs[(1)]{Hl}{}$}
&\colorbox{SkyBlue}{$\wcs[(3)]{Hl}{}$}
&
&
&
&\colorbox{SkyBlue}{$\wcs[(3)]{lq}{}$}
&
&
&\\
&$\wcs[(1)]{Hl}{}$
&
&\colorbox{RedOrange}{ $\wcs[(1)]{Hl}{}$}
& & &
&\colorbox{Green}{$\wcs[(1)]{lq}{}$}
&
&
&
&
\\ 
&$\wcs[(3)]{Hl}{}$
&\colorbox{SkyBlue}{$\wcs[]{ll}{}$}
&
&\colorbox{RedOrange}{$\wcs[(3)]{Hl}{}$}
&
&\colorbox{SkyBlue}{$\wcs[(3)]{Hq}{}$}
&
&\colorbox{Green}{$\wcs[(3)]{lq}{}$}
&
&
&
&
\\ 
&$\wcs[(1)]{Hq}{}$
&
&
&
&\colorbox{RedOrange}{$\wcs[(1)]{Hq}{}$}
&\colorbox{Green}{$\wcs[(3)]{Hq}{}$}
&
&
&\colorbox{Green}{$\wcs[(1)]{qq}{}$}
&
&
&
&
\\ 
&$\wcs[(3)]{Hq}{}$
&
&
&\colorbox{Green}{$\wcs[(3)]{Hl}{}$}
&\colorbox{BurntOrange}{$\wcs[(1)]{Hq}{}$}
&\colorbox{RedOrange}{$\wcs[(3)]{Hq}{}$}
&
&\colorbox{SkyBlue}{$\wcs[(3)]{lq}{}$}
&\colorbox{SkyBlue}{$\wcs[(1)]{qq}{}$}
&\colorbox{Green}{$\wcs[(3)]{qq}{}$}
&
&
&
\\
&$\wcs[(1)]{lq}{}$
&
&\colorbox{BurntOrange}{$\wcs[(1)]{Hl}{}$}
&
&
&
&\colorbox{RedOrange}{$\wcs[(1)]{lq}{}$}
&\colorbox{Green}{$\wcs[(3)]{lq}{}$}
&
\\ 
&$\wcs[(3)]{lq}{}$
&\colorbox{Green}{$\wcs[]{ll}{}$}
&\colorbox{SkyBlue}{$\wcs[(1)]{Hl}{}$}
&\colorbox{BurntOrange}{$\wcs[(3)]{Hl}{}$}
&
&\colorbox{SkyBlue}{$\wcs[(3)]{Hq}{}$}
&\colorbox{BurntOrange}{$\wcs[(1)]{lq}{}$}
&\colorbox{RedOrange}{$\wcs[(3)]{lq}{}$}
&
&\colorbox{SkyBlue}{$\wcs[(3)]{qq}{}$}
\\
&$\wcs[(1)]{qq}{}$
&
&
&
&\colorbox{RedOrange}{$\wcs[(1)]{Hq}{}$}
&\colorbox{Green}{$\wcs[(3)]{Hq}{}$}
&\colorbox{SkyBlue}{$\wcs[(1)]{lq}{}$}
&\colorbox{SkyBlue}{$\wcs[(3)]{lq}{}$}
&\colorbox{RedOrange}{$\wcs[(1)]{qq}{}$}
&\colorbox{BurntOrange}{$\wcs[(3)]{qq}{}$}
&
&
&
\\ 
&$\wcs[(3)]{qq}{}$
&
&
&\colorbox{SkyBlue}{$\wcs[(3)]{Hl}{}$}
&\colorbox{Green}{$\wcs[(1)]{Hq}{}$}
&\colorbox{BurntOrange}{$\wcs[(3)]{Hq}{}$}
&\colorbox{SkyBlue}{$\wcs[(1)]{lq}{}$}
&\colorbox{Green}{$\wcs[(3)]{lq}{}$}
&\colorbox{RedOrange}{$\wcs[(1)]{qq}{}$}
&\colorbox{RedOrange}{$\wcs[(3)]{qq}{}$}
&
&\\ 
\bottomrule
\end{tabular}
\end{adjustbox}
\caption{\small The pattern of operator mixing for SMEFT operators of {\bf set A}. Other details are the same as for Tab.~\ref{tab:mixing-bosonic-color}.}
\label{tab:mixing-bosonic-fermionic-color2new}
\end{center}
\end{table}

Tabs.~\ref{tab:mixing-bosonic-color}-\ref{tab:mixing-fermionic-color4} give us an impression of how a given WC at the NP scale $\Lambda$ affects through RG evolution WCs at the EW scale. These tables correspond precisely to the tables of operators in the Warsaw basis given in App.~\ref{Warsawbasis}. It is evident that the mixing in question acts across different classes of operators. In order to get a better insight into the involved mixing it is useful to divide the 59 Warsaw basis operators into different sets in which on the one hand the operators in a given set mix dominantly among themselves, and on the other hand describe specific processes. The operator mixing in a given set is then described by a matrix analogous to the ADM.

Such sets of phenomenologically relevant operators are collected in Tabs.~\ref{tab:mixing-bosonic-fermionic-color2new}-\ref{tab:mixing-bosonic-colornew6}. They are defined as follows:

{\bf Set A:} This set plays an important role in semi-leptonic and leptonic decays of mesons and the corresponding matrix is given in Tab.~\ref{tab:mixing-bosonic-fermionic-color2new}.

{\bf Set B:} This set plays an important role in non-leptonic decays and quark mixing and the corresponding matrix is given in Tab.~\ref{tab:mixing-fermionic-color3new}.

{\bf Set C:} This set plays an important role in lepton flavour violating decays and the corresponding matrix is given in Tab.~\ref{tab:mixing-fermionic-color4new}.

{\bf Set D:} This set plays an important role for EDMs and Magnetic Moments of leptons and the corresponding matrix is given in Tab.~\ref{tab:mixing-bosonic-fermionic-color1new1}.

{\bf Set E:} This set plays a very important role in the cases of EDMs of quarks and hadrons and the corresponding matrix is given in Tab.~\ref{tab:mixing-bosonic-fermionic-color1new5}.

{\bf Set F:} This set plays a very important role for EW precision tests and the corresponding matrix is given in Tab.~\ref{tab:mixing-bosonic-colornew6}.

\begin{table}[H]
\centering
\setlength{\tabcolsep}{2pt}
\begin{center}
\begin{adjustbox}{width=0.7\textwidth}
\begin{tabular}{l@{\ }|c|ccccccccccccccccccc{c}}
\toprule
\multicolumn{13}{c}{Operator-mixing for the SMEFT operators of {\bf set B}}  & \\ 
\multicolumn{13}{c}{ \colorbox{RedOrange}{$\lambda$}, \colorbox{BurntOrange}{$\lambda^2$},
\colorbox{Green}{$\lambda^3$}, \colorbox{SkyBlue}{$\lambda^4$}}  & \\ \hline
\rowcolor{Grayd}
&    &     \multicolumn{4}{l}{} &  & &\multicolumn{6}{l}{} \\
\midrule
&$\wcs[(1)]{qu}{}$
&\colorbox{RedOrange}{$\wcs[(1)]{qu}{}$}
& \colorbox{SkyBlue}{$\wcs[(1)]{qd}{}$}    
&\colorbox{RedOrange}{$\wcs[(8)]{qu}{}$}
&\colorbox{SkyBlue}{$\wcs[(8)]{qd}{}$}
&\colorbox{Green}{$\wcs[]{uu}{}$}
&        
 &     
& \colorbox{SkyBlue}{$\wcs[(8)]{ud}{}$}
&\colorbox{BurntOrange}{$\wcs[]{Hu}{ }$}
&
\\ 
&$\wcs[(1)]{qd}{}$
&\colorbox{SkyBlue}{$\wcs[(1)]{qu}{}$}
&\colorbox{RedOrange}{$\wcs[(1)]{qd}{}$}
&\colorbox{SkyBlue}{$\wcs[(8)]{qu}{}$}
&\colorbox{RedOrange}{$\wcs[(8)]{qd}{}$}
&
& \colorbox{SkyBlue}{$\wcs[]{dd}{}$}
&\colorbox{Green}{$\wcs[(1)]{ud}{}$}
&\colorbox{SkyBlue}{$\wcs[(8)]{ud}{}$}
&
&\colorbox{BurntOrange}{$\wcs[]{Hd}{ }$}
&
&
&
&
&
&
\\ 
&$\wcs[(8)]{qu}{}$
&\colorbox{BurntOrange}{$\wcs[(1)]{qu}{}$}
&
&\colorbox{RedOrange}{$\wcs[(8)]{qu}{}$}
&\colorbox{Green}{$\wcs[(8)]{qd}{}$}
&\colorbox{Green}{$\wcs[]{uu}{}$}
&
&
&\colorbox{Green}{$\wcs[(8)]{ud}{}$}
& \colorbox{SkyBlue}{$\wcs[]{Hu}{ }$}
&
&
&
&
&
&
\\ 
&$\wcs[(8)]{qd}{}$
&
&\colorbox{BurntOrange}{$\wcs[(1)]{qd}{}$}
&\colorbox{Green}{$\wcs[(8)]{qu}{}$}
&\colorbox{RedOrange}{$\wcs[(8)]{qd}{}$}
&
&\colorbox{Green}{$\wcs[]{dd}{}$}
&
&\colorbox{Green}{$\wcs[(8)]{ud}{}$}
&
&\colorbox{SkyBlue}{$\wcs[]{Hd}{ }$}
&
\\
&$\wcs[]{uu}{}$
&\colorbox{Green}{$\wcs[(1)]{qu}{}$}
&
&\colorbox{BurntOrange}{$\wcs[(8)]{qu}{}$}
&
&\colorbox{RedOrange}{$\wcs[]{uu}{}$}
&
&\colorbox{SkyBlue}{$\wcs[(1)]{ud}{}$}
&\colorbox{BurntOrange}{$\wcs[(8)]{ud}{}$}
&\colorbox{RedOrange}{$\wcs[]{Hu}{ }$}
&
&
\\ 
&$\wcs[]{dd}{}$
&
&
&
&\colorbox{BurntOrange}{$\wcs[(8)]{qd}{}$}
&
&\colorbox{RedOrange}{$\wcs[]{dd}{}$}
&\colorbox{SkyBlue}{$\wcs[(1)]{ud}{}$}
&\colorbox{BurntOrange}{$\wcs[(8)]{ud}{}$}
&
&\colorbox{SkyBlue}{$\wcs[]{Hd}{ }$}
&
&
\\
&$\wcs[(1)]{ud}{}$
&
&\colorbox{Green}{$\wcs[(1)]{qd}{}$}
&\colorbox{SkyBlue}{$\wcs[(8)]{qu}{}$}
&\colorbox{SkyBlue}{$\wcs[(8)]{qd}{}$}
&\colorbox{SkyBlue}{$\wcs[]{uu}{}$}
&\colorbox{SkyBlue}{$\wcs[]{dd}{}$}
&\colorbox{RedOrange}{$\wcs[(1)]{ud}{}$}
&\colorbox{RedOrange}{$\wcs[(8)]{ud}{}$}
&
&\colorbox{BurntOrange}{$\wcs[]{Hd}{ }$}
&
&
&
\\ 
&$\wcs[(8)]{ud}{}$
&
&
&\colorbox{Green}{$\wcs[(8)]{qu}{}$}
&\colorbox{Green}{$\wcs[(8)]{qd}{}$}
&\colorbox{SkyBlue}{$\wcs[]{uu}{}$}
&\colorbox{SkyBlue}{$\wcs[]{dd}{}$}
&\colorbox{BurntOrange}{$\wcs[(1)]{ud}{}$}
&\colorbox{RedOrange}{$\wcs[(8)]{ud}{}$}
&
&\colorbox{SkyBlue}{$\wcs[]{Hd}{ }$}
&
&
\\
&$\wcs[]{Hu}{ }$
&\colorbox{Green}{$\wcs[(1)]{qu}{}$}
&
&\colorbox{SkyBlue}{$\wcs[(8)]{qu}{}$}
&
&\colorbox{Green}{$\wcs[]{uu}{}$}
&
&
&
&\colorbox{RedOrange}{$\wcs[]{Hu}{ }$}
&
&
&
&
&
&
\\ 
&$\wcs[]{Hd}{ }$
&
&\colorbox{Green}{$\wcs[(1)]{qd}{}$}
&
&\colorbox{SkyBlue}{$\wcs[(8)]{qd}{}$}
&
&
&\colorbox{Green}{$\wcs[(1)]{ud}{}$}
&\colorbox{SkyBlue}{$\wcs[(8)]{ud}{}$}
&
& \colorbox{RedOrange}{$\wcs[]{Hd}{ }$}
&
\\ 
&$\wcs[]{Hud}{}$
&
&
&
&
&
&
&
&
&
&
&\colorbox{RedOrange}{$\wcs[]{Hud}{}$}
&
&\\
\bottomrule
\end{tabular}
\end{adjustbox}
\caption{\small The pattern of operator mixing for the {\bf set B} SMEFT operators. Other details are the same as for Tab.~\ref{tab:mixing-bosonic-color}.}
\label{tab:mixing-fermionic-color3new}
\end{center}
\end{table}
\begin{table}[H]
\centering
\setlength{\tabcolsep}{5pt}
\begin{center}
\begin{adjustbox}{width=0.7\textwidth}
\begin{tabular}{l@{\ }|c|cccccccc{c}}
\toprule
\multicolumn{10}{c}{Operator-mixing for the SMEFT operators of {\bf set C}} & \\ 
\multicolumn{10}{c}{ \colorbox{RedOrange}{$\lambda$}, \colorbox{BurntOrange}{$\lambda^2$},
\colorbox{Green}{$\lambda^3$}, \colorbox{SkyBlue}{$\lambda^4$}}  & \\ \hline
\rowcolor{Grayd}
 &    &     \multicolumn{3}{l}{} & & &\multicolumn{4}{l}{} \\
        \midrule

&$\wcs[]{le}{}$
&\colorbox{RedOrange}{$\wcs[]{le}{}$}
&\colorbox{SkyBlue}{$\wcs[]{lu}{}$}
 &
&\colorbox{SkyBlue}{$\wcs[]{ee}{}$}
&\colorbox{SkyBlue}{$\wcs[]{eu}{}$}
&
&
&
\\ 
&$\wcs[]{lu}{}$
&\colorbox{SkyBlue}{$\wcs[]{le}{}$}
&\colorbox{RedOrange}{$\wcs[]{lu}{}$}
&\colorbox{SkyBlue}{$\wcs[]{ld}{}$}
&
&\colorbox{SkyBlue}{$\wcs[]{eu}{}$}
&
&
&
&
\\ 
&$\wcs[]{ld}{}$
&\colorbox{SkyBlue}{$\wcs[]{le}{}$}
&\colorbox{SkyBlue}{$\wcs[]{lu}{}$}
&\colorbox{RedOrange}{$\wcs[]{ld}{}$}
&
&
&\colorbox{SkyBlue}{$\wcs[]{ed}{}$}
&
&
&\\
&$\wcs[]{ee}{}$
&\colorbox{SkyBlue}{$\wcs[]{le}{}$}
&
&
&\colorbox{RedOrange}{$\wcs[]{ee}{}$}
&\colorbox{SkyBlue}{$\wcs[]{eu}{}$}
&\colorbox{SkyBlue}{$\wcs[]{ed}{}$}
&\colorbox{SkyBlue}{$\wcs[]{qe}{}$}
&\colorbox{SkyBlue}{$\wcs[]{He}{ }$}
\\ 
&$\wcs[]{eu}{}$
&\colorbox{SkyBlue}{$\wcs[]{le}{}$}
&
&
&\colorbox{Green}{$\wcs[]{ee}{}$}
&\colorbox{RedOrange}{$\wcs[]{eu}{}$}
&\colorbox{SkyBlue}{$\wcs[]{ed}{}$}
&\colorbox{Green}{$\wcs[]{qe}{}$}
&\colorbox{BurntOrange}{$\wcs[]{He}{ }$}
\\
&$\wcs[]{ed}{}$
&&&
&\colorbox{SkyBlue}{$\wcs[]{ee}{}$}
&\colorbox{SkyBlue}{$\wcs[]{eu}{}$}
&\colorbox{RedOrange}{$\wcs[]{ed}{}$}
&
&
\\ 
&$\wcs[]{qe}{}$
&&&
&\colorbox{SkyBlue}{$\wcs[]{ee}{}$}
&\colorbox{Green}{$\wcs[]{eu}{}$}
&
&\colorbox{RedOrange}{$\wcs[]{qe}{}$}
&\colorbox{BurntOrange}{$\wcs[]{He}{ }$}
\\
&$\wcs[]{He}{ }$
&&&
&\colorbox{SkyBlue}{$\wcs[]{ee}{}$}
&\colorbox{Green}{$\wcs[]{eu}{}$}
&
&\colorbox{Green}{$\wcs[]{qe}{}$}
&\colorbox{RedOrange}{$\wcs[]{He}{ }$}
\\
\bottomrule
\end{tabular}
\end{adjustbox}
\caption{\small The pattern of operator mixing for the {\bf set C} SMEFT operators. Other details are the same as for Tab.~\ref{tab:mixing-bosonic-color}.}
\label{tab:mixing-fermionic-color4new}
\end{center}
\end{table}

\begin{table}[H]
\centering
\setlength{\tabcolsep}{5pt}
\begin{center}
\begin{adjustbox}{width=0.7\textwidth}
\begin{tabular}{l@{\ }|c|ccccccc{c}}
\toprule
\multicolumn{8}{c}{Operator-mixing for the SMEFT operators of {\bf set D}} & \\ 
\multicolumn{8}{c}{ \colorbox{RedOrange}{$\lambda$}, \colorbox{BurntOrange}{$\lambda^2$},
\colorbox{Green}{$\lambda^3$}, \colorbox{SkyBlue}{$\lambda^4$}}  & \\ \hline
\rowcolor{Grayd}
&&&&&&&& \\ 
\hline
&$\wcs[]{eW}{}$
&\colorbox{RedOrange}{$\wcs[]{eW}{}$}
&\colorbox{SkyBlue}{$\wcs[]{eB}{}$}
&\colorbox{SkyBlue}{$\wcs[(1)]{lequ}{}$}
&\colorbox{Green}{$\wcs[(3)]{lequ}{}$}
&\colorbox{Green}{$\wcs[]{eH}{}$}
&
&
\\
&$\wcs[]{eB}{}$
&
&\colorbox{RedOrange}{$\wcs[]{eB}{}$}
&
&\colorbox{SkyBlue}{$\wcs[(3)]{lequ}{}$}
&
&
&
\\
&$\wcs[(1)]{lequ}{}$
&&
&\colorbox{RedOrange}{$\wcs[(1)]{lequ}{}$}
&\colorbox{SkyBlue}{$\wcs[(3)]{lequ}{}$}
&\colorbox{BurntOrange}{$\wcs[]{eH}{}$}
&
&
\\ 
&$\wcs[(3)]{lequ}{}$
&\colorbox{BurntOrange}{$\wcs[]{eW}{}$}
&\colorbox{BurntOrange}{$\wcs[]{eB}{}$}
&\colorbox{RedOrange}{$\wcs[(1)]{lequ}{}$}
&\colorbox{RedOrange}{$\wcs[(3)]{lequ}{}$}
&\colorbox{Green}{$\wcs[]{eH}{}$}
&
&
\\
&$\wcs[]{eH}{}$
&&&&
&\colorbox{RedOrange}{$\wcs[]{eH}{}$}
&
&
\\
&$\wcs[]{ledq}{}$
&&&&&
&\colorbox{RedOrange}{$\wcs[]{ledq}{}$}
&
\\
\bottomrule
\end{tabular}
\end{adjustbox}
\caption{\small The mixing pattern of {\bf set D} operators playing an important role in EDMs and Magnetic Moments of leptons. Other details are the same as for Tab.~\ref{tab:mixing-bosonic-color}.}
\label{tab:mixing-bosonic-fermionic-color1new1}
\end{center}
\end{table}

\begin{table}[H]
  \centering
    \setlength{\tabcolsep}{6pt}
\begin{center}
\begin{adjustbox}{width=\textwidth}
  \begin{tabular}{l@{\ }|c|ccccccccccccccccc{c}}
 \toprule
\multicolumn{16}{c}{Operator-mixing for the SMEFT operators of {\bf set E}} & \\ 
\multicolumn{16}{c}{ \colorbox{RedOrange}{$\lambda$}, \colorbox{BurntOrange}{$\lambda^2$},
\colorbox{Green}{$\lambda^3$}, \colorbox{SkyBlue}{$\lambda^4$}}  & \\ \hline
\rowcolor{Grayd}
 &&&&&&&&&&&&&&&& \\ 
\hline 
&$\wcs[]{G}{}$
&\colorbox{RedOrange}{$\wcs[]{G}{}$}
&
&\colorbox{BurntOrange}{$\wcs[]{uG}{}$}
&\colorbox{SkyBlue}{$\wcs[]{uW}{}$}
&\colorbox{SkyBlue}{$\wcs[]{uB}{}$}
&\colorbox{SkyBlue}{$\wcs[]{dG}{}$}
& &&&
&\colorbox{BurntOrange}{$\wcs[]{uH}{}$}
&
&\colorbox{SkyBlue}{$\wcs[]{HG}{}$}
&\\
&$\wcs[]{\widetilde G }{}$
&
&\colorbox{RedOrange}{$\wcs[]{\widetilde G }{}$}
&\colorbox{BurntOrange}{$\wcs[]{uG}{}$}
&\colorbox{SkyBlue}{$\wcs[]{uW}{}$}
&\colorbox{SkyBlue}{$\wcs[]{uB}{}$}
&\colorbox{SkyBlue}{$\wcs[]{dG}{}$}
&&&&
&\colorbox{BurntOrange}{$\wcs[]{uH}{}$}
&&&
\colorbox{SkyBlue}{$\wcs[]{H\widetilde G}{}$}
\\
&$\wcs[]{uG}{}$
&
&
&\colorbox{RedOrange}{$\wcs[]{uG}{}$}
&\colorbox{Green}{$\wcs[]{uW}{}$}
&\colorbox{Green}{$\wcs[]{uB}{}$}
&&&
&\colorbox{SkyBlue}{$\wcs[(1)]{quqd}{}$}
&\colorbox{SkyBlue}{$\wcs[(8)]{quqd}{}$}
&\colorbox{RedOrange}{$\wcs[]{uH}{}$}
&
&\colorbox{BurntOrange}{$\wcs[]{HG}{}$}
\\ 
&$\wcs[]{uW}{}$
&
&
&\colorbox{BurntOrange}{$\wcs[]{uG}{}$}
&\colorbox{RedOrange}{$\wcs[]{uW}{}$}
&\colorbox{SkyBlue}{$\wcs[]{uB}{}$}
&&&&
& \colorbox{SkyBlue}{$\wcs[(8)]{quqd}{}$}
&\colorbox{BurntOrange}{$\wcs[]{uH}{}$}
&
&\colorbox{SkyBlue}{$\wcs[]{HG}{}$}
&
\\
&$\wcs[]{uB}{}$
&&
&\colorbox{Green}{$\wcs[]{uG}{}$}
&
&\colorbox{RedOrange}{$\wcs[]{uB}{}$}
&&&&&
&\colorbox{Green}{$\wcs[]{uH}{}$}
&
&\\ 
&$\wcs[]{dG}{}$
&&&&&
&\colorbox{RedOrange}{$\wcs[]{dG}{}$}
&\colorbox{Green}{$\wcs[]{dW}{}$}
&\colorbox{SkyBlue}{$\wcs[]{dB}{}$}
&\colorbox{RedOrange}{$\wcs[(1)]{quqd}{}$}
&\colorbox{RedOrange}{$\wcs[(8)]{quqd}{}$}
&
&\colorbox{SkyBlue}{$\wcs[]{dH}{}$}
\\ 
&$\wcs[]{dW}{}$
&&&&&
&\colorbox{BurntOrange}{$\wcs[]{dG}{}$}
&\colorbox{RedOrange}{$\wcs[]{dW}{}$}
&\colorbox{Green}{$\wcs[]{dB}{}$}
&\colorbox{BurntOrange}{$\wcs[(1)]{quqd}{}$}
&\colorbox{RedOrange}{$\wcs[(8)]{quqd}{}$}
&
&\colorbox{BurntOrange}{$\wcs[]{dH}{}$}
&\\ 
&$\wcs[]{dB}{}$
&&&&&
&\colorbox{SkyBlue}{$\wcs[]{dG}{}$}
&\colorbox{SkyBlue}{$\wcs[]{dW}{}$}
&\colorbox{RedOrange}{$\wcs[]{dB}{}$}
&\colorbox{Green}{$\wcs[(1)]{quqd}{}$}
&\colorbox{BurntOrange}{$\wcs[(8)]{quqd}{}$}
&
&\colorbox{SkyBlue}{$\wcs[]{dH}{}$}
\\ 
&$\wcs[(1)]{quqd}{}$
&&&&&
&\colorbox{Green}{$\wcs[]{dG}{}$}
&\colorbox{SkyBlue}{$\wcs[]{dW}{}$}
&\colorbox{SkyBlue}{$\wcs[]{dB}{}$}
&\colorbox{RedOrange}{$\wcs[(1)]{quqd}{}$}
&\colorbox{RedOrange}{$\wcs[(8)]{quqd}{}$}
&
&\colorbox{RedOrange}{$\wcs[]{dH}{}$} 
\\ 
&$\wcs[(8)]{quqd}{}$
&&&&&
&\colorbox{SkyBlue}{$\wcs[]{dG}{}$}
&\colorbox{SkyBlue}{$\wcs[]{dW}{}$}
&\colorbox{SkyBlue}{$\wcs[]{dB}{}$}
&\colorbox{BurntOrange}{$\wcs[(1)]{quqd}{}$}
&\colorbox{RedOrange}{$\wcs[(8)]{quqd}{}$}
&
&\colorbox{Green}{$\wcs[]{dH}{}$}
\\
&$\wcs[]{uH}{}$
&&&&&&&&&&
&\colorbox{RedOrange}{$\wcs[]{uH}{}$}
\\ 
&$\wcs[]{dH}{}$
&&&&&&&&&&&
&\colorbox{RedOrange}{$\wcs[]{dH}{}$}
\\
&$\wcs[]{HG}{}$
&&
&\colorbox{BurntOrange}{$\wcs[]{uG}{}$}
&&&&&&&
&\colorbox{RedOrange}{$\wcs[]{uH}{}$}
&\colorbox{Green}{$\wcs[]{dH}{}$}
&\colorbox{RedOrange}{$\wcs[]{HG}{}$}
&\\
&$\wcs[]{H\widetilde G}{}$
&&
&\colorbox{BurntOrange}{$\wcs[]{uG}{}$}
&&&&&&&
&\colorbox{RedOrange}{$\wcs[]{uH}{}$}
&\colorbox{Green}{$\wcs[]{dH}{}$}
&
&\colorbox{RedOrange}{$\wcs[]{H\widetilde G}{}$}
&\\
\bottomrule
\end{tabular}
\end{adjustbox}
\caption{\small The mixing pattern of {\bf set E} operators playing an important role in EDMs of quarks and hadrons. Other details are the same as for Tab.~\ref{tab:mixing-bosonic-color}.}
\label{tab:mixing-bosonic-fermionic-color1new5}
\end{center}
\end{table}

Inspecting Tabs.~\ref{tab:mixing-bosonic-color}-\ref{tab:mixing-fermionic-color4} we find 400 entries while Tabs.~\ref{tab:mixing-bosonic-fermionic-color2new}-\ref{tab:mixing-bosonic-colornew6} have 256 entries. Evidently 144 entries present in Tabs.~\ref{tab:mixing-bosonic-color}-\ref{tab:mixing-fermionic-color4} but absent in Tabs.~\ref{tab:mixing-bosonic-fermionic-color2new}-\ref{tab:mixing-bosonic-colornew6} describe the connections between the sets A-F generated {through} operator mixing. In this manner one can imagine that the sets A-F represent six cities and the connections represent the roads between them.

However, one should keep in mind that a given correlation between two WCs with two Lorentz structures as given in the tables above includes only the leading correlation among all possible flavour combinations. Therefore some correlations with higher powers than the leading one are not included in the counting of correlations given below. While this is a limitation,
it allows to present the general pattern of correlations in a transparent way. Additional insight will be provided through the charts in PART II of our review and with the help of $\rho$ and $\eta$ parameters introduced in next section.

\begin{table}[H]
\centering
\setlength{\tabcolsep}{1.5pt}
\begin{center}
\begin{tabular}{l@{\ }|c|ccccccccccccccccc{c}}
\toprule
\multicolumn{14}{c}{Operator-mixing for the SMEFT operators of {\bf set F}}  \\ 
\multicolumn{14}{c}{ \colorbox{RedOrange}{$\lambda$}, \colorbox{BurntOrange}{$\lambda^2$},   
\colorbox{Green}{$\lambda^3$}, \colorbox{SkyBlue}{$\lambda^4$}}  \\ 
\midrule
\rowcolor{Grayd}  &&&&&&&&&&&&& \\
\midrule
&$\wcs[]{W}{}$
&\colorbox{RedOrange}{$\wcs[]{W}{}$}
&&&&&&
&\colorbox{BurntOrange}{$\wcs[]{HW}{}$}
&\colorbox{SkyBlue}{$\wcs[]{H W B}{}$}
&
&
&
&&&&&&\\ 
&$\wcs[]{\widetilde W }{}$
&
&\colorbox{RedOrange}{$\wcs[]{\widetilde W }{}$}
&&&&&&&&
&\colorbox{BurntOrange}{$\wcs[]{H\widetilde W}{}$}
&\colorbox{SkyBlue}{$\wcs[]{H  \widetilde W B}{}$}
&
&
\\ 
&$\wcs[]{H}{}$
&&
&\colorbox{RedOrange}{$\wcs[]{H}{}$}
&&&&&&&&&\\ 
&$\wcs[]{H\Box}{}$
&&
&\colorbox{Green}{$\wcs[]{H}{}$}
&\colorbox{RedOrange}{$\wcs[]{H\Box}{}$}
&\colorbox{SkyBlue}{$\wcs[]{HD}{}$}
&
&
&
&
&
\\ 
&$\wcs[]{HD}{}$
&&
&\colorbox{SkyBlue}{$\wcs[]{H}{}$}
&\colorbox{SkyBlue}{$\wcs[]{H\Box}{}$}
&\colorbox{RedOrange}{$\wcs[]{HD}{}$}
&
&
&
&
&
&
&
\\ 
&$\wcs[]{HB}{}$
&&&&&&
&\colorbox{RedOrange}{$\wcs[]{HB}{}$}
&
&\colorbox{SkyBlue}{$\wcs[]{H W B}{}$}
&
&
&
&
&&&&\\ 
&$\wcs[]{HW}{}$
&&
&\colorbox{SkyBlue}{$\wcs[]{H}{}$}
&&&&
&\colorbox{RedOrange}{$\wcs[]{HW}{}$}
&\colorbox{SkyBlue}{$\wcs[]{H W B}{}$}
&
&
&&\\ 
&$\wcs[]{H W B}{}$
&&&&&&
&\colorbox{Green}{$\wcs[]{HB}{}$}
&\colorbox{SkyBlue}{$\wcs[]{HW}{}$}
&\colorbox{RedOrange}{$\wcs[]{H W B}{}$}
&
&
&
&
&
&
\\ 
&$\wcs[]{H\widetilde B}{}$
&&&&&&&&&
&\colorbox{RedOrange}{$\wcs[]{H\widetilde B}{}$}
&
&\colorbox{SkyBlue}{$\wcs[]{H  \widetilde W B}{}$}
&\\ 
&$\wcs[]{H\widetilde W}{}$
&&&&&&&&&&
&\colorbox{RedOrange}{$\wcs[]{H\widetilde W}{}$}
&\colorbox{SkyBlue}{$\wcs[]{H  \widetilde W B}{}$}
&&&\\ 
&$\wcs[]{H  \widetilde W B}{}$
&&&&&&&&&
&\colorbox{Green}{$\wcs[]{H\widetilde B}{}$}
&\colorbox{SkyBlue}{$\wcs[]{H\widetilde W}{}$}
&\colorbox{RedOrange}{$\wcs[]{H  \widetilde W B}{}$}
&\\ 

\bottomrule
\end{tabular}
\caption{\small The mixing pattern of {\bf set F} operators playing an important role for EW precision tests. Other details are the same as for Tab.~\ref{tab:mixing-bosonic-color}.}
\label{tab:mixing-bosonic-colornew6}
\end{center}
\end{table}

Keeping in mind this limitation it is instructive to find the roads {of} $\mathcal{O}(\lambda)$ representing motorways and the ones {of} $\mathcal{O}(\lambda^4)$ being small roads. The result of this search is presented in Tabs.~\ref{tab:impact A}-\ref{tab:impact CF}. We observe the following
\begin{itemize}
\item
As seen in Tab.~\ref{tab:impact A} set A has the largest impact on other sets followed by the set B with the impact shown in Tab.~\ref{tab:impact B}. The impact of the remaining sets C-F on other sets is individually much smaller and has been collected in a single Tab.~\ref{tab:impact CF}
\item
We find 6 $\mathcal{O}(\lambda)$ connections, {22} $\mathcal{O}(\lambda^2)$, 30 $\mathcal{O}(\lambda^3)$ and {87} $\mathcal{O}(\lambda^4)$.
\item
We find that the set D is almost fully isolated. It is not affected by other sets and has only $\mathcal{O}(\lambda^4)$ impact on set C.
\item
Also set C has very small impact on other operator sets, mainly on sets A and B. On the other hand it is affected by several operators in sets A and B.
\item
Finally, sets E and F are significantly more active than the sets C and D. They are significantly affected by both sets A and B.
\end{itemize}

\begin{table}[H]
\centering
\setlength{\tabcolsep}{2pt}
\begin{center}
\begin{adjustbox}{width=0.7\textwidth}
\begin{tabular}{l@{\ }|c|ccccccccccccccccccc{c}}
\toprule
\multicolumn{13}{c}{Impact of {\bf Set A}  on other sets} & \\ 
\multicolumn{13}{c}{ \colorbox{RedOrange}{$\lambda$}, \colorbox{BurntOrange}{$\lambda^2$},
\colorbox{Green}{$\lambda^3$}, \colorbox{SkyBlue}{$\lambda^4$}}  & \\ \hline
\rowcolor{Grayd}
& $C_i(\Lambda)$   &     \multicolumn{4}{l}{} &$C_j(\muEW)$ & &\multicolumn{6}{l}{} \\
\midrule
&$\wcs[]{ll}{}$
&\colorbox{Green}{$\wcs[]{le}{}$~{\bf (C)}}
&\colorbox{SkyBlue}{$\wcs[]{lu}{}$~{\bf (C)}}
&\colorbox{SkyBlue}{$\wcs[]{ld}{}$~{\bf (C) }}
&
&&&&\\
&$\wcs[(1)]{Hl}{}$
&\colorbox{Green}{$\wcs[]{lu}{}$~{\bf (C)}}
&\colorbox{SkyBlue}{$\wcs[]{HD}{}$~{\bf (F)}}
&\colorbox{SkyBlue}{$\wcs[]{He}{}$~{\bf (C)}}
&
&
&
&
&
&
&
&
&
\\ 
&$\wcs[(3)]{Hl}{}$
&\colorbox{Green}{$\wcs[]{H\Box}{}$~{\bf (F)}}
&\colorbox{SkyBlue}{$\wcs[]{uH}{}$~{\bf (E)}}
&\colorbox{SkyBlue}{$\wcs[]{H}{}$~{\bf (F)}}
&
&\\
&$\wcs[(1)]{Hq}{}$
&\colorbox{RedOrange}{$\wcs[]{HD}{}$~{\bf (F)}}
&\colorbox{BurntOrange}{$\wcs[]{H\Box}{}$~{\bf (F)}}
&\colorbox{Green}{$\wcs[(1)]{qu}{}$~{\bf (B)}}
&\colorbox{Green}{$\wcs[]{uH}{}$~{\bf (E)}}
&\colorbox{Green}{$\wcs[]{Hu}{ }$~{\bf (B)}}
&\colorbox{SkyBlue}{$\wcs[(8)]{qu}{}$~{\bf (B)}}
&\colorbox{SkyBlue}{$\wcs[]{H}{}$~{\bf (F)}}
\\&
&\colorbox{SkyBlue}{$\wcs[]{He}{ }$~{\bf (C)}}
&
&
&
&
&
\\ 
&$\wcs[(3)]{Hq}{}$
&\colorbox{RedOrange}{$\wcs[]{H\Box}{}$~{\bf (F)}}
&\colorbox{BurntOrange}{$\wcs[]{uH}{}$~{\bf (E)}}
&\colorbox{BurntOrange}{$\wcs[]{H}{}$~{\bf (F)}}
&\colorbox{Green}{$\wcs[]{HD}{}$~{\bf (F)}}
&\colorbox{SkyBlue}{$\wcs[]{Hu}{ }$~{\bf (B)}}
&\colorbox{SkyBlue}{$\wcs[(1)]{qu}{}$~{\bf (B)}}
&
&
\\ 
&$\wcs[(1)]{lq}{}$
&\colorbox{Green}{$\wcs[]{lu}{}$~{\bf (C)}}
&\colorbox{SkyBlue}{$\wcs[]{le}{}$~{\bf (C)}}
&\colorbox{SkyBlue}{$\wcs[]{qe}{}$~{\bf (C)}}
&\colorbox{SkyBlue}{$\wcs[(1)]{qu}{}$~{\bf (B)}}
&\\
&$\wcs[(1)]{qq}{}$
&\colorbox{BurntOrange}{$\wcs[(1)]{qu}{}$~{\bf (B)}}
&\colorbox{BurntOrange}{$\wcs[(8)]{qu}{}$~{\bf (B)}}
&\colorbox{BurntOrange}{$\wcs[(8)]{qd}{}$~{\bf (B)}}
&\colorbox{BurntOrange}{$\wcs[]{HD}{}$~{\bf (F)}}
&\colorbox{Green}{$\wcs[]{H\Box}{}$~{\bf (F)}}
      &\colorbox{SkyBlue}{$\wcs[]{qe}{}$~{\bf (C)}}
&\colorbox{SkyBlue}{$\wcs[]{Hu}{ }$~{\bf (B)}}
\\&
&\colorbox{SkyBlue}{$\wcs[]{uH}{}$~{\bf (E)}}
&\colorbox{SkyBlue}{$\wcs[(8)]{quqd}{}$~{\bf (E)}}
&\colorbox{SkyBlue}{$\wcs[(8)]{ud}{}$~{\bf (B)}}
&\colorbox{SkyBlue}{$\wcs[(1)]{quqd}{}$~{\bf (E)}}
&
&
&
&
\\ 
&$\wcs[(3)]{qq}{}$
&\colorbox{RedOrange}{$\wcs[(8)]{qu}{}$~{\bf (B)}}
&\colorbox{RedOrange}{$\wcs[(8)]{qd}{}$~{\bf (B)}}
&\colorbox{Green}{$\wcs[]{H\Box}{}$~{\bf (F)}}
&\colorbox{Green}{$\wcs[(8)]{quqd}{}$~{\bf (E)}}
&\colorbox{Green}{$\wcs[(1)]{qu}{}$~{\bf (B)}}
&\colorbox{Green}{$\wcs[]{HD}{}$~{\bf (F)}}
&\colorbox{SkyBlue}{$\wcs[]{uH}{}$~{\bf (E)}}
\\&
&\colorbox{SkyBlue}{$\wcs[(8)]{ud}{}$~{\bf (B)}}
&\colorbox{SkyBlue}{$\wcs[(1)]{qd}{}$~{\bf (B)}}
&\colorbox{SkyBlue}{$\wcs[(1)]{quqd}{}$~{\bf (E)}}
&\colorbox{SkyBlue}{$\wcs[]{H}{}$~{\bf (F)}}
&\colorbox{SkyBlue}{$\wcs[]{dd}{}$~{\bf (B)}}
&\colorbox{SkyBlue}{$\wcs[]{uu}{}$~{\bf (B)}}
&\colorbox{SkyBlue}{$\wcs[]{qe}{}$~{\bf (C)}}
&
&
\\
\bottomrule
\end{tabular}
\end{adjustbox}
\caption{\small The impact of the set A on the operator sets B, C, E and F. Other details are the same as for Tab.~\ref{tab:mixing-bosonic-color}.}
\label{tab:impact A}
\end{center}
\end{table}

\begin{table}[H]
\centering
\setlength{\tabcolsep}{2pt}
\begin{center}
\begin{adjustbox}{width=0.7\textwidth}
\begin{tabular}{l@{\ }|c|ccccccccccccccccccc{c}}
\toprule
\multicolumn{13}{c}{Impact of {\bf Set B}  on other sets} & \\ 
\multicolumn{13}{c}{ \colorbox{RedOrange}{$\lambda$}, \colorbox{BurntOrange}{$\lambda^2$},
\colorbox{Green}{$\lambda^3$}, \colorbox{SkyBlue}{$\lambda^4$}}  & \\ \hline
\rowcolor{Grayd}
& $C_i(\Lambda)$   &     \multicolumn{4}{l}{} &$C_j(\muEW)$ & &\multicolumn{6}{l}{} \\
\midrule
&$\wcs[(1)]{qu}{}$
&\colorbox{BurntOrange}{$\wcs[]{uH}{}$~{\bf (E)}}
&\colorbox{BurntOrange}{$\wcs[(1)]{Hq}{}$~{\bf (A)}}
&\colorbox{Green}{$\wcs[]{HD}{}$~{\bf (F)}}
&\colorbox{Green}{$\wcs[(1)]{qq}{}$~{\bf (A)}}
&\colorbox{Green}{$\wcs[]{H}{}$~{\bf (F)}}
&\colorbox{SkyBlue}{$\wcs[]{H\Box}{}$~{\bf (F)}}
\\&
&\colorbox{SkyBlue}{$\wcs[]{qe}{}$~{\bf (C)}}
&\colorbox{SkyBlue}{$\wcs[]{eu}{}$~{\bf (C)}}
&\colorbox{SkyBlue}{$\wcs[(1)]{lq}{}$~{\bf (A)}}
&\colorbox{SkyBlue}{$\wcs[(3)]{Hq}{}$~{\bf (A)}}
&\colorbox{SkyBlue}{$\wcs[(8)]{quqd}{}$~{\bf (E)}}
&
&
\\ 
&$\wcs[(1)]{qd}{}$
&\colorbox{SkyBlue}{$\wcs[]{qe}{}$~{\bf (C)}}
&\colorbox{SkyBlue}{$\wcs[]{ed}{}$~{\bf (C)}}
&\colorbox{SkyBlue}{$\wcs[(8)]{quqd}{}$~{\bf (E)}}
&
&
&
&
&
&
\\ 
&$\wcs[(8)]{qu}{}$
&\colorbox{BurntOrange}{$\wcs[]{uH}{}$~{\bf (E)}}
&\colorbox{Green}{$\wcs[]{H}{}$~{\bf (F)}}
&\colorbox{SkyBlue}{$\wcs[(3)]{qq}{}$~{\bf (A)}}
&\colorbox{SkyBlue}{$\wcs[(1)]{qq}{}$~{\bf (A)}}
&\colorbox{SkyBlue}{$\wcs[(1)]{Hq}{}$~{\bf (A)}}
&
&
&
&
&
&
\\ 
&$\wcs[(8)]{qd}{}$
&\colorbox{SkyBlue}{$\wcs[(3)]{qq}{}$~{\bf (A)}}
&\colorbox{SkyBlue}{$\wcs[(1)]{qq}{}$~{\bf (A)}}
&
&
&
\\
&$\wcs[]{uu}{}$
&\colorbox{BurntOrange}{$\wcs[]{HD}{}$~{\bf (F)}}
&\colorbox{Green}{$\wcs[]{eu}{}$~{\bf (C)}}
&\colorbox{Green}{$\wcs[]{H\Box}{}$~{\bf (F)}}
&\colorbox{SkyBlue}{$\wcs[]{lu}{}$~{\bf (C)}}
&\colorbox{SkyBlue}{$\wcs[]{uH}{}$~{\bf (E)}}
&\colorbox{SkyBlue}{$\wcs[(1)]{Hq}{}$~{\bf (A)}}
&
&
&
&
&
&
\\ 
&$\wcs[]{dd}{}$
&\colorbox{SkyBlue}{$\wcs[]{ed}{}$~{\bf (C)}}
&\colorbox{SkyBlue}{$\wcs[]{ld}{}$~{\bf (C)}}
&
&
&
&
&\\
&$\wcs[(1)]{ud}{}$
&\colorbox{SkyBlue}{$\wcs[]{ed}{}$~{\bf (C)}}
&\colorbox{SkyBlue}{$\wcs[]{eu}{}$~{\bf (C)}}
&\colorbox{SkyBlue}{$\wcs[]{ld}{}$~{\bf (C)}}
&\colorbox{SkyBlue}{$\wcs[(8)]{quqd}{}$~{\bf (E)}}
&
&
&
&
&
&
\\ 
&$\wcs[]{Hu}{ }$
&\colorbox{RedOrange}{$\wcs[]{HD}{}$~{\bf (F)}}
&\colorbox{BurntOrange}{$\wcs[]{H\Box}{}$~{\bf (F)}}
&\colorbox{Green}{$\wcs[]{uH}{}$~{\bf (E)}}
&\colorbox{SkyBlue}{$\wcs[(1)]{Hq}{}$~{\bf (A)}}
&\colorbox{SkyBlue}{$\wcs[]{He}{ }$~{\bf (C)}}
&\colorbox{SkyBlue}{$\wcs[(1)]{Hl}{}$~{\bf (A)}}
&
&
&
&
\\ 
&$\wcs[]{Hd}{ }$
&\colorbox{SkyBlue}{$\wcs[]{HD}{}$~{\bf (F)}}
&\colorbox{SkyBlue}{$\wcs[]{He}{ }$~{\bf (C)}}
&
&
&
&
&
\\ 
&$\wcs[]{Hud}{}$
&\colorbox{Green}{$\wcs[]{dH}{}$~{\bf (E)}}
&\colorbox{SkyBlue}{$\wcs[]{HD}{}$~{\bf (F)}}
&\colorbox{SkyBlue}{$\wcs[]{H\Box}{}$~{\bf (F)}}
&\colorbox{SkyBlue}{$\wcs[]{uH}{}$~{\bf (E)}}
&
&
&&&&\\ 
\bottomrule
\end{tabular}
\end{adjustbox}
\caption{\small The impact of the set B on the operator sets A, C, E and F. Other details are the same as for Tab.~\ref{tab:mixing-bosonic-color}.}
\label{tab:impact B}
\end{center}
\end{table}

\begin{table}[H]
\centering
\setlength{\tabcolsep}{12pt}
\begin{center}
\begin{adjustbox}{width=0.8\textwidth}
\begin{tabular}{l@{\ }|c|ccccccc{c}}
\toprule
\multicolumn{8}{c}{The Impact of {\bf Sets C-F}  on other operator sets} & \\ 
\multicolumn{8}{c}{ \colorbox{RedOrange}{$\lambda$}, \colorbox{BurntOrange}{$\lambda^2$},
\colorbox{Green}{$\lambda^3$}, \colorbox{SkyBlue}{$\lambda^4$}}  & \\ \hline
\rowcolor{Grayd}
& $C_i(\Lambda)$   &     \multicolumn{1}{l}{} &$C_j(\muEW)$ & &\multicolumn{4}{l}{} \\
\midrule
&$\wcs[]{lu}{}~{\bf (C)}$
&\colorbox{BurntOrange}{$\wcs[(1)]{Hl}{}$~{\bf (A)}}
&\colorbox{Green}{$\wcs[(1)]{lq}{}$~{\bf (A)}}
&\colorbox{SkyBlue}{$\wcs[]{ll}{}$~{\bf (A)}}
&\colorbox{SkyBlue}{$\wcs[]{uu}{}$~{\bf (B)}}
&
&
 \\
 &$\wcs[]{ld}{}~{\bf (C)}$
&\colorbox{SkyBlue}{$\wcs[(1)]{ud}{}$~{\bf (B)}}
&
&
 \\
&$\wcs[]{eu}{}~{\bf (C)}$
&\colorbox{SkyBlue}{$\wcs[(1)]{uu}{}$~{\bf (B)}}
&
 \\
 &$\wcs[]{ed}{}~{\bf (C)}$
&\colorbox{SkyBlue}{$\wcs[(1)]{ud}{}$~{\bf (B)}}
&
\\
    &$\wcs[]{qe}{}~{\bf (C)}$
&\colorbox{SkyBlue}{$\wcs[(1)]{qu}{}$~{\bf (B)}}
&
 \\    
&$\wcs[]{He}{}~{\bf (C)}$
&\colorbox{SkyBlue}{$\wcs[]{HD}{}$~{\bf (F)}}
&
&
&
&
&
\\
\hline
& $\wcs[(3)]{lequ}{}~{\bf (D)}$
&\colorbox{SkyBlue}{$\wcs[]{eu}{}$~{\bf (C)}}
&\colorbox{SkyBlue}{$\wcs[]{qe}{}$~{\bf (C)}}
&\colorbox{SkyBlue}{$\wcs[]{lu}{}$~{\bf (C)}}
        &&&\\
\hline
&$\wcs[]{G}{}~{\bf (E)}$
&\colorbox{SkyBlue}{$\wcs[]{H}{}$~{\bf (F)}}
&
&
&
& &
\\ 
 &$\wcs[]{uG}{}~{\bf (E)}$
&\colorbox{BurntOrange}{$\wcs[]{H}{}$~{\bf (F)}}
&\colorbox{SkyBlue}{$\wcs[]{H W B}{}$~{\bf (F)}}
        &
&
&
&\\
 &$\wcs[]{uW}{}~{\bf (E)}$
&\colorbox{BurntOrange}{$\wcs[]{HW}{}$~{\bf (F)}}
&\colorbox{BurntOrange}{$\wcs[]{HWB}{}$~{\bf (F)}}
&\colorbox{SkyBlue}{$\wcs[]{H}{}$~{\bf (F)}}
&
&
&
\\
&$\wcs[]{uB}{}~{\bf (E)}$
&\colorbox{BurntOrange}{$\wcs[]{H W B}{}$~{\bf (F)}}
&\colorbox{BurntOrange}{$\wcs[]{HB}{}$~{\bf (F)}}
&\colorbox{SkyBlue}{$\wcs[]{H}{}$~{\bf (F)}}
&
&
&
\\ 
&$\wcs[(1)]{quqd}{}~{\bf (E)}$
&\colorbox{SkyBlue}{$\wcs[(8)]{qd}{}$~{\bf (B)}}
&\colorbox{SkyBlue}{$\wcs[(8)]{qu}{}$~{\bf (B)}}
&
&
&
&
\\ 
&$\wcs[]{uH}{}~{\bf (E)}$
&\colorbox{RedOrange}{$\wcs[]{H}{}$~{\bf (F)}}
&
&
&&&
\\
&$\wcs[]{HG}{}~{\bf (E)}$
&\colorbox{BurntOrange}{$\wcs[]{H}{}$~{\bf (F)}}
&
&
&
&&
\\
\hline
&$\wcs[]{W}{}~{\bf (F)}$
&\colorbox{SkyBlue}{$\wcs[]{uH}{}$~{\bf (E)}}
&
&
\\
&$\wcs[]{\tilde W}{}~{\bf (F)}$
&\colorbox{SkyBlue}{$\wcs[]{uH}{}$~{\bf (E)}}
&
&
\\
&$\wcs[]{H\Box}{}~{\bf (F)}$
&\colorbox{BurntOrange}{$\wcs[]{uH}{}$~{\bf (E)}}
&\colorbox{Green}{$\wcs[]{Hu}{ }$~{\bf (B)}}
&\colorbox{SkyBlue}{$\wcs[(1)]{Hq}{}$~{\bf (A)}}
&\colorbox{SkyBlue}{$\wcs[(3)]{Hq}{}$~{\bf (A)}}
&
&
\\ 
&$\wcs[]{HD}{}~{\bf (F)}$
&\colorbox{Green}{$\wcs[]{Hu}{ }$~{\bf (B)}}
&\colorbox{SkyBlue}{$\wcs[]{uH}{}$~{\bf (E)}}
&\colorbox{SkyBlue}{$\wcs[(1)]{Hq}{}$~{\bf (A)}}
&
&
&
\\
&$\wcs[]{HB}{}~{\bf (F)}$
&\colorbox{Green}{$\wcs[]{uH}{}$~{\bf (E)}}
&\colorbox{SkyBlue}{$\wcs[]{uB}{}$~{\bf (E)}}
&
&
&
&
\\ 
&$\wcs[]{HW}{}~{\bf (F)}$
&\colorbox{BurntOrange}{$\wcs[]{uH}{}$~{\bf (E)}}
&\colorbox{Green}{$\wcs[]{uW}{}$~{\bf (E)}}
&
&
&
&
\\
&$\wcs[]{H W B}{}~{\bf (F)}$
&\colorbox{Green}{$\wcs[]{uB}{}$~{\bf (E)}}
&\colorbox{SkyBlue}{$\wcs[]{uW}{}$~{\bf (E)}}
&\colorbox{SkyBlue}{$\wcs[]{uH}{}$~{\bf (E)}}
&
&
&
\\
&$\wcs[]{H\widetilde B}{}~{\bf (F)}$
&\colorbox{Green}{$\wcs[]{uH}{}$~{\bf (E)}}
&\colorbox{SkyBlue}{$\wcs[]{uB}{}$~{\bf (E)}}
&
&
&
&
\\ 
&$\wcs[]{H\widetilde W}{}~{\bf (F)}$
&\colorbox{BurntOrange}{$\wcs[]{uH}{}$~{\bf (E)}}
&\colorbox{Green}{$\wcs[]{uW}{}$~{\bf (E)}}
&
&
&
&
\\ 
&$\wcs[]{H  \widetilde W B}{}~{\bf (F)}$
&\colorbox{Green}{$\wcs[]{uB}{}$~{\bf (E)}}
&\colorbox{SkyBlue}{$\wcs[]{uW}{}$~{\bf (E)}}
&\colorbox{SkyBlue}{$\wcs[]{uH}{}$~{\bf (E)}}
&
&
&
\\ 
\bottomrule
\end{tabular}
\end{adjustbox}
\caption{\small The impact of the sets C-F on other operator sets. Other details are the same as for Tab.~\ref{tab:mixing-bosonic-color}.}
\label{tab:impact CF}
\end{center}
\end{table}

The pattern of mixings in Tabs.~\ref{tab:impact A}-\ref{tab:impact CF} is summarized in the collection of tables in Tab.~\ref{tab:grandpattern}, where the diagonal elements indicate the number of non-vanishing entries in the Tabs.~\ref{tab:mixing-bosonic-fermionic-color2new}-\ref{tab:mixing-bosonic-colornew6}. The non-diagonal entries represent the numbers in Tabs.~\ref{tab:impact A}-\ref{tab:impact CF}. The four tables represent the four different orders in $\lambda$: $\mathcal{O}(\lambda)$, $\mathcal{O}(\lambda^2)$  $\mathcal{O}(\lambda^3)$ and $\mathcal{O}(\lambda^4)$ as explained in the table caption. All the qualitative statements made above are given now in explicit terms.

\begin{table}[H]
  \centering
  \colorbox{RedOrange}{
\begin{tabular}{|c||c|c|c|c|c|c|}
\toprule
\hline
& A & B & C & D & E & F 
\\
\hline
A & 11  & 2  &  &  &  & 2 \\
\hline
B &  & 15 &  &  &  & 1 \\
\hline
C &  &   & 8  &  &  &  \\
\hline
D &  &  &  & 7 &  &  \\
\hline
E &  &  &  &  & 22 & 1 \\
\hline
F &  &  &  &  &  & 11 \\
\hline 
\bottomrule
\end{tabular}
}
\hspace{7mm}
\colorbox{BurntOrange}{
\begin{tabular}{|c||c|c|c|c|c|c|}
\toprule
\hline
& A & B & C & D & E & F 
\\
\hline
A & 6 & 3 &  &  & 1 & 3 \\
\hline
B & 1 & 10 &  &  & 2 & 2 \\
\hline
C & 1 &  & 2 &  &  &  \\
\hline
D &  &  &  & 3 &  &  \\
\hline
E &  &  &  &  & 14 & 6 \\
\hline
F &  &  &  &  & 3 & 2 \\
\hline 
\bottomrule
\end{tabular}
}
\colorbox{Green}{
\begin{tabular}{|c||c|c|c|c|c|c|}
\toprule
\hline
& A & B & C & D & E & F 
\\
\hline
A & 11 & 3 & 3 &  & 2 & 5 \\
\hline
B & 1 & 16 & 1 &  & 2 & 4 \\
\hline
C & 1 &   & 5 &  &  &  \\
\hline
D &  &  &  & 3 &  &  \\
\hline
E &  &  &  &  & 11 &  \\
\hline
F &  & 2 &  &  & 6 & 3 \\
\hline 
\bottomrule
\end{tabular}
}
\hspace{7mm}
\colorbox{SkyBlue}{
\begin{tabular}{|c||c|c|c|c|c|c|}
\toprule
\hline
& A & B & C & D & E & F 
\\
\hline
A & 14 & 10 & 8 &  & 6 & 4 \\
\hline
B & 10 & 22 & 12 &  & 5 & 4 \\
\hline
C & 1 & 5  & 20 &  &  & 1 \\
\hline
D &  &  & 3 & 4 &  &  \\
\hline
E &  & 2 &  &  & 23 & 4 \\
\hline
F & 3 &  &  &  & 9 & 12 \\
\hline 
\bottomrule
\end{tabular}
}
\caption{The pattern of mixings within each set and between different sets: top-left ($\mathcal{O}(\lambda)$), top-right ($\mathcal{O}(\lambda^2)$), bottom-left ($\mathcal{O}(\lambda^3)$) and bottom-right ($\mathcal{O}(\lambda^4)$).}
\label{tab:grandpattern}
\end{table}

\begin{figure}[tbp]%
\centering
\includegraphics[clip, trim=0cm 13cm 6cm 0cm, width=0.7\textwidth]{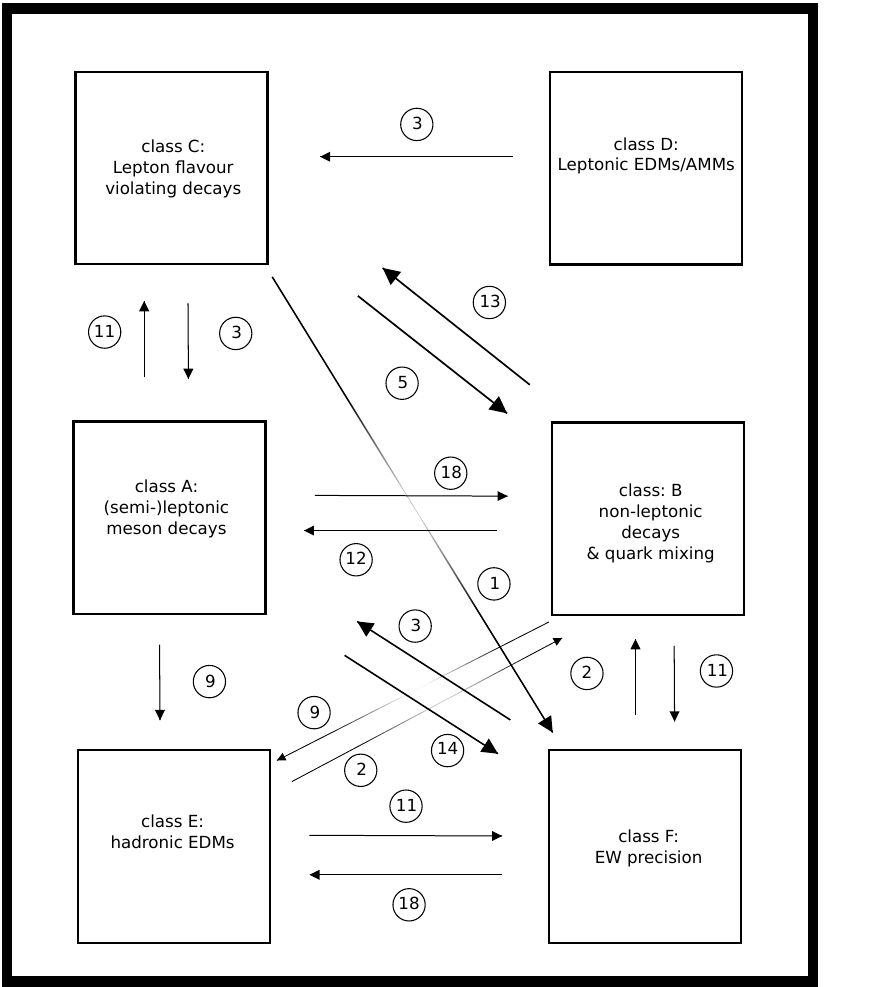}
\caption{ Cartoon of the operator mixing between the operator sets collected in Tab.~\ref{tab:grandpattern}. See text for explanations.}%
\label{fig:cartoon}%
\end{figure}

Finally, the pattern of mixing presented in Tab.~\ref{tab:grandpattern} is illustrated with a cartoon in Fig.~\ref{fig:cartoon} with the arrows presenting the impact of a given set on another set and the numbers in circles represent the sum of all connections between two sets collected in Tab.~\ref{tab:grandpattern}.

\section{Mapping Operators to Observables}\label{sec:11}
\subsection{Preliminaries}
To connect the SMEFT operators triggered by the UV physics to the low energy observables as classified in Sec.~\ref{sub:obs-classes}, one has to resort to the RG running governed by gauge as well as Yukawa interactions. As discussed there, where the flavour indices were omitted for simplicity, the operator mixing introduced by the running is complicated. This becomes even more complex once the flavour indices are restored which obscures the connection between UV theories to a given lower energy  observable, unlike in WET, where the electroweak and Yukawa interactions do not play any role in the operator mixing. Also, the phenomenological impact of the WET RG running is understood very well by now and has been systematically documented in \cite{Buchalla:1995vs,Buras:2011we} and in the appropriate references listed in Section~\ref{RGMA}. Therefore, the SMEFT running will be the main focus in the present work.

The upcoming sections aim to systematically group the list of SMEFT operators contributing to our 10 classes of observables which can play an important role in constraining the NP structure by studying a huge reservoir of data collected by various experiments. We will provide a compendium of SMEFT operators which contribute to each class at the EW scale as well as at the NP scale -- which is dictated by the SMEFT RG running. This information will serve as a comprehensive guide for the phenomenology within the SMEFT framework.

The set of tree-level SMEFT operators contributing at the EW as well as at the NP scale will be given in the shorthand notation e.g. $\ops[(1)]{qq}{ijkl}$. The explicit definition for these operators can be found in Tabs.~\ref{tab:no4fermsmeft} and \ref{tab:4fermsmeft}.

\subsection{CWET, JMS and SMEFT Bases}\label{3bases}
The low-energy WET operator basis for flavour changing transitions as found often in the literature has been constructed by extending the basis of the SM operators. This is in particular the case of operators in classes 1-3. This basis differs from the JMS and SMEFT bases introduced in previous sections. The basic formulae for numerous  observables in classes 1-3
have been given for decades in this basis, first within the SM \cite{Buchalla:1995vs}, subsequently in several BSM analyses \cite{Altmannshofer:2008dz,Altmannshofer:2009ma,Buras:2014fpa} and recently in a book \cite{Buras:2020xsm}. Also, many fits have been performed in this basis. To distinguish it from the popular WET basis these days, the JMS basis, we will denote this basis by Classic WET (CWET). Two examples of CWET are the operator bases for non-leptonic $\Delta F=2$ and $\Delta F=1$ processes, which is the classes 1 and 3, that carry the name of the BMU bases introduced in \cite{Buras:2000if}.

For WET, from the phenomenological point of view, in particular for classes 1-3, this is an important advantage of CWET basis over the JMS basis, simply because in the CWET both hadronic
matrix elements of operators and also the WCs entering various formulae in the SM is given in this basis. However, eventually, we would like to express the WCs of the operators in the CWET basis in terms of the WCs of the SMEFT operators. In this manner, the impact of NP on the observables can be studied very conveniently.

It is precisely this reason why the role of the JMS basis  \cite{Jenkins:2017jig} in reaching this goal should not be underestimated. It allows to match very efficiently the WET operators in this basis with the SMEFT ones. The matching in question is known by now both at tree-level \cite{Jenkins:2017jig} and one-loop level \cite{Dekens:2019ept}. Previous partial results can be found, for example, in \cite{Aebischer:2015fzz,Bobeth:2016llm,Bobeth:2017xry,Hurth:2019ula, Endo:2018gdn,Grzadkowski:2008mf}. In this manner the operators and WCs in the JMS basis
are given in terms of SMEFT operators and WCs at the EW scale $\muEW$. In fact in the recent literature, the JMS basis is often used for phenomenology from the start. This is the case of classes
{4-10.} We will therefore use the notation WET=JMS in what follows.

Having these results at hand we can proceed to reach our goal as follows:

{\bf Classes 1-3}

For classes 1-3 we can express the WCs in the CWET basis in terms of the ones in the JMS basis at the  EW scale, $\muEW$ and use the results of \cite{Jenkins:2017jig,Dekens:2019ept} so that finally one obtains formulae for observables in the CWET basis with the WCs directly dependent on the SMEFT WCs and consequently on the parameters of NP models.

{\bf Classes 4-10}

For the remaining classes we can directly use the JMS basis and use the formulae of \cite{Jenkins:2017jig,Dekens:2019ept} to relate it to the SMEFT.

Unless specified we will work in the down-basis of the SMEFT. The issues related to various bases in the SMEFT are discussed in detail in Sec.~\ref{sec:smeft-weak-basis}.

\subsection{SMEFT Charts}\label{SMEFTcharts}   
Due to the SMEFT running, new operators can contribute to a given process and the CKM rotations in the ADMs as well as the tree-level WET-SMEFT matching bring in new flavour structures into the game. As a result, the set of contributing SMEFT operators can be quite large. Therefore we will present the so-called {\em SMEFT Charts} for each process, showing the SMEFT operators at {the} scale $\Lambda$ which through mixing imply non-vanishing WCs of the set of tree-level operators at the EW scale. Typically, there are two kinds of operator mixing effects in the charts shown.
\begin{itemize}
  \item
First, there is {\em flavour-independent} operator mixing. In this terminology, the mixing between flavour-conserving parts of the same or different operators involving a change of generation of the fermions  (e.g. $\ops[(1)]{qq}{1211} \to \ops[(1)]{qq}{1233}$ or $\ops[(1)]{qq}{1211} \to \ops[(3)]{qq}{1233}$) is counted as flavour-independent, because in such cases the RG running does not generate flavour-violating operators from the flavour-conserving ones.   
\item
Second, representing {\em flavour-mixed} operator mixing that relates the operators at the two scales having different flavour structures. Here the mixing of the kind $\ops[(1)]{qq}{1211} \to \ops[(1)]{qq}{1212}$ or $\ops[(1)]{qq}{1211} \to \ops[(3)]{qq}{1212}$ takes place. 
\end{itemize}

In the light of the above observations, we divide the SMEFT operators into two categories based on their flavour indices:
\begin{equation*}
\textrm{\bf  Trivial Flavour Structures}\,, \quad  \textrm{\bf  Non-Trivial Flavour Structures}\,.
\end{equation*}
Here 'trivial' refers to a flavour structure that matches with the flavour structure of the corresponding WET operators in the mass basis, which in turn is fixed by the observable of interest.
All other flavour structures of the SMEFT operators stemming from the CKM rotations in whatever way are called non-trivial.

Typically, the first type of mixing is caused by the EW and QCD gauge interactions, while the second happens due to the Yukawa couplings. Further, we will use the following colour coding.
\begin{itemize}
\item
Operator mixing due to the strong, electroweak and top-Yukawa couplings will be indicated by the {\em dashed green}, {\em dashed red} and {\em solid black} lines in the RGE charts, respectively. 
\item
For simplicity, in the charts we omit the self-mixing since this only leads to a shift in the initial SMEFT WCs.
\end{itemize}

\subsection{The Impact of UV Physics on IR Observables}\label{GImpact}
To describe the impact of UV physics expressed in terms of the values of the SMEFT WCs at $\Lambda$ on the IR observables such as branching ratios which are functions of the  WET WCs as well as the SM parameters, we employ the SMEFT charts accompanied by two sets of $\rho_{ki}$ {and} $\eta_{km}$ coefficients. The general structure of the SMEFT charts has been already described in Sec.~\ref{SMEFTcharts} and here we will give general expressions for the coefficients $\rho_{ki}$ and  $\eta_{km}$. In {the following sections} we will use these formulae while discussing the different classes listed above.

\boldmath 
\subsubsection{$\rho_{ki}$ {and} $\eta_{km}$ Coefficients}\label{sec:rhoeta}
\unboldmath
The $\rho_{ki}$ coefficients describe the NP impact on the WET through the matching of the SMEFT on the WET at $\muEW$. In what follows we will deal with tree-level matching but it can be generalized to one-loop matching in a straightforward manner. On the other hand the $\eta_{km}$ coefficients parameterize the impact of RG running effects on IR quantities in terms of a finite number of parameters for a given class of observables.
 
To define both coefficients we will derive in what follows an analytic formula that describes the impact of NP  on the WET coefficients through tree-level matching at $\muEW$ and RG evolution within the SMEFT from $\Lambda$ to $\muEW$ in the first leading log approximation.

To this end, the tree-level matching conditions can be generically written as
\be\label{tree0}
 C^{\rm WET}_k(\muEW)=\sum_i \rho_{ki}\, C^{\rm SMEFT}_i(\muEW)\,.
\ee
As the WCs on both sides have the same dimension, the coefficients $\rho_{ki}$  are {\em dimensionless}. In most cases they are $\pm 1$ but generally could have different values. 

The next step is to express the $C^{\rm SMEFT}_i(\muEW)$ in terms of SMEFT WCs at $\Lambda$. The latter WCs can be divided into two sets:
\begin{itemize}
\item
$C^{\rm SMEFT}_j(\Lambda)$ with the index $j$  indicating that after RG running to $\muEW$ they enter the tree level matching relations, as the ones in \eqref{tree0}. Here the index was changed from $i$ to $j$ to be able to describe transparently the mixing under renormalization between the corresponding operators $\mathcal{O}^{\rm SMEFT}_j$ in this set, including self-mixing.
\item
$C^{\rm SMEFT}_r(\Lambda)${, denoting} operators which do not enter tree-level relations at $\muEW$ but can have an impact on the values of the $C^{\rm SMEFT}_i(\muEW)$ that enter
these relations through RG evolution from $\Lambda$ to $\muEW$.
\end{itemize}

Including these contributions we arrive at
\be\label{tree+RG}
 C^{\rm WET}_k(\muEW)=
\sum_{i,j,r} \rho_{ki}\left[\left(\delta_{ij}+
\frac{\beta_{ij}}{16\pi^2} \ln(\frac{\muEW}{\Lambda})\right) C^{\rm SMEFT}_j(\Lambda)+\frac{\beta_{ir}}{16\pi^2} \ln(\frac{\muEW}{\Lambda}) C^{\rm SMEFT}_r(\Lambda)\right]\,.
\ee
While performing the sums one should remember that the index $r$ differs from $i$ and $j$ but $i=j$ should be included.

This formula can finally be cast into
\be\label{ABKfinal}
C^{\rm WET}_k(\muEW)= 
\sum_{i} \rho_{ki} C_i^{\rm SMEFT}(\Lambda) +   \sum_{m=(j,r)} \eta_{km} C^{\rm SMEFT}_m(\Lambda)
\ln(\frac{\muEW}{\Lambda})\,,
\ee
where
\be\label{rhoeta}
\eta_{km}=\sum_{i}\rho_{ki}\frac{\beta_{im}}{16\pi^2}{=:C^{\rm WET}_k(C^{\rm SMEFT}_m)}
\,.
 \ee
 Here we work with the first LLA so that the coupling constants in $\beta_{im}$ are evaluated at $\muEW$. The last expression in \eqref{rhoeta} allows to present transparently the values of $\eta_{km}$ for different classes of observables without getting lost in indices.

We will present the numerical values of the $\rho_{ki}$ and $\eta_{km}$ coefficients class by class in App.~\ref{App:etas}. In this manner the SMEFT charts in association with the numerical values of $\rho_{ki}$ and $\eta_{km}$ will allow us to get a better insight into the pattern of the impact of NP on the WET WCs than a purely numerical analysis with the help of existing codes. In particular, they answer the fundamental question:
\begin{center}
{\bf What are all UV scale operators relevant for the observables in a given class?}
\end{center}

Further, the usefulness of the SMEFT charts increases when studying them in combination with the numerical values of $\rho_{ki}$ and $\eta_{km}$ for the following reasons.
\begin{enumerate}
\item With the help of the SMEFT charts we will be able to identify all UV scale operators which could contribute to a given class of observables.
\item
Numerical values of $\rho_{ki}$ tell us which SMEFT WCs have a direct impact on WET WCs without RG effects. 
\item On the other hand the $\eta_{km}$ coefficients describe the RG running effects including non-trivial back rotations for a given class. Their values allow for a quantitative description of the RG running effects, unlike the charts, which are purely qualitative. In particular, the hierarchy in the $\eta_{km}$ values will allow us to isolate the operators at the UV scale, beyond those involved directly in the matching on the WET, which could be responsible for a possible low-energy deviation from the SM expectation found in the future. 
\item
Moreover, some of the  $\eta_{km}$ coefficients with $m=r$ could be very suppressed so that the corresponding operators would not play any role in phenomenology, although being present in the charts. 
\item Finally, the presence of the same significant $\rho_{ki}$ and $\eta_{km}$ coefficients appearing in different classes indicate a strong correlation between the observables in those classes. In particular, the $\eta_{km}$ coefficients identify correlations that one would miss without including RG effects.
\end{enumerate}

Needless to say, once the general pattern of NP  has been understood it can be further improved at the quantitative level by using numerical codes that can sum all logarithms , thereby improving to the first leading log approximation in \eqref{ABKfinal}. Moreover in principle such codes can take NLO RG running and one-loop matching effects into account.

We are now well prepared to discuss the classes of observables one-by-one. In the processes of presenting the classes we will encounter for each class:
\paragraph{A:} The tree-level matching between the WET and the SMEFT operators involved at the electroweak scale.
\paragraph{B:} The SMEFT operators involved in this matching.
\paragraph{C:} Additional operators at the NP scale that through one-loop RG evolution down to electroweak scale affect those SMEFT WCs in a given class that enter tree-level matching on to the WET at the electroweak scale. In this context we stress  again that the SMEFT charts and the numerical values of $\eta_{ij}$ coefficients are based on the first leading log approximation. This is in contrast to Tables~\ref{tab:mixing-bosonic-color}-\ref{tab:mixing-fermionic-color4} in which summation of all leading logs is performed. Comparing these tables with the SMEFT charts one finds that quite generally our SMEFT charts give a satisfactory representation of the LO RG effects.
\vspace{5mm}

\begin{table}[H]
\centering  
\renewcommand{\arraystretch}{1.3}
\begin{adjustbox}{width=\textwidth}
\begin{tabular}{|ll|l|l|l|}
\hline
Class  &  Obs.    & A   &    B & C  \\
\hline
\hline
1 & $\Delta F=2$  & \eqref{eq:left-smeft-down}, \eqref{eq:left-smeft-up}  & \eqref{eq:df2smefttree}, \eqref{eq:df2upsmefttree}     & \eqref{class1A-smeftopsL}, \eqref{class1B-smeftopsL}  
\\
2 & FCNC  & \eqref{class2A-match}, \eqref{class2B-match}, \eqref{class2C-match}  &  \eqref{eq:class2A-smeft-tree}, \eqref{eq:class2B-smeft-tree}, \eqref{eq:class2C-smeft-tree}    & \eqref{eq:class2A-smeft-loop}  \\
3 & Non-Leptonic & \eqref{AJB1}-\eqref{eq:wet-match-srr} & \eqref{eq:smeft-class3-tree} & \eqref{eq:class3-smeft-loopl} \\
4 &$Z$-pole  & \eqref{eq:zcouplingsa}-\eqref{eq:gztree} & \eqref{eq:class4-smeft-tree}     & \eqref{class4-smeftopsL} \\
5 & Leptonic LFV  & \eqref{eq:class5A-tree-match}, \eqref{class5A-match2}, \eqref{class5B-match}, \eqref{eq:class5-match-C}   & \eqref{treeLFV}, \eqref{eq:class5B-tree}, \eqref{eq:class5C-smeft-tree}, \eqref{eq:class5D-smeft-tree}     & \eqref{eq:class5:smeft-lambda1}-\eqref{class5D-smeftopsL} \\
6 & SemiLep. LFV&  \eqref{class6A-match1}, \eqref{class6A-match2}, \eqref{eq:class6B-match-tree-down}, \eqref{eq:class6B-match-tree-up} 
& \eqref{class6A-smeftopsT}, \eqref{class6B-smeftopsT} & \eqref{class6A-smeftopsL}, \eqref{class6B-smeftopsL}  \\
7 & EDMs $\&$ MDMs &\eqref{eq:class7-tree-match1}, \eqref{eq:class7-tree-match2}  & \eqref{class7A-smeftopsT}, \eqref{class7B-smeftopsT}   & \eqref{class7A-smeftopsL}, \eqref{class7B-smeftopsL} \\
8 & Semilep. Charge & \eqref{class8-match1}, \eqref{class8-match2} & \eqref{eq:ops-cctree}, \eqref{GF-tree}   & \eqref{eq:class8-smeft-loop}, \eqref{eq:class8GF-smeft-loop1}\\
9 & Higgs & -- & \eqref{eq:class9-smeft-tree}  & \eqref{eq:class9:smeft-loop}  \\
10 &  High-$p_T$ & --& -- & \eqref{eq:smeft-10A}, \eqref{eq:smeft-10BC}, \eqref{eq:smeft-10D}   \\    \hline
\end{tabular}
\end{adjustbox}
\caption{\small The guide to the tree-level matching  equations of the SMEFT on to the WET (A) and to the lists of operators at $\muEW$ (B) and $\muNP$ (C). }
\label{tab:equations}
\end{table}

In Table~\ref{tab:equations} we guide the reader to the corresponding equations. Moreover, the reader will find in each class a multitude of RG equations that describe the RG evolution of the WCs within the SMEFT. They are rather complicated and some readers may  wonder why we present them at all as they can be found in several existing computer  codes. Yet, they are an important part of the SMEFT and in our view they have to be exposed if the goal is to see the landscape beyond the SM in its complexity and beauty. They simply constitute the SMEFT ATLAS and do not have to be read one by one like common reviews. Similar to a common atlas, each section can be consulted independently by readers who want to explore a particular part of the landscape beyond the SM.

\begin{table}[H]
\centering
\begin{tabular}{|c|c|}
\hline
\hfill $m_{B_s} = 5366.8(2) \mev$  \cite{Zyla:2020zbs}   &  $m_{B_d}=5279.58(17)\mev$\hfill\cite{Zyla:2020zbs}\\
$\Delta M_s = 17.749(20) \,\text{ps}^{-1}$\hfill \cite{Zyla:2020zbs}    &  $\Delta M_d = 0.5065(19) \,\text{ps}^{-1}$\hfill \cite{Zyla:2020zbs}\\
$\Delta M_K = 0.005292(9) \,\text{ps}^{-1}$\hfill \cite{Zyla:2020zbs} &
{$m_{K^0}=497.61(1)\mev$}\hfill \cite{Zyla:2020zbs}\\
$S_{\psi K_S}= 0.699(17)$\hfill\cite{Zyla:2020zbs}
                &  {$F_K=155.7(3)\mev$\hfill  \cite{Aoki:2019cca}}\\
        $|V_{us}|=0.2253(8)$\hfill\cite{Zyla:2020zbs} &
 $|\eps_K|= 2.228(11)\cdot 10^{-3}$\hfill\cite{Zyla:2020zbs}\\
$F_{B_s}$ = $230.3(1.3)\mev$ \hfill \cite{Aoki:2021kgd} & $F_{B_d}$ = $190.0(1.3)\mev$ \hfill \cite{Aoki:2021kgd}  \\
$F_{B_s} \sqrt{\hat B_s}=256.1(5.7) \mev$\hfill  \cite{Dowdall:2019bea}&
$F_{B_d} \sqrt{\hat B_d}=210.6(5.5) \mev$\hfill  \cite{Dowdall:2019bea}
\\
 $\hat B_s=1.232(53)$\hfill\cite{Dowdall:2019bea}        &
 $\hat B_d=1.222(61)$ \hfill\cite{Dowdall:2019bea}
\\
{$m_t(m_t)=162.83(67)\GeV$\hfill\cite{Brod:2021hsj} }  & {$m_c(m_c)=1.279(13)\GeV$} \\
{$S_{tt}(x_t)=2.303$} & {$S_{ut}(x_c,x_t)=-1.983\times 10^{-3}$} \\
    $\eta_{tt}=0.55(2)$\hfill\cite{Brod:2019rzc} & $\eta_{ut}= 0.402(5)$\hfill\cite{Brod:2019rzc}\\
$\kappa_\varepsilon = 0.94(2)$\hfill \cite{Buras:2010pza}       &
$\eta_B=0.55(1)$\hfill\cite{Buras:1990fn,Urban:1997gw}\\
$\tau_{B_s}= 1.515(4)\,\text{ps}$\hfill\cite{Amhis:2016xyh} & $\tau_{B_d}= 1.519(4)\,\text{ps}$\hfill\cite{Amhis:2016xyh}
\\
\hline
\end{tabular}
\caption {\textit{Input values used in $\DF=2$ and other flavour changing processes. For future 
updates see FLAG \cite{Aoki:2021kgd}, PDG \cite{Zyla:2020zbs} and HFLAV \cite{Aoki:2019cca}. 
}}
\label{tab:parameters}
\end{table}

{\boldmath
\section{Meson-Antimeson Mixing (Class 1)}
\label{class1}}
\noindent
This class is devoted to meson-antimeson mixing processes in the SMEFT. The most prominent observables in this class are
\begin{itemize}
\item
The mass differences 
\be
\Delta M_d\,, \qquad\Delta M_s\,,\qquad \Delta M_K\,,\qquad \Delta M_D\,,
\ee
in $B^0_d-\bar B^0_d$, $B^0_s-\bar B^0_s$, $K^0-\bar K^0$ and $D^0-\bar D^0$ mixing, respectively.
\item
The measures of mixing induced CP violation corresponding to these mixings, in particular
\be
S_{\psi K_S}\,,\qquad S_{\psi\phi}\,,\qquad \varepsilon_K\,, \qquad S_f\,.
\ee
The first two and the last one are CP asymmetries in decays of $B_d$, $B_s$ and $D^0$ to a CP-eigenstate $f$. $\varepsilon_K$ can be measured in $K_L\to\pi^+\pi^-$ decays. 
\end{itemize}

The observables in this class played a prominent role in the tests of the SM and the search for NP for the last five decades. Being by now very precisely measured they play a very important role in the selection of the viable extensions of the SM. Details on these observables within the SM can be found in Chapters 8 and 11 in \cite{Buras:2020xsm}. In the following we will present a detailed analysis of SMEFT operators contributing to such processes through RG running from the NP scale $\Lambda$ down to the electroweak scale $\muEW$ and through tree-level matching of the SMEFT onto the WET at the latter scale. First, we provide the list of WET operators in the mass basis contributing to the processes that violate quark flavour by two units, such as meson-antimeson mixing in the $B_{s,d}$, $K^0$ or $D^0$ systems. In this case NLO QCD corrections are known both for the WET \cite{Buras:2000if} and the SMEFT \cite{Aebischer:2020dsw,Aebischer:2022anv}.

The presentation of the effective Hamiltonians for $\DF=2$ processes is simpler than for processes discussed subsequently, in particular compared to Class 2 and Class 3. The reason is that in the meson-mixing case, the number of operators involved is sufficiently small so that many details can still be presented in explicit terms. These details will illustrate in a simple setting what happens in the case of more complicated processes.

In Tab.~\ref{tab:parameters} we give the set of input parameters required to compute $\DF=2$ observables. These parameters are also useful for processes belonging to other classes discussed in this review. As far as the parameter $\hat B_K$, relevant for $\varepsilon_K$ and $\Delta M_K$, is concerned the most accurate result {is from the} RBC-UKQCD collaboration \cite{Boyle:2024gge}, which after the inclusion of NNLO QCD corrections reads $\hat B_K=0.7600(53)$ \cite{Gorbahn:2024qpe}. This should be compared with the DQCD result $\hat B_K=0.73\pm0.02$ \cite{Buras:2014maa} obtained ten years earlier. References to other LQCD results are given in \cite{Boyle:2024gge,Gorbahn:2024qpe} and \cite{Buras:2024per}. The explanation why these results are so close to $\hat B_K=3/4$ obtained in the strict large $N$ limit has been so far only provided in the DQCD \cite{Buras:2014maa}.

\subsection{\boldmath CWET and WET for Class 1 at $\muEW$}
We begin by listing the CWET operators which govern $\DF=2$ transitions. Several operator bases were employed in the past. One of them is the so-called SUSY basis, which was used for the first two-loop ADMs of BSM operators \cite{Ciuchini:1997bw}. In our discussion, we will focus on the so-called BMU basis \cite{Buras:2000if}, which consists of $(5 + 3) = 8$ physical operators. In this basis, the operators are given as follows:
\begin{eqnarray}   \label{eq:BMU-basis} 
& \OpL[ij]{\text{VLL}} & = (\bar{d}_i \gamma_\mu P_L d_j)(\bar{d}_i \gamma^\mu P_L d_j)\,,   \quad \quad
 \OpL[ij]{\text{LR},1}  = (\bar{d}_i \gamma_\mu P_L d_j)(\bar{d}_i \gamma^\mu P_R d_j)\,, \notag \\[2mm]
 &\OpL[ij]{\text{SLL},1} & = (\bar{d}_i P_L d_j)(\bar{d}_i P_L d_j)\,,   \quad\quad\quad\quad
 \OpL[ij]{\text{LR},2}   = (\bar{d}_i P_L d_j)(\bar{d}_i P_R d_j) \,,  \notag \\[2mm]
& \OpL[ij]{\text{SLL},2} & = -(\bar{d}_i \sigma_{\mu\nu} P_L d_j)(\bar{d}_i \sigma^{\mu\nu} P_L d_j) \,.
\end{eqnarray}
The chirality-flipped operators $\OpL[ij]{\text{VRR}}$ and $\OpL[ij]{\text{SRR},1,2}$ are obtained by interchanging $P_L\leftrightarrow P_R$ in their left-handed counter-parts. This basis choice is favourable for DQCD calculations \cite{Buras:2018lgu, Aebischer:2018rrz}, due to the operator structure, containing only colour-singlet currents. Finally, we note that the minus sign in the tensor operator $\OpL[ij]{\text{SLL},2}$ arises from different definitions of $\tilde{\sigma}_{\mu\nu} \equiv [\gamma_\mu,\, \gamma_\nu]/2$ in \cite{Buras:2000if} w.r.t. $\sigma_{\mu\nu} = i \tilde{\sigma}_{\mu\nu}$ used in this review. The complete NLO QCD ADM in the BMU basis was calculated in \cite{Buras:2000if}, and numerical solutions for $ij = ds, bd, bs$ can be found in \cite{Buras:2001ra}.

While the BMU basis is useful for QCD calculations, the JMS basis \cite{Jenkins:2017jig} is more practical for matching the WET to the SMEFT. The complete JMS basis is given in App.~\ref{app:jmsbasis}. We divide the operators of this class into two sub-classes, dependently on whether they contribute to {\em down-type} $\DF=2$ transitions (Class 1A) or {\em up-type} $\DF=2$ transitions (Class 1B) transitions. The set of JMS operators for Class 1A is given by 
\begin{center}
{\bf WET-1A}
\end{center}
\be
\begin{aligned}
\opL[V,LL]{dd}{ijij}\,, \,\,
\opL[V1,LR]{dd}{ijij}\,, \,\,
\opL[V8,LR]{dd}{ijij}\,, \,\,
\opL[S1,RR]{dd}{ijij}\,, \,\,
\opL[S8,RR]{dd}{ijij}\,, \,\,
\opL[V,RR]{dd}{ijij}\,, \,\,
(\opL[S1,RR]{dd}{jiji})^\dag\,, \,\,
(\opL[S8,RR]{dd}{jiji})^\dag\,.
\end{aligned}
\ee
Dependently on a given NP scenario, they can be generated similar to the SM first at the one-loop level but in contrast to the SM also at tree-level as in the case of $Z^\prime$ models. Here we express them in terms of the operators in the BMU basis. We have
\begin{align} \label{eq:jmsbmu}
&  \opL[V,LL]{dd}{ijij}  = \OpL[ij]{\text{VLL}} \,,\quad  \opL[VRR]{dd}{ijij}  = \OpL[ij]{\text{VRR}} \,,  \quad
\opL[V1,LR]{dd}{ijij}  = \OpL[ij]{\text{LR},1} \,, \quad    \opL[V8,LR]{dd}{ijij} 
= -\frac{1}{6}\OpL[ij]{\text{LR},1} -\OpL[ij]{\text{LR},2} \,, \notag
\\[2mm]
&  (\opL[S1,RR]{dd}{jiji})^\dagger  = \OpL[ij]{\text{SLL},1}\,, \quad
(\opL[S8,RR]{dd}{jiji})^\dagger  =  -\frac{5}{12} \OpL[ij]{\text{SLL},1} + \frac{1}{16}\OpL[ij]{\text{SLL},2} \,,
\\[2mm]
&  \opL[S1,RR]{dd}{ijij}  = \OpL[ij]{\text{SRR},1}\,, \quad
\opL[S8,RR]{dd}{ijij} =  -\frac{5}{12} \OpL[ij]{\text{SRR},1} + \frac{1}{16}\OpL[ij]{\text{SRR},2} \,. \notag
\end{align}

Using the hadronic matrix elements for the operators in the BMU basis from LQCD, these expressions allow us to calculate the corresponding hadronic matrix elements in the JMS basis at the hadronic scale and, using RG evolution, at the electroweak scale $\muEW$. Subsequently, the WCs in the JMS basis can be expressed in terms of the SMEFT ones as discussed for $\DF=2$ transitions below and for other classes in the subsequent sections.
 
Alternatively, one can use hadronic matrix elements in the BMU basis and express the WCs in this basis in terms of the ones in the JMS one and subsequently in terms of the SMEFT ones. 

As already discussed in Part I of our review, in general in all these transformations at the 1-loop level one has to take care of evanescent operators to cancel the scheme dependence of hadronic matrix elements at NLO in QCD. However, the evanescent operators for Class 1 vanish at one-loop level in QCD \cite{Aebischer:2020dsw}. 

Using \eqref{eq:jmsbmu} we can derive relationships between WCs in the BMU basis and the ones in the JMS basis. The transformation matrix for the WCs is given by the inverse transposed of the corresponding matrix for the operators \eqref{eq:jmsbmu}.

For Class 1B, the WET operators contributing to $\DDbar$ mixing are
\begin{center}
{\bf WET 1B}
\end{center}
\be 
\begin{aligned}
\opL[V,LL]{uu}{1212}\,, \,\,
\opL[V1,LR]{uu}{1212}\,, \,\,
\opL[V8,LR]{uu}{1212}\,, \,\,
\opL[S1,RR]{uu}{1212}\,, \,\,
\opL[S8,RR]{uu}{1212}\,, \,\,
\opL[V,RR]{uu}{1212}\,, \,\,
(\opL[S1,RR]{uu}{2121})^\dag\,, \,\,
(\opL[S8,RR]{uu}{2121})^\dag\,.
\end{aligned}
\ee
The above sets of WCs in WET-1A and WET-1B are separately closed under WET QCD and QED RG running.

\subsection{\boldmath SMEFT Operators for Class 1 at $\muEW$}
The most important SMEFT operators that govern $\DF=2$ processes are \cite{Aebischer:2020dsw,Aebischer:2022anv}

\begin{center}
{\boldmath \bf SMEFT-Tree 1A}
\end{center}
\be 
\begin{aligned}
\label{eq:df2smefttree}
\ops[(1)]{qq}{ijij}  \,, \quad
\ops[(3)]{qq}{ijij}  \,,  \quad
\ops[(1)]{qd}{ijij}  \,, \quad
\ops[(8)]{qd}{ijij} \,,  \quad
\ops[ ]{dd}{ijij}.
\end{aligned}
\ee
Here {\em tree} indicates that these operators match onto the WET operators listed above at tree-level. Next we give the WET WCs at $\muEW$ in the JMS basis in terms of SMEFT WCs at $\muEW$. This will allow us to construct the SMEFT Lagrangian at the scale $\Lambda$, as well as the relevant RGEs and charts showing the mapping between corresponding SMEFT WCs at $\muEW$ and $\Lambda$, relevant for $\DF=2$ transitions. 

At tree-level one finds for $\BBbar$ and $\KKbar$ mixing the following matching 
conditions between WET and SMEFT WCs, at~$\muEW$, in the down-basis
\hfill
\begin{equation}
\label{eq:left-smeft-down}
\begin{aligned}
\wcL[V,LL]{dd}{ijij} &
=  \wc[(1)]{qq}{ijij} + \wc[(3)]{qq}{ijij} \,, \quad
&
\wcL[V1,LR]{dd}{ijij} & =  \wc[(1)]{qd}{ijij} \,,
\\
\wcL[V,RR]{dd}{ijij}  & =  \wc{dd}{ijij}      \,,
&
\wcL[V8,LR]{dd}{ijij} & =  \wc[(8)]{qd}{ijij}.
\end{aligned}
\end{equation}
The leading order SMEFT matching for scalar WET operators only arises 
at the dim-8 level \cite{Hamoudou:2022tdn}. For $\DDbar$ mixing in the up-basis, the relevant operators are
\begin{center}
{\boldmath \bf SMEFT-Tree 1B}
\end{center}
\be
\label{eq:df2upsmefttree}
\begin{aligned}
\opup[(1)]{qq}{1212}  \,, \quad
\opup[(3)]{qq}{1212}  \,,  \quad
\opup[(1)]{qu}{1212}  \,, \quad
\opup[(8)]{qu}{1212} \,,  \quad
\opup[ ]{uu}{1212}.
\end{aligned}
\ee
Their matching onto the WET reads
\be
  \label{eq:left-smeft-up}
\begin{aligned}
\wcL[V,LL]{uu}{1212} &
=  \wcup[(1)]{qq}{1212} + \wcup[(3)]{qq}{1212} \,, \quad
&
\wcL[V1,LR]{uu}{1212} & =  \wcup[(1)]{qu}{1212} \,,
\\
\wcL[V,RR]{uu}{1212}  & =  \wcup{uu}{1212}      \,,
&
\wcL[V8,LR]{uu}{1212} & =  \wcup[(8)]{qu}{1212} \,.
\end{aligned}
\ee
{Here the tilde notation is used for the up-basis SMEFT WCs.}
Note that we use (contrary to \cite{Aebischer:2022anv}) the Lagrangian to define our WET {WCs} as done in \cite{Jenkins:2017jig,Dekens:2019ept}. Otherwise, there would be additional minus signs in front of WET coefficients. Finally, the relation between the down- and up-basis WCs is discussed in Sec.~\ref{sec:smeft-weak-basis}.

For Class 1, it is straightforward to read off the $\rho$ parameters from the matching conditions:
\be
\wcL[V,LL]{dd}{ijij} [\wc[(1)]{qq}{ijij}] =
\wcL[V,LL]{dd}{ijij} [\wc[(3)]{qq}{ijij}] =
\wcL[V1,LR]{dd}{ijij} [\wc[(1)]{qd}{ijij}] =
\wcL[V8,LR]{dd}{ijij} [\wc[(8)]{qd}{ijij}] =
\wcL[V,RR]{dd}{ijij} [\wc[]{dd}{ijij}] = 1\,. 
\ee

Likewise, for the up-sector all $\rho$ parameters are 1. The most important are the $\eta$ 
{coefficients  which quantify the RG} running {effects}. 

\subsection{ \boldmath SMEFT Operators for Class 1 at $\Lambda$}

Next we find according to the four step procedure of Sec.~\ref{Bottom-Up} the set of contributing operators at the scale $\Lambda$ that through RG evolution from $\Lambda$ to $\muEW$ affect the WCs of SMEFT operators at $\muEW$ listed above. To this end we use the relevant ADMs including gauge and Yukawa couplings.

In the gauge sector, the most sizable ADMs that govern the RGEs are those due to the strong coupling $4 \pi \alpha_s = g_s^2 \approx 1.4$ and less so for $\text{SU(2)}_L \times \text{U}(1)_Y$ gauge couplings. The SMEFT four-quark operators in \eqref{eq:df2smefttree} and \eqref{eq:df2upsmefttree} undergo under gauge-interactions only self-mixing or mix among themselves.\footnote{A strict use of the term ``self-mixing'' implies that the flavour structure of the $\DF=2$ WC is conserved to be $\wc{b}{ijij}$. {On the other hand, mixing with flavour variant operators is not referred to as self mixing in this review.}} This can be seen in the chart in Fig.~\ref{chart:df2-down}.\footnote{In the chart, for simplicity we only show mixing between different WCs and have suppressed the self-mixing.}

We also {present} here the anatomy of RGEs for Class 1 operators. To 
{simplify the complex structure }  of the ADMs, we {organize} 
the {RGEs} according to both {operator classes} 
and underlying interactions that constitute various components of the relevant ADMs. \\

{As a first step, we examine the gauge coupling dependent RGEs.}

\noindent
\underline{\bf \boldmath $f^4 \to f^4$ (Gauge)}:

{The 4f to 4f operator mixing is given by}

\be
\begin{aligned}
{\dotwc[(1)]{qq}{ijij}}
&= \left(\frac{g_1^2}{3}+g_s^2\right) \wc[(1)]{qq}{ijij}+9 \left(g_2^2+g_s^2\right) \wc[(3)]{qq}{ijij}\,, \\
{\dotwc[(3)]{qq}{ijij}}  &= 
3 \left(g_2^2+g_s^2\right) \wc[(1)]{qq}{ijij}+\frac{1}{3} \left(g_1^2-3 \left(6 g_2^2+5 g_s^2\right)\right) \wc[(3)]{qq}{ijij} \,, \\
{\dotwc[(1)]{qd}{ijij}} & = \frac{2}{3} \left(g_1^2 \wc[(1)]{qd}{ijij}-4 g_s^2 \wc[(8)]{qd}{ijij}\right)\,, \\
{\dotwc[(8)]{qd}{ijij}} & = \frac{2}{3} \left(g_1^2-21 g_s^2\right) \wc[(8)]{qd}{ijij}-12 g_s^2 \wc[(1)]{qd}{ijij}\,,\\
{\dotwc[]{dd}{ijij}} & = \frac{4}{3} \left(g_1^2+3 g_s^2\right) \wc[]{dd}{ijij}\,.
\end{aligned}
\ee

More interesting are Yukawa interactions that bring new operators into the game. 
The most sizable ADMs from Yukawa-mixing that govern RGEs are due to the top-Yukawa 
coupling $y_t \approx 1$ and $y_b $ of relatively lesser strength. 
The impact of this evolution on the SMEFT WCs at $\muEW$ is seen in the charts in 
Fig.~\ref{chart:df2-down} and \ref{chart:df2-up}. 

To illustrate this mixing, the explicit Yukawa RGEs are reported in 
the following, where on the r.h.s only new WCs that satisfy our 
Criteria-I {(as defined by \eqref{eq:criteriaI})} are kept. Furthermore, vanishing RGEs have been omitted. For this class, the new operators are found to be only of non-leptonic (NL) nature containing two ($f^2$) or four fermions ($f^4$). \\

{The Yukawa dependent RGEs are given below.}

\noindent
\underline{\bf \boldmath $f^2 H^2 D \to f^4$ (Yukawa)}:

\be
\begin{aligned}
{\dotwc[(1)]{qq}{ijij}} &
=  y_t^2 V^*_{3i} V_{3j} \wc[(1)]{Hq}{ij}\,, \quad
{\dotwc[(3)]{qq}{ijij}}  = 
-y_t^2 V^*_{3i} V_{3j} \wc[(3)]{Hq}{ij}  \,,\quad
{\dotwc[(1)]{qd}{ijij}}  = 
y_t^2 V^*_{3i} V_{3j} \wc[]{Hd}{ij} \,. \label{eq:Cqd1Hd}
\end{aligned}
\ee


\noindent
\underline{\bf \boldmath $f^4 \to f^4$ (Yukawa)}: 

{The operator mixing between 4f vector operators, excluding self-mixing, is given by}
\be
\begin{aligned}\label{eq:f4f4yukcl1}
{\dotwc[(1)]{qq}{ijij}}
&= -\frac{1}{12} y_t^2 V^*_{3i} V_{3j} (12 \wc[(1)]{qu}{ij33}+\wc[(8)]{qu}{ij33})\,, \\
{\dotwc[(3)]{qq}{ijij}}  &= 
-\frac{1}{4} y_t^2 V^*_{3i} V_{3j} \wc[(8)]{qu}{ij33} \,, \\
{\dotwc[(1)]{qd}{ijij}} & = \frac{4}{9} y_b^2 \wc[(8)]{qd}{3j3j} \delta_{i3}+\frac{4}{9} y_b^2 \wc[(8)]{qd}{i3i3} \delta_{j3}-y_t^2 V^*_{3i} V_{3j} \wc[(1)]{ud}{33ij}\,, \\
{\dotwc[(8)]{qd}{ijij}} & = 2 y_b^2 \wc[(1)]{qd}{3j3j} \delta_{i3}+2 y_b^2 \wc[(1)]{qd}{i3i3} \delta_{j3}-y_t^2 V^*_{3i} V_{3j} \wc[(8)]{ud}{33ij}\,.
\end{aligned}
\ee
The RGEs dictating mixing between the scalar and vector operators are given as
\begin{align}\label{eq:cl1scalarmix}
{\dotwc[(1)]{qq}{ijij}} &= 
\frac{1}{12} y_b y_t (-3 \wc[(1)]{quqd}{i3i3} \delta_{j3} V_{3j}+\delta_{i3} V^*_{3i} (2 \wc[(8)*]{quqd}{j3j3}-3 \wc[(1)*]{quqd}{j3j3})+2 \wc[(8)]{quqd}{i3i3} \delta_{j3} V_{3j})\,, \notag\\
\dotwc[(1)]{qd}{ijij} & = 
\frac{1}{3} y_b y_t \delta_{i3} V_{3j} \wc[(1)]{quqd}{33ij}+\frac{1}{2} y_b y_t \delta_{i3} V_{3j} \wc[(1)]{quqd}{i33j}+\frac{1}{3} y_b y_t \delta_{j3} V^*_{3i} \wc[(1)*]{quqd}{33ji}+\frac{1}{2} y_b y_t \delta_{j3} V^*_{3i} \wc[(1)*]{quqd}{j33i}\notag\\
&+\frac{4}{9} y_b y_t \delta_{i3} V_{3j} \wc[(8)]{quqd}{33ij}+\frac{4}{9} y_b y_t \delta_{j3} V^*_{3i} \wc[(8)*]{quqd}{33ji}\,,\notag\\
{\dotwc[(8)]{qd}{ijij}} & = 2 y_b y_t \delta_{i3} V_{3j} \wc[(1)]{quqd}{33ij}+2 y_b y_t \delta_{j3} V^*_{3i} \wc[(1)*]{quqd}{33ji}-\frac{1}{3} y_b y_t \delta_{i3} V_{3j} \wc[(8)]{quqd}{33ij}+\frac{1}{2} y_b y_t \delta_{i3} V_{3j} \wc[(8)]{quqd}{i33j}\notag\\
&-\frac{1}{3} y_b y_t \delta_{j3} V^*_{3i} \wc[(8)*]{quqd}{33ji}+\frac{1}{2} y_b y_t \delta_{j3} V^*_{3i} \wc[(8)*]{quqd}{j33i}\,.
\end{align}
The expression for $\dotwc[(3)]{qq}{ijij}$ is the same as for $\dotwc[(1)]{qq}{ijij}$, up to a sign. Note that in these equations we neglected $\mathcal{O}(Y_d Y_d)$ and $\mathcal{O}(Y_d Y_u)$ terms relative to $\mathcal{O}(Y_u Y_u)$, unless they are the only contributions.

Here, to get the exact flavour structure of the operators which contribute due to operator mixing {we need the} $\hat Y_u$. In the down-basis, it is given by\footnote{For illustration we neglect here the dimension-six terms to the mass matrix, but they can be included into numerics using \tt{wilson}.}
\begin{equation}
\hat Y_u = \frac{\sqrt 2}{v} \hat V^\dagger \hat M_u^{\rm diag} \,.
\end{equation}
{The flavour mixing induced by the Yukawa couplings thus originates from the off-diagonal elements of the CKM matrix. Further details are provided}  in Sec.~\ref{sec:smeft-weak-basis}.

Although at first sight numerically suppressed, {the mixing due to light Yukawas}  can be phenomenologically relevant, depending on the UV completion and also on the SM suppression factors for the observable under consideration. 

In {\bf the up-basis}, as the up-type Yukawa matrix 
\be
\hat Y_u = \frac{\sqrt 2 \,M_u^\text{diag}}{v}\,,
\ee
 is diagonal, all flavour-changing mixing terms in the ADMs 
due to $\hat Y_u$ disappear. But the mixing due to gauge interactions 
remains unaltered.

Using the above information, we find additional SMEFT operators for Class 1A that complement the ones in \eqref{eq:df2smefttree}
at the scale $\Lambda$:
\begin{center}
{\bf SMEFT-Loop 1A}
\end{center}
\be \label{class1A-smeftopsL}
\begin{aligned}
\textrm{\bf Yukawa-mixing:} 
&~ 
\ops[(1)]{Hq}{}\,, \quad
\ops[(3)]{Hq}{}\,, \quad
\ops[]{Hd}{}\,, \quad
\ops[(1)]{qu}{}\,, \quad
\ops[(8)]{qu}{}\,, \\
&~
\ops[(1)]{ud}{}\,,  \quad
\ops[(8)]{ud}{}\,, \quad
\ops[(1)]{quqd}{}\,,\quad
\ops[(8)]{quqd}{}\,.
 \\
\textrm{\bf Gauge-mixing:} &~  \textrm{No new {operators.}}
\end{aligned}
\ee
We conclude therefore, that not only four-fermion operators {can}  have an 
impact on {meson}-mixing but also the Higgs-quark operators 
through top-Yukawa-RG evolution.

{The quantitaive impact of these operators in Class 1 can be assessed
 using  corresponding $\eta$ coefficients (see \eqref{ABKfinal} for the definition). This provides a first indication of 
which RG effects are most significant\footnote{It should be noted that the RG 
effects of the operators already present at the EW scale are generally larger 
than these ones.}. For the $B_s$-mixing,  
restricting to contributions having $|\eta| \ge 6 \cdot 10^{-6} $,} we find
\be \begin{aligned} &
\qquad \quad 
\wcs[(3)]{qq}{}\to \wcL[V,LL]{dd}{ }\,, \quad
\wcs[(1)]{qq}{}\to \wcL[V,LL]{dd}{ }\,, \quad
\wcs[(1)]{qu}{}\to \wcL[V,LL]{dd}{ }\,, \quad
\wcs[(1)]{Hq}{}\to \wcL[V,LL]{dd}{ }\,, \\ 
& 
\qquad \quad 
\wcs[(3)]{Hq}{}\to \wcL[V,LL]{dd}{ }\,, \quad
\wcs[(8)]{qu}{}\to \wcL[V,LL]{dd}{ }\,, \quad
\wcs[]{dd}{}\to \wcL[V,RR]{dd}{ }\,, \quad
\wcs[(8)]{qd}{}\to \wcL[V1,LR]{dd}{ }\,, \\ 
& 
\qquad \quad 
\wcs[(1)]{qd}{}\to \wcL[V1,LR]{dd}{ }\,, \quad
\wcs[(1)]{ud}{}\to \wcL[V1,LR]{dd}{ }\,, \quad
\wcs[]{Hd}{ }\to \wcL[V1,LR]{dd}{ }\,, \quad
\wcs[(1)]{qd}{}\to \wcL[V8,LR]{dd}{ }\,, \\ 
& 
\qquad \quad 
\wcs[(8)]{qd}{}\to \wcL[V8,LR]{dd}{ }\,, \quad
\wcs[(8)]{ud}{}\to \wcL[V8,LR]{dd}{ }\,, \quad
\wcs[(1)]{quqd}{}\to \wcL[V8,LR]{dd}{ }. 
\end{aligned} \ee
Note that the remaining operators such as $\wc[(8)]{quqd}{}$ 
could still contribute to Class 1, but have smaller values of $\eta$.
{The tree-level operators seen in this equation are 
actualy their flavour variants.}
For the Kaon-mixing we found exact same pattern.
See Tabs.~\ref{tab:etaclass1Bs}- \ref{tab:etaclass1K} for more details.

The strong constraints on NP contributions from mixing therefore imply indirectly 
also constraints on these {new}  operators. This issue is extensively discussed in \cite{Bobeth:2017xry,Bobeth:2016llm,Endo:2016tnu,Endo:2018gdn,Aebischer:2020dsw}. For instance, if at the NP scale the coefficients $\wc{H d}{ij}$ with $i,j=d,s,b$ are non-vanishing, neutral flavour violating right-handed (RH) quark currents, represented here by $\wcs[(1)]{qd}{}$, are generated with profound implications for $\DF=2$ transitions and their specific correlations with $\DF=1$ observables. Concrete models with this structure are models with flavour-violating $Z$-couplings and in particular models with vector-like quarks. Indeed, $\wc{Hd}{ij}(\muNP) \neq 0$ generates through Yukawa RG effects a leading-logarithmic contribution to the LR operator $\OpL[ij]{\text{LR},1}$ at the electroweak scale $\muEW$, as can be seen in \eqref{eq:Cqd1Hd}. Its phenomenological implications are discussed in detail in \cite{Bobeth:2017xry}. There it was found that the RG contribution given above is, for $\muNP$ sufficiently larger than $\muEW$, not only the most important NP effect in this scenario but also contributes with an opposite sign compared to non-logarithmic effects at the matching scale.

On the other hand, if at the NP scale $\Lambda$ only neutral flavour violating left-handed (LH) quark currents are generated by NP, the implications are different, in particular for the correlation between $\DF=2$ and $\DF=1$ transitions. As this time left-left operators are generated, the impact on the $\DF=2$ process is generally smaller than in the previous scenario in which left-right operators with large hadronic matrix elements are generated.

Let us now {examine} the RGE running of $\DF=2$ operators in more detail. 
The operator mixing for the down-type $\DF=2$ processes are illustrated in the 
charts of Fig.~\ref{chart:df2-down}, while for $\DDbar$ are shown in Fig.~\ref{chart:df2-up}. 

\noindent
For Fig.~\ref{chart:df2-down}, the following restrictions apply:

\begin{figure}[tb]
\centering
\includegraphics[clip, trim=1cm 12cm 0.5cm 11.5cm, width=1\textwidth]{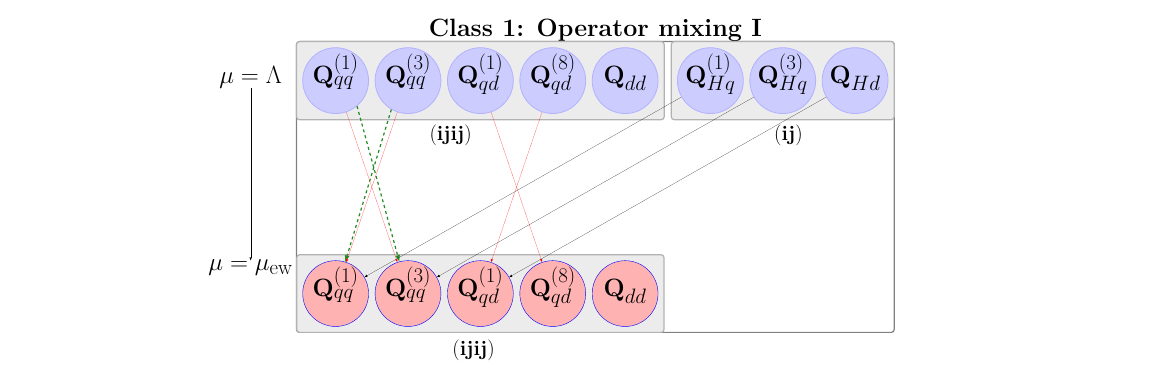}
\includegraphics[clip, trim=4.5cm 12.8cm 0.2cm 12cm, width=1.3\textwidth]{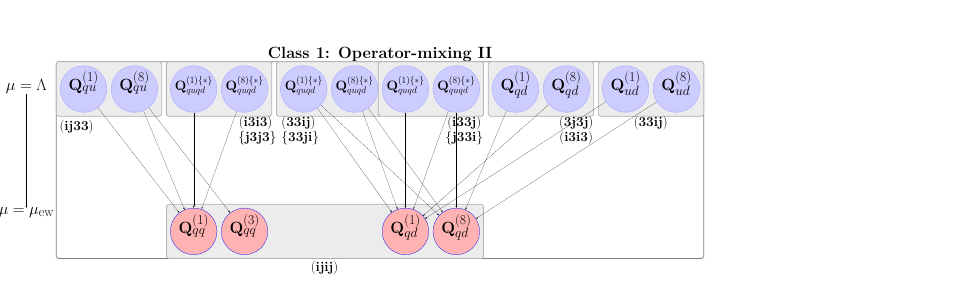}

\caption{\small Class 1: Operator mixing as per criteria-I relevant for $\DF=2$ observables for the $\KKbar$ and $\BBbar$ mixing in the Warsaw down-basis. The solid red, dashed green and solid black lines indicate the mixing due to strong, electroweak and third generation-Yukawa couplings, respectively. Self-mixing is not depicted. The top panel is also valid for $\DDbar$ mixing in the up-basis by replacing the RH down-quark $d$ with a RH up-quark $u$ in the corresponding operators. In contrast, all lines in the bottom panel are absent in the up-basis. The indices in round and curly brackets correspond to the non-conjugate and conjugate operators, respectively.}
\label{chart:df2-down}
\end{figure}

\begin{itemize}
\item In the top panel, the black lines showing mixings of $\wc[(1)]{qd}{ijij}$ and $\wc[(8)]{qd}{ijij}$ into $\wc[(8)]{qd}{ijij}$ and $\wc[(1)]{qd}{ijij}$, respectively, exist only if some of the indices on the r.h.s. among $i$ and $j$ are equal to 3, as can be seen from \eqref{eq:f4f4yukcl1}. 
\item In the bottom panel, the mixings of $\wc[(1)]{quqd}{ijij}$ and $\wc[(8)]{quqd}{ijij}$ {into} $\wc[(1)]{qq}{ijij}$ and $\wc[(3)]{qq}{ijij}$ involving $y_b$ exist only if some of the indices on the r.h.s. among $i$ and $j$ {are} equal to 3 (see \eqref{eq:cl1scalarmix}).
\item In the bottom panel, the mixings of $\wc[(1)]{quqd}{33ji}$ and $\wc[(8)]{quqd}{33ji}$ {into} $\wc[(1)]{qd}{ijij}$ and $\wc[(8)]{qd}{ijij}$ involving $y_b$ exist only if some of the indices on the r.h.s. among $i$ and $j$ {are} equal to 3, as can be seen in \eqref{eq:cl1scalarmix}. 
\end{itemize}

\begin{figure}[tb]
\centering
\includegraphics[trim={3.2cm 11.5cm 4cm 11cm},clip, width=0.57\textwidth]{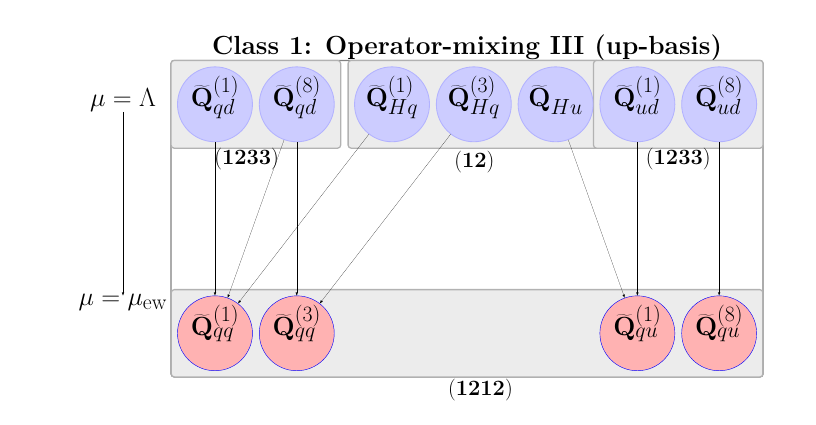}
\caption{\small Class 1: Operator mixing relevant for $\DDbar$-mixing in the Warsaw up-basis. The solid black lines indicate the mixing due to bottom-Yukawa couplings. The self-mixing is not shown here.}
\label{chart:df2-up}
\end{figure}

At the EW scale, the operators contributing to $\DDbar$ mixing are given in \eqref{eq:df2upsmefttree}. The set of relevant operators at the scale $\Lambda$ due to bottom Yukawa coupling operator mixing can be read from Fig.~\ref{chart:df2-up}. 

\begin{center}
\textrm{\bf SMEFT-Loop: 1B}
\end{center}
\be \label{class1B-smeftopsL}
\begin{aligned}
\textrm{\bf Yukawa-mixing:} &~ 
\opup[(1)]{Hq}{}\,, \quad 
\opup[(3)]{Hq}{}\,, \quad 
\opup[]{Hu}{}\,, \quad 
\opup[(1)]{qd}{}\,, \quad 
\opup[(8)]{qd}{}\,, \quad 
\opup[(1)]{ud}{}\,, \quad 
\opup[(8)]{ud}{}\,. 
 \\
\textrm{\bf Gauge-mixing:} &~  
\textrm{No new {operators}.}
\end{aligned}
\ee
Like in Class 1A, there are no new operators due to gauge-coupling ADMs. 
We observe that also in this case not only four-fermion operators have an impact 
on {meson} mixing but also the Higgs-quark operators.

\subsection{$\text{SU(2)}_L$ Correlations}

{To investigate correlations between Class 1 and other classes, we identify the 
SMEFT operators that contribute at tree-level to two different classes simultaneously.}

The WET operators in other classes can be generated from Class 1 SMEFT operators either 
due to $\text{SU(2)}_L$ symmetry at the level of tree-level matching \eqref{eq:left-smeft-down} 
and from non-trivial flavour structures emerging from CKM rotations when changing from 
the SMEFT down-basis to the mass-basis. 

In both cases, the presence of left-handed fermion fields is a necessary condition. In Tab.~\ref{tab:class1-su2}, we list such tree-level correlations to other classes, where we only consider operators containing quark-doublet fields $q$. 

In the top-down approach one usually starts with a model in which a specific set of SMEFT WCs are generated. For example, a $Z^\prime$ model could generate Class 1 WCs $\wc[(1)]{qq}{2323}$. Using the relation in this table, we find then
\be
\wcL[V1,LL]{ud}{mn23} = V_{m2} \wc[(1)]{qq}{2323} V^*_{3n}.
\ee
This implies due to $\text{SU(2)}_L$-symmetry, that several other WET operators are also generated in addition to $\wcL[V1,LL]{dd}{2323}$. Therefore, the WC $\wc[(1)]{qq}{2323}$ can be constrained by other processes, in addition to Class 1 processes. These are indicated in the third column in Tab.~\ref{tab:class1-su2}.

{
It is important to note that the actual phenomenological significance of these correlations depends on details such as their numerical size  of the WCs, which is beyond the scope of this review and will be addressed in future work.}

Interestingly, these correlations can also be interpreted in a bottom-up manner. For instance, one can examine how the effects in certain low-energy processes, generated by specific WET operators that match onto Class 1 SMEFT WCs 
can be constrained by meson mixing observables, assuming that SMEFT is the valid effective theory above the EW scale.

For this, it is instructive to expand Class 1 SMEFT WCs in powers of $\lambda$ as functions of {other} 
WET WCs.  Inverting the matching conditions in Tab.~\ref{tab:class1-su2}, using the unitarity of the CKM, 
we can find {other WET operators} correlated to Class 1.

For example, writing $(\overline L L)(\overline L L)$ SMEFT WCs in terms of $\wcL[V1,LL]{ud}{}$, we obtain

\be
\begin{aligned}
\wcL[(1)]{qq}{2323} & \supset
\lambda^2 \wcL[V1,LL]{ud}{2223} 
+\lambda^3 (
\wcL[V1,LL]{ud}{1223} +
\wcL[V1,LL]{ud}{2123})
+ \lambda^4 \wcL[V1,LL]{ud}{1123} \,,  \\
\wc[(1)]{qq}{1212} & \supset
\wcL[V1,LL]{ud}{1212} +
\lambda (
\wcL[V1,LL]{ud}{1112}-
\wcL[V1,LL]{ud}{2212}) 
-\lambda^2
\wcL[V1,LL]{ud}{2112} \,, \\
\wc[(1)]{qq}{1313} & \supset
\lambda^2 \wcL[V1,LL]{ud}{1213} 
+ \lambda^3(\wcL[V1,LL]{ud}{1113} -\wcL[V1,LL]{ud}{2213})
-\lambda^4 \wcL[V1,LL]{ud}{2113}.
\end{aligned}
\ee
The expansion for the other SMEFT operators have a similar form. Since, on the RHS the WET operators are in the mass basis, they can be directly associated to physical processes such as
\be
b \to s c \bar c: \mathcal{O}(\lambda^2) \,, \quad  s \to d u \bar u: \mathcal{O}(\lambda)\,, \quad 
b \to d u \bar u: \mathcal{O}(\lambda^3).
\ee
Here $\mathcal{O}(\lambda^n)$ indicates how strongly WET WCs related to these processes will contribute to Class 1 processes. In addition $pp \to jj$ processes are possible.

\begin{table}[tb]
\begin{center}
\renewcommand*{\arraystretch}{1.0}
\resizebox{0.9\textwidth}{!}{
\begin{tabular}{ |c|c|c|c| }
\hline
\multicolumn{4}{|c|}{$\text{SU}(2)_L$ Correlations for Class 1} \\
\hline
Class 1 WCs at $\muEW$& $\text{SU(2)}_L$  correlated WET WCs   & processes & class \\
\hline
$\wc[(1)]{qq}{ijij}, \wc[(3)]{qq}{ijij}$ & $\wcL[V,LL]{uu}{mnrs}
=  V_{mi}  V_{ri'}(\wc[(1)]{qq}{iji'j'}+ \wc[(3)]{qq}{iji'j'}) { V^*_{nj} 
 V^*_{sj'}}$ &  $ pp \to $ jets  & 10 \\
& &   $ c \to  u u \bar u$  & --  \\
& &   $ \DDbar$  & 1 \\
\hline 
$\wc[(1)]{qq}{ijij}, \wc[(3)]{qq}{ijij}$ & $\wcL[V1,LL]{ud}{mnrs}
	=  V_{mi} (\wc[(1)]{qq}{ijrs}-\frac{1}{3} \wc[(3)]{qq}{ijrs}) { V^*_{nj}}$  & $d_j \to d_i u_m \bar u_m$  & 3\\
& &   $ pp \to $ jets  & 10\\
\hline 
$\wc[(3)]{qq}{ijij}$ & $\wcL[V8,LL]{ud}{mnrs}
= 4 V_{mi} \wc[(3)]{qq}{ijrs} { V^*_{nj} }$ &  $d_j \to d_i u_m \bar u_m$  & 3 \\
& &   $ pp \to $ jets  & 10\\
\hline 
$\wc[(1)]{qd}{ijij}$ & $\wcL[V1,LR]{ ud}{mnij} =  V_{mi'} \wc[(1)]{qd}{i'j'ij}{  V^*_{nj'}} $ & $d_j \to d_i u_m \bar u_m$  & 3 \\
& &   $ pp \to $ jets  & 10\\
\hline 
$\wc[(8)]{qd}{ijij}$ & $\wcL[V8,LR]{ ud}{mnij} =  V_{mi'} \wc[(8)]{qd}{i'j'ij} { V^*_{nj'}} $ & $d_j \to d_i u_m \bar u_m$  & 3 \\
& &   $ pp \to $ jets  & 10\\
\hline 
\end{tabular}
}
\caption{\small Summary of $\text{SU(2)}_L$ correlations for Class 1 operators at tree-level in the WET and SMEFT. The first column shows Class 1 operators at $\muEW$. The second column lists the other WET operators generated by the operators in the first column, and the third column shows processes that are generated by the WET operators in the second column. The last column shows the classes that are correlated to Class 1. We {have adopted} the down-basis for the SMEFT.} 
\label{tab:class1-su2}
\end{center}
\end{table}

Since $\DF=2$ observables typically place strong constraints on Class 1 SMEFT WCs, 
sizable effects  in observables correlated to $\DF=2$ transitions 
are generally disfavoured \footnote{See for instance \cite{Silvestrini:2018dos}.} 
unless the interplay of left-handed and right-handed couplings allows to suppress 
NP contributions to $\Delta F=2$ processes as pointed 
out in \cite{Buras:2014sba,Buras:2014zga,Crivellin:2015era}. 

For a most recent analysis of rare $B$ decays in the context of $Z^\prime$ models using such a strategy see \cite{Buras:2024mnq}. In the case of rare $K$ decays the constraint from $\varepsilon_K$ can be practically eliminated within $Z^\prime$ models by choosing the $Z^\prime$ couplings to quarks to be almost imaginary \cite{Aebischer:2023mbz}. Yet, such choices have further implications which are indirect consequences of the $\text{SU(2)}_L$ correlations discussed here.

{\boldmath
\section{FCNC Decays of Mesons (Class 2)}
\label{class2}
}

The main goal of this section is to discuss the SMEFT Lagrangian responsible for the semileptonic, leptonic and radiative FCNC decays of mesons. The non-leptonic meson decays are discussed in the next section. In Class 2A, we consider following semileptonic decays:
\begin{center}
\textrm{\bf Class 2A: Semileptonic Decays}
\end{center}
\be
\begin{aligned}
&B\to K^{(*)} \ell^+\ell^-\,,\quad  
B\to X_{s,d} \ell^+\ell^-\,,  \quad  B\to K^{(*)}\nu\bar\nu\,,         \\ 
& B\to X_{s,d}\nu\bar\nu \,,\quad B\to \pi \nu \bar \nu\,,\quad B\to \varrho \nu \bar \nu\,,\\
&\kpn, \quad \klpn\,, \quad  K_L \to \pi^0 \ell^+\ell^-\,,
\end{aligned}
\ee
which are governed at the quark level by $d_j \to d_i \ell^+ \ell^-$ and $d_j \to d_i \nu \bar \nu$ transitions. There are also other semileptonic decays, in particular charm decays \cite{Gambino:2010jz,deBoer:2016dcg,DeBoer:2018pdx,Fael:2019umf,Golz:2022alh,Gisbert:2024kob} but we concentrate here on $K$ and $B$ decays. 

{Class 2B is dedicated to leptonic decays.}

\begin{center}
\textrm{\bf Class 2B: Leptonic Decays}
\end{center}
\be
B_{s,d}\to\ell^+\ell^-\,, \quad  K_{L,S}\to \ell^+\ell^-\,.
\ee
The Class 2B could in principle be {combined} together with Class 2A, because the operators responsible for leptonic FCNC decays of mesons are among the operators of Class 2A. But within the SM, the  decays like $B_{s,d}\to\mu^+\mu^-$ and $ K_{L,S}\to\mu^+\mu^-$ receive only contributions from the operator $\OpL[]{10}$ listed below. This is the consequence of the vanishing of the relevant hadronic matrix elements of vector currents so that the operator $\OpL[]{9}$ does not contribute. Correspondingly, this is also the case for $\OpL[']{10}$ and $\OpL[']{9}$ which involve right-handed FCNCs, that are absent within the SM.

The processes of interest in Class 2C are the radiative decays {$d_j \to d_i \gamma$ and $d_j \to d_i g$}:
\begin{center}
\textrm{\bf Class 2C: Radiative Decays}
\end{center}
\be
 B \to X_s \gamma\,, \quad  K \to \pi \gamma\,, \quad B \to X_s g\,, \quad K\to \pi g\,.
\ee
Similar to the observables in Class 1 these observables played a prominent role in the tests of the SM and the search for NP for the last five decades. Being strongly suppressed within the SM they play a very important role in the selection of the viable extensions of the SM. However, in contrast to processes of Class 1 several of them are still poorly measured and for several of them only experimental upper bounds are known. Details on these observables within the SM can be found in Chapter 9 in \cite{Buras:2020xsm}. In the following we discuss the CWET and WET Lagrangians that govern these processes and subsequently we will devote this section to their description within the SMEFT.

{\boldmath
\subsection{CWET and WET for Class 2 at $\muEW$}
}

\subsubsection{Semileptonic Decays (2A)}

The low-energy (e.g. at $\mu \sim m_b$ of $B$-decays and $\mu \sim 1$ GeV for $K$-decays) CWET operator basis for $d_j \to d_i \ell^+_p \ell^-_p$ transitions as usually found in the literature is given by the effective Lagrangian
\begin{equation}\label{eq:Leffqll}
\mathcal{L}_{\rm CWET}^{6A} = \mathcal{N}\, {
 \sum_{a \in \{7-10, S,P\}} (\WCL[]{a} \OpL[]{a}  + \WCL[\prime ]{a} \OpL[\prime ]{a}   + h.c.)} \, .
\end{equation}
Here the normalization depends upon the specific flavour transition, for $j\to i$
\begin{equation}\label{eq:Heff_normalization}
\mathcal{N}_{ji} =\frac{4\,G_F}{\sqrt{2}}\frac{\alpha}{4\pi}V_{ti}^* V_{tj}\,.
\end{equation}
This ensures that the WCs in the CWET are dimensionless as used in the literature. When relating them to the dimensionfull WET and SMEFT WCs, this factor has to be taken into account.

For the evaluation of these factors we can use the results of \cite{Buras:2022wpw}
\be\label{CKMoutput}
\boxed{\vcb=42.6(4)\times 10^{-3}\,, \,\,
\gamma=64.6(16)^\circ\,, \,\ \beta=22.2(7)^\circ\,, \,\, \vub=3.72(11)\times 10^{-3}\,,\,}
\ee
and consequently \cite{Buras:2022qip}
\be\label{CKMoutput2}
\boxed{\vts=41.9(4)\times 10^{-3}\,, \,\, \vtd=8.66(14)\times 10^{-3}\,,\,\,
{\im}\lambda_t=1.43(5)\times 10^{-4}\,,}
\ee
\be\label{CKMoutput3}
\boxed{\bar\varrho=0.164(12),\qquad \bar\eta=0.341(11)\,,}
\ee
where $\lambda_t=V_{ts}^*V_{td}$. They have been obtained from $\Delta F=2$ processes using the input values in Tab.~\ref{tab:parameters}. We find then
\be\label{Nfactors}
\mathcal{N}_{sb} = -8.6 \times 10^{-4} e^{i\beta_s}{\rm TeV^{-2}}\,,\,\, \mathcal{N}_{db} = 1.8 \times 10^{-4} e^{i\beta}{\rm TeV^{-2}}\,,\,\, \mathcal{N}_{ds} = -7.4 \times 10^{-6} e^{i(\beta-\beta_s)}{\rm TeV^{-2}}\,,
\ee
with the phases $\beta$ and $\beta_s$ defined through
\be
V_{td}=\vtd e^{-i\beta}\,,\qquad V_{ts}=-\vts e^{-i\beta_s}\,.
\ee

It is instructive to compare the factors $\mathcal{N}_{ij}$ with the corresponding $1/\Lambda^2$ factor in the SMEFT WCs by defining
\be
|\mathcal{N}_{ij}|=\frac{1}{[\Lambda^{\rm eff}_{ij}]^2}\,.
\ee
Then
\be
\Lambda^{\rm eff}_{sb}=34 {\rm TeV}\,,\qquad \Lambda^{\rm eff}_{db}=75 {\rm TeV}\,,\qquad
\Lambda^{\rm eff}_{ds}=366 {\rm TeV}\,.
\ee
This shows that $K$-meson decays are more powerful than $B$-meson decays in testing very high energy scales. This is demonstrated explicitly in \cite{Buras:2014zga} and applies also to $K^0-\bar K^0$ mixing when compared with $B^0_{s,d}-\bar B^0_{s,d}$ mixings \cite{Isidori:2010kg}.

The operators in \eqref{eq:Leffqll} are defined as
\be
\label{eq:wet-bsll}
\begin{aligned} 
\OpL[]{7} & =   \frac{m_{k}}{e} (\bar d_i \sigma^{\mu\nu} P_R d_j) F_{\mu\nu}\,, \quad
\OpL[\prime ]{7}  =   \frac{m_{k}}{e} (\bar d_i \sigma^{\mu\nu} P_L d_j) F_{\mu\nu}\,,\\ 
\OpL[]{8} & =   m_{k}\frac{g_s}{e^2} (\bar d_i \sigma^{\mu\nu}T^A P_R d_j) G^{A}_{\mu\nu}\,, \quad
\OpL[\prime ]{8}  =   m_{k}\frac{g_s}{e^2} (\bar d_i \sigma^{\mu\nu}T^A P_L d_j)  G^{A}_{\mu\nu}\,, \\ 
\OpL[]{9} &=   (\bar d_i \gamma_\mu P_L d_j) (\bar \ell_p \gamma^\mu \ell_p)\,, \quad
\OpL[\prime ]{9} =   (\bar d_i \gamma_\mu P_R d_j) (\bar \ell_p \gamma^\mu \ell_p)\,,  \\
\OpL[]{10} &=   (\bar d_i \gamma_\mu P_L d_j) (\bar \ell_p \gamma^\mu \gamma_5 \ell_p)\,, \quad
\OpL[\prime ]{10} =   (\bar d_i \gamma_\mu P_R d_j) (\bar \ell_p \gamma^\mu \gamma_5 \ell_p)\,,  \\
\OpL[]{S} &=  m_{k} (\bar d_i P_R d_j) (\bar \ell_p  \ell_p)\,, \quad
\OpL[\prime ]{S} =  m_{k} (\bar d_i P_L d_j) (\bar \ell_p \ell_p)\,,  \\
\OpL[]{P} &=   m_{k} (\bar d_i  P_R d_j) (\bar \ell_p \gamma_5 \ell_p)\,, \quad
\OpL[\prime ]{P} = {m_{k}}  (\bar d_i P_L d_j) (\bar \ell_p \gamma_5 \ell_p)\,. 
\end{aligned}
\ee

Due to the disparity of the masses of the involved quarks, in $B$ decays $m_k=m_b(\mu)$ and in $K$ decays $m_k=m_s(\mu)$ are usually used. The inclusion of these factors in dipole and scalar operators is a convention which simplifies the RG analysis of the WCs as then the diagonal evolution of these operators is absent. Since the scalar operators are strongly constrained by leptonic decays like $B_s \to \mu^+ \mu^-$ and $K_L\to\mu^+\mu^-$, we discuss them separately in Class 2B.

In the case of $d_j \to d_i \nu_p \bar\nu_p $ observables, the relevant set of low energy operators is given by
\be\label{eq:Leffnunu}
\mathcal{L}_{\rm CWET}^{6A} =  \mathcal{N} 
(\WCL[]{L}\OpL[]{L} + \WCL[]{R}\OpL[]{R}) + h.c. \, ,
\ee
with
\be\label{eq:bsnunu_ops}
\OpL[]{L}  = (\bar d_i \gamma_\mu P_L d_j)  
(\bar \nu_p \gamma^\mu  (1-\gamma_5) \nu_p)\,, \quad \OpL[]{R} =
  (\bar d_i \gamma_\mu P_R d_j)  
(\bar \nu_p \gamma^\mu  (1-\gamma_5) \nu_p)\,.
\ee
Only the first operator is present in the SM.

As already stated before, the CWET basis is commonly used in phenomenological analyses of decays belonging to Class 2. The basic formulae for related observables have been given for decades in this basis, first within the SM \cite{Buchalla:1995vs}, subsequently in several BSM analyses \cite{Altmannshofer:2008dz,Altmannshofer:2009ma,Buras:2014fpa} and recently in a book \cite{Buras:2020xsm}. Furthermore, many fits have been performed in this basis. A review of the status of semileptonic $B$ anomalies can be found in \cite{London:2021lfn, Capdevila:2023yhq}.

In the following subsections, we will establish the relation between CWET WCs at $\muEW$ and the SMEFT operators at the scales $\muEW$ and $\Lambda$. For the latter, the 1-loop contributions due to the SMEFT evolution from $\Lambda$ to $\muEW$ will be considered. The matching relations will be given at the tree-level \cite{Aebischer:2015fzz}. One-loop matching effects can be found for instance in \cite{Dekens:2019ept}.

It is also useful to present Class 2A in terms of the JMS basis because of its simple matching to the Warsaw basis. The relevant set of JMS operators is
\begin{center}
\textrm{\bf WET-Tree 2A}
\end{center}
\be
\begin{aligned}
&
\opL[V,LL]{ed}{ppij}\,, \quad
\opL[V,RR]{ed}{ppij}\,, \quad
\opL[V,LR]{ed}{ppij}\,, \quad
\opL[V,LR]{de}{ijpp}\,, \quad
\opL[S,RR]{ed}{ppij}\,, \quad
\opL[S,RL]{ed}{ppij}\,, \quad
\opL[T,RR]{ed}{ppij}\,, \\
&
\opL[V,LL]{\nu d}{ppij} \,, \quad
\opL[V,LR]{\nu d}{ppij} \,, \quad
\opL[]{d\gamma}{ij}\,, \quad
{\opL[]{dG}{ij}}\,.
\end{aligned}
\ee
Here the index $p$ stands for the external leptons which can take values 1-3. $j$ and $i$ are quark flavour indices in the process $j\to i$, for $b\to s: j=3, i=2 $, $s\to d: j=2, i=1$ and $b\to d: j=3, i=1$. The last two operators belong to Class 2C but also play a role in Class 2A decays. Furthermore, we have neglected LFV operators which will be discussed in Sec.~\ref{class6}.

\subsubsection{Leptonic Decays (2B)}
Beyond the SM the following operators contribute to the leptonic meson decays without LFV within the CWET:
\be
\begin{aligned} \label{eq:wet-bsmumu}
\OpL[]{10} &=   (\bar d_i \gamma_\mu P_L d_j) (\bar \ell_p \gamma^\mu \gamma_5 \ell_p)\,, \quad
\OpL[']{10} =   (\bar d_i \gamma_\mu P_R d_j) (\bar \ell_p \gamma^\mu \gamma_5 \ell_p)\,, \\
\OpL[]{S} &=  m_j (\bar d_i P_R d_j) (\bar \ell_p  \ell_p)\,, \quad
\OpL[']{S} =  m_j (\bar d_i P_L d_j) (\bar \ell_p \ell_p)\,, \\
\OpL[]{P} &=   m_j (\bar d_i  P_R d_j) (\bar \ell_p \gamma_5 \ell_p)\,, \quad
\OpL[' ]{P} = m_j  (\bar d_i P_L d_j) (\bar \ell_p \gamma_5 \ell_p)\,.
\end{aligned}
\ee
For the $B$ or $K$ decays of this class, flavour indices have to be chosen accordingly. In the SM only the first operator is present.

The WET and the SMEFT analyses of these decays are similar to the decays of Class 2A except that they are simpler given the smaller number of contributing  operators.

As an example, the explanation of the observed anomalies in $B\to K(K^*)\mu^+\mu^-$ decays requires at least a NP contribution to the WC of the operator $\OpL[bs\ell\ell]{9}$. This in turn generates, through RG effects related to the top-Yukawa coupling, non-vanishing NP contributions to the WC of $\OpL[bs\ell\ell]{10}$, affecting the decay $B_s\to\mu^+\mu^-$. A recent analysis in \cite{Buras:2024mnq} demonstrates this within a $Z^\prime$ model in explicit terms.

It should also be recalled that the decays in Class 2B are not only loop suppressed within the SM, like decays in Class 2A. As the final state is purely leptonic and the initial state is a pseudoscalar, the decays in question are also strongly helicity suppressed in view of the smallness of $m_\mu$. Even stronger suppression is found for $e^+e^-$ in the final state. This suppression is lifted through the contributions of the operators $\OpL[bs\ell\ell]{S}$, $\OpL[\prime bs\ell\ell]{S}$, $\OpL[bs\ell\ell]{P}$ and $\OpL[\prime bs\ell\ell]{P}$. In principle the decays of this Class could be strongly affected by such operators. The fact however, that the SM prediction for the $B_s\to \mu^+\mu^-$ branching ratio is presently within $1\sigma$ of the experimental data tells us that either the relevant couplings are small or the masses of scalar and pseudoscalar particles generating such operators are large. In fact, as demonstrated in \cite{Buras:2014zga}, they could be as high as a few hundred TeV. For the most recent phenomenological analysis of this class of decays in correlation with $B\to K(K^*)\nu\bar\nu$ and $B\to K(K^*)\mu^+\mu^-$ decays see \cite{Bause:2021cna,He:2021yoz,Bause:2022rrs,Becirevic:2023aov,Bause:2023mfe,Allwicher:2023xba,Wang:2023trd, Altmannshofer:2023hkn, Gabrielli:2024wys, Hou:2024vyw, He:2024iju, Bolton:2024egx,Marzocca:2024hua,Buras:2024ewl,Buras:2024mnq,Altmannshofer:2024kxb,Hati:2024ppg,Bolton:2025fsq,He:2025zfy,Crivellin:2025qsq}. Also correlations with the decay $\kpn$ explored by NA62 are discussed in some of these papers.

The WET basis for Class 2B is given as
\begin{center}
\textrm{\bf WET-Tree 2B}
\end{center}
\be
\begin{aligned}
&
\opL[V,LL]{ed}{ppij}\,, \quad
\opL[V,RR]{ed}{ppij}\,, \quad
\opL[V,LR]{ed}{ppij}\,, \quad
\opL[V,LR]{de}{ijpp}\,, \quad
\opL[S,RR]{ed}{ppij}\,, \quad
\opL[S,RL]{ed}{ppij}\,, \quad
\opL[T,RR]{ed}{ppij}\,, \quad\\
&
\opL[]{d\gamma}{ij}\,, \quad
{\opL[]{dG}{ij}}\,.
\end{aligned}
\ee
Comparing this basis with the corresponding CWET operators, we find that only certain linear combinations of the above operators matter for 2B. For example, for the vector operators only the pairs $\wc[V,LL]{ed}{ppij}, \wcL[V,LR]{de}{ijpp}$ and $\wcL[V,RR]{ed}{ppij}, \wcL[V,LR]{ed}{ppij}$ map onto $Q_{10}$ and $Q_{10}^\prime$ operators while the remaining two combinations do not appear in 2B.\footnote{This can easily be deduced from \eqref{class2A-matchWETCWET}.} Similarly, linear combinations of $\opL[S,RR]{ed}{ppij}, \opL[S,RL]{ed}{ppij}$ along with their hermitian conjugates map onto $Q_{S}^{(\prime)}$ and $Q_{P}^{(\prime)}$.

\subsubsection{Radiative Decays (2C)}
The low-energy CWET operators for Class 2C are \cite{Buchalla:1995vs} given by the first four operators in \eqref{eq:wet-bsll}. For 2C the WET dipoles are denoted as
\begin{center}
\textrm{\bf WET-Tree 2C}
\end{center}
\be
\begin{aligned}
\opL[]{d\gamma}{ij} \,, \quad \opL[]{dG}{ij}.
\end{aligned}
\ee
In the WET basis the hermitian conjugate of $\opL[]{d\gamma}{{ji}}$ and $\opL[]{dG}{{ji}}$ serve as the primed dipole operators in CWET. 

Further, it is worth mentioning that at the 1-loop level the scalar and tensor four-fermion WET operators can mix with the dipole operators. The relevant 1-loop WET operators are $\opL[T,RR]{ed}{ppij}$, $\opL[S1,RR]{dd}{{ippj}}$, $\opL[S8,RR]{dd}{{ippj}}$, $\opL[S1,RR]{uddu}{{pjip}}$, and $\opL[S8,RR]{uddu}{{pjip}}$. The electromagnetic and chromomagnetic dipole operators do also mix with each other.

The size of these effects would be of the following order: (size of 1-loop RGE in WET) $\times $ (size of 1-loop RGE in SMEFT). Such effects could be relevant in certain cases. However, for the discussion at hand we will restrict ourselves to the tree-level WET operators (i.e. we will ignore RG or matching generated new operators within WET) given above.

\subsection{SMEFT Operators for Class 2 at $\muEW$ }

In this section, we discuss the tree-level SMEFT Lagrangian at the EW scale.

\subsubsection{Semileptonic Decays}
Suppressing the flavour indices, the tree-level SMEFT operators contributing to Class 2A processes are given by \cite{Buras:2014fpa,Celis:2017doq,Datta:2019zca}
\begin{center}
\textrm{\bf SMEFT-Tree 2A}
\end{center}
\be
\label{eq:class2A-smeft-tree}
\begin{aligned} 
&
\ops[(1)]{\ell q}{}   \,,  \quad
\ops[(3)]{\ell q}{}  \,,  \quad
\ops[]{\ell d}{}   \,,  \quad
\ops[]{ed}{}  \,,  \quad
{\ops[]{qe}{}}    \,, \quad
\ops[]{Hd}{} \,, \quad
\ops[(1)]{Hq}{} \,, \quad
\ops[(3)]{Hq}{}\,.
\end{aligned}
\ee

Their matching onto WET at $\muEW$ reads \cite{Jenkins:2017jig,Dekens:2019ept}
\be
\label{class2A-match}
\begin{aligned}
\wcL[V,LL]{ed}{ppij} & = \zeta_1  C_Z^q + \wc[(1)]{l q}{ppij} + \wc[(3)]{l q}{ppij}\,, \quad \wcL[V,RR]{ed}{ppij}=2 s_w^2 C_Z^d + \wc[]{ed}{ppij}\,, \\
\wcL[V,LR]{ed}{ppij}  &=  \zeta_1 C_Z^d +\wc[]{l d}{ppij}   \,, \quad \wcL[V,LR]{de}{{ijpp}}  = 2 s_w^2 C_Z^q +\wc[]{qe}{{ijpp}}\,, \\
\wcL[V,LL]{\nu d}{ppij} & =  C_Z^q + \wc[(1)]{l q}{ppij} - \wc[(3)]{l q}{ppij}\,, \quad 
\wcL[V,LR]{\nu d}{ppij}=  C_Z^d + \wc[]{ld}{ppij}\,.
\end{aligned}
\ee
Here $\zeta_1 = 1-2c_w^2$. 
The parameters $C_Z^{q(d)}$ contribute at the same order $( \sim 1 / v^2 \times v^2 / \Lambda^2 = 1 / \Lambda^2)$ in the power counting as the four-fermion operators:
\begin{equation}
\label{cZ}
C_Z^q = (\wc[(1)]{Hq}{ij}   + \wc[(3)]{Hq}{ij}){\delta_{pp}} \,,
\qquad
C_Z^d = \wc[]{Hd}{ij}{\delta_{pp}}\,.
\end{equation}
The CWET WCs $\WCL[ijpp]{9,10}$, $\WCL[\prime, ijpp]{9,10}$, and $\WCL[ijpp]{L,R}$ defined through \eqref{eq:wet-bsll}-\eqref{eq:bsnunu_ops} at $\muEW$ are related to the WCs of the SMEFT operators in \eqref{eq:class2A-smeft-tree} at tree-level as follows: 
\be
\begin{aligned} \label{eq:SMEFT2WET_C9_C10}
2\, \,\mathcal{N}_{ji}\,C_{9}^{ijpp} &= \wc[]{qe}{ijpp}  + \wc[(1)]{\ell q}{ppij} 
+ \wc[(3)]{\ell q}{ppij}  - \zeta_2 \, C_Z^q\,, 
\\
2\, \,\mathcal{N}_{ji}\,C_{10}^{ijpp} &=  \wc[]{qe}{ijpp}  - \wc[(1)]{\ell q}{ppij} - \wc[(3)]{\ell q}{ppij} + C_Z^q\,, \\
2\, \,\mathcal{N}_{ji}\, C^{ijpp}_L &= \wc[(1)]{\ell q}{ppij}   - \wc[(3)]{\ell q}{ppij}  + C_Z^q\,,\\
2\, \,\mathcal{N}_{ji}\,C_{9}^{\prime, ijpp} &= \wc[]{ed}{ppij}  +  \wc[]{\ell d}{ppij}  -\zeta_2 \, C_Z^d\,,
\\
2\, \,\mathcal{N}_{ji}\,C_{10}^{\prime, ijpp} &= \wc[]{ed}{ppij}  -\wc[]{\ell d}{ppij}   + C_Z^d\,,
\\
2\, \,\mathcal{N}_{ji}\,C^{ijpp}_R &=\wc[]{\ell d}{ppij}  + C_Z^d\,,
\end{aligned}
\ee
where $\mathcal{N}_{ji}$ is defined in (\ref{eq:Heff_normalization}). $\zeta_2 = 1- 4 s_w^2\approx 0.08$ is an accidentally small vector coupling of the $Z$ to the charged leptons. The equations in (\ref{eq:SMEFT2WET_C9_C10}) are consistent with the ones in \cite{Buras:2014fpa}. The missing factors of $1/2$ in (\ref{cZ}) relative to the corresponding expressions in \cite{Buras:2014fpa} are due to symmetry factors that occur when a non-redundant basis is used in the matching. We refer to appendix A in \cite{Aebischer:2018iyb} for further details.

As these days the phenomenology is performed usually in terms of CWET WCs, it is useful to express WET WCs in terms of the CWET ones. This can simply be done by inserting \eqref{SMEFTCWET1} and \eqref{SMEFTCWET2} into \eqref{class2A-match}. We find
\be
\label{class2A-matchWETCWET}
\begin{aligned}
\wcL[V,LL]{ed}{ppij} & ={\mathcal{N}_{ji}}\left[C_{9}^{ijpp}-C_{10}^{ijpp}\right]
\,, \quad \wcL[V,RR]{ed}{ppij}=
{\mathcal{N}_{ji}}
  \left[C_{9}^{\prime, ijpp}+C_{10}^{\prime, ijpp}\right]\,,\\
\wcL[V,LR]{ed}{ppij}  &={\mathcal{N}_{ji}}
  \left[C_{9}^{\prime, ijpp}-C_{10}^{\prime, ijpp}\right]\,, \quad
\wcL[V,LR]{de}{ijpp}  =\mathcal{N}_{ji}\left[C_{9}^{ijpp}+C_{10}^{ijpp}\right]
\,, \\
\wcL[V,LL]{\nu d}{ppij} & =2\mathcal{N}_{ji}C^{ijpp}_L
\,, \quad 
\wcL[V,LR]{\nu d}{ppij}= 2\mathcal{N}_{ji}C^{ijpp}_R\,. 
\end{aligned}
\ee
Note that $ C_Z^q$ and $C_Z^d$ do not enter these relations as it should be.

Appropriate formulae for $B_d$, $B^+$ and $K$ mesons can be obtained by properly choosing the flavour indices. In writing these formulae we omitted the dipole operators which will be discussed in more detail in Class 2C. Obviously the WCs $\wc[ ]{ed}{ppij} $ and $\wc[]{qe}{ijpp}$ do not contribute to the $d_j \to d_i \nu \bar \nu$ processes.

\subsubsection{Leptonic Decays (2B)}
The SMEFT operators for Class 2B are the same as in Class 2A, given in \eqref{eq:class2A-smeft-tree}. Additionally we have, scalar and pseudoscalar operators which receive matching contribution from
{
\begin{center}
\textrm{\bf SMEFT-Tree 2B}
\end{center}
}
\be \label{eq:class2B-smeft-tree}
\ops[]{\ell e d q}{ppij} \,.
\ee
The SMEFT matching for $C_{10}$ and $C_{10}^{\prime }$ can be found in \eqref{eq:SMEFT2WET_C9_C10}. For the scalar operators, we have
\be
{2m_k}\mathcal{N}_{ji}\,C_{S}^{\prime, ijpp} = \wc[]{\ell e dq }{ppij}\,, \quad {2m_k}\mathcal{N}_{ji}C_{S}^{ijpp} = (\wc[]{\ell e dq }{ppji})^*\,.
\ee
The $\text{SU(2)}_L$ symmetry imposes \cite{Alonso:2014csa}
\be
C_S^{} = -C_P^{}\,, \quad  C_S^{\prime } = C_P^{\prime }\,. 
\ee
For the WET, the matching reads
\be
\label{class2B-match}
\wcL[S,RL]{ed}{ppij} = \wc[]{ledq}{ppij}\,.
\ee
The matching conditions for SRR and TRR operators vanish at the dim-6 level in the SMEFT, i.e. 
\be 
\wcL[S,RR]{ed}{ppij}  = 0 \,, \quad \wcL[T,RR]{ed}{ppij}  = 0\,,
\ee
but is non-zero when dim-8 contributions are considered \cite{Hamoudou:2022tdn}.

\subsubsection{Radiative Decays (2C)}
The SMEFT operators for Class 2C are 
\begin{center}
\textrm{\bf SMEFT-Tree 2C}
\end{center}
\be
\label{eq:class2C-smeft-tree}
\begin{aligned}
\ops[]{dW}{}   \,,  \quad \ops[]{dB}{}  \,,  \quad \ops[]{dG}{}\,.
\end{aligned}
\ee
The SMEFT-to-WET {matching}  at $\muEW$ is given by
\be \label{class2C-match}
\begin{aligned}
\wcL[]{{d\gamma}}{ij} & = {v \over \sqrt{2}} (-\wc[]{dW}{ij} s_w
+ \wc[]{dB}{ij} c_w) \,,\quad
\wcL[]{{dG}}{ij}  =  {v \over \sqrt{2}} \wc[]{dG}{ij}\,. 
\end{aligned}
\ee
The CWET dipoles differ only by prefactors from the WET dipoles. It would be straightforward to extend this analyses to up-sector where $\wc[]{uW}{}$, $\wc[]{uB}{}$ and $\wc[]{uG}{}$ will come into play.

\subsection{SMEFT Operators for Class 2 at $\Lambda$}
\label{sec:class2Lambda}

Having the knowledge of the SMEFT Lagrangian at $\muEW$, the next step is to find the SMEFT operators at the scale $\Lambda$ that through operator mixing in the RG evolution from the NP scale $\Lambda$ down to the EW scale $\muEW$ contribute to the WCs of the operators shown in \eqref{eq:class2A-smeft-tree}, \eqref{eq:class2B-smeft-tree} and \eqref{eq:class2C-smeft-tree}. The anatomy of the SMEFT ADMs reveals the associated operators to be
\begin{center}
\textrm{\bf SMEFT-Loop 2}
\end{center}
\be
\label{eq:class2A-smeft-loop}
\begin{aligned} 
\textrm{\bf Gauge-mixing}:& ~
\ops[(1)]{q q}{}  \,,\quad
\ops[(3)]{q q}{}  \,, \quad
\ops[(1)]{q d}{}  \,, \quad
\ops[(1)]{q u}{} \,,\quad
\ops[]{dd}{}  \,, \quad 
\ops[(1)]{ud}{}.\\
\textrm{\bf Yukawa-mixing}: &~
\ops[(1)]{qq}{}\,, \quad
\ops[(3)]{qq}{}\,, \quad
\ops[(1)]{qd}{}\,, \quad
\ops[(1)]{qu}{}\,, \quad
\ops[]{dd}{}\,, \quad
\ops[(1)]{ud}{}\,, \\
& ~
\ops[]{l u}{}   \,,  \quad
\ops[]{e u}{} \,, \quad
\ops[(1)]{Hl}{}\,, \quad
\ops[(3)]{Hl}{} \,, \quad
\ops[]{Hud}{} \,, \quad
\ops[]{He}{} \,, \\
&~
\ops[]{Hu}{} \,, \quad
\ops[]{uX}{} \,, \quad
\ops[]{H\Box}{} \,, \quad
\ops[]{HD}{}\,, \quad
{\ops[(1)]{quqd}{}}\,, \quad
{\ops[(8)]{quqd}{}}\,.
\end{aligned}
\ee

Here the relevant flavour indices for operators have been suppressed which can be found in the SMEFT charts or the explicit RGEs to be given below.

Setting $|\eta|\ge 1 {\rm TeV}^{2}$, we can get a quantitative sense of which RG effects in question are most significant. In Class 2A and 2B, for $b\to s \mu^+ \mu^-$ and $b\to s \nu_\mu \bar \nu_\mu$ decays within CWET, these turn out to be 
\be \begin{aligned} 
& \wcs[(3)]{qq}{}\to C_{9bs}^{\mu\mu}\,, \quad
\wcs[(1)]{qu}{}\to C_{9bs}^{\mu\mu}\,, \quad
\wcs[(1)]{qq}{}\to C_{9bs}^{\mu\mu}\,, \quad
\wcs[(1)]{lq}{}\to C_{9bs}^{\mu\mu}\,, 
\\ 
&  \wcs[]{qe}{}\to C_{9bs}^{\mu\mu}\,, \quad
\wcs[(1)]{qd}{}\to C_{9bs}^{\mu\mu}\,, \quad
\wcs[(1)]{qu}{}\to C_{10bs}^{\mu\mu}\,, \quad
\wcs[(1)]{qq}{}\to C_{10bs}^{\mu\mu}\,, 
\\ 
&  \wcs[(3)]{qq}{}\to C_{10bs}^{\mu\mu}\,, \quad
\wcs[(1)]{qd}{}\to C_{9bs}^{\prime \mu\mu}\,, \quad
\wcs[(1)]{ud}{}\to C_{9bs}^{\prime \mu\mu}\,, \quad
\wcs[]{dd}{}\to C_{9bs}^{\prime \mu\mu}\,, 
\\ 
&  \wcs[]{ld}{}\to C_{9bs}^{\prime \mu\mu}\,, \quad
\wcs[]{ed}{}\to C_{9bs}^{\prime \mu\mu}\,, \quad
\wcs[(1)]{ud}{}\to C_{10bs}^{\prime \mu\mu}\,, \quad
\wcs[(1)]{qd}{}\to C_{10bs}^{\prime \mu\mu}\,, 
\\ & 
\wcs[(1)]{qu}{}\to C_{Lbs}^{\nu_\mu\nu_\mu}\,, \quad
\wcs[(1)]{qq}{}\to C_{Lbs}^{\nu_\mu\nu_\mu}\,, \quad
\wcs[(3)]{qq}{}\to C_{Lbs}^{\nu_\mu\nu_\mu}\,, \quad
\wcs[(1)]{ud}{}\to C_{Rbs}^{\nu_\mu\nu_\mu}\,, 
\\ 
&  \wcs[(1)]{qd}{}\to C_{Rbs}^{\nu_\mu\nu_\mu}.
\end{aligned} 
\ee
See Tabs.~\ref{tab:2A1}-\ref{tab:2A2} for more details.

In Class 2A and 2B for $s \to d \mu^- \mu^+$ and  $s \to d \nu_\mu \bar \nu_\mu$, we use WET. For $|\eta| \ge  10^{-3}$, the dominant operators are given by
\be \begin{aligned} &
\wcs[(3)]{qq}{}\to \wcL[V,LL]{ed}{ }\,, \quad
\wcs[(1)]{qq}{}\to \wcL[V,LL]{ed}{ }\,, \quad
\wcs[(1)]{qu}{}\to \wcL[V,LL]{ed}{ }\,, \quad
\wcs[(1)]{ud}{}\to \wcL[V,RR]{ed}{ }\,, 
\\ & 
\wcs[(1)]{qd}{}\to \wcL[V,RR]{ed}{ }\,, \quad
\wcs[]{dd}{}\to \wcL[V,RR]{ed}{ }\,, \quad
\wcs[(1)]{qd}{}\to \wcL[V,LR]{ed}{ }\,, \quad
\wcs[(1)]{ud}{}\to \wcL[V,LR]{ed}{ }\,, 
\\ & 
\wcs[]{dd}{}\to \wcL[V,LR]{ed}{ }\,, \quad
\wcs[(1)]{qu}{}\to \wcL[V,LR]{de}{ }\,, \quad
\wcs[(1)]{qq}{}\to \wcL[V,LR]{de}{ }\,, \quad
\wcs[(3)]{qq}{}\to \wcL[V,LR]{de}{ }\,, 
\\ & 
\wcs[(1)]{qu}{}\to \wcL[V,LL]{\nu d}{ }\,, \quad
\wcs[(3)]{qq}{}\to \wcL[V,LL]{\nu d}{ }\,, \quad
\wcs[(1)]{qq}{}\to \wcL[V,LL]{\nu d}{ }\,, \quad
\wcs[(1)]{ud}{}\to \wcL[V,LR]{\nu d}{ }\,, 
\\ & 
\wcs[(1)]{qd}{}\to \wcL[V,LR]{\nu d}{ }.
\end{aligned}
\ee
Finally, in Class 2C, for $s \to d \gamma$ {for $|\eta| \ge 10^{-3}$}, we find
\be
\begin{aligned}
&
\wcs[]{dW}{}\to \wcL[]{d \gamma }{ }\,, \quad
\wcs[]{dG}{}\to \wcL[]{d \gamma }{ }\,, \quad
\wcs[(8)]{quqd}{}\to \wcL[]{d \gamma }{ }\,,  \quad
\wcs[(1)]{quqd}{}\to \wcL[]{d \gamma }{ }\,, \quad
\\ & 
\wcs[]{dB}{}\to \wcL[]{d \gamma }{ }\,, \quad
\wcs[(1)]{quqd}{}\to \wcL[]{d G }{ }\,, \quad
\wcs[]{dW}{}\to \wcL[]{d G }{ }\,, \quad
\wcs[]{dB}{}\to \wcL[]{d G }{ }\,, \quad
\\ & 
\wcs[(8)]{quqd}{}\to \wcL[]{d G }{ }\,, \quad
\wcs[]{dG}{}\to \wcL[]{d G }{ }.
\end{aligned} 
\ee
See Tabs.~\ref{tab:class2-3}-\ref{tab:class2-5} for more details. Using this approach the dominant set of WCs for other processes in 
this class can be found.

\begin{figure}[H]
\centering
\includegraphics[clip, trim=3cm 11cm 4cm 11cm, width=0.6\textwidth]{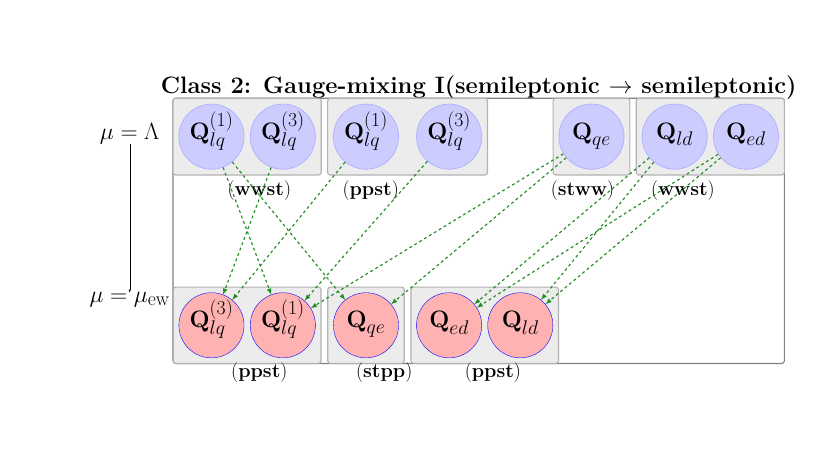}
\hspace{-15mm}
\includegraphics[clip, trim=4cm 11.5cm 0.4cm 12.5cm, width=1.3\textwidth]{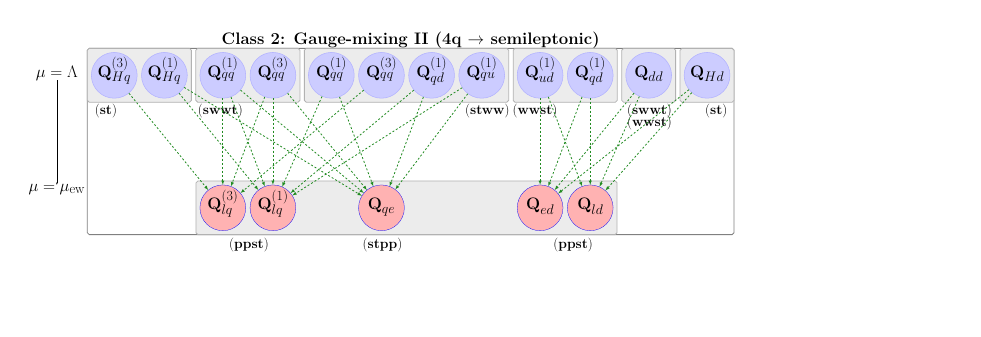}
\includegraphics[clip, trim=4.4cm 13cm 1.3cm 13cm, width=1.3\textwidth]{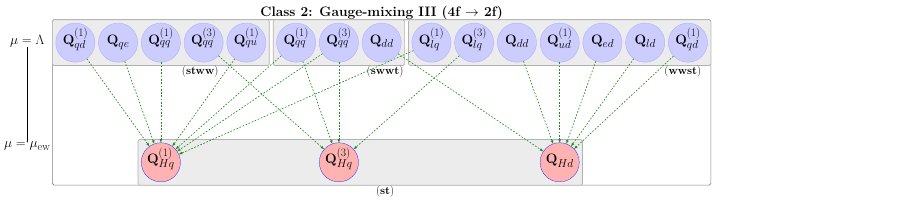}
\caption{\small Class 2: Gauge coupling dependent operator mixing for $\DF=1$ transitions such as $b \to s\ell^+ \ell^-$, $b\to s\nu \bar \nu$ and $s\to d \nu \bar \nu$ in the Warsaw down-basis. The dashed green lines indicate the mixing due to electroweak couplings. The index $w$ is summed over 1-3. All other indices are 
fixed by the {external states involved in the}  specific process of interest. The self-mixing is omitted.}%
\label{chart:df1-gauge}%
\end{figure}

\begin{figure}[H]
\hspace{-19mm}
\vspace{-0.9cm}
\includegraphics[clip, trim=4cm 13.2cm 0.2cm 13cm, width=1.5\textwidth]{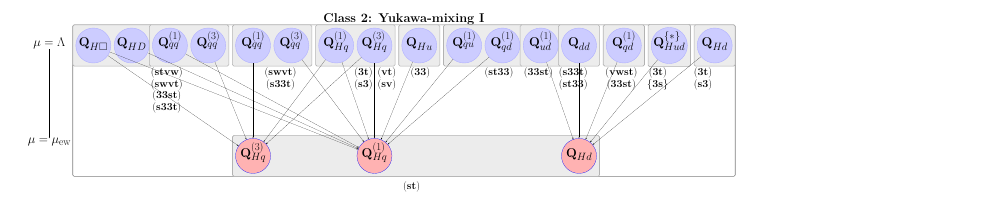}
\vspace{-0.7cm}
\includegraphics[clip, trim=4cm 12.5cm 0.7cm 12.0cm, width=1.2\textwidth]{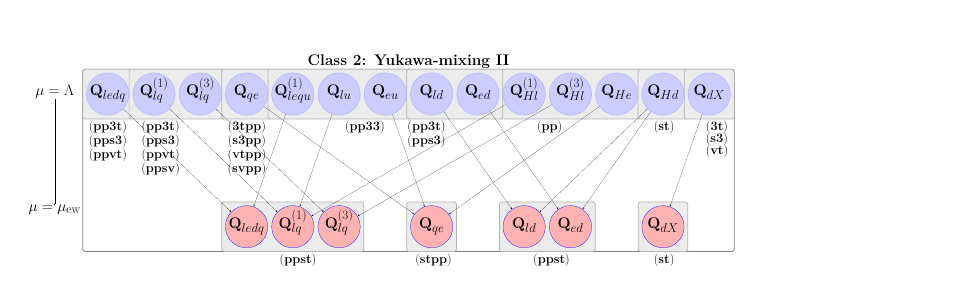}
\caption{\small Class 2: Yukawa dependent operator mixing relevant for $\DF=1$ observables such as $b \to s\ell^+ \ell^-$, $b\to s\nu \bar \nu$ and $s\to d \nu \bar \nu$ in the Warsaw down-basis. The solid black lines indicate the mixing due to top and bottom Yukawa couplings. Here the indices $v,w$ are summed over 1-3. For $\ops[]{dX}{}$ and $\ops[]{uX}{}$, $X=B,W$ or $G$. These operators are relevant for $\DF=1$ radiative processes (Class 2C) such as $b \to s\gamma(g),\, b \to d \gamma(g)$, and $s\to d \gamma(g)$. The self-mixing is omitted. } \label{chart:df1-yukawa}
\end{figure}

On the other hand, qualitatively the operator mixing can be visualized by the SMEFT charts that depend on the class of processes considered. These are displayed in Figs.~\ref{chart:df1-gauge} and \ref{chart:df1-yukawa} for gauge and Yukawa coupling dependent operator mixing, respectively. These particular charts visualize the action of the operators in \eqref{eq:class2A-smeft-loop} on the sets of operators in \eqref{eq:class2A-smeft-tree} in the process of RG evolution from the NP scale $\Lambda$ down to the EW scale $\muEW$. The charts in Fig.~\ref{chart:df1-gauge} exhibit the following features of the operator mixing
\begin{itemize}
\item Upper panel depicts the mixing of semileptonic with semileptonic operators.
\item
In the middle panel four-quark and two-quark operators generate semi-leptonic operators implying correlations between non-leptonic decays and semi-leptonic decays. Indeed this leads to correlations between hadronic and semileptonic $B$-decays \cite{Datta:2024zrl}.
\item
In the bottom panel, four-quark and four-lepton operators generate quark-Higgs operators which in turn can also generate semi-leptonic operators. This implies additional correlations between non-leptonic and semi-leptonic decays.
\item
It should also be emphasized that the generation of quark-Higgs operators has also an impact on quark mixing as seen in the chart in Fig.~\ref{chart:df2-down}. The strong constraints on NP contributions from mixing therefore imply also indirectly constraints for semi-leptonic decays. Using the same approach, correlation between different classes can be identified. 
\end{itemize}

It is useful to report the results of the anatomy of the relevant RGEs behind the SMEFT charts in question. It can be obtained from \cite{Alonso:2013hga} for gauge couplings and from \cite{Jenkins:2013wua} for Yukawa couplings. But in order to be consistent with the non-redundant basis proposed in \cite{Aebischer:2017ugx}, some terms in the formulae of these papers have to be {adjusted} as discussed in Sec.~\ref{NONRED}.

Finally, we present the explicit RGEs. The ones involving only gauge couplings are as follows.

\noindent
\underline{\bf \boldmath $f^4 \to f^4$ (Gauge)}:

Limiting to mixing between four-fermion operators, we have
\be
\begin{aligned}
\dotwc[(1)]{lq}{ppst}   
&= \frac{2}{9} g_1^2 (3  \wc[(1)]{lq}{wwst}+3  \wc[(1)]{qd}{stww}
+3  \wc[]{qe}{stww}-3  \wc[(1)]{qq}{stww}- \frac{1}{2} \wc[(1)]{qq}{swwt}-\frac{3}{2}  \wc[(3)]{qq}{swwt}-6  \wc[(1)]{qu}{stww})\delta_{pp} \\
&-g_1^2 \wc[(1)]{lq}{ppst}  +9 g_2^2 \wc[(3)]{lq}{ppst}\,, \nn
\end{aligned}
\ee
\vspace{-0.2cm}
\be
\begin{aligned}
\dotwc[(3)]{lq}{ppst}
 &=
 3 g_2^2 \wc[(1)]{lq}{ppst}+\frac{2}{3} g_2^2 ( \wc[(3)]{lq}{wwst}+\frac{1}{2} \wc[(1)]{qq}{swwt}+3  \wc[(3)]{qq}{stww}- \frac{1}{2}\wc[(3)]{qq}{swwt})\delta_{pp}-\left(g_1^2+6 g_2^2\right) \wc[(3)]{lq}{ppst}\,, \\
\dotwc[]{qe}{stpp} &=  
-\frac{2}{9} g_1^2 (-6  \wc[(1)]{lq}{wwst}-6  \wc[(1)]{qd}{stww}-6  \wc[]{qe}{stww}
+6  \wc[(1)]{qq}{stww}+  \wc[(1)]{qq}{swwt}+3  \wc[(3)]{qq}{swwt}+12  \wc[(1)]{qu}{stww})\delta_{pp} \\
& +2g_1^2 \wc[]{qe}{stpp}\,, \\
\dotwc[]{ld}{ppst} &=  
\frac{2}{9} g_1^2 (3  \wc[]{dd}{wwst}+\wc[]{dd}{swwt}+3  \wc[]{ed}{wwst}+3  \wc[]{ld}{wwst}
-3  \wc[(1)]{qd}{wwst}-6  \wc[(1)]{ud}{wwst})\delta_{pp} -2g_1^2 \wc[]{ld}{ppst}\,, \\
\dotwc[]{ed}{ppst} &=                
\frac{4}{9} g_1^2 (3  (\wc[]{dd}{wwst}+\wc[]{ed}{wwst}+\wc[]{ld}{wwst}-\wc[(1)]{qd}{wwst}-2 \wc[(1)]{ud}{wwst})+\wc[]{dd}{swwt})\delta_{pp}+4g_1^2 \wc[]{ed}{ppst}\,.
\end{aligned}
\ee
As we can see both semileptonic and four-quark operators can mix with semileptonic operators.

Next we report the RGEs that involve only gauge couplings for the mixing of two-fermion operators into four-fermion ones.

\noindent
\underline{\bf \boldmath $f^2 H^2 D \to f^4$ (Gauge)}:

\be
\begin{aligned}\label{eq:f2H2Dtof4gaugecl2}
  \dotwc[(1)]{lq}{ppst}   &
= - \frac{1}{3} g_1^2 \wc[(1)]{Hq}{st}\delta_{pp}
\,,  \quad
\dotwc[(3)]{lq}{ppst}   =
\frac{1}{3} g_2^2 \wc[(3)]{Hq}{st}\delta_{pp}
\,, \\
\dotwc[]{qe}{stpp} &=  
-\frac{2}{3} g_1^2 \wc[(1)]{Hq}{st}\delta_{pp}
\,,\quad
\dotwc[]{ed}{ppst} = -\frac{2}{3} g_1^2 \wc[]{Hd}{st}\delta_{pp}\,,\\
\dotwc[]{ld}{ppst} &=  -\frac{1}{3} g_1^2  \wc[]{Hd}{st}\delta_{pp}\, .
\end{aligned}
\ee
The four-fermion operators can also mix with two-fermion operators of Class 2. The gauge coupling dependent RGEs governing such operator mixing are given below.

\noindent
\underline{\bf \boldmath $f^4 \to  f^2 H^2 D $ (Gauge)}:

\be \label{eq:class2-hq1hq3hd-f4f2-gauge1}
\begin{aligned}
\dotwc[(1)]{Hq}{st} &= 
-\frac{2}{3} g_1^2 \left(\wc[(1)]{lq}{wwst}+\wc[(1)]{qd}{stww}+\wc[]{qe}{stww}-\wc[(1)]{qq}{stww}-\frac{1}{6} \wc[(1)]{qq}{swwt}-\frac{1}{2} \wc[(3)]{qq}{swwt}-2 \wc[(1)]{qu}{stww}\right) \,, \\
\dotwc[(3)]{Hq}{st} &= 
\frac{1}{3} g_2^2 (2 \wc[(3)]{lq}{wwst}+\wc[(1)]{qq}{swwt}+6 \wc[(3)]{qq}{stww}-\wc[(3)]{qq}{swwt})\,, \\
\dotwc[]{Hd}{st} &= 
-\frac{2}{3} g_1^2 \left(\wc[]{dd}{stww}+\frac{1}{3} \wc[]{dd}{swwt}+\wc[]{ed}{wwst}+\wc[]{ld}{wwst}-\wc[(1)]{qd}{wwst}-2 \wc[(1)]{ud}{wwst}\right)\,.
\end{aligned}
\ee
Since we require quark flavour violation in this class, the $f^2 H^2 D$ operators do not mix with each other through gauge interactions (up to self-mixing).

Next, we have operator mixing due to Yukawa interactions. The charts in Fig.~\ref{chart:df1-yukawa} have only black lines, stressing that the mixing is caused by Yukawa couplings only. The ones involving only four fermion operators are given as follows.
\\

\noindent
\underline{\bf \boldmath $f^4 \to f^4$ (Yukawa)}:

\be \label{eq:class2-gauge-lequ13-f4-f4}
\begin{aligned}
\dotwc[(1)]{lq}{ppst} &=
\frac{1}{2} y_b^2 \delta_{s3} \wc[(1)]{lq}{pp3t}+\frac{1}{2} y_b^2 \delta_{t3} \wc[(1)]{lq}{pps3}+\frac{1}{2} y_t^2 V^*_{3s} V_{3v} \wc[(1)]{lq}{ppvt}+\frac{1}{2} y_t^2 V_{3t} V^*_{3v} \wc[(1)]{lq}{ppsv}-y_t^2 \wc[]{lu}{pp33} V^*_{3s} V_{3t}  \,, \\
\dotwc[(3)]{lq}{ppst} &=
\frac{1}{2} y_b^2 \delta_{s3} \wc[(3)]{lq}{pp3t}+\frac{1}{2} y_b^2 \delta_{t3} \wc[(3)]{lq}{pps3}+\frac{1}{2} y_t^2 V^*_{3s} V_{3v} \wc[(3)]{lq}{ppvt}+\frac{1}{2} y_t^2 V_{3t} V^*_{3v} \wc[(3)]{lq}{ppsv}  \,, \\
\dotwc[]{qe}{stpp} &=
-y_t^2 \wc[]{eu}{pp33} V^*_{3s} V_{3t}+\frac{1}{2} y_b^2 \delta_{s3} \wc[]{qe}{3tpp}+\frac{1}{2} y_b^2 \delta_{t3} \wc[]{qe}{s3pp}+\frac{1}{2} y_t^2 V^*_{3s} V_{3v} \wc[]{qe}{vtpp}+\frac{1}{2} y_t^2 V_{3t} V^*_{3v} \wc[]{qe}{svpp} \,,\\
\dotwc[]{ld}{ppst} &=
y_b^2 \delta_{s3} \wc[]{ld}{pp3t}+y_b^2 \delta_{t3} \wc[]{ld}{pps3}  \,, \quad
\dotwc[]{ed}{ppst} =
y_b^2 \delta_{s3} \wc[]{ed}{pp3t}+y_b^2 \delta_{t3} \wc[]{ed}{pps3} \,, \\
\dotwc[]{ledq}{ppst} &=\frac{1}{2} y_b^2 \delta_{s3} \wc[]{ledq}{pp3t}+y_b^2 \delta_{t3} \wc[]{ledq}{pps3}+\frac{1}{2} y_t^2 V^*_{3s} V_{3v} \wc[]{ledq}{ppvt}+2 y_b y_t \wc[(1)]{lequ}{pp33} \delta_{s3} V_{3t} \,.
 \end{aligned}
\ee

Then we have $f^2 H^2 D$ type operators mixing into EW operators.

\noindent
\underline{\bf \boldmath $ f^2 H^2 D \to f^4$ (Yukawa)}:

\be \label{eq:class2-f2H2D-to-f4-yukawa1}
\begin{aligned}
\dotwc[(1)]{lq}{ppst} &=
y_t^2 \wc[(1)]{Hl}{pp} V^*_{3s} V_{3t}  \,, \quad
\dotwc[(3)]{lq}{ppst} =
-y_t^2 \wc[(3)]{Hl}{pp} V^*_{3s} V_{3t} \,, \\
\dotwc[]{qe}{stpp} &=
y_t^2 \wc[]{He}{pp} V^*_{3s} V_{3t} \,,\quad
\dotwc[]{ld}{ppst} =
-y_\tau^2 \delta_{p3} \wc[]{Hd}{st} \,, \\
\dotwc[]{ed}{ppst} &=
2 y_\tau^2 \delta_{p3} \wc[]{Hd}{st}\,.
\end{aligned}
\ee
\\

\noindent
\underline{\bf \boldmath $f^4 \to f^2 H^2 D$ (Yukawa)}:

The operator mixing between four-fermion and two-fermion operators is given by
\be \label{eq:class2-hq1hq3hd-f4f2-yuk}
\begin{aligned}
  \dotwc[(1)]{Hq}{st} &=
y^2_t\left(6 V_{3v} V^*_{3w} \wc[(1)]{qq}{stvw}+V_{3v} V^*_{3w} \wc[(1)]{qq}{swvt}
+3 V_{3v} V^*_{3w} \wc[(3)]{qq}{swvt}-6 \wc[(1)]{qu}{st33}\right) \\
&+ y^2_b\left(6\wc[(1)]{qd}{st33}-6\wc[(1)]{qq}{33st}- \wc[(1)]{qq}{s33t}-3 \wc[(3)]{qq}{s33t}\right)\,, \\
\dotwc[(3)]{Hq}{st} &=  -y_t^2 \left(V_{3v} V^*_{3w} \wc[(1)]{qq}{swvt}+
6 V_{3v} V^*_{3w} \wc[(3)]{qq}{stvw}- V_{3v} V^*_{3w} \wc[(3)]{qq}{swvt}\right)
-y_b^2\left(\wc[(1)]{qq}{s33t}-\wc[(3)]{qq}{s33t}+6 \wc[(3)]{qq}{33st}\right) \,, \\
\dotwc[]{Hd}{st} &=6y^2_t\left(V_{3v} V^*_{3w} \wc[(1)]{qd}{vwst}- \wc[(1)]{ud}{33st} \right)+
2 y_b^2 \left(\wc[]{dd}{s33t}+3 \wc[]{dd}{st33}-3  \wc[(1)]{qd}{33st}\right) \,.
\end{aligned}
\ee

\noindent
\underline{\bf \boldmath $ f^2 H^2 D \to f^2 H^2 D$ (Yukawa)}:

The operator mixing between two-fermion operators reads
\be \label{eq:class2-hq1hq3hd-f2f2-yuk}
\begin{aligned}
\dotwc[(1)]{Hq}{st} &=y_t^2\left(2 V^*_{3s} V_{3v} \wc[(1)]{Hq}{vt}+
2 V_{3t} V^*_{3v} \wc[(1)]{Hq}{sv}+{6}\wc[(1)]{Hq}{st}-\frac{9}{2} V^*_{3s} V_{3v} \wc[(3)]{Hq}{vt}-\frac{9}{2} V_{3t} V^*_{3v} \wc[(3)]{Hq}{sv}- \wc[]{Hu}{33} V^*_{3s} V_{3t}\right)\\&
+y_b^2\left(2 \wc[(1)]{Hq}{3t} \delta_{s3}+2 \wc[(1)]{Hq}{s3}\delta_{t3} +\frac{9}{2} \wc[(3)]{Hq}{3t} \delta_{s3}
+\frac{9}{2} \wc[(3)]{Hq}{s3} \delta_{t3}\right)\,,
\\
\dotwc[(3)]{Hq}{st} &=y_t^2\left(-\frac{3}{2}V^*_{3s} V_{3v} \wc[(1)]{Hq}{vt}-\frac{3}{2}V_{3t} V^*_{3v} \wc[(1)]{Hq}{sv}+
V^*_{3s} V_{3v} \wc[(3)]{Hq}{vt}+V_{3t} V^*_{3v} \wc[(3)]{Hq}{sv}+6\wc[(3)]{Hq}{st}
\right)\\&
+y^2_b\left(\frac{3}{2} \wc[(1)]{Hq}{3t} \delta_{s3}+\frac{3}{2} \wc[(1)]{Hq}{s3} \delta_{t3}+ \wc[(3)]{Hq}{3t} \delta_{s3}+\wc[(3)]{Hq}{s3} \delta_{t3}\right), \\
\dotwc[]{Hd}{st} &=6 y_t^2 \wc[]{Hd}{st}-y_b y_t \left(\wc[]{Hud}{3t} \delta_{s3}+\wc[*]{Hud}{3s} \delta_{t3}\right)
+4 y_b^2 \left(\wc[]{Hd}{3t} \delta_{s3}+\wc[]{Hd}{s3} \delta_{t3}\right)\,.
\end{aligned}
\ee
We also have non-fermionic operators mixing into EW scale operators of Class 2.
\\

\noindent
\underline{\bf \boldmath $H^4 D^2 \to f^2 H^2 D $ (Yukawa)}:

\be \label{eq:class2-hq1hq3hd-f0f2-yuk}
\begin{aligned}
\dotwc[(1)]{Hq}{st} &= \frac{1}{2} y_t^2 (\wc[]{H\square}{}+\wc[]{HD}{}) V^*_{3s} V_{3t}\,, \\
\dotwc[(3)]{Hq}{st} &= -\frac{1}{2} \wc[]{H\square}{} y_t^2 V^*_{3s} V_{3t}\,.
\end{aligned}
\ee

Note that the subleading $\mathcal{O}(\hat Y_d \hat Y_d)$ and $\mathcal{O}(\hat Y_d \hat Y_u)$ terms have been dropped relative to $\mathcal{O}(\hat Y_u \hat Y_u)$, unless being the sole contribution.

Finally, the gauge and Yukawa dependent RGEs for the dipole operators relevant mainly for Class 2C are presented.

\noindent
\underline{\bf \boldmath $f^2 X H \to f^2 XH  $ (Gauge)}:

\be
\begin{aligned}
\dotwc[]{dG}{st} &=
-\frac{2}{3} g_1 g_s \wc[]{dB}{st}-\frac{1}{36} \left(31 g_1^2+81 g_2^2+204 g_s^2\right) \wc[]{dG}{st}+6 g_2 g_s \wc[]{dW}{st}\,, \\
\dotwc[]{dW}{st}& = \frac{5}{6} g_1 g_2 \wc[]{dB}{st}+\frac{8}{3} g_2 g_s \wc[]{dG}{st}-\frac{1}{36} \left(31 g_1^2+33 g_2^2-96 g_s^2\right) \wc[]{dW}{st}\,, \\
\dotwc[]{dB}{st} &= -\frac{1}{36} \left(-253 g_1^2+81 g_2^2-96 g_s^2\right) \wc[]{dB}{st}-\frac{8}{9} g_1 g_s \wc[]{dG}{st}+\frac{5}{2} g_1 g_2 \wc[]{dW}{st}\,.
\end{aligned}
\ee

\noindent
\underline{\bf \boldmath $f^2 X H \to f^2 XH  $ (Yukawa)}:

\be \label{eq:class2yukawa-dX-all}
\begin{aligned}
\dotwc[]{dG}{st} &=
\frac{5}{2} y_b^2 \wc[]{dG}{3t} \delta_{s3}+2 y_b^2 \wc[]{dG}{s3} \delta_{t3}-\frac{3}{2} y_t^2 V^*_{3s} V_{3v} \wc[]{dG}{vt}+3 y_t^2 \wc[]{dG}{st}-y_b y_t \wc[]{uG}{s3} \delta_{t3}\,, \\
\dotwc[]{dW}{st}& =\frac{1}{2} y_b^2 \wc[]{dW}{3t} \delta_{s3}+2 y_b^2 \wc[]{dW}{s3} \delta_{t3}+\frac{5}{2} y_t^2 V^*_{3s} V_{3v} \wc[]{dW}{vt}+3 y_t^2 \wc[]{dW}{st}-y_b y_t \wc[]{uW}{s3} \delta_{t3}\,, \\
\dotwc[]{dB}{st} &= \frac{5}{2} y_b^2 \wc[]{dB}{3t} \delta_{s3}+2 y_b^2 \wc[]{dB}{s3} \delta_{t3}-\frac{3}{2} y_t^2 V^*_{3s} V_{3v} \wc[]{dB}{vt}+3 y_t^2 \wc[]{dB}{st}-y_b y_t \wc[]{uB}{s3} \delta_{t3}\,.
\end{aligned}
\ee

\noindent
\underline{\bf \boldmath $f^4  \to f^2 XH  $ (Gauge-Yukawa)}:

\be \label{eq:class2yukawa-dX-all}
\begin{aligned}
\dotwc[]{dG}{st} &= \frac{1}{6} g_s y_t V_{3v} (\wc[(8)]{quqd}{s3vt}-6 \wc[(1)]{quqd}{s3vt})\,, \\
\dotwc[]{dW}{st}& = \frac{1}{12} g_2 y_t V_{3v} (3 \wc[(1)]{quqd}{s3vt}+4 \wc[(8)]{quqd}{s3vt})\,, \\
\dotwc[]{dB}{st} &= -\frac{5}{36} g_1 y_t V_{3v} (3 \wc[(1)]{quqd}{s3vt}+4 \wc[(8)]{quqd}{s3vt})\,.
\end{aligned}
\ee
The $\rho$ and $\eta$ parameter values are collected in Tabs.~\ref{tab:2A1} and \ref{tab:2A2} in App.~\ref{App:etas}.

\subsection{$\text{SU(2)}_L$ Correlations}
\label{CORR2su2}
In this section we briefly discuss the $SU(2)_L$ correlations between various WET WCs due to $\text{SU(2)}_L$ imposed by SMEFT. These relations directly follow from the matching condition between SMEFT and WET, as derived in \cite{Jenkins:2017jig} (up to dim-6) and \cite{Hamoudou:2022tdn} (up to dim-8) level. Nevertheless, we emphasize that even when 
limited to dim-6 and tree-level, validity of $\text{SU(2)}_L$ relations relies on assumptions regarding the presence (or absence) of specific operators in SMEFT. 

Under the assumption of dominance of certain SMEFT Class 2 operators we can work out correlations of Class 2 observables with other observables. For example, assuming that the NP produces only  
a single WC such as such as $\wc[(1)]{lq}{}$, $\wc[]{qe}{}$ or $\wc[]{ledq}{}$, the matching condition \eqref{class2A-match} can be inverted. For the $b\to s$ case, we have

\be
\begin{aligned}
\wc[(1)]{lq}{2223} & =\lambda^2 \wcL[V,LL]{eu}{2222}  + \lambda^3( \wcL[V,LL]{eu}{2212} + \wcL[V,LL]{eu}{2221} ) +\lambda^4 \wcL[V,LL]{eu}{2211} \,, \\
\wc[]{qe}{2322} & =\lambda^2 \wcL[V,LR]{ue}{2222} + \lambda^3( \wcL[V,LR]{ue}{1222} + \wcL[V,LR]{ue}{2122} ) 
+\lambda^4 \wcL[V,LR]{ue}{1122} \,, \\
\wc[]{ledq}{2223} & =\lambda^2 \wcL[S,RL]{\nu e du}{2222}  + \lambda^3 \wcL[S,RL]{\nu e du}{2221} .
\end{aligned}
\ee
The WCs on the l.h.s. can be directly fixed by measurements of the observables in Class 2. On the other hand the WET WCs on the r.h.s. contribute to processes beyond Class 2, leading to correlations among them.

Similarly, for the $s\to d$ operators, we find
\be
\begin{aligned}
\wc[(1)]{lq}{2212} & =\wcL[V,LL]{eu}{2212} +  \lambda (\wcL[V,LL]{eu}{2211}-\wcL[V,LL]{eu}{2222} )
- \lambda^2 \wcL[V,LL]{eu}{2221} \,, \\
\wc[]{qe}{1222} & =\wcL[V,LR]{ue}{1222} + \lambda^2
( \wcL[V,LR]{ue}{1122}  -  \wcL[V,LR]{ue}{2222})
-\lambda^2   \wcL[V,LR]{ue}{2122}  \,, \\
\wc[]{ledq}{2212} & = \wcL[S,RL]{\nu e du}{2212} 
+\lambda  \wcL[S,RL]{\nu e du}{2211} .
\end{aligned}
\ee
These relations can be used to see the relationship between a number of low energy processes associated with WET operators on the r.h.s. with the $b\to s $ and $s\to d$ processes in Class 2, 
assuming the SMEFT to be the correct theory above the weak scale. It is worth to mention that such correlations can be significantly affected by the RG running discussed in this class 
or the dim-8 SMEFT effects on the WET\cite{Hamoudou:2022tdn}.

A summary of tree-tree $\text{SU(2)}_L$ correlations is given in Tab.~\ref{tab:class2-su2}. Depending on which WCs in the first column are non-zero, different  
patterns of correlations among the WET WCs will emerge.
\begin{table}[tbp]
\begin{center}
\renewcommand*{\arraystretch}{1.0}
\resizebox{0.9\textwidth}{!}{
\begin{tabular}{ |c|c|c|c| }
\hline
\multicolumn{4}{|c|}{$\text{SU}(2)_L$ Correlations for Class 2} \\
\hline
Class 2 WCs&  correlated WET WCs   & processes & classes \\
\hline
$\wc[(1)]{lq}{ppji}, \wc[(3)]{lq}{ppji}$ & $\wcL[V,LL]{e u}{ppmn}=V_{mj}(\wc[(1)]{lq}{ppji}- \wc[(3)]{lq}{ppji}) {V^*_{ni}}$ & $pp\to \ell^+\ell^-$  & 10\\
&& $c\to u \ell^+ \ell^-$ & \\
$\wc[(1)]{lq}{ppji}, \wc[(3)]{lq}{ppji}$ & $\wcL[V,LL]{\nu u}{ppmn}=V_{mj}(\wc[(1)]{lq}{ppji} + \wc[(3)]{lq}{ppji}){V^*_{ni}}$ &
$p p \to \nu \bar \nu$  & 10 \\
&& $c\to u \nu \bar \nu$  & \\
{$\wc[(3)]{lq}{ppji}$} & {$\wcL[V,LL]{\nu edu}{ppmn}=2V_{mj}\wc[(3)]{lq}{ppji}V^*_{ni}$} &
{$b\to c\ell\bar\nu$}  & {8} \\
$\wc[]{qe}{jipp}$ & $ \wcL[V,LR]{u e}{mnpp} = V_{mj} \wc[]{qe}{jipp} {V^*_{ni}} $ & $pp\to \ell^+ \ell^-$  & 10\\ && $c\to u \ell^+ \ell^-$   &\\
$\wc[]{ledq}{ppji}$   & $\wcL[S,RL]{\nu e du}{ppjn} = \wc[]{ledq}{ppji} {V^*_{ni}} $ & $b\to c \ell \bar \nu$   & {8} \\
$\wc[(1)]{Hq}{ji}, \wc[(3)]{Hq}{ji}$ & $\wcL[V,LL]{\nu u}{ppmn} =V_{mj} (\wc[(1)]{Hq}{ji} - \wc[(3)]{Hq}{ji}){V^*_{ni}\delta_{pp}}$ & $c\to u \nu \bar \nu$   & \\
$\wc[(1)]{Hq}{ji}, \wc[(3)]{Hq}{ji}$ & $\wcL[V,LL]{e u}{ppmn} ={\zeta_1} V_{mj} (\wc[(1)]{Hq}{ji} {-} \wc[(3)]{Hq}{ji}){V^*_{ni}\delta_{pp}}$ & $c \to u \ell^+ \ell^-$   & \\
&& $p p \to \ell^+ \ell^-$ & 10, 4\\
&& $Z \to q  \bar q$ & 10, 4\\
\hline
\end{tabular}
}
\caption{\small Summary of $\text{SU(2)}_L$ correlations for Class 2 operators at tree-level. The first column shows Class 2 operators at $\muEW$, the second column lists the other WET operators {(beyond Class 2)} generated by the operators in the first column, and the third column shows processes that are generated by the WET operators in the second column. Finally in the last column classes that are correlated to Class 2 are collected. {Recall that} the SMEFT down-basis was adopted, and the PMNS matrix is set to unity.} \label{tab:class2-su2}
\end{center}
\end{table}
{Moreover}, two-quark WCs such as $\wc[(1)]{Hq}{{ji}}$, $\wc[(3)]{Hq}{{ji}}$, and $\wc[]{Hd}{{ji}}$ match onto many other WET operators after integrating out the $Z$-boson, such as $\wcL[V,LL]{dd}{}$, $\wcL[V1,LL]{ud}{}$, $\wcL[V8,LL]{ud}{}$, $\wcL[V,RR]{dd}{}$, $\wcL[V1,RR]{ud}{}$, $\wcL[V8,RR]{ud}{}$, $\wcL[V1,LR]{dd}{}$, implying many more correlations with hadronic $\Delta F= 1$ processes. We will see this explicitly in the {context of } Class 3.

\subsection{{RGE Correlations}}\label{CORR2rge}
Let us consider a few examples of correlations between various processes that are induced both by the RG evolution presented above and by $\text{SU(2)}_L$ gauge symmetry. In this class, dominated by semi-leptonic operators, there are, as easily deduced from the charts, correlations not only among the semi-leptonic processes of Class 2 but also correlations with purely leptonic (Class 5) and non-leptonic (Class 3) processes among others.

{\bf Example 1}

At the high NP scale $\Lambda$ only semi-leptonic operators are present as found for example in leptoquark models. RG evolution in the presence of Yukawa couplings and gauge interactions has then an impact on lepton-flavour violating lepton decays (Class 5) and implies modifications of SM couplings. This case is important in the context of $B$ physics anomalies, which require enhanced contributions from semi-leptonic operators. These can in turn have an impact on lepton decays, implying enhancements of the latter. The experimental bounds on lepton decays can in turn bound the required enhancements of semi-leptonic $B$ decays as stressed in particular in \cite{Feruglio:2016gvd,Feruglio:2017rjo}.

For instance LFU breaking effects in $\tau \to \ell \bar\nu \nu$ (with $\ell_{1,2}=e,\mu$) are described by the observables
\begin{align}
\!\!\!\!\!R^{\tau/\ell_{1,2}}_\tau = \frac{\mathcal{B}(\tau \to \ell_{2,1} \nu\bar\nu)_{\rm exp}/\mathcal{B}(\tau \to \ell_{2,1} \nu\bar\nu)_{\rm SM}}{\mathcal{B}(\mu \to e \nu\bar\nu)_{\rm exp}/\mathcal{B}(\mu \to e \nu\bar\nu)_{\rm SM}} \,,
\end{align}
and are experimentally tested at the few permille level~\cite{Pich:2013lsa}
\begin{align}
\!\!\!R^{\tau/\mu}_\tau = 1.0022 \pm 0.0030 \,,~~ R^{\tau/e}_\tau = 1.0060 \pm 0.0030 \,.
\label{eq:tau_LFU_data}
\end{align}

Similarly, modifications of the leptonic $Z$ couplings are constrained by LEP measurements of the $Z$ decay width as well as the left-right and forward-backward asymmetries. According to the PDG one finds for the lepton-universal couplings
\begin{align}
\frac{v_\tau}{v_e} = 0.959\; (29)\,, \qquad
\frac{a_\tau}{a_e} = 1.0019\; (15)\,,
\label{eq:zpole_pdg}
\end{align}
where $v_\ell$ and $a_\ell$ are the vector and axial-vector $Z$-couplings to leptons, respectively. These two examples show how careful one has to be in order not to violate the existing bounds. Further examples can be found in \cite{Feruglio:2016gvd,Feruglio:2017rjo}.

To demonstrate the occurrence of these correlations we consider semi-leptonic $(\bar LL)(\bar LL)$ operators, so that the operators $\Op[(1,3)]{\ell q}$ are the driving force. The complex-valued coefficients of these operators are
\be
\label{eq:SMEFT-wilson-coeffsIV}
{\wc[(1)]{\ell q}{abij}, \qquad \wc[(3)]{\ell q}{abij},}
\ee
where lepton-flavour indices are denoted by $ab$ and quark-flavour indices by $ij$.

Keeping only $Y_u$ effects we find (summation over repeated indices $ww$ is understood)

\begin{align}
\label{eq:ADM-Yuk-lq1lq1}
\dotwc[(1)]{l q}{abij} & 
= \frac{1}{2}y_t^2 V^*_{3i}V_{3w} \wc[(1)]{l q}{abwj}+ \frac{1}{2}
y_t^2 V^*_{3w}V_{3j}
\wc[(1)]{l q}{abiw}\,,
\\
\label{eq:ADM-Yuk-lq3lq3}
\dotwc[(3)]{l q}{abij} & 
= \frac{1}{2} y_t^2 V^*_{3i}V_{3w} \wc[(3)]{l q}{abwj}+ \frac{1}{2}
y_t^2 V^*_{3w}V_{3j}\wc[(3)]{l q}{abiw}\,,
\end{align}

\noindent
which shows that these two WCs in this approximation evolve independently of each other. Yet, when gauge couplings are included they start to mix 
\begin{align}
\label{eq:ADM-Yuk-lq1lq1gauge}
\dotwc[(1)]{l q}{kkij} & 
=  g_1^2 ( \frac{2}{3}\wc[(1)]{l q}{wwij}\delta_{kk}- \wc[(1)]{l q}{kkij}) +9 g_2^2\wc[(3)]{l q}{kkij}\,,
\\
\label{eq:ADM-Yuk-lq3lq3gauge}
\dotwc[(3)]{l q}{kkij} & 
= g_2^2  (\frac{2}{3}\wc[(3)]{l q}{wwij}\delta_{kk}+3  \wc[(1)]{l q}{kkij}) -6(g_2^2+\frac{1}{6}g_1^2)\wc[(3)]{l q}{kkij}\,.
\end{align}

However, what is more important is the generation of purely leptonic operators via electroweak interactions:\footnote{Note that now there are two new lepton indices {\em cd} and the summation is only over quark indices.} 
\begin{align}\label{purelep}
 \dotwc[]{\ell \ell}{abcd}& =-\frac{1}{3} g_1^2 ( \wc[(1)]{\ell q}{abww}\delta_{cd}+\wc[(1)]{\ell q}{cdww}\delta_{ab} )- g_2^2 (\wc[(3)]{\ell q}{abww}\delta_{cd}+\wc[(3)]{\ell q}{cdww}\delta_{ab}
-2 \wc[(3)]{\ell q}{cbww}\delta_{ad}-2\wc[(3)]{\ell q}{adww}\delta_{bc} )\,,
\end{align}
\noindent
which can contribute to purely leptonic lepton decays, thereby putting significant bounds on the coefficients of semi-leptonic operators as stressed in particular in \cite{Feruglio:2016gvd, Feruglio:2017rjo}. We refer to these papers for phenomenological implications of these equations.

{\bf Example 2}

Again, as in the first example, at the high scale only semi-leptonic operators are present but this time we study the impact of RG evolution on $\Delta F=1$ non-leptonic decays, like $K_L\to\pi\pi$ and the ratio $\epe$ that belong to Class 3. This case is again interesting in the context of leptoquark models with more details found in \cite{Bobeth:2017ecx}. In what follows we list RG equations which govern the generation of non-leptonic (NL)-$f^4$ coefficients from semi-leptonic (SL)-$f^4$ ones. They can be derived from \cite{Alonso:2013hga} and have been presented in detail in \cite{Bobeth:2017ecx}. Only electroweak interactions are involved here. Adjusting these results to our notation we find \\

\noindent
{\boldmath NL-$f^4$} $(\overline{L}L)(\overline{L}L)$:
\begin{align}
\label{eq:4-quark-RGE-qq1}
\dotwc[(1)]{qq}{ijkk} & 
= - \frac{1}{9} g^2_1 (\wc[(1)]{l q}{wwij} + \wc{qe}{ijww}) \delta_{kk}\,, \qquad
\dotwc[(3)]{qq}{ijkk} = \frac{1}{3} g^2_2 \wc[(3)]{lq}{wwij} \delta_{kk}\,.
\end{align}
{\boldmath NL-$f^4$} $(\overline{R}R)(\overline{R}R)$:
\begin{align}
\dotwc[]{uu}{ijkk} & 
= - \frac{4}{9} g^2_1 (\wc{eu}{wwij}+\wc{l u}{wwij})\delta_{kk}\,,\qquad
\dotwc[]{dd}{ijkk} =  \frac{2}{9} g^2_1 (\wc{ed}{wwij}+\wc{ld}{wwij}) \delta_{kk}\,,
\\
\label{eq:4-quark-RGE-ud1}
\dotwc[(1)]{ud}{ijkk} & 
= \frac{4}{9} g^2_1 (\wc{eu}{wwij} + \wc{l u}{wwij}) \delta_{kk}\,,\qquad
\dotwc[(1)]{ud}{kkij} = - \frac{8}{9} g^2_1 (\wc{l d}{wwij}+\wc{ed}{wwij})\delta_{kk}\,.
\end{align}
{\boldmath NL-$f^4$} $(\overline{L}L)(\overline{R}R)$:
\begin{align}
\label{eq:4-quark-RGE-qu1}
\dotwc[(1)]{qu}{ijkk} & 
= - \frac{8}{9} g^2_1 (\wc[(1)]{l q}{wwij}+\wc{qe}{ijww}) \delta_{kk}\,,\qquad
\dotwc[(1)]{qu}{kkij} = - \frac{2}{9} g^2_1 (\wc{l u}{wwij}+\wc{eu}{wwij}) \delta_{kk}\,,
\\
\label{eq:4-quark-RGE-qd1}
\dotwc[(1)]{qd}{ijkk} & 
= \frac{4}{9} g^2_1 ( \wc[(1)]{l q}{wwij} +\wc{qe}{ijww}) \delta_{kk}\,,\qquad
\dotwc[(1)]{qd}{kkij} = - \frac{2}{9} g^2_1 ( \wc{l d}{wwij} +\wc{ed}{wwij} )\delta_{kk}\,.
\end{align}
\noindent
Finally for all other NL-$f^4$ operators we have
\begin{equation}
\begin{aligned}
\dotwc[(8)]{ud}{ijkk} & = \dotwc[(8)]{ud}{kkij} = 0\,, &
\dotwc[(8)]{qu}{ijkk} & =\dotwc[(8)]{qu}{kkij} = 0\,, &
\dotwc[(1)]{quqd}{ijkk} & =  \dotwc[(1)]{quqd}{kkij}= 0\,, 
\\
\dotwc[(8)]{qd}{ijkk} & =\dotwc[(8)]{qd}{kkij}= 0\,, &
\dotwc[(8)]{quqd}{ijkk} & = \dotwc[(8)]{quqd}{kkij}= 0\,.
\end{aligned}
\end{equation}

We observe that the SM gauge-mixing of SL-$f^4$ into NL-$f^4$ operators within the SMEFT generates in 1stLLA only $(\bar{L}L)(\bar{L}L)$, $(\bar{L}L)(\bar{R}R)$ and $(\bar{R}R)(\bar{R}R)$ NL-$f^4$ operators from the corresponding semi-leptonic classes.

Further details on this scenario can be found in \cite{Bobeth:2017ecx}, where the ratio $\epe$ was studied in leptoquark models. The results of this analysis are summarized in Section~14.7 in \cite{Buras:2020xsm} and in Sec.~\ref{class3loopcorr} in the context of Class 3.

{\bf Example 3}

In this example we consider only operators relevant for $b\to s\nu\bar\nu$ and $b\to s\ell^+\ell^-$ transitions with the goal to identify some correlations between them that follow from SM gauge invariance. Indeed the $b\to s\nu\bar\nu$ transition is closely related to the $b\to s\ell^+\ell^-$ transition because the neutrinos and left-handed charged leptons are related by ${\rm SU(2)}_L$ symmetry. 

The same comments apply to the $d\to s\nu\bar\nu$ and $d\to s\ell^+\ell^-$ transitions. While, the correlations between these transitions imply stringent constraints on parameters of specific NP scenarios, significant long-distance (LD) uncertainties in $d\to s\ell^+\ell^-$ and no real plans for measuring $K_L\to\pi^0 \ell^+\ell^-$ in the near future make these connections less interesting than the $B$ physics case at present. Yet, there exist dedicated analyses of such correlations in the literature and we will refer to them as we proceed.

The implication of SM gauge invariance for flavour observables have been discussed in many papers, in particular in \cite{Fajfer:2012jt,Alonso:2014csa,Glashow:2014iga,Bhattacharya:2014wla,Buras:2014fpa,Alonso:2015sja,Calibbi:2015kma,Buras:2024mnq}. Here, as an example, we will just inspect the general formulae in \eqref{eq:SMEFT2WET_C9_C10}, which allow to study correlations on the one hand between $b\to s\nu\bar\nu$ and $b\to s\ell^+\ell^-$ transitions and on the other hand between $d\to s\nu\bar\nu$ and $d\to s\ell^+\ell^-$ transitions.

Inspecting equations in \eqref{eq:SMEFT2WET_C9_C10} we observe that the number of operators in the SMEFT is in general larger than in the low-energy effective Hamiltonian. Hence, for a completely model-independent basis no general correlations can be derived. Indeed, from \eqref{eq:SMEFT2WET_C9_C10} it is clear that in complete generality, the size of NP effects in $b\to s\nu\bar\nu$ is not constrained by the $b\to s\ell^+\ell^-$ measurements. First, the decays with charged leptons are only sensitive to the combination $\Wc[(1)]{\ell q} +\Wc[(3)]{\ell q}$, while the decays with neutrinos in the final state probe $\Wc[(1)]{\ell q} -\Wc[(3)]{\ell q}$.
Second, even if the WCs $\Wc[(3)]{\ell q}$ vanish, cancellations between the operators with left- and right-handed charged leptons can lead to small deviations from the SM in $b\to s\ell^+\ell^-$ transitions, even when large effects are present in $b\to s\nu\bar\nu$. However, in concrete NP models, often only a subset of operators is relevant and in these cases correlations characteristic for these NP scenarios are obtained. We remark that the present bounds on $b\to s\nu\bar\nu$ transitions can in certain models be problematic for the explanation of the anomalies in $b\to s\ell^+\ell^-$ measurements.

A solution to the latter problem is the equality $\Wc[(1)]{\ell q}=\Wc[(3)]{\ell q}$ which removes the contributions not only to the $b\to s\nu\bar\nu$ transition but if valid for all down-quark flavours, also to $d\to s\nu\bar\nu$ transitions, that is $\kpn$ and $\klpn$. However, as emphasized in \cite{Feruglio:2016gvd, Feruglio:2017rjo}, this equality can be valid only at one scale. RG evolution from the NP to the EW scale breaks this relation. See \cite{Bobeth:2017ecx} for an explicit analysis of such RG effects in the context of leptoquark models.

While such explicit models are considered in PART III of our review, here we just illustrate general correlations on two examples presented in \cite{Buras:2014fpa}. We consider first the general case of $Z^\prime$ models in which a single $Z^\prime$ gauge boson with left-handed and right-handed couplings dominates the scene. In this case one finds at the NP scale $\Wc[(3)]{\ell q}=0$, together with the relations
\be\label{CORRZprime}
C_{L,{\rm NP}}^{baji}=\frac{C_{9,{\rm NP}}^{baji}- C_{10,{\rm NP}}^{baji}}{2}\,, \qquad
C_{R,{\rm NP}}^{baji}=\frac{C_{9',{\rm NP}}^{baji}- C_{10',{\rm NP}}^{baji}}{2}\,.
\ee

However, if NP contributions to the processes considered are fully dominated by induced FCNC couplings of the SM $Z$ boson -- as is for instance the case in the MSSM, in models with partial compositeness and models with vector-like quarks -- one finds
\be\label{CORRZ}
C_{L,{\rm NP}}^{baji}=C_{10,{\rm NP}}^{baji}\,,\quad C_{9,{\rm NP}}^{baji}=-\zeta_2 C_{10,{\rm NP}}^{baji}\,,\quad  C_{R,{\rm NP}}^{baji}= C_{10',{\rm NP}}^{baji}\,,\quad
C_{9',{\rm NP}}^{baji}=-\zeta_2 C_{10',{\rm NP}}^{baji}\,,
\ee
\noindent
where $\zeta_2 = 1- 4 s_w^2\approx 0.08$ is the accidentally small vector coupling of the $Z$ to charged leptons. Evidently the correlations in \eqref{CORRZprime} and \eqref{CORRZ} differ from each other. In particular the sign in front of $C_{10,{\rm NP}}^{baji}$ in $C_{L,{\rm NP}}^{baji}$ is different and similar for the right-handed coefficients. In this manner the $Z^\prime$ and $Z$ scenarios can be distinguished from each other. The plots in \cite{Buras:2014fpa} illustrate these differences.

Finally, if the presence of a $Z^\prime$ induces FCNC couplings of the $Z$ through $Z-Z^\prime$ mixing, the relations given above are modified and depend on the size of the $Z-Z^\prime$ mixing. This mixing is clearly model-dependent and the resulting correlations can vary from model to model. They were illustrated in the context of 331 models in \cite{Buras:2014yna}.

Of particular interest are scenarios which imply correlations between $b\to s\nu\bar\nu$ transitions studied at BELLE II and $ d\to s \nu\bar\nu$ investigated at NA62 and KOTO. In Minimal Flavour Violation (MFV) scenarios and simplified NP scenarios these correlations are very stringent \cite{Buras:2001af,Buras:2015yca}. Involving two meson systems they test the flavour structure of different models. Several recent analyses of $B\to K(K^*)\nu\bar\nu$ decays \cite{Bause:2021cna,He:2021yoz,Bause:2022rrs,Becirevic:2023aov,Bause:2023mfe,Allwicher:2023xba,Dreiner:2023cms,He:2023bnk,Hou:2024vyw,Buras:2024mnq,Chen:2025npb}, motivated by the last Belle~II data~\cite{Belle-II:2023esi}, studied such correlations.

{\boldmath
\section{Non-Leptonic Decays of Mesons (Class 3)}
 \label{class3}
}
We next present Class 3 in which the non-leptonic decays $K\to\pi\pi$, $B\to\pi\pi$, $B\to\pi K$, $B\to KK$, $D\to\pi\pi$, $D\to\pi K$ and $D\to KK$ played an important role for decades. While not as suppressed as the decays of Class 2, they are subject to significant non-perturbative QCD uncertainties so that the tests of the SM and of its extensions through observables of this class require not only precise measurements but also control of the non-perturbative uncertainties. Details on these observables within the SM can be found in Chapters 7, 8 and 10 in \cite{Buras:2020xsm}. In the following we discuss the CWET and WET Lagrangians that govern these processes and subsequently we will devote this section to their description within the SMEFT.

\subsection{CWET and WET Operators for Class 3}

We recall first the well known basis of operators within the SM \cite{Buchalla:1995vs} here specified to describe the $B_s$ system:

{\bf Current--Current:}
\begin{equation}\label{O1} 
Q_1 = (\bar c_{\alpha} b_{\beta})_{V-A}\;(\bar s_{\beta} c_{\alpha})_{V-A}
\,,~~~~~Q_2 = (\bar c b)_{V-A}\;(\bar s c)_{V-A} \,.
\end{equation}

{\bf QCD--Penguins:}
\begin{equation}\label{O2}
Q_3 = (\bar s b)_{V-A}\sum_{q=u,d,s,c,b}(\bar qq)_{V-A}\,,~~~~~   
Q_4 = (\bar s_{\alpha} b_{\beta})_{V-A}\sum_{q=u,d,s,c,b}(\bar q_{\beta} 
      q_{\alpha})_{V-A}\,,
\end{equation}
\begin{equation}\label{O3}
 Q_5 = (\bar s b)_{V-A} \sum_{q=u,d,s,c,b}(\bar qq)_{V+A}\,,~~~~  
 Q_6 = (\bar s_{\alpha} b_{\beta})_{V-A}\sum_{q=u,d,s,c,b}
       (\bar q_{\beta} q_{\alpha})_{V+A}\,.
\end{equation}

{\bf Electroweak Penguins:}
\begin{equation}\label{O4} 
Q_7 = \frac{3}{2}\;(\bar s b)_{V-A}\sum_{q=u,d,s,c,b}e_q\;(\bar qq)_{V+A} 
\,,~~~~ Q_8 = \frac{3}{2}\;(\bar s_{\alpha} b_{\beta})_{V-A}\sum_{q=u,d,s,c,b}e_q
        (\bar q_{\beta} q_{\alpha})_{V+A}\,,
\end{equation}
\begin{equation}\label{O5} 
 Q_9 = \frac{3}{2}\;(\bar s b)_{V-A}\sum_{q=u,d,s,c,b}e_q(\bar q q)_{V-A}
\,,~~~~Q_{10} =\frac{3}{2}\;
(\bar s_{\alpha} b_{\beta})_{V-A}\sum_{q=u,d,s,c,b}e_q\;
       (\bar q_{\beta}q_{\alpha})_{V-A}\,. 
\end{equation}
Here, $\alpha,\beta$ denote colours and $e_q$ denote the electric quark charges reflecting the electroweak origin of $Q_7,\ldots,Q_{10}$. Finally, $(\bar c b)_{V-A}\equiv \bar c_\alpha\gamma_\mu(1-\gamma_5) b_\alpha$. 

Beyond the SM the number of operators increases to such an extent that it is not possible to present them all in this review in explicit terms. The first studies including NLO corrections go back to \cite{Ciuchini:1997bw,Buras:2000if}. In particular in \cite{Buras:2000if} the BMU basis was introduced that we already encountered in Class 1. But the most general non-leptonic $\Delta F=1$ WET at the NLO in QCD, involving both BMU and JMS bases was presented only recently \cite{Aebischer:2021raf}. In light of a recent revisited study of two-loop ADMs \cite{Morell:2024aml} four entries related to the left-left operator $Q_{11}$ were modified relative to \cite{Buras:2000if}, but this modification is expected to have a marginal effect on phenomenology of non-leptonic decays in view of potential long-distance uncertainties. In the following we summarise what can be found in the latter papers and related ones.
The complete list of BMU operators that capture most generic BSM effects is given by 

\begin{center}
\textrm{\bf CWET-3}
\end{center}
\be
\begin{aligned}
\text{VLL} & :
\big\{ \OpL{1},\; \OpL{2},\;
\OpL{3},\; \OpL{4},\; \OpL{9},\; \OpL{10},\;
\OpL{11},\; \OpL{14}
\big\}\,,  \\
\text{VLR} & :
\big\{ \OpL{5},\; \OpL{6},\; \OpL{7},\; \OpL{8},\;
\OpL{12},\; \OpL{13}, \; \OpL{15}, \ldots , \OpL{24}
\big\}\,,  \\
\text{SRR} & :
\big\{ \OpL{25}, \ldots , \OpL{40}
\big\}\,.
\end{aligned}
\ee
While in the SM only $\OpL{2}$ is generated at tree-level, most of the operators can be generated at tree-level beyond the SM, in particular in models containing $Z^\prime$ and $G^\prime$ gauge bosons.\footnote{In the SM $\OpL{1}$ and $\OpL{3}-\OpL{10}$ are generated at loop level via $W^\pm$ boson and gluon exchanges.}
In the chirality-flipped sectors (VRR, VRL and SLL), the ordering is analogous, up to shifting the operator subscripts ($Q_i \to Q_{40 + i})$. The explicit form of the BSM operators $Q_{11}-Q_{40}$ can be found in App.~A2 of \cite{Aebischer:2021raf}. 

In general, we can have different kinds of quark-flavour violating transitions in Class 3. Possible quark level transitions are, for $i \ne j \in \{ d,s,b\}$, 
\be
\begin{aligned}
d_i &\to d_j d_k \bar  d_k\,,  \quad (i \ne k, j \ne k) \,,\\
d_i &\to d_j d_i \bar  d_i \,,\\
d_i &\to d_j d_j  \bar  d_j \,,\\
d_i &\to d_j u_k \bar u_k  \,, \quad    ( {u_k} \in \{ u,c  \}   )\,.
\end{aligned}
\ee
Few examples of corresponding hadron level transitions are 
$B \to K \pi$, $B \to \pi \pi$ and $K \to \pi \pi$.
The complete list of 80 WET operators to describe such processes is given by:\footnote{Note that {sometime in} this review the index $k$ has been kept to represent the repeated index summed over 1-3, but in Class 3 $k$ is used as a fixed index that depends on the process considered.}
\begin{center}
\textrm{\bf WET-3}
\end{center}
\be
\begin{aligned}
\text{VLL} & :
\big\{ \opL[V1,LL]{ud}{11ji},\; \opL[V8,LL]{ud}{11ji},\;
\opL[V1,LL]{ud}{22ji},\; \opL[V8,LL]{ud}{22ji},
\opL[V,LL]{dd}{jikk}, \; \opL[V,LL]{dd}{jkki},\;
\opL[V,LL]{dd}{jiii}, \; \opL[V,LL]{dd}{jijj}
\big\}\,, \\
\text{VLR} & :
\big\{ \opL[V1,LR]{du}{ji11},\; \opL[V8,LR]{du}{ji11},\;
\opL[V1,LR]{du}{ji22},\; \opL[V8,LR]{du}{ji22},
\opL[V1,LR]{dd}{jikk}, \; \opL[V8,LR]{dd}{jikk},\;
\opL[V1,LR]{dd}{jiii}, \; \opL[V8,LR]{dd}{jiii},\; \nn \\
&\phantom{:\big\{\;}    
\opL[V1,LR]{dd}{jijj}, \; \opL[V8,LR]{dd}{jijj},
\opL[V1,LR]{uddu}{1ji1}^\dagger,\; \opL[V8,LR]{uddu}{1ji1}^\dagger, \;
\opL[V1,LR]{uddu}{2ji2}^\dagger,\; \opL[V8,LR]{uddu}{2ji2}^\dagger,
\opL[V1,LR]{dd}{jkki}, \; \opL[V8,LR]{dd}{jkki}
\big\}\,, \nn \\
\text{SRR} & :
\big\{ \opL[S1,RR]{dd}{jiii},\; \opL[S8,RR]{dd}{jiii},\;
\opL[S1,RR]{dd}{jijj},\; \opL[S8,RR]{dd}{jijj},\;
\opL[S1,RR]{ud}{11ji},\;   \opL[S8,RR]{ud}{11ji},\;
\opL[S1,RR]{uddu}{1ij1},\; \opL[S8,RR]{uddu}{1ij1}, \notag
\\ & \phantom{: \big\{ \;}
\opL[S1,RR]{ud}{22ji},\;   \opL[S8,RR]{ud}{22ji},\;
\opL[S1,RR]{uddu}{2ij2},\; \opL[S8,RR]{uddu}{2ij2},
\opL[S1,RR]{dd}{jikk},\; \opL[S8,RR]{dd}{jikk},\;
\opL[S1,RR]{dd}{jkki},\; \opL[S8,RR]{dd}{jkki}
\big\}\,, \\
\text{VRR} & :
\big\{ \opL[V1,RR]{ud}{11ji},\; \opL[V8,RR]{ud}{11ji},\;
\opL[V1,RR]{ud}{22ji},\; \opL[V8,RR]{ud}{22ji}, 
\opL[V,RR]{dd}{jikk}, \; \opL[V,RR]{dd}{jkki},\;
\opL[V,RR]{dd}{jiii}, \; \opL[V,RR]{dd}{jijj}
\big\}\,,
\\
\text{VRL} & :
\big\{  \opL[V1,LR]{ud}{11ji},\; \opL[V8,LR]{ud}{11ji},\;
\opL[V1,LR]{ud}{22ji},\; \opL[V8,LR]{ud}{22ji}, \;
\opL[V1,LR]{dd}{kkji}, \; \opL[V8,LR]{dd}{kkji},\;
\opL[V1,LR]{dd}{iiji}, \; \opL[V8,LR]{dd}{iiji},
\\
&   \opL[V1,LR]{dd}{jjji}, \; \opL[V8,LR]{dd}{jjji}, \;
\opL[V1,LR]{uddu}{1ij1},\; \opL[V8,LR]{uddu}{1ij1}, \;
\opL[V1,LR]{uddu}{2ij2},\; \opL[V8,LR]{uddu}{2ij2}, \;
\opL[V1,LR]{dd}{kijk}, \; \opL[V8,LR]{dd}{kijk} 
\big\}\,.
\end{aligned}
\ee
Note that the SLL sector operators are simply hermitian conjugates of the SRR operators. It is evident that the number of operators in this class is very large and we are only in the position to give the reader a general look at the operator structure relevant for these processes and list five papers which taken together summarize the present status of Class 3. These are as follows.
\begin{itemize}
\item
The most general WET analysis of non-leptonic decays of mesons including LO and NLO QCD corrections was presented in \cite{Aebischer:2021raf}. Both the CWET basis (the BMU basis) and JMS basis were discussed and relations between them were presented. A brief look at this paper shows that in view of large ADMs the RG analysis is rather involved. In particular the WCs in the JMS basis are expressed in terms of the BMU ones as
\be
\vec C^T_{\rm JMS}=\vec C^T_{\rm BMU}\hat R\,,
\ee
with the matrix $\hat R$ given in Section 3 of \cite{Aebischer:2021raf}.
\item
The application of these results for the ratio $\varepsilon^\prime/\varepsilon$ in  $K_L \to \pi \pi$ decays was presented in \cite{Aebischer:2021hws} where the complete list of WET operators relevant for this ratio can be found.
\item
As far as the SMEFT is concerned only the LO QCD analysis including also top-Yukawa effects is known \cite{Aebischer:2018csl}. In this paper, devoted to the ratio $\epe$, the relevant operators in the CWET, WET and SMEFT bases are listed as well as relations between them. Also RG evolution is discussed in some detail. Moreover, various NP scenarios representing a particular set of operators are discussed. We will summarise the main result of this analysis at the end of this section.
\item
Finally, a detailed analysis of $\epe$, $\Delta M_K$ and $K\to\pi\nu\bar\nu$ in $Z^\prime$ models within the SMEFT was presented in \cite{Aebischer:2020mkv} and recently in \cite{Aebischer:2023mbz}.
\item
It should be emphasized that to be able to perform these analyses, the hadronic matrix elements of the involved BSM operators have to be known. Until now they are only known within the DQCD approach \cite{Aebischer:2018rrz}.
\end{itemize}
Examples of the charts involving $\Delta F=1$ non-leptonic operators and also their impact on $\Delta F=2$ operators can be found in Figs.~1-3 in \cite{Aebischer:2020mkv}. The phenomenology in the latter paper concentrates on the Kaon system within $Z^\prime$ models. Finally, see \cite{Datta:2024zrl} for a LO analysis of $B\to K\pi$ within the SMEFT, where the emphasis is given to the correlations between semileptonic $B$-decays due to SMEFT RG running effects. To our knowledge no other SMEFT analysis of non-leptonic $B$ decays exists in the literature.

\subsection{SMEFT Operators for Class 3 at $\muEW$}
The connection between the hadronic processes of this class and the UV physics can be achieved through transforming the CWET operators in the BMU basis to the JMS basis, which in turn can be matched onto the Warsaw basis. The transformation matrix for the former can be found in Eq.~(17) of \cite{Aebischer:2023mbz}. The SMEFT operators for the non-leptonic FCNC $\Delta F=1$ $K$ and $B$ decays that match onto CWET and JMS operators at the EW scale are \cite{Aebischer:2018csl} 
\begin{center}
\textrm{\bf SMEFT-Tree 3}
\end{center}
\be
\begin{aligned} \label{eq:smeft-class3-tree}
&
\ops[(1)]{Hq}{}  \,,  \quad 
\ops[(3) ]{Hq}{}   \,,  \quad
\ops[]{Hd}{}   \,, \quad  
\ops[]{Hud}{} \,,  \quad
\ops[(1)]{q q}{}  \,, \quad
\ops[(3)]{q q}{}  \,,  \quad
\ops[]{dd}{}      \,,  \\
&
\ops[(1)]{qu}{}  \,, \quad
\ops[(8)]{qu}{}  \,,  \quad
\ops[(1) ]{qd}{}   \,,  \quad 
\ops[(8)]{qd}{}  \,,  \quad
\ops[(1) ]{quqd}{}  \,,  \quad 
\ops[(8)]{quqd}{}  \,, \quad
\ops[(1)]{ud}{}\,, \\
&
\ops[(8)]{ud}{}\,.
\end{aligned}
\ee
Here we have suppressed the flavour indices, which become evident in the tree-level matching. It is given for the VLL WET WCs and the Warsaw-down basis as follows: 

\be\label{AJB1}
\begin{aligned}
\wcL[V1,LL]{ud}{ppji} &=-\frac{2}{N_c} (V_{pm}  \wc[(3)]{Hq}{mi} {V_{pj}^*} + V_{pi} { \wc[(3)*]{Hq}{mj}} V^*_{pm})
- {1\over 3} {\zeta_3} (\wc[(1)]{Hq}{ji}  + \wc[(3)]{Hq}{ji})\delta_{pp} \\
&+  V_{pm} (\wc[(1)]{qq}{mnji} - \wc[(3)]{qq}{mnji}   
+     {2\over N_c} \wc[(3)]{qq}{mijn} ) {V^*_{pn}}\,,  \\
\wcL[V8,LL]{ud}{ppji} & =
-4(V_{pm} \wc[(3)]{Hq}{mi} {V_{pj}^*} + V_{pi}  {\wc[(3)*]{Hq}{mj}} V_{pm}^*)
+4 V_{pm}  \wc[(3)]{qq}{mijn} {V^*_{pn}}\,, \\
\wcL[V,LL]{dd}{jikk} & = -{{1\over 6}(1+ 2c_w^2)} (\wc[(1)]{Hq}{ji} + \wc[(3)]{Hq}{ji})\delta_{kk}
+  (\wc[(1)]{qq}{jikk} + \wc[(3)]{qq}{jikk})\,, \\
\wc[V,LL]{dd}{jkki} & =  \wc[(1)]{qq}{jkki} + \wc[(3)]{qq}{jkki}\,,
\end{aligned}
\ee
with $\zeta_3 = 1-4c_w^2$. Since in down-basis, the up-type quarks in the SMEFT are not in the mass basis, we have to use the following relation to transform a quark bilinear from the SMEFT to the WET in the mass basis:
\be \label{eq:upquark-downbasis}
\wcL[\rm WET]{kl}{} \,\  \bar u^k \gamma_\mu P_L u^l 
=   V_{ki} \wc[\rm SMEFT]{ij}{} V^\dagger_{jl} \,\ \bar q^i  \gamma_\mu q^j=   {V_{ki} \wc[\rm SMEFT]{ij}{} V^*_{lj} \,\ \bar q^i  \gamma_\mu q^j}\, .
\ee 
We prefer to keep $V$ in the expressions, instead of $V^\dagger$ and therefore express the latter in terms of the former. The matching for the VRR sector is given by 
\be\label{AJB2}
\begin{aligned}
\wcL[V1,RR]{ud}{ppji} & =- {4\over 3} s_w^2 \wc[]{Hd}{ji}\delta_{pp} + \wc[(1)]{ud}{ppji}\,, \quad 
\wcL[V8,RR]{ud}{ppji}  = \wc[(8)]{ud}{ppji}\,,  \\
\wcL[V,RR]{dd}{jikk} & = {1\over 3} s_w^2 \wc[]{Hd}{ji}\delta_{kk} +  \wc[]{dd}{jikk}\,,\quad
\wcL[V,RR]{dd}{jkki} =   \wc[]{dd}{jkki}\,.
\end{aligned}
\ee
For the VLR operators, we have
\be
\begin{aligned}\label{AJB3}
\wcL[V1,LR]{du}{jipp} &= -\frac{4}{3} s_w^2  (\wc[(1)]{Hq}{ji} +\wc[(3)]{Hq}{ji})\delta_{pp} + 
  \wc[(1)]{qu}{jipp}\,, \quad
\wcL[V8,LR]{du}{jipp} =  \wc[(8)]{qu}{jipp}\,, \\
\wcL[V1,LR]{dd}{jikk} & = {2\over 3} s_w^2 (\wc[(1)]{Hq}{ji} + \wc[(3)]{Hq}{ji})\delta_{kk}
 + \wc[(1)]{qd}{jikk}\,, \quad
\wc[V8,LR]{dd}{jikk}  = \wc[(8)]{qd}{jikk} \,, \\
\wc[V1,LR]{uddu}{pjip} & =  -\wc[]{Hud}{ip} V_{pj}\,, \quad 
\wc[V1,LR]{dd}{jkki}  = \wc[(1)]{qd}{jkki}\,, \quad
\wc[V8,LR]{dd}{jkki}  = \wc[(8)]{qd}{jkki}\,.
\end{aligned}
\ee
Similarly, for the VRL sector, we find
\be \label{eq:wet-match-vrl}
\begin{aligned}
\wcL[V1,LR]{ud}{ppji} & = -{1\over 3} {\zeta_3} \wc[]{Hd}{ji}\delta_{pp} 
+  V_{pm} \wc[(1)]{qd}{mnji} {V^*_{pn}}     \,, 
\quad \wcL[V8,LR]{ud}{ppji} =  V_{pm} \wc[(8)]{qd}{mnji}  {V^*_{pn}} \,, \\
\wcL[V1,LR]{dd}{kkji} & = -{1\over 3} (1+2c_w^2) \wc[]{Hd}{ji}\delta_{kk}
+ \wc[(1)]{qd}{kkji}\,, \quad
\wcL[V8,LR]{dd}{kkji}  = \wc[(8)]{qd}{kkji} \,, \\
\wcL[V1,LR]{uddu}{ppji}  &=\wc[]{Hud}{ji} V_{pp} \,, \quad
\wcL[V1,LR]{dd}{kijk}  = \wc[(1)]{qd}{kijk}\,, \quad
\wcL[V8,LR]{dd}{kijk} = \wc[(8)]{qd}{kijk}\,.
\end{aligned}
\ee
Finally, for the SRR sector, we find
\be
\begin{aligned}\label{eq:wet-match-srr}
\wc[S1,RR]{ud}{ppji} & = V_{pm} \wc[(1)]{quqd}{mpji}\,, \quad
\wc[S8,RR]{ud}{ppji}  = V_{pm} \wc[(8)]{quqd}{mpji}\,, \\
\wc[S1,RR]{uddu}{pijp} & = - V_{pm} \wc[(1)]{quqd}{jpmi}\,, \quad
\wc[S8,RR]{uddu}{pijp}  =  - V_{pm} \wc[(8)]{quqd}{jpmi}\,.
\end{aligned}
\ee
The index $p$, $i,j$ or $k$ is not summed over and depend on the external states. The matching for the operators $\wc[V8,LR]{uddu}{pjip}$, $\wc[S1,RR]{dd}{}$ and $\wc[S8,RR]{dd}{}$ vanish at dim-6 level in SMEFT.

\subsection{\boldmath SMEFT Operators for Class 3 at $\Lambda$}
The operator mixing due to ADMs bring in more operators at $\Lambda$ that through RG evolution have an effect on the WCs of the previously listed operators at $\muEW$. A careful analysis of relevant RGEs at 1st LL level reveals these to be:
\begin{center}
\textrm{\bf SMEFT-Loop 3}
\end{center}
\be
\label{eq:class3-smeft-loopl}
\begin{aligned} 
  \textrm{\bf Gauge-mixing:}& \,\
 \ops[(1)]{lq}{}  \,, \,\
\ops[(3)]{lq}{}  \,,  \,\
\ops[]{qe}{}   \,,  \,\
\ops[]{ld}{}\,,  \,\
\ops[]{ed}{} \,, \,\
\ops[]{eu}{} \,, \\
\textrm{\bf Yukawa-mixing:}& \,\ 
\ops[]{H\Box}{}  \,,\,\ 
\ops[]{HD}{}  \,,  \,\
\ops[]{Hu}{}\,,  \,\
\ops[]{dB}{}\,,  \,\ 
\ops[]{dW}{}\,,  \,\ 
\ops[]{dG}{}\,,  \,\  \\
&
\ops[]{uB}{}\,,  \,\ 
\ops[]{uW}{}\,,  \,\ 
\ops[]{uG}{}\,.  \,\ 
\end{aligned}
\ee
Note that the tree-level operator structures having non-trivial flavour structures can also contribute at the 1-loop level. However, in the above list, we include only new operator structures. 
We follow this rule for all other classes as well.

Some of the RGEs for this class are in common with other classes. The details of references to appropriate equations is as follows\\

\noindent
\underline{\bf \boldmath $f^4 \to f^2 H^2 D$ (Gauge)}:

\be
\begin{aligned}
\wc[(1)]{Hq}{}\,, \wc[(3)]{Hq}{}\,, \wc[]{Hd}{} &: \eqref{eq:class2-hq1hq3hd-f4f2-gauge1}\,.
\end{aligned}
\ee

The WC $\wc[]{Hud}{}$ {exhibits} only self-mixing due to gauge couplings.

\noindent
\underline{\bf \textrm{Non-fermion}, \boldmath $f^2 H^2 D \to f^2 H^2 D$ (Gauge)}: 

\noindent
\newline
This type of operator mixing happens only for flavour diagonal operators which in Class 3 are generated through CKM rotations in the tree-level matching conditions. However, the flavour conserving $f^2 H^2 D$ type operators which modify the $W$ and $Z$ boson couplings to fermions are trivial operators for Class 4
\be
\begin{aligned}
\wc[(1)]{Hq}{}\,, \wc[(3)]{Hq}{}\,, \wc[]{Hd}{}  &: \eqref{eq:class4-hq1hq3huhd-f2f2-gauge}\,. 
\end{aligned}
\ee

\noindent
\underline{\bf \boldmath $f^4 \to f^2 H^2 D$ (Yukawa) }:
\be
\begin{aligned}
\wc[(1)]{Hq}{}\,, \wc[(3)]{Hq}{}\,, \wc[]{Hd}{}  &: \eqref{eq:class2-hq1hq3hd-f4f2-yuk}\,.
\end{aligned}
\ee

\noindent
\underline{\bf \boldmath $f^2 H^2 D \to f^2 H^2 D$ (Yukawa)}:

\noindent
\newline
Unlike for the case of gauge coupling operator mixing \eqref{eq:class4-hq1hq3huhd-f2f2-gauge}, in this case even flavour violating operators can have operator mixing via Yukawa interactions. This is similar to Class 2 which also involves flavour violating $f^2 H^2 D$ operators
\be
\begin{aligned}
\wc[(1)]{Hq}{}\,, \wc[(3)]{Hq}{}\,, \wc[]{Hd}{}  &: \eqref{eq:class2-hq1hq3hd-f2f2-yuk}\,.
\end{aligned}
\ee

\noindent
\underline{\bf \boldmath $H^4 D^2  \to f^2 H^2 D$ (Yukawa)}:
\be
\begin{aligned}
\wc[(1)]{Hq}{}\,, \wc[(3)]{Hq}{}   &: \eqref{eq:class2-hq1hq3hd-f0f2-yuk}\,.
\end{aligned}
\ee

Next we move on to four-quark operators as many of them contribute already at the tree-level to this class. In Class 1, we also have $\Delta F=2$ four-quark operators which however play the role of non-trivial operators for Class 3, since they come into play only via CKM rotations at the level of tree-level matching conditions. \\

\noindent
\underline{\bf \boldmath $f^4 \to f^4$ (Gauge)}:

\noindent
\newline
In the following RGEs, only the repeated index $ww$ is summed over.

\be
\begin{aligned}
\dotwc[(1)]{qq}{prst}  &=  
\frac{1}{54} \delta_{pr} \left(-6 g_1^2 \wc[(1)]{lq}{wwst}-6 g_1^2 \wc[(1)]{qd}{stww}-3 g_s^2 (\wc[(8)]{qd}{stww}+2 \wc[(1)]{qq}{swwt}+6 \wc[(3)]{qq}{swwt}+\wc[(8)]{qu}{stww}) \right.\\
&\left.+g_1^2 (-6 \wc[]{qe}{stww}+6 \wc[(1)]{qq}{stww}+\wc[(1)]{qq}{swwt}+3 \wc[(3)]{qq}{swwt}+12 \wc[(1)]{qu}{stww})\right)\\
&+\frac{1}{54} \delta_{st} \left(-6 g_1^2 \wc[(1)]{lq}{wwpr}-6 g_1^2 \wc[(1)]{qd}{prww}-3 g_s^2 (\wc[(8)]{qd}{prww}+2 \wc[(1)]{qq}{pwwr}+6 \wc[(3)]{qq}{pwwr}+\wc[(8)]{qu}{prww})\right.\\
&\left.+g_1^2 (-6 \wc[]{qe}{prww}+6 \wc[(1)]{qq}{prww}+\wc[(1)]{qq}{pwwr}+3 \wc[(3)]{qq}{pwwr}+12 \wc[(1)]{qu}{prww})\right)+\frac{1}{12} g_s^2 \delta_{pt} (\wc[(8)]{qd}{srww}+2 \wc[(1)]{qq}{swwr}+6 \wc[(3)]{qq}{swwr}+\wc[(8)]{qu}{srww})\\
&+\frac{1}{12} g_s^2 \delta_{rs} (\wc[(8)]{qd}{ptww}+2 \wc[(1)]{qq}{pwwt}+6 \wc[(3)]{qq}{pwwt}+\wc[(8)]{qu}{ptww})+3 g_s^2 (\wc[(1)]{qq}{ptsr}+3 \wc[(3)]{qq}{ptsr})+\frac{1}{3} \left(g_1^2-6 g_s^2\right) \wc[(1)]{qq}{prst}+9 g_2^2 \wc[(3)]{qq}{prst}\,,
\end{aligned}
\ee

\be
\begin{aligned}
\dotwc[(3)]{qq}{prst}   &=  
\frac{1}{6} g_2^2 \delta_{pr} (2 \wc[(3)]{lq}{wwst}+\wc[(1)]{qq}{swwt}+6 \wc[(3)]{qq}{stww}-\wc[(3)]{qq}{swwt})+\frac{1}{6} g_2^2 \delta_{st} (2 \wc[(3)]{lq}{wwpr}+\wc[(1)]{qq}{pwwr}+6 \wc[(3)]{qq}{prww}-\wc[(3)]{qq}{pwwr})\\
&+\frac{1}{12} g_s^2 \delta_{pt} (\wc[(8)]{qd}{srww}+2 \wc[(1)]{qq}{swwr}+6 \wc[(3)]{qq}{swwr}+\wc[(8)]{qu}{srww})+\frac{1}{12} g_s^2 \delta_{rs} (\wc[(8)]{qd}{ptww}+2 \wc[(1)]{qq}{pwwt}+6 \wc[(3)]{qq}{pwwt}+\wc[(8)]{qu}{ptww})\\
&+3 g_2^2 \wc[(1)]{qq}{prst}+3 g_s^2 \wc[(1)]{qq}{ptsr}+\frac{1}{3} \left(g_1^2-6 \left(3 g_2^2+g_s^2\right)\right) \wc[(3)]{qq}{prst}-3 g_s^2 \wc[(3)]{qq}{ptsr}\,,
\end{aligned}
\ee

\be
\begin{aligned}
\dotwc[]{dd}{prst}   &= 
\frac{1}{54} \delta_{pr} \left(\left(4 g_1^2-6 g_s^2\right) \wc[]{dd}{swwt}+12 g_1^2 \wc[]{dd}{stww}+12 g_1^2 (\wc[]{ed}{wwst}+\wc[]{ld}{wwst}-\wc[(1)]{qd}{wwst}-2 \wc[(1)]{ud}{wwst})\right.\\
&\left.-3 g_s^2 (2 \wc[(8)]{qd}{wwst}+\wc[(8)]{ud}{wwst})\right)+\frac{1}{6} g_s^2 \delta_{rs} (2 (\wc[]{dd}{pwwt}+\wc[(8)]{qd}{wwpt})+\wc[(8)]{ud}{wwpt})+\left(\frac{4 g_1^2}{3}-2 g_s^2\right) \wc[]{dd}{prst}\\
&+\frac{1}{54} \delta_{st} \left(\left(4 g_1^2-6 g_s^2\right) \wc[]{dd}{pwwr}+12 g_1^2 \wc[]{dd}{prww}+12 g_1^2 (\wc[]{ed}{wwpr}+\wc[]{ld}{wwpr}-\wc[(1)]{qd}{wwpr}-2 \wc[(1)]{ud}{wwpr})\right.\\
&\left.-3 g_s^2 (2 \wc[(8)]{qd}{wwpr}+\wc[(8)]{ud}{wwpr})\right)+\frac{1}{6} g_s^2 \delta_{pt} (2 (\wc[]{dd}{swwr}+\wc[(8)]{qd}{wwsr})+\wc[(8)]{ud}{wwsr})+6 g_s^2 \wc[]{dd}{ptsr}\,,
\end{aligned}
\ee

\be
\begin{aligned}
\dotwc[(1)]{qu}{prst} &= -\frac{2}{9} g_1^2 \delta_{pr}\left(\wc[]{eu}{wwst}
+\wc[]{lu}{wwst}-\wc[(1)]{qu}{wwst}+\wc[(1)]{ud}{stww}-2\wc[]{uu}{stww}-
\frac{2}{3}\wc[]{uu}{swwt}\right)\\&
-\frac{8}{9} g_1^2 \delta_{st}\left(\wc[(1)]{lq}{wwpr}+\wc[(1)]{qd}{prww}
+\wc[]{qe}{prww}-\frac{1}{6}\wc[(1)]{qq}{pwwr}-\frac{1}{2}\wc[(3)]{qq}{pwwr}
-\wc[(1)]{qq}{prww}-2\wc[(1)]{qu}{prww}\right)\\&
-\frac{4}{3} g_1^2 \wc[(1)]{qu}{prst} 
-\frac{8}{3} g_s^2 \wc[(8)]{qu}{prst}\,,
 \end{aligned}
\ee

\be
\begin{aligned}
\dotwc[(8)]{qu}{prst} &= \frac{2}{3}g_s^2 \delta_{st} \left(\wc[(8)]{qd}{prww}+2 \wc[(1)]{qq}{pwwr}+6 \wc[(3)]{qq}{pwwr}+\wc[(8)]{qu}{prww}\right)
+\frac{2}{3}g_s^2 \delta_{pr} \left(2 \wc[(8)]{qu}{wwst} +\wc[(8)]{ud}{stww}
+2 \wc[]{uu}{swwt}\right)
\\&
-12g_s^2 \wc[(1)]{qu}{prst}
-14 g_s^2 \wc[(8)]{qu}{prst}-\frac{4}{3} g_1^2 \wc[(8)]{qu}{prst}\,,
\end{aligned}
\ee

\be
\begin{aligned}
\dotwc[(1)]{qd}{prst} &= -\frac{2}{9} g_1^2 \delta_{pr} \left(\wc[]{dd}{stww}+
\frac{1}{3}\wc[]{dd}{swwt}+ \wc[]{ed}{wwst}+\wc[]{ld}{wwst}-\wc[(1)]{qd}{wwst} -2 \wc[(1)]{ud}{wwst}\right)\\ 
&+\frac{4}{9} g_1^2 \delta_{st} \left( \wc[(1)]{lq}{wwpr}+ \wc[(1)]{qd}{prww}+ \wc[]{qe}{prww}- \wc[(1)]{qq}{prww}-\frac{1}{6}\wc[(1)]{qq}{pwwr} 
- \wc[(3)]{qq}{pwwr}-2\wc[(1)]{qu}{prww}\right)\\&
+\frac{2}{3} g_1^2 \wc[(1)]{qd}{prst}-\frac{8}{3} g_s^2 \wc[(8)]{qd}{prst}\,,
  \end{aligned}
\ee

\be
\begin{aligned}
\dotwc[(8)]{qd}{prst} &= \frac{2}{3} g_s^2 \delta_{pr}\left(2 \wc[]{dd}{swwt}+2 \wc[(8)]{qd}{wwst}+\wc[(8)]{ud}{wwst}\right)-12  g_s^2\wc[(1)]{qd}{prst}
-14 g_s^2\wc[(8)]{qd}{prst}+\frac{2}{3}g_1^2 \wc[(8)]{qd}{prst}\,,
\\&
+\frac{2}{3} g_s^2\delta_{st} \left(\wc[(8)]{qd}{prww}+2 \wc[(1)]{qq}{pwwr}
+6\wc[(3)]{qq}{pwwr}+\wc[(8)]{qu}{prww}\right)\,,
 \end{aligned}
\ee

\be
\begin{aligned}\label{eq:quqd1G}
 { \dotwc[(1)]{quqd}{prst}} &= -\frac{1}{2} \left(\frac{11 g_1^2}{9}+3 g_2^2+32 g_s^2\right) \wc[(1)]{quqd}{prst}-\frac{1}{3} \left(-\frac{5 g_1^2}{9}-3 g_2^2+\frac{64 g_s^2}{3}\right)\wc[(1)]{quqd}{srpt} \\ 
 & -\frac{4}{9} \left(-\frac{5 g_1^2}{9}-3 g_2^2+\frac{28 g_s^2}{3}\right) \wc[(8)]{quqd}{srpt} 
 +\frac{16}{9} g_s^2 \wc[(8)]{quqd}{prst}\,,
 \end{aligned}
\ee

\be
\begin{aligned}\label{eq:quqd8G}
{\dotwc[(8)]{quqd}{prst}} &= \left(\frac{10 g_1^2}{9}+6 g_2^2+\frac{16 g_s^2}{3}\right) \wc[(1)]{quqd}{srpt}+8 g_s^2 \wc[(1)]{quqd}{prst}+\left(-\frac{11 g_1^2}{18}-\frac{3 g_2^2}{2}+\frac{16 g_s^2}{3}\right)\wc[(8)]{quqd}{prst} \\ 
& -\frac{1}{3} \left(\frac{5 g_1^2}{9}+3 g_2^2+\frac{44 g_s^2}{3}\right) \wc[(8)]{quqd}{srpt}\,,
\end{aligned}
\ee

\be
\begin{aligned}
  \dotwc[(1)]{ud}{prst}    &=  \frac{2}{27} g_1^2 \delta_{pr} (2 \wc[]{uu}{swwt}-3 (\wc[]{eu}{wwst}+\wc[]{lu}{wwst}-\wc[(1)]{qu}{wwst}+\wc[(1)]{ud}{stww}-2 \wc[]{uu}{stww}))\\
  &-\frac{4}{27} g_1^2 \delta_{st} (6 \wc[(1)]{lq}{wwpr}+6 \wc[(1)]{qd}{prww}+6 \wc[]{qe}{prww}-6 \wc[(1)]{qq}{prww}-\wc[(1)]{qq}{pwwr}-3 (\wc[(3)]{qq}{pwwr}+4 \wc[(1)]{qu}{prww}))\\
  &-\frac{4}{3} \left(g_1^2 \wc[(1)]{qu}{prst}+2 g_s^2 \wc[(8)]{qu}{prst}\right)\,,
\end{aligned}
\ee

\be
\begin{aligned}
  \dotwc[(8)]{ud}{prst}    &=  \frac{2}{3} g_s^2 \delta_{st} (\wc[(8)]{qd}{prww}+2 \wc[(1)]{qq}{pwwr}+6 \wc[(3)]{qq}{pwwr}+\wc[(8)]{qu}{prww})-\frac{2}{3} \left(18 g_s^2 \wc[(1)]{qu}{prst}+\left(2 g_1^2+21 g_s^2\right) \wc[(8)]{qu}{prst}\right)\\
  &+\frac{2}{3} g_s^2 \delta_{pr} (2 \wc[(8)]{qu}{wwst}+\wc[(8)]{ud}{stww}+2 \wc[]{uu}{swwt})\,.
\end{aligned}
\ee

\noindent
\underline{\bf \boldmath $f^2 H^2 D  \to f^4$ (Gauge)}:

\be
\begin{aligned}
  \dotwc[(1)]{qq}{prst}  &=  
  \frac{1}{18} g_1^2 \wc[(1)]{Hq}{st} \delta_{pr}+\frac{1}{18} g_1^2 \wc[(1)]{Hq}{pr} \delta_{st} \,, 
  \quad
\dotwc[(3)]{qq}{prst}   &=  
\frac{1}{6} g_2^2 \wc[(3)]{Hq}{st} \delta_{pr}+\frac{1}{6} g_2^2 \wc[(3)]{Hq}{pr} \delta_{st}\,,
\\  
\dotwc[(1)]{qu}{prst}    &=  \frac{4}{9} g_1^2 \wc[(1)]{Hq}{pr} \delta_{st}+\frac{1}{9} g_1^2 \wc[]{Hu}{st} \delta_{pr}\,, \quad
  \dotwc[(1)]{qd}{prst}    &=  \frac{1}{9} g_1^2 \wc[]{Hd}{st} \delta_{pr}-\frac{2}{9} g_1^2 \wc[(1)]{Hq}{pr} \delta_{st}\,,
\\  
\dotwc[]{dd}{prst}  &=  -\frac{1}{9} g_1^2 \wc[]{Hd}{st} \delta_{pr}-\frac{1}{9} g_1^2 \wc[]{Hd}{pr} \delta_{st}\,, \quad
  \dotwc[(1)]{ud}{prst}    &=  \frac{4}{9} g_1^2 \wc[]{Hd}{st} \delta_{pr}-\frac{2}{9} g_1^2 \wc[]{Hu}{pr} \delta_{st}\,.
\end{aligned}
\ee
{The} WCs $\wc[(8)]{qu}{}$, $\wc[(8)]{qd}{}$, $\wc[(8)]{ud}{}$, $\wc[(1)]{quqd}{}$ and $\wc[(8)]{quqd}{}$ do not exhibit operator mixing of this type.

In the current and some of the subsequent classes the RGs due to Yukawa mixing turn out to be very lengthy. Therefore, in such cases we present only the contributing operators, suppressing the flavour indices but showing the dependence on Yukawa couplings in order of importance. 

We use the functions $F_i$ to indicate the additional factors in front of the corresponding WCs, like CKM factors. Typically, a given entry in $F_i$ represents several operators distinguished by different flavour indices. In certain cases also the dependence on gauge couplings has been suppressed because the hierarchy in Yukawa couplings is much stronger than in the gauge couplings. Such simplified RGEs reflect basis independent patterns in the operator mixing, since the dependence on the CKM elements is controlled by the choice of the SMEFT flavour basis. Simpler structures are however shown explicitly. \\

With these simplifications one finds:

\noindent
\underline{\bf \boldmath Dipoles $\to f^4$ (Yukawa-Gauge)}: 

\noindent
\newline
The dipole operators mix with the scalar four-quark operators.

\be
\begin{aligned}
\dotwc[(1)]{quqd}{}    &=  y_tF_1(\wc[]{dB}{},\wc[]{dG}{},\wc[]{dW}{})
+y_b F_2(\wc[]{uB}{},\wc[]{uG}{},\wc[]{uW}{})\,, \\
\dotwc[(8)]{quqd}{}    &= y_tF_1( \wc[]{dB}{},\wc[]{dG}{},\wc[]{dW}{})+y_bF_2(\wc[]{uB}{},\wc[]{uG}{},\wc[]{uW}{})\,. 
\end{aligned}
\ee

\noindent
\underline{\bf \boldmath $f^4  \to f^4$ (Yukawa)}:

\be
\begin{aligned}
\dotwc[(1)]{qq}{}  &=  
y_t^2 F_1(\wc[(1)]{qq}{},\wc[(1)]{qu}{},\wc[(8)]{qu}{})+
y_b y_tF_2(\wc[(1)]{quqd}{},\wc[(8)]{quqd}{}{,\wc[(1)*]{quqd}{},\wc[(8)*]{quqd}{}})+{y_b^2 F_3(\wc[(1)]{qd}{},\wc[(8)]{qd}{},\wc[(1)]{qq}{})}\,, 
\\
\dotwc[(3)]{qq}{}   &=  
y_t^2 F_1(\wc[(3)]{qq}{},\wc[(8)]{qu}{})
+y_by_tF_2(\wc[(1)]{quqd}{},\wc[(8)]{quqd}{}{,\wc[(1)*]{quqd}{},\wc[(8)*]{quqd}{}})+{y_b^2 F_3(\wc[(8)]{qd}{},\wc[(3)]{qq}{})}\,.
\end{aligned}
\ee

\be
\begin{aligned}
\dotwc[(1)]{qu}{}    &=  y_t^2 F_1(\wc[(1)]{qq}{},\wc[(3)]{qq}{},\wc[(1)]{qu}{},\wc[(8)]{qu}{},{\wc[(1)*]{qu}{},\wc[(8)*]{qu}{},}\wc[]{uu}{})+y_b y_t F_2(\wc[(1)]{quqd}{},\wc[(8)]{quqd}{}{,\wc[(1)*]{quqd}{},\wc[(8)*]{quqd}{}})\\
&+{y_b^2 F_3(\wc[(1)]{qu}{})+y_ty_\tau F_4(\wc[(1)]{lequ}{},\wc[(1)*]{lequ}{})}\,, \\
\dotwc[(8)]{qu}{}    &= y_t^2 F_1(\wc[(1)]{qq}{},\wc[(3)]{qq}{},
\wc[(1)]{qu}{},\wc[(8)]{qu}{},{\wc[(1)*]{qu}{},\wc[(8)*]{qu}{},}\wc[]{uu}{})
+y_b y_tF_2(\wc[(1)]{quqd}{},\wc[(8)]{quqd}{}{,\wc[(1)*]{quqd}{},\wc[(8)*]{quqd}{}})\\
&+{y_b^2 F_3(\wc[(8)]{qu}{})+y_ty_\tau F_4(\wc[(1)]{lequ}{},\wc[(1)*]{lequ}{})}\,, \\
\dotwc[(1)]{qd}{}    &=   y_t^2 F_1({\wc[(1)]{qd}{}},\wc[(1)]{ud}{})+
y_b y_t F_2(\wc[(1)]{quqd}{},\wc[(8)]{quqd}{}{,\wc[(1)*]{quqd}{},\wc[(8)*]{quqd}{}})\\  
&+y_b^2 F_3(\wc[(1)]{qd}{},{\wc[(8)]{qd}{},\wc[(1)*]{qd}{},\wc[(8)*]{qd}{}}{,\wc[(1)]{qq}{},\wc[(3)]{qq}{}})
+{y_by_\tau F_4(\wc[]{ledq}{},\wc[*]{ledq}{})}\,, \\
\dotwc[(8)]{qd}{}    &=   
y_t^2 F_1(\wc[(8)]{qd}{},\wc[(8)]{ud}{})+y_b y_t F_2(\wc[(1)]{quqd}{},\wc[(8)]{quqd}{}{,\wc[(1)*]{quqd}{},\wc[(8)*]{quqd}{}}) \\
&+ y_b^2 F_3(\wc[(1)]{qd}{},\wc[(8)]{qd}{}{,\wc[(1)*]{qd}{},\wc[(8)*]{qd}{}}{,\wc[(1)]{qq}{},\wc[(3)]{qq}{}})
+{y_by_\tau F_4(\wc[]{ledq}{},\wc[*]{ledq}{})}\,. \\
\end{aligned}
\ee

\be
\begin{aligned}\label{eq:quqd18Yuk}
\dotwc[(1)]{quqd}{}    &= y_t^2F_1(\wc[(1)]{quqd}{},\wc[(8)]{quqd}{})+y_b y_t F_2(\wc[(1)]{qd}{},\wc[(8)]{qd}{}, 
\wc[(1)]{qq}{},\wc[(3)]{qq}{},\wc[(1)]{qu}{},\wc[(8)]{qu}{},\wc[(1)]{ud}{},
\wc[(8)]{ud}{})\\&
+  y_b^2 F_3(\wc[(1)]{quqd}{},\wc[(8)]{quqd}{}) {+  y_t y_\tau F_4(\wc[*]{ledq}{})+  y_b y_\tau F_5(\wc[(1)]{lequ}{})}\,,\\
\dotwc[(8)]{quqd}{}    &= y_t^2 F_1(\wc[(8)]{quqd}{})+y_b y_t F_2(\wc[(1)]{qq}{},\wc[(3)]{qq}{},\wc[(1)]{qu}{},\wc[(8)]{qu}{},\wc[(1)]{ud}{},\wc[(8)]{ud}{}) {+y_b^2 F_3(\wc[(8)]{quqd}{})}\,.
\end{aligned}
\ee

\be
\begin{aligned}
\dotwc[]{dd}{}   &= 
y_b^2 F_1(\wc[]{dd}{},\wc[(1)]{qd}{},\wc[(8)]{qd}{})\,,\\ 
\dotwc[(1)]{ud}{} &=  y_t^2F_1(\wc[(1)]{ud}{})+
y_b y_t  F_2(\wc[(1)]{quqd}{},\wc[(8)]{quqd}{}{,\wc[(1)*]{quqd}{},\wc[(8)*]{quqd}{}}){+ y_b^2F_3(\wc[(1)]{qu}{},\wc[(1)]{ud}{})}\,, \\
\dotwc[(8)]{ud}{}    &= y_t^2F_1(\wc[(8)]{ud}{})
+y_b y_t F_2(\wc[(1)]{quqd}{},\wc[(8)]{quqd}{}{,\wc[(1)*]{quqd}{},\wc[(8)*]{quqd}{}}){+ y_b^2F_3(\wc[(8)]{qu}{},\wc[(8)]{ud}{})}\,. 
\end{aligned}
\ee
After this shorthand presentations of the rather long RG equations above it is useful to exhibit the full dependences in the following simpler equations. For completeness the full version of above RGEs is given in the supplemental material of this review.\\

\noindent
\underline{\bf \boldmath $f^2 H^2 D  \to f^4$ (Yukawa)}:

\be
\begin{aligned}
\dotwc[(1)]{qq}{ijkk}  &=  
\frac{1}{2} y_t^2 \wc[(1)]{Hq}{kk} V^*_{3i} V_{3j}+\frac{1}{2} y_t^2 V_{3k} V^*_{3k} \wc[(1)]{Hq}{ij} \,, 
\\
\dotwc[(3)]{qq}{ijkk}   &=  
-\frac{1}{2} y_t^2 \wc[(3)]{Hq}{kk} V^*_{3i} V_{3j}-\frac{1}{2} y_t^2 V_{3k} V^*_{3k} \wc[(3)]{Hq}{ij}\,,
\end{aligned}
\ee

\be
\begin{aligned}
\dotwc[]{dd}{ijkk}  &= y_b^2 \delta_{k3} \wc[]{Hd}{ij}\,,\quad
\dotwc[(1)]{qu}{ijkk} =  -2 y_t^2 \delta_{k3} \wc[(1)]{Hq}{ij}+y_t^2 \wc[]{Hu}{kk} V^*_{3i} V_{3j}\,, \quad
\dotwc[(1)]{qu}{kkij}  =  y_t^2 V_{3k} V^*_{3k} \wc[]{Hu}{ij}\,,
\end{aligned}
\ee

\be
\begin{aligned}
\dotwc[(1)]{qd}{ijkk}    &=  y_t^2 \wc[]{Hd}{kk} V^*_{3i} V_{3j}\,, \qquad
\dotwc[(1)]{qd}{kkij}    =  y_t^2 V_{3k} V^*_{3k} \wc[]{Hd}{ij}\,,
\end{aligned}
\ee

\be
\begin{aligned}
\dotwc[(1)]{ud}{ijkk}    &=  \frac{2}{3} y_b y_t \delta_{j3} \delta_{k3} \wc[]{Hud}{ik}+\frac{2}{3} y_b y_t \delta_{i3} \delta_{k3} \wc[*]{Hud}{jk}\,, \\
\dotwc[(1)]{ud}{kkij}    &=  -2 y_t^2 \delta_{k3} \wc[]{Hd}{ij}+\frac{2}{3} y_b y_t \delta_{i3} \delta_{k3} \wc[]{Hud}{kj}+\frac{2}{3} y_b y_t \delta_{j3} \delta_{k3} \wc[*]{Hud}{ki}\,, \\
\dotwc[(8)]{ud}{ijkk}    &=  4 y_b y_t \delta_{j3} \delta_{k3} \wc[]{Hud}{ik}+4 y_b y_t \delta_{i3} \delta_{k3} \wc[*]{Hud}{jk}\,, \quad
\dotwc[(8)]{ud}{kkij}    =  4 y_b y_t \delta_{i3} \delta_{k3} \wc[]{Hud}{kj}+4 y_b y_t \delta_{j3} \delta_{k3} \wc[*]{Hud}{ki}\,.
\end{aligned}
\ee
The WCs $\wc[(8)]{qd}{}$, $\wc[(1)]{quqd}{}$ and $\wc[(8)]{quqd}{}$ do not {exhibit} operator mixing in this case.

The above selection of RGEs will be crucial for the one-loop analyses of hadronic decays as well as their corrlelations with other classes. However, detailed numerical studies would be necessary for different type of processes. For a partial set of operators encountered in this class, such as $\wc[(1)]{Hq}{ji}$, $\wc[(3)]{Hq}{ji}$ and $\wc[]{Hd}{ji}$, the RGE charts (common with Class 2) for $j\ne i$ can be found in Class 2 in  Fig.~\ref{chart:df1-gauge} (for gauge-couplings) and in Fig.~\ref{chart:df1-yukawa} (for top and bottom Yukawas). Moreover, in Class 2, the same operators with $j =i$ also contribute due to CKM rotations. Such flavour conserving operators directly contribute to $Z$-pole measurements by modifying its couplings after EW symmetry breaking. The corresponding SMEFT charts can be found in Class 4, see Fig.~\ref{chart:ewp-gauge-2}-\ref{chart:ewp-gauge-1} (for gauge-couplings) and \ref{chart:ewp-yukawa-1}-\ref{chart:ewp-yukawa-2} (for top and bottom Yukawas). However, we refrain from presenting the detailed SMEFT charts for Class 3 because these are found to be very complicated and less illuminating.

\subsection{\boldmath {$\text{SU(2)}_L$ and RGE Correlations}}\label{class3loopcorr}
Most of the operators relevant for Class 3 are of four-quark nature involving either two down quark currents or a down and an up-type current. Given that in the mass basis the corresponding WET operators are of $\Delta F=1$ nature,  the SMEFT operators in the down basis having non-trivial flavour structures can also contribute to Class 3 and other processes such as of $\Delta F=0,2$ type, e.g. $pp\to $jets (Class 10) and meson-antimeson mixing (Class 1), respectively.

The $f^2 H^2 D$ type $\Delta F=1$ operators can also contribute to various $\Delta F=0$ and $\Delta F=1$ semileptonic processes, e.g. $d_j \to d_i \ell^+ \ell^-$ (Class 2) and $pp\to \ell^+ \ell^-$ (Class 10), respectively. 

As an illustration, in Tab~\ref{tab:class3-su2} we present a few examples of additional WET operators generated due to $\text{SU(2)}_L$ symmetry and CKM rotations, together with processes beyond Class 3 to which they contribute. The correlations for the remaining WCs such as $\wc[(1),(3)]{Hq}{}$ (see also Tab.~\ref{tab:class2-su2} in Class 2), $\wc[(1),(3)]{qq}{},\wc[(1),(8)]{qd}{}$, and $\wc[(1),(8)]{quqd}{}$ can be predicted following similar steps.

\begin{table}[tbp]
\begin{center}
\renewcommand*{\arraystretch}{1.0}
\resizebox{0.8\textwidth}{!}{
\begin{tabular}{ |c|c|c|c| }
\hline
\multicolumn{4}{|c|}{$\text{SU}(2)_L$ correlations for Class 3} \\
\hline
Class 3 WCs& correlated WET WCs   & processes & classes \\
\hline
$\wc[(1)]{qu}{jipp}$ & $\wcL[V1,LR]{uu}{mnpp} = V_{mj}  \wc[(1)]{qu}{jipp} {V^*_{ni}}$ &  $ pp \to$ {jets}  & 10 \\
& &   $ c \to  u u \bar u$  & --  \\
$\wc[(8)]{qu}{jipp}$ & $\wcL[V8,LR]{uu}{mnpp} = V_{mj}  \wc[(8)]{qu}{jipp} {V^*_{ni}}$ &  $ pp \to $ {jets}  & 10 \\
& &   $ c \to  u u \bar u$  & --  \\
\hline
\end{tabular}
}
\caption{$\text{SU(2)}_L$ correlations for Class 3 operators, arising at tree-level in the WET and SMEFT. The first column shows Class 3 operators at $\muEW$. The second column lists the other WET operators generated by the operators in the first column, and the third column shows processes that are generated by the WET operators in the second column. The last column shows the classes that are correlated to Class 3. We assume the SMEFT in the down-basis.}
\label{tab:class3-su2}
\end{center}
\end{table}

In Sec.~\ref{CORR2su2} we have mentioned the correlation between the processes in Class 2 and Class 3. There are two main differences between these two classes:
\begin{itemize}
\item
The processes in Class 2 are theoretically cleaner than the ones of Class 3.
\item
On the other hand, while many interesting decay branching ratios of Class 2 have not yet been measured and only experimental upper bounds on them are known, several of the non-leptonic branching ratios and other observables like the ratio $\epe$ have been measured with sufficient precision but due to non-perturbative QCD uncertainties in many cases it is still not clear whether the SM is capable of explaining the data or not.
\end{itemize}

Most spectacular in the latter context are the $\Delta I=1/2$ rule in $K\to\pi\pi$ decays and the ratio $\epe$. Detailed reviews on this topic can be found in \cite{Buras:2020xsm,Buras:2022cyc,Buras:2023qaf}. Here we want to summarize first the status of $\epe$ within the SMEFT.

As discussed in detail in \cite{Buras:2020xsm,Buras:2022cyc,Buras:2023qaf}, while the estimates of this ratio from the RBC-UKQCD lattice collaboration \cite{RBC:2020kdj} and Chiral Perturbation Theory \cite{Cirigliano:2019ani} agree within very large uncertainties with the experimental value from the NA48 \cite{Batley:2002gn} and KTeV \cite{AlaviHarati:2002ye,Abouzaid:2010ny} collaborations 
\be\label{EXP}
(\epe)_\text{exp}=(16.6\pm 2.3)\times 10^{-4} \,,
\ee
the estimates using the Dual QCD (DQCD) approach \cite{Bardeen:1986vz,Buras:2014maa} indicate the SM value of $\epe$ to be by a factor of 2 to 3 below the data \cite{Buras:2015xba,Buras:2016fys,Buras:2020wyv}. We refer to this finding as $\epe$-anomaly.

In \cite{Aebischer:2018csl} the first model-independent anatomy of the ratio $\epe$ in the context of the SMEFT has been presented. This was only possible thanks to the 2018 calculations of the $K\to\pi\pi$ matrix elements of BSM operators, namely of the chromo-magnetic dipole operators by lattice QCD \cite{Constantinou:2017sgv} and DQCD \cite{Buras:2018evv} and in particular through the calculation of matrix elements of all four-quark BSM operators, including scalar and tensor operators, in DQCD \cite{Aebischer:2018rrz}. Even if the latter calculation was performed in the chiral limit, it offers for the first time some insight into the world of BSM operators contributing to $\epe$.

The main messages from \cite{Aebischer:2018csl} are as follows:
\begin{itemize}
\item
Tree-level vector exchanges, like $Z^\prime$, $W^\prime$ and $G^\prime$ contributions and vector-like quarks \cite{Bobeth:2016llm} can be responsible for the observed $\epe$-anomaly but generally one has to face important constraints from $\Delta S=2$ and $\Delta C=2$ transitions as well as direct searches and often some fine tuning is required. Here the main role is played by the electroweak operator $Q_8$ with its WC significantly modified by NP.
\item
Models with tree-level exchanges of heavy colourless or coloured scalars are a new avenue, opened with the results for BSM operators from DQCD in \cite{Aebischer:2018rrz}. In particular scalar and tensor operators, having chirally enhanced matrix elements, are candidates for the explanation of the anomaly in question. Moreover, some of these models, in contrast to the ones with tree-level $Z^\prime$ and $G^\prime$ exchanges, are free from both $\Delta S=2$ and $\Delta C=2$ constraints. The EDM of the neutron is an important constraint for these models, depending on the couplings, but does not preclude sizable NP effects in $\epe$. See Sec.~\ref{class7} for related comments.
\item
Models with modified $W^\pm$ or $Z^0$ couplings can induce sizable effects in $\epe$ without appreciable constraints from semi-leptonic decays such as $\kpn$ or $\klpll$. In the case of a SM singlet $Z'$ that mixes with the $Z$, sizable $Z$-mediated contributions are disfavoured by electroweak precision tests. Yet, as discussed in \cite{Buras:2018wmb} and in references therein, also such models could contribute to our understanding of the role of NP in $\epe$. This is in particular the case for models with vector-like quarks \cite{Bobeth:2016llm}.
\item
Finally, as already mentioned in Sec.~\ref{classification}, the attempt to obtain significant enhancement of the ratio $\epe$ (Class 3) within leptoquark models \cite{Bobeth:2017ecx} is very much restricted by the decays of Class 2.
\end{itemize}

The future of $\epe$ in the SM and in the context of searches for NP will depend on how accurately it can be calculated. This requires improved lattice calculations not only of the matrix elements of SM operators but also of the BSM ones, which are known presently only from the DQCD approach in the chiral limit. In any case, in the coming years the ratio $\epe$ is expected to play a significant role in the search for NP. In this respect, the results presented in \cite{Aebischer:2018quc,Aebischer:2018csl} will be helpful in disentangling potential models of new CP violating sources beyond the SM as well as constraining the magnitude of their effects.

As far as the $\Delta I=1/2$ rule \cite{GellMann:1955jx,GellMann:1957wh}
\be\label{N1a}
R=\frac{{\rm Re}A_0}{{\rm Re}A_2}=22.35\,,
\ee
is concerned it was demonstrated already in 1986 in the DQCD framework \cite{Bardeen:1986vz}, with some improvements in 2014 \cite{Buras:2014maa}, that this rule is dominated by low energy QCD dynamics in hadronic matrix elements of current-current operators. This finding has been confirmed more than 30 years later by the RBC-UKQCD collaboration \cite{RBC:2020kdj} although the modest accuracy of both approaches still allows for some NP contributions at the level of $15\%$. While no detailed SMEFT analysis of this rule exists, a pilot analysis in \cite{Buras:2014sba} suggests that heavy $G^\prime$ gauge bosons are superior to a heavy $Z^\prime$ in this case. See \cite{Buras:2022cyc} for the most recent summary.

{\boldmath
\section{Electroweak Precision Measurements (Class 4)}
\label{class4}}
In Class 4 we look at the electroweak precision (EWP) measurements constituted by 
the $Z$ and {$W$-pole} observables:
{
\be
m_Z, \quad m_W, \quad  \Gamma_Z, \quad \sigma_{\rm had}^0, \quad R^0_f, \quad  \rm A_{FB, f}^0, \quad A_f.
\ee
Here $f$ stands for fermions and quarks. We refer to \cite{ALEPH:2005ab, ALEPH:2013dgf, ALEPH:2006bhb} for 
their definition and measurements. 
}

\begin{figure}[tbp]
\centering
\includegraphics{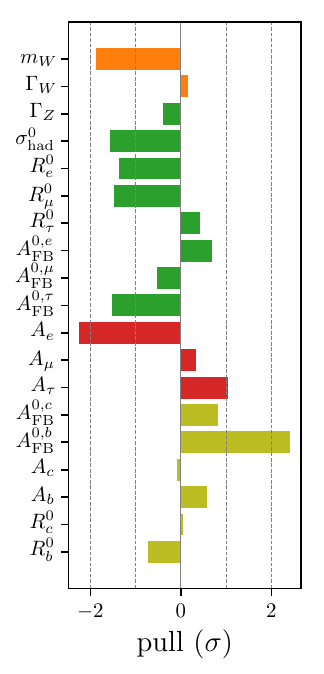}
\caption{\small The SM pulls for the EWP observables, from Ref.~\cite{Aebischer:2018iyb}. }
\label{ewpulls}
\end{figure}

These are mainly controlled by the $W$ and $Z$ boson couplings. These measurements serve as important tests of the SM and at the same time impose stringent constraints on BSM possibilities. The $2\to 2$ scattering processes are also an important component of EWP Physics at LEP \cite{ALEPH:2013dgf}, but these will be discussed in a separate class in Sec.~\ref{class10} dedicated to scattering processes. The list of EWP measurements and their pulls from the SM are displayed in Fig.~\ref{ewpulls}. The observables shown in there can put stringent constraints on the SMEFT WC space. Essentially any SMEFT WC that gives dim-6 corrections to $W$/$Z$ couplings at $\muEW$ is constrained by these observables.

\subsection{Anomalous Gauge Boson Couplings in SMEFT}
The SMEFT effects on the anomalous gauge boson couplings to fermions enter via two routes (a) the higher dimensional shifts in the SM parameters, and (b) via direct contributions from higher dimensional operators to $Z$ and $W$ boson vertices\cite{Berthier:2015oma}. In the chiral basis, we can express the NP shifts in the couplings between fermions and gauge boson by \cite{Kumar:2021yod}
\be
\begin{aligned} \label{gauge-couplings}
\mathcal{L}_{Z,W}^{\rm NP} & = { g_{Z,eff} }  \bar f_i \gamma^\mu 
\left ( \Delta g_{ij, L}^{ffZ} P_L + \Delta g_{ij, R}^{ffZ} P_R    \right ) f_j Z_\mu \\
& - \frac{{\sqrt{2\pi\hat\alpha_e}}}{s_{\hat w}} 
\left ( \Delta g_{ij}^{\nu eW}  \bar \nu_i \gamma_\mu P_L e_j  +
\Delta g_{ij}^{udW} \bar u_i \gamma_\mu P_L d_j  \right ) W^\mu + h.c.
\end{aligned}
\ee
Here, the summation over the fermionic fields $f\in \{u,d,e,{\nu}\}$ is implied. We have used the $\{\hat \alpha_{e}, \hat G_F, \hat m_Z\}$ scheme, so that these act as input parameters taken from the measured values. Here $ g_{Z,eff}$ is given by \cite{Berthier:2015oma}
\be
\begin{aligned}
 g_{Z,eff} & = -2\sqrt{\sqrt{2} \hat G_F} \hat m_Z\,.
\end{aligned}
\ee
We have used the hat (unhat) notation for the measured (SM parameters in the absence of higher dimensional operators) parameters \cite{Berthier:2015oma}. 

$\Delta g^{ffZ}_{ij,L}$, $\Delta g^{ffZ}_{ij,R}$ and $\Delta g_{ij}^{\nu eW}$ are the effective couplings receiving direct and indirect contributions from dim-6 SMEFT operators. The direct contributions originate due to dim-6 shifts in the $Z$ and $W$ vertices from two-fermion operators. The indirect effects stem from SMEFT shifts in the SM parameters such as $g_Z$, $s_w$ (that is in the limit of vanishing higher dimensional SMEFT operators). Both effects have to be included simultaneously, through \cite{Berthier:2015oma, Kumar:2021yod}
\be \label{gauge-couplings-full}
\begin{aligned} 
\Delta g^{ffZ}_{ij,X} & = \Delta g_Z ~  g^{ffZ,{\rm SM}}_{ij,X}  
{+} {Q_f} ~\Delta s^2_w \delta_{ij} + g^{ffZ, {\rm dir}}_{ij,X} \,, \\
 \Delta g_{ij}^{\nu eW} & =    \frac{\Delta s^2_w}{2s^2_{\hat w}}{\delta_{ij}}
+ \Delta g_{ij}^{\nu eW, {\rm dir}}\,, \\
 \Delta g_{ij}^{udW} & =  \frac{\Delta s^2_w}{2s^2_{\hat w}}{\delta_{ij}}
+ \Delta g_{ij}^{udW, {\rm dir}}.
\end{aligned}
\ee 
Here $X \in \{L,R \}$ stands for the chirality of fermions and $Q_f$ denote their electric charges. At dim-6 level, the NP shifts through the indirect contributions  $\Delta g_Z$, ${\Delta s^2_w}$, and direct contributions $\Delta g^{ffZ,{\rm dir}}_{ij,X}$, $\Delta g_{ij}^{\nu eW,{\rm dir}}$, $\Delta g_{ij}^{udW,{\rm dir}}$ can be expressed as linear functions of the SMEFT WCs at $\muEW$, as detailed in the next subsection.

\subsection{SMEFT Operators for Class 4 at $\muEW$}
At the tree-level, the precision $Z$-pole measurements are controlled by the following list of SMEFT operators at the EW scale 
\begin{center}
\textrm{\bf SMEFT-Tree 4}
\end{center}
\be \label{eq:class4-smeft-tree}
\begin{aligned}
&Q_{HD} \,, \quad 
Q_{H W B}  \,, \quad
\ops[(1)]{H l}{} \,, \quad 
\ops[(3)]{H l}{} \,,\quad 
\ops[]{He}{} \,, \\
&\ops[]{ll}{}\,, \quad
\ops[(1)]{Hq}{}\,, \quad 
\ops[(3)]{Hq}{}\,,\quad
\ops[]{Hu}{}\,,\quad
\ops[]{Hd}{}\,.
\end{aligned}
\ee

After EW symmetry breaking, these operators give dim-6 corrections to the anomalous gauge boson couplings \eqref{gauge-couplings-full}. The relationship between the shifts in the $Z$ boson quark couplings and SMEFT WCs read \cite{Berthier:2015oma,Kumar:2021yod}
\be
\begin{aligned}
\label{eq:zcouplingsa}
\Delta g_{ij, L}^{uuZ, \textrm{dir}} &= - {1 \over {2 \sqrt 2 \hat G_F}} V_{im}  
 ( \wc[(1)]{H q}{mn}  - \wc[(3)]{H q}{mn} ) {{V^*_{jn}}}\,, \quad
\Delta g_{ij,R}^{uuZ,\textrm{dir}} = - {1 \over {2 \sqrt 2 \hat G_F}}  \wc[]{Hu}{ij}\,,   \\
\Delta g_{ij,L}^{ddZ, \textrm{dir}} & =  - {1 \over {2 \sqrt 2 \hat G_F}}  ( \wc[(1)]{H q}{ij}  + \wc[(3)]{H q}{ij} )\,,  \qquad
\Delta g_{ij,R}^{ddZ, \textrm{dir}} = - {1 \over {2 \sqrt 2 \hat G_F}} \wc[]{H d}{ij}\,.
\end{aligned}
\ee
The corresponding shifts in leptonic couplings read
\be
\label{eq:zcouplingsb}
\begin{aligned}   
\Delta g_{ij,L}^{eeZ, \textrm{dir}} & =  - {1 \over {2 \sqrt 2 \hat G_F}}   ( \wc[(1)]{Hl}{ij}  + \wc[(3)]{Hl}{ij} )\,,\quad
\Delta g_{ij,R}^{eeZ, \textrm{dir}}  = - {1 \over {2 \sqrt 2 \hat G_F}}  \wc[]{H e}{ij}\,, \\
\Delta g_{ij,L}^{\nu \nu Z,\textrm{dir}}  &= - {1 \over {2 \sqrt 2 \hat G_F}}  ( \wc[(1)]{Hl}{ij}  - \wc[(3)]{Hl}{ij}  )\,.
\end{aligned}
\ee
The shifts in $W$ boson couplings can be analogously parameterized as
\be
\Delta g_{ij}^{\nu eW, \textrm{dir}} = {1 \over { \sqrt 2 \hat G_F}} \wc[(3)]{Hl}{ij}  \,, \quad
\Delta g_{ij}^{udW, \textrm{dir}} =  {1 \over {\sqrt 2 \hat G_F}} V_{im}  \wc[(3)]{H q}{mj} \,.
\ee
The up-type gauge couplings depend on non-trivial flavour structures of SMEFT WCs due to CKM rotations. 

The anomalous $Wtb$ does not directly enter the EWP observables, but it can be indirectly  constrained by the EWP constraints due to operator mixing. Indirect effects of $Wtb$ were considered long ago in $\bar B\to X_s\gamma  $ process \cite{Grzadkowski:2008mf}. Therefore, as already stated in Sec.~\ref{sub:obs-classes} the anomalous $Wtb$ couplings can also been studied 
efficiently in processes of Class 1, 2, 8 and 9. More recent global fits related to EWP or $Wtb$ can be found in Sec.~\ref{sec:11}. However, here we focus on the $W$ and $Z$ couplings directly affecting the EWP observables.

Finally, the SMEFT corrections to the SM parameters $g_Z$ and $s^2_w$ read \cite{Berthier:2015oma, Kumar:2021yod} 
\be
\begin{aligned} 
\label{eq:gztree}
\Delta g_Z &=  - \frac{1}{2 \sqrt 2 \hat G_F}  \left({ \wc[(3)]{Hl}{11} +\wc[(3)]{Hl }{22} - \frac{1}{2}\wc[]{ll}{1221}}+\frac{1}{2}\wc[]{HD}{} +(\sqrt{\frac{2\sqrt{2}\pi\hat \alpha}{M_Z^2 \hat G_F}}-2s_{\hat w}c_{\hat w} )\wc[]{HWB}{}\right) \,,\\
\Delta s^2_w &= -\frac{1}{2 \sqrt 2 \hat G_F} \frac{ s^2_{2 \hat w}}{2 c_{2 \hat w}}
(\frac{1}{2}\wc[]{HD}{} +  {2 \over s_{2 \hat w}} \wc[]{HWB}{} + {2(\wc[(3)]{Hl}{11}+ \wc[(3)]{Hl}{22})
- \wc[]{ll}{1221})}\,. 
\end{aligned}
\ee
The information obtained in this subsection, together with analysis of the SMEFT ADMs, can identify the contributing WCs at $\Lambda$.

\subsection{SMEFT Operators for Class 4 at $\Lambda$}
At the UV scale many four-fermion as well as other operators come into operation because of operator mixing while running from $\Lambda$ to $\muEW$. We identify all such operators that contribute via this mechanism:
\begin{center}
\textrm{\bf SMEFT-Loop 4}
\end{center}
\be \label{class4-smeftopsL}
\begin{aligned}
\textrm{ \bf Gauge-mixing:} ~ & 
\ops[(1)]{lq}{}\,, \quad
\ops[(3)]{lq}{}\,, \quad
\ops[]{qe}{}\,, \quad
\ops[(1)]{qu}{}\,, \quad
\ops[(1)]{qd}{}\,, \quad
\ops[(1)]{ud}{}\,, \quad
\ops[]{HW}{} \,, \quad
\\
&
\ops[]{HB}{} \,,\quad
\ops[]{W}{} \,, \quad
\ops[]{ld}{}\,, \quad
 \ops[]{lu}{} \,, \quad
\ops[]{le}{}\,, \quad
\ops[]{ed}{}\,, \quad
\ops[]{eu}{}\,, \quad
\ops[]{ee}{}\,, 
\\
&
\ops[]{uu}{}\,, \quad
\ops[]{dd}{}\,, \quad
\ops[(1)]{qq}{}\,, \quad
\ops[(3)]{qq}{}\,,  \quad
{\ops[]{H\Box}{}}\,, 
\\
 \textrm{ \bf Yukawa-mixing:}  ~&
\ops[(1)]{lq}{}\,, \quad
\ops[(3)]{lq}{}\,, \quad
\ops[]{qe}{}\,, \quad
\ops[]{ld}{}\,, \quad
\ops[]{lu}{}\,, \quad
\ops[]{ed}{}\,, \quad
\ops[]{eu}{}\,, \quad
\ops[]{le}{}\,, \\
&
\ops[]{ee}{}\,, \quad
\ops[(1)]{qu}{}\,, \quad
\ops[(1)]{ud}{}\,, \quad
\ops[]{uu}{}\,, \quad
\ops[]{Hud}{}\,, \quad
\ops[]{dB}{}\,, \quad
\ops[]{dW}{}\,, \quad
\ops[]{uB}{}\,, \\
&
\ops[]{uW}{}\,, \quad
\ops[]{eB}{}\,, \quad
\ops[]{eW}{}\,. \quad
\ops[]{H\Box}{}\,. \\
\end{aligned}
\ee

The specific flavour structures of these operators will be defined in the explicit RGEs and SMEFT charts to be given below.
 
First, we present the RGEs for the SMEFT-tree level operators collected in \eqref{eq:class4-smeft-tree}. In the Yukawa dependent RGEs, the subdominant terms $\mathcal{O}(Y_d^2)$ were dropped as 
compared to $\mathcal{O}(Y_u^2)$ unless the latter contribution is absent. 

\begin{figure}[tbp]
\centering
\begin{minipage}{1.0\linewidth}%
\includegraphics[clip, trim=0.1cm 8cm 0.1cm 8cm, width=0.91\textwidth]{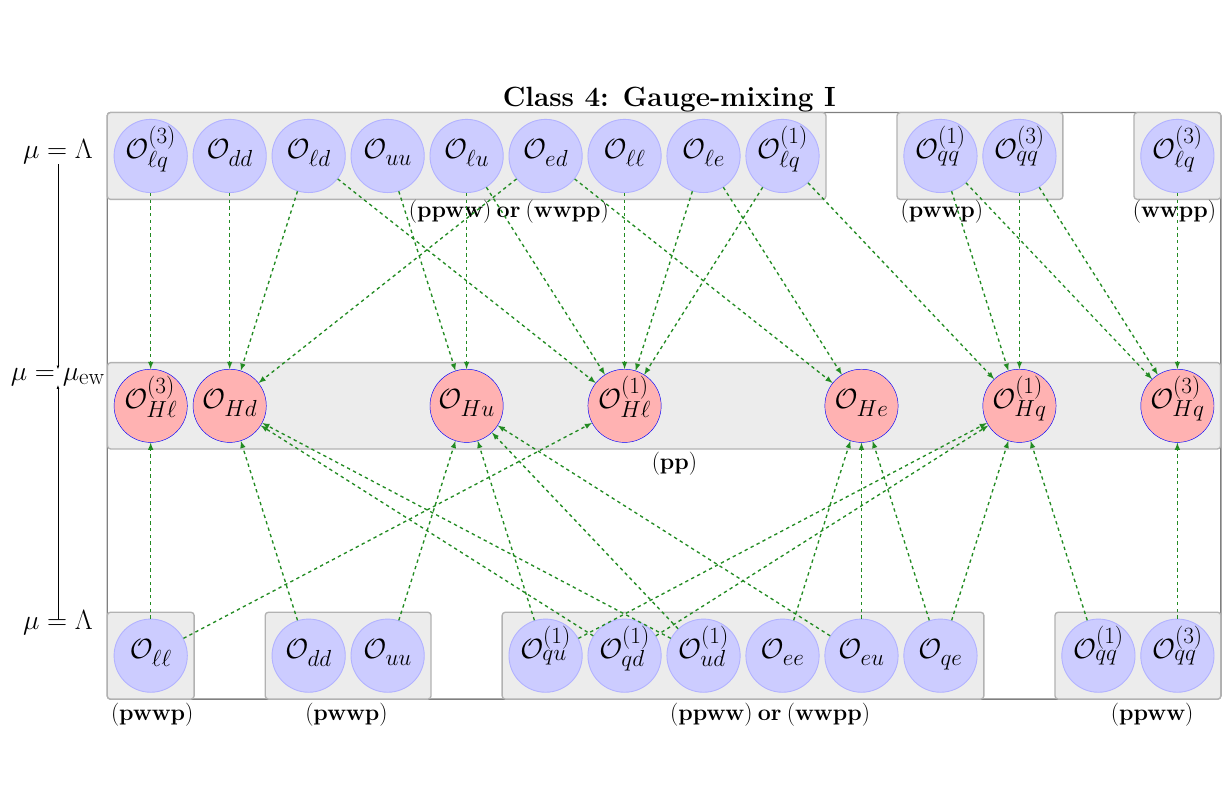}
\end{minipage}
\vspace{-0.8cm}
\caption{\small Class 4: Operator mixing relevant for $Z$-pole observables. Mixing due to the weak gauge coupling is indicated by dashed green lines. Here the index $w$ is summed over $1-3$. Depending upon the closed fermion loop only one flavour structure out of $(wwpp)$ and $(ppww)$ is allowed at $\Lambda$. For example, $\ops[]{He}{pp}$ and $\ops[]{Hu}{pp}$ mix only with $\ops[]{eu}{ppww}$ and $\ops[]{eu}{wwpp}$, respectively. For $\wc[(1)]{Hq}{}$, $\wc[(3)]{Hq}{}$ and $\wc[]{Hd}{}$, the operator-mixing for the flavour violating case is not shown here. It can be found in the RGEs \eqref{eq:class2-hq1hq3hd-f4f2-gauge1}.}
\label{chart:ewp-gauge-2}
\end{figure}

\begin{figure}[H] 
\begin{minipage}{1.0\linewidth}%
\includegraphics[clip, trim=4.5cm 12.5cm 0.3cm 12cm, width=1.3\textwidth]{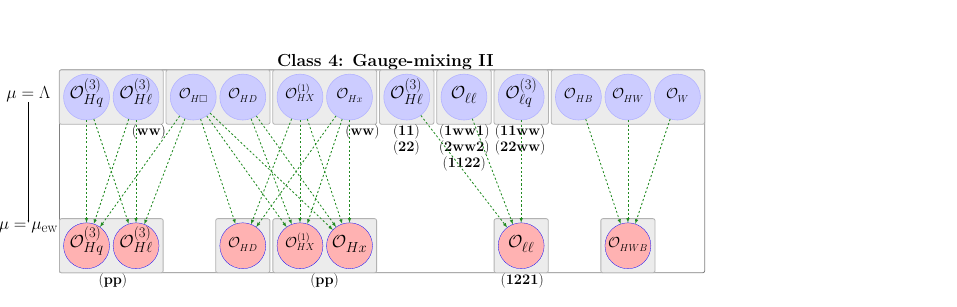}
\end{minipage}
\caption{\small Class 4: Operator mixing relevant for the $Z$-pole observables. Mixing due to the weak gauge coupling is indicated by dashed green lines. Here the index $w$ is summed over $1$-$3$. Note {that} $\ops[(3)]{Hq}{pq}$ for $p\ne q$ only exhibits self-mixing. Here the $X$ in $\wc[(1)]{HX}{}$ is a variable, which is summed over $(l, q )$ at $\Lambda$. Similarly, for $\wc[]{Hx}{}$, $x \in (u,d,e)$ is summed over.}
\label{chart:ewp-gauge-1}
\end{figure}

Depending upon  the type of SMEFT operator involved in the operator mixing, we can have several categories of RGEs for Class 4. The gauge couplings dependent RGEs are given in the following. The corresponding SMEFT charts are shown in Figs.~\ref{chart:ewp-gauge-2}, \ref{chart:ewp-gauge-1}.

\noindent
\underline{\bf \boldmath $f^4 \to f^4$ (Gauge)}: 

The only four-fermion operator which contributes Class 4 at tree-level is $\ops[]{ll}{1221}$:
\be
\begin{aligned}\label{eq:ll1221}
\dotwc[]{ll}{1221} &= 3 \wc[]{ll}{1221} \left(g_1^2-g_2^2\right)+\frac{1}{3} g_2^2 \left(\wc[]{ll}{1ww1}+\wc[]{ll}{2ww2}\right)+2 g_2^2\left(\wc[(3)]{lq}{11ww}
+\wc[(3)]{lq}{22ww}\right)+6g_2^2 \wc[]{ll}{1122} .
\end{aligned}
\ee

\noindent
\underline{\bf \boldmath $f^2 H^2 D \to f^4$ (Gauge)}:
\be
\begin{aligned}\label{eq:cllHl}
\dotwc[]{ll}{1221} &= \frac{1}{3} g_2^2 (\wc[(3)]{Hl}{11}+\wc[(3)]{Hl}{22})\,.
\end{aligned}
\ee

\noindent
\underline{\bf \boldmath $f^4 \to f^2 H^2 D$ (Gauge)}:

\noindent
Ignoring the PMNS matrix, the two lepton operators can only have two identical flavour indices, as evident from the matching conditions. The relevant RGEs for these operators read: 
\be
\begin{aligned}\label{eq:2lep_class4}
\dotwc[(1)]{Hl}{pp} &= -\frac{2}{3} g_1^2 \left(\wc[]{ld}{ppww}+\wc[]{le}{ppww}+{\frac{1}{2} \wc[]{ll}{pwwp}}+\wc[]{ll}{ppww}  -\wc[(1)]{lq}{ppww}-2 \wc[]{lu}{ppww}\right)\,,\\
\dotwc[(3)]{Hl}{pp} &= {\frac{1}{3} g_2^2 \wc[]{ll}{pwwp}}+2 g_2^2 \wc[(3)]{lq}{ppww}\,,\\
\dotwc[]{He}{pp} &= -\frac{2}{3} g_1^2 (\wc[]{ed}{ppww}+\wc[]{ee}{ppww}
-2 \wc[]{eu}{ppww}+\wc[]{le}{wwpp}-\wc[]{qe}{wwpp})\,,
\end{aligned}
\ee
where we have used Fierz identities to remove the WC $\wc[]{ee}{pwwp}$.
The two quark operators containing LH fermion currents can have non-diagonal indices. The RGEs for such operators are already presented in Class 2:
\be
\begin{aligned}
\wc[(1)]{Hq}{}\,, \wc[(3)]{Hq}{} &: \eqref{eq:class2-hq1hq3hd-f4f2-gauge1}\,.
\end{aligned}
\ee
However, within Class 4 the two quark operators with RH fermion currents can have only diagonal indices, for which the corresponding RGEs are given by
\be
\begin{aligned}
\dotwc[]{Hu}{pp} &=-\frac{2}{3} g_1^2 \left(\wc[]{eu}{wwpp}+\wc[]{lu}{wwpp}-\wc[(1)]{qu}{wwpp}+\wc[(1)]{ud}{ppww}-2 \wc[]{uu}{ppww}-\frac{2}{3} \wc[]{uu}{pwwp}\right)\,, \\
\wc[]{Hd}{pp} &: \eqref{eq:class2-hq1hq3hd-f4f2-gauge1}~ {\rm setting}\,\,\, s=t=p\,. 
\end{aligned}
\ee

\noindent
\underline{\bf Non-Fermion, \boldmath $f^2 H^2 D \to f^2 H^2 D$ (Gauge)}:

\noindent
\newline
The gauge-dependent mixing of the non-fermion, and $f^2 H^2 D$ sectors into $f^2 H^2 D$ is given for the two-lepton operators as follows

\be
\begin{aligned}\label{eq:Hl_gauge}
\dotwc[(1)]{Hl}{pr} &= \frac{1}{6} g_1^2 (2 \wc[(1)]{Hl}{pr}-\xi  \delta_{pr})\,,\quad
\dotwc[]{He}{pr} = \frac{1}{3} g_1^2 (\wc[]{He}{pr}-\xi  \delta_{pr})\,, \\
\dotwc[(3)]{Hl}{pr} &= \frac{1}{6} g_2^2 (\delta_{pr} (4 \wc[(3)]{Hl}{ww}+ 
12 \wc[(3)]{Hq}{ww}+\wc[]{H\square}{})-34 \wc[(3)]{Hl}{pr})\,,
\end{aligned}
\ee
and for two-quark operators as
\be \label{eq:class4-hq1hq3huhd-f2f2-gauge}
\begin{aligned}
\dotwc[(1)]{Hq}{pr} &= \frac{1}{18} g_1^2 (6 \wc[(1)]{Hq}{pr}+\xi  \delta_{pr})\,,\\
\dotwc[(3)]{Hq}{pr} &= \frac{1}{6} g_2^2 (\delta_{pr} (4 \wc[(3)]{Hl}{ww}+12 \wc[(3)]{Hq}{ww}+\wc[]{H\square}{})-34 \wc[(3)]{Hq}{pr})\,,\\
\dotwc[]{Hu}{pr} &= \frac{1}{9} g_1^2 (3 \wc[]{Hu}{pr}+2 \xi  \delta_{pr})\,,\quad 
\dotwc[]{Hd}{pr} = \frac{1}{9} g_1^2 (3 \wc[]{Hd}{pr}-\xi  \delta_{pr})\,,
\end{aligned}
\ee
where we defined
\be
\xi=-4 \wc[]{Hd}{ww}-4 \wc[]{He}{ww}-4 \wc[(1)]{Hl}{ww}+4 \wc[(1)]{Hq}{ww}+8 \wc[]{Hu}{ww}+\wc[]{H\square}{}+\wc[]{HD}{}\,.
\ee

\noindent
\underline{\bf Non-Fermion, \boldmath $f^2 H^2 D\to $  Non-Fermion (Gauge)}: 

\noindent
\newline
Finally, for purely bosonic operators, we find

\be
\begin{aligned}\label{eq:CHD}
\dotwc[]{HD}{} & = -\frac{8}{3} g_1^2( \wc[]{Hd}{vv}+
\wc[]{He}{vv}+\wc[(1)]{Hl}{vv}-\wc[(1)]{Hq}{vv}-2\wc[]{Hu}{vv}-\frac{5}{2}\wc[]{H\square}{}+\frac{5}{16} \wc[]{HD}{} )
+\frac{9}{2} \wc[]{HD}{} g_2^2\,,
\\
\dotwc[]{HWB}{} & = 2  g_1 g_2(\wc[]{HB}{} +\wc[]{HW}{})+\frac{19}{3}g_1^2 \wc[]{HWB}{} +\frac{4}{3}g_2^2 \wc[]{HWB}{} 
+3 \wc[]{W}{} g_1 g_2^2\,.
\end{aligned}
\ee

\begin{figure}[H]
\begin{minipage}{1.0\linewidth}%
\hspace{-1.5cm}
\includegraphics[clip, trim=0.0cm 11.5cm 0.2cm 11.5cm, width=1.1\textwidth]{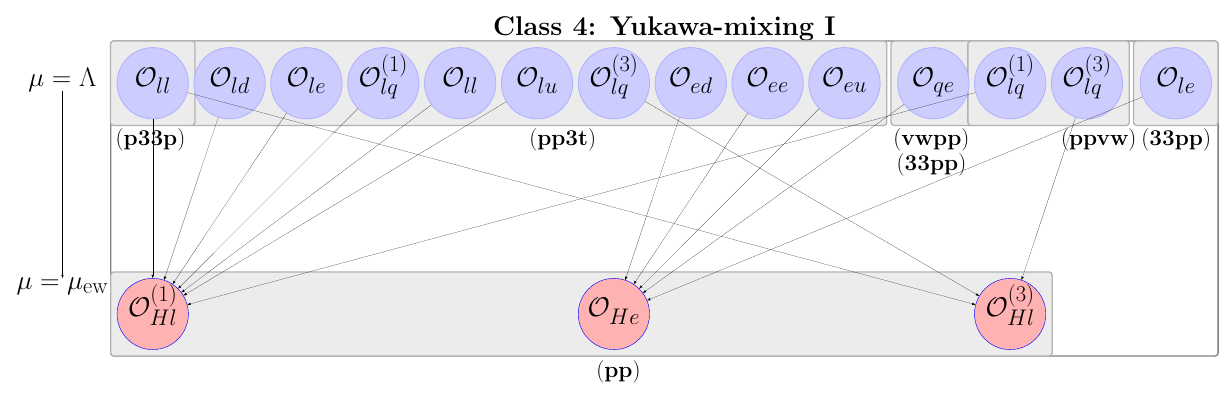}
\end{minipage}
\caption{\small Class 4: Operator mixing due to top Yukawa as indicated by solid black lines for the set of operators relevant for $Z$-pole observables in the Warsaw down-basis. The indices $v,w$ should be summed over $1$-$3$, but not $pp$. See text for the Yukawa operator mixing of $\wc[(1)]{Hq}{}, \wc[(3)]{Hq}{}, \wc[]{Hu}{}, \wc[]{Hd}{}$, $\wc[]{HD}{}$ and $\wc[]{HWB}{}$.}\label{chart:ewp-yukawa-1}
\end{figure}

\begin{figure}[H]
  \centering 
\begin{minipage}{1.0\linewidth}%
  \hspace{4cm}
\includegraphics[clip, trim= 0.3cm 9cm 3cm 9cm, width=0.4\textwidth]{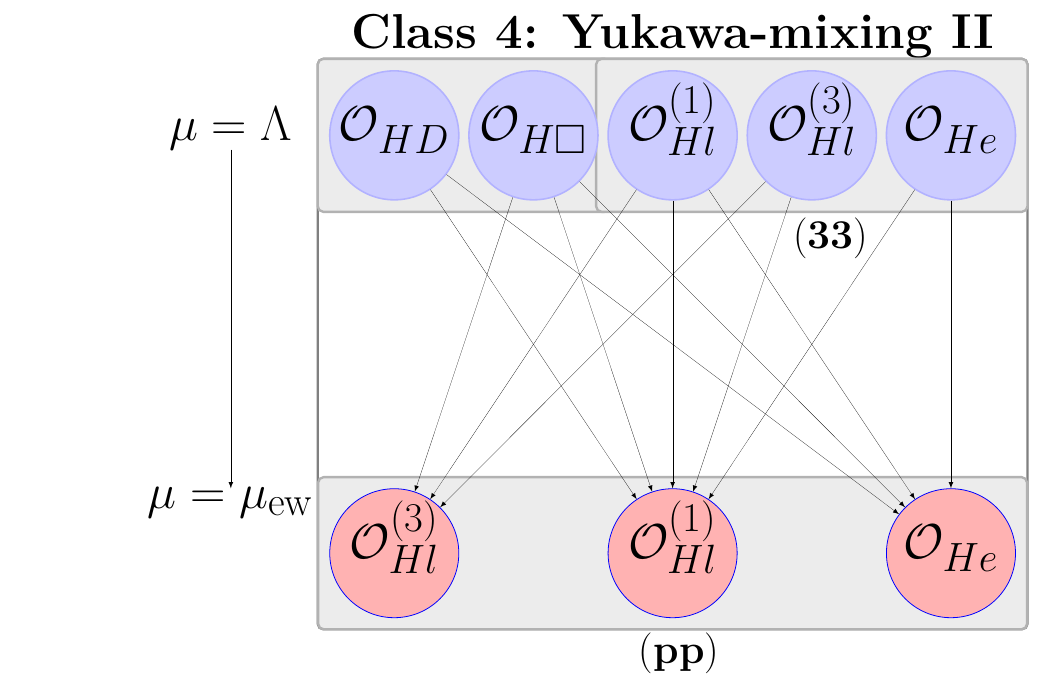}
\end{minipage}
\caption{\small Class 4: Operator mixing due to Yukawas as indicated by solid black lines for the set of operators relevant for $Z$-pole observables in the Warsaw down-basis. The repeated index $p$ should not be summed over. See text for the Yukawa operator mixing of $\wc[(1)]{Hq}{}, \wc[(3)]{Hq}{}, \wc[]{Hu}{}, \wc[]{Hd}{}$, $\wc[]{HD}{}$ and $\wc[]{HWB}{}$.}
\label{chart:ewp-yukawa-2}
\end{figure}

Next we present the Yukawa dependent RGEs for Class 4 and refer to Figs.~\ref{chart:ewp-yukawa-1} and \ref{chart:ewp-yukawa-2} for the corresponding RGE charts.

\noindent
\underline{\bf \boldmath $f^4 \to f^2 H^2 D$ (Yukawa)}:
For the two lepton operators one finds:
\be
\begin{aligned}\label{eq:2lep_yuk_class4}
\dotwc[(1)]{Hl}{pp} &= 6 y_b^2 \wc[]{ld}{pp33}+2 y_\tau^2 \wc[]{le}{pp33}-2y_\tau^2 \wc[]{ll}{pp33}-y_\tau^2 \wc[]{ll}{p33p}+6 y_t^2 V_{3v} V^*_{3w} \wc[(1)]{lq}{ppvw}-6 y_b^2 \wc[(1)]{lq}{pp33}-6 y_t^2 \wc[]{lu}{pp33}\,,\\
\dotwc[(3)]{Hl}{pp} &= -y_\tau^2 \wc[]{ll}{p33p}-6 y_t^2 V_{3v} V^*_{3w} \wc[(3)]{lq}{ppvw}-6 y_b^2 \wc[(3)]{lq}{pp33}\,,\\
\dotwc[]{He}{pp} &= 6 y_b^2 \wc[]{ed}{pp33}+2 y_\tau^2 \wc[]{ee}{pp33}-6 y_t^2 \wc[]{eu}{pp33}-2 y_\tau^2 \wc[]{le}{33pp}+6 y_t^2 V_{3v} V^*_{3w} \wc[]{qe}{vwpp}-6 y_b^2 \wc[]{qe}{33pp}\,.
\end{aligned}
\ee
For three of the two-quark operators, the RGEs were already given in Class 2 in Sec.~\ref{class2}.
\be
\begin{aligned}
\wc[(1)]{Hq}{}\,, \wc[(3)]{Hq}{}\,, \wc[]{Hd}{} &: \eqref{eq:class2-hq1hq3hd-f4f2-yuk}\,.
\end{aligned}
\ee
As the corresponding RGE charts are also given in Fig.~\ref{chart:df1-yukawa} in Class 2, we do not repeat them here. The remaining one is given by:
\be \label{eq:class4-hq1hq3hd-f4f2-yuk}
\begin{aligned}
\dotwc[]{Hu}{pp} &= 2 y_\tau^2 \wc[]{eu}{33pp}-2 y_\tau^2 \wc[]{lu}{33pp}+6 y_t^2 V_{3v} V^*_{3w} \wc[(1)]{qu}{vwpp}-6 y_b^2 \wc[(1)]{qu}{33pp}+6 y_b^2 \wc[(1)]{ud}{pp33}-2 y_t^2 \wc[]{uu}{p33p}-6 y_t^2 \wc[]{uu}{pp33}\,.
\end{aligned}
\ee

\noindent
\underline{\bf \boldmath $f^2 H^2 D \to f^2 H^2 D$ (Yukawa)}:

\noindent
\newline
Further, two-fermion operators can also mix with two-fermion operators with different flavour indices. For the case of two lepton operators one finds:
\be
\begin{aligned}\label{eq:Hlyuk}
\dotwc[(1)]{Hl}{pp} &= -y_\tau^2 \wc[]{He}{33} \delta_{p3}+6 y_t^2 \wc[(1)]{Hl}{pp}+4 y_\tau^2 \wc[(1)]{Hl}{33}+9 y_\tau^2 \wc[(3)]{Hl}{33}\,,\\
\dotwc[(3)]{Hl}{pp} &= 3 y_\tau^2 \wc[(1)]{Hl}{33}+6 y_t^2 \wc[(3)]{Hl}{pp}+2 y_\tau^2 \wc[(3)]{Hl}{33}\,,\\
\dotwc[]{He}{pp} &= 6 y_t^2 \wc[]{He}{pp}+8 y_\tau^2 \wc[]{He}{33}-2 y_\tau^2 \wc[(1)]{Hl}{33} \delta_{p3}\,.
\end{aligned}
\ee
For some of the two-quark operators the RGEs were already given in Sec.~\ref{class2}.
\be
\begin{aligned}
\wc[(1)]{Hq}{}, \wc[(3)]{Hq}{}, \wc[]{Hd}{} &: \eqref{eq:class2-hq1hq3hd-f2f2-yuk}\,.
\end{aligned}
\ee
For the remaining two-quark operators one finds:
\be \label{eq:class4-hq1hq3hd-f4f2-yuk}
\begin{aligned}
\dotwc[]{Hu}{pp} &= -2 y_t^2 \delta_{p3} V_{3v} V^*_{3w} \wc[(1)]{Hq}{vw}+6 y_t^2 \wc[]{Hu}{pp}+8 y_t^2 \wc[]{Hu}{33}+y_b y_t \wc[]{Hud}{33}+y_b y_t \wc[*]{Hud}{33}\,,\\
\dotwc[]{Hud}{pp} &= -2 y_b y_t \wc[]{Hd}{33}+2 y_b y_t \wc[]{Hu}{33}+6 y_t^2 \wc[]{Hud}{pp}+3 y_t^2 \wc[]{Hud}{33}\,,
\end{aligned}
\ee
where we have dropped $y_b^2$ contributions with respect to $y_t^2$ terms.

\noindent
\underline{\bf \boldmath \textrm{Non-Fermion} $\to f^2 H^2 D$ (Yukawa)}: 

\noindent
\newline
Some of the non-fermion operators can mix with two-fermion operators via top and bottom Yukawas. The RGEs for the two-lepton operators are given by:
\be
\begin{aligned}
\dotwc[(1)]{Hl}{pp} &= -\frac{1}{2} y_\tau^2 (\wc[]{H\square}{}+\wc[]{HD}{}) \delta_{p3}\,,\qquad
\dotwc[(3)]{Hl}{pp} = -\frac{1}{2} \wc[]{H\square}{} y_\tau^2 \delta_{p3}\,,\qquad
\dotwc[]{He}{pp} = y_\tau^2 (\wc[]{H\square}{}+\wc[]{HD}{}) \delta_{p3}\,.
\end{aligned}
\ee
The RGEs for the other two-quark operators are given by:
\be \label{eq:class4-hq1hq3hd-f0f2-yuk}
\begin{aligned}
\dotwc[(1)]{Hq}{pp} &= -\frac{1}{2} (\wc[]{H\square}{}+\wc[]{HD}{}) \left(y_b^2 \delta_{p3}-y_t^2 V_{3p} V^*_{3p}\right)\,,\\
\dotwc[(3)]{Hq}{pp} &= -\frac{1}{2} \wc[]{H\square}{} \left(y_b^2 \delta_{p3}+y_t^2 V_{3p} V^*_{3p}\right)\,,\\
\dotwc[]{Hd}{pp} &= y_b^2 (\wc[]{H\square}{}+\wc[]{HD}{}) \delta_{p3}\,,\ \quad
\dotwc[]{Hu}{pp} = -y_t^2 (\wc[]{H\square}{}+\wc[]{HD}{}) \delta_{p3}\,.
\end{aligned}
\ee

\noindent
\underline{\bf Non-Fermion, \boldmath $f^2 H^2 D\to $ Non-Fermion (Yukawa)}:

\noindent
\newline
Finally, the operator mixing for the non-fermion operators is given by the RGEs
\be
\begin{aligned}\label{eq:CHDyuk}
\dotwc[]{HD}{} & = 24 y_b^2 (\wc[]{Hd}{33}-\wc[(1)]{Hq}{33})+12 y_t^2 (2 V_{3v} V^*_{3w} \wc[(1)]{Hq}{vw}-2 \wc[]{Hu}{33}+\wc[]{HD}{})+8 y_\tau^2 (\wc[]{He}{33}-\wc[(1)]{Hl}{33})\\
&-12 y_b y_t (\wc[]{Hud}{33}+\wc[*]{Hud}{3v} V_{3v})+6 \wc[]{HD}{} \lambda\,,
\end{aligned}
\ee
\be
\begin{aligned}\label{eq:CHWByuk}
  \dotwc[]{HWB}{} & = y_b (-3 g_2 \wc[]{dB}{33}-3 g_2 \wc[*]{dB}{33}+g_1 \wc[]{dW}{33}+g_1 \wc[*]{dW}{33})+y_\tau (-g_2 \wc[]{eB}{33}-g_2 \wc[*]{eB}{33}+3 g_1 \wc[]{eW}{33}+3 g_1 \wc[*]{eW}{33})\\
  &+y_t (3 g_2 \wc[]{uB}{v3} V_{3v}+3 g_2 \wc[*]{uB}{w3} V^*_{3w}+5 g_1 \wc[]{uW}{v3} V_{3v}+5 g_1 \wc[*]{uW}{w3} V^*_{3w})+2 \wc[]{HWB}{} \lambda+6 \wc[]{HWB}{} y_t^2.
\end{aligned}
\ee

Using the formulae presented here, it is straightforward to compute the $\eta$ parameters for the effective $W$ and $Z$ couplings, thereby, identifying the most significant SMEFT WCs for Class 4. 

\subsection{\boldmath {$\textrm{SU(2)}_L$} Correlations} 
The EWP observables probe flavour-conserving gauge couplings to fermions in the mass basis. However, in the SMEFT down basis, the gauge couplings involving LH up-type quarks become functions of non-trivial flavour dim-6 operators, which could be flavour violating as well. 

For example, expressing the up-type LH couplings in terms of SMEFT WCs and expanding them in Wolfenstein parameter $\lambda$, we find
\be
\begin{aligned}
\Delta g_{11,L}^{uuZ, \textrm{dir}} & \propto 
(\wc[(1)]{Hq}{11} -  \wc[(3)]{Hq}{11} ) +
\lambda(\wc[(1)]{Hq}{12} -  \wc[(3)]{Hq}{12} +\wc[(1)]{Hq}{21} -  \wc[(3)]{Hq}{21}) \\
&+
\lambda^2(\wc[(1)]{Hq}{22} -  \wc[(3)]{Hq}{22} ) +
\lambda^3(\wc[(1)]{Hq}{13} -  \wc[(3)]{Hq}{13} +\wc[(1)]{Hq}{31} -  \wc[(3)]{Hq}{31}) 
\\
& +
\lambda^4(\wc[(1)]{Hq}{23} -  \wc[(3)]{Hq}{23} +\wc[(1)]{Hq}{32} -  \wc[(3)]{Hq}{32}) +
\lambda^6(\wc[(1)]{Hq}{33} -  \wc[(3)]{Hq}{33} ) \,, \\
\Delta g_{22,L}^{uuZ, \textrm{dir}} & \propto
(\wc[(1)]{Hq}{22} -  \wc[(3)]{Hq}{22} ) -
\lambda(\wc[(1)]{Hq}{12} -  \wc[(3)]{Hq}{12} +\wc[(1)]{Hq}{21} -  \wc[(3)]{Hq}{21}) \\
& +  \lambda^2(\wc[(1)]{Hq}{11} -  \wc[(3)]{Hq}{11}  + \wc[(1)]{Hq}{23} - \wc[(3)]{Hq}{23} 
+ \wc[(1)]{Hq}{32} - \wc[(3)]{Hq}{32}) \\
& - \lambda^3(\wc[(1)]{Hq}{13} -  \wc[(3)]{Hq}{13} +\wc[(1)]{Hq}{31} -  \wc[(3)]{Hq}{31}) +
\lambda^4(\wc[(1)]{Hq}{33} -  \wc[(3)]{Hq}{33} )\,.
\end{aligned}
\ee
In a similar manner we also expand the gauge boson couplings to left-handed charged fermion currents
\be
\begin{aligned}
\Delta g_{11,L}^{udW, \textrm{dir}} & \propto
\wc[(3)]{Hq}{11} + \lambda \wc[(3)]{Hq}{21} +\lambda^3  \wc[(3)]{Hq}{31}   \,,\\              
\Delta g_{22,L}^{udW, \textrm{dir}} & \propto
\wc[(3)]{Hq}{22} -  \lambda \wc[(3)]{Hq}{12}+  \lambda^2 \wc[(1)]{Hq}{32} .
\end{aligned}
\ee
Observe that the r.h.s exhibits non-trivial flavour structures with hierarchical suppressions governed by powers of $\lambda$: the leading terms are of $\mathcal{O}(1)$, while 
the smallest ones of $\mathcal{O}(\lambda^6)$. Additionally, there are intermediate terms whose magnitude may be comparable to SMEFT RGE effects. 

As seen in \eqref{eq:zcouplingsa}, the operators on the r.h.s. in the above relations also match onto gauge couplings with down-type fermions reflecting $\textrm{SU(2)}_L$ correlations between up- and down-type fermions in the SMEFT.

The flavour conserving gauge couplings can be directly probed via EWP measurements. However, the flavour-violating WCs on the r.h.s. also contribute to other observables after integrating out the $W/Z$ gauge bosons.

Here is a list of 4f WET operators involving down-type quarks 
\be
\begin{aligned}
\wcL[V,LL]{\nu d}{ppij} & = {\wc[(1)]{lq}{ppij}-\wc[(3)]{lq}{ppij}} +(\wc[(1)]{Hq}{ij} + \wc[(1)]{Hq}{ij}){\delta_{pp}}\,,  \\
\wcL[V,LL]{e d}{ppij} & = {\wc[(1)]{lq}{ppij}+\wc[(3)]{lq}{ppij}} +(1-2c_w^2)  (\wc[(1)]{Hq}{ij} + \wc[(3)]{Hq}{ij}){\delta_{pp}}\,, \\
\wcL[V,LL]{\nu edu }{ppij} & = 2\wc[(3)]{lq}{ppin}V^*_{nj}-2\wc[(3)]{Hl}{pp} V^*_{ji}-
{2\wc[(3)*]{Hq}{ki} V^*_{jk}  \delta_{pp} } \,.
\end{aligned}
\ee
This suggests correlations between the EWP measurements and neutral and charged current meson decays $d_j \to d_i \nu \bar \nu$ (Class 2) and $d_i \to u_j \ell \bar \nu$ (Class 8) \cite{Kumar:2021yod}. The same holds for the down-type hadronic decays which depend on $\wcL[V,LL]{dd}{}$. Similarly, Class 4 operators, after EWSB also generate 4f operators in WET 
involving up-type quarks. 

On the other hand, the leptonic operators $\wc[(1)]{Hl}{pp}$, $\wc[(3)]{Hl}{pp}$ and $\wc[]{He}{pp}$ contribute to many other semileptonic and pure leptonic processes (through $\wcL[V,LL]{\nu\nu}{}$,  $\wcL[V,LL]{ee}{}$, $\wcL[V,LL]{\nu e}{}$ and $\wcL[V,RR]{ee}{}$, $\wcL[V,RR]{eu}{}$, $\wcL[V,RR]{dd}{}$). This suggests that the EWP observables can be correlated with many other observables through $\rm SU(2)_L$ and CKM rotations.

\section{Charged  LFV Decays (Class 5)}
\label{class5}

In this section, we discuss the effective Lagrangians of the WET and the SMEFT relevant for the charged lepton flavour violating (cLFV) observables. For a review of the experimental status see \cite{Davidson:2022jai}, where further references can be found. No lepton flavour violating decays have been observed until now but the non-vanishing neutrino masses give a hint that they must be present. The observation of any cLFV decay would be a smoking gun for NP at work, since simply adding neutrino masses to the SM is not sufficient to enhance them at the observable level. There is a large interest in this class, because the upper bounds on many of the corresponding processes will be improved in this and next decade by several orders of magnitude. Very importantly most of these decays are theoretically very clean. Selected references to phenomenological analyses in the context of the SMEFT and the WET are listed in Sec.~\ref{sec:obs}.
 
In this class we discuss only the non-hadronic cLFV processes. There are four types:
\be
\begin{aligned}
\textrm{(A)}~ {\Delta F=1}~ \textrm{Decays} &:  {l_i^- \to l_j^- l_p^-  l_p^+} \,, \quad 
\textrm{(B)}~ {\Delta F=2~ \textrm{Decays} : l_i^- \to l_j^- l_j^- l_p^+}\,, \\
\textrm{(C)}~ {\Delta F=1}~  \textrm{Decays} & :  l_i^- \to l_j^- \gamma\,,\quad  \quad 
\textrm{(D)}~ {\Delta F=1}~ \textrm{Decays}: Z\to l_j^+ l_i^- \,. 
\end{aligned}
\ee
$l_i^- \to l_j^- l_p^-  l_p^+$ include processes with three leptons of the same flavour for $(p = j)$ or two distinguishable leptons for $(p \ne j)$, in the final state.\footnote{Obviously, the case $p=i$ is forbidden by kinematics.} Examples of such processes include $\mu \to 3e$ and $\tau \to \mu ee $, respectively. These processes violate lepton flavour by one unit, i.e. $\Delta F=1$. On the other hand, $l_i^- \to l_j^- l_j^-  l_p^+$ processes violate lepton flavour by two units for $(p\ne j)$. An example is $\tau^- \to e^- e^- \mu^+ $. A simultaneous $\tau \to e$ and $\mu \to e$ conversion makes it a $\Delta F=2$ type process. 

A useful collection of branching ratios for such processes can be found in \cite{Crivellin:2013hpa} and its compact version in Chapter~17 of \cite{Buras:2020xsm}.

{\boldmath
\subsection{WET Operators for Class 5 at $\muEW$}
}
The complete list of WET operators at the one-loop level governing  $l_i^- \to l_j^- l_p^-  l_p^+$ ($j=p$ or $j \ne p$) processes in Class 5A at the EW scale is \\
\begin{center}
{\bf WET-5A}
\end{center}
\be 
\label{eq:class5-WETA}
\begin{aligned}
&
\opL[V,LL]{ee}{jipp} \,, \quad
\opL[V,RR]{ee}{jipp}  \,, \quad
\opL[V,LR]{ee}{jipp} \,,  \quad 
\opL[V,LR]{ee}{ppji}\,, \quad
\opL[V,LR]{ee}{jppi} \,, \quad
\opL[V,LR]{ee}{pijp} \,, \\
& 
\opL[S,RR]{ee}{jipp}\,, \quad
\opL[S,RR]{ee}{ijpp}\,, \quad
\opL[S,RR]{ee}{jppi}\,, \quad
\opL[S,RR]{ee}{pjip}\,, \quad 
\underline{\opL[V,LL]{ee}{jikk}} \,, \quad
\underline{\opL[V,RR]{ee}{jikk}}\,, \\
&
\underline{\opL[V,LR]{ee}{jikk}} \,, \quad %
\underline{\opL[V,LR]{ee}{kkji}} \,, \quad
\underline{\opL[V,LL]{ed}{jikk}} \,, \quad 
\underline{\opL[V,LL]{eu}{jikk}} \,, \quad
\underline{\opL[V,LR]{ed}{jikk}} \,, \quad 
\underline{\opL[V,LR]{eu}{jikk}} \,,  \\
&
\underline{\opL[V,LR]{de}{kkji}} \,, \quad
\underline{\opL[V,RR]{ed}{jikk}} \,, \quad
\underline{\opL[V,LR]{ue}{kkji}} \,, \quad
\underline{\opL[V,RR]{eu}{jikk}} \,. 
\end{aligned}
\ee
Here $j,i$, and $p$ are fixed by the external states, but the index $k$ should be summed over 1-3. 

\noindent
Within the WET, the underlined operators contribute at 1-loop via operator mixing with the remaining operators entering the amplitudes for the cLFV decays at tree-level. In what follows we will designate the underlined operators to be {\em 1-loop WET operators}. Since the index $k$ has to be summed over 1-3, the 1-loop operators with all fermion generations can contribute to a given LFV decay.

\noindent
We stress that for operators satisfying the relation $$\mathcal{O}^\dagger \neq\mathcal{O},$$ we also need to add hermitian conjugate partners as per \texttt{WCxf} convention, which is always followed in this work, see \eqref{Lag-WCxf}. Further, the dipole operators such as, $\opL[]{e\gamma}{}$, could be listed above as they contribute to class 5A via $l^+l^-$ emission by the photon, however in \eqref{eq:class5-WETA} we restrict to four-quark operators because dipole operators primarily belong to class 5C.

The $l_i^- \to l_j^- l_j^- l_p^+$ processes ($j \ne p$) in class 5B are driven by
\begin{center}
{{\bf WET-5B}}
\end{center}
\be \label{eq:class5-WETB}
\begin{aligned}
&
\opL[V,LL]{ee}{jijp} \,, \quad
\opL[V,RR]{ee}{jijp}  \,, \quad
\opL[V,LR]{ee}{jijp} \,, \quad
\opL[V,LR]{ee}{jpji} \,, \quad
\opL[S,RR]{ee}{jijp} \,,\quad
\opL[S,RR]{ee}{jpji} \,, \\
&
\opL[S,RR]{ee}{ijpj} \,, \quad
{\opL[S,RR]{ee}{pjij}}\,.
\end{aligned}
\ee

To better understand the structure of the operators in \eqref{eq:class5-WETB}, it is illuminating to look at the Lagrangian in its full form. For example,\footnote{In this context the collection of Fierz identities in Appendix A.3 in \cite{Buras:2020xsm} is useful.} consider
\be
\begin{aligned}
\mathcal{L}_{\rm WET}^{\rm 5B} & \supset 
\wcL[V,LR]{ee}{jijp} \opL[V,LR]{ee}{jijp} +
\wcL[V,LR]{ee}{jpji} \opL[V,LR]{ee}{jpji} +
\wcL[V,LR]{ee}{jijp}^* \opL[V,LR]{ee}{jijp}^\dagger +
\wcL[V,LR]{ee}{jpji}^* \opL[V,LR]{ee}{jpji}^\dagger \\
&=
\wcL[V,LR]{ee}{jijp} \opL[V,LR]{ee}{jijp} -
2\wcL[V,LR]{ee}{jpji} \opL[S,RL]{ee}{jijp} +
\wcL[V,LR]{ee}{jijp}^* \opL[V,LR]{ee}{jijp}^\dagger -
2\wcL[V,LR]{ee}{jpji}^* \opL[S,RL]{ee}{jijp}^\dagger \,.
\end{aligned}
\ee
Here, due to the \texttt{WCxf} convention of our Lagrangian in \eqref{Lag-WCxf}, for all non-hermitian operators, one also needs to add the corresponding hermitian conjugated terms. The reason why it was necessary to include both $\opL[V,LR]{ee}{jijp}$ and $\opL[V,LR]{ee}{jpji}$ in \eqref{eq:class5-WETB}, is simply because the Fierzed conjugate of the latter is a scalar operator with $jijp$ flavour structure, which must be included to complete the basis.

Since the case of scalar operators is subtle, let us consider an explicit example of it. The operator $\opL[S,RR]{ee}{jijp}$ directly contributes to the process of our interest. Like in the VLR case, the Fierzed conjugate of it generates $\opL[T,RR]{ee}{jijp}$, hence $\opL[S,RR]{ee}{jpji}$ must be included in the basis \eqref{eq:class5-WETB}. Furthermore, hermitian conjugation of scalar operators flips chiralities along with flavour indices, i.e.,
\be
\opL[S,RR]{ee}{ijpj}^\dagger  = \opL[S,LL]{ee}{jijp}\,. 
\ee
Therefore, $\opL[S,RR]{ee}{ijpj}$ must be included as well. Further, the Fierz conjugate of $\opL[S,LL]{ee}{jpji}$ leads to a new operator $\opL[T,LL]{ee}{jijp}$ with the appropriate flavor structure. This explains why $\opL[S,RR]{ee}{pjij}$ are included in the basis \eqref{eq:class5-WETB}, since
\be 
\opL[S,RR]{ee}{pjij}^\dagger  = \opL[S,LL]{ee}{jpji}\,.
\ee
In class 5B, the operator mixing in the WET does not lead to any new operators simply because QCD and QED, being unbroken symmetries of the WET, conserve the lepton flavour.

For radiative decays $l_i^- \to l_j^- \gamma$ in class 5C, we have
\begin{center}
{\bf WET-5C}
\end{center}
\be \label{eq:class5-WETC}
\begin{aligned}
\opL[]{e \gamma}{ji}\,, \quad 
\underline{\opL[T,RR]{ed}{jikk}}\,, \quad 
\underline{\opL[T,RR]{eu}{jikk}}\,, \quad 
\underline{\opL[S,RR]{ee}{jkki}}\,.
\end{aligned}
\ee
Here only the dipole operator enters at tree-level whereas the other operators contribute only at the one-loop level via operator mixing. 

We will discuss the class 5D Lagrangian only in the SMEFT context as the $Z$ boson is already integrated out below $\muEW$. Furthermore, it is important to emphasize that in this class  only first leading-log solutions for the RGEs have been used to isolate the relevant WET operators at the one-loop level (using underlined notation) rather than the full LO contribution of the RG improved perturbation theory. But, since LFV decays are in general sensitive to very high scales, the inclusion of operators due to higher powers of logs could be essential. We will discuss such effects in Sec.~\ref{class7} in the context of electric dipole moments \cite{Kumar:2024yuu}, where operator mixing taking place even in multiple steps is found to be important.

\subsection{SMEFT Operators for Class 5 at $\muEW$}
Since the processes of type A, B, C and D depend upon different sets of SMEFT operators, below we discuss them separately.

{\boldmath
\subsubsection{$l_i^- \to l_j^- l_p^-  l_p^+$ Decays ($\Delta F=1$)}
}
For $\Delta F=1$ LFV processes the set of contributing tree-level SMEFT operators at the EW scale is found to be
\begin{center}
{\bf SMEFT-Tree 5A}
\end{center}
\be
\begin{aligned}\label{treeLFV}
&
\ops[]{ll}{} \,,   \quad
\ops[]{l e}{} \,,  \quad 
\ops[]{ee}{}  \,,  \quad 
\ops[(1)]{H l}{}  \,, \quad
\ops[(3)]{H l}{} \,, \quad 
\ops[]{He}{} \,,  \quad
\ops[]{eH}{} \,, 
\\
& 
\underline{ \ops[]{ld}{}}\,, \quad
\underline{\ops[]{lu}{}}\,, \quad 
\underline{\ops[]{qe}{}}\,, \quad 
\underline{\ops[]{ed}{}}\,, \quad 
\underline{\ops[]{eu}{}}\,, \quad 
\underline{\ops[(1)]{lq}{}}\,, \quad 
\underline{\ops[(3)]{lq}{}}.
\end{aligned}
\ee
Here the underlined semileptonic operators contribute to cLFV due to their tree-level matching onto one-loop WET operators. Interestingly, the SMEFT 1-loop RGEs (discussed in Sec:~\ref{sec:SMEFTLambda}) can also give rise to the mixing between semileptonic and leptonic operators, leading to similar effects. However, due to different values of the running gauge couplings in the WET and SMEFT regimes, as well as different structures of the anomalous dimension, the numerical size of these two competing effects within the two theories could differ. 

For the tree-level operators in \eqref{eq:class5-WETA}, the SMEFT-WET matching at $\muEW$ reads 
\be \label{eq:class5A-tree-match}
\begin{aligned} 
\wcL[V,LL]{ee}{jipp} &= \wc[]{ll}{jipp}  + 
\frac{1}{4} {\zeta_1}  C^l_Z {\delta_{pp}}\,, \qquad
\wcL[V,RR]{ee}{jipp} = \wc[]{ee}{jipp}  +  
{1\over 4}(1+  {\zeta_1})  C_Z^{e} {\delta_{pp}}\,, \\
\wcL[V,LR]{ee}{jipp} &= \wc[]{l e}{jipp}  +  (1+  {\zeta_1})  C^l_Z {\delta_{pp}} \,, \quad
\wcL[V,LR]{ee}{ppji} = \wc[]{l e}{ppji}  +  {\zeta_1}  C_Z^{e} {\delta_{pp}} \,, \quad
\wcL[V,LR]{ee}{jppi} = \wc[]{l e}{jppi}  \,, \\
\wcL[S,RR]{ee}{jipp}& = - {[\hat Y_e]_{pp} \over 4 }\frac{v^2}{m_h^2}  \wc[]{eH}{ji}\,.
\end{aligned}
\ee
The remaining operators $\wc[S,RR]{ee}{jppi}$ and $\wc[S,RR]{ee}{ippj}$ being flavour violating in both fermionic currents cannot be generated within the SMEFT at the dim-6 level. Here $C^l_Z$ and $C_Z^{e}$ are defined by
\begin{equation}\label{cZtilde}
C^l_Z = \wc[(1)]{Hl}{ji}   + \wc[(3)]{Hl}{ji} \,,
\qquad
C_Z^{e} = \wc[]{He}{ji} \,,
\end{equation}
and $ {\zeta_1 = 1-2 c_w^2}$. Similarly, for the semileptonic operators in \eqref{eq:class5-WETA}, the matching is given by
\be \label{class5A-match2}
\begin{aligned}
\wcL[V,LL]{ed}{jikk} & = -{1 \over 3} (2-\zeta_1) C_Z^l {\delta_{kk}} 
+ \wc[(1)]{lq}{jikk} + \wc[(3)]{lq}{jikk}  \,, \nn \\
\wcL[V,LL]{eu}{jikk}  &= -{1\over 3} {\zeta_3} C_Z^l {\delta_{kk}} 
+ V_{km} (\wc[(1)]{lq}{jimn} {-}  \wc[(3)]{lq}{jimn}) V^*_{kn} \,, 
\end{aligned}
\ee
\be
\begin{aligned}
\wcL[V,LR]{ed}{jikk}&={1\over 3} (1+ {\zeta_1}) C_Z^l {\delta_{kk}} + \wc[]{ld}{jikk} \,, \qquad 
\wcL[V,LR]{eu}{jikk}=-  {{2\over 3} (1+\zeta_1)}  C_Z^l {\delta_{kk}} + \wc[]{lu}{jikk}\,,   \\
\wcL[V,LR]{de}{kkji} &=-{1\over 3} (2-\zeta_1) {C_Z^e\delta_{kk}}  + \wc[]{qe}{kkji}    \,, \quad
\wcL[V,RR]{ed}{jikk} =  {\frac{1}{3}(1+\zeta_1)C^e_Z\delta_{kk}}  + \wc[]{ed}{jikk}  \,,\\
\wcL[V,LR]{ue}{kkji} & =-{1\over 3} {\zeta_3} C_Z^e\delta_{kk} + {V_{km}} \wc[]{qe}{mnji} {V^*_{kn}}\,, 
\quad 
\wcL[V,RR]{eu}{jikk} = - {\frac{2}{3}(1+\zeta_1)C^e_Z\delta_{kk}} + \wc[]{eu}{jikk}\,,
\end{aligned}
\ee
where $\zeta_3=1-4c_w^2$. In the next subsection, we will not discuss the SMEFT RGEs for semileptonic operators appearing on the r.h.s. of these relations, as the corresponding WET operators on the l.h.s. contribute to cLFV only at loop level. In principle the SMEFT operators running onto one-loop WET operators can affect cLFV, but the net effect is expected to be small, given the double loop suppression. However, it might be worth to investigate such effects case by case which is beyond the scope of our analyses here.

{\boldmath
\subsubsection{$l_i^- \to l_j^- l_p^+  l_j^-$ Decays $(\Delta F=2)$ }
}
The tree-level SMEFT operators governing the $\Delta F=2$ LFV decays, such as $\tau^-\to e^-\mu^+ e^-$ and $\tau^-\to \mu^-e^+ \mu^-$, are found to be
\begin{center}
{\bf SMEFT-Tree 5B}
\end{center}
\be \label{eq:class5B-tree}
\begin{aligned}
\ops[]{ ll}{} \,,   \quad
\ops[]{l e}{} \,,  \quad 
\ops[]{ee}{}\,.
\end{aligned}
\ee
The corresponding SMEFT-WET matching relations at $\muEW$ are given by
\be
\begin{aligned} \label{class5B-match}
\wcL[V,LL]{ee}{jijp} &=  \wc[]{ll}{jijp} \,, \quad \quad
\wcL[V,RR]{ee}{jijp} =  \wc[]{ee}{jijp} \,, \\
\wcL[V,LR]{ee}{jijp} &= \wc[]{l e}{jijp}  \,, \qquad \quad
\wcL[V,LR]{ee}{jpji} = \wc[]{l e}{jpji} \,.
\end{aligned}
\ee
We remind that here $p\ne i$ and $p \ne j$. All other scalar operators in \eqref{eq:class5-WETB} vanish at dim-6 level in the SMEFT.

{\boldmath
\subsubsection{$l_i^- \to l_j^- \gamma$ Decays (Radiative)}
}
The radiative LFV decays like $\mu^- \to e^- \gamma$ are controlled by dim-6 SMEFT dipole and semileptonic tensor operators at tree-level:
\begin{center}
{\bf SMEFT-Tree 5C}
\end{center} 
\be  \label{eq:class5C-smeft-tree}
\begin{aligned}
\ops[]{eW}{}\,, \quad
\ops[]{eB}{}\,, \quad
\underline{\ops[(3)]{lequ}{}}\,.
\end{aligned}
\ee
They match onto dim-5 WET dipoles after EW symmetry-breaking
\be \label{eq:class5-match-C}
\wcL[]{e\gamma}{ji} = \frac{v }{\sqrt{2}} (c_w\wc[]{eB}{ji} - s_w\wc[]{eW}{ji})\,, \quad
\wcL[T,RR]{eu}{jipp} = -\wc[(3)]{lequ}{jipp}\,. 
\ee
The remaining WCs $\wcL[T,RR]{ed}{jikk}$, $\wcL[S,RR]{ee}{jkki}$ do not obtain a matching contribution in the SMEFT at the dim-6 level. Furthermore, we will neglect SMEFT RGE effects for $\wc[(3)]{lequ}{jipp}$, because $\wcL[T,RR]{eu}{jipp}$ is a one-loop WET operator.

{\boldmath
\subsubsection{$Z \to f_j \bar f_i$ Decays}
}
The generic parameterization of $Z$-fermion-fermion couplings is given in \eqref{eq:zcouplingsa} (for quarks) and \eqref{eq:zcouplingsb} (for leptons). They depend upon the following SMEFT set of operators at $\muEW$:
\\

\begin{center}
{\bf SMEFT-Tree 5D}
\end{center}
\be 
\label{eq:class5D-smeft-tree}
\begin{aligned}
\ops[(1)]{Hl}{ji}\,, \quad
\ops[(3)]{Hl}{ji}\,, \quad
\ops[]{He}{ji}\,.
\end{aligned}
\ee
Here the lepton flavour indices can take values 1-3. It would be trivial to generalize this class to non-leptonic LFV decays of the $Z$ boson. In that case the operators $\ops[(1)]{Hq}{ji}\,, \ops[(3)]{Hq}{ji}\,,$ and $ \ops[]{Hu}{ji}$ have to be included. In case of non-leptonic decays, given that we work in the down-basis, WCs with $j=i$ can give rise to $Z\to u_j \bar u_i$, due to a rotation from the weak to the mass basis in the SMEFT (see \eqref{eq:zcouplingsa}). However, for the leptonic case, discussed exclusively here, we assume the PMNS matrix to be unity, hence $j\ne i$ is essential for cLFV $Z$ transitions.

\subsection{SMEFT Operators for Class 5 at $\Lambda$}
\label{sec:SMEFTLambda}
At loop level, the WCs of the tree-level operators identified in \eqref{treeLFV}, \eqref{eq:class5B-tree}, \eqref{eq:class5C-smeft-tree} and \eqref{eq:class5D-smeft-tree} can mix with other SMEFT operators. In this subsection, we systematically discuss the RG running and operator mixing within SMEFT for classes 5A-5D. This will allow us to categorized all possible SMEFT operator that give rise to cLFV decays at low energies. 

As far as gauge coupling running is concerned, for $\Delta F=2$ LFV processes (5B), the SMEFT operators in \eqref{eq:class5B-tree}, do not exhibit a rich operator mixing structure due to their unique flavour structure. In 5C, $\wc[]{eW}{ji}$ and $\wc[]{eB}{ji}$ mix only among themselves. As a result, for $\Delta F=2$ type processes (5B) and radiative decays (5C), we do not find new operators due to 1-loop RGEs. However, the situation is different when including Yukawa running. As mentioned before, we do not discuss the RGEs of the semileptonic SMEFT operators because they are generated only at the 1-loop level in the WET.

Coming to the SMEFT charts, because of the limited operator mixing for 5B, 5C and 5D, the SMEFT charts are pretty simple. Moreover, in class 5D, the leptonic operators are in common with 5A, so SMEFT charts presented  for the latter will partly be applicable for the former. Finally, had we  considered also the non-leptonic processes in 5D, $\ops[(1)]{Hq}{ji}$ and $\ops[(3)]{Hq}{ji}$ {would come into the picture}, however for these the new operators at $\Lambda$ as well as SMEFT charts are already given in Sec.~\ref{sec:class2Lambda}.

Below, we give the simplified form of the RGEs tailored to this class. First, the mixing of four-fermion operators into two-fermion ones is given below. $\wc[(1)]{Hl}{ji}, \wc[(1)]{Hl}{ji}$, and $\wc[]{He}{ji}$ are involved in class 5A and 5D. The corresponding RGEs are given by

\noindent
\underline{\bf \boldmath $f^4 \to f^2 H^2 D$ (Gauge)}:
\be\label{eq:class5-Hl1-Hl3-He}
\begin{aligned}
\dotwc[(1)]{Hl}{ji} &= -\frac{1}{3} g_1^2 (2 \wc[]{ld}{jiww}+2 \wc[]{le}{jiww}+2 \wc[]{ll}{jiww}+\wc[]{ll}{jwwi}-2 \wc[(1)]{lq}{jiww}-4 \wc[]{lu}{jiww})\,,\\
\dotwc[(3)]{Hl}{ji} &= \frac{1}{3} g_2^2 (\wc[]{ll}{jwwi}+6 \wc[(3)]{lq}{jiww})\,,\\
\dotwc[]{He}{ji} &=-\frac{2}{3} g_1^2 (\wc[]{ed}{jiww}+\wc[]{ee}{jiww}-2 \wc[]{eu}{jiww}+\wc[]{le}{wwji}-\wc[]{qe}{wwji})\,.
\end{aligned}
\ee
Here the index $w$ must be summed over 1-3. The running of these operators can in principle be deduced from the flavour conserving case in class 4 given by \eqref{eq:2lep_class4}. The gauge coupling dependent RGEs do not induce mixing of two-fermion $\Delta F=1$ (leptonic) operators with each other except for the  self-mixing. 

The Yukawa dependent operator mixing is given by:

\noindent
\underline{\bf \boldmath $f^4 \to f^2 H^2 D$ (Yukawa)}:
  \be\label{eq:class5-Hl1-Hl3-He-yuk}
\begin{aligned}
\dotwc[(1)]{Hl}{ji} &= 6 y_b^2 (\wc[]{ld}{ji33}-\wc[(1)]{lq}{ji33})+{y_\tau^2 (2 \wc[]{le}{ji33}-\wc[]{ll}{j33i}-2 \wc[]{ll}{ji33})} \\
& +y_t^2 (6 V_{3v} V^*_{3w} \wc[(1)]{lq}{jivw}-6 \wc[]{lu}{ji33})\,, \\
  \dotwc[(3)]{Hl}{ji} &= {-y_\tau^2 \wc[]{ll}{j33i}}-6 y_t^2 V_{3v} V^*_{3w} \wc[(3)]{lq}{jivw}-6 y_b^2 \wc[(3)]{lq}{ji33}\,,\\
\dotwc[]{He}{ji} &= 6 y_b^2 (\wc[]{ed}{ji33}-\wc[]{qe}{33ji})+{2 y_\tau^2 (\wc[]{ee}{ji33}-\wc[]{le}{33ji})}\\
& +y_t^2 (6 V_{3v} V^*_{3w} \wc[]{qe}{vwji}-6 \wc[]{eu}{ji33})\,.
\end{aligned}
\ee

The Yukawa mixing within the same operator class reads

\noindent
\underline{\bf \boldmath {$f^2 H^2 D \to f^2 H^2 D$} (Yukawa)}:

\be \label{eq:f2h2d-f2h2d-yukawa}
\begin{aligned}
 \dotwc[(1)]{Hl}{ji} &= \frac{1}{2} y_\tau^2 (\delta_{j3} (4 \wc[(1)]{Hl}{3i}+9 \wc[(3)]{Hl}{3i})+\delta_{i3} (4 \wc[(1)]{Hl}{j3}+9 \wc[(3)]{Hl}{j3}))+6y_t^2\wc[(1)]{Hl}{ji}\,,\\
\dotwc[(3)]{Hl}{ji} &= \frac{1}{2} y_\tau^2 (\delta_{j3} (3 \wc[(1)]{Hl}{3i}+2 \wc[(3)]{Hl}{3i})+\delta_{i3} (3 \wc[(1)]{Hl}{j3}+2 \wc[(3)]{Hl}{j3}))+6y_t^2 \wc[(3)]{Hl}{ji}\,,\\
\dotwc[]{He}{ji} &=4 y_\tau^2 (\wc[]{He}{3i} \delta_{j3}+\wc[]{He}{j3} \delta_{i3})+6y_t^2\wc[]{He}{ji}\,.
\end{aligned}
\ee

Next, we give the gauge RGEs for the four lepton operators. The allowed flavour indices are $\wc[]{ll}{jipp}$ and $\wc[]{ll}{jppi}$ (Class 5A) 
$\wc[]{ll}{jijp}$ (Class 5B): The gauge dependent RGEs are given by

\noindent
\underline{\bf \boldmath $f^4 \to f^4$ (Gauge)}: 

\be \label{eq:llgauge}
\begin{aligned}
  \dotwc[]{ll}{jipp} &=
\frac{1}{3} g_1^2 \delta_{pp}(\wc[]{ld}{jiww}+\wc[]{le}{jiww}+
\frac{1}{2} \wc[]{ll}{jwwi}+\wc[]{ll}{jiww}-\wc[(1)]{lq}{jiww}-2 \wc[]{lu}{jiww})\\&
+\frac{1}{3} g_2^2 \delta_{jp}(\wc[]{ll}{pwwi}+6\wc[(3)]{lq}{piww})
+\frac{1}{3} g_2^2 \delta_{pi} (\wc[]{ll}{jwwp}+6 \wc[(3)]{lq}{jpww})
\\&
-\frac{1}{6}g_2^2 \delta_{pp} (\wc[]{ll}{jwwi}+6\wc[(3)]{lq}{jiww})
+3 \left(g_1^2-g_2^2\right) {\wc[]{ll}{jipp}}
+6 g_2^2 \wc[]{ll}{jppi}\,,
\\  
\dotwc[]{ll}{jijp} &= 3 \left(g_1^2-g_2^2\right) \wc[]{ll}{jijp}.
\end{aligned}
\ee

The relevant RGEs for $\wc[]{ee}{jipp}$ (Class 5A) and $\wc[]{ee}{jijp}$ (class 5B) are given by:

\be
\begin{aligned}
{\dotwc[]{ee}{jipp}} &=\frac{1}{3} g_1^2 \delta_{pp}(\wc[]{ed}{jiww}+\wc[]{ee}{jiww}-2\wc[]{eu}{jiww}+\wc[]{le}{wwji}-\wc[]{qe}{wwji})
\\&
+\frac{1}{3} g_1^2 \delta_{jp} (\wc[]{ed}{piww}+\wc[]{ee}{piww}-2\wc[]{eu}{piww}+\wc[]{le}{wwpi}-\wc[]{qe}{wwpi})
\\&
+\frac{1}{3} g_1^2 \delta_{pi} (\wc[]{ed}{jpww}+\wc[]{ee}{jpww}-2\wc[]{eu}{jpww}+\wc[]{le}{wwjp}-\wc[]{qe}{wwjp})
+12g_1^2 \wc[]{ee}{jipp}\,,
\\
\dotwc[]{ee}{jijp} &= 12 g_1^2 \wc[]{ee}{jijp}.
\end{aligned}
\ee

RGEs for $\wc[]{le}{jipp}$, $\wc[]{le}{ppji}$ (Class 5A), and $\wc[]{le}{jijp}$, $\wc[]{le}{jpji}$ (Class 5B) are given by

\be
\begin{aligned}
\dotwc[]{le}{jipp} &= \frac{4}{3} g_1^2 \delta_{pp} (\wc[]{ld}{jiww}+\wc[]{le}{jiww}+\frac{1}{2}\wc[]{ll}{jwwi}+ \wc[]{ll}{jiww}-\wc[(1)]{lq}{jiww}-2 \wc[]{lu}{jiww})-6 g_1^2 \wc[]{le}{jipp}\,,\\
\dotwc[]{le}{ppji} &= \frac{2}{3} g_1^2 \delta_{pp} (\wc[]{ed}{jiww}+\wc[]{ee}{jiww}-2 \wc[]{eu}{jiww}+\wc[]{le}{wwji}-\wc[]{qe}{wwji})-6 g_1^2 \wc[]{le}{ppji}\,,
\\
\dotwc[]{le}{jijp} &= -6 g_1^2 \wc[]{le}{jijp}\,, \quad
\dotwc[]{le}{jpji} = -6 g_1^2 \wc[]{le}{jpji}\,.
\end{aligned}
\ee

\noindent
\underline{\bf \boldmath $ f^2 H^2 D \to f^4$ (Gauge)}:

\be
\begin{aligned}
\dotwc[]{ll}{jipp} &=
-\frac{1}{6} \delta_{pp} \left(g_1^2 \wc[(1)]{Hl}{ji}+g_2^2 \wc[(3)]{Hl}{ji}\right)+\frac{1}{3} g_2^2 \wc[(3)]{Hl}{pi} \delta_{jp}+\frac{1}{3} g_2^2 \wc[(3)]{Hl}{jp} \delta_{ip}\,,
\\
{\dotwc[]{ee}{jipp}} &= -\frac{1}{6} g_1^2 (\wc[]{He}{jp} \delta_{pi}+\delta_{pp} \wc[]{He}{ji})-\frac{1}{6} g_1^2 \wc[]{He}{pp} \delta_{jp}\,,
\\
\dotwc[]{le}{jipp} &= -\frac{2}{3} g_1^2 \delta_{pp} \wc[(1)]{Hl}{ji}\,,\quad
\dotwc[]{le}{ppji} = -\frac{1}{3} g_1^2 \delta_{pp} \wc[]{He}{ji}\,.
\end{aligned}
\ee

The Yukawa dependence for the four-fermion operators goes as

\noindent
\underline{\bf \boldmath $f^2 H^2 D \to f^4$ (Yukawa)}:

\be
\begin{aligned}
\dotwc[]{ll}{jipp} &=
-\frac{1}{2} y_\tau^2 \delta_{p3} (\wc[(1)]{Hl}{ji}+2 \delta_{j3} \wc[(3)]{Hl}{pi}+2 \delta_{i3} \wc[(3)]{Hl}{jp}-\wc[(3)]{Hl}{ji})\,,
\\
{\dotwc[]{ee}{jipp}} &= y_\tau^2 \delta_{p3} \wc[]{He}{ji}\,, \quad
\dotwc[]{le}{jipp} = 2 y_\tau^2 \delta_{p3} \wc[(1)]{Hl}{ji}\,,\quad
\dotwc[]{le}{ppji} = -y_\tau^2 \delta_{p3} \wc[]{He}{ji}\,.
\end{aligned}
\ee

\noindent
\underline{\bf \boldmath $f^4 \to f^4$ (Yukawa)}:

\be
\begin{aligned}
\label{eq:lleele-yukawa}
\dotwc[]{ll}{jipp} &= \frac{1}{2} y_\tau^2 (\delta_{p3} (-\wc[]{le}{ji33}+\wc[]{ll}{3pji}+\wc[]{ll}{p3ji})+\delta_{j3} \wc[]{ll}{3ipp}+\delta_{i3} \wc[]{ll}{j3pp})\,,\\
{\dotwc[]{ll}{jijp}} &= \frac{1}{2} y_\tau^2 (\delta_{i3} \wc[]{ll}{j3jp}+\delta_{j3} \wc[]{ll}{3ijp}+\delta_{j3} \wc[]{ll}{ji3p}+\delta_{p3} \wc[]{ll}{jij3})\,,\\
\dotwc[]{ee}{jipp} &= y_\tau^2 (\delta_{p3} (\wc[]{ee}{3pji}+\wc[]{ee}{p3ji}-\wc[]{le}{33ji})+\delta_{j3} \wc[]{ee}{3ipp}+\delta_{i3} \wc[]{ee}{j3pp})\,,\\
{\dotwc[]{ee}{jijp}} &= y_\tau^2 (\delta_{i3} \wc[]{ee}{j3jp}+\delta_{j3} \wc[]{ee}{3ijp}+\delta_{j3} \wc[]{ee}{ji3p}+\delta_{p3} \wc[]{ee}{jij3})\,, \\
\dotwc[]{le}{}&=y_\tau y_tF_1(\wc[(1)]{lequ}{},\wc[(1)*]{lequ}{})+y_\tau y_b F_2(\wc[]{ledq}{},\wc[*]{ledq}{}) \\
&+
y_\tau^2 F_3( \wc[]{le}{}, \wc[*]{le}{},\wc[]{ll}{},\wc[]{ee}{})\,,
\end{aligned}
\ee
where the running of the last WC is given in full detail in the supplemental material.

\noindent
\underline{\bf \boldmath $f^2XH \to f^2XH$ (Gauge)}:

$\wc[]{eB}{ji}$ and $\wc[]{eW}{ji}$ belong to Class 5C. The corresponding RGEs are

\be
\begin{aligned}\label{eq:eBeWgauge}
\dotwc[]{eB}{ji} &= \frac{151}{12} g_1^2 \wc[]{eB}{ji}-\frac{9}{4} g_2^2 \wc[]{eB}{ji}-\frac{3}{2} g_1 g_2 \wc[]{eW}{ji}\,,\\
\dotwc[]{eW}{ji} &= -\frac{1}{2} g_1 g_2 \wc[]{eB}{ji}+\frac{1}{4} g_1^2 \wc[]{eW}{ji}-\frac{11}{12} g_2^2 \wc[]{eW}{ji}\,.
\end{aligned}
\ee

\noindent
\underline{\bf \boldmath $f^4 \to f^2XH$ (Yukawa)}:

\be
\begin{aligned}\label{eq:eBeWyuk4F}
\dotwc[]{eB}{ji} &= 10 g_1 y_t V_{3v} \wc[(3)]{lequ}{jiv3}\,,\quad
\dotwc[]{eW}{ji} = -6 g_2 y_t V_{3v} \wc[(3)]{lequ}{jiv3}\,.\\
\end{aligned}
\ee

\noindent
\underline{\bf \boldmath non-fermion $\to f^2XH$ (Yukawa):}

\be
\begin{aligned}\label{eq:eBeQyukno4F}
\dotwc[]{eB}{} &= y_t^2 F_1(\wc[]{eB}{})+
y_\tau F_2( \wc[]{H B}{},\wc[]{H\tilde B}{},\wc[]{HWB}{},\wc[]{H\tilde WB}{})+
y_\tau^2 F_3(\wc[]{eB}{})\,, \\
\dotwc[]{eW}{} &= y_t^2F_1(\wc[]{eW}{})+
y_\tau F(\wc[]{HW}{},\wc[]{H\tilde W}{},\wc[]{HWB}{},\wc[]{H\tilde WB}{})+
y_\tau^2 F(\wc[]{eW}{})\,.
\end{aligned}
\ee

The RGE for $\wc[]{eH}{ip}$ in Class 5A is given by

\noindent
\underline{\bf \boldmath $f^2XH \to f^2H^3$ (Gauge)}:

\begin{equation}
\dotwc[]{eH}{ip} = 9 g_1^3 \wc[]{eB}{ip}-3 g_1 g_2^2 \wc[]{eB}{ip}+9 g_1^2 g_2 \wc[]{eW}{ip}-9 g_2^3 \wc[]{eW}{ip}\,.
\end{equation}

\noindent
\underline{\bf \boldmath $f^4  \to f^2 H^3 $ (Yukawa)}:

\begin{align}
\dotwc[]{eH}{ip} &=  2 \left(\left(4 y_\tau^3-2 \lambda y_\tau\right) \wc[]{le}{i33p}+3 y_b \left(\lambda-2 y_b^2\right) \wc[]{ledq}{ip33}+3y_t(2y_t^2 - \lambda ) V_{3v} \wc[(1)]{lequ}{ipv3}\right)\, .
\label{CeH}
\end{align}

Inspecting the RG equations above, we can identify the all new operators at $\Lambda$ that have an impact on the WCs at $\muEW$. For $\Delta F=1$ (5A) LFV decays these are:
\begin{center}
\textrm{\bf SMEFT-Loop 5A}
\end{center}
\be \label{eq:class5:smeft-lambda1}
\begin{aligned} 
\textrm{\bf Yukawa-mixing:}&~ 
{ \ops[]{ld}{}}\,, \quad
{\ops[]{lu}{}}\,, \quad
{\ops[]{qe}{}}\,, \quad
{\ops[]{ed}{}}\,, \quad
{\ops[]{eu}{}}\,, \quad
{\ops[(1)]{lq}{}}\,, \quad
{\ops[(3)]{lq}{}}\,, \\
&
\ops[(1)]{\ell e qu }{}\,, \,\
\ops[]{ ledq }{}.
\\
\textrm{\bf Gauge-mixing:} 
& ~
{ \ops[]{ld}{}}\,, \quad
{\ops[]{lu}{}}\,, \quad
{\ops[]{qe}{}}\,, \quad
{\ops[]{ed}{}}\,, \quad
{\ops[]{eu}{}}\,, \quad
{\ops[(1)]{lq}{}}\,, \quad
{\ops[(3)]{lq}{}}.
\\
\end{aligned}
\ee
For 5B and 5C we have:
\begin{center}
{\bf SMEFT-Loop 5B }
\end{center}
\be \label{class5B-smeftopsL}
\begin{aligned}
\textrm{\bf Yukawa-mixing: } 
& ~
\ops[]{ledq}{}\,, \,\
\ops[(1)]{lequ}{} \,.
\\
\textrm{\bf Gauge-mixing:} & ~\textrm{No new {operators}} .
\end{aligned}
\ee

\begin{center}
{\bf SMEFT-Loop 5C}
\end{center}
\be \label{class5C-smeftopsL}
\begin{aligned}
{\textrm{\bf Yukawa-mixing: }} & ~ 
\ops[]{ HW }{}\,, \,\
\ops[]{ H \widetilde W }{}\,, \,\
\ops[]{ H B}{}\,, \,\
\ops[]{ H \widetilde B}{}\,, \,\
\ops[]{ H W B}{}\,, \,\
\ops[]{ H \widetilde W B}{}\,, \,\
\ops[(3)]{lequ}{}.\\
{\textrm{\bf Gauge-mixing:}} & ~\textrm{No new {operators}} .
\end{aligned}
\ee
For 5D, we find following new operators at $\Lambda$:
\begin{center}
{\bf SMEFT-Loop 5D}
\end{center}
\be \label{class5D-smeftopsL}
\begin{aligned}
{\textrm{\bf Yukawa-mixing: }}  ~ &
\ops[(1)]{lq}{}\,, \quad
\ops[(3)]{lq}{}\,, \quad
\ops[]{qe}{}\,, \quad
\ops[]{lu}{}\,, \quad
\ops[]{ld}{}\,, \quad
\ops[]{eu}{}\,, \quad
\ops[]{ed}{}\,, \\
&
\ops[]{ll}{}\,, \quad
\ops[]{le}{}\,, \quad
\ops[]{ee}{}\,, 
\\
{\textrm{\bf Gauge-mixing:}}  ~ &
\textrm{The same {operators} as for the Yukawa-mixing}.
\end{aligned}
\ee

For $|\eta| \ge 5\cdot 10^{-6}$, the largest RG running effects are exhibited by following SMEFT operators in 5A and 5C for $\tau \to 3\mu$ and $\tau \to  \mu \gamma$ ($\Delta F=1$  decays):

\be \begin{aligned} &
\wcs[(3)]{lq}{}\to \wcL[V,LL]{ee}{ }\,, \quad
\wcs[(1)]{lq}{}\to \wcL[V,LL]{ee}{ }\,, \quad
\wcs[]{lu}{}\to \wcL[V,LL]{ee}{ }\,, \quad
\wcs[]{ll}{}\to \wcL[V,LL]{ee}{ }\,, 
\\ & 
\wcs[]{le}{}\to \wcL[V,LL]{ee}{ }\,, \quad
\wcs[]{ld}{}\to \wcL[V,LL]{ee}{ }\,, \quad
\wcs[]{eu}{}\to \wcL[V,RR]{ee}{ }\,, \quad
\wcs[]{qe}{}\to \wcL[V,RR]{ee}{ }\,, 
\\ & 
\wcs[]{ee}{}\to \wcL[V,RR]{ee}{ }\,, \quad
\wcs[]{ed}{}\to \wcL[V,RR]{ee}{ }\,, \quad
\wcs[]{le}{}\to \wcL[V,RR]{ee}{ }\,, \quad
\wcs[]{qe}{}\to \wcL[V,LR]{ee}{ }\,, 
\\ & 
\wcs[]{eu}{}\to \wcL[V,LR]{ee}{ }\,, \quad
\wcs[]{ee}{}\to \wcL[V,LR]{ee}{ }\,, \quad
\wcs[]{le}{}\to \wcL[V,LR]{ee}{ }\,, \quad
\wcs[]{ed}{}\to \wcL[V,LR]{ee}{ }\,, 
\\ & 
\wcs[(1)]{lequ}{}\to \wcL[V,LR]{ee}{ }\,, \quad
\wcs[]{lu}{}\to \wcL[V,LR]{ee}{ }\,, \quad
\wcs[(1)]{lq}{}\to \wcL[V,LR]{ee}{ }\,, \quad
\wcs[(3)]{lq}{}\to \wcL[V,LR]{ee}{ }\,, 
\\ & 
\wcs[]{ll}{}\to \wcL[V,LR]{ee}{ }\,, \quad
\wcs[]{ld}{}\to \wcL[V,LR]{ee}{ }\,, \quad
\wcs[]{le}{}\to \wcL[V,LR]{ee}{ }\,, \quad
\wcs[(3)]{lequ}{}\to \wcL[]{e \gamma }{ }\,, \quad
\\ & 
\wcs[]{eW}{}\to \wcL[]{e \gamma }{ }\,, \quad
\wcs[]{eB}{}\to \wcL[]{e \gamma }{ }
. \end{aligned} \ee
The flavour indices are suppressed here for simplicity.
In 5B, for $\Delta F=2$ decays $\tau^- \to \mu^- e^+ \mu^-$, we do not find any relevant operator mixing for the chosen value of the $\eta$. We refer to the $\rho-\eta$ Tabs.~\ref{tab:rhoeta5-1} -\ref{tab:rhoeta5-3} for more details.

\begin{figure}[H]%
\includegraphics[clip, trim=2.cm   12.6cm 1.5cm 12cm, width=1.0\textwidth]{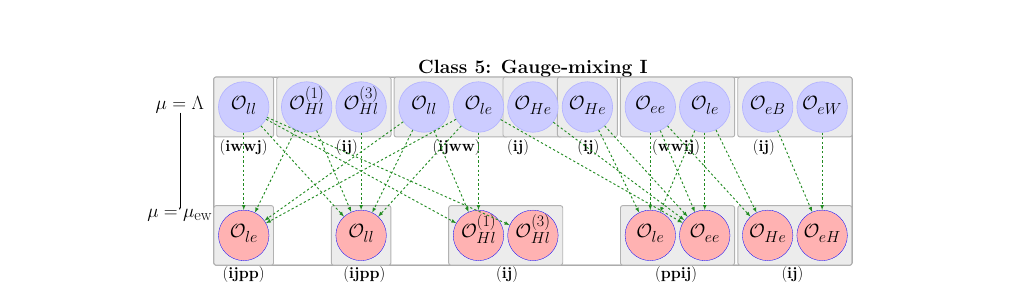}
\includegraphics[clip, trim=2.cm   12.5cm 1.5cm 12.5cm, width=1.0\textwidth]{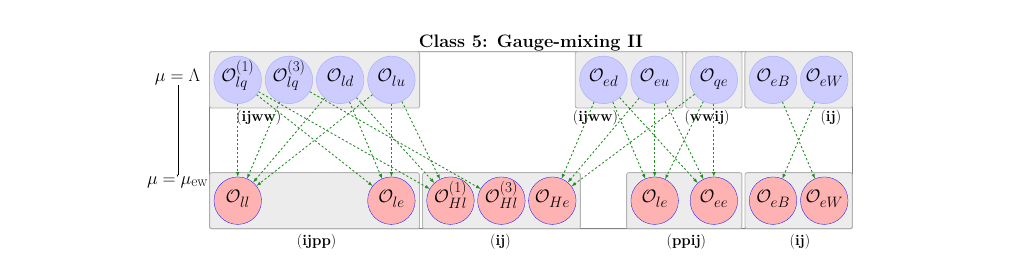}
\caption{ \small Class 5 -- Operator mixing for leptonic LFV processes of $\Delta F=1$ type, in the Warsaw down-basis. Dashed green lines show mixing due to electroweak gauge couplings. The self-mixing has been omitted. These charts are {valid}  for the SMEFT operators entering at tree-level in WET given {by} the first line of \eqref{treeLFV}. For some WCs at $\Lambda$ we have suppressed the flavour indices for simplicity, {they} can be found in the corresponding RGEs.}
\label{chart:class5-gauge1} 
\end{figure}

\begin{figure}[htb]%
\hspace{-45mm}
\includegraphics[clip, trim=1.5cm 13.2cm 1.2cm 13.0cm, width=1.5\textwidth]{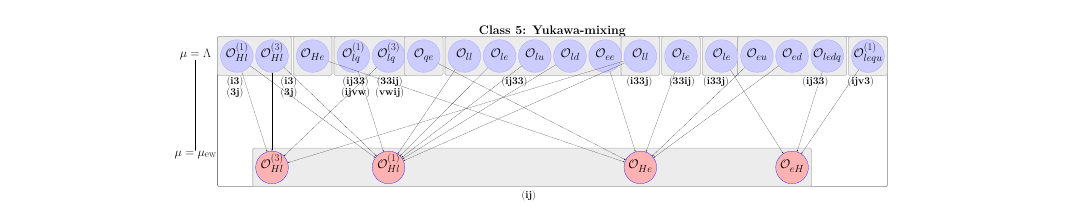}
\caption{ \small Class 5 -- Yukawa mixing of operators governing the leptonic LFV processes of $\Delta F=1$ type, in the Warsaw down-basis. The black lines indicate the mixing due to the top, bottom and tau Yukawa couplings. The self-mixing is not shown. This chart is {also valid only}  for the SMEFT operators that match at tree-level onto the WET \eqref{treeLFV}. Since the four-fermion and SMEFT dipole operators exhibit a complicated operator mixing pattern, we do not show them here. We refer to {the} corresponding RGEs given in the text {as well as to the} supplemental material. See text for more details. }%
\label{chart:class5-yuk}%
\end{figure}

The explicit flavour structures of the corresponding WCs can also be found in the RGEs given above or the SMEFT charts. 
The charts are shown in Fig.~\ref{chart:class5-gauge1} (for gauge couplings) and in 
Fig.~\ref{chart:class5-yuk} (for top and bottom Yukawa couplings). A few important comments about the charts are in order:
\begin{itemize}
\item The charts in Figs.~\ref{chart:class5-gauge1} and \ref{chart:class5-yuk} are applicable only to the SMEFT operators that contribute to the matching at tree-level, corresponding to the first line in \eqref{treeLFV}, leaving aside SMEFT operators that match onto 1-loop WET operators. 
\item For this class we do not find any operator mixing due to QCD.
\end{itemize}
Finally, we note that the semileptonic SMEFT operators can enter class 5 via two routes (i) tree-level SMEFT operators matching onto 1-loop WET operators (see second line of \eqref{treeLFV}) (ii) the SMEFT operators {running} into SMEFT operators at $\muEW$ that match at tree-level onto Class 5 WET operators, which are listed in \eqref{eq:class5:smeft-lambda1}, \eqref{class5B-smeftopsL}, \eqref{class5C-smeftopsL}, and \eqref{class5D-smeftopsL}. The latter operators can further mix with other SMEFT operators which is an NLL effect, but might be relevant given the highly sensitive nature of LFV processes.

Until this point, our discussion was focused on the qualitative behaviour of RGE running and operator mixing. In order to get quantitative estimates of various effects, in Tabs.~\ref{tab:rhoeta5-1}, \ref{tab:rhoeta5-2} and \ref{tab:rhoeta5-3}, we show the values of $\rho$ and $\eta$ parameters for Class 5 operators in the JMS basis.
{\boldmath
\subsection{$\text{SU(2)}_L$ and RG Correlations}
}

A summary of some interesting $\text{SU(2)}_L$ correlations of Class 5 with other processes is given in Tab.~\ref{tab:class5-su2}. Apart from this, Class 5 SMEFT WCs also match onto many other WET WCs due to $\text{SU(2)}_L$. For example,
$\wcL[V,RR]{eu}{} $,
$\wcL[V,RR]{ed}{} $,
$\wcL[V,LL]{\nu u}{} $,
$\wcL[V,LL]{\nu d}{} $,
$\wcL[V,LL]{\nu e  d u}{} $,
$\wcL[V,LR]{\nu e}{} $,
$\wcL[V,LR]{\nu u}{} $, and
$\wcL[V,LR]{\nu d}{} $, etc.

At the end, which correlations will turn out to be phenomenologically important will depend upon the experimental precision of the involved processes and specific scenarios. 
A detailed numerical study of them  is beyond the scope of the present work and will be investigated elsewhere.

There are many analyses of the decays discussed in this class. In the context of the SMEFT they can be found in Sec.~\ref{sec:obs}, in particular in Tab.~\ref{tab:LOSMEFTobsflavor}. Here we just list selected papers which are particularly useful for $\mu\to e$ transitions including $\mu\to e$ conversion in nuclei and correlations with other processes. First we mention \cite{Crivellin:2017rmk}, where a RG improved analyses of $\mu\to e$ processes in a systematic EFT was performed. More recent analyses can be found in \cite{Davidson:2020ord,Davidson:2020hkf,Cirigliano:2022ekw,Haxton:2022piv,Ardu:2023yyw,Ardu:2024bua,Haxton:2024lyc,Delzanno:2024ooj}. See also \cite{Buras:2021btx} where the global analysis of leptophilic $Z^\prime$ bosons was performed. Here we only  discuss briefly the case of RG effects in models with vector-like leptons.

Among possible NP scenarios are the ones with vector-like leptons instead of vector-like quarks mentioned briefly in Class 1. This time we deal with neutral flavour violating RH and LH lepton currents, with implications through RG evolution on semi-leptonic operators. Now the operators $\ops[(1,3)]{Hl}{}$ and $\ops[]{He}{}$ are the driving force. The complex-valued coefficients of these operators are
\be
\label{eq:SMEFT-wilson-coeffs}
\wc[(1)]{Hl}{ab}\,, \qquad \wc[(3)]{Hl}{ab}\,, \qquad \wc{He}{ab}\,,
\ee
where the indices $a,b = 1,2,3$ denote the different generations of up- and down-type leptons, that is neutrinos and charged leptons.

Through RG Yukawa interactions, these operators generate contributions to semileptonic 4-fermion operators. Keeping only $Y_u$ effects one finds (following \eqref{eq:class6-f2H2D-to-f4-yukawa1})

\begin{table}[H]
\begin{center}
\renewcommand*{\arraystretch}{1.0}
\resizebox{1.0\textwidth}{!}{
\begin{tabular}{ |c|c|c| }
\hline
\multicolumn{3}{|c|}{$\text{SU}(2)_L$ correlations for Class 5} \\
\hline
Class 5 WC at $\muEW$ &  $\text{SU}(2)_L$ correlation   & correlated processes  \\
\hline
$\wc[]{ll}{jipp}$ &$\wcL[V,LL]{\nu\nu}{jipp} = \wc[]{ll}{jipp}$ &  $\nu \nu \to \nu \nu$   \\
"  &$\wcL[V,LL]{\nu e}{jipp} = \wc[]{ll}{jipp}$ & $\nu \ell \to \nu \ell$   \\
$\wc[]{le}{jipp}$ &$\wcL[V,LR]{\nu e}{jipp} = \wc[]{le}{jipp}$ & $\nu \ell \to \nu \ell$  \\
$\wc[(1)]{Hl}{ji}, \wc[(3)]{Hl}{ji}$  &$ \wcL[V,LL]{\nu \nu}{jipp} = {1\over 4}(\wc[(1)]{Hl}{ji} -\wc[(3)]{Hl}{ji}){\delta_{pp}} $ & $\nu \nu \to \nu \nu$    \\
" &  $ \wcL[V,LL]{\nu e}{jipp} = \zeta_1 (\wc[(1)]{Hl}{ji} -\wc[(3)]{Hl}{ji}){\delta_{pp}} $ &  $\nu \ell \to \nu \ell$   \\
" &  $ \wcL[V,LL]{\nu e}{ppji} = (\wc[(1)]{Hl}{ji} +\wc[(3)]{Hl}{ji}){\delta_{pp}} $ & $\ell_i \to \ell_j \nu \bar \nu$    \\
" &  $ \wcL[V,LL]{ed}{jipp} = -{{1+2c_w^2} \over 3} (\wc[(1)]{Hl}{ji} +\wc[(3)]{Hl}{ji}){\delta_{pp}} $ & $\ell_i \to \ell_j d \bar d$   \\
" &  $ \wcL[V,LL]{eu}{jipp} = -{\zeta_2 \over 3} (\wc[(1)]{Hl}{ji} +\wc[(3)]{Hl}{ji}){\delta_{pp}} $ & $\ell_i \to \ell_j u \bar u$    \\
\hline
\end{tabular}
}
\caption{Summary of $\text{SU(2)}_L$ correlations for Class 5 at tree-level in the WET and SMEFT. The first column shows Class 5 SMEFT operators at $\muEW$, the second column lists the other WET 
operators generated by the operators in the first column. The third column shows processes that are generated by the WET operators in the second column. Here, $\zeta_1=1-2c_w^2$. We assume the SMEFT down-basis together with the PMNS matrix to be set to unity.}
\label{tab:class5-su2}
\end{center}
\end{table}

{\boldmath $f^4$} $(\overline{L}L)(\overline{L}L)$:
\begin{align}
  \label{eq:ADM-Yuk-lq1}
  \dotwc[(1)]{lq}{abij} & 
  = y_t^2  V^*_{3i}V_{3j} \wc[(1)]{Hl}{ab}\,,
\qquad
  \dotwc[(3)]{lq}{abij}  
  = -y_t^2  V^*_{3i}V_{3j}\wc[(3)]{Hl}{ab}\,, 
\end{align}

{\boldmath $f^4$} $(\overline{R}R)(\overline{R}R)$:
\begin{align}
  \dotwc{eu}{abij} & 
  = -2 y_t^2\,\delta_{i3}\delta_{j3}  \wc{He}{ab}\,, 
\end{align}

{\boldmath $f^4$} $(\overline{L}L)(\overline{R}R)$:
\begin{align}
  \dotwc{\ell u}{abij} & 
  = -2 y_t^2\,\delta_{i3}\delta_{j3}  \wc[(1)]{H\ell}{ab}\,, 
\qquad
  \dotwc{q e}{ijab} 
  = y_t^2 V^*_{3i}V_{3j}  \wc{He}{ab}\,.
\end{align}
These RG effects, being enhanced by the top-Yukawa coupling imply important constraints on models with vector-like leptons through semi-leptonic decays.

\section{Semileptonic LFV Decays (Class 6)} 
\label{class6}

In this section, we discuss two types of semileptonic LFV decays: 
\begin{enumerate}[(A)]
\item quark flavour-conserving processes, such as $\tau \to P \ell$ or $\tau \to V \ell$ for $P =\pi$, $V=\rho$.
\item quark flavour-violating processes, such as $B\to K \ell \ell'$, $B \to K^* \ell \ell'$ and $K\to\pi \ell \ell'$.
\end{enumerate}
There is a rich literature on these decays. A collection of references to phenomenological analyses in NP models like $Z^\prime$ models and Littlest Higgs Model with T-parity (LHT) and the current limits on the branching ratios can be found in Section 17.2 of \cite{Buras:2020xsm}.

{\boldmath 
\subsection{WET Operators for Class 6 at $\muEW$}}

Both quark flavour-conserving and violating LFV processes are governed by 
the following WET operators \cite{Kumar:2018kmr, Aebischer:2018iyb}
\begin{center}
{\bf WET-Tree 6A, 6B}
\end{center}
\begin{equation}\label{eq:class6-wet}
\begin{aligned} 
&
\opL[V,LL]{ed}{jiqp} \,,  \quad
\opL[V,LR]{ed}{jiqp} \,, \quad 
\opL[V,LR]{de}{qpji}  \,,  \quad
\opL[V,RR]{ed}{jiqp} \,,  \quad
\opL[T,RR]{ed}{jiqp} \,,  \quad 
\opL[S,RL]{ed}{jiqp} \,, \quad 
\opL[S,RR]{ed}{jiqp}\,, \quad
\opL[]{e\gamma}{ji} \,, \\
&
\opL[V,LL]{eu}{jiqp} \,,  \quad
\opL[V,LR]{eu}{jiqp} \,, \quad 
{\opL[V,LR]{ue}{qpji}}  \,,  \quad
\opL[V,RR]{eu}{jiqp} \,,  \quad
\opL[T,RR]{eu}{jiqp} \,,  \quad 
\opL[S,RL]{eu}{jiqp} \,, \quad 
\opL[S,RR]{eu}{jiqp}\,.
\end{aligned}
\end{equation}
In Class 6A, the tensor and dipole operators do not contribute the $\tau \to  P \ell$ processes and the scalar operators do not contribute to $\tau \to V \ell$ \cite{Aebischer:2018iyb}.

In Class 6B, the semileptonic decays violate quark and lepton flavour simultaneously. At the quark-level they involve $l_i \to l_j q_q \bar q_p$ or $q_p \to q_q l_j \bar l_i$ transitions. The governing WET operators are the same as given in \eqref{eq:class6-wet} but 
with $q \ne p$.

Note that the hermitian conjugate of some operators results in chirality flipped operators. For example, the chirality flipped SLL operator can be expressed in terms of a SRR operator using  
\be
\opL[S,LL]{ed}{jiqp} = \opL[S,RR]{ed}{ijpq}^\dagger\,.
\ee
Therefore, such operators must be included in the Class 6 basis to fully describe the LFV decays. This applies to dipole, scalar, and tensor operators. Specifically, \eqref{eq:class6-wet} should be supplemented by the following operators:

\begin{center}
{\bf WET-Tree 6A, 6B continue}
\end{center}
\be \label{eq:class6AB-wettree2}
\begin{aligned} 
&
\opL[T,RR]{ed}{ijpq} \,,  \quad 
\opL[S,RL]{ed}{ijpq} \,, \quad 
\opL[S,RR]{ed}{ijpq}\,, \quad
\opL[]{e\gamma}{ij} \,, \\
&
\opL[T,RR]{eu}{ijpq} \,,  \quad 
\opL[S,RL]{eu}{ijpq} \,, \quad 
\opL[S,RR]{eu}{ijpq}\,.
\end{aligned}
\ee
On the other hand the vector operators in \eqref{eq:class6-wet} do not flip chirality under hermitian conjugation. Given the fact that Class 6B operators violate both quark and lepton flavours simultaneously, the 1-loop QCD+QED WET RGEs do not lead to non-trivial operator mixing except the self-mixing. In Class 6A, operator mixing and running within WET can be important. 

\subsection{SMEFT Operators for Class 6 at $\muEW$ }
Having discussed the WET Lagrangian, in this section, we construct the most general SMEFT Lagrangian governing Class 6 at the tree-level.

\subsubsection{Lepton Flavour Violation {(Class 6A)}}
Class 6A processes are controlled by the following SMEFT operators at $\muEW$
\begin{center}
{\bf SMEFT-Tree 6A}
\end{center}
\be \label{class6A-smeftopsT}
\begin{aligned}
&
\ops[(1)]{l q}{} \,,  \quad
\ops[(3)]{ l q}{} \,,  \quad
\ops[]{ l u}{}  \,, \quad
\ops[]{ l d}{}     \,,   \quad
\ops[]{ed}{}  \,,\quad 
\ops[]{eu}{} \,, \quad
\ops[]{qe}{}   \,,  \\
&
\ops[]{ l edq}{}   \,, \quad
\ops[(1)]{ l equ}{} \,,  \quad
\ops[(3)]{ l e qu}{}  \,,  \quad
\ops[(1)]{H l}{}\,,\quad
\ops[(3)]{H l}{}\,, \quad
\ops[]{He}{}\,.
\end{aligned}
\ee
This list can be found from the matching conditions. For the operators involving down-quarks, we find
\be \label{class6A-match1}
\begin{aligned}
\wcL[V,LL]{ed}{jipp} & = {{1 \over 3} (\zeta_1-2)} \wc[l]{Z}{} {\delta_{pp}} 
+ \wc[(1)]{l q}{jipp}+\wc[(3)]{l q}{jipp} \,, \quad \quad
\wcL[V,RR]{ed}{jipp}  =  {{1 \over 3} (\zeta_1+1)}  \wc[e]{Z}{} {\delta_{pp}} +  \wc[]{ed}{jipp}   \,, \\
\wcL[V,LR]{ed}{jipp} & = {{1 \over 3} (\zeta_1+1)} \wc[l]{Z}{} {\delta_{pp}} +  \wc[]{ld}{jipp} \,, \quad \quad
\wcL[V,LR]{de}{ppji}  = {{1 \over 3} (\zeta_1-2)}\wc[e]{Z}{} {\delta_{pp}} + \wc[]{qe}{ppji}\,, \quad
\wcL[S,RL]{ed}{jipp} = \wc[]{ l e dq}{jipp}\,. 
\end{aligned}
\ee
Here $\zeta_1= 1-2 c_w^2$, and 
\be 
\wc[l]{Z}{}  = \wc[(1)]{H l}{ji} + \wc[(3)]{H l}{ji}\,, \quad 
\wc[e]{Z}{} = \wc[]{He}{ji}\,.
\ee 
The WCs $\wcL[T,RR]{ed}{jipp}$ and $\wcL[SRR]{ed}{jipp}$ have a vanishing matching conditions at the dim-6 level. The matching for the dipole operator can be found in \eqref{eq:class5-match-C}. For the up-quark operators, the matching reads
\be \label{class6A-match2}
\begin{aligned}
\wcL[V,LL]{eu}{jipp} & = - {{\zeta_3} \over 3} \wc[l]{Z}{} {\delta_{pp}}
+  V_{pm} (\wc[(1)]{l q}{jimn}-\wc[(3)]{l q}{jimn}) {V^*_{pn}}  \,, \quad \quad
\wcL[V,RR]{eu}{jipp}  = {-{2 \over 3} (\zeta_1+1)} \wc[e]{Z}{} {\delta_{pp}} +   \wc[]{eu}{jipp}   \,, \\
\wcL[V,LR]{eu}{jipp} & = {-{2 \over 3} (\zeta_1+1)} \wc[l ]{Z}{} {\delta_{pp}} +  \wc[]{l u}{jipp} \,, \quad \quad 
\wcL[V,LR]{ue}{ppji}  = - {{\zeta_3} \over 3} \wc[e]{Z}{} {\delta_{pp}} + V_{pm}\wc[]{qe}{mnji} {V^*_{pn}}  \,, \\
\wcL[S,RR]{eu}{jipp} & = - V_{pm}  \wc[(1)]{lequ}{jimp} \,, \quad
\wcL[T,RR]{eu}{jipp}  = - V_{pm}  \wc[(3)]{lequ}{jimp}\,. 
\end{aligned}
\ee
Here $\zeta_3= 1-4 c_w^2$. The WC $\wcL[SRL]{eu}{jipp}$ has a vanishing matching condition at the dim-6 level. We remind the reader that the CKM matrix always acts on the flavour index of the LH quark doublet denoted here by $q$. As evident from \eqref{class6A-match2} in addition to flavour diagonal operators, through CKM mixing the ones with off-diagonal quark flavour index also match onto the contributing WET operators. Such operators can also trigger the LFV decays of mesons belonging to Class 6B.

By adjusting the flavour indices appropriately, it is trivial to obtain the matching for the additional operators in \eqref{eq:class6AB-wettree2}.

\subsubsection{Quark and Lepton Flavour Violation (Class 6B)}

Class 6B processes are controlled by the following SMEFT operators at $\muEW$
\begin{center}
{\bf SMEFT-Tree 6B}
\end{center}
\be \label{class6B-smeftopsT}
\begin{aligned}
&
\ops[(1)]{l q}{} \,,  \quad
\ops[(3)]{ l q}{} \,,  \quad
\ops[]{ l u}{}  \,, \quad
\ops[]{ l d}{}     \,,   \quad
\ops[]{ed}{}  \,,\quad
\ops[]{eu}{} \,, \quad
\ops[]{qe}{}   \,,  \\
&
\ops[]{ l edq}{}   \,, \quad
\ops[(1)]{ l equ}{} \,,  \quad
\ops[(3)]{ l e qu}{}\,.
\end{aligned}
\ee
While some of the operator structures are in common with Class 6A, their flavour indices differ. In this case, the matching onto the WET down-quark operators, for $p \ne q$ reads 
\be \label{eq:class6B-match-tree-down}
\begin{aligned}
\wcL[V,LL]{ed}{jiqp} & =  \wc[(1)]{lq}{jiqp}+\wc[(3)]{lq}{jiqp} \,, \quad \quad
\wcL[V,RR]{ed}{jiqp}  =   \wc[]{ed}{jiqp}   \,, \\
\wcL[V,LR]{ed}{jiqp} & =  \wc[]{ld}{jiqp} \,, \quad \quad
\wcL[V,LR]{de}{qpji}  =    \wc[]{qe}{qpji}  \,, \quad
\wcL[S,RL]{ed}{jiqp} & = \wc[]{ledq}{jiqp}\,.
\end{aligned}
\ee
As for the case of 6A, in 6B the WC $\wcL[T,RR]{ed}{jiqp}$ and $\wcL[SRR]{ed}{jiqp}$ have vanishing matching conditions at the dim-6 level.  

For the up-quark operators, the matching {is given by}
\be \label{eq:class6B-match-tree-up}
\begin{aligned}
\wcL[V,LL]{eu}{jiqp} & =   V_{qm} (\wc[(1)]{l q}{jimn}-\wc[(3)]{l q}{jimn}) {V^*_{pn}}  \,, \quad \quad
\wcL[V,RR]{eu}{jiqp}  =    \wc[]{eu}{jiqp}   \,, \\
\wcL[V,LR]{eu}{jiqp} & =   \wc[]{l u}{jiqp} \,, \quad \quad
\wcL[V,LR]{ue}{qpji}  =    V_{qm}    \wc[]{qe}{mnji} {V^*_{pn}}  \,, \\
\wcL[S,RR]{eu}{jiqp} & = - V_{qm}  \wc[(1)]{l e qu}{jimp}  \,, \quad
\wcL[T,RR]{eu}{jiqp}  = - V_{qm}  \wc[(3)]{l e qu}{jimp}\,.
\end{aligned}
\ee
Unlike 6A, in 6B due to violation of quark flavour, these relations do not involve $\wc[(1)]{Hq}{ji}$, $\wc[(3)]{Hq}{ji}$ and $\wc[]{He}{ji}$ WCs. Below we explore the SMEFT Lagrangian at 1-loop level including effects due to RGEs.

\subsection{SMEFT Operators for Class 6 at $\Lambda$}
Now we move on to construct the SMEFT Lagrangian at 1-loop level for Classes 6A and 6B. Interestingly, pure leptonic operators also come into the picture via RG running from $\Lambda$ down to $\muEW$ and affect the SMEFT WCs at $\muEW$ just listed. This becomes clear from the explicit RGEs. 

They are given below for Classes 6A and 6B with the following distinction:
\be
\text{Class 6A}:  s=t, \qquad \text{Class 6B}: s\neq t\,.
\ee
Therefore for Class 6A, the dominant operators have $\delta_{s t}=1$ while for Class 6B $\delta_{s t}=0$ so that only few terms survive. However, $s \neq t$ flavour violating operators can also subdominantly affect Class 6A due to CKM rotations in the tree-level matching as discussed before. 

We remind that these RGEs are tailored to non-redundant basis as adopted in this review. 

\noindent
\underline{\bf \boldmath $f^4 \to f^4$ (Gauge)}: First we list the 4f to 4f operator mixing.

For the $(LL)(LL)$ operators, we have
\be
\begin{aligned}\label{eq:lq3_gauge_full}
\dotwc[(1)]{lq}{prst} &= -\frac{2}{9} g_1^2 \delta_{st} (\wc[]{ld}{prww}+\wc[]{le}{prww}+\wc[]{ll}{prww}+\frac{1}{2}\wc[]{ll}{pwwr} -\wc[(1)]{lq}{prww}-2 \wc[]{lu}{prww})-g_1^2 \wc[(1)]{lq}{prst}+9 g_2^2 \wc[(3)]{lq}{prst}\,,\\
\dotwc[(3)]{lq}{prst} &= \frac{2}{3} g_2^2 \delta_{st} (\frac{1}{2}\wc[]{ll}{pwwr}+3 \wc[(3)]{lq}{prww})+3 g_2^2 \wc[(1)]{lq}{prst}-\left(g_1^2+6 g_2^2\right) \wc[(3)]{lq}{prst}\,.\\
\end{aligned}
\ee
For the $(LL)(RR)$ operators, we have

\be
\begin{aligned}
\dotwc[]{lu}{prst} &= \frac{4}{9} g_1^2 (9 \wc[]{lu}{prst}-2 \delta_{st} (\wc[]{ld}{prww}+\wc[]{le}{prww}+\wc[]{ll}{prww}+\frac{1}{2}\wc[]{ll}{pwwr}-\wc[(1)]{lq}{prww}-2 \wc[]{lu}{prww}))\,,\\
\dotwc[]{ld}{prst} &= -\frac{2}{9} g_1^2 (9 \wc[]{ld}{prst}-2 \delta_{st} (\wc[]{ld}{prww}+\wc[]{le}{prww}+\wc[]{ll}{prww}+\frac{1}{2}\wc[]{ll}{pwwr}-\wc[(1)]{lq}{prww}-2 \wc[]{lu}{prww}))\,,\\
\dotwc[]{qe}{stpr} &= -\frac{2}{9} g_1^2 (\delta_{st} (\wc[]{ed}{prww}+\wc[]{ee}{prww}-2 \wc[]{eu}{prww}+\wc[]{le}{wwpr}-\wc[]{qe}{wwpr})-9 \wc[]{qe}{stpr})\,.\\
\end{aligned}
\ee
For the $(RR)(RR)$ operators the RGEs are given by
\be
\begin{aligned}
\dotwc[]{ed}{prst} &= \frac{4}{9} g_1^2 (\delta_{st} (\wc[]{ed}{prww}+\wc[]{ee}{prww}
-2 \wc[]{eu}{prww}+\wc[]{le}{wwpr}-\wc[]{qe}{wwpr})+9 \wc[]{ed}{prst})\,,\\
\dotwc[]{eu}{prst} &= -\frac{8}{9} g_1^2 (\delta_{st} (\wc[]{ed}{prww}+\wc[]{ee}{prww}
-2 \wc[]{eu}{prww}+\wc[]{le}{wwpr}-\wc[]{qe}{wwpr})+9 \wc[]{eu}{prst})\,.\\
\end{aligned}
\ee
The scalar operators only undergo self-mixing or mix among themselves:

\be
\begin{aligned}\label{eq:scal_gauge}
\dotwc[]{ledq}{prst} &= -\frac{8}{3} \left(g_1^2+3 g_s^2\right) \wc[]{ledq}{prst}\,,\\
\dotwc[(1)]{lequ}{prst} &= \left(-\frac{11 g_1^2}{3}-8 g_s^2\right) \wc[(1)]{lequ}{prst}+6 \left(5 g_1^2+3 g_2^2\right) \wc[(3)]{lequ}{prst}\,,\\
\dotwc[(3)]{lequ}{prst} &= \frac{1}{8} \left(5 g_1^2+3 g_2^2\right) \wc[(1)]{lequ}{prst}+\left(\frac{2 g_1^2}{9}-3 g_2^2+\frac{8 g_s^2}{3}\right) \wc[(3)]{lequ}{prst}\,.\\
\end{aligned}
\ee

\noindent
\underline{\bf \boldmath {$f^2 H^2 D \to f^4$} (Gauge)}:

\noindent
\newline
Since the lepton flavour must be violated in this class, only leptonic $f^2 H^2 D$ operators can mix with four-fermi operators in contrast to \eqref{eq:class2-f2H2D-to-f4-yukawa1} in Class 2 where two-quark operators were involved. Recall that semileptonic operators involving up-quarks also contribute to Class 6 unlike Class 2.  

\be \label{eq:class6-f2H2D-to-f4-gauge}
\begin{aligned}
\dotwc[(1)]{lq}{prst} &= \frac{1}{9} g_1^2 \wc[(1)]{Hl}{pr} \delta_{st}\,, \quad
\dotwc[(3)]{lq}{prst} =  \frac{1}{3} g_2^2 \wc[(3)]{Hl}{pr} \delta_{st}\,,\\
\dotwc[]{lu}{prst} &=  \frac{4}{9} g_1^2 \wc[(1)]{Hl}{pr} \delta_{st}\,, \quad
\dotwc[]{ld}{prst} = -\frac{2}{9} g_1^2 \wc[(1)]{Hl}{pr} \delta_{st}\,, \\
\dotwc[]{qe}{stpr} &=  \frac{1}{9} g_1^2 \wc[]{He}{pr} \delta_{st}\,, \quad
\dotwc[]{ed}{prst} =  -\frac{2}{9} g_1^2 \wc[]{He}{pr} \delta_{st}\,, \\
\dotwc[]{eu}{prst} &=  \frac{4}{9} g_1^2 \wc[]{He}{pr} \delta_{st}\,.
\end{aligned}
\ee

\noindent
\underline{\bf \boldmath {$f^4 \to f^2 H^2 D$} (Gauge)}:

\noindent
\newline
This running can be deduced from results in Class 5:
\be
\begin{aligned}
\wc[(1)]{Hl}{pr}\,, \wc[(3)]{Hl}{pr}\,, \wc[]{He}{pr} &:{ \eqref{eq:class5-Hl1-Hl3-He}}\,.
\end{aligned}
\ee

\noindent
\underline{\bf \boldmath {$f^2 H^2 D \to f^2 H^2 D$} (Gauge)}:

\noindent
\newline
The corresponding RGEs can again be deduced from results in Class 4
\be
\begin{aligned}
\wc[(1)]{Hl}{pr}\,, \wc[(3)]{Hl}{pr}\,, \wc[]{He}{pr} &: {\eqref{eq:Hl_gauge}}\,.
\end{aligned}
\ee

\begin{figure}[H]%
\includegraphics[clip, trim=4.2cm 12cm 1.2cm 12.5cm, width=1.2\textwidth]{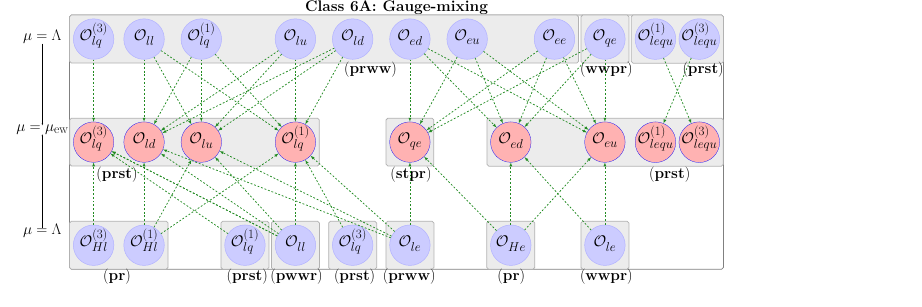}
\caption{\small Class 6A -- The mixing of four-fermion and two-fermion operators with semileptonic operators. The dashed green lines indicate the mixing due to electroweak gauge couplings. QCD does not cause any operator mixing in this case. Moreover, for Class 6A we have quark flavour diagonal WCs at $\muEW$, i.e. $s=t$. For the $s\ne t$ case (Class 6B), the mixing due to gauge couplings is very limited due to the presence of $\delta_{st}$ in the RGEs. See \eqref{eq:lq3_gauge_full}-\eqref{eq:class6-f2H2D-to-f4-gauge} for more details. The index $w$ is summed over 1-3, while all other indices are fixed by the external states in a process. The self-mixing has been omitted.}
\label{chart:lfv-qfc-gauge}
\end{figure}

Next, we look at the Yukawa dependent RGEs which do not bring in any new Lorentz structure.

\noindent
\underline{\bf \boldmath $f^4 \to f^4$ (Yukawa)}:

\noindent
\\
Similar to Classes 3 and 5 the expressions in the current subsection turn out to be very long. Therefore, we present only the contributing operators, suppressing the flavour indices but showing the dependence on Yukawa couplings in order of importance. The full RGEs with indices can be found in the appended supplemental material. 

As in Classes 3 and 5 the functions $F_i$ indicate that there are additional factors in front of the WCs like CKM factors, with the structure explicitly shown in simpler equations. Typically, a given entry in $F_i$ represents several operators with different flavour indices. We have then

\be
\begin{aligned}\label{eq:lq3_yuk}
\dotwc[(1)]{lq}{} &= y_t^2 F_1(\wc[(1)]{lq}{},\wc[]{lu}{})+
y_t y_\tau F_2(\wc[(1)]{lequ}{},\wc[(3)]{lequ}{}{,\wc[(1)*]{lequ}{},\wc[(3)*]{lequ}{}})\\
&+ y_b^2F_3(\wc[]{ld}{},\wc[(1)]{lq}{}) 
 +y_b y_\tau F_4(\wc[]{ledq}{}{,\wc[*]{ledq}{}})
+y_\tau^2 F_5(\wc[(1)]{lq}{})\,,\\
\dotwc[(3)]{lq}{} &=y_t^2F_1(\wc[(3)]{lq}{})+
y_t y_\tau F_2(\wc[(1)]{lequ}{},\wc[(3)]{lequ}{},\wc[(1)*]{lequ}{},\wc[(3)*]{lequ}{})  \\
& +y_b^2F_3(\wc[(3)]{lq}{})+
y_b y_\tau F_4(\wc[]{ledq}{}{,\wc[*]{ledq}{}}) +
y_\tau^2 F_5(\wc[(3)]{lq}{})\,,
\end{aligned}
\ee
\be
\begin{aligned}
\dotwc[]{qe}{} &=y_t^2F_1(\wc[]{eu}{},\wc[]{qe}{})+
y_t y_\tau F_2(\wc[(1)]{lequ}{},\wc[(3)]{lequ}{}{,\wc[(1)*]{lequ}{},\wc[(3)*]{lequ}{}})
\\
&
+y_b^2F_3(\wc[]{ed}{}, \wc[]{qe}{})
+
y_b y_\tau F_4(\wc[]{ledq}{},\wc[*]{ledq}{})
+ y_\tau^2 F_5(\wc[]{qe}{})\,,\\
\dotwc[]{lu}{} &=y_t^2 F_1(\wc[(1)]{lq}{},\wc[]{lu}{})+
y_\tau y_t F_2(\wc[(1)]{lequ}{},\wc[(3)]{lequ}{}{,\wc[(1)*]{lequ}{},\wc[(3)*]{lequ}{}})+
y_\tau^2 F_3(\wc[]{lu}{})\,,\\
\dotwc[]{ld}{} &=y_b y_\tau F_1(\wc[]{ledq}{}{,\wc[*]{ledq}{}})+
y_b^2 F_2(\wc[]{ld}{},\wc[(1)]{lq}{})+
y_\tau^2 F_3(\wc[]{ld}{})\,,\\
\dotwc[]{ed}{} &= y_b^2 F_1(\wc[]{ed}{},\wc[]{qe}{})+ y_b y_\tau F_2(\wc[]{ledq}{}{,\wc[*]{ledq}{}}) +  y_\tau^2 F_3(\wc[]{ed}{})\,,\\
\dotwc[]{eu}{} &= y_t^2 F_1(\wc[]{eu}{},\wc[]{qe}{})+
y_\tau y_t F_2(\wc[(1)]{lequ}{},\wc[(3)]{lequ}{}{,\wc[(1)*]{lequ}{},\wc[(3)*]{lequ}{}})+
y_\tau^2F_3(\wc[]{eu}{})\,.
\end{aligned}
\ee

The running for the scalar and tensor operators is given by:
\be
\begin{aligned}\label{eq:scal_yuk}
\dotwc[]{ledq}{} &=y_t^2 F_1(\wc[]{ledq}{})+y_b y_t F_2(\wc[(1)]{lequ}{})+ y_b^2 F_4(\wc[]{ledq}{})\\
&+y_b y_\tau F_3(\wc[]{ed}{},\wc[]{ld}{},\wc[]{le}{},\wc[(1)]{lq}{},\wc[(3)]{lq}{},\wc[]{qe}{})+ y_\tau^2 F_5(\wc[]{ledq}{})\,,\\
\dotwc[(1)]{lequ}{} &=y_t^2F_1(\wc[(1)]{lequ}{})+ y_b y_tF_2(\wc[]{ledq}{})+  y_t y_\tau F_3(\wc[]{eu}{},\wc[]{le}{},\wc[(1)]{lq}{}, \wc[(3)]{lq}{},\wc[]{lu}{},\wc[]{qe}{})\\&
+  y_b^2 F_4(\wc[(1)]{lequ}{})+y_\tau^2F_5(\wc[(1)]{lequ}{})\,,\\
\dotwc[(3)]{lequ}{} &= y_t^2  F_1(\wc[(3)]{lequ}{})+
y_t y_\tau F_2(\wc[]{eu}{},\wc[(1)]{lq}{},\wc[(3)]{lq}{},\wc[]{lu}{},
\wc[]{qe}{})+  y_b^2 F_3(\wc[(3)]{lequ}{}) +
y_\tau^2 F_4(\wc[(3)]{lequ}{})\,.
\end{aligned}
\ee

\noindent
\underline{\bf \boldmath {$f^2 H^2 D \to f^4$} (Yukawa)}:

\be \label{eq:class6-f2H2D-to-f4-yukawa1}
\begin{aligned}
\dotwc[(1)]{lq}{prst} &= y_t^2 V^*_{3s} V_{3t} \wc[(1)]{Hl}{pr}\,, \quad
\dotwc[(3)]{lq}{prst} = -y_t^2 V^*_{3s} V_{3t} \wc[(3)]{Hl}{pr} \,,\\
\dotwc[]{lu}{prst} &=  -2 y_t^2 \delta_{s3} \delta_{t3} \wc[(1)]{Hl}{pr}\,, \quad
\dotwc[]{ld}{prst} = 2 y_b^2 \delta_{s3} \delta_{t3} \wc[(1)]{Hl}{pr}\,, \\
\dotwc[]{qe}{stpr} &= y_t^2 V^*_{3s} V_{3t} \wc[]{He}{pr} \,, \quad
\dotwc[]{ed}{prst} = 2 y_b^2 \delta_{s3} \delta_{t3} \wc[]{He}{pr} \,, \\
\dotwc[]{eu}{prst} &=  -2 y_t^2 \delta_{s3} \delta_{t3} \wc[]{He}{pr}\,.
\end{aligned}
\ee

Note that $f^2 H^2 D$ operators do not mix with scalar semileptonic operators at LO. Also, we observe that  $\wc[]{ld}{}, \wc[]{lu}{}, \wc[]{eu}{}$ and $\wc[]{ed}{}$ operators can only have flavour diagonal indices in the above RGEs.  

Given that operators on r.h.s. can directly contribute to Class 6A at the tree-level, the matching effects are likely to win over 1-loop RG running.

\begin{figure}[H]%
\includegraphics[clip, trim=2.0cm 10.7cm 1.0cm 11.cm, width=1.0\textwidth]{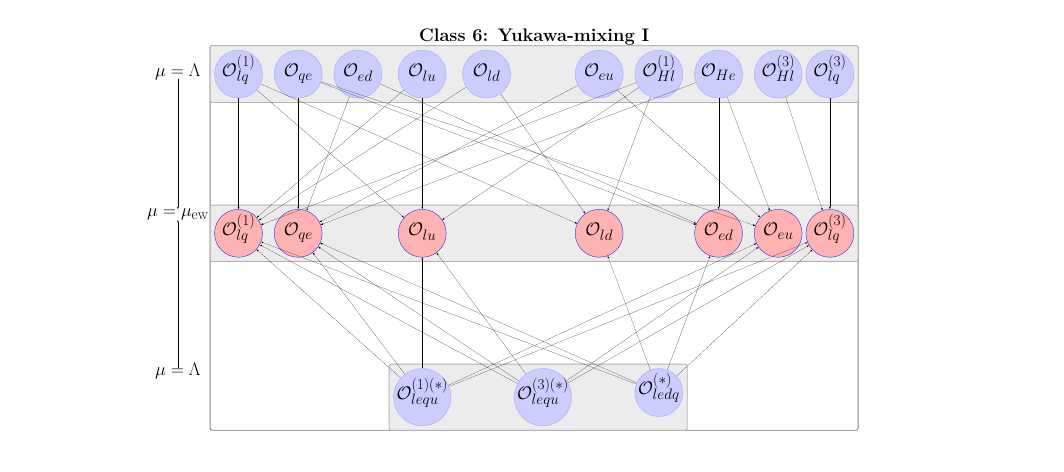}
\caption{ \small Class 6A: Operator mixing for semileptonic LFV processes, which are quark flavour conserving. The black lines indicate the mixing due to top and bottom Yukawa couplings. The self-mixing is not shown here. For simplicity we have suppressed flavour indices which can be found in the full RGEs given in the supplemental material.}%
\label{chart:lfv-qfc-yukawa1}%
\end{figure}

\noindent
\underline{\bf \boldmath $f^4 \to f^2 H^2 D$ (Yukawa)}:

\noindent
\newline
This running can be deduced from results in Class 5
\be
\begin{aligned}
\wc[(1)]{Hl}{pr}\,, \wc[(3)]{Hl}{pr}\,, \wc[]{He}{pr} &: \eqref{eq:class5-Hl1-Hl3-He-yuk}\,.
\end{aligned}
\ee

\noindent
\underline{\bf \boldmath {$f^2 H^2 D \to f^2 H^2 D$} (Yukawa)}:

\be
\begin{aligned}
\wc[(1)]{Hl}{pr}, \wc[(3)]{Hl}{pr}, \wc[]{He}{pr} &: \eqref{eq:f2h2d-f2h2d-yukawa}\,.
\end{aligned}
\ee
This anatomy of ADMs enables the identification of new operator structures that are generated purely through RGE evolution. For Class 6A, the new operators are summarized as:
\begin{center}
{\bf SMEFT-Loop 6A}
\end{center}
\be \label{class6A-smeftopsL}
\begin{aligned}
{\textrm{\bf Yukawa-mixing: }} & ~ \ops[]{ll}{}\,, 
\quad \ops[]{le}{}\,, 
\quad \ops[]{ee}{}\,. 
 \\
{\textrm{\bf Gauge-mixing:}} &
~ \ops[]{l l}{} 
 \,,  \quad
\ops[]{l e}{}\,, \quad
\ops[]{ee}{}\,.
\end{aligned}
\ee
Similar to Class 3 there are very few new one-loop operators. 

\begin{figure}[tbp]%
\includegraphics[clip, trim=2cm 10cm 0.0cm 10cm, width=1.0\textwidth]{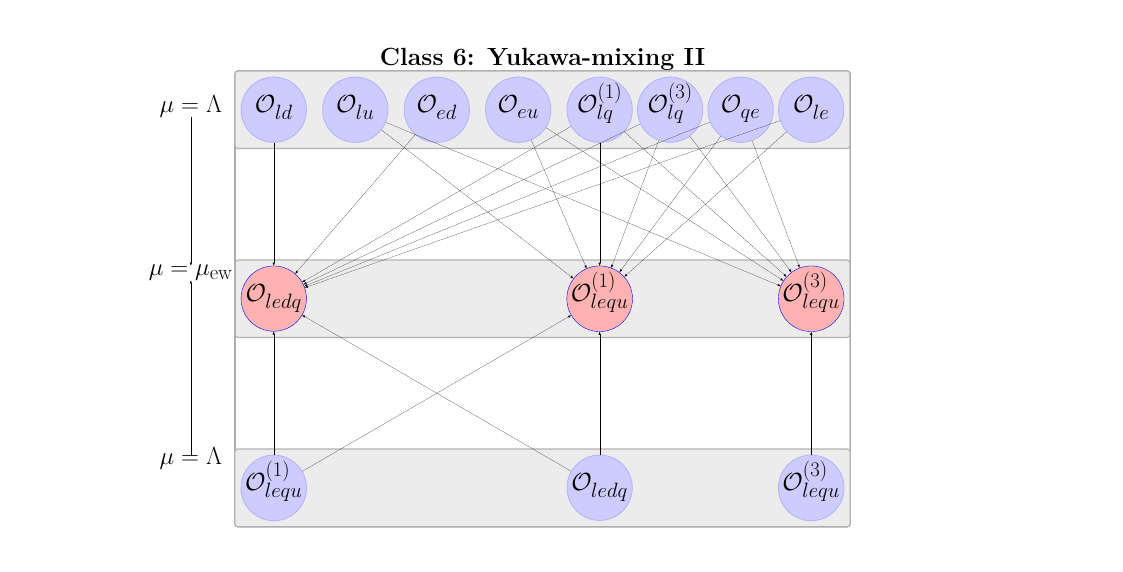}
\caption{ \small Class 6: Operator mixing for LFV processes. The black lines indicate the mixing due to the top and bottom Yukawa couplings. The self-mixing is not shown here. Also, or simplicity we have suppressed flavour indices which can be found in the corresponding RGEs.}%
\label{chart:lfv-qfc-yukawa2}%
\end{figure}

The flavour structure of the operators at $\muEW$ emerging beyond the tree-level can be found in the corresponding SMEFT charts or full RGEs. The operators contributing at the NP scale are shown in Figs.~\ref{chart:lfv-qfc-gauge} and \ref{chart:lfv-qfc-yukawa1}, \ref{chart:lfv-qfc-yukawa2}. In Fig.~\ref{chart:lfv-qfc-gauge} for the gauge couplings, for Class 6A, we need to set $s=t$ in the WCs at $\muEW$ in the corresponding RGEs. The SMEFT charts for $\wc[(1)]{Hl}{ji}$, $\wc[(3)]{Hl}{ji}$ and $\wc[]{He}{ji}$\footnote{Note the different choice of dummy flavour indices indicating the flavour violation: in the matching conditions (ji) and charts as well RGEs (pr).} are not shown here since these are already given in other classes, for instance see Class 5 Figs.~\ref{chart:class5-gauge1} (gauge) and \ref{chart:class5-yuk} (Yukawa). 

Now let us summarize the 1-loop operators for Class 6B, in which simultaneously quark and lepton flavours are violated. Interestingly, in this case, no new structures 
are introduced at $\Lambda$ due to gauge operator mixing.
\begin{center}
{\bf SMEFT-Loop 6B}
\end{center}
\be \label{class6B-smeftopsL}
\begin{aligned}
\textrm{\bf Yukawa-mixing: } & ~ 
\ops[]{le}{}\,.
\\
{\textrm{\bf Gauge-mixing:}} & ~\textrm{No new structures} . 
\end{aligned}
\ee
Indeed in Class 6B, due to gauge interactions, we only have self-mixing in most cases, apart from a few exceptions. For example, for $p \ne q$
\begin{itemize}
\item Apart from self mixing, only $\wc[(1)]{lq}{jiqp}$ run into $\wc[(3)]{lq}{jiqp}$ due to gauge couplings and vice versa.
\item Apart from self mixing, only $\wc[(1)]{lequ}{jiqp}$ run into $\wc[(3)]{lequ}{jiqp}$ due to gauge couplings and vice versa.
\item All other lines in Fig.~\ref{chart:lfv-qfc-gauge} are absent. 
\item All lines in Fig.~\ref{chart:lfv-qfc-yukawa1} remain unchanged.
\end{itemize}
Also, for top and bottom Yukawa mixing there are some differences with respect to Class 6A. 

Finally, in order to identify the most significant operators  which contribute to Class 6A at 1-loop level, using the condition $|\eta| \ge 10^{-6}$, for $\tau \to \mu (u \bar u) $, we find
\be \begin{aligned} &
\wcs[(3)]{lq}{}\to \wcL[V,LL]{eu}{ }\,, \quad
\wcs[(1)]{lq}{}\to \wcL[V,LL]{eu}{ }\,, \quad
\wcs[]{lu}{}\to \wcL[V,LL]{eu}{ }\,, \quad
\wcs[]{ll}{}\to \wcL[V,LL]{eu}{ }\,, 
\\ & 
\wcs[]{le}{}\to \wcL[V,LL]{eu}{ }\,, \quad
\wcs[]{ld}{}\to \wcL[V,LL]{eu}{ }\,, \quad
\wcs[]{lu}{}\to \wcL[V,LR]{eu}{ }\,, \quad
\wcs[(1)]{lq}{}\to \wcL[V,LR]{eu}{ }\,, 
\\ & 
\wcs[(3)]{lq}{}\to \wcL[V,LR]{eu}{ }\,, \quad
\wcs[]{ll}{}\to \wcL[V,LR]{eu}{ }\,, \quad
\wcs[]{ld}{}\to \wcL[V,LR]{eu}{ }\,, \quad
\wcs[]{le}{}\to \wcL[V,LR]{eu}{ }\,, 
\\ & 
\wcs[]{qe}{}\to \wcL[V,LR]{ue}{ }\,, \quad
\wcs[]{eu}{}\to \wcL[V,LR]{ue}{ }\,, \quad
\wcs[]{ee}{}\to \wcL[V,LR]{ue}{ }\,, \quad
\wcs[]{le}{}\to \wcL[V,LR]{ue}{ }\,, 
\\ & 
\wcs[]{ed}{}\to \wcL[V,LR]{ue}{ }\,, \quad
\wcs[]{eu}{}\to \wcL[V,RR]{eu}{ }\,, \quad
\wcs[]{qe}{}\to \wcL[V,RR]{eu}{ }\,, \quad
\wcs[]{ee}{}\to \wcL[V,RR]{eu}{ }\,, 
\\ & 
\wcs[]{ed}{}\to \wcL[V,RR]{eu}{ }\,, \quad
\wcs[]{le}{}\to \wcL[V,RR]{eu}{ }\,, \quad
\wcs[(1)]{lequ}{}\to \wcL[T,RR]{eu}{ }\,, \quad
\wcs[(3)]{lequ}{}\to \wcL[S,RR]{eu}{ }
. \end{aligned} \ee

Likewise, in Class 6B, for $b\to s \tau^+ \mu^-$ decays, we find:

\be \begin{aligned} &
\wcs[(1)]{lq}{}\to \wcL[V,LL]{ed}{ }\,, \quad
\wcs[(3)]{lq}{}\to \wcL[V,LL]{ed}{ }\,, \quad
\wcs[]{lu}{}\to \wcL[V,LL]{ed}{ }\,, \quad
\wcs[(1)]{Hl}{}\to \wcL[V,LL]{ed}{ }\,, 
\\ & 
\wcs[(3)]{Hl}{}\to \wcL[V,LL]{ed}{ }\,, \quad
\wcs[]{ld}{}\to \wcL[V,LR]{ed}{ }\,, \quad
\wcs[]{qe}{}\to \wcL[V,LR]{de}{ }\,, \quad
\wcs[]{eu}{}\to \wcL[V,LR]{de}{ }\,, 
\\ & 
\wcs[]{He}{ }\to \wcL[V,LR]{de}{ }\,, \quad
\wcs[(3)]{lequ}{}\to \wcL[V,LR]{de}{ }\,, \quad
\wcs[(1)]{lequ}{}\to \wcL[V,LR]{de}{ }\,, \quad
\wcs[]{ed}{}\to \wcL[V,RR]{ed}{ }\,, 
\\ & 
\wcs[]{ledq}{}\to \wcL[S,RL]{ed}{ }\,, \quad
\wcs[(1)]{lequ}{}\to \wcL[S,RL]{ed}{ }\,, \quad
\wcs[(3)]{lq}{}\to \wcL[S,RL]{ed}{ }\,, \quad
\wcs[(1)]{lq}{}\to \wcL[S,RL]{ed}{ }\,, 
\\ & 
\wcs[]{ledq}{}\to \wcL[S,RL]{ed}{ }\,, \quad
\wcs[(1)]{lequ}{}\to \wcL[S,RL]{ed}{ }\,, \quad
\wcs[]{qe}{}\to \wcL[S,RL]{ed}{ }. \end{aligned} \ee

We remind that the tree-level operators appearing here actually have different flavour indices as compared to the operators involved in the matching conditions 
\eqref{class6A-match1}-\eqref{class6A-match2}. Therefore, technically they are different operators which comes into existence only through loop corrections.

The values for the $\rho$ and $\eta$ parameters are collected in Tabs.~\ref{tab:6A-1}, \ref{tab:6A-2} and \ref{tab:6B} in App.~\ref{App:etas}. These are useful to get a quantitative sense of the 1-loop SMEFT effects on the LFV decays.
{\boldmath
\subsection{$\text{ SU(2)}_L$ Correlations }
}
The general concept of $\text{SU(2)}_L$ correlation has already been discussed in earlier classes. Here we provide a non-exhaustive summary of tree-level $\text{SU(2)}_L$ correlations 
relevant to Class 6A and 6B, presented in in Tabs.~\ref{tab:class6A-su2} and \ref{tab:class6B-su2}, respectively. It would be particularly interesting to examine the resulting phenomenological implications which is beyond the scope of this review.
\begin{table}[H]
\begin{center}
\renewcommand*{\arraystretch}{1.0}
\resizebox{1.0\textwidth}{!}{
\begin{tabular}{ |c|c|c| }
\hline
\multicolumn{3}{|c|}{$\text{SU}(2)_L$ correlations for Class 6A} \\
\hline
Class 6A WC at $\muEW$ &  $ \text{SU}(2)_L$ correlations   & correlated processes  \\
\hline
$\wc[(1)]{lq}{jipp}, \wc[(3)]{lq}{jipp}$ & $\wcL[V,LL]{\nu d}{jipp} = \wc[(1)]{lq}{jipp}- \wc[(3)]{lq}{jipp}$ & $\nu d \to \nu d$  \\
$\wc[(1)]{lq}{jipp}, \wc[(3)]{lq}{jipp}$ & $\wcL[V,LL]{\nu u}{jipp} =V_{pm} (\wc[(1)]{lq}{jimn} + \wc[(3)]{lq}{jimn}){V^*_{pn}}$ & $\nu u \to \nu u$   \\
$\wc[]{ld}{jipp}$ & $\wcL[V,LR]{\nu d}{jipp} =\wc[]{ld}{jipp} $ & $\nu d \to \nu d$  \\
$\wc[]{lu}{jipp}$ & $\wcL[V,LR]{\nu u}{jipp} = \wc[]{lu}{jipp}$ &  $\nu u \to \nu u$   \\
\hline
$\wc[]{ledq}{jipp}$ & $\wcL[S,RL]{\nu e du}{jipp} = \wc[]{ledq}{jipn}{V^*_{pn}}$ & charged currents   \\
$\wc[(1)]{lequ}{jipp}$ & $ \wcL[S,RR]{\nu e du}{jipp} = {V_{pm}}\wc[(1)]{lequ}{jimp} $ &  charged currents   \\
$\wc[(3)]{lequ}{jipp}$ &  $ \wcL[T,RR]{\nu e du}{jipp} = {V_{pm}}\wc[(3)]{lequ}{jimp}  $ & charged currents   \\
\hline 
\end{tabular}
}
\caption{$\text{SU(2)}_L$ correlations for Class 6A at tree-level. The first column shows Class 6A operators at $\muEW$, the second column lists the other WET operators {(beyond Class 6A)} generated by the operators in the first column, and the third column shows processes that are generated by the WET operators in the second column. We assume the SMEFT down-basis with the PMNS set to unity.}
\label{tab:class6A-su2}
\end{center}
\end{table}

\begin{table}[H]
\begin{center}
\renewcommand*{\arraystretch}{1.0}
\resizebox{1.0\textwidth}{!}{
\begin{tabular}{ |c|c|c| }
\hline
\multicolumn{3}{|c|}{$\text{SU}(2)_L$ correlations for Class 6B} \\
\hline
Class 6B WC at $\muEW$ &  $\text{SU}(2)_L$ correlation   & correlated processes  \\
\hline
$\wc[(1)]{lq}{jiqp}, \wc[(3)]{lq}{jiqp}$ & $\wcL[V,LL]{\nu d}{jiqp} = \wc[(1)]{lq}{jiqp}- \wc[(3)]{lq}{jiqp}$ 
& $ d_p \to  d_q \nu \bar \nu$ \\
$\wc[(1)]{lq}{jiqp}, \wc[(3)]{lq}{jiqp}$ & $\wcL[V,LL]{\nu u}{jiqp} ={V_{qm}} (\wc[(1)]{lq}{jimn} + \wc[(3)]{lq}{jimn}){V^*_{pn}}$ & $ u_p \to u_q \nu \bar \nu$  \\
$\wc[]{ld}{jiqp}$     & $\wcL[V,LR]{\nu d}{jiqp} =\wc[]{ld}{jiqp} $ &
$ d_p  \to d_q  \nu \bar \nu$    \\
$\wc[]{lu}{jiqp}$     & $\wcL[V,LR]{\nu u}{jiqp} =\wc[]{lu}{jiqp} $ & $u_p \to u_q \nu \bar \nu$   \\
\hline
$\wc[]{ledq}{jiqp}$   & $\wcL[S,RL]{\nu e du}{jiqp} = \wc[(1)]{ledq}{jiqn} {V^*_{pn}} $ & charged currents   \\
$\wc[(1)]{lequ}{jiqp}$ & $ \wcL[S,RR]{\nu e du}{jiqp} = {V_{qm}} \wc[(1)]{lequ}{jimp} $ &  charged currents  \\
$\wc[(3)]{lequ}{jiqp}$ &  $ \wcL[T,RR]{\nu e du}{jiqp} = {V_{qm}} \wc[(3)]{lequ}{jimp}  $ & charged currents   \\
\hline 
\end{tabular}
}
\caption{\small {Same as for Tab.~\ref{tab:class6A-su2}, but for $q\ne p$, for Class 6B.}}
\label{tab:class6B-su2}
\end{center}
\end{table}

{\boldmath
\section{Electric and Magnetic Dipole Moments (Class 7)}
\label{class7}
}
EDMs of leptons, hadrons,  atoms as well as molecules can play a very important role in the search for NP because they are very strongly suppressed within the SM. Any measurement of an EDM would signal not only NP but also a new source of CP violation beyond the SM that in this case takes place in flavour-conserving processes.

While EDMs are CP-violating but flavour conserving, Magnetic Dipole Moments (MDM) are both CP and flavour conserving. Here the MDMs of leptons play a very important role represented by $(g-2)_\ell$ with $\ell=e,\mu,\tau$. A very recent simple review on EDMs and ADMs intended for PhD students and anyone entering the field for the first time is given in \cite{Vives:2025clr}.

{We divide Class 7 into two subclasses 
\begin{equation}
{\rm {\bf Class~ 7A}: EDMs\,, \quad {\bf Class~ 7B}: MDMs.}
\end{equation}
}
Details on EDMs can be found in the review \cite{Engel:2013lsa} and a summary is given in Section 17.3 of \cite{Buras:2020xsm}. Here we will concentrate our discussion on the WET and the SMEFT. In addition to the two references just cited that discuss various aspects of EDMs within the SMEFT several papers discuss details of EDMs in the SMEFT and WET.

This is, in particular, the case of rather recent analyses in \cite{Aebischer:2021uvt,Kley:2021yhn} where different aspects of the RG evolution and of the matching at tree and one-loop level have been discussed in detail. Both papers contain a very valuable list of references to earlier literature.

As far as MDMs are concerned they became in particular popular due to the $(g-2)_\mu$ anomaly. There is a very reach literature on this topic. We refer here only to a selection of papers where further references can be found. These are \cite{Isidori:2021gqe,Aebischer:2021uvt,Fajfer:2021cxa,Allwicher:2021jkr,Cirigliano:2021peb}. For the most recent update see \cite{Aliberti:2025beg}.

For a review of possible NP explanations with many references see \cite{Athron:2021iuf}. The first detailed study in the WET and SMEFT can be found in \cite{Aebischer:2021uvt}. It was shown that below the EW scale the most dominant contribution to $(g-2)_\mu$ is given by the dipole operator, followed by semileptonic tensor operators. This picture remains true in the SMEFT, where the contributions come from the WCs $\wc[]{eB}{},\wc[]{eW}{}$, followed by $\wc[(3)]{lequ}{}$.

The NLO and NNLO mixing of dipole operators has already been discussed in connection with the decay $B\to X_s\gamma$. Here the papers \cite{Misiak:1994zw} and \cite{Gorbahn:2005sa} should be mentioned.

More recent NLO studies of dipole operators in connection with electric dipole moments including also the neutron EDM in the context of the WET and the SMEFT can be found in \cite{Panico:2018hal,Brod:2018lbf,Brod:2018pli} and very recently in~\cite{Brod:2022bww,Brod:2023wsh}. These analyses play a significant role in constraining the CP-violating phases of Yukawa couplings. In particular, the correlations between EDM and LHC data are a powerful tool in the search for NP. Very recently a comprehensive WET analysis of various EDMs due to heavy and light fermion generation operators is presented in \cite{Kumar:2024yuu}. See also Section 17.3.6 in \cite{Buras:2020xsm}.

Finally, NP in CP-violating and flavour changing quark dipole transitions has been recently analyzed within the SMEFT in \cite{Fajfer:2023gie}. Bounds on the WCs of dipole operators resulting from CPV observables in non-leptonic and radiative $B$, $D$, and $K$ decays as well as from the neutron and electron EDMs were presented.
\subsection{WET Operators for Class 7}
The complete list of WET operators in the JMS basis for the EDM observables was recently presented in \cite{Kumar:2024yuu}. It includes the effects of heavy fermion generations that are captured through renormalization group running and matching contributions within the WET with different numbers of flavour (e.g. between WET-5 and WET-4, WET-4 and WET-3). In particular the following effects were taken into account
\begin{itemize}
\item
1-loop WET ADMs, called LL in that paper.
\item
2-loop WET ADMs or 1-loop WET threshold corrections, called NLL there.
\item
A combination of 2-loop ADMs and 1-loop threshold corrections in the WET, called NNLL there.
\end{itemize}
Next, we sketch the effective operators up to dim-6 level constituting the WET Lagrangian for EDMs. At dim-$4$, we have the $\theta$-term (like in the SMEFT case~\eqref{eq:Lyuk1}):
\begin{equation}
\mathcal L_\theta = \bar \theta\, \frac{g_s^2}{32\pi^2} G^A_{\mu \nu} \tilde G^{A\, \mu \nu}\,,
\end{equation}
with $\bar\theta$ having an unnaturally small value $\bar \theta\le 10^{-10}$ unless it can be removed by some symmetries. This very popular topic is beyond the scope of our review.

Then, at dim-5, we have electromagnetic and chromo-dipole operators
\be
\begin{aligned}
\opL[]{u\gamma}{ii}\,, \quad
 \opL[]{d\gamma}{ii}\,,  \quad
\opL[]{uG}{ii}\,, \quad
\opL[]{dG}{ii}\,, \quad
\opL[]{ e \gamma}{ii}\,,
\end{aligned}
\ee
where $i$ denotes the flavour. Down-quark operators having $ii=11,22$, and operators with up-quark with $ii=11$ directly contribute to the neutron and atomic EDMs. Dipole operators with $ii=33,22$ i.e. $b$ and $c$ quarks can contribute at the loop level. See Tab.~\ref{tab:wet5} for a summary at which level in RG improved perturbation theory specific operators are generated. In what follows, we closely follow \cite{Kumar:2024yuu}.

In the leptonic sector, $ii=11$, $22$, and $33$ contribute directly to electron, muon, and tau EDMs, respectively. Since the limits on $d_\mu$ and $d_\tau$ are much weaker than on the electron EDMs, it is important to consider the matching and running of heavier leptons into leptonic and semileptonic operators that contribute to molecular EDMs.

The CP-violating Weinberg operator is defined as
\be
\opL[]{\widetilde G}{} =  f^{ABC} \widetilde G_\mu^{A\nu} G_\nu^{B\rho} G_\rho^{C\mu}\,.
\ee
Its WC $\wc[]{\widetilde G}{}$ can be real but the operator itself is CP violating implying the so-called strong CP problem.\footnote{Solutions to the strong CP problem in the context of Nelson-Barr models within the SMEFT were studied in \cite{Alves:2025owr}.} 

In the next category, there are many four-fermion WET operators. The ones falling within the $n_f = 3+1$ quark and lepton structure contribute directly to the EDMs.

Suppressing the flavour indices, the complete list of operators is given by
\begin{center}
\textrm{\bf WET-Loop 7A}
\end{center}
\be
\label{eq:wet-loop-edm}
\begin{aligned} 
&\opL[V1,LR]{uddu}{}\,, \quad
\opL[V8,LR]{uddu}{}\,, \quad   
\opL[V1,LR]{dd}{} \,, \quad
\opL[V8,LR]{dd}{}  \,, \quad
\underline{\opL[V1,LR]{uu}{}} \,, \quad
\underline{\opL[V8,LR]{uu}{}}  \,, \\
&\opL[S1,RR]{dd}{} \,, \quad
\opL[S8,RR]{dd}{} \,, \quad
{\opL[S1,RR]{uu}{}} \,, \quad
{\opL[S8,RR]{uu}{}} \,, \quad
\opL[S1,RR]{ud}{} \,, \quad
\opL[S8,RR]{ud}{} \,, \\
&
\opL[S1,RR]{uddu}{} \,, \quad
\opL[S8,RR]{uddu}{} \,, \quad
\opL[S,RL]{eu}{}  \,, \quad
\opL[S,RL]{ed}{} \,, \quad
\opL[S,RR]{eu}{} \,, \quad
\opL[S,RR]{ed}{} \,, \\
&
\opL[T,RR]{eu}{} \,, \quad
\opL[T,RR]{ed}{} \,, \quad 
\underline{\opL[S,RR]{ee}{}}\,, \quad
\underline{\opL[V,LR]{ee}{}}\,, \quad
\opL[]{f\gamma}{}\,, \quad
\opL[]{\widetilde G}{}\,,\quad
\bar \theta\,.
\end{aligned}
\ee
\begin{table}[H] 
\begin{center}
 \renewcommand*{\arraystretch}{1.5}
\begin{tabular}{ |cccccc| } 
\hline 
\multicolumn{6}{|c|}{Four-fermion Operators for EDMs with $n_l+n_q$ flavours} \\
\hline
\hline
\multicolumn{6}{|c|}{ $e, u, d, s$ operators $\subset$ 1+3 flavour WET} \\
\hline \hline
$ [O_{uddu}^{V1,LR}]_{1111}$  & $ [O_{uddu}^{V8,LR}]_{1111}$   & $[O_{uddu}^{V1,LR}]_{1221}$ &  $[O_{uddu}^{V8,LR}]_{1221}$ &
$[O_{dd}^{V1,LR}]_{1221}$  & $[O_{dd}^{V8,LR}]_{1221}$  \\ 
$[O_{uu}^{S1,RR}]_{1111}$  & $[O_{uu}^{S8,RR}]_{1111}$ &  $[O_{dd}^{S1,RR}]_{1111}$  & $[O_{dd}^{S8,RR}]_{1111}$   & $[O_{dd}^{S1,RR}]_{2222}$ &  $[O_{dd}^{S8,RR}]_{2222}$ \\  
$[O_{dd}^{S1,RR}]_{1122}$  & $[O_{dd}^{S8,RR}]_{1122}$  & $[O_{dd}^{S1,RR}]_{1221}$  & $[O_{dd}^{S8,RR}]_{1221}$ & 
$[O_{ud}^{S1,RR}]_{1111}$  & $[O_{ud}^{S8,RR}]_{1111}$  \\ 
$[O_{ud}^{S1,RR}]_{1122}$  & $[O_{ud}^{S8,RR}]_{1122}$ &  $[O_{uddu}^{S1,RR}]_{1111}$  & $[O_{uddu}^{S8,RR}]_{1111}$  & $[O_{uddu}^{S1,RR}]_{1221}$  & $[O_{uddu}^{S8,RR}]_{1221}$\\  \hline 
$[O_{eu}^{S,RL}]_{1111}$  & $[O_{eu}^{S,RR}]_{1111}$ & 
$[O_{ed}^{S,RL}]_{1111}$ & $[O_{ed}^{S,RR}]_{1111}$ & 
$[O_{ed}^{S,RL}]_{1122}$ &  $[O_{ed}^{S,RR}]_{1122} $\\
$[O_{eu}^{T,RR}]_{1111}$  & $[O_{ed}^{T,RR}]_{1111}$ 
& $[O_{ed}^{T,RR}]_{1122}$  &    $[O_{ee}^{S,RR}]_{1111}^*$     && \\  \hline
\hline
\multicolumn{6}{|c|}{$c$ operators $\subset$ 1+4 flavour WET  } \\
\hline \hline
$[O_{uu}^{S1,RR}]_{2222}^*$  & $[O_{uu}^{S8,RR}]_{2222}^*$ &  $[O_{uu}^{S1,RR}]_{1122}$  & $[O_{uu}^{S8,RR}]_{1122}$ &
$[O_{ud}^{S1,RR}]_{2211}$  & $[O_{ud}^{S8,RR}]_{2211}$  \\
$[O_{ud}^{S1,RR}]_{2222}$  & $[O_{ud}^{S8,RR}]_{2222}$ & $[O_{uddu}^{S1,RR}]_{2112}$  & $[O_{uddu}^{S8,RR}]_{2112}$  & $[O_{uddu}^{S1,RR}]_{2222}$  & $[O_{uddu}^{S8,RR}]_{2222}$\\ 
$[O_{uu}^{S1,RR}]_{1221}$  & $[O_{uu}^{S8,RR}]_{1221}$ &  
$[O_{uddu}^{V1,LR}]_{2112}^*$  & $[O_{uddu}^{V8,LR}]_{2112}^*$   & $[O_{uddu}^{V1,LR}]_{2222}^*$ &  $[O_{uddu}^{V8,LR}]_{2222}^*$ \\  
$[O_{uu}^{V1,LR}]_{1221}^*$  & $[O_{uu}^{V8,LR}]_{1221}^*$  &  &    & & \\ 
\hline
$[O_{eu}^{S,RR}]_{1122}$  &   $[O_{eu}^{T,RR}]_{1122}$ & &    & & \\ 
\hline \hline
\multicolumn{6}{|c|}{ $\mu, \tau, b$ operators $\subset$ 3+5 flavour WET } \\
\hline \hline
$[O_{dd}^{S1,RR}]_{3333}^*$  & $[O_{dd}^{S8,RR}]_{3333}^*$   & $[O_{dd}^{S1,RR}]_{1133}$  & $[O_{dd}^{S8,RR}]_{1133}$ & 
$[O_{dd}^{S1,RR}]_{2233}$  & $[O_{dd}^{S8,RR}]_{2233}$  \\ 
$[O_{dd}^{S1,RR}]_{1331}$  & $[O_{dd}^{S8,RR}]_{1331}$ & 
$[O_{dd}^{S1,RR}]_{2332}$  & $[O_{dd}^{S8,RR}]_{2332}$ & $[O_{ud}^{S1,RR}]_{1133}$  & $[O_{ud}^{S8,RR}]_{1133}$ \\
$[O_{uddu}^{S1,RR}]_{1331}$  & $[O_{uddu}^{S8,RR}]_{1331}$  &  $[O_{ud}^{S1,RR}]_{2233}^*$  & $[O_{ud}^{S8,RR}]_{2233}^*$  & 
$[O_{uddu}^{S1,RR}]_{2332}^*$  & $[O_{uddu}^{S8,RR}]_{2332}^*$  \\
$[O_{uddu}^{V1,LR}]_{1331}^*$  & $[O_{uddu}^{V8,LR}]_{1331}^*$ &  
$[O_{dd}^{V1,LR}]_{1331}^*$  & $[O_{dd}^{V8,LR}]_{1331}^*$ &   $[O_{dd}^{V1,LR}]_{2332}^*$  & $[O_{dd}^{V8,LR}]_{2332}^*$  \\
$[O_{uddu}^{V1,LR}]_{2332}^{**}$  & $[O_{uddu}^{V8,LR}]_{2332}^{**}$   &   &   &   &      \\
\hline
$[O^{T,RR}_{ed}]_{1133}$  & $[O^{T,RR}_{ed}]_{2211}$   & $[O_{ed}^{T,RR}]_{2222}$  & $[O_{ed}^{T,RR}]_{2233}^*$   &$[O_{ed}^{T,RR}]_{3311}$    &  $[O_{ed}^{T,RR}]_{3322}$    \\
$[O^{T,RR}_{ed}]_{3333}^*$  & $[O^{T,RR}_{eu}]_{2211}$   & $[O_{eu}^{T,RR}]_{2222}^*$    &$[O_{eu}^{T,RR}]_{3311}$    &  $[O_{eu}^{T,RR}]_{3322}^*$   & \\

$[O^{S,RR}_{ed}]_{1133}$  & $[O^{S,RR}_{ed}]_{2211}$   & $[O_{ed}^{S,RR}]_{2222}$  & $[O_{ed}^{S,RR}]_{2233}^*$   &$[O_{ed}^{S,RR}]_{3311}$    &  $[O_{ed}^{S,RR}]_{3322}$    \\
$[O^{S,RR}_{ed}]_{3333}^*$  & $[O^{S,RR}_{eu}]_{2211}$   & $[O_{eu}^{S,RR}]_{2222}^*$    &$[O_{eu}^{S,RR}]_{3311}$    &  $[O_{eu}^{S,RR}]_{3322}^*$   & \\

$[O_{ee}^{V,LR}]_{1221}^*$  & $[O_{ee}^{V,LR}]_{1331}^*$   &$[O_{ee}^{S,RR}]_{1221}$    &   $[O_{ee}^{S,RR}]_{1331}$  & $[O_{ee}^{S,RR}]_{1122}$  & $[O_{ee}^{S,RR}]_{1133}$   \\ \hline
\end{tabular}
\caption{Four-fermion flavour conserving CP-violating operators contributing to EDMs in the $n_q+ n_\ell$ =5+3 flavour WET. None, single and double asterisk signs denote LL, NLL, and NNLL contributions, respectively. This table is taken from \cite{Kumar:2024yuu}.
}
\label{tab:wet5}
\end{center}
\end{table}

Here the underlined operators contribute only at the 1- or 2-loop level. The remaining operators containing flavour indices of the first two fermion generations contribute at tree-level. On the other hand, operators with heavier fermions do so at loop level. The complete list of contributing operators with explicit flavour indices in the ${5+3}$ WET can be found in Tab.~\ref{tab:wet5}. In this table the LL, NLL, and NNLL operators are indicated by no, single and double asterisk signs, respectively. 

Its worth to mention that the VLR operators with heavier fermion generations contribute to EDMs only through 2-loop ADMs, which in principle would require 1-loop matching onto the SMEFT. The other Dirac structures such as SRR with heavier generations contribute via 1-loop running or threshold corrections at $m_b$, $m_c$, or $m_\mu$. More details about different operators and the mechanisms through which they generate EDMs are discussed in \cite{Kumar:2024yuu}.

For the magnetic moments, a much smaller set of operators is relevant \cite{Aebischer:2021uvt}
\begin{center}
\textrm{\bf WET-Loop 7B}
\end{center}
\be \label{eq:wet-mm}
\begin{aligned}
&\opL[]{e\gamma}{ii}\,, \quad
\underline{\opL[V,LR]{ee}{immi}}\,, \quad
\underline{\opL[S,RR]{ee}{immi}}\,, \quad
\underline{\opL[T,RR]{ed}{iimm}}\,, \quad
\underline{\opL[T,RR]{eu}{iimm}}\,, \quad
\underline{\opL[S,RR]{eu}{iimm}}\,, \\
&
\underline{\opL[S,RR]{ed}{iimm}}\,, \quad
\underline{\opL[S,RL]{ed}{iimm}}\,, \quad
\underline{\opL[S,RL]{eu}{iimm}}\,.
\end{aligned}
\ee
Here the index $m$ is summed over 1-3 and $i$ is fixed by the lepton for which the magnetic moment is computed. The underlined operators contribute at 1-loop.

\subsection{SMEFT Operators for Class 7 at $\muEW$}

The complete list of tree-level operators (suppressing flavour indices) in the SMEFT at the EW scale are
\begin{center}
\textrm{\bf SMEFT-Tree 7A} 
\end{center}
\be \label{class7A-smeftopsT}
\begin{aligned}
& \ops[]{Hud}{} \,,\quad 
\underline{\ops[]{le}{}}\,, \quad
\ops[(1)]{qd}{}\,,  \quad
\ops[(8)]{qd}{} \,, \quad 
\underline{\ops[(1)]{qu}{}}\,, \quad
\underline{\ops[(8)]{qu}{}}\,,\quad 
\ops[(1)]{quqd}{}\,, \quad  
\ops[(8)]{quqd}{}\,,\quad\\
&\ops[]{ledq}{}\,,   \quad
\ops[(1)]{lequ}{}\,, \quad 
\ops[(3)]{lequ}{}\,, \quad
\ops[]{fB}{}\,, \quad
\ops[]{fW}{}\,, \quad
\ops[]{fG}{}\,, \quad
\ops[]{\widetilde G}{}\,.
\end{aligned}
\ee
Here the underlined operators match at tree-level onto WET-loop operators. The SMEFT to WET matching is given by
\be \label{eq:class7-tree-match1}
\begin{aligned}
&\wcL[V1,LR]{uddu}{ijji}   =- V_{ij} {\wc[*]{Hud}{ij}} \,, \qquad 
\wcL[V1,LR]{dd}{ijji}   =  \wc[(1)]{qd}{ijji} \,, \qquad 
\wcL[V8,LR]{dd}{ijji} = \wc[(8)]{qd}{ijji}\,,\\
\wcL[V1,LR]{uu}{ijji} &  = V_{im}  \wc[(1)]{qu}{mnji} {V_{jn}^*} \,, \qquad 
\wcL[V8,LR]{uu}{ijji} =   V_{im}  \wc[(8)]{qu}{mnji} {V_{jn}^*} \,,\qquad
\wcL[S1,RR]{ud}{iijj}   = V_{im}    \wc[(1)]{quqd}{mijj}  \,,\\
\wcL[S8,RR]{ud}{iijj} &= V_{im} \wc[(8)]{quqd}{mijj}\,,\qquad\quad
\wcL[S1,RR]{uddu}{ijji}   =  - V_{im} \wc[(1)]{quqd}{jimj}  \,, \qquad 
\wcL[S8,RR]{uddu}{ijji} = -V_{im} \wc[(8)]{quqd}{jimj} \,,\\
\wcL[S,RR]{eu}{iijj}  &= -V_{jm}  \wc[(1)]{lequ}{iimj} \,,\quad 
\wcL[S,RL]{ed}{iijj}  =  \wc[]{ledq}{iijj} \,, \quad 
\wcL[T,RR]{eu}{iijj}   =  - V_{jm} \wc[(3)]{lequ}{iimj}\,, \quad
\wcL[V,LR]{ee}{ijji}   =  \wc[]{le}{ijji}\,.
\end{aligned}
\ee
Here only the indices $m$ or $n$ are summed over. The remaining operators such as $\wcL[V8,LR]{uddu}{}$, $\wcL[S1(8),RR]{uu}{}$, $\wcL[S1(8),RR]{dd}{}$, $\wcL[S,RR]{ed}{} $, $\wcL[S,RL]{eu}{}$, $\wcL[T,RR]{ed}{}$, and $\wcL[S,RR]{ee}{}$ match first at dim-8 level to the SMEFT \cite{Jenkins:2017jig,Burgess:2021ylu}. 

The matching conditions for the dipoles read
\be \label{eq:class7-tree-match2}
\begin{aligned}
\wcL[]{u\gamma}{ii} & = \frac{v}{\sqrt 2} V_{im} (c_w \wc[]{uB}{mi} + s_w \wc[]{uW}{mi})\,, \\
\wcL[]{d\gamma}{ii} & = \frac{v}{\sqrt 2} (c_w \wc[]{dB}{ii} - s_w \wc[]{dW}{ii})\,, \\
\wcL[]{e\gamma}{ii} & = \frac{v}{\sqrt 2} (c_w \wc[]{eB}{ii} - s_w \wc[]{eW}{ii})\,, \\
\wcL[]{uG}{ii} & = \frac{v}{\sqrt 2} V_{im}  \wc[]{uG}{mi}\,, \quad
\wcL[]{dG}{ii}   = \frac{v}{\sqrt 2} \wc[]{dG}{ii}\,.
\end{aligned}
\ee
Here too the repeated index $m$ is summed over whereas $i$ is fixed. For the down-type quark operators, at the tree-level, we can have $ii=11,22$, for up-quark operators $ii=11$, and 
the leptonic operators admit $ii=11,22,33$. 

The WCs of the Weinberg operator in the WET and SMEFT are simply equal to each other, i.e. 
\begin{equation}\label{eq:tG_match}
\wcL[]{\widetilde G}{} = \wc[]{\widetilde G}{}\,.
\end{equation}
Next, we discuss the SMEFT operators for MDMs. The matching to the tree-level operators in \eqref{eq:wet-mm} involve only the two leptonic SMEFT dipole operators
\begin{center}
\textrm{\bf SMEFT-Tree 7B}
\end{center}
\be \label{class7B-smeftopsT}
\begin{aligned}
\ops[]{eB}{}\,, \quad
\ops[]{eW}{}\,, \quad
\underline{\ops[]{le}{}}\,, \quad
\underline{\ops[]{ledq}{}}\,,   \quad
\underline{\ops[(1)]{lequ}{}}\,, \quad
\underline{\ops[(3)]{lequ}{}}\,.
\end{aligned}
\ee
The additional (underlined) operators in this class match onto 1-loop WET operators (underlined) in \eqref{eq:wet-mm}. The specific flavour indices can be inferred from the matching conditions in \eqref{eq:class7-tree-match1} and \eqref{eq:class7-tree-match2}. 

\subsection{SMEFT Operators for Class 7 at $\Lambda$}
Now we explore the structure of the 1-loop SMEFT Lagrangian relevant for EDMs and MDMs. In this regard, we consider the relevant anomalous dimension matrices that renormalize the tree-level SMEFT operators for Class 7 as identified in the previous subsections. This allows to identify all 1-loop operators relevant for Class 7: 
\begin{center}
\textrm{\bf SMEFT-Loop 7A}
\end{center}
\be \label{class7A-smeftopsL}
\begin{aligned} 
\textrm{\bf Yukawa-Mixing}: \quad&
\ops[]{Hd}{}\,, \quad
\ops[(1)]{ud}{}\,, \quad
\ops[(8)]{ud}{}\,, \quad
\\
\textrm{\bf Gauge-Mixing}: \quad &  \textrm{No additional {operators}.}
\end{aligned}
\ee
Even though operator mixing due to Yukawa interactions brings in three new operators into the picture, in this class, the operator mixing due to gauge couplings does not introduce any new operator structures.

The flavour structures of these SMEFT operators at $\Lambda$ are determined by the WET flavours appearing in the tree-level matching conditions, \eqref{eq:class7-tree-match1} and \eqref{eq:class7-tree-match2}, together with the structure of the ADMs. 

Recall that throughout this review we do not consider the operators due to RGEs that are suppressed by light Yukawa (i.e. $y_u, y_c,  y_d, y_s$) at the ADM level. However, for the WET framework such terms are known to be important due to the sensitivity of EDM observables to high new physics scales \cite{Kumar:2024yuu}.

A similar effect maybe present in the SMEFT, motivating a careful numerical study that include these contributions. These effects can be incorporated automatically using tools such as {\tt wilson} and {\tt DsixTools}.

For the MDMs one finds
\begin{center}
\textrm{\bf SMEFT-Loop 7B}
\end{center}
\be \label{class7B-smeftopsL}
\begin{aligned}
\textrm{\bf Yukawa-Mixing}: \quad & \textrm{No additional {operators}.} \\
\textrm{\bf Gauge-Mixing}: \quad & \textrm{No additional {operators}.}
\end{aligned}
\ee
In the following, we present the RGEs for the SMEFT operators at present at $\muEW$ that match with, at tree-level, to EDM and MDM operators in WET. 
\\

\noindent
\underline{\bf \boldmath $f^4 \to f^4$ (Gauge)}:

The gauge coupling dependent 4f to 4f operator mixing is given by

\be
\begin{aligned}
\dotwc[(1)]{qd}{ijji} &= \frac{2}{3} g_1^2 \wc[(1)]{qd}{ijji}-\frac{8}{3} g_s^2 \wc[(8)]{qd}{ijji}\,, \\
\dotwc[(8)]{qd}{ijji} &= -12  g_s^2\wc[(1)]{qd}{ijji}-14 g_s^2\wc[(8)]{qd}{ijji}+\frac{2}{3}g_1^2 \wc[(8)]{qd}{ijji}\,.
 \end{aligned}
\ee
We do not show RGEs for $\wc[(1)]{qu}{}$, $\wc[(8)]{qu}{}$ and $\wc[]{le}{}$ because the corresponding WET operators contribute to EDMs only at loop level, see \eqref{class7A-smeftopsT}. Moreover, the RGEs for the following operators have been presented in Class 3 and Class 6:

\be
\begin{aligned}
\wc[(1)]{quqd}{}&: \eqref{eq:quqd1G}\,, \\
\wc[(8)]{quqd}{}&: \eqref{eq:quqd8G}\,,\\
\wc[(1)]{lequ}{}\,,\,\wc[(3)]{lequ}{}\,,\,\wc[]{ledq}{}&: \eqref{eq:scal_gauge}\,.
\end{aligned}
\ee

\noindent
\underline{\bf \boldmath $f^2H^2D \to f^4$ (Gauge)}:
\\

No relevant RGEs are found in this case.\\

\noindent
\underline{\bf \boldmath $f^2H^2D \to f^2H^2D$ (Gauge)}:

\be
\begin{aligned}\label{eq:Hudgauge}
  \dotwc[]{Hud}{pr} &= -3 g_1^2 \wc[]{Hud}{pr}\,.
\end{aligned}
\ee

\noindent
\underline{\bf \boldmath $f^2XH \to f^2XH$ (Gauge)}:

\be
\begin{aligned}
\dotwc[]{uB}{pr} &= \frac{313}{36} g_1^2 \wc[]{uB}{pr}-\frac{9}{4} g_2^2 \wc[]{uB}{pr}+\frac{8}{3} g_s^2 \wc[]{uB}{pr}+\frac{40}{9} g_1 g_s \wc[]{uG}{pr}-\frac{1}{2} g_1 g_2 \wc[]{uW}{pr}\,,\\
\dotwc[]{uW}{pr} &= -\frac{1}{6} g_1 g_2 \wc[]{uB}{pr}+\frac{8}{3} g_2 g_s \wc[]{uG}{pr}-\frac{19}{36} g_1^2 \wc[]{uW}{pr}-\frac{11}{12} g_2^2 \wc[]{uW}{pr}+\frac{8}{3} g_s^2 \wc[]{uW}{pr}\,,\\
\dotwc[]{uG}{pr} &=\frac{10}{3} g_1 g_s \wc[]{uB}{pr}-\frac{19}{36} g_1^2 \wc[]{uG}{pr}-\frac{9}{4} g_2^2 \wc[]{uG}{pr}-\frac{17}{3} g_s^2 \wc[]{uG}{pr}+6 g_2 g_s \wc[]{uW}{pr} \,,\\
\dotwc[]{dB}{pr} &= \frac{253}{36} g_1^2 \wc[]{dB}{pr}-\frac{9}{4} g_2^2 \wc[]{dB}{pr}+\frac{8}{3} g_s^2 \wc[]{dB}{pr}-\frac{8}{9} g_1 g_s \wc[]{dG}{pr}+\frac{5}{2} g_1 g_2 \wc[]{dW}{pr}\,,\\
\dotwc[]{dW}{pr} &= \frac{5}{6} g_1 g_2 \wc[]{dB}{pr}+\frac{8}{3} g_2 g_s \wc[]{dG}{pr}-\frac{31}{36} g_1^2 \wc[]{dW}{pr}-\frac{11}{12} g_2^2 \wc[]{dW}{pr}+\frac{8}{3} g_s^2 \wc[]{dW}{pr}\,,\\
\dotwc[]{dG}{pr} &= -\frac{2}{3} g_1 g_s \wc[]{dB}{pr}-\frac{31}{36} g_1^2 \wc[]{dG}{pr}-\frac{9}{4} g_2^2 \wc[]{dG}{pr}-\frac{17}{3} g_s^2 \wc[]{dG}{pr}+6 g_2 g_s \wc[]{dW}{pr}\,.
\end{aligned}
\ee
The RGEs for the following two operators have already been presented in {Class 5}:
\be
\wc[]{eB}{}\,,\wc[]{eW}{}: \eqref{eq:eBeWgauge}\,.
\ee

\begin{figure}[tbp]
\hspace{-0.5cm}
\includegraphics[clip,trim=4.0cm  12.5cm 3.5cm 12.6cm, width=1.0\textwidth]{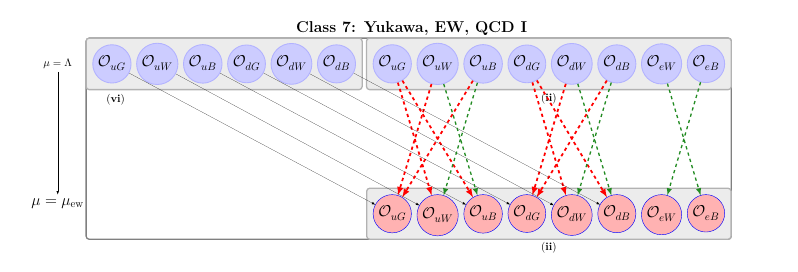}
\vspace{-0.3cm}
\caption{\small Class 7 -- Mixing of operators in the Warsaw down-basis, relevant for the EDMs and MDMs. The dashed green, solid black and dashed red lines indicate the mixing due to electroweak, Yukawa, and strong couplings, respectively. The index $v=1,2,3$ originates from CKM rotations in the ADMs and is always summed over, whereas index $i$ is fixed as per tree-level matching onto the WET. See text for more details.}
\label{chart:class7-1}
\end{figure}

\noindent
\underline{\bf \boldmath $X^3 \to X^3$ (Gauge)}:

\be
\begin{aligned}
  \dotwc[]{\widetilde G}{} &= 15 \wc[]{\widetilde G}{} g_s^2\,.
\end{aligned}
\ee
The SMEFT chart in Fig.~\ref{chart:class7-1} provides a graphical view of the operator mixing discussed above. 

Interestingly, the Yukawa RGEs are significantly simplified due to unique nature of this class. The simplification arises from the fact that only WET operators involving light fermion flavours $(u,d,s,e)$ directly contribute to the EDMs (see, for example, \cite{Kumar:2024yuu} for the WET expressions for various EDMs) and that we retain only dominant terms proportional to $y_t$, $y_b$ and $y_\tau$ in the Yukawa sector.

The Yukawa-dependent RGEs are analysed below.

\noindent
\underline{\bf \boldmath $f^4 \to f^4$ (Yukawa)}
\\

In principle, {the WCs} $\wc[(1)]{qd}{1221}$ and $\wc[(8)]{qd}{1221}$ can mix with many other WCs but only through light quark Yukawas, which we have neglected here.

For $\wc[(1)]{quqd}{},\wc[(8)]{quqd}{}$, using their SMEFT matching conditions and allowed flavour indices of corresponding WET WCs listed in Tab.~\ref{tab:wet5}, the possible flavour combinations are 

\begin{equation}
{m111\,, m122}\,, 11m1\,, 21m2\,,
\end{equation} 
where $m$ can take any value between $1-3$. The RGEs for these operators have already been presented in Sec.~\ref{class3}:
\be
\begin{aligned}
\wc[(1)]{quqd}{}\,,\,\wc[(8)]{quqd}{}&: \eqref{eq:quqd18Yuk}\,.
\end{aligned}
\ee
The only relevant terms {in the mixing are} the ones proportional to 
\begin{equation}
\wc[(1)]{ud}{}\,, \wc[(8)]{ud}{}\,, \wc[(1)]{quqd}{}\,, \wc[(8)]{quqd}{}\,.
\end{equation}

$\wc[(1)]{lequ}{11m1}, \wc[(3)]{lequ}{11m1}$, and $\wc[]{ledq}{1111}, \wc[]{ledq}{1122}$ also exhibit  limited operator mixing. Due to involvement of third generation Yukawa couplings in the RGEs, and only specific flavour indices allowed in this class, these WC predominantly mix with themselves. Their relevant RGEs can be found in the supplemental material.

\noindent
\underline{\bf \boldmath $f^2H^2D \to f^4$ (Yukawa)}:

\be
\begin{aligned}
\dotwc[(1)]{qd}{prst} &= \wc[]{Hd}{st} y_t^2 V^*_{3p} V_{3r}\,.\\
\end{aligned}
\ee

\noindent
\underline{\bf \boldmath $f^4 \to f^2H^2D$ (Yukawa)}:

\be
\begin{aligned}\label{eq:Hudf4}
\dotwc[]{Hud}{pr} &= \frac{4}{3} y_b y_t (3 \wc[(1)]{ud}{p33r}+4 \wc[(8)]{ud}{p33r})\,.
\end{aligned}
\ee

\noindent
\underline{\bf \boldmath ${f^2 H^2 D} \to f^2H^2D$ (Yukawa)}:

\be
\begin{aligned}\label{eq:Hudnon4f}
  \dotwc[]{Hud}{pr} &= 6 y_t^2\wc[]{Hud}{pr}\,.
\end{aligned}
\ee

\noindent
\underline{\bf \boldmath $f^4 \to f^2XH$ (Yukawa)}:

\be
\begin{aligned}
\dotwc[]{uB}{pr} &= \frac{1}{36} g_1 y_b (3 \wc[(1)]{quqd}{3rp3}+4 \wc[(8)]{quqd}{3rp3})-6 g_1 y_\tau \wc[(3)]{lequ}{33pr}\,,\\
\dotwc[]{uW}{pr} & = \frac{1}{12} g_2 y_b (3 \wc[(1)]{quqd}{3rp3}+4 \wc[(8)]{quqd}{3rp3})-2 g_2 y_\tau \wc[(3)]{lequ}{33pr}\,,\\
\dotwc[]{uG}{pr} &= \frac{1}{6} g_s y_b (\wc[(8)]{quqd}{3rp3}-6 \wc[(1)]{quqd}{3rp3})\,,\quad
\dotwc[]{dB}{pr} =-\frac{5}{36} g_1 y_t V_{3v} (3 \wc[(1)]{quqd}{p3vr}+4 \wc[(8)]{quqd}{p3vr}) \,,\\
\dotwc[]{dW}{pr} &= \frac{1}{12} g_2 y_t V_{3v} (3 \wc[(1)]{quqd}{p3vr}+4 \wc[(8)]{quqd}{p3vr})\,,\quad
\dotwc[]{dG}{pr} = \frac{1}{6} g_s y_t V_{3v} (\wc[(8)]{quqd}{p3vr}-6 \wc[(1)]{quqd}{p3vr})\,.
\end{aligned}
\ee
The RGEs for the following operators have already been presented in Class 5:
\be
\wc[]{eB}{},\wc[]{eW}{}:\eqref{eq:eBeWyuk4F}\,.
\ee
{The WCs} $\wc[]{eB}{11}$, and $\wc[]{eW}{11}$ can only have self-mixing.

\noindent
\underline{\bf \boldmath non-{fermion} $\to f^2XH$ (Yukawa)}:
 
There is no operator mixing in this category to the considered order.
\\

\noindent
\underline{\bf \boldmath {$f^2 XH $}$\to f^2XH$ (Yukawa)}:

\noindent
\newline
The allowed indices for the dipole WCs in Class 7 are 
\begin{equation}\label{eq:cl7_yukdipoles}
\wc[]{uB}{m1}\,, \,\wc[]{uW}{m1}\,, \,\wc[]{uG}{m1}\,, \,\wc[]{dB}{11}\,, \,\wc[]{dW}{11}\,, \,\wc[]{dG}{11}\,, 
\,\wc[]{dB}{22}\,, \,\wc[]{dW}{22}\,, \,\wc[]{dG}{22}\,,  \,\wc[]{eB}{11}\,, \,\wc[]{eW}{11}\,, \, \wc[]{eB}{22}\,, \,\wc[]{eW}{22}\,.  
\end{equation}
Here WCs $\wc[]{eB}{33}\,, \,\wc[]{eW}{33}$ are omitted because they contribute to tau EDMs for which the current limits are rather weak. 

The full RGEs with explicit index dependence can be found in the appended supplemental material. The RGEs for the following operators have already been presented in Class 5:
\be
\begin{aligned}
\wc[]{eB}{}\,, \wc[]{eW}{}: \eqref{eq:eBeQyukno4F}\,.
\end{aligned}
\ee

\begin{figure}[H]
\hspace{2.4cm}
\includegraphics[clip,trim=2cm  11cm 3.5cm 11.2cm, width=0.6\textwidth]{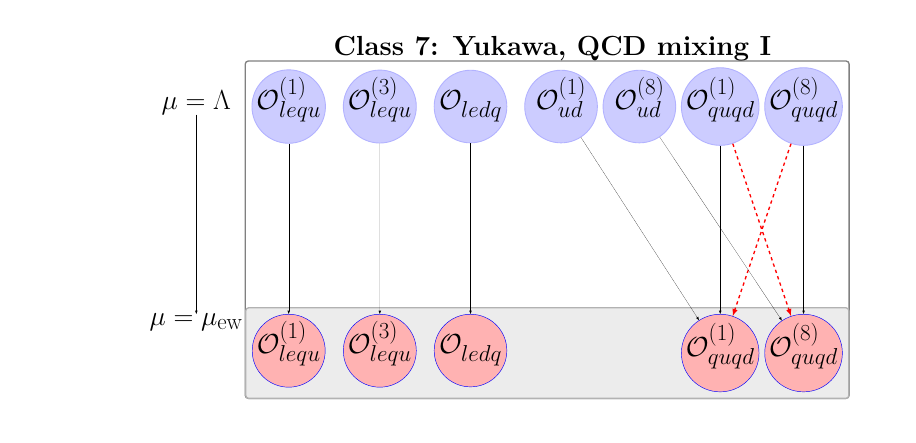}
\vspace{-0.4cm}
\caption{\small Class 7 -- Mixing pattern of four-quark operators relevant for the EDMs, in the Warsaw down-basis. Solid black and dashed red lines indicate the mixing due to Yukawa and strong interactions, respectively.}
\label{chart:class7-2}
\end{figure}

\begin{figure}[H]
\hspace{-1.0cm}
\includegraphics[clip,trim=2cm 12.5cm 3.5cm 13cm, width=1\textwidth]{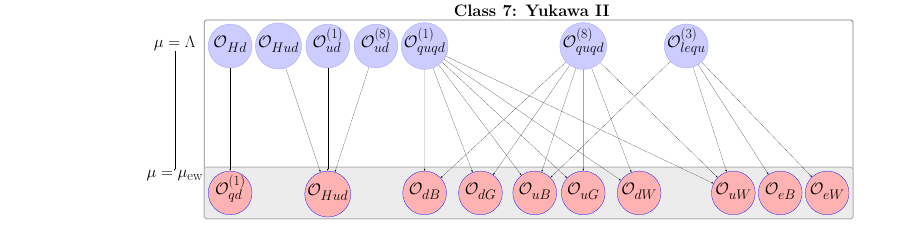}
\vspace{-0.4cm}
\caption{\small Class 7 -- Mixing pattern {of} four-quark and non-leptonic operators relevant for EDMs, in the Warsaw down-basis. Solid black lines indicate the mixing due to Yukawa interactions.}
\label{chart:class7-3}
\end{figure}

An overview of the Yukawa operator mixing pattern is given by the SMEFT charts in Figs.~\ref{chart:class7-1}-\ref{chart:class7-3}. The relevant SMEFT operators at $\muEW$, as shown in red at the bottom, carry flavour indices $ii, jiij, iijj, ijji, ji$ (suppressed there).  The specific flavour indices can be fixed by the observables of interest.
 
{\boldmath
\subsection{$\text{SU(2)}_L$ and RG Correlations}
}

In this subsection, we highlight key  correlations between Class 7 and other classes arising due to $ \text{SU(2)}_L$ symmetry at tree-level and RG running effects at 1-loop. By examining the matching conditions \eqref{eq:class7-tree-match1} one can identify $\text{SU(2)}_L$-induced correlations involving Class 7 .\footnote{We remind that since we adopted the down-basis, the down-quark SMEFT WCs directly match onto the WET ones without involvement of the CKM rotations i.e. the SMEFT WCs carry the original WET indices. The up-type SMEFT WCs match onto WET WCs with non-trivial flavour structures.} 
The key idea is to examine Class 7 WCs appearing on r.h.s of \eqref{eq:class7-tree-match1}, which due to $\text{SU(2)}_L$ simultaneously match onto WET
operators beyond Class 7 that contribute to other processes. The summary of all such correlations is given in Tab.~\ref{tab:class7-su2}. In particular we note the correlations with non-leptonic decays (Class 3) that have been analysed 
in detail in \cite{Fajfer:2023gie}.

We take a closer look at some correlations induced by {$\text{SU(2)}_L$} and  CKM rotations. For example, consider an EDM operator expressed in terms of the SMEFT operators at $\muEW$:
\be \label{eq:class7-corr1a}
\wcL[V1,LR]{uu}{1221}  = 
\wc[(1)]{qu}{1221}+ \lambda (\wc[(1)]{qu}{2221}-\wc[(1)]{qu}{1121}) +\lambda^2 (\wc[(1)]{qu}{1321} - \wc[(1)]{qu}{2121}) +\mathcal{O}(\lambda^3)
   \,. 
\ee
Here the leading term contributes to EDM, while $\mathcal{O}(\lambda)$ term contributes to $c\to u d \bar d$ and $c\to u s \bar s$ processes, because due to $\text{SU(2)}_L$:
{
\be \label{eq:class7-corr1b}
\begin{aligned}
\wcL[V1,LR]{du}{1121}  &= \wc[(1)]{qu}{1121} \,, \quad
\wcL[V1,LR]{du}{2221}  = \wc[(1)]{qu}{2221}\,.
\end{aligned}
\ee}
Given that $\wcL[V1,LR]{uu}{1221}$ is constrained by the neutron EDM, the relations in \eqref{eq:class7-corr1a} and \eqref{eq:class7-corr1b} imply stringent bounds on CP violation in the charm decays, barring cancellations. Another possible way to evade this correlation would be additional operator contributions to $d_n$. The above discussion also applies to $\wcL[V8,LR]{uu}{1212}$.

As a second example, consider
\be \label{eq:class7-corr1c}
\begin{aligned}
\wcL[V1,LR]{ud}{1112} &= \wc[(1)]{qd}{1112} + \lambda (\wc[(1)]{qd}{2112} + \wc[(1)]{qd}{1212}) 
+ \lambda^2  \wc[(1)]{qd}{2212} + {\mathcal{O}(\lambda^3)}  \,,\quad\\
\wcL[V1,LR]{ud}{2223}  &=\wc[(1)]{qd}{2223}  -  
\lambda(\wc[(1)]{qd}{1223}+\wc[(1)]{qd}{2123})
+ \lambda^2( \wc[(1)]{qd}{2323}  +\wc[(1)]{qd}{3223} + \wc[(1)]{qd}{1123}  ) 
+\mathcal{O}(\lambda^3)\,.
\end{aligned}
\ee 
Here the two operators on the l.h.s. generate $s \to d u \bar u $ and $b\to s  c \bar c$, whereas, on the r.h.s, some of the operators contribute to the EDMs, because
\be
\wcL[V1,LR]{dd}{2112}  = \wc[(1)]{qd}{2112} \,,\quad
\wcL[V1,LR]{dd}{3223}  =  \wc[(1)]{qd}{3223}  \,,
\ee
resulting in correlations between these two type of observables. 

A third example would be the case of two EDM operators
\be
\wcL[S,RR]{eu}{1111}  = -\wc[(1)]{lequ}{1111}+ \mathcal{O}(\lambda)  \,,\quad
\wcL[S,RR]{eu}{1122}  =  -\wc[(1)]{lequ}{1122} + \mathcal{O}(\lambda) \,,
\ee
here the operators on the r.h.s. generate charged currents through $\text{SU(2)}_L$ symmetry, because
\be
\wcL[S,RR]{\nu e du}{1111}  = \wc[(1)]{lequ}{1111}  \,,\quad
\wcL[S,RR]{\nu e du}{1122}  =  \wc[(1)]{lequ}{1122} \,.
\ee
This induces correlations between EDMs and $d\to u e^- \bar \nu$ or $c \to s e^- \bar \nu$ processes. 

As a fourth example, we consider the EDM operators  
\be
\wcL[V,LR]{ee}{1221}  = \wc[]{le}{1221}   \,,\quad
\wcL[V,LR]{ee}{1331}  =  \wc[]{le}{1331} .
\ee
{However, due to $\text{SU(2)}_L$:}
\be
\wcL[V,LR]{\nu e}{1221}  = \wc[]{le}{1221}  \,,\quad
\wcL[V,LR]{\nu e}{1331}  =  \wc[]{le}{1331}  \,,
\ee
suggesting correlation between EDMs and $\mu \to e \nu \bar \nu$ or $\tau \to e \nu \bar \nu$. 

Finally, take the EDM operators 
\be
\wcL[T,RR]{eu}{1111}  = -(\wc[(3)]{lequ}{1111}+  
\lambda \wc[(3)]{lequ}{1121}+ \lambda^3 \wc[(3)]{lequ}{1131}) \,,\quad
\wcL[T,RR]{eu}{1122}  = -(\wc[(3)]{lequ}{1122} -\lambda \wc[(3)]{lequ}{1112} + 
\lambda^2 \wc[(3)]{lequ}{1132})    \,,
\ee
where some of the operators on the r.h.s. generate charged currents 
due to $\text{SU(2)}_L$: 
\be
\wcL[T,RR]{\nu e d u}{1121}  = \wc[(3)]{lequ}{1121}  \,,\quad
\wcL[T,RR]{\nu e d u}{1132}  = \wc[(3)]{lequ}{1132}  \,.
\ee
These contribute to $s \to u {e} \bar \nu$ and $b\to c {e} \bar\nu$ processes at $\mathcal{O}(\lambda)$ and $\mathcal{O}(\lambda^2)$, according to this expansion. 

Similarly, $\wcL[S,RL]{\nu e du}{}$ and $\wc[S,RL]{ed}{}$ get correlated via the SMEFT WC $\wc[]{ledq}{}$, introducing correlations between EDMs and charged currents. Some of these $\text{SU(2)}_L$ correlations have been pointed out earlier \cite{Cirigliano:2017tqn, Aebischer:2018csl}.

\begin{table}[H]
\begin{center}
\renewcommand*{\arraystretch}{1.0}
\resizebox{1.0\textwidth}{!}{
\begin{tabular}{ |c|c|c|c| }
\hline
\multicolumn{4}{|c|}{$\text{SU}(2)_L$ Correlations for Class 7} \\
\hline
Class 7 WC at $\muEW$ &  $\text{SU}(2)_L$ correlations   & correlated processes  & correlated classes \\
\hline
$\wc[(1)]{qu}{vwji}$ &$\wcL[V1,LR]{du}{vwji} = \wc[(1)]{qu}{vwji}$ & $c \to u d \bar d (s\bar s)$    & {3} \\
$\wc[(8)]{qu}{vwji}$ &$\wcL[V8,LR]{du}{vwji} = \wc[(8)]{qu}{vwji}$ & $c \to u d \bar d (s\bar s)$    & {3} \\
$\wc[(1)]{qd}{ijji}$ &$\wcL[V1,LR]{ud}{mnji} = V_{mv} \wc[(1)]{qd}{vwji} {V_{nw}^*}$ &
$s \to d u \bar u,~  b\to s s \bar s $    & {3} \\
$\wc[(8)]{qd}{ijji}$ &$\wcL[V8,LR]{ud}{mnji} = V_{mv} \wc[(8)]{qd}{vwji} {V_{nw}^*}$ &
$s \to d u \bar u,~  b\to s s \bar s $    & {3} \\
$\wc[]{le}{ijji}$ &$\wcL[V,LR]{\nu e}{ijji} = \wc[]{le}{ijji}$ & $\mu \to e \nu \bar \nu,~ \tau \to e \nu \bar \nu$    &  {--}\\
$\wc[]{ledq}{iijj}$ &$\wcL[S,RL]{\nu edu}{iijj} = {\wc[]{ledq}{iijm}V_{jm}^*}$ &  $s\to u  l \nu,~ {b\to c l \nu} $   & {8} \\
$\wc[(1)]{lequ}{iivj}$ &$\wcL[S,RR]{\nu e du}{iivj} = \wc[(1)]{lequ}{iivj}$ & $s\to u  l \nu,~ {b\to c l \nu} $     & {8} \\
$\wc[(3)]{lequ}{iivj}$ &$\wcL[T,RR]{\nu edu}{iivj} = \wc[(3)]{lequ}{iivj}$ & $s\to u  l \nu,~ {b\to c l \nu} $    & {8} \\
\hline
\end{tabular}
}
\caption{A summary of $\text{SU(2)}_L$ correlations for Class 7. The first column shows EDM operators, the second column lists the other WET operators generated by the operators in the first column and the third column shows processes generated by operators in the second column. The last column shows the classes that are correlated with Class 7.}
\label{tab:class7-su2}
\end{center}
\end{table}

\section{Semileptonic Charged Current Decays (Class 8)}\label{class8}

In this section, we will discuss charged current processes such as $B$-decays: $\bar B\to D \ell\bar\nu_l$, $\bar B\to D^* \ell\bar\nu_l$, $B_c\to J/\psi\tau\bar\nu$, $B_c^+\to \tau^+\nu_\tau$, $B^+\to \tau^+\nu_\tau$, Kaon decays: $K\to l\nu \,,$ Pion-decays: $\pi^- \to l^- \bar \nu$, neutron decay, and nuclear beta-decay, etc. At the quark level all these processes are governed by $d_q \to u_p l^-_i\bar\nu_i$ underlying transitions.

\subsection{WET Operators for Class 8}
For this class we will use the JMS basis. The effective Lagrangian is usually written as 
\be \label{lag:cc}
\mathcal{L}_{\rm WET}^{CC} = \wcL[V,LL]{\nu e du}{iiqp} \opL[V,LL]{\nu e du}{iiqp} 
+\sum_{I}  \wcL[I]{\nu e du}{iiqp} {\opL[I]{\nu edu}{iiqp}} +~ h.c. 
\ee
In the SM only $ \opL[V,LL]{\nu e du}{iiqp}$ is present, while the rest of the operators can be generated only by NP. Yet, NP can also enter indirectly through the Fermi-constant. At tree-level, all possible WET operators contributing directly to semileptonic charged currents are
\begin{center}
$\textrm{\bf WET 8}$
\end{center}
\be \label{eq:class8-wet-tree} 
\begin{aligned}
\opL[V,LL]{\nu e du}{iiqp}  \,, \quad
\opL[V,LR]{\nu e du}{iiqp}   \,,   \quad
\opL[S,RL]{\nu e du}{iiqp} \,, \quad
\opL[S,RR]{\nu e du}{iiqp}  \,,   \quad
\opL[T,RR]{\nu edu}{iiqp}. 
\end{aligned}
\ee
However, $d_q \to u_p l^-_i \bar\nu_i$ processes are actually governed by the hermitian conjugate versions of these operators. Also, since the neutrino flavour cannot be tagged experimentally, the leptonic flavour violating operators with indices $jiqp, i \neq j$ can in principle contribute to charged current processes.

In the SM we have 
\be \label{eq:vllSM}
\wcL[V,LL]{\nu e d u}{iiqp}  =  -\frac{4G_F}{\sqrt 2}  V_{pq}^*\,.
\ee
Here $V_{pq}$ denote the elements of the CKM matrix. Since the most precise determination of the Fermi constant $G_F$ is obtained from muon decay, any operator affecting muon decay can also contribute to the charged current decays. The WET operators governing the muon decay are \cite{Jenkins:2017jig}

\begin{center}
$\textrm{\bf WET 8 via $G_F$}$
\end{center}
\be
\begin{aligned} \label{eq:fermi-operators-wet}
\opL[V,LL]{{\nu}e}{1221}\,, \quad \opL[V,LR]{{\nu}e}{1221}.
\end{aligned}
\ee
They will be termed as {\em $G_F$-operators} in what follows. The lists of WET operators shown in \eqref{eq:class8-wet-tree} and \eqref{eq:fermi-operators-wet} fully characterize many charged-current decays for decaying particles involving down-type quarks, that are of the form $d_q \to u_p l^-_i \bar\nu_i$.

\subsection{SMEFT Operators for Class 8 at $\muEW$}
Suppressing flavour indices, at tree-level the relevant SMEFT operators at $\muEW$ are found to be
\begin{center}
$\textrm{\bf SMEFT-Tree 8}$
\end{center}
\be
\begin{aligned}
\label{eq:ops-cctree}
\ops[(3)]{l q}{} \,, \quad  
\ops[(3)]{Hl}{} \,,  \quad 
\ops[(3)]{H q}{} \,, \quad 
\ops[]{Hud}{} \,,   \quad
\ops[ ]{l edq}{} \,, \quad  
\ops[(1)]{l e qu}{}  \,,   \quad 
\ops[ (3)]{l e qu}{}\,.
\end{aligned}
\ee
This does not include the operators that enter Class 8 through the effective Fermi-constant $G_F$ in the SMEFT. The tree-level matching of \eqref{eq:ops-cctree} onto the WET at $\muEW$, excluding the SM contribution in \eqref{eq:vllSM}, reads
\be \label{class8-match1}
\begin{aligned}
\wcL[V,LL]{\nu e du}{iiqp} & = -2\wc[(3)]{Hl}{ii}  V^*_{pq} -2 \wc[(3)*]{Hq}{mq}\delta_{ii} V_{pm}^* + 2\wc[(3)]{lq}{iiqm} {V_{pm}^*}  \,,  \quad 
\wcL[V,LR]{\nu e du}{iiqp}  = -{\wc[*]{Hud}{pq}\delta_{ii}} \,,  \\
\wcL[S,RL]{\nu e du}{iiqp} & = \wc[]{ledq}{iiqm} {V_{pm}^*}  \,,  \quad
\wcL[S,RR]{\nu e du}{iiqp}  =  \wc[(1)]{lequ}{iiqp}  \,,\quad 
\wcL[T,RR]{\nu e du}{iiqp}  =\wc[(3)]{l e qu}{iiqp}\,.
\end{aligned}
\ee
The CKM elements appear due to rotation to mass-basis from the down-basis, but the rotation due to the PMNS matrix has been neglected. 

Here $q$ and $p$ are the fixed flavour indices of the external quarks. For example, setting ${p=2} ({c}\textrm{-quark}), q=3 (b\textrm{-quark})$ and $i=j=3 (\tau)$ the WET operators describe $b\to c \tau \bar \nu $ processes so that the leading flavour indices for the relevant four-fermion and two-fermion operators are $3332$ and $32$ or $23$, respectively. 

Analogously, for other processes one can appropriately choose the WET flavour indices. For the WET operators involving LH $u$-quarks the leading index $2$ is generalized to $m=1-3$ on the SMEFT side within the down basis. Consequently, they can have non-trivial flavour structures at $\muEW$. 

Notably, by construction at the dim-6 level $\wcL[V,LR]{\nu e du}{iiqp}$ is lepton-flavour universal within SMEFT, assuming a unit PMNS matrix. However, this restriction can be lifted at the dim-8 level~\cite{Burgess:2021ylu}. 

As already pointed out before, for the complete description of the charged current processes, it is essential to consider additional SMEFT operators affecting the muon decay through \eqref{eq:fermi-operators-wet}. The effective Fermi constant within SMEFT exhibits dependence on the following new operators:
\begin{center}
$\textrm{\bf SMEFT-Tree 8 via $G_F$}$
\end{center}
\be
\label{GF-tree}
\ops[(3)]{H l}{}\,, \quad \ops[]{le}{}\,, \quad \ops[]{ll}{}\,.
\ee
Their matching onto the WET in the non-redundant basis, including the SM contribution, reads \cite{Jenkins:2017jig}
\be \label{class8-match2}
\wcL[V,LL]{\nu e}{1221}  = -{2 \over v_T^2} - 2 \wc[(3)]{Hl}{11} - 2 \wc[(3)]{Hl}{22} + \wc[]{ll}{1221}\,,  \quad
\wcL[V,LR]{\nu e}{1221}  = \wc[]{l e}{1221}\,.
\ee
In the SM only $\wcL[V,LL]{\nu e}{1221}$ is generated through $W$-boson exchange which is given by the first term here.

Having discussed the tree-level SMEFT operators, we are now in a position to explore the one-loop structure of the SMEFT Lagrangian.

\subsection{SMEFT Operators for Class 8 at $\Lambda$}
The relevant operators at $\Lambda$ can be inferred from the RGEs for the operators in \eqref{eq:ops-cctree} and \eqref{GF-tree}. We begin by presenting the structure of gauge coupling dependent RGEs. In particular, we dissect the RGEs according to classes of SMEFT operators involved.
\\

\noindent
\underline{\bf \boldmath $f^4 \to f^4$ (Gauge)}:

\be
{\dotwc[(3)]{lq}{iipq}} = {3 g_2^2 \wc[(1)]{lq}{iipq}+\frac{1}{3} g_2^2 \delta_{ii} (2 \wc[(3)]{lq}{wwpq}+\wc[(1)]{qq}{pwwq}+6 \wc[(3)]{qq}{pqww}-\wc[(3)]{qq}{pwwq})-\left(g_1^2+6 g_2^2\right) \wc[(3)]{lq}{iipq}\,.} 
\ee
The running of the scalar and tensor operators was already discussed in Class 6:
\be
\wc[]{ledq}{}\,,\,\wc[(1)]{lequ}{}\,,\,\wc[(3)]{lequ}{} : \eqref{eq:scal_gauge}\,.
\ee
For the remaining four-fermion operators giving dim-6 corrections to $G_F$, we have
\be
\begin{aligned}\label{eq:cll4g}
\wc[]{ll}{1221} &= \eqref{eq:ll1221} \,,\\
\dotwc[]{le}{1221} &= -6 g_1^2 \wc[]{le}{1221}\,. 
\end{aligned}
\ee

\noindent
\underline{\bf \boldmath {$f^2 H^2 D \to f^4$} (Gauge)}:

\be
\begin{aligned}
\wc[(3)]{lq}{} &: {\eqref{eq:f2H2Dtof4gaugecl2}}\,, \\
\wc[]{ll}{1221} &= \eqref{eq:cllHl}\,.
\end{aligned}
\ee

\noindent
\underline{\bf \boldmath {$f^4 \to f^2H^2D$} (Gauge)}:

\noindent
\newline
With the appropriate changes one can also find the RGEs for this sector in the previous sections:
\be
\begin{aligned}
\wc[(1)]{Hl}{}\,,\wc[(3)]{Hl}{} &: \eqref{eq:2lep_class4}\,, \\
\wc[(3)]{Hq}{} &: \eqref{eq:class2-hq1hq3hd-f4f2-gauge1}\,.
\end{aligned}
\ee

\noindent
\underline{\bf \boldmath $f^2H^2D \to f^2H^2D$ (Gauge)}:

\noindent
\newline
One can also find the RGEs for this sector in the previous sections:
\be
\begin{aligned}
\wc[(1)]{Hl}{}\,,\wc[(3)]{Hl}{} &: \eqref{eq:Hl_gauge}\,, \\
\wc[(3)]{Hq}{} &: \eqref{eq:class4-hq1hq3huhd-f2f2-gauge}\,, \\
\wc[]{Hud}{} &: \eqref{eq:Hudgauge}\,. \\
\end{aligned}
\ee

Next we present the components of Yukawa dependent RGEs relevant for Class 8.

\noindent
\underline{\bf \boldmath $f^4 \to f^4$ (Yukawa)}:

\be
\begin{aligned}\label{eq:scal_yuk_Cl8}
\dotwc[]{ledq}{} &= y_t^2 F_1(\wc[]{ledq}{})+y_b y_t F_2(\wc[(1)]{lequ}{})+ y_b^2 F_3(\wc[]{ledq}{})\\
&+y_b y_\tau F_4(\wc[]{ed}{},\wc[]{ld}{},\wc[]{le}{},\wc[(1)]{lq}{},\wc[(3)]{lq}{},\wc[(1)*]{qd}{},\wc[(8)*]{qd}{},\wc[]{qe}{})+y_\tau y_t F_5(\wc[(1)*]{quqd}{},\wc[(8)*]{quqd}{})+ y_\tau^2 F_6(\wc[]{ledq}{})\,,\\
\dotwc[(1)]{lequ}{} &=y_t^2F_1(\wc[(1)]{lequ}{})+ y_b y_tF_2(\wc[]{ledq}{})+  y_t y_\tau F_3(\wc[]{eu}{},\wc[]{le}{},\wc[(1)]{lq}{}, \wc[(3)]{lq}{},\wc[]{lu}{},\wc[]{qe}{}{,\wc[(1)]{qu}{},\wc[(8)]{qu}{}})\\&
+  y_b^2 F_4(\wc[(1)]{lequ}{}){+y_\tau y_bF_5(\wc[(1)]{quqd}{},\wc[(8)]{quqd}{})}+
y_\tau^2F_6(\wc[(1)]{lequ}{})\,.
\end{aligned}
\ee
The flavour dependence in these RGEs can be found in the supplemental material. The running for the remaining Wilson coefficients can be deduced from other classes:

\be
\begin{aligned}
\wc[(3)]{lq}{} &: {\eqref{eq:class2-gauge-lequ13-f4-f4}}\,, \\
\wc[(3)]{lequ}{} &: {\eqref{eq:scal_yuk}}\,, \\
\wc[]{ll}{}  &: \eqref{eq:lleele-yukawa}\,, \\
{\wc[]{le}{}} &: \eqref{eq:lleele-yukawa}\,.
\end{aligned}
\ee
The full Yukawa dependence for the $\wc[(1)]{lequ}{iiqp}$ and $\wc[(3)]{lequ}{iiqp}$ is given in the supplemental material.

\noindent
\underline{\bf \boldmath $f^2H^2D \to f^4$ (Yukawa)}:
\be
\dotwc[(3)]{lq}{iiqp} = -y_t^2 \wc[(3)]{Hl}{ii} V_{3p} V^*_{3q}\,.
\ee

\noindent
\underline{\bf \boldmath $f^4 \to f^2H^2D$ (Yukawa)}:

\noindent
\newline
With the appropriate changes one can find the RGEs for this sector in the previous sections:
\be
\begin{aligned}
\wc[(1)]{Hl}{}\,,\wc[(3)]{Hl}{} &: \eqref{eq:2lep_yuk_class4}\,, \\
\wc[(3)]{Hq}{} &: \eqref{eq:class2-hq1hq3hd-f4f2-yuk}\,, \\
\wc[]{Hud}{} &: \eqref{eq:Hudf4}\,.
\end{aligned}
\ee

\noindent
\underline{\bf \boldmath $f^2H^2D \to f^2H^2D$ (Yukawa)}:

\noindent
\newline
Again, the RGEs for this sector can be derived from the previous sections:
\be
\begin{aligned}
\wc[(1)]{Hl}{}\,,\wc[(3)]{Hl}{} &: \eqref{eq:Hlyuk}\,, \\
\wc[(3)]{Hq}{} &: \eqref{eq:class2-hq1hq3hd-f2f2-yuk}\,, \\
\wc[]{Hud}{} &: \eqref{eq:Hudnon4f}\,, \\
\end{aligned}
\ee

Building on our discussion of RGEs, we are now in position to list, at 1-loop level, operators relevant for the charged current processes including both trivial and non-trivial flavours. At the scale $\Lambda$, we identify following new operator structures:

\begin{center}
$\textrm{\bf SMEFT-Loop 8}$
\end{center}
\be
\begin{aligned} \label{eq:class8-smeft-loop}
\textrm{\bf Yukawa-Mixing}: ~&
\ops[(1)]{qq}{} \,,  \quad
\ops[(3)]{qq}{}\,, \quad 
{\ops[(1)]{lq}{}} \,,  \quad
{\ops[]{le}{}} \,,  \quad
{\ops[]{ld}{}} \,,  \quad
\ops[]{lu}{} \,,  \\
&
\ops[]{qe}{} \,,  \quad
\ops[]{HD}{} \,,  \quad
\ops[]{H \Box}{}\,, \quad  
\ops[(1)]{Hq}{} \,, \quad 
\ops[]{Hu}{} \,,  \quad
\ops[]{Hd}{} \,,  \\
&
\ops[]{He}{} \,,  \quad
{\ops[]{ed}{}} \,,  \quad
{\ops[]{eu}{}} \,,  \quad
{\ops[]{ee}{}} \,, \quad 
\ops[(1)]{ud}{} \,,  \quad
\ops[(8)]{ud}{} \,,  
\\
\textrm{\bf Gauge-Mixing}: & ~
\ops[]{H \Box}{}\,,\quad
{\ops[]{le}{}}\,,\quad
\ops[(1)]{lq}{}\,,\quad
\ops[]{He}{} \,,  \quad
\ops[(1)]{qq}{}\,,\quad
\ops[(3)]{qq}{}\,.
\end{aligned}
\ee
Some of these operators contribute to $G_F$ at tree-level, cf. \eqref{GF-tree}. 

\begin{figure}[H]%
\hspace{-0.5cm}
\includegraphics[clip, trim=1.5cm 12.cm 1.2cm 12cm, width=1\textwidth]{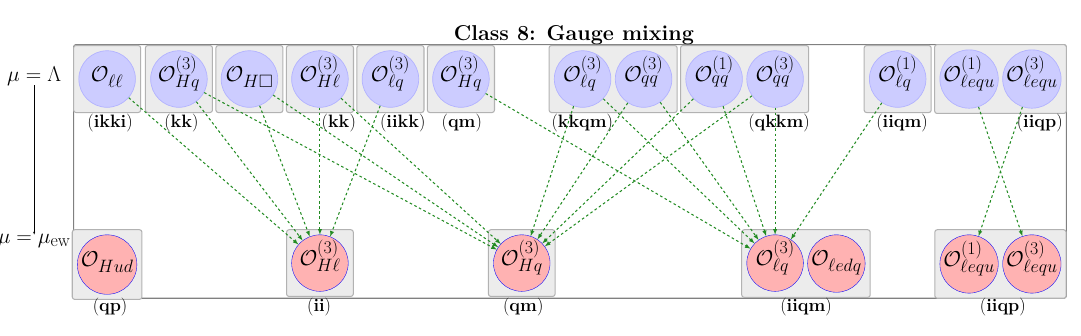}
\vspace{-0.2cm}
\caption{ \small Operator mixing relevant for $\DF=1$ charged current processes in the Warsaw down-basis. The leptonic flavour indices are fixed to $ii$ which can be 11, 22 or 33. The quark flavour indices $q$, $p$ and $m$ are fixed by tree-level matching conditions (i.e. by the observables of interest). The index $k$ should be summed over 1-3 for all occurrences. The dashed green lines indicate the operator mixing due to electroweak couplings.  $\wc[]{Hud}{}$ and $\wc[]{ledq}{}$ do not mix with any other operator at 1-loop under gauge interactions. The self-mixing is not shown here.}%
\label{chart:df1-bclnu-gauge}%
\end{figure}

\begin{figure}[H]%
\hspace{-0.5cm}
\includegraphics[clip, trim=1.2cm 12.cm 1.0cm 12.5cm, width=1\textwidth]{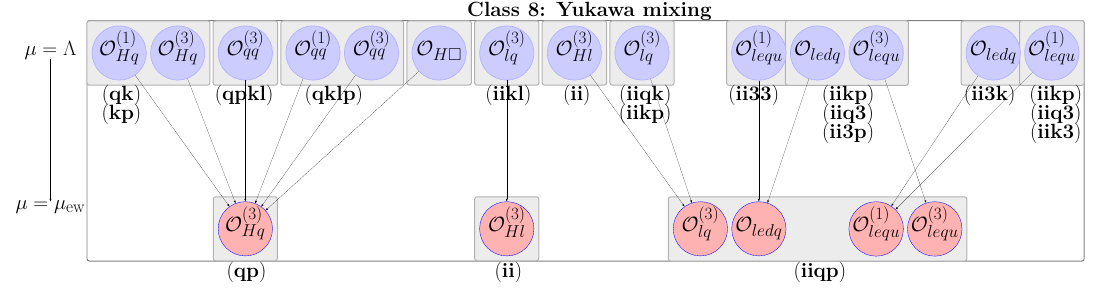}
\caption{ \small  Operator mixing relevant for charged current processes in the Warsaw down-basis. The leptonic flavour indices are fixed to $ii=11,22,33$. The indices $k$ and $l$ have to be summed. The quark flavour indices $q$ and $p$ are fixed by tree-level matching conditions (i.e. observables of interest). The black lines indicate the mixing due to the top or bottom Yukawa couplings. More details can be found in the text. Note that $\wc[(1)]{lequ}{iiqp}$ and $\wc[(3)]{lequ}{iiqp}$ also mix with many other WCs such as $\wc[]{eu}{}, \wc[]{ledq}{}, \wc[(1)]{lq}{}, \wc[(3)]{lq}{}, \wc[]{lu}{}, \wc[]{qe}{}$. These can be found in \eqref{eq:scal_yuk}. Also the self-mixing is not shown here.}
\label{chart:df1-bclnu-yukawa}%
\end{figure}

For the pure charged current operators at $\muEW$, (excluding $G_F$-operators) having trivial or non-trivial flavour structures, the RG evolution of the SMEFT operators at $\Lambda$ is shown in the charts displayed in Figs.~\ref{chart:df1-bclnu-gauge} and \ref{chart:df1-bclnu-yukawa}. It is useful to identify the 1-loop operators which specifically mix with $G_F$ tree-level operators \eqref{GF-tree}. These are listed below.

\begin{center}
{\boldmath
$\textrm{\bf SMEFT-Loop via} ~G_F$ }
\end{center}
\be
\begin{aligned} \label{eq:class8GF-smeft-loop1}
\textrm{\bf Yukawa-Mixing:}&~ \,
\ops[]{l u}{}\,, \quad 
{\ops[]{ld}{}}\,,  \quad
{\ops[(1)]{lq}{}}\,,  \quad
{\ops[(3)]{lq}{}}\,,  \quad
{\ops[]{ee}{}}\,,  \\
\textrm{\bf Gauge-Mixing:}&~\,
\ops[]{ld}{}\,,\quad 
{\ops[]{lu}{}}\,, \quad    
\ops[]{HD}{} \,, \quad
{\ops[]{H \Box}{}}\,, \quad 
\ops[]{He}{}\,, \\
&\,\,\,
\ops[]{Hd}{}\,, \quad 
\ops[]{Hu}{}\,, \quad 
{\ops[(1)]{lq}{}}\,, \quad
{\ops[(3)]{Hq}{}}\,.
\end{aligned}
\ee
Note that $\ops[(3)]{lq}{}$, and $\ops[(3)]{Hq}{}$ operators already contribute at tree-level to processes in Class 8 (see \eqref{eq:ops-cctree}). These operators are kept in \eqref{eq:class8GF-smeft-loop1} because these one-loop $G_F$ could also be relevant for other classes, e.g. Class 4. 

Utilizing the $\eta$ parameters defined in \eqref{ABKfinal} we can get a first estimate of which RG effects are most significant. Suppressing the flavour indices, and considering  $|\eta| \ge 10^{-4} $, we find following dominant contributions for $d\to u e \bar \nu$ processes
\be \begin{aligned} &
\wcs[(3)]{lq}{}\to \wcL[V,LL]{\nu edu}{ }\,, \quad
\wcs[(3)]{qq}{}\to \wcL[V,LL]{\nu edu}{ }\,, \quad
\wcs[]{ll}{}\to \wcL[V,LL]{\nu edu}{ }\,, \quad
\wcs[(1)]{lq}{}\to \wcL[V,LL]{\nu edu}{ } 
\,,\\ & 
\wcs[(1)]{qq}{}\to \wcL[V,LL]{\nu edu}{ }\,, \quad
\wcs[(3)]{Hl}{}\to \wcL[V,LL]{\nu edu}{ }\,, \quad
\wcs[(1)]{Hq}{}\to \wcL[V,LL]{\nu edu}{ }\,, \quad
\wcs[(8)]{ud}{}\to \wcL[V,LR]{\nu edu}{ } 
\,,\\ & 
\wcs[(1)]{ud}{}\to \wcL[V,LR]{\nu edu}{ }\,, \quad
\wcs[(3)]{lequ}{}\to \wcL[S,RR]{\nu edu}{ }\,, \quad
\wcs[(1)]{lequ}{}\to \wcL[T,RR]{\nu edu}{ } .
\end{aligned} \ee
All WCs appearing in this equation contribute only at 1-loop level via RG running. The presence of some operators that also appear at the tree-level should not be surprising, as they carry different flavour indices from those involved in the tree-level matching. The exact flavour structure can be found in the $\rho-\eta$ table \ref{tab:8-1}.

Similarly, for $b\to c \tau \bar \nu$ processes we find
\be \begin{aligned} &
\wcs[(3)]{qq}{}\to \wcL[V,LL]{\nu edu}{ }\,, \quad
\wcs[(1)]{lq}{}\to \wcL[V,LL]{\nu edu}{ }\,, \quad
\wcs[(1)]{Hq}{}\to \wcL[V,LL]{\nu edu}{ }\,, \quad
\wcs[(1)]{qq}{}\to \wcL[V,LL]{\nu edu}{ } 
\,,\\ & 
\wcs[(3)]{lq}{}\to \wcL[V,LL]{\nu edu}{ }\,, \quad
\wcs[]{ll}{}\to \wcL[V,LL]{\nu edu}{ }\,, \quad
\wcs[(3)]{lequ}{}\to \wcL[V,LL]{\nu edu}{ }\,, \quad
\wcs[(3)]{Hq}{}\to \wcL[V,LL]{\nu edu}{ } 
\,,\\ & 
\wcs[(8)]{ud}{}\to \wcL[V,LR]{\nu edu}{ }\,, \quad
\wcs[(1)]{ud}{}\to \wcL[V,LR]{\nu edu}{ }\,, \quad
\wcs[]{Hu}{ }\to \wcL[V,LR]{\nu edu}{ }\,, \quad
\wcs[(1)]{quqd}{}\to \wcL[S,RL]{\nu edu}{ } 
\,,\\ & 
\wcs[(3)]{lequ}{}\to \wcL[S,RR]{\nu edu}{ }\,, \quad
\wcs[(8)]{qu}{}\to \wcL[S,RR]{\nu edu}{ }\,, \quad
\wcs[(1)]{qu}{}\to \wcL[S,RR]{\nu edu}{ }\,, \quad
\wcs[(1)]{lequ}{}\to \wcL[S,RR]{\nu edu}{ } 
\,,\\ & 
\wcs[]{eu}{}\to \wcL[S,RR]{\nu edu}{ }\,, \quad
\wcs[]{lu}{}\to \wcL[S,RR]{\nu edu}{ }\,, \quad
\wcs[(1)]{lequ}{}\to \wcL[T,RR]{\nu edu}{ }\,, \quad
\wcs[(3)]{lequ}{}\to \wcL[T,RR]{\nu edu}{ }. 
\end{aligned} \ee
This demonstrates that different type of charged-current interactions exhibit distinct patterns of operator mixing, for a fixed threshold on $|\eta|$. For further details on flavour indices, we refer the reader to Tab.~\ref{tab:8-2}.

The SMEFT charts for the $G_F$-operators are given in Fig.~\ref{chart:ewp-gauge-2}, \ref{chart:ewp-gauge-1}, \ref{chart:ewp-yukawa-1}.

Finally, we note following restrictions that apply for the SMEFT chart shown in Fig.~\ref{chart:df1-bclnu-gauge}:
\begin{itemize}
\item The operator mixing  $\wc[(3)]{Hl}{kk}, \wc[(3)]{Hq}{kk} \to \wc[(3)]{Hq}{qm}$ due to gauge coupling takes place only for $q=m$. The same applies for $\wc[]{H\Box}{} \to  \wc[(3)]{Hq}{qm}$. 
\end{itemize}
Therefore, some of the lines in the chart in Fig.~\ref{chart:df1-bclnu-gauge} will disappear for $q\ne m$. 

Even though our selection criteria (i.e. keeping only $y_\tau, y_b, y_t$ dependent terms in the RGEs) for the operators at $\Lambda$ treat those with indices 12 differently from operators with 
indices 13 and 23, still in this class we do not observe any difference in the operator mixing in the Yukawa sector.
The values of $\rho$ and $\eta$ parameters are collected in Tabs.~\ref{tab:8-1} and \ref{tab:8-2} in App.~\ref{App:etas}.

{\boldmath
\subsection{$\text{SU(2)}_L$ Correlations}
}
The charged currents involve both up- and down-type quarks. The contributing WET operators involving LH up-quarks match with many SMEFT operators having non-trivial flavour structure due to 
the change of basis from the weak to the mass basis. However, at low energies those SMEFT operators can also be determined directly by measuring purely down-type processes, for which the CKM rotations are not involved when using the down-basis. 

In what follows we will identify various down quark processes for all such SMEFT operators containing LH up-quarks which control charged currents. This, in turn, also implies correlations between various other processes and Class 8. 

To find possible correlations we take all WET operators containing LH up-quarks and expand them in terms of SMEFT WCs in powers of the Wolfenstein parameter $\lambda$. For instance, for $b\to c \tau \bar \nu$ operators, we have
\be \label{eq:class8-parent}
\begin{aligned}
\wcL[V,LL]{\nu edu}{3332 } & =   -2\lambda^2 \wc[(3)]{Hl}{33}
-2 (\wc[(3)*]{Hq}{23}  -\lambda  \wc[(3)*]{Hq}{13} + \lambda^2 \wc[(3)*]{Hq}{33} )
+ 2( \wc[(3)]{lq}{3332} -\lambda  \wc[(3)]{lq}{3331} +  \lambda^2 \wc[(3)]{lq}{3333}  )\,,
\\
\wcL[S,RL]{\nu edu}{3332 } & =  
 \wc[]{ledq}{3332}  -\lambda  \wc[]{ledq}{3331} + \lambda^2 \wc[]{ledq}{3333}\,. 
\end{aligned}
\ee
Now we look at the $\text{SU(2)}_L$ WET counterparts of the SMEFT operators on the r.h.s. For the {four-fermi} operators we find
\be \label{eq:class8-counterparts}
\begin{aligned} 
\wcL[V,LL]{\nu d}{3332 } & =  -\wc[(3)]{lq}{3332}\,,  \quad \wcL[V,LL]{e d}{3332 }  =   \wc[(3)]{lq}{3332}\,, 
\quad
\wcL[V,LL]{\nu d}{3331 }  =  -\wc[(3)]{lq}{3331}\,,  \quad \wcL[V,LL]{e d}{3331 }  =   \wc[(3)]{lq}{3331}\,, \\
\wcL[V,LL]{\nu d}{3333 } & =  -\wc[(3)]{lq}{3333}\,,  \quad \wcL[V,LL]{e d}{3333 }  =   \wc[(3)]{lq}{3333}\,, \quad
{\wcL[S,RL]{e d}{3333 }}  =  \wc[]{ledq}{3333}\,,  \quad \wcL[S,RL]{e d}{3331 }  =   \wc[]{ledq}{3331}\,, \\
{\wcL[S,RL]{e d}{3332 }} & =  \wc[]{ledq}{3332}\,.
\end{aligned}
\ee
Similarly, two-fermion operators in \eqref{eq:class8-parent} can contribute to other processes via $W$ exchange, because after integrating out the $W$, these operators match onto various four-fermion operators. 

The $\text{SU(2)}_L$ relations in \eqref{eq:class8-counterparts} imply that $b\to c \tau \bar \nu$ processes are correlated with neutral current processes involving $b\to s \nu \bar \nu$, $b\to s \tau^+ \tau^-$ at $\mathcal{O}(1)$ and $b\to d \nu \bar \nu$, $b\to d \tau^+ \tau^-$ at $\mathcal{O}(\lambda)$ at the matching level. Such $\text{SU(2)}_L$ correlations were used to explain neutral and charged current anomalies in \cite{Bhattacharya:2014wla, Kumar:2018kmr}. The flavour conserving operator $\wc[(3)]{lq}{3333}$ contributes to the EWP observables at one-loop \cite{Kumar:2021yod}.

The remaining two WET operators in Class 8 contain LH down-quarks and RH up-quarks. These operators receive tree-level matching contributions from $\wc[(1)]{lequ}{}$ and $\wc[(3)]{lequ}{}$ (see \eqref{class8-match1}). Due to $\text{SU(2)}_L$ invariance they generate neutral current WET operators:
\be
\begin{aligned}
\wc[S,RR]{eu}{iiqp} = -V_{qm} \wc[(1)]{lequ}{iimp}\,, \\
\wc[T,RR]{eu}{iiqp} = -V_{qm}  \wc[(3)]{lequ}{iimp}\,.
\end{aligned}   
\ee
The procedure outlined above can be easily generalized to other charged current processes.

{\boldmath
\subsection{The Role of Tensor Operators and Related RG Effects}
}
It has been pointed out in \cite{Biancofiore:2013ki} that tensor operators could play a significant role in the explanation of the $R(D)$ and $R(D^*)$ anomalies. Subsequently, tensor operators in this context and in the context of the tensions between exclusive and inclusive determinations of $\vcb$ have been considered in \cite{Sakaki:2014sea,Duraisamy:2014sna,Freytsis:2015qca,Becirevic:2016hea,Alonso:2016gym,Colangelo:2016ymy,Li:2016vvp,Bardhan:2016uhr,Bhattacharya:2016zcw,Ivanov:2017mrj,Chen:2017hir}. In particular the role of leptoquarks in the explanation of $R(D)$ and $R(D^*)$ anomalies through tensor operators was analysed in these papers. However, RG effects have not been included in these analyses. But, as pointed out in \cite{Gonzalez-Alonso:2017iyc} the large RG mixing of tensor operator ${\cal O}_{lequ}^{(3)}$ into the (pseudo)scalar ones through top Yukawa coupling, also found in \cite{Aebischer:2017gaw}, has important phenomenological implications on these analyses and generally on charged current processes. Fig.~2 of \cite{Gonzalez-Alonso:2017iyc} demonstrates how the inclusion of these effects and also of QED effects in the WET set a strong bound on the size of the WCs of tensor operators. Subsequently, detailed implications of this finding have been studied in particular in \cite{Feruglio:2018fxo} with additional implications for $(g-2)_\mu$ and Yukawa couplings. See also the analysis in the context of leptoquark models in \cite{Becirevic:2018afm}. A recent review on $R(D)$ and $R(D^*)$ anomalies can be found in \cite{Capdevila:2023yhq}.

{\boldmath
\section{Higgs Measurements (Class 9)}
}\label{class9}
The SM Higgs boson couples to fermions through Yukawa interactions and to the $W$ and $Z$ bosons via the kinetic terms in the SM Lagrangian. In addition, it interacts with gluons and photons at 1-loop level, with the loops mediated by fermions and $W$ bosons.  As a result, both production mechanisms  and decays channels of the Higgs boson are governed by these interactions.

The 3rd generation Yukawa couplings ($t,b,\tau$) have already been measured at the level of $10$-$20\%$ precision \cite{CMS:2022dwd, ATLAS:2022vkf}. Very recently an upper bound on $h\to c \bar c$ has been set \cite{CMS:2025qmm, ATLAS:2024yzu}. The couplings to light fermions are yet to be  measured.\footnote{See e.g. Ref.~\cite{CMS:2025xkn} and \href{https://cds.cern.ch/record/2917791/files/ATL-PHYS-SLIDE-2024-588.pdf}{ATL-PHYS-SLIDE-2024-588}.} As of now the Higgs couplings to $W, Z, \gamma$ and gluons are known with a precision better than $10\%$ \cite{CMS:2022dwd, ATLAS:2022vkf}. We can also expect much improvements in their measurements at future experiments such as the HL-LHC~\cite{Cepeda:2019klc, Mlynarikova:2023bvx}. 

The most important Higgs measurements consist of the so-called signal strength parameters defined by 
\be
\mu_i^j ={   \sigma_i \times {\rm BR_j}  \over  \sigma_i({\rm SM}) \times {\rm BR_j{\rm (SM)}} }\,, 
\ee
where, $\sigma_i $ refers to the production cross-section in different modes such as gluon fusion, associated production with $W$ and $Z$, vector boson fusion or associated production with a quark and antiquark. ${\rm BR_j}$ denotes the decays modes: $f\bar f$, $WW, ZZ, \gamma\gamma$ and $gg$. The Higgs boson can also decay invisibly.

Deviations in the Higgs signal strength due to BSM physics can arise from modification of Higgs coupling induced by new particles or new interactions. In the following, we provide theoretical tools required to analyze such BSM effects within the SMEFT framework at the 1-loop level. With advent of high precision measurements of Higgs couplings, these effects maybe experimentally accessible. Moreover, an SMEFT based study of the Higgs sector provides values insights into the potential correlations between Higgs physics and other sectors.

Alternatively, BSM effects in the Higgs signal strength can also be parameterized through the $\kappa$-framework \cite{LHCHiggsCrossSectionWorkingGroup:2012nn}. In this approach, the basic idea is to dress the Higgs decay widths and production cross-sections by $\kappa_i$ factors representing the BSM physics, with $i$ representing different production and decay channels. 

In addition to the signal strength parameters, there are also other important Higgs observables such as measurements of kinematic distributions \cite{Dittmaier:2012vm}.

\subsection{Non-standard Higgs Couplings} 
Apart from uncovering correlations between Higgs measurements and other precision observables, including those from the EWP sector, the formulation of Higgs interactions in terms of the SMEFT also enables us a systematic treatment of 1-loop effects, taking advantage of fully known ADMs at the 1-loop level.

First we list the CP-even and CP-odd non-standard single Higgs interactions parameterized in terms of the Higgs basis. The Higgs boson couplings with vector boson pairs, $hVV$, where $V=W, Z, g, \gamma$, can be written as in ~\cite{LHCHiggsCrossSectionWorkingGroup:2016ypw} 
\be
\label{eq:hvv}
\begin{aligned}
 {\cal L}_{  h V \bar V} 
& ~ =  \left (1 +  \delta c_W \right )~  {g^2 v^2 \over 2} W_\mu^+ W^{\mu - } { h\over v} \\
&~  +    \left (1 +  \delta c_Z \right )~  {(g^2+g^{\prime 2}) v^2 \over 4} Z_\mu Z^\mu {h \over v}\\
& ~  + c_{WW} ~ {g^2 \over  2} W_{\mu \nu}^+  W^{\mu\nu -} {h \over v}  ~+~ \widetilde c_{WW}~  {g^2 \over  2} 
W_{\mu \nu}^+   \widetilde W^{\mu\nu -} {h \over v} \\
&~ + c_{W \Box}~ g^2 \left (W_\mu^- \partial_\nu W^{\mu \nu -} { h\over v} + {\mathrm h.c.} \right )   \\ 
& ~ +  c_{gg} {g_s^2 ~\over 4 } G_{\mu \nu}^A G^{A\mu \nu} {h\over v}  
 ~+~ c_{\gamma \gamma} ~{e^2 \over 4} A_{\mu \nu} A^{\mu \nu} { h \over v} \\
&~ + c_{Z \gamma} ~{e \sqrt{g^2 + g'{}^2}  \over  2} Z_{\mu \nu} A^{\mu\nu}  {h  \over v}
 ~+~ c_{ZZ} ~{g^2 + g'{}^2 \over  4} Z_{\mu \nu} Z^{\mu\nu} { h\over v}  \\
&~ + c_{Z \Box}~ g^2 Z_\mu \partial_\nu Z^{\mu \nu} {h \over v}
 ~+~ c_{\gamma \Box}~ g g' Z_\mu \partial_\nu A^{\mu \nu} {h \over v}  \\ 
&~ +  \widetilde c_{gg}~ {g_s^2 \over 4} G_{\mu \nu}^A \widetilde G^{A\mu \nu }  { h \over v}
 ~+~ \widetilde c_{\gamma \gamma}~ {e^2 \over 4} A_{\mu \nu} \widetilde A^{\mu \nu}{ h \over v}  \\
&~ + \widetilde c_{Z \gamma} ~{e \sqrt{g^2 + g'{}^2} \over  2} Z_{\mu \nu} \widetilde A^{\mu\nu} { h \over v}
 ~+~ \widetilde c_{ZZ} ~ {g^2 + g'{}^2  \over  4} Z_{\mu \nu} \widetilde Z^{\mu\nu}{ h \over v}\,,
\end{aligned}
 \ee
where $A_{\mu\nu} = \partial_\mu A_\nu - \partial_\nu A_\mu$\, etc. Here $h$ is the physical Higgs boson, in contrast to the symbol $H$ used for the Higgs doublet in this review. Likewise, the non-standard Yukawa couplings $hf\bar f$ can be parameterized as 
\begin{eqnarray}
\label{eq:hff}
{\cal L}_{ h f \bar f} &= & 
-  \sum_{f \in u,d,e}  
\sqrt{m_{f_i} m_{f_j}} \left ( \delta_{ij} 
+ [\delta y_f]_{ij}  e^{i [\phi_f]_{ij}}   \right ) 
\bar f_{R,i} f_{L,j} { h\over v} + {\rm h.c.}
\end{eqnarray} 
Here $i,j$ are the flavour indices in the mass-basis. At tree-level the Higgs measurements are only sensitive to the flavour-conserving interactions. 

In addition, many four-fermion operators can be relevant for the LHC Higgs phenomenology \cite{Azatov:2022kbs}.

Interestingly, not all of the couplings discussed above are independent. Upon suppressing flavour indices, only 16 independent single-Higgs CP-even and CP-odd couplings remain that affect the Higgs production and decay rates, but do not affect the EW precision observables \cite{LHCHiggsCrossSectionWorkingGroup:2016ypw}. These are:
\be \label{eq:}
\begin{aligned} 
\delta c_{Z}\,, & \quad c_{gg}\,,  \quad c_{\gamma\gamma}\,,\quad c_{Z\gamma}\,, \quad 
c_{ZZ}\,,\quad c_{Z \Box}\,, \quad
[\delta y_{u}]_{ij}\,, \quad [\delta y_d]_{ij}\,, \quad [\delta y_e]_{ij}\,, \\
&
\widetilde c_{gg}\,, \quad \widetilde c_{\gamma \gamma}\,, \quad \widetilde c_{Z\gamma}\,, \quad \widetilde c_{ZZ} \,,  \quad
 [\delta \phi_u]_{ij}\,, \quad [\delta \phi_d]_{ij}\,, \quad [\delta \phi_e]_{ij}\,.
\end{aligned}
\ee
The couplings in the first (second) line are of CP-even (CP-odd) in nature. Apart from these, gauge boson couplings to fermions \eqref{gauge-couplings} can also affect the Higgs rates at the tree-level. However, since they are primarily constrained by the $Z$-pole observables, they are grouped under Class 4 (Sec.~\ref{class4}). This also implies the presence of strong correlations between the EWP and the Higgs sectors. 

\subsection{SMEFT Operators for Class 9 at $\muEW$}
Now we will discuss the relationship between the non-standard interactions in the Higgs basis and the Warsaw basis. At the tree-level, non-fermion, 2f as well as 4f operators 
can contribute to the single-Higgs couplings. The list is as follows
\begin{center}
{\bf SMEFT-Tree 9}
\end{center}
\be \label{eq:class9-smeft-tree}
\begin{aligned}
& \ops[]{H}{} \,, \quad
Q_{H\Box}  \,, \quad      
Q_{H B} \,, \quad
Q_{HW}  \,,\quad
Q_{H \widetilde B} \,, \quad
Q_{H \widetilde W} \,,  \\
& Q_{H G} \,, \quad
Q_{H \widetilde G}  \,, \quad
Q_{H W B} \,, \quad
Q_{H \widetilde W B}\,, \quad
Q_{HD}\,, \quad
\ops[(3)]{H\ell }{pp} \,,  \quad \\
& \ops[]{uH}{pq} \,, \quad 
\ops[]{dH}{pp}  \,,   \quad
\ops[]{eH}{pp}\,, \quad
\ops[]{\ell \ell}{1221}.
\end{aligned}
\ee
Here the WCs $\wc[]{ll}{1221}$ and $\wc[(3)]{H l}{pp}$ contribute through the effective Fermi constant within the SMEFT, therefore, $pp$ =$11$ or $22$ is allowed. The complete list of other four-fermion operators relevant for the LHC Higgs phenomenology are given in Eq.~(4.13) of \cite{Azatov:2022kbs}. 

The $\wc[]{xH}{}$ operators give direct corrections to Yukawa couplings after EW symmetry breaking. The flavour violating operator $\ops[]{uH}{pq}$ contributes to up-type quark couplings due to misalignment of up quarks in the down-basis. 

The actual relations between these operators and the Higgs couplings in \eqref{eq:hvv}-\eqref{eq:hff} can be found in \cite{Azatov:2022kbs}. See also this reference for the mapping of Higgs couplings to other SMEFT bases, such as the SILH basis.

Here we have not considered operators contributing simultaneously to Higgs and EWP observables discussed in Class 4. With the current experimental sensitivities the latter observables are expected to put stronger constraints on such operators. 
 
\subsection{SMEFT Operators for Class 9 at $\Lambda$}

Having identified the tree-level SMEFT operators for the Higgs signal strength, we now turn to the relevant pieces of the ADMs for this class. This will then allow to determine the complete
set of SMEFT operators that contribute at 1-loop level. 

Let us recall, that we assume the dominance of third generation Yukawas, hence the Yukawa RG running effects in the first and second generation operators are expected to be suppressed. Under this assumption, we find that the operators in \eqref{eq:class9-smeft-tree} can be generated from several other operators due to top, bottom, tau Yukawas and gauge coupling dependent RG running. They are given as follows:
\begin{center}
{\bf SMEFT-Loop 9}
\end{center}
\be \label{eq:class9:smeft-loop}
\begin{aligned}
\textrm{\bf Yukawa:}&\,\,\, 
\ops[(1)]{Hq}{}  \,,  \quad
\ops[(3)]{Hq}{}  \,,  \quad
\ops[]{Hd}{} \,, \quad
\ops[]{Hu}{} \,, \quad
\ops[]{Hud}{} \,, \quad
\ops[(1)]{Hl}{} \,,  \\
&
{\ops[]{He}{} } \,,  \quad
\ops[(1)]{quqd}{}  \,, \quad
\ops[(8)]{quqd}{}   \,, \quad
\ops[(1)]{qu}{}  \,, \quad
\ops[(8)]{qu}{}  \,, \quad 
\ops[(1)]{qd}{}  \,, \quad \\
&
\ops[(8)]{qd}{}  \,, \quad
\ops[(1) ]{\ell e qu}{} \,, \quad
{\ops[]{ledq}{}}\,,  \quad
{\ops[]{le}{}}.
\\
\textrm{\bf Gauge:} &\,\,\,
{\ops[(1)]{Hq}{}}\,, \quad 
\ops[(3)]{Hq}{}\,, \quad 
\ops[]{Hd}{} \,, \quad
\ops[]{Hu}{} \,, \quad
\ops[(1)]{Hl}{} \,, \quad
\ops[]{He}{} \,, \\
&
\ops[]{uB}{}  \,, \quad
\ops[]{uW}{} \,, \quad
\ops[]{eW}{} \,, \quad
\ops[]{eB}{} \,, \quad
\ops[]{dB}{}\,, \quad
\ops[]{dW}{}  \,, \\
&
\ops[]{W}{}\,, \quad
\ops[]{\widetilde W}{}\,.
\end{aligned}
\ee
The tree-level operators are not displayed as the RG running only introduces small shifts in their values at $\muEW$. However, flavour variants for these operators through running can appear and must be included in a realistic analysis.

Next we present the explicit RGEs used to identify the above operators. Some of the equations are already given in the previous classes. For example, the RG running of the $G_F$ operator $\wc[]{ll}{1221}$ can be found in Class 4, see \eqref{eq:ll1221} and \eqref{eq:cllHl}. For $\wc[(3)]{Hl}{}$ the relevant RGEs are already given in \eqref{eq:2lep_class4}, \eqref{eq:Hl_gauge}, \eqref{eq:2lep_yuk_class4} and \eqref{eq:Hlyuk}. 

We now present the RGEs of the remaining operators in Class 9.
\\

{
\noindent
\underline{\bf \boldmath {\bf non-fermion} $\to$ {\bf non-fermion} (Gauge)}:
In this category we find}
\be
\begin{aligned}
\dotwc[]{H}{} &=-\frac{9}{2}(g_1^2+3 g_2^2) \wc[]{H}{} -3 (g_1^4+ g_1^2 g_2^2)\wc[]{HB}{} 
-\frac{3}{4}(g_1^2+g_2^2)^2 \wc[]{HD}{}
\\
&-3 (g_1^2 g_2^2 + 3g_2^4)\wc[]{HW}{}
-3( g_1g_2^3+g_1^3 g_2)\wc[]{HWB}{}\,,
\end{aligned}
\ee
\be
\begin{aligned}
\dotwc[]{HB}{} & = \frac{85}{6}g_1^2 \wc[]{HB}{}-\frac{9}{2}g_2^2 \wc[]{HB}{}
+3g_1 g_2\ \wc[]{HWB}{}\,,\\
\dotwc[]{HW}{} & = 
-\frac{3}{2}g_1^2 \wc[]{HW}{}-\frac{53}{6}g_2^2 \wc[]{HW}{} +g_1 g_2\wc[]{HWB}{} -15 \wc[]{W}{} g_2^3\,,\\
\dotwc[]{HG}{} & = -\frac{1}{2}( 3g_1^2+9 g_2^2+28g_s^2) \wc[]{HG}{} \,,
\end{aligned}
\ee
\be
\begin{aligned}
\dotwc[]{H\widetilde B}{} & = 
\frac{85}{6} g_1^2 \wc[]{H\widetilde B}{} -\frac{9}{2} g_2^2\wc[]{H\widetilde B}{}  +3 g_1 g_2 \wc[]{H\widetilde WB}{}\,, \\
\dotwc[]{H\widetilde W}{} & = 
-\frac{3}{2}g_1^2 \wc[]{H\widetilde W}{}-\frac{53}{6}g_2^2 \wc[]{H\widetilde W}{} +g_1 g_2\wc[]{H\widetilde WB}{} -15 g_2^3 \wc[]{\widetilde W}{} \,,\\
\dotwc[]{H\widetilde G}{} & = -\frac{1}{2}( 3g_1^2+9 g_2^2+28g_s^2) \wc[]{H\widetilde G}{}\,,\\
\dotwc[]{H\widetilde WB}{} & = 2g_1 g_2(\wc[]{H\widetilde B}{}+ \wc[]{H\widetilde W}{})+\frac{19}{3}g_1^2\wc[]{H\widetilde WB}{} +\frac{4}{3}g_2^2 \wc[]{H\widetilde WB}{} +3g_1 g_2^2 \wc[]{\widetilde W}{}.
\end{aligned}
\ee

{
\noindent
\underline{\bf \boldmath {\bf non-fermion, 2f} $\to$ {\bf non-fermion} (Gauge)}:
}
\be
\wc[]{HD}{}: \eqref{eq:CHD}\,, \quad
\wc[]{HWB}{}: \eqref{eq:CHD}\,,
\ee

\be
\begin{aligned}
\dotwc[]{H\Box}{} &= -\frac{2}{3} g_1^2(\wc[]{Hd}{vv}+\wc[]{He}{vv}+\wc[(1)]{Hl}{vv}
-\wc[(1)]{Hq}{vv}-2\wc[]{Hu}{vv}+2\wc[]{H\square}{}-\frac{5}{2}\wc[]{HD}{}) \\
&
+2 g_2^2( \wc[(3)]{Hl}{vv}+3\wc[(3)]{Hq}{vv}-2\wc[]{H\square}{}).
\end{aligned}
\ee

{
\noindent
\underline{\bf \boldmath  $f^2 H^3, f^2 X H \to f^2 H^3  $ (Gauge)}:
}
\be
\begin{aligned}
\dotwc[]{dH}{pr} &=\left(g_1^3-3 g_1 g_2^2\right) \wc[]{dB}{pr}-\frac{1}{12} \left(23 g_1^2+81 g_2^2+96 g_s^2\right) \wc[]{dH}{pr}+g_2 \left(g_1^2-9 g_2^2\right) \wc[]{dW}{pr}\,,\\
\dotwc[]{uH}{pr} & = \left(3 g_1 g_2^2-5 g_1^3\right) \wc[]{uB}{pr}-\frac{1}{12} \left(35 g_1^2+81 g_2^2+96 g_s^2\right) \wc[]{uH}{pr}+g_2 \left(5 g_1^2-9 g_2^2\right) \wc[]{uW}{pr}\,,\\
\dotwc[]{eH}{pr} & = \left(9 g_1^3-3 g_1 g_2^2\right) \wc[]{eB}{pr}-\frac{3}{4} \left(7 g_1^2+9 g_2^2\right) \wc[]{eH}{pr}+9 g_2 \left(g_1^2-g_2^2\right) \wc[]{eW}{pr}.
\end{aligned}
\ee
\begin{figure}[tbp]
\centering
\begin{minipage}{1.0\linewidth}%
\hspace{-4.2cm}
\includegraphics[clip, trim=0.5cm 13.2cm 0.5cm 13.2cm, width=1.5\textwidth]{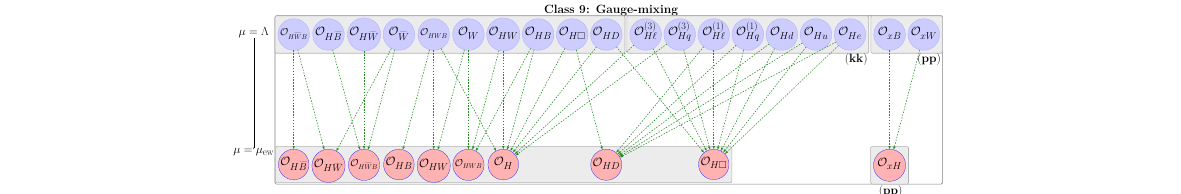}
\end{minipage}
\caption{\small Electroweak operator mixing for the Higgs observables (Class 9) in the Warsaw down-basis. Here $x=u,d,e$. The dashed green lines indicate the mixing due to electroweak interactions. The index $k$ is summed over the values 1-3. Similar SMEFT charts for $\wc[(3)]{H\ell}{}$ and $\wc[]{\ell \ell}{1221}$ are already given in Figs.~\ref{chart:ewp-gauge-1} and \ref{chart:ewp-gauge-2}, respectively. Self-mixing is not shown.} 
\label{chart:class13-gauge}
\end{figure}

For the case of Yukawa interactions the operator mixing is much more involved, which is illustrated by the following RGEs:
\\

\noindent
\underline{\bf \boldmath $f^4 \to f^2H^3$}: 
Within this category following structures are identified:
\be
\begin{aligned}
\dotwc[]{dH}{pr} & = \frac{1}{3} \left(6 y_\tau \left(\lambda-2 y_\tau^2\right) \wc[*]{ledq}{33rp}-12 y_b \left(\lambda-2 y_b^2\right) \wc[(1)]{qd}{p33r}-16 \lambda y_b \wc[(8)]{qd}{p33r} \right.\\ 
&+32 y_b^3 \wc[(8)]{qd}{p33r}+3 \lambda y_t V_{3v} \wc[(1)]{quqd}{p3vr}+18 \lambda y_t V_{3v} \wc[(1)]{quqd}{v3pr}-6 y_t^3 V_{3w} \wc[(1)]{quqd}{p3wr}-36 y_t^3 V_{3w} \wc[(1)]{quqd}{w3pr} \\ 
&\left.+4 \lambda y_t V_{3v} \wc[(8)]{quqd}{p3vr}-8 y_t^3 V_{3w} \wc[(8)]{quqd}{p3wr}\right)\,,
\end{aligned}
\ee
\be
\begin{aligned}
\dotwc[]{uH}{pr} & = -2 \lambda y_\tau \wc[(1)]{lequ}{33pr}+4 y_\tau^3 \wc[(1)]{lequ}{33pr}-4 \lambda y_t V^*_{3v} \wc[(1)]{qu}{pv3r}+8 y_t^3 V^*_{3v} \wc[(1)]{qu}{pv3r}-\frac{16}{3} \lambda y_t V^*_{3v} \wc[(8)]{qu}{pv3r} \\ 
&+\frac{32}{3} y_t^3 V^*_{3v} \wc[(8)]{qu}{pv3r}+\lambda y_b \wc[(1)]{quqd}{3rp3}+6 \lambda y_b \wc[(1)]{quqd}{pr33}-2 y_b^3 \wc[(1)]{quqd}{3rp3}-12 y_b^3 \wc[(1)]{quqd}{pr33}+\frac{4}{3} \lambda y_b \wc[(8)]{quqd}{3rp3} \\ 
&-\frac{8}{3} y_b^3 \wc[(8)]{quqd}{3rp3}\,,
\end{aligned}
\ee
\be
\begin{aligned}
\dotwc[]{eH}{pr} & = 2 \left(\left(4 y_\tau^3-2 \lambda y_\tau\right) \wc[]{le}{p33r}+3 y_b \left(\lambda-2 y_b^2\right) \wc[]{ledq}{pr33}-3 \lambda y_t V_{3v} \wc[(1)]{lequ}{prv3}+6 y_t^3 V_{3w} \wc[(1)]{lequ}{prw3}\right)\,. 
\end{aligned}
\ee
Several of the remaining RGEs are even more involved and we use the shorthand presentation like in Classes 3, 6 and 7. 
\\

{
\noindent
\underline{\bf \boldmath {non-fermion, 2f} $\to$ {\bf non-fermion}}:
}
\be
\begin{aligned}
\dotwc[]{H}{} &= y_t^3F_1(\wc[]{uH}{},\wc[*]{uH}{})+ 
y_t^2 F_2(\wc[(3)]{Hq}{},\wc[]{H}{})
+
y_t F_3(\wc[]{uH}{},\wc[*]{uH}{}) 
+
y_b F_4(\wc[]{dH}{},\wc[*]{dH}{})
+
y_b^2 F_5(\wc[(3)]{Hq}{})\\
&+
y_b^3 F_6(\wc[]{dH}{},\wc[*]{dH}{})
+
y_\tau F_7(\wc[]{eH}{},\wc[*]{eH}{})
+
y^2_\tau F_8(\wc[(3)]{Hl}{})
+
y^3_\tau F_9 (\wc[]{eH}{},\wc[*]{eH}{})
+
y_b y_t F(\wc[]{Hud}{},\wc[*]{Hud}{})\\
&
+\frac{8}{3} g_2^2 \lambda \wc[(3)]{Hl}{}+8 g_2^2 \lambda \wc[(3)]{Hq}{}
+54 \wc[]{H}{} \lambda
+6 \wc[]{HB}{} g_1^2 \lambda+\frac{20}{3} \wc[]{H\square}{} g_2^2 \lambda
-40 \wc[]{H\square}{} \lambda^2\\
&+3 \wc[]{HD}{} g_1^2 \lambda
-3 \wc[]{HD}{} g_2^2 \lambda+12 \wc[]{HD}{} \lambda^2
+18 \wc[]{HW}{} g_2^2 \lambda+6 \wc[]{HWB}{} g_1 g_2 \lambda\,,
\end{aligned}
\ee
\be
\begin{aligned}
\dotwc[]{H\Box}{} &=
 y_t^2F_1(\wc[(1)]{Hq}{},\wc[(3)]{Hq}{},\wc[]{Hu}{},\wc[]{H\square}{})+y_b y_t F_2(\wc[]{Hud}{}{,\wc[*]{Hud}{}})+y_b^2F_3(\wc[]{Hd}{},\wc[(1)]{Hq}{},\wc[(3)]{Hq}{})\\&
+ y_\tau^2 F_4(\wc[]{He}{}, \wc[(1)]{Hl}{},\wc[(3)]{Hl}{})+12 \lambda \wc[]{H\square}{} \,,
\end{aligned}
\ee

\be
\begin{aligned}
\dotwc[]{HD}{} & = \eqref{eq:CHDyuk}\,,
\end{aligned}
\ee

\be
\begin{aligned}
\dotwc[]{HB}{} & = y_t^2 F_1(\wc[]{HB}{})+g_1 y_t F_2({\wc[]{uB}{},\wc[*]{uB}{}})+g_1 y_b F_3(\wc[]{dB}{}{,\wc[*]{dB}{}})
+g_1 y_\tau F_4(\wc[]{eB}{}{,\wc[*]{eB}{}})+ 6 \lambda \wc[]{HB}{}\,,  \\
\dotwc[]{HW}{} & =y_t^2 F_1(\wc[]{HW}{})+g_2 y_\tau F_2(\wc[]{eW}{}{,\wc[*]{eW}{}})
+ g_2 y_b F_3(\wc[]{dW}{}{,\wc[*]{dW}{}}) +g_2 y_t F_4(\wc[]{uW}{}{,\wc[*]{uW}{}}) +6 \lambda\wc[]{HW}{}\,, \\
\dotwc[]{HG}{} & = y_t^2F_1(\wc[]{HG}{})+g_s y_b F_2(\wc[]{dG}{}{,\wc[*]{dG}{}})+ g_s y_t F_3(\wc[]{uG}{}{,\wc[*]{uG}{}}) +6 \lambda \wc[]{HG}{}\,, 
\end{aligned}
\ee

\be
\begin{aligned}
\dotwc[]{HWB}{} & = \eqref{eq:CHWByuk}\,, 
\end{aligned}
\ee
\be
\begin{aligned}
\dotwc[]{H\widetilde B}{} & = g_1 y_b F_1(\wc[]{dB}{}{,\wc[*]{dB}{}})+ g_1 y_\tau F_2(\wc[]{eB}{}{,\wc[*]{eB}{}})+ g_1 y_t F_3(\wc[]{uB}{}{,\wc[*]{uB}{}})
+6 \lambda \wc[]{H\widetilde B}{} +6y_t^2  \wc[]{H\widetilde B}{}\,, \\
\dotwc[]{H\widetilde W}{} & = g_2 y_b F_1(\wc[]{dW}{}{,\wc[*]{dW}{}}) + g_2 y_\tau F_2(\wc[]{eW}{}{,\wc[*]{eW}{}})
+g_2 y_t F_3(\wc[]{uW}{}{,\wc[*]{uW}{}})
+6\lambda  \wc[]{H\widetilde W}{} +6y_t^2 \wc[]{H\widetilde W}{}\,, \\
\dotwc[]{H\widetilde G}{} & = g_s y_b F_1(\wc[]{dG}{}{,\wc[*]{dG}{}})+g_s y_t F_2(\wc[]{uG}{}{,\wc[*]{uG}{}}) 
+6 \lambda \wc[]{H\widetilde G}{} +6 y_t^2 \wc[]{H\widetilde G}{} \,, \\
\dotwc[]{H\widetilde WB}{} & = g_2 y_b F_1(\wc[]{dB}{}{,\wc[*]{dB}{}})+g_1 y_b F_2(\wc[]{dW}{}{,\wc[*]{dW}{}})
+g_2 y_\tau F_3(\wc[]{eB}{}{,\wc[*]{eB}{}})+
g_1 y_\tau F_4(\wc[]{eW}{}{,\wc[*]{eW}{}})\\&
+g_2y_t F_5(\wc[]{uB}{}{,\wc[*]{uB}{}}) +g_2y_t F_6(\wc[]{uW}{}{,\wc[*]{uW}{}}) 
+2\lambda \wc[]{H\widetilde WB}{} 
+6y_t^2 \wc[]{H\widetilde WB}{}\,. 
\end{aligned}
\ee

\noindent
\underline{\bf \boldmath $f^2H^2D \to f^2H^3$}:

\be
\begin{aligned}
\dotwc[]{dH}{} & =  y_t^3 F_1(\wc[]{Hud}{})+
y_tF_2(\wc[]{Hud}{})+y_b y_t^2 F_3(\wc[(3)]{Hq}{})
+y_b y_t F_4(\wc[]{uH}{}{,\wc[*]{uH}{}})+
y_b F_5(\wc[]{Hd}{},\wc[(3)]{Hl}{},\wc[(1)]{Hq}{},\wc[(3)]{Hq}{})\\&
+ y_b^3 F_6(\wc[]{Hd}{},\wc[(1)]{Hq}{},\wc[(3)]{Hq}{})
+y_b y_\tau^2 F_7(\wc[(3)]{Hl}{})
+ y_b^2 y_t F_8(\wc[]{Hud}{}{,\wc[*]{Hud}{}})
+6 \lambda y_b \wc[(3)]{Hq}{} -2 \lambda y_t \wc[]{Hud}{}\,,
\end{aligned}
\ee

\be
\begin{aligned}
\dotwc[]{uH}{} & =y_t^3F_1(\wc[(1)]{Hq}{},\wc[(3)]{Hq}{},\wc[]{Hu}{})+
y_t^2 F_2({\wc[*]{uH}{}})+
y_t F_3(\wc[(3)]{Hl}{},\wc[(1)]{Hq}{},\wc[(3)]{Hq}{},\wc[]{Hu}{})
+
y_t y_\tau F_4({\wc[*]{eH}{}})\\&
+
 y_b y_t F_5({\wc[*]{dH}{}})
+
y_b y_t^2 F_6(\wc[]{Hud}{}{,\wc[*]{Hud}{}})
+
y_b^2 y_t F_7(\wc[(3)]{Hq}{})
+
y_t y_\tau^2F_8( \wc[(3)]{Hl}{})
+
y_b F_9(\wc[]{Hud}{})
\\&
+
y_b^3F_{10}({\wc[*]{Hud}{}})+6 \lambda y_t \wc[(3)]{Hq}{}
+2 \lambda y_t \wc[]{Hu}{} 
-2 \lambda y_b \wc[*]{Hud}{}\,,
\end{aligned}
\ee

\be
\begin{aligned}
\dotwc[]{eH}{} & = y_t^2 y_\tau F_1(\wc[(3)]{Hq}{})
+
y_b y_t y_\tau F_2(\wc[]{Hud}{}{,\wc[*]{Hud}{}})
+
 y_t y_\tau F_3({\wc[*]{uH}{}})
 +
y_\tau F_4(\wc[]{He}{},\wc[(1)]{Hl}{},\wc[(3)]{Hl}{})\\
& +
y_\tau^3 F_5(\wc[]{He}{},\wc[(1)]{Hl}{},\wc[(3)]{Hl}{})
+ y_b^2 y_\tau F_6(\wc[(3)]{Hq}{}) \,.
\end{aligned}
\ee

\noindent
\underline{\bf \boldmath $f^2H^3 \to f^2H^3$}:

\be
\begin{aligned}
\dotwc[]{dH}{} &= y_t^2 F_1(\wc[]{dH}{})
+ y_b y_t F_2(\wc[]{uH}{}{,\wc[*]{uH}{}})
+ y_b^2 F_3( \wc[]{dH}{})
+y_b y_\tau F_4(\wc[]{eH}{})
+12 \lambda \wc[]{dH}{}\,,
\\
\dotwc[]{uH}{} & =y_t^2 F_1(\wc[]{uH}{}{,\wc[*]{uH}{}})
+y_b y_t F_2(\wc[]{dH}{}{,\wc[*]{dH}{}})
+ y_b^2 F_3( \wc[]{uH}{})
+y_t y_\tau F_4({\wc[]{eH}{}},{\wc[*]{eH}{}})
+
12 \lambda \wc[]{uH}{}\,,
\\
\dotwc[]{eH}{} & =
 y_t^2 F_1(\wc[]{eH}{}) +y_t y_\tau F_2(\wc[]{uH}{}{,\wc[*]{uH}{}})
+ y_b y_\tau F_3(\wc[]{dH}{})+y_\tau^2 F_4(\wc[]{eH}{},{\wc[*]{eH}{}})
+
12 \lambda \wc[]{eH}{}\,.
\end{aligned}
\ee

\noindent
\underline{\bf \boldmath Bosonic $\to f^2H^3$}:

\be
\begin{aligned}
\dotwc[]{dH}{} & =
y_b F_1(\wc[]{HB}{},\wc[]{H\square}{},\wc[]{H\widetilde B}{}, \wc[]{HD}{},\wc[]{HG}{},\wc[]{H\widetilde G}{},\wc[]{HW}{},\wc[]{HWB}{},\wc[]{H\widetilde W}{},\wc[]{H\widetilde WB}{})\\&
+\lambda y_b F_2(\wc[]{H\square}{}, \wc[]{HD}{} )+
y_b^3 F_3(\wc[]{H\square}{}, \wc[]{HD}{})\,,
\end{aligned}
\ee
\be
\begin{aligned}
\dotwc[]{uH}{} & = y_t F_1(\wc[]{HB}{},\wc[]{H\square}{},\wc[]{H\widetilde B}{},
\wc[]{HD}{},\wc[]{HG}{},\wc[]{H\widetilde G}{}, \wc[]{HW}{},\wc[]{HWB}{},\wc[]{H\widetilde W}{},\wc[]{H\widetilde WB}{})\\&
+\lambda y_t F_2(\wc[]{H\square}{},\wc[]{HD}{})+
y_t^3 F_3(\wc[]{H\square}{},\wc[]{HD}{})\,,
\end{aligned}
\ee

\be
\begin{aligned}
\dotwc[]{eH}{} & = y_\tau F_1(\wc[]{HB}{}, \wc[]{H\square}{},\wc[]{H\widetilde B}{}
,\wc[]{HD}{},\wc[]{HW}{},\wc[]{HWB}{},\wc[]{H\widetilde W}{},\wc[]{H\widetilde WB}{})\\&
+ \lambda y_\tau F_2(\wc[]{H\square}{},\wc[]{HD}{})
+y_\tau^3 F_3(\wc[]{H\square}{},\wc[]{HD}{})\,.
\end{aligned}
\ee
The operator mixing due to gauge interactions is shown in the SMEFT chart in Fig.~\ref{chart:class13-gauge}. The explicit flavour dependent Yukawa operator mixing can be found in the supplemental material.

To quantify these effects, one can compute the $\eta$ parameters treating the Higgs couplings at $\muEW$ as IR quantities. Specifically, $\eta$ coefficients for this class can be defined by:
\be
\eta^{(9)}_{km} = c_{k}(C_m^{\rm SMEFT})  = \frac{c_{k}(C_{\rm SMEFT}(\muEW))}{C_m^{\rm SMEFT} (\Lambda)}.
\ee
Here $c_{k}(C_{\rm SMEFT}(\muEW))$ are the Higgs couplings defined in \eqref{eq:hvv}, which are functions of SMEFT WCs at $\muEW$.

{\boldmath
\subsection{$ \text{SU(2)}_L$ Correlations}
}

As seen in \eqref{eq:class9-smeft-tree}, this class predominantly involves non-fermion operators at the tree-level. The only fermionic operators that appear are of $\psi^2 H^3$ type, which induce only dim-6 corrections to the SM Yukawa couplings after EWSB, but do not match onto additional operators in WET at dim-6 level. Consequently, there are not many $\text{SU(2)}_L$ correlations in this class. 

However, double insertions of Yukawa couplings generate contributions to flavour changing mesonic processes and electric dipole moments, thereby establishing correlation with Class 9 \cite{Kumar:2024yuu}.

{\boldmath
\section{\boldmath High-$p_T$ Scattering Processes (Class 10)}
\label{class10}
}
With an increased center of mass energy $(\sqrt{\hat s})$ and {significantly} higher 
luminosities achievable at the LHC and future colliders, direct searches will be {capable of probing} 
heavy new resonances in the multi-TeV regime. Furthermore, processes {involving} 
different type of initial states such as $ep,pp, ee$ and final states containing 
leptons, jets and/or missing transverse energy can be studied. {Although no} BSM particles 
have been observed at the LHC {so far, such discoveries could still occur in the future}.

The tails of invariant mass distributions in high-$p_T$ scattering processes can be {modified} 
by the presence of effective operators \cite{Greljo:2017vvb}. Therefore, the LHC data from direct 
{searches can be used} to constrain WCs. These constraints are complementary to 
{those obtained from the first nine classes, which primarily involved} low energy processes. 
In this respect, this class is fundamentally different from the first nine classes 
{as} processes in this class occur at {energy} scale well above 
the weak scale.  

Note that the choice of $\mu$ is not unique and depends on the specific scattering processes. 
{It is often set equal to the center of mass energy or to characteristic momentum transfer.} 
Typically, one has
\begin{equation}
\mu \approx \sqrt{\hat s} \ll \Lambda \,. 
\end{equation}
Here the second inequality needs to be ensured for the EFT expansion to remain valid. 
In this class, we {will not discuss the RG running, as the number of operators is  too large 
and most of the RGEs are already contained in previous classes}. 
It would be easy include it following same strategy used in other classes. 

{The RGEs, however, would be} necessary, when studying correlations with 
low-energy observables, discussed in the previous classes. {To find these correlations, } 
 one has to compare the list of operators at $\mu \approx \sqrt{\hat s}$ for class 10 with 
the corresponding operators in other classes at $\Lambda$.

We will mainly focus on the scattering processes of four types {based on their} 
initial and final states. The first category is the purely leptonic scattering \cite{Falkowski:2015krw}:
\\

\begin{center}
\textrm{\bf \boldmath Class 10A:  $\ell^+ \ell^- \to \ell^{'+} \ell^{'-}\,.$}
\end{center}
Here $\ell^{'}$ and $\ell$ can be equal or different. Examples include processes 
with leptonic final states at $e^+ e^-$ colliders, such as LEP. 
Then second class entails hadronic final states at lepton colliders \cite{Falkowski:2017pss},
\\

\begin{center}
\textrm{\bf \boldmath Class 10B:  $\ell^+ \ell^- \to  q \bar q'$}. 
\end{center}
The LHC processes involve protons in the initial states. In case of leptonic final states one has Drell-Yan  processes \cite{Allwicher:2022gkm}
\begin{center}
\textrm{\bf  \boldmath Class 10C:  $pp \to \ell^{+} \ell^-, \,\ p p \to \ell^- \bar \nu$}.
\end{center}
Finally, hadronic final states at proton colliders are characterized by the following class
\\

\begin{center}
\textrm{\bf  \boldmath Class 10D:  $pp \to  q \bar q'$}.
\end{center}
Clearly, the processes having gauge bosons and Higgs particles in the final states are beyond class 10, 
 {but they fall under the same category}. There are numerous SMEFT studies of high $p_T$-processes, which are summarized in Tabs.~\ref{tab:LOSMEFTobsHiggs}-\ref{tab:LOSMEFTobsqqll} and \ref{tab:NLOSMEFTobsHiggs}-\ref{tab:NLOSMEFTobsneut}.

Ignoring the running from $\Lambda$ to $\mu$, for this class one can directly put constraints on the SMEFT WCs at {the NP scale} $\Lambda$. Consequently, we do not discuss the corresponding operators at $\muEW$.

\subsection{\boldmath SMEFT Operators for Class 10 at $\boldmath \Lambda$ }
Suppressing flavour indices, at the scale $\Lambda$ the relevant tree-level operators at mass dimension six involve two or more fermion fields:

\begin{center}
\textrm{\bf SMEFT-Tree 10A}
\end{center}
\be
\label{eq:smeft-10A}
\begin{aligned}
\ops[]{ll}{}\,, \quad
\ops[]{le}{}\,, \quad
\ops[]{ee}{}\,, \quad
\ops[(1)]{Hl}{}\,, \quad
\ops[(3)]{Hl}{}\,, \quad
\ops[]{He}{}\,, \quad
\ops[]{eW}{}\,, \quad
\ops[]{eB}{}\,.
\end{aligned}
\ee

\begin{center}
\textrm{\bf SMEFT-Tree 10B, 10C}
\end{center}
\be
\label{eq:smeft-10BC}
\begin{aligned}
&\ops[(1)]{lq}{}\,, \quad
\ops[(3)]{lq}{}\,, \quad
\ops[]{lu}{}\,, \quad
\ops[]{ld}{}\,, \quad
\ops[]{qe}{}\,, \quad
\ops[]{eu}{}\,, \quad
\ops[]{ed}{}\,, \quad
\ops[(1)]{lequ}{}\,, \quad
\ops[(3)]{lequ}{}\,, \quad
\\
&
\ops[]{ledq}{}\,, \quad
\ops[(1)]{Hl}{}\,, \quad
\ops[(3)]{Hl}{}\,, \quad
\ops[]{He}{}\,, \quad
{\ops[]{eH}{}}\,, \quad
\ops[(1)]{Hq}{}\,, \quad
\ops[(3)]{Hq}{}\,, \quad
\ops[]{Hu}{}\,, \quad
\ops[]{Hd}{}\,, \\\
&
\ops[]{Hud}{}\,, \quad
\ops[]{eW}{}\,, \quad
\ops[]{eB}{}\,, \quad
\ops[]{uB}{}\,, \quad
\ops[]{uW}{}\,, \quad
\ops[]{dB}{}\,, \quad
\ops[]{dW}{}\,.
\end{aligned}
\ee

\begin{center}
\textrm{\bf SMEFT-Tree 10D}
\end{center}
\be 
\label{eq:smeft-10D}
\begin{aligned}
&\ops[(1)]{qq}{}\,, \quad
\ops[(3)]{qq}{}\,, \quad
\ops[(1)]{qu}{}\,, \quad
\ops[(8)]{qu}{}\,, \quad
\ops[(1)]{qd}{}\,, \quad
\ops[(8)]{qd}{}\,, \quad
\ops[]{uu}{}\,, \quad
\ops[]{dd}{}\,, \\
& \ops[(1)]{ud}{}\,, \quad
\ops[(8)]{ud}{}\,, \quad
\ops[(1)]{quqd}{}\,, \quad
\ops[(8)]{quqd}{}\,, \quad
\ops[(1)]{Hq}{}\,, \quad
\ops[(3)]{Hq}{}\,, \quad
\ops[]{Hu}{}\,, \quad
 \ops[]{Hd}{}\,, \\
&\ops[]{uB}{}\,, \quad
\ops[]{uW}{}\,, \quad
\ops[]{uG}{}\,, \quad
\ops[]{dB}{}\,, \quad
\ops[]{dW}{}\,, \quad
\ops[]{dG}{}\,.
\end{aligned}
\ee

We stress that, since the scattering cross-section is $\propto |\mathcal{A}|^2$, it consists of terms which scale as $\mathcal{O}(1/\Lambda^2)$ as well as $\mathcal{O}(1/\Lambda^4)$. The latter terms can originate from squared contributions of dim-6 operators or from genuine dim-8 operators interfering with the SM. Therefore, to be consistent with the EFT power counting, the latter must be included as well. In \cite{Murphy:2020rsh} the complete list of dim-8 SMEFT operators is provided. Further, see \cite{Allwicher:2022gkm} for a discussion on dim-8 effects in semileptonic high-energy processes.

Depending upon the specific high-energy process of interest, at the parton level the flavour indices have to be properly chosen. As evident from the above lists of operators, in general plenty of SMEFT operators can affect class 10A-10D processes, correlating class 10 to many other classes. This is evident from Tabs.~\ref{tab:cor-bosonic-fermionic} and \ref{tab:cor-fermionic}.

The practical challenge that one faces in class 10 like in other classes is the choice of weak basis. At the high scale the SMEFT is defined only in the weak basis, which could be chosen to be either the down or up-basis. However, scattering processes involve particles in the mass basis for both up- and down-type fermions. As a result a given SMEFT operator in the up- or down basis can yield several different {type} of parton level transitions due to CKM rotations. This leads to quantitatively different constraints from high-$p_T$ distributions in the two bases. See for example \cite{Allwicher:2022gkm, Allwicher:2022mcg} for a discussion on the impact of basis choice on the resulting constraints from Drell-Yan processes.

As already mentioned, we have ignored RG running effects in this class, so we do not need to look at the RGEs, SMEFT charts etc. But we stress that if the NP scale $\Lambda$ is much larger than $\sqrt{\hat s}$, such effects might be important numerically.

\subsection{\boldmath $\textrm{SU(2)}_L$ Correlations}
As evident from \eqref{eq:smeft-10A}-\eqref{eq:smeft-10D}, classes 10A-10D are correlated at the tree-level due to common WCs. Further, correlations within a class can also arise due to $\textrm{SU(2)}_L$ symmetry. For example, $u \bar u \to \mu^+ \mu^- $ and $d \bar d \to \mu^+ \mu^-$ processes within class 10C become correlated in the presence of $\wc[(1)]{lq}{2211}$. Moreover, correlations to other classes can arise. For instance, the WC $\wc[(1)]{lq}{2222}$ contributes to $c\bar c (s\bar s) \to \mu^+ \mu^-$ as well as $b\to s \mu^+ \mu^-$ (due to back rotation and/or SMEFT RG running effects). Another example is the correlation between nuclear beta decay ($d \to u e \bar \nu$) and $u \bar u \to e^+ e^-$ (or $e^+ e^- \to u \bar u$) and $d \bar d \to e^+ e^-$ (or $e^+ e^- \to d \bar d$) in the presence of $\wc[(3)]{lq}{1111}$. 

It is worth to keep in mind that the phenomenological importance of these correlations will depend on the specific values of the NP scales to which different processes involved in the correlations are sensitive to. The presence of large hierarchies could result in severe constraints from one of the process precluding any sizeable deviations in the other observable sensitive to the same WC. A good example in this respect are flavour violating processes that usually lead to stronger constraints on SMEFT WCs as compared to direct searches. It is however possible that these two types of observables can give comparable constraints \cite{Faroughy:2016osc,Aebischer:2022oqe}.

The characteristic scale for class 10 is much larger than the typical scale for other classes because most of them deal with low-energy observables. Therefore, the tree-level operators in \eqref{eq:smeft-10A}-\eqref{eq:smeft-10D}, when evolved to lower scales can affect other classes at loop-level. This will introduce {additional} RGE correlations. To deduce such correlations, one simply has to {identify} the overlapping sets of WCs from other classes and the ones given in \eqref{eq:smeft-10A}-\eqref{eq:smeft-10D}. A more focused study of such effects will be presented elsewhere.

\newpage
\part{SMEFT to Particles}

\section{Classification of BSM Particles}\label{sec:5}

\subsection{Preliminaries}\label{PRE27}
Until now our presentation of the SMEFT was dominated by the operators and their WCs. This is clearly the proper language for any effective field theory. Yet, eventually our main goal is to find out which UV completion is responsible for possible anomalies observed in low-energy experiments. With the technology developed until now at hand we can begin to consider specific NP scenarios with the goal to identify their specific signatures that could be observed in low-energy experiments. Quite generally, having a specific NP model or some simplified NP scenario with new heavy scalars, fermions and vectors, the implications of this extension of the SM on low-energy observables can be found systematically by performing six steps which we summarized in Sec.~\ref{sec:GV}.\footnote{A detailed discussion on light new particles called Loryons, which are non-decoupling can be found in \cite{Banta:2021dek}, fractionally charged particles are discussed in \cite{Koren:2025utp} and an analysis of non-decoupling scalars at future colliders is given in \cite{Crawford:2024nun}.}

We would like to emphasize that our goal in this section is not the presentation of numerical analyses of these NP scenarios which can be found in the literature to which we will provide extensive lists of references. We will rather consider a number of simplified NP scenarios with the goal to find out which operators are generated at the NP scale $\Lambda$ when a given new heavy particle is integrated out. Furthermore, we will determine the resulting WCs of these operators in terms of the NP parameters in the tree-level approximation. In this context the complete tree-level dictionary of this kind presented in \cite{deBlas:2017xtg} turned out to be very useful. Indeed, the complete matching of any NP model onto the SMEFT at tree-level has been calculated in \cite{deBlas:2017xtg}. Moreover, the classification of the new fields relevant for this dictionary, presented before in a series of papers for new quarks \cite{delAguila:2000rc}, new leptons \cite{delAguila:2008pw}, new vectors \cite{delAguila:2010mx} and new scalars \cite{deBlas:2014mba}, has been summarized and generalized to non-renormalizable UV-completions. Starting with a general Lagrangian involving scalars, fermions and vector bosons the authors have computed WCs of the SMEFT at the NP scale in terms of couplings and masses of {the} new particles. In Tables~1-3 of \cite{deBlas:2017xtg} one can find the list of new scalar bosons, new vector-like fermions and new vector bosons contributing to the SMEFT at tree-level. In the corresponding Tables~7-9 of \cite{deBlas:2017xtg} a list of SMEFT operators generated by each of these fields is given. Finally, in Appendix D of \cite{deBlas:2017xtg} the tree-level SMEFT WCs as functions of masses and couplings of all new fields are collected. Even if the formulae in this paper are very complicated, this paper is very useful both for the top-down approach and also for the bottom-up one. We will see it in the rest of our review.

However, given the increasing precision arising in experimental physics and the fact that several observables receive their leading contributions at loop-level (for instance the anomalous magnetic moment of the muon) the one-loop matching of a given UV completion onto the SMEFT should be included. This is also required for a consistent NLO analysis as already mentioned previously. An important step toward a complete one-loop dictionary was made recently in \cite{Guedes:2023azv,Guedes:2024vuf}. It includes an arbitrary number of heavy scalars and fermions. The code \texttt{SOLD}, listed in Sec.~\ref{SMEFTtools}, is very helpful in this context.

In Tab.~\ref{TAB3} we collect references to papers in which tree-level and one-loop matching results for various NP models can be found. We have separated Leptoquarks (LQs) which are either scalar or vector particles. Some of these results will be used below.

\begin{table}[tb]
\begin{center}
\begin{tabular}{|l|l|l|}
\hline
\bf \phantom{XXXXX} NP Scenario &  {\bf Tree} & {\bf One-Loop}  \\
\hline
\hline
New Quarks & \cite{delAguila:2000rc,deBlas:2017xtg}    & \cite{Bobeth:2016llm,Bobeth:2017xry,Endo:2016tnu,Endo:2018gdn}   \\
New Leptons & \cite{delAguila:2008pw,deBlas:2017xtg,Cepedello:2024ogz}    &  \\
New Vectors & \cite{delAguila:2010mx,deBlas:2017xtg,Dawson:2024ozw}    & \cite{Buras:2012fs,Buras:2012gm,Brivio:2021alv}\\
New Scalars &  \cite{deBlas:2014mba,Henning:2014gca,deBlas:2017xtg,Ellis:2023zim,Dawson:2022cmu,Dawson:2023ebe}   & \cite{Ellis:2017jns,Buras:2012fs,Buras:2012gm,Haisch:2020ahr,Jiang:2018pbd,DasBakshi:2024krs} \\
Leptoquarks  &\cite{Buras:2020xsm}   &  \cite{Bobeth:2017ecx,Dekens:2018bci,Gherardi:2020det}\\
\hline
\end{tabular}
\end{center}
\vspace{-0.5cm}
\caption{{SMEFT matching in} various New Physics scenarios. \label{TAB3}}
\end{table}

In this context several useful papers should be mentioned:
\begin{itemize}
\item
The covariant derivative expansion for the SMEFT is explained in \cite{Henning:2014wua,deBlas:2017xtg}. 
\item General aspects of tree-level matching onto the SMEFT were studied in \cite{Jiang:2016czg}, stressing that always more than one operator is generated when considering complete UV theories.
\item
The one-loop matching of all scalar LQ representations onto dipole operators can be found in~\cite{Dekens:2018bci}.
\item
The complete one-loop SMEFT matching for $S_1$, $S_3$ LQs is given in~\cite{Gherardi:2020det}.
\item
The one-loop matching of scalars onto the SMEFT, using the UOLEA approach can be found in \cite{Ellis:2017jns}. The first one-loop matching computation involving a scalar onto SMEFT is given in \cite{Jiang:2018pbd} and the full matching including all contributions in \cite{Haisch:2020ahr}.
\item 
A matching of the Two-Higgs-doublet model (2HDM) onto the SMEFT can be found in \cite{Banta:2023prj}, {and an analysis of EDMs in the decoupling limit of the 2HDM was performed in \cite{Davila:2025goc}}.
\item
Recently, the complete one-loop matching of the full $R$-parity conserving MSSM onto the SMEFT was worked out in \cite{Kraml:2025fpv}. The matching was performed with the \texttt{Matchete} package and includes all correlations among the different SMEFT Wilson coefficients that are governed by supersymmetry.

\item
One-loop QCD corrections to tree-level $Z^\prime$ and neutral scalar exchanges for $\Delta F=1$ and $\Delta F=2$ processes have been calculated in \cite{Buras:2012fs,Buras:2012gm}. This allows to perform the NLO QCD analysis in models with these tree-level exchanges.
\item 
A complete UV/IR dictionary for BSM models matched onto the SMEFT was obtained in \cite{Li:2023cwy} for dim-5, dim-6 and dim-7 operators, using on-shell methods.
\item 
A complete UV dictionary for dim-5, dim-6, dim-7 and dim-8 operators was obtained in \cite{Li:2022abx,Li:2023pfw}, using the $J$-basis approach, mentioned in Sec.~\ref{sec:obs}.
\item 
The matching of type-I, II and III see-saw models onto the SMEFT were performed in \cite{Zhang:2021jdf,Li:2022ipc,Li:2023ohq}.
\item 
Some progress has been made lately on the functional method side, where the one-loop effective action for non-degenerate scalars and fermions was derived in \cite{Banerjee:2023xak}.
\item 
The matching of Froggatt-Nielsen theories onto the SMEFT was discussed in \cite{Loisa:2024xuk}.
\item 
Diagrammatic approaches to find UV completions of the SMEFT were studied in \cite{DasBakshi:2021xbl,Naskar:2022rpg,Cepedello:2022pyx}.
\item 
Possible UV completions for the SMEFT were studied from the WET perspective in \cite{Chakrabortty:2020mbc}.
\item 
The ability of heavy mediators, transforming under $\mathbb{Z}_6$ to test the faithful SM gauge group is investigated in \cite{Li:2024nuo}.
\item 
Linear extensions of the SMEFT including fermion and scalar fields were studied at the one-loop level in the context of the Tera-$Z$ run of the FCC-ee in \cite{Gargalionis:2024jaw}.
\item 
The two-loop effective action up to dimension six resulting from integrating out heavy scalars in the Heat-Kernel method was presented in \cite{Adhikary:2025pbb}.
\item 
A study of various discrete lepton flavour symmetries and their connections to tree-level mediators is discussed in \cite{Palavric:2024gvu}.
\item 
Tree-level UV-completions for $\Delta L =2$ dim-7 operators were identified in \cite{Fridell:2024pmw}.
\end{itemize}

Our strategy will be to extract from \cite{deBlas:2017xtg} information relevant for simplified scenarios which in our view are expected to play an important role in phenomenology. Our expectations are based on various model constructions presented in the literature in the context of attempts to explain several anomalies. While our presentation is by definition not as exhaustive as the listings in~\cite{deBlas:2017xtg}, we hope that it could provide additional insight into the very lengthy formulae presented in \cite{deBlas:2017xtg}. While these authors state that their results are insufficient for the study of NP which enters first at the one-loop level, this statement applies really only to the matching. As we have seen in numerous SMEFT charts in PART II of our review, NP operators can be generated through RG evolution at the one-loop level. Moreover, already with the tree-level matching one can get a good impression of what a given NP scenario implies, in particular in combination with the SMEFT charts presented in Part II of our review.

\subsection{$Z^\prime$, $W^\prime$, $G^\prime$ and $Z_1^\prime$, $W_1^\prime$, $G_1^\prime$}
We consider first popular new vector bosons that are listed in Tab.~\ref{t:vectors}. In addition to $Z^\prime$, $W^\prime$ and $G^\prime$ that have zero hypercharge, we consider the corresponding $Z_1^\prime$, $W_1^\prime$ and $G_1^\prime$ with $Y=1$ in order to illustrate the change of matching conditions, which depend on the hypercharge. 

In Tab.~\ref{tab:topdown_vectors} we list the four-fermion operators generated by these gauge bosons at tree-level. We note that ${W_1^\prime}$ cannot generate four-fermion operators at tree-level. In Tab.~\ref{tab:topdown_phi} we list the dim-6 operators involving a Higgs field $H$, together with the dim-4 Higgs self-interaction operator $\mathcal{O}_{H^4}$, generated by these gauge bosons. For this case we note that ${G^\prime}$ and ${G_1^\prime}$ cannot generate such operators at tree-level, whereas ${W_1^\prime}$ can.

\begin{table}[tb]
\begin{center}
{\small
\begin{tabular}{lcccccc} 
\hline
      Name &
${ Z^\prime}$ &
${ Z_1^\prime}$ &
${ W^\prime}$ &
${ W_1^\prime}$ &
${ G^\prime}$ &
${ G_1^\prime}$ 
\\
Irrep &
$\left(1,1\right)_0$ &
$\left(1,1\right)_1$ &
$\left(1,3\right)_0$ &
$\left(1,3\right)_1$ &
$\left(8,1\right)_0$ &
$\left(8,1\right)_1$ 
\\[1.3mm]
\hline
    \end{tabular}
}
\caption{Selected new vector bosons contributing to the dimension-six SMEFT at tree level.}
\label{t:vectors}
\end{center}
\end{table}

\begin{table}[tbp]
\begin{center}
\begin{tabular}{cl}
\hline
Fields & Operators 
\\
\hline
$Z^\prime$ &
$\mathcal{O}_{ll}$, $\mathcal{O}^{(1)}_{qq}$,
$\mathcal{O}^{(1)}_{lq}$, $\mathcal{O}_{ee}$,
$\mathcal{O}_{dd}$, $\mathcal{O}_{uu}$,
$\mathcal{O}_{ed}$, $\mathcal{O}_{eu}$,
$\mathcal{O}^{(1)}_{ud}$, $\mathcal{O}_{le}$,
$\mathcal{O}_{ld}$, $\mathcal{O}_{lu}$,
$\mathcal{O}_{qe}$, $\mathcal{O}^{(1)}_{qu}$,
$\mathcal{O}^{(1)}_{qd}$ \\
$Z^\prime_1$ &
$\mathcal{O}^{(1)}_{ud}$,
$\mathcal{O}^{(8)}_{ud}$\\
$W^\prime$ &
$\mathcal{O}_{ll}$,
$\mathcal{O}^{(3)}_{qq}$, $\mathcal{O}^{(3)}_{lq}$
\\
$G^\prime$ & 
$\mathcal{O}^{(1)}_{qq}$, $\mathcal{O}^{(3)}_{qq}$,
$\mathcal{O}_{dd}$, $\mathcal{O}_{uu}$,
$\mathcal{O}^{(8)}_{ud}$, $\mathcal{O}^{(8)}_{qu}$,
$\mathcal{O}^{(8)}_{qd}$ \\
$G^\prime_1$ & 
$\mathcal{O}^{(1)}_{ud}$, $\mathcal{O}^{(8)}_{ud}$ \\
&\\[-0.5cm]
\hline
\end{tabular}
\caption{Four-fermion operators generated by the heavy vector bosons listed in Tab.~\ref{t:vectors}. Based on \cite{deBlas:2017xtg}.}
\label{tab:topdown_vectors}
\end{center}
\end{table}

\begin{table}[tbp]
\begin{center}
\begin{tabular}{cl}
\hline
Fields & Operators 
\\
\hline
${Z^\prime}$ &
$\mathcal{O}_{H D}$, $\mathcal{O}_{H \square}$,
$\mathcal{O}_{eH}$, $\mathcal{O}_{dH}$,
$\mathcal{O}_{uH}$, $\mathcal{O}^{(1)}_{H l}$,
$\mathcal{O}^{(1)}_{H q}$, $\mathcal{O}_{H e}$,
$\mathcal{O}_{H d}$, $\mathcal{O}_{H u}$ \\
${Z_1^\prime}$ &
$\mathcal{O}_{H^4}$, $\mathcal{O}_{H}$,
$\mathcal{O}_{H D}$, $\mathcal{O}_{H \square}$,
$\mathcal{O}_{eH}$, $\mathcal{O}_{dH}$,
$\mathcal{O}_{uH}$, $\mathcal{O}_{H ud}$ \\
${W^\prime}$ &
$\mathcal{O}_{H^4}$, 
$\mathcal{O}_{H}$, $\mathcal{O}_{H D}$,
$\mathcal{O}_{H \square}$, $\mathcal{O}_{eH}$,
$\mathcal{O}_{dH}$, $\mathcal{O}_{uH}$,
$\mathcal{O}^{(3)}_{H l}$, $\mathcal{O}^{(3)}_{H q}$ \\
${W_1^\prime}$ & 
$\mathcal{O}_{H^4}$, $\mathcal{O}_{H}$,
$\mathcal{O}_{H D}$, $\mathcal{O}_{H \square}$,
$\mathcal{O}_{eH}$, $\mathcal{O}_{dH}$,
$\mathcal{O}_{uH}$ \\
& \\[-0.5cm]
\hline
\end{tabular}
\caption{Operators involving Higgs field $H$ generated by the heavy vector bosons presented in Tab.~\ref{t:vectors}. Based on \cite{deBlas:2017xtg}.}
\label{tab:topdown_phi}
\end{center}
\end{table}

The non-vanishing tree-level contributions of these gauge bosons to WCs of four-fermion operators are given as follows \cite{deBlas:2017xtg}
\begin{align}
\wc[]{ll}{ijkl}= &
- \frac{(g^l_{{Z^\prime}})_{rkl}
(g^l_{{Z^\prime}})_{rij}}{
2  M_{{Z^\prime_r}}^{2}}- \frac{
(g^l_{{W^\prime}})_{rkj}
(g^l_{{W^\prime}})_{ril}}{
4  M_{{W^\prime_r}}^{2}}
+ \frac{(g^l_{{W^\prime}})_{rkl}
(g^l_{{W^\prime}})_{rij}}{
8  M_{{W^\prime_r}}^{2}}\,,
\label{eq:Cll}
\\[5mm]
\wc[(1)]{qq}{ijkl}= &
- \frac{
(g^q_{{Z^\prime}})_{rkl}
(g^q_{{Z^\prime}})_{rij}}{
2  M_{{Z^\prime_r}}^{2}}
- \frac{
(g^q_{{G^\prime}})_{rkj}
(g^q_{{G^\prime}})_{ril}}{
8  M_{{G^\prime_r}}^{2}}
+ \frac{
(g^q_{{G^\prime}})_{rkl}
(g^q_{{G^\prime}})_{rij}}{
12  M_{{G^\prime_r}}^{2}}\,,
\label{eq:Cqq1}
\\[5mm]
\wc[(3)]{qq}{ijkl}= &  - \frac{
(g^q_{{W^\prime}})_{rkl}
(g^q_{{W^\prime}})_{rij}}{
8  M_{{W^\prime_r}}^{2}}- \frac{(g^q_{{G^\prime}})_{rkj}
(g^q_{{G^\prime}})_{ril}}{
8  M_{{G^\prime_r}}^{2}}\,,
  \label{eq:Cqq3}
\\[5mm]
\wc[(1)]{lq}{ijkl}= &
- \frac{
(g^q_{{Z^\prime}})_{rkl}
(g^l_{{Z^\prime}})_{rij}}{
M_{{Z^\prime_r}}^{2}}, \quad   \wc[(3)]{lq}{ijkl} = - \frac{(g^q_{{W^\prime}})_{rkl}
(g^l_{{W^\prime}})_{rij}}{4  M_{{W^\prime_r}}^{2}}\,,
\label{eq:Clq1}
\quad
\wc[]{ee}{ijkl} = 
- \frac{
(g^e_{{Z^\prime}})_{rkl}
(g^e_{{Z^\prime}})_{rij}}{
2  M_{{Z^\prime_r}}^{2}}\,,
\\[5mm]
\wc[]{dd}{ijkl}= &
- \frac{
(g^d_{{Z^\prime}})_{rkl}
(g^d_{{Z^\prime}})_{rij}}{
2  M_{{Z^\prime_r}}^{2}}- \frac{
(g^d_{{G^\prime}})_{rkj}
(g^d_{{G^\prime}})_{ril}}{
4  M_{{G^\prime_r}}^{2}}
+ \frac{
(g^d_{{G^\prime}})_{rkl}
(g^d_{{G^\prime}})_{rij}}{
12  M_{{G^\prime_r}}^{2}}\,,
\\[5mm]
\wc[]{uu}{ijkl}= &
- \frac{
(g^u_{{Z^\prime}})_{rkl}
(g^u_{{Z^\prime}})_{rij}}{
2  M_{{Z^\prime_r}}^{2}}- \frac{
(g^u_{{G^\prime}})_{rkj}
(g^u_{{G^\prime}})_{ril}}{
4  M_{{G^\prime_r}}^{2}}
+ \frac{
(g^u_{{G^\prime}})_{rkl}
(g^u_{{G^\prime}})_{rij}}{
12  M_{{G^\prime_r}}^{2}}\,,
\\[5mm]
\wc[]{ed}{ijkl}= &
- \frac{
(g^d_{{Z^\prime}})_{rkl}
(g^e_{{Z^\prime}})_{rij}}{
M_{{Z^\prime_r}}^{2}}\,,\quad
\wc[]{eu}{ijkl}= 
- \frac{
(g^u_{{Z^\prime}})_{rkl}
(g^e_{{Z^\prime}})_{rij}}{
M_{{Z^\prime_r}}^{2}}\,,
\\[5mm]
\wc[(1)]{ud}{ijkl}= &
- \frac{
(g^u_{{Z^\prime}})_{rij}
(g^d_{{Z^\prime}})_{rkl}}{
M_{{Z^\prime_r}}^{2}}
- \frac{
(g^{du}_{{Z^\prime_1}})^*_{rli}
(g^{du}_{{Z^\prime_1}})_{rkj}}{
3  M_{{Z^\prime_{1r}}}^{2}}
- \frac{
4   (g_{{G^\prime_1}})^*_{rli}
(g_{{G^\prime_1}})_{rkj}}{
9  M_{{G^\prime_{1r}}}^{2}}\,,
\\[5mm]
\wc[(8)]{ud}{ijkl}= &
- \frac{
2   (g^{du}_{{Z^\prime_1}})^*_{rli}
(g^{du}_{{Z^\prime_1}})_{rkj}}{
M_{{Z^\prime_{1r}}}^{2}} - \frac{
(g^d_{{G^\prime}})_{rkl}
(g^u_{{G^\prime}})_{rij}}{
M_{{G^\prime_r}}^{2}}+ \frac{
(g_{{G^\prime_1}})^*_{rli}
(g_{{G^\prime_1}})_{rkj}}{
3  M_{{G^\prime_{1r}}}^{2}}\,,
\\[5mm]
\wc[]{le}{ijkl}= &
- \frac{
(g^e_{{Z^\prime}})_{rkl}
(g^l_{{Z^\prime}})_{rij}}{
M_{{Z^\prime_r}}^{2}}\,,\quad 
\wc[]{ld}{ijkl}  = 
- \frac{
(g^d_{{Z^\prime}})_{rkl}
(g^l_{{Z^\prime}})_{rij}}{
M_{{Z^\prime_r}}^{2}}\,,\quad \wc[]{lu}{ijkl}= 
- \frac{
(g^u_{{Z^\prime}})_{rkl}
(g^l_{{Z^\prime}})_{rij}}{
M_{{Z^\prime_r}}^{2}}\,,
\\[7.5mm]
\wc[]{qe}{ijkl}=& 
- \frac{
(g^e_{{Z^\prime}})_{rkl}
(g^q_{{Z^\prime}})_{rij}}{
M_{{Z^\prime_r}}^{2}}\,,
\quad
\wc[(1)]{qu}{ijkl}= 
- \frac{
(g^u_{{Z^\prime}})_{rkl}
(g^q_{{Z^\prime}})_{rij}}{
M_{{Z^\prime_r}}^{2}}\,,\quad \wc[(8)]{qu}{ijkl}=   - \frac{
(g^u_{{G^\prime}})_{rkl}
(g^q_{{G^\prime}})_{rij}}{
M_{{G^\prime_r}}^{2}}\,,
\\[5mm]
\wc[(1)]{qd}{ijkl}= &
- \frac{
(g^d_{{Z^\prime}})_{rkl}
(g^q_{{Z^\prime}})_{rij}}{
M_{{Z^\prime_r}}^{2}}\,,\quad \wc[(8)]{qd}{ijkl}=   - \frac{
(g^d_{{G^\prime}})_{rkl}
(g^q_{{G^\prime}})_{rij}}{
M_{{G^\prime_r}}^{2}}\,.
\end{align}
The index ``$r$'' allows to include several gauge bosons of the same kind that could have different masses and couplings. Therefore, summation over this index is understood.

These formulae give an impression of how the tree-matching looks like and how it depends on the particle involved, in particular which operator is generated. The corresponding expressions for operators containing Higgs fields listed in Tab.~\ref{tab:topdown_phi} have a similar structure and can be found in Appendix D in \cite{deBlas:2017xtg}.

Among these heavy particles the most prominent role in phenomenology until now played $Z^\prime$ models, not only in the context of the so-called $B$ physics anomalies but also generally in models with an extended gauge group. In particular extensive analyses of flavour physics in 331 models are given in \cite{Buras:2012dp,Buras:2013dea,Buras:2014yna,Buras:2015kwd,Buras:2016dxz,Buras:2021rdg,Buras:2023ldz,Escalona:2025rxu} and for more general $Z^\prime$ models in \cite{Buras:2012jb,Buras:2013qja,Buras:2014zga,Aebischer:2019blw,Buras:2014fpa}. The most recent phenomenological SMEFT analyses in $Z^\prime$ models can be found in \cite{Aebischer:2020mkv,Aebischer:2022vky,Aebischer:2020dsw,Aebischer:2023mbz,Buras:2023ldz,Colangelo:2024sbf,Buras:2024mnq}.

\subsection{Scalars}
Next we consider the most popular new heavy scalars, listed in Tab.~\ref{t:scalars}. The dim-6 operators, together with the dim-4 Higgs self-interaction operator $\mathcal{O}_{H^4}$ generated by them are listed in Tab.~\ref{tab:topdown_scalars}. The non-vanishing tree-level contributions of these scalars to four-fermion WCs are given as follows \cite{deBlas:2017xtg} 

\begin{table}[tbp]
\begin{center}
{\small
\begin{tabular}{lccccc}
\hline
Name &
${\cal \phi}_0$ &
${\cal  \phi}_1$ &
${\cal \phi}_2$ &
$S$ & $\Phi$ \\
Irrep &
$\left(1,1\right)_0$ &
$\left(1,1\right)_1$ &
$\left(1,1\right)_2$ &
$\left(1,2\right)_{\frac 12}$ &    $\left(8,2\right)_{\frac 12}$ \\[1.3mm]
\hline
\end{tabular}
}
\caption{Selected new scalar bosons contributing to the dimension-six SMEFT at tree-level.}
\label{t:scalars}
\end{center}
\end{table}

\begin{table}[tbp]
\centering
\begin{tabular}{cl}
\hline
Fields & Operators \\
\hline
$\mathcal{ \phi}_0$ &
$\mathcal{O}_{H^4}$, $\mathcal{O}_{H}$,
$\mathcal{O}_{H \square}$, $\mathcal{O}_{H B}$,
$\mathcal{O}_{H \tilde{B}}$, $\mathcal{O}_{H W}$,
$\mathcal{O}_{H \tilde{W}}$, $\mathcal{O}_{H G}$,
$\mathcal{O}_{H \tilde{G}}$, $\mathcal{O}_{eH}$,
$\mathcal{O}_{dH}$, $\mathcal{O}_{uH}$ \\
$\mathcal{\phi}_1$ &
$\mathcal{O}_{ll}$ \\
$\mathcal{ \phi}_2$ &
$\mathcal{O}_{ee}$ \\
$S$ &
$\mathcal{O}_{le}$, $\mathcal{O}^{(1)}_{qu}$,
$\mathcal{O}^{(8)}_{qu}$, $\mathcal{O}^{(1)}_{qd}$,
$\mathcal{O}^{(8)}_{qd}$, $\mathcal{O}_{ledq}$,
$\mathcal{O}^{(1)}_{quqd}$, $\mathcal{O}^{(1)}_{lequ}$,
$\mathcal{O}_{H}$, $\mathcal{O}_{eH}$,
$\mathcal{O}_{dH}$, $\mathcal{O}_{uH}$ \\
$\Phi$ &  $\mathcal{O}^{(1)}_{qu}$, $\mathcal{O}^{(8)}_{qu}$, $\mathcal{O}^{(1)}_{qd}$,     $\mathcal{O}^{(8)}_{qd}$,  $\mathcal{O}^{(8)}_{quqd}$\\
& \\[-0.4cm]
\hline
\end{tabular}
\caption{Operators generated by the heavy scalar fields introduced in Tab.~\ref{t:scalars}. The results are based on \cite{deBlas:2017xtg}.} 
\label{tab:topdown_scalars}
\end{table}

\begin{align}
\wc[]{ll}{ijkl}= &
\frac{
(y_{\mathcal{\phi}_1})^*_{rjl}
(y_{\mathcal{\phi}_1})_{rik}}{
M_{\mathcal{\phi}_{1r}}^{2}}\,,\quad
\wc[]{ee}{ijkl} = 
\frac{
(y_{\mathcal{\phi}_2})_{rki}
(y_{\mathcal{\phi}_2})^*_{rlj}}{
2  M_{\mathcal{\phi}_{2r}}^{2}}\,,
\\[5mm]
\wc[]{le}{ijkl} = &
- \frac{
(y^e_{S})^*_{rli}
(y^e_{S})_{rkj}}{
2  M_{S_r}^{2}}\,,\quad
\wc[(1)]{qu}{ijkl}= 
- \frac{
(y^u_{S})^*_{rjk}
(y^u_{S})_{ril}}{
6  M_{S_r}^{2}}
- \frac{
2   (y^{qu}_{\Phi})^*_{rjk}
(y^{qu}_{\Phi})_{ril}}{
9  M_{\Phi_{r}}^{2}}\,,
\\[5mm]
\wc[(8)]{qu}{ijkl}= &
- \frac{
(y^u_{S})^*_{rjk}
(y^u_{S})_{ril}}{
M_{S_r}^{2}}
+ \frac{
(y^{qu}_{\Phi})^*_{rjk}
(y^{qu}_{\Phi})_{ril}}{
6  M_{\Phi_{r}}^{2}}\,,
\\[5mm]
\wc[(1)]{qd}{ijkl}= &
- \frac{
(y^d_{S})^*_{rli}
(y^d_{S})_{rkj}}{
6  M_{S_r}^{2}}
- \frac{
2   (y^{dq}_{\Phi})^*_{rli}
(y^{dq}_{\Phi})_{rkj}}{
9  M_{\Phi_{r}}^{2}}\,,
\\[5mm]
\wc[(8)]{qd}{ijkl}= &
- \frac{
(y^d_{S})^*_{rli}
(y^d_{S})_{rkj}}{
M_{S_r}^{2}}
+ \frac{
(y^{dq}_{\Phi})^*_{rli}
(y^{dq}_{\Phi})_{rkj}}{
6  M_{\Phi_{r}}^{2}}\,, 
\\[5mm]
\wc[]{ledq}{ijkl}= &
\frac{
(y^d_{S})_{rkl}
(y^e_{S})^*_{rji}}{
M_{S_r}^{2}}\,,\quad
\wc[(1)]{quqd}{ijkl} = 
- \frac{
(y^u_{S})_{rij}
(y^d_{S})^*_{rlk}}{
M_{S_r}^{2}}\,,
\\[5mm]
\wc[(8)]{quqd}{ijkl}= &
- \frac{
(y^{dq}_{\Phi})^*_{rlk}
(y^{qu}_{\Phi})_{rij}}{
M_{\Phi_{r}}^{2}}\,,\quad
\wc[(1)]{lequ}{ijkl}= 
\frac{
(y^{e}_{S})^*_{rji}
(y^{u}_{S})_{rkl}}{
M_{S_{r}}^{2}}\,.
\end{align}

Selective phenomenological SMEFT analyses with scalars are listed in Tab.~\ref{TAB3} and described briefly in Sec.~\ref{PRE27}.

\subsection{Vector-Like Quarks}
Next, we discuss vector-like quark (VLQ) models. There are 7 renormalizable VLQ representations \cite{Ishiwata:2015cga}. All are $\text{SU(3)}_C$ triplets and transform under $\text{SU(2)}_{\rm L}$ either as singlets, doublets and triplets
\begin{equation}\label{tab:VLQ}
\begin{aligned}
\mbox{singlets} & : & D(1, -1/3)\,, & \,U(1,2/3)\,,   &  {\rm (V,VI)~~(LH)}
\\
\mbox{doublets} & : & Q_V(2, +1/6)\,, & \, Q_d(2, -5/6)\,, Q_u(2,7/6)\,, &\, {\rm(IX,XI,X)~~(RH)}
\\
\mbox{triplets} & : & T_d(3, -1/3)\,,  & \,T_u(3, +2/3)\,, &  {\rm (VII,VIII)~~(LH)}
\end{aligned}
\end{equation}
\noindent
where the transformation properties are indicated as $(\text{SU(2)}_{\rm L}, \text{U(1)}_{\rm Y}$). The representations $D$, $U$, $Q_V$, $Q_d$, $Q_u$ $T_d$, $T_u$ correspond to the models V, VI, IX, XI, X, VII, VIII introduced in~\cite{Ishiwata:2015cga}, respectively. We indicated which models bring in new LH currents and which ones RH currents.

We list in Tab.~\ref{tab:topdown_fermions} the operators generated by VLQs. As seen in Tab.~\ref{tab:topdown_phi} several of these operators can also be generated by vector bosons. Among them an important role in these models is played by the $f^2H^2D$ operators,
\be
\mathcal{O}^{(1)}_{H q},\quad
\mathcal{O}^{(3)}_{H q},\quad\mathcal{O}_{H d},\quad  \mathcal{O}_{H u},\quad \mathcal{O}_{H ud}\,.
\ee
The first two generate left-handed quark currents, the remaining ones right-handed ones. On the other hand new operators can be generated that are not present in the case of vector bosons and scalars. These are the dipole operators $f^2XH$
\be
\mathcal{O}_{dB},\quad  \mathcal{O}_{dW},\quad
\mathcal{O}_{dG}, \quad \mathcal{O}_{uB},\quad
\mathcal{O}_{uW}, \quad \mathcal{O}_{uG}\,.
\ee
As we have seen in Sec.~\ref{class7} they play an important role for EDMs.

\begin{table}[tb]
\centering
\begin{tabular}{cl}
\hline
Fields & Operators 
\\
\hline
$D$ & 
$\mathcal{O}_{dH}$, $\mathcal{O}_{dB}$,
$\mathcal{O}_{dG}$, $\mathcal{O}^{(1)}_{H q}$,
$\mathcal{O}^{(3)}_{H q}$ \\
$U$ & 
$\mathcal{O}_{uH}$, $\mathcal{O}_{uB}$,
$\mathcal{O}_{uG}$, $\mathcal{O}^{(1)}_{H q}$,
$\mathcal{O}^{(3)}_{H q}$ \\
$Q_V$ & 
$\mathcal{O}_{dH}$, $\mathcal{O}_{uH}$,
$\mathcal{O}_{dB}$,  $\mathcal{O}_{dW}$,
$\mathcal{O}_{dG}$, $\mathcal{O}_{uB}$,
$\mathcal{O}_{uW}$, $\mathcal{O}_{uG}$,
$\mathcal{O}_{H d}$, $\mathcal{O}_{H u}$,
$\mathcal{O}_{H ud}$ \\
$Q_d$ & 
$\mathcal{O}_{dH}$, $\mathcal{O}_{H d}$ \\
$Q_u$ & 
$\mathcal{O}_{uH}$, $\mathcal{O}_{H u}$ \\
$T_d$ & 
$\mathcal{O}_{dH}$, $\mathcal{O}_{uH}$,
$\mathcal{O}_{dW}$, $\mathcal{O}^{(1)}_{H q}$,
$\mathcal{O}^{(3)}_{H q}$ \\
$T_u$ &
$\mathcal{O}_{dH}$, $\mathcal{O}_{uH}$,
$\mathcal{O}_{uW}$, $\mathcal{O}^{(1)}_{H q}$,
$\mathcal{O}^{(3)}_{H q}$\\
& \\[-0.5cm]
\hline
\end{tabular}
\caption{Operators generated by the heavy vector-like quarks in \eqref{tab:VLQ}. The results are based on \cite{deBlas:2017xtg}.}
\label{tab:topdown_fermions}
\end{table}

The VLQs interact with the SM quarks ($q_L,\, u_R,\, d_R$) via Yukawa interactions 
\begin{equation}
\label{eq:Yuk:H}
\begin{aligned}
- {\cal L}_{\rm Yuk}(H) & =  
\left( \lambda_{ri}^D \, H^\dagger \overline{D}_{Rr}
+   {\lambda_{ri}^U \, \widetilde{H}^\dagger \overline{U}_{Rr}}
+ {\frac{1}{2}\lambda_{ri}^{T_d} \, H^\dagger \overline{T}^I_{dRr}\tau^I}
+ \frac{1}{2}\lambda_{ri}^{T_u} \,\widetilde{H}^\dagger \overline{T}^I_{uRr}\tau^I \right) q_L^i     
\\ 
&  
+ {\bar{u}_R^i\left(\lambda_{ri}^{V_u}  \widetilde{H}^\dagger Q_{VLr} + \lambda_{ri}^{Q_u} \, {H}^\dagger Q_{uLr} \right) }
+ \bar{d}_R^i \left( \lambda_{ri}^{V_d} \,  H^\dagger Q_{VLr}
+ \lambda_{ri}^{Q_d} \, \widetilde{H}^\dagger Q_{dLr} \right) 
+ \mbox{h.c.}
\end{aligned}
\end{equation}
\noindent
where $\widetilde H =\varepsilon_{jk}(H^k)^*$ and $r$ distinguishes between different representations of the same type. The complex-valued Yukawa couplings $\lambda_{ri}^{\rm VLQ}$ with $i=1,2,3$, or equivalently either $i=d,s,b$ or $i=u,c,t$, give rise to mixing with the SM quarks and consequently to flavour-changing $Z$-couplings, which have been worked out in detail in \cite{Ishiwata:2015cga} and confirmed up to a sign in the $T_u$ model in \cite{Bobeth:2016llm}. We refer to Sec.~16.3 in \cite{Buras:2020xsm} for a compact description of VLQ models. Here we just list the tree-level matching conditions at $\Lambda$ for models relevant for $K$, $B_d$ and $B_s$ meson systems that have been first found in \cite{delAguila:2000rc} and confirmed in \cite{Bobeth:2016llm}. 
\begin{equation}
\label{eq:GSM:matching:SMEFT:psi2H2D}
\begin{aligned}
D : &&
\left[\Wc[(1)]{Hq}\right]_{ij} & = \left[\Wc[(3)]{Hq}\right]_{ij}
= - \frac{1}{4} \frac{\lambda_i^\ast \lambda_j}{M^2}\,,
\\
T_d : &&
\left[\Wc[(1)]{Hq}\right]_{ij} & = - 3\, \left[\Wc[(3)]{Hq}\right]_{ij} 
= - \frac{3}{8} \frac{\lambda_i^\ast \lambda_j}{M^2}\,,
\\
T_u : &&
\left[\Wc[(1)]{Hq}\right]_{ij} & = 3\, \left[\Wc[(3)]{Hq}\right]_{ij} 
= \frac{3}{8} \frac{\lambda_i^\ast \lambda_j}{M^2}\,,
\\
Q_d : &&
\left[\Wc{Hd}\right]_{ij} &  
= - \frac{1}{2} \frac{\lambda_i \lambda_j^\ast}{M^2}\,, 
\\
Q_V : &&
\left[\Wc{Hd}\right]_{ij} &  
= \frac{1}{2} \frac{\lambda_i^{V_d} \lambda_j^{V_d\ast}}{M^2}\,, \quad
\left[\Wc{Hu}\right]_{ij} 
= - \frac{1}{2} \frac{\lambda_i^{V_u} \lambda_j^{V_u\ast}}{M^2}\,, \quad
\left[\Wc{Hud}\right]_{ij} 
= \frac{\lambda_i^{V_u} \lambda_j^{V_d\ast}}{M^2}\,,
\end{aligned}
\end{equation}
where $\lambda$ generically denotes the Yukawa coupling of the corresponding VLQ model. The matching of $U$ and $Q_u$ relevant for the $D$-system can be found in \cite{deBlas:2017xtg}.

Having these WCs at the NP scale $\Lambda$ at hand, one can with the help of the tables in Secs.~\ref{classification} and \ref{sec:10} and the charts in Part II of our review investigate the implications of VLQ models for rare decays of mesons. There is a rich literature on FCNCs implied by the presence of VLQs. See in particular \cite{Nir:1990yq, Branco:1992wr,delAguila:2000rc, Barenboim:2001fd, Buras:2009ka, Botella:2012ju, Fajfer:2013wca,Buras:2013td, Altmannshofer:2014cfa, Alok:2015iha, Ishiwata:2015cga, Arnan:2016cpy,Bobeth:2016llm,Cepedello:2024qmq,Erdelyi:2024sls}, which analyzed the patterns of flavour violation in a number of VLQ models listed above.

An important effect of the presence of VLQs is the generation of FCNCs of the $Z$ boson. Let us define the FC quark couplings of the $Z$ by
\begin{align}  
\label{eq:Zcouplings}
\mathcal{L}_{\psi\bar\psi Z}^{\rm NP} & 
= Z_{\mu} \sum_{\psi = u,d} \bar \psi_i \, \gamma^{\mu} \left( 
[\Delta_L^{\psi}(Z)]_{ij} \, P_L 
\,+\, [\Delta_R^{\psi}(Z)]_{ij} \, P_R \right) \psi_j {\,\,+\,\, h.c.}
\end{align}
with $\psi=u,d$ distinguishing between $up-$ and $down-$ quark couplings. These complex-valued couplings are related to the WCs of the SMEFT in (\ref{eq:GSM:matching:SMEFT:psi2H2D}) through \cite{Bobeth:2016llm,Bobeth:2017xry} 
\be
\label{eq:Z-Deltas:dim-6-WC}
\phantom{x}[\Delta^u_L(Z)]_{ij}  
= -\frac{g_Z}{2} v^2 \left[\Wc[(1)]{Hq} - \Wc[(3)]{Hq}\right]_{ij} \,, \qquad 
[\Delta^u_R(Z)]_{ij} 
= -\frac{g_Z}{2} v^2 \left[\Wc{Hu}\right]_{ij} \,,
\ee
\be
[\Delta^d_L(Z)]_{ij}  
= -\frac{g_Z}{2} v^2 \left[\Wc[(1)]{Hq} + \Wc[(3)]{Hq}\right]_{ij} \,, \qquad 
[\Delta^d_R(Z)]_{ij} 
= -\frac{g_Z}{2} v^2 \left[\Wc{Hd}\right]_{ij}\,,
\ee
where $g_Z=\sqrt{g_1^2+g_2^2}$. But what is most important in this context are the RG effects related to the top Yukawa coupling that in the case of models like $Q_d$ and $Q_V$, that have non-vanishing WCs $\left[\Wc{Hd}\right]_{ij}$, cause the generation of left-right operators contributing to $\Delta F=2$ processes. This can be seen by inspecting the charts in Part II of our review. These issues including numerical analyses are discussed in detail in \cite{Bobeth:2016llm,Bobeth:2017xry,Endo:2016tnu,Endo:2018gdn} and summarized briefly in Section 15.5.2 in \cite{Buras:2020xsm}. VLQs play also an important role in the explanation of the so-called Cabibbo Anomaly (CAA), which is nicely reviewed in \cite{Crivellin:2022ctt}. Additional useful references in this context are \cite{Belfatto:2019swo,Belfatto:2021jhf,Botella:2021uxz,Crivellin:2022rhw}. The most recent extensive review on vector-like quarks can be found in \cite{Alves:2023ufm} and in Chapter 16.3 of \cite{Buras:2020xsm}.

\subsection{Vector-Like Leptons}
We next discuss vector-like leptons (VLLs), which are singlets under $\text{SU(3)}_{\rm C}$. There are four Dirac vector-like lepton representations \cite{Ishiwata:2015cga} that transform under $\text{SU(2)}_{\rm L}$ as singlets, doublets and triplets as well as two Majorana vector-like leptons, $N$ and $T_0$ that transform as singlet and triplet:
\begin{equation}\label{tab:VLL}
\begin{aligned}
\mbox{singlets} & : & N(1,0)\,,\,\,\, & E(1, -1)\,,    & &  {\rm (I)~~{(LH)}}
\\
\mbox{doublets} & : & \Delta_1(2, -1/2)\,,\,\,\, & \Delta_3(2, -3/2)\,, & & {\rm(III,IV)~~{(RH)}}
\\
\mbox{triplets} & : & T_0(3,0)\,,\,\,\, & T_e(3, -1)\,.  & & {\rm (II)~~{(LH)}}
\end{aligned}
\end{equation}
\noindent
Their transformation properties are indicated as $(\text{SU(2)}_{\rm L}, \text{U(1)}_{\rm Y}$). The representations $E$, $\Delta_1$, $\Delta_3$, $T_e$ correspond to the models I, III, IV, II introduced in~\cite{Ishiwata:2015cga}. We indicated which models bring in new LH currents and which ones RH currents.

\begin{table}[tb]
\centering
\begin{tabular}{cl}
\hline
Fields & Operators 
\\
\hline
$N$ & 
$\mathcal{O}_{\nu\nu}$, $\mathcal{O}^{(1)}_{H l}$,
$\mathcal{O}^{(3)}_{H l}$ \\
$E$ & 
$\mathcal{O}_{eH}$, $\mathcal{O}_{eB}$,
$\mathcal{O}^{(1)}_{H l}$, $\mathcal{O}^{(3)}_{H l}$ \\
$\Delta_1$ & 
$\mathcal{O}_{eH}$, $\mathcal{O}_{eB}$,
$\mathcal{O}_{eW}$, $\mathcal{O}_{H e}$ \\
$\Delta_3$ & 
$\mathcal{O}_{eH}$, $\mathcal{O}_{H e}$ \\
$T_0$ & 
$\mathcal{O}_{\nu\nu}$, $\mathcal{O}_{eH}$,
$\mathcal{O}^{(1)}_{H l}$, $\mathcal{O}^{(3)}_{H l}$ \\
$T_e$ & 
$\mathcal{O}_{eH}$, $\mathcal{O}_{eW}$,
$\mathcal{O}^{(1)}_{H l}$, $\mathcal{O}^{(3)}_{H l}$ \\
& \\[-0.5cm]
\hline
\end{tabular}
\caption{Operators generated by the heavy vector-like leptons in \eqref{tab:VLL}. The results are based on \cite{deBlas:2017xtg}.}
\label{tab:topdown_leptons}
\end{table}

The VLLs interact with SM leptons ($l_L,\, e_R$) via Yukawa interactions
\begin{equation}
\label{eq:Yuk:H}
\begin{aligned}
- {\cal L}_{\rm Yuk}(H) & =  
\left( \lambda^N_{ri} \overline{N}_{Rr}\widetilde{H}^\dagger+\lambda_{ri}^E \, H^\dagger \overline{E}_{Rr}{+\frac{1}{2} \lambda_{ri}^{T_0} \,\widetilde H^\dagger \overline{T}^I_{0Rr} \tau^I}
+\frac{1}{2} \lambda_{ri}^{T_e} \, H^\dagger \overline{T}^I_{eRr} \tau^I\right) l_L^i
\\ 
&  
+ \bar{e}_R^i \left( \lambda_{ri}^{\Delta_1} \,  H^\dagger \Delta_{1Lr}
+ \lambda_{ri}^{\Delta_3} \, \widetilde{H}^\dagger \Delta_{3Lr} \right) 
+ \mbox{h.c.}
\end{aligned}
\end{equation}
\noindent
where $\widetilde H =\varepsilon_{jk}(H^k)^*$ and $r$ distinguishes between different representations of the same type. 

The complex-valued Yukawa couplings $\lambda_{ri}^{\rm VLL}$ with $i=1,2,3$, or equivalently either $i=e,\mu,\tau$ or $i=\nu_\e,\nu_\mu,\nu_\tau$, give rise to mixing with the SM leptons and consequently to flavour-changing $Z$-couplings, which have been worked out in detail in \cite{Ishiwata:2015cga}. The list of tree-level matchings to the SMEFT can be found in \cite{deBlas:2017xtg}.

In Tab.~\ref{tab:topdown_leptons} we list the dim-5 and dim-6 SMEFT operators generated by the heavy vector-like leptons. Again, using the tables in Secs.~\ref{classification} and \ref{sec:10} and the charts in Part II of our review one can investigate the implications of vector-like lepton models for rare decays of leptons.

The literature on these models is not as rich as for VLQs and is dominated by the motivation to explain the $(g-2)_\mu$ anomaly \cite{Dermisek:2013gta,Sierra:2015fma,Belanger:2015nma,Altmannshofer:2016oaq,Kowalska:2017iqv,Darme:2018hqg,Crivellin:2020ebi}. They can also play a role in the explanation of the CAA anomaly \cite{Crivellin:2020ebi}, and for collider signatures see e.g. \cite{Falkowski:2013jya,Bissmann:2020lge}. VLLs were studied in the context of the FCC-ee and the electron Yukawa coupling in \cite{Erdelyi:2025axy}. We can also recommend the analysis in \cite{Kawamura:2019rth}, where a model with a complete vector-like fourth family and additional $\text{U(1)}^\prime$ symmetry has been proposed. One can then see VLQs, VLLs and a $Z^\prime$ together in action with the goal to explain the $(g-2)_\mu$ anomaly and $b\to s\ell^+\ell^-$ anomalies in the context of a rich flavour structure.

\subsection{Leptoquarks}
Finally, we discuss leptoquarks that are reviewed in \cite{Buchmuller:1986zs,Davies:1990sc,Davidson:1993qk,Dorsner:2016wpm,Bobeth:2017ecx} and in Section 16.4 in \cite{Buras:2020xsm}.

In Tab.~\ref{tab:LQ-q-numbers} we list all 10 leptoquark models, together with their quantum numbers. The stars indicate to which processes a given LQ contributes at tree-level. The relevant Lagrangians are discussed in detail in Section 16.4 of \cite{Buras:2020xsm}.

LQ contributions are dominated by semi-leptonic operators like
\be 
(\bar d^i\gamma_\mu P_L l^j)(\bar l^l\gamma^\mu P_L d^k)\,, \qquad 
(\bar d^i\gamma_\mu P_R l^j)(\bar l^l\gamma^\mu P_R d^k)\,,\qquad
(\bar d^i\gamma_\mu P_L l^j)(\bar l^l\gamma^\mu P_R d^k)\,.
\ee
In order to find the WCs of the operators consisting of products of quark and lepton currents, as given in the Warsaw basis, Fierz transformations have to be performed. One finds then respectively
\be 
(\bar d^i\gamma_\mu P_L d^k)(\bar l^l\gamma^\mu P_L l^j)\,, \qquad 
(\bar d^i\gamma_\mu P_R d^k)(\bar l^l\gamma^\mu P_R l^j)\,, \qquad
-2 (\bar d^i P_R d^k)(\bar l^l  P_L l^j)\,.
\ee
This means that in the presence of both LH and RH vector couplings SLR operators contribute to various observables. On the other hand in some models, like $S_3$, operators with LH and RH scalar couplings and charge conjugated fields are present. The corresponding Fierz transformations, that can be found in Appendix~A.3 of \cite{Buras:2020xsm}, imply VLR operators.

The four-fermion operators contributing to both $b\to s\nu\bar\nu$ and $b\to s\ell^+\ell^-$ transitions are 
\be
 {Q_{lq}^{(1)} = (\bar \ell \gamma_\mu \ell)(\bar q \gamma^\mu q) \,,
\qquad 
 Q_{lq}^{(3)} = (\bar \ell \gamma_\mu \tau^I \ell)(\bar q \gamma^\mu \tau^I q) \,,
\qquad
 Q_{ld} = (\bar \ell\gamma_\mu \ell)(\bar d \gamma^\mu d) \,,}
\label{eq:ops}
\ee
and the ones contributing to $b\to s\ell^+\ell^-$ but {\em not} to $b\to s\nu\bar\nu$ are
\begin{align}
 Q_{ed} &= {(\bar e\gamma_\mu e)(\bar d \gamma^\mu d) }
\,,&
 Q_{qe} &= {(\bar q \gamma_\mu q) (\bar e\gamma^\mu e)
\,.}
\label{eq:ops2}
\end{align}

We illustrate the tree-level matching in specific leptoquark models that are relevant for both $b\to s\nu\bar\nu$ and $b\to s\ell^+\ell^-$ decays \cite{Bobeth:2017ecx}:
\begin{align}
S_1:&\qquad \wc[(1)]{\ell q}{ijkl} = - \wc[(3)]{\ell q}{ijkl} =  \frac{1}{4}\frac{\lambda_{j l} \lambda^*_{i k}}{m_\phi^2}\,, 
\\
S_3:& \qquad \wc[(1)]{\ell q}{ijkl} =  3\wc[(3)]{\ell q}{ijkl} = \frac{3}{4} \frac{\lambda_{j l} \lambda^*_{i k}}{m_\phi^2}\,,
\\
\widetilde{R}_2:&\qquad \wc[]{\ell d}{ijkl} = -\frac{1}{2} \frac{\lambda_{i l} \lambda^*_{j k}}{m_\phi^2}\,,
\end{align}

Here $m_\phi$ is the leptoquark mass and $\lambda_{ij}$ the relevant couplings. The remaining four leptoquark models $\tilde S_1$, $R_2$, $\tilde U_1$ and $\tilde V_2$ are less popular and their properties are briefly summarized in \cite{Buras:2020xsm}.

There is a huge number of such analyses, in particular in the context of $B$ physics anomalies. Selected references can be found in Table 16.5 of \cite{Buras:2020xsm}. Here we refer to more recent papers. In particular to~\cite{Gherardi:2020qhc,Marzocca:2021miv,Marzocca:2024hua} where low-energy phenomenology of scalar leptoquarks, for $B$, $K$, $D$ physics and also for lepton flavour violating processes were presented. See also an extensive analysis in the $U_1$ model \cite{Cornella:2021sby} with $\text{U(2)}^5$ symmetry. Most recently correlations between $R(D^{(*)})$ anomalies and $K\to\pi\nu\bar\nu$ and $B\to K\nu\bar\nu$ via leptoquarks have been investigated in \cite{Crivellin:2025qsq}.

\begin{table}[H]
\renewcommand{\arraystretch}{1.5}
\centering
{\begin{tabular}{|c|ccccc|c|c|c|c|}
\hline
LQ                & $\text{SU(3)}_c$ & $\text{SU(2)}_L$ & $\text{U(1)}_Y$ & $T_3$  & $Q_{\rm em}$ &
  $d_i$, $\ell^+\ell^-$ & $d_i$, $\nu\bar \nu$ & $u_i$, $\ell^+\ell^-$ & $u_i$, $\nu\bar \nu$ \\
\hline \hline
$S_1$             & $3^\ast$  & 1         & $1/3$    & 0      & $1/3$ & & $*$ & $*$ & 
\\[0.1cm]
$\tilde{S}_1$     & $3^\ast$  & 1         & $4/3$    & 0      & $4/3$ & $*$ & & & 
\\[0.1cm]
$R_2$             & $3\;\;$   & 2         & $7/6$    & $+1/2$ & $5/3$ 
& $$ &  &$* $& 
\\
&           &           &          & $-1/2$ & $2/3$
& $*$ & & & $*$
\\[0.1cm]
$\tilde{R}_2$     & $3\;\;$   & 2         & $1/6$    & $+1/2$ & $2/3$
& $*$ & $$ & &
\\
&           &           &          & $-1/2$ & $-1/3\;\;\;$
& & $*$ && 
\\[0.1cm]
$S_3$             & $3^\ast$  & 3         & $1/3$    & $+1$   & $4/3$
&$*$ & && 
\\
&         &       &       & $\;\;\;0$ & $1/3$
&& $*$ &$*$ &
\\
&       &       &       & $-1$  & $-2/3\;\;\;$
&&&& $*$
\\ 
\hline
$U_1$         & $3\;\;$  & 1      & $2/3$  & 0    & $2/3$
& $*$ & & & $*$
\\[0.1cm]
$\tilde{U}_1$   & $3\;\;$  & 1      & $5/3$ & 0    & $5/3$
& & &$*$ &
\\[0.1cm]
$V_2$         & $3^\ast$ & 2      & $5/6$  & $+1/2$ & $4/3$
&$*$& & & 
\\
&       &       &      & $-1/2$ & $1/3$
& & $*$ & $*$ &
\\[0.1cm]
$\tilde{V}_2$   & $3^\ast$ & 2      & $-1/6\;\;\;$  & $+1/2$ & $1/3$
&&& $*$ & 
\\
&       &       &      & $-1/2$ & $-2/3\;\;\;$
&&&& $*$
\\[0.1cm]
$U_3$         & $3\;\;$  & 3      & $2/3$  & $+1$  & $5/3$
&&&$*$&
\\
&       &       &       & $\;\;\;0$ & $2/3$
&$*$ & & & $*$
\\
&       &       &       & $-1$ & $-1/3\;\;\;$
&&$*$ &&
\\ 
\hline
\end{tabular}
\caption{Quantum numbers of interactions of LQs. The stars indicate to which processes a given LQ contributes at tree-level with $d_i$ representing $K$ or $B$ mesons, $u_i$ D mesons and $\ell^+\ell^-$ and $\nu\bar\nu$ lepton pair in the final state. Note that antitriplets $3^\ast$ interact with charge conjugated quark fields. From \cite{Buras:2020xsm}.
}
\label{tab:LQ-q-numbers}}
\end{table}
\begin{align}
U_1:&\qquad \wc[(1)]{\ell q}{ijkl}  = \wc[(3)]{\ell q}{ijkl} = -\frac{1}{2}\frac{\lambda_{i l} \lambda^*_{j k}}{m_\phi^2}\,,
\\
U_3:&\qquad \wc[(1)]{\ell q}{ijkl} = - 3 \wc[(3)]{\ell q}{ijkl} = - \frac{3}{2}\frac{\lambda_{i l} \lambda^*_{j k}}{m_\phi^2}\,,
\\
V_2:&\qquad \wc[]{\ell d}{ijkl} = \frac{\lambda_{i l} \lambda^*_{j k}}{m_\phi^2}\,.
\end{align}

\section{Bottom-Up Approach at Work} \label{sec:6}

\subsection{Preliminaries}
In the previous section we have presented a number of NP models which can be confronted with experiment using the top-down approach. Having constructed a UV completion and using the full machinery of the SMEFT and WET in which RG evolution plays the crucial role, it is possible to obtain WCs evaluated at the low-energy hadronic scale $\muLow$ which are then given in terms of the parameters of the given BSM scenario. By performing a global fit to a multitude of low energy observables one can determine these parameters and check whether this model can describe the data. In particular whether it can explain possible anomalies.

The construction of a given BSM scenario can generally be motivated by the open questions that the SM cannot answer. Yet, the number of models one could construct without any hints from low-energy data is large and to reach this goal in an efficient manner it is crucial to get some information from low-energy data, in particular from the pattern of possible anomalies observed there. These are described in terms of $C_i(\muLow)$. But to find out which NP is responsible for these anomalies what we really need are $C_k(\muNP)$. However, due to the mixing under renormalization, not only due to gauge interactions but in particular because of Yukawa interactions, the values of the non-vanishing $C_k(\muNP)$ and in particular their number will differ from the ones of $C_i(\muLow)$.

This is evident from our presentation of different classes of processes in Part II of our review. There, in fact, following the four-step procedure summarized in Sec.~\ref{Bottom-Up} we already employed the bottom-up approach to identify for a given class the most important SMEFT operators at $\Lambda$ including those which were absent in the tree-level matching of the WET on to SMEFT. However, eventually, to identify more suitable UV completions, it would be beneficial to determine the values of $C_k(\muNP)$ following from the measured values of $C_i(\muLow)$. The goal of this section is to develop a procedure which would in principle allow us to find in a systematic manner $C_k(\muNP)$ from $C_i(\muLow)$. We limit our discussion to the LO case of the RG improved perturbation theory as done in PART II, leaving a NLO analysis for the future. Due to simplicity, we will showcase the method for pure QCD only. 

To this end we recall the RG formulae in the top-down approach that we already encountered in Sec.~\ref{RSDEP}, keeping only the leading terms. These are
\begin{align}
\label{eq:fullEvol+}
\vec{C}_\text{JMS}(\muLow) &
= \hat U_\text{JMS}^{(0)}(\muLow, \muEW) \;
\hat K^{(0)}(\muEW) \;
\hat U_\text{SMEFT}^{(0)}(\muEW,\muNP) \;
\vec{\mathcal{C}}_\text{SMEFT}(\muNP),
\end{align}
where the matrix $\hat K^{(0)}(\muEW)$ represents the tree-level matching of the SMEFT onto the WET
\be
\label{eq:SMEFTWET+}
\vec{\mathcal{C}}_\text{JMS}(\muEW) 
= \hat K^{(0)}(\muEW) \; \vec{\mathcal{C}}_\text{SMEFT}(\muEW)\,.
\ee
$\hat U_i^{(0)}$ with $i$ distinguishing between JMS and SMEFT are LO evolution matrices which are given generally as follows
\begin{equation}\label{u0vdP} 
\hat U_i^{(0)}(\mu_1,\mu_2)= \hat V_i
\left(
{\left[\frac{\as(\mu_2)}{\as(\mu_1)}\right]}^{{\vec\gamma^{(0)}_i\over 2\beta_0}}
\right)_D \hat V_i^{-1}, 
\end{equation}
\noindent
where $\hat V_i$ diagonalizes ${\hat\gamma_i^{(0)T}}$
\begin{equation}
\label{ga0dP} 
\hat\gamma^{(0)}_{iD}=\hat V_i^{-1} {\hat \gamma_i^{(0)T}} 
\hat V_i\,,
\end{equation}
\noindent
and $\vec\gamma_i^{(0)}$ is the vector containing the diagonal elements of the diagonal matrix $\hat\gamma^{(0)}_{iD}$.

\subsection{Bottom-Up Running: QCD}
At first sight it appears that our goal can be reached in a straightforward manner by just inverting \eqref{eq:fullEvol+}. As
\be\label{invU}
\left[\hat U_i^{(0)}(\mu_1,\mu_2)\right]^{-1}=\hat U_i^{(0)}(\mu_2,\mu_1)\,,
\ee     
we find
\begin{align}
\label{eq:Bottom-Up}
\vec{C}_\text{SMEFT}(\muNP) &
= \hat U^{(0)}_\text{SMEFT}(\muNP, \muEW) \;
\left[\hat K^{(0)}(\muEW)\right]^{-1}\,  \;
\hat U^{(0)}_\text{JMS}(\muEW, \muLow)\; 
\vec{\mathcal{C}}_\text{JMS}(\muLow) \,,
\end{align}
with all matrices defined above. Note that the arguments in the evolution matrices are interchanged relatively to the top-down approach as given in \eqref{eq:fullEvolX} and \eqref{eq:fullEvol+}, because RG evolution takes place now in the opposite direction.

Next, due to \eqref{invU} the RG running itself consists of two steps:

\paragraph{Step 1:}
\be
\label{eq:Step 1}
\vec{C}_\text{JMS}(\muEW) 
= \hat U^{(0)}_\text{JMS}(\muEW, \muLow)\; 
\vec{\mathcal{C}}_\text{JMS}(\muLow)\,.
\ee
This evolution will modify the values of $\vec{\mathcal{C}}_\text{JMS}(\muLow)$ and could generate non-vanishing values of new operators so that the pattern of the values of $\vec{C}_\text{JMS}(\muEW)$ will be generally different from the one of $\vec{\mathcal{C}}_\text{JMS}(\muLow)$.

\paragraph{Step 2:}
\begin{align}
\label{eq:Step 2}
\vec{C}_\text{SMEFT}(\muNP) &
= \hat U^{(0)}_\text{SMEFT}(\muNP, \muEW) \;
\vec{\mathcal{C}}_\text{SMEFT}(\muEW)\,.
\end{align}
Also in this step new SMEFT operators are generated at $\Lambda$ that were absent at $\muEW$. We have already seen this in Part II when searching for operators at $\Lambda$ in a given class. This marks Step 3 of the four-step procedure of Sec.~\ref{Bottom-Up}.

What is left to do is to invert (\ref{eq:SMEFTWET+})
\be
\vec{\mathcal{C}}_\text{SMEFT}(\muEW) 
= [\hat K^{(0)}(\muEW)]^{-1} \; \vec{\mathcal{C}}_\text{JMS}(\muEW)\,.
\ee

However, the inspection of tree-level matching conditions presented in Part II 
demonstrates very clearly that generally the inverse of the matrix $\hat K^{(0)}(\muEW)$ does not exist. 
{Indeed the SMEFT to WET matching is not \emph{bijective} at the EW threshold. 
Exactly due to this reason the RG running can not be performed across the EW threshold. 
}
The reason is very simple. The number of SMEFT operators involved in the matching is generally different from the number of WET operators.

Thus, it looks like generally it is not possible without any additional model assumptions to calculate the coefficients $\vec{C}_\text{SMEFT}(\muNP)$ on the basis of the experimentally measured $\vec{\mathcal{C}}_\text{JMS}(\muLow)$. While this is a clear disappointment, let us illustrate this issue on a few examples which also will inform us under which circumstances our goals can still be reached.

{\bf Class 1}

Inspecting the matching conditions in \eqref{eq:left-smeft-down} we find that in the absence of LL operators our goal can be reached:
\be
\wc[(1)]{qd}{ijij}(\muEW)=\wcL[V1,LR]{dd}{ijij}(\muEW)\,,
\quad 
\wc[(8)]{qd}{ijij}(\muEW)=\wcL[V8,LR]{dd}{ijij}(\muEW)\,,\quad
\wc{dd}{ijij}(\muEW)=\wcL[V,RR]{dd}{ijij}(\muEW)\,.
\ee
Indeed, the values of the three SMEFT WCs at the EW scale can be determined from experiment and knowing them and using \eqref{eq:Step 2} we can find their values at $\muNP$ and also the values of the additional WCs generated in the process of RG running.

On the other hand for LL operators this is not possible because in this case we have
\be
\wc[(1)]{qq}{ijij}(\muEW)+ \wc[(3)]{qq}{ijij}(\muEW)= \wcL[V,LL]{dd}{ijij}(\muEW)\,,
\ee
so that only the sum of the two SMEFT coefficients is constrained by the data. In certain NP scenarios one of them is very small and then under this assumption our goal can be reached. This is the case for $Z^\prime$ scenarios in which at the NP scale the coefficients $\wc[(3)]{qq}{ijij}(\muNP)=0$ at tree-level. Even if they are generated through RG running at $\muEW$ they are much smaller than the $\wc[(1)]{qq}{ijij}(\muEW)$. Such hierarchies do however not hold in general.

{\bf Class 2}

A more interesting situation takes place for Class 2. Here it is favourable to use the CWET operator basis instead of the JMS one. Inspecting then the matching conditions in \eqref{eq:SMEFT2WET_C9_C10} we again find that in the RH sector our goal can be reached. Suppressing $\muEW$ we indeed find at the EW scale
\be
\begin{aligned}\label{SMEFTCWET1}
\wc[]{\ell d}{ppij} &= \frac{2\, \,\mathcal{N}_{ji}}{(3+\zeta_2)}
\left[C_{9}^{\prime, ijpp}-C_{10}^{\prime, ijpp}+(1+\zeta_2)\,C^{ijpp}_R \right]\,,
\\
\wc[]{ed}{ppij} &=\frac{ 2\, \,\mathcal{N}_{ji}}{(3+\zeta_2)}
\left[2C_{9}^{\prime, ijpp}+(1+\zeta_2)C_{10}^{\prime, ijpp}+(\zeta_2-1)\,C^{ijpp}_R \right]\,,\\
C_Z^d &=\frac{ 2\, \,\mathcal{N}_{ji}}{(3+\zeta_2)}
\left[C_{10}^{\prime, ijpp}-C_{9}^{\prime, ijpp}+2\,C^{ijpp}_R \right]\,,
\end{aligned}
\ee
so that the values of the three SMEFT coefficients can be determined at the EW scale and subsequently at $\muNP$.

Again, for the case of LH currents our goal cannot be reached without further assumptions. This time we find
\be
\begin{aligned}\label{SMEFTCWET2}
\wc[(1)]{\ell q}{ppij} &=\frac{\mathcal{N}_{ji}}{2}\left[C_{9}^{ijpp}-C_{10}^{ijpp}+2C^{ijpp}_L\right]-{\frac{1}{4}}(1-\zeta_2)C_Z^q\,,
\\
\wc[(3)]{\ell q}{ppij} &=\frac{\mathcal{N}_{ji}}{2}\left[C_{9}^{ijpp}-C_{10}^{ijpp}-2C^{ijpp}_L\right]+{\frac{1}{4}(3+\zeta_2)}C_Z^q\,,
\\
\wc[]{qe}{ijpp} &=\mathcal{N}_{ji}\left[C_{9}^{ijpp}+C_{10}^{ijpp}\right]-\frac{1}{2}(1-\zeta_2)C_Z^q\,.
\end{aligned}
\ee
Thus, unless we make some model-dependent assumption on $C_Z^q$ our goal cannot be reached. This WC is often related to generated flavour changing $Z$ couplings through RG effects and/or $Z-Z^\prime$ mixing and could be small but this clearly depends on the NP scenario.

Similar derivations can be made for other classes. The main message from this analysis is the following one: While in certain cases, dictated by the experimental data, it is possible to evaluate the WCs at the NP scale, this is not always the case. To our knowledge this finding is new and for sure has not been emphasized enough in the literature.

Still an analysis of that type can give useful information on the kind of NP at work. Having the lists of operators from Part II for each class one can then search for explicit models that generate them. We will elaborate on it in Sec.~\ref{Searching}. Moreover, with explicit expressions given above we know that in the absence of new left-handed currents not only the contributing operators at $\Lambda$ but also their WCs can be determined.

It should be noticed that Step 1, which includes only QCD and QED RG effects, is a good approximation because the top Yukawa does not enter the evolution. This is not the case for Step 2, where top Yukawa effects have to be included. This is what we will discuss next.

\subsection{Bottom-Up Running: Top Yukawa Effects}
The case of Yukawa evolution is more involved. Fortunately such effects are only important within the SMEFT due to the presence of the top quark, which is absent in the WET. Due to effects of the mixing through Yukawa couplings and the need for back-rotation, as discussed in Sec.~\ref{sec:ren-gro-run}, analytic expressions are much more involved and will not be presented here. Some insight into their structure can be obtained by inspecting numerous formulae in \cite{Buras:2018gto}. While there top-down evolution was discussed, the corresponding expressions for bottom-up running can be simply obtained by interchanging $\mu$ and $\Lambda$. A short summary of this method can be found in Appendix E of \cite{Buras:2020xsm}.

\section{Hunting New Particles}\label{Searching}

In the presence of anomalies one can identify various sets of WCs that can give a satisfactory description of the observed data. Presently, having only a small number of rare branching ratios that have been measured with respectable precision and which simultaneously have small theoretical uncertainties, many sets are possible. However, in the future the situation will certainly improve and the number of sets of operators with non-vanishing WCs capable of describing existing anomalies, not only in weak decays but also in other processes represented by classes 4, 7, 9 and 10, will be reduced. They would represent several promising avenues towards the construction of viable NP scenarios to be explored. In this context the accuracy of the data and also the theoretical uncertainties will play a crucial role. One should hope that at the end of this decade, when not only more accurate data will be available but also non-perturbative long distance uncertainties will be under a better control than they are now, such an approach will lead to definite conclusions. In this context we should emphasize that not only the uncertainties in hadronic matrix elements of SM operators should be decreased but also in the ones of NP operators. Here still definite progress should be made.

Last, but certainly not least, new models solving existing open questions could be proposed and analyzed in detail. One should not forget that although in the last decade particular attention of theorists was payed to models explaining LFUV, the confirmation of $\mu-e$ universality in some $b\to s\ell^+\ell^-$ observables changed significantly the strategies, and models with $\mu-e$ universality are again in the game. Whether $\mu-\tau$ universality breakdown will survive more accurate data is an open question. However, as pointed out in \cite{Fleischer:2023zeo}, in the presence of CPV phases the fact that the ratios $R(K)$ and $R(K^*)$ are in agreement with the SM does not necessarily imply $\mu-e$ universality in all $b\to s\ell^+\ell^-$ observables, in particular not in CP asymmetries in neutral and charged $B\to K\ell^+\ell^-$ decays.

\begin{table}[tbp]
\begin{center}
\renewcommand{\arraystretch}{1.3}
\begin{tabular}{|c|c|c||c|c|c|}
\hline
WET & SMEFT &Class& WET & SMEFT &Class\\
\hline

$\wcL[V,LL]{\nu u}{}$     & $\wc[(1)]{l q}{},\, \wc[(3)]{l q}{} $  &{6}&
$\wcL[V,LR]{\nu u}{}$         & $\wc[]{lu}{}$  & {6} \\
$\wcL[V,LL]{\nu d}{}$       & $\wc[(1)]{Hq}{} , \wc[(3)]{Hq}{},  \wc[(1)]{l q}{},\, \wc[(3)]{l q}{} $   &2   &
$\wcL[V,LR]{\nu d}{}$         &  $ \wc[]{Hd}{},\,\wc[]{l d}{} $ &2   \\
$\wcL[V,LL]{eu}{}$      & $\wc[(1)]{Hl}{} , \wc[(3)]{Hl}{},\,  \wc[(1)]{l q}{}, \wc[(3)]{l q}{}$ &  6 &
$\wcL[V,LR]{eu}{}$     & $\wc[(1)]{Hl}{} , \wc[(3)]{Hl}{}, \wc[]{l u}{}$  & 6   \\
$\wcL[V,LL]{ed}{}$       & $\wc[(1)]{Hl}{} , \wc[(3)]{Hl}{} ,\wc[(1)]{Hq}{} , \wc[(3)]{Hq}{},\, \wc[(1)]{l q}{}, \wc[(3)]{l q}{} $  &2,4,6 &
$\wcL[V,LR]{ed}{}$        & $ \wc[(1)]{Hl}{} , \wc[(3)]{Hl}{}, \wc[]{Hd}{},\,\wc[]{l d}{} $      &2,6  \\
$\wcL[V,LL]{\nu edu}{}$  
& $\wc[(3)]{lq}{},\, \wc[(3)]{Hl}{}, \wc[(3)]{Hq}{}$ &  8 &
$\wcL[V,LR]{ue}{}$ 
      &$   \wc[]{He}{}, \wc[]{qe}{}$  &6  \\
  $\wcL[V,RR]{eu}{}$     & $ \wc[]{He}{},  \wc[]{eu}{} $  &6 &
$\wcL[V,LR]{de}{}$         & $ \wc[]{He}{},\wc[]{Hd}{}, \wc[(1)]{Hq}{} , \wc[(3)]{Hq}{},  \wc[]{qe}{}$  &2,6       \\
$\wcL[V,RR]{ed}{}$     & $\wc[]{He}{},\wc[]{Hd}{},\,\wc[]{ed}{}$   &2, 6     &
$\wcL[V,LR]{\nu edu}{}$        & $\wc[]{Hud}{}$  & 8  \\
$\wcL[S,RR]{eu}{}$   & $\wc[(1)]{lequ}{}$ &6,7 & 
$\wcL[S,RL]{eu}{}$ & 0  & {6,7} \\
$\wcL[T,RR]{eu}{}$ & $\wc[(3)]{lequ}{}$ & 6,7  &
$\wcL[S,RL]{ed}{}$
&$\wc[]{ledq}{}$  &2,6,7 \\
$\wcL[S,RR]{ed}{}$  & 0  & {2,6,7}&
$\wcL[S,RL]{\nu edu}{}$  
 & $\wc[]{ledq}{}$   & 8\\
$\wcL[T,RR]{ed}{}$ & 0 & {2,6,7} && &  \\
$\wcL[S,RR]{\nu edu}{}$  
 & $\wc[(1)]{lequ}{}$  &8&&  &\\
$\wcL[T,RR]{\nu edu}{}$  
 &  $\wc[(3)]{lequ}{}$  &8&&  &  \\
\hline
\end{tabular}
\caption{{C}onnections of semi-leptonic four-fermion WET WCs to SMEFT WCs {and various classes, at the tree-level}.}
\label{tab:wetsemilepSMEFT}
\end{center}
\end{table}

While our review was dominated by SMEFT WCs the phenomenology at hadronic scales is performed in terms of WET WCs. Therefore, possible anomalies in low-energy processes will be first signalled by the values of these coefficients. The connections of these coefficients to the SMEFT ones are given by the numerous matching conditions which can be easily found with the help of Tab.~\ref{tab:equations}. In order to get an idea  what the anomaly given in terms of WET WCs implies for SMEFT WCs, it is useful to have the Tabs.~\ref{tab:wetsemilepSMEFT}-\ref{tab:wetdipSMEFT} at hand, summarizing the matching conditions at the EW scale.\footnote{The ``0'' means that there is no tree-level matching at dimension 6 level. Such WET operators are sometimes referred to as non-SMEFT operators. If an anomaly is seen in one of these WET WCs below $\muEW$, it can point us towards physics beyond the SMEFT or a dim-8 effect within SMEFT \cite{Burgess:2021ylu}. As emphasized in this reference, after having evidence for such a signal, a systematic exploration of ultraviolet completions would be necessary. } 
{They} give us hints which SMEFT WCs could be used to describe the observed anomalies that were identified using the WET \footnote{{Note that Classes 4,9 and 10 absent in Tabs.~\ref{tab:wetsemilepSMEFT}-\ref{tab:wetdipSMEFT} can also be correlated though vertex corrections to EW particles and various four-fermion operators.   }}. As discussed in previous sections these hints are not unique but still reduce the number of possibilities.

Indeed, as seen in these tables, an anomaly in a given WET WC can have an impact on several SMEFT WCs and also on different classes of decays. Even though flavour indices are not shown in these tables, the inspection of the classes affected by a given anomaly together with the relevant experimental data allows to make first steps in the phenomenological analysis that should be improved by including RG effects and the one-loop matching.

\begin{table}[tbp]
\begin{center}
\renewcommand{\arraystretch}{1.3}
\begin{tabular}{|c|c|c||c|c|c|}
  \hline
  WET & SMEFT& Class & WET & SMEFT& Class\\
\hline
$\wcL[V,LL]{uu}{}       $ & $  \wc[(1)]{qq}{}\,,\wc[(3)]{qq}{}$  &1    &
$\wcL[V1,LR]{uu}{}       $&  $\wc[(1)]{qu}{}$ &1  \\
$ \wcL[V,LL]{dd}{}  $ &  $ \wc[(1)]{qq}{}\,, \wc[(3)]{qq}{}, \wc[(1)]{Hq}{}, 
 \wc[(3)]{Hq}{}$  &1,3 &
$\wcL[V8,LR]{uu}{}      $& $ \wc[(8)]{qu}{}$  &1  \\
$\wcL[V1,LL]{ud}{}      $& $\wc[(1)]{Hq}{},\,\wc[(3)]{Hq}{},\,\wc[(1)]{qq}{},\, \wc[(3)]{qq}{} $   &3 &
 $\wcL[V1,LR]{ud}{}      $& $\wc[]{Hd}{},\, \wc[(1)]{qd}{}$ &3 \\
$\wcL[V8,LL]{ud}{}  $& $\wc[(3)]{Hq}{},\wc[(3)]{qq}{} $ &3 &
$\wcL[V8,LR]{ud}{}      $& $\wc[(8)]{qd}{} $   &3    \\
$\wcL[V,RR]{uu}{}    $&  $\wc{uu}{}$ &1 & 
$\wcL[V1,LR]{du}{}     $& $\wc[(1)]{Hq}{},\,\wc[(3)]{Hq}{}, \wc[(1)]{qu}{}$ &3\\
$ \wcL[V,RR]{dd}{}     $& $\wc{dd}{}, \wc[]{Hd}{}$ &1,3  & 
$\wcL[V8,LR]{du}{}      $& $\wc[(8)]{qu}{}$ &3  \\
$\wcL[V1,RR]{ud}{}      $& $\wc[]{Hd}{}, \wc[(1)]{ud}{}$ &3  & 
$ \wcL[V1,LR]{dd}{}     $&     $\wc[]{Hd}{}, \wc[(1)]{qd}{}, \wc[(1)]{Hq}{},\,\wc[(3)]{Hq}{}$  &1,3,7 \\
$ \wcL[V8,RR]{ud}{}  $& $ \wc[(8)]{ud}{}$ &3  &
$ \wcL[V8,LR]{dd}{}  $&  $\wc[(8)]{qd}{}$  &1,3,7  \\
 $\wc[S1,RR]{uu}{}$  &0  & {1,7} &
$ \wc[V1,LR]{uddu}{}   $  & $\wc[]{Hud}{}$ &3,7  \\
 $\wc[S8,RR]{uu}{}$  &0  & {1,7} &
$ \wc[V8,LR]{uddu}{}  $& 0   & {3,7} \\
$\wc[S1,RR]{ud}{}   $& $\wc[(1)]{quqd}{}$  &3,7 &
$\wc[S8,RR]{ud}{}   $& $ \wc[(8)]{quqd}{}$  &3,7    \\
 $\wc[S1,RR]{dd}{}$    & 0 & {1,3,7}&
      $\wc[S8,RR]{dd}{}$     & 0 & {1,3,7}\\
$\wc[S1,RR]{uddu}{} $& $\wc[(1)]{quqd}{}$  &3,7 &
$\wc[S8,RR]{uddu}{}   $& $\wc[(8)]{quqd}{}$ &3,7 \\
\hline
\end{tabular}
\caption{{C}onnections of non-leptonic four-fermion WET WCs to SMEFT WCs {and various classes, at the tree-level}.} \label{tab:wetnonlepSMEFT}
\end{center}
\end{table}

\begin{table}[tbp]
\begin{center}
\renewcommand{\arraystretch}{1.3}
\begin{tabular}{|c|c|c|c|c|c|}
\hline
WET & SMEFT & Class & WET & SMEFT & Class\\
\hline
$\wcL[V,LL]{\nu\nu}{} $
& $\wc[]{ll}{}$&  5&
$\wcL[V,RR]{ee}{} $
&  $\wc[]{ee}{},\,   \wc[]{He}{}  $ & 5\\
$\wcL[V,LL]{ee}{} $ & $\wc[]{ll}{},\,\wc[(1)]{Hl}{} , \wc[(3)]{Hl}{}  $
& 5, 8 & $\wcL[V,LR]{\nu e}{} $ & $\wc[]{l e}{},   \wc[(1)]{Hl}{} , \wc[(3)]{Hl}{} $ &  8 \\
$\wcL[V,LL]{\nu e}{} $
& $\wc[]{ll}{}$, $\wc[(3)]{Hl}{}$ &  8
& $\wcL[V,LR]{ee}{}$  & $\wc[]{l e}{},  \wc[]{He}{}, \wc[(1)]{Hl}{} , \wc[(3)]{Hl}{}$ & 5,7,  \\
$\wcL[S,RR]{ee}{} $ & $\wc[]{eH}{}$\,& 5 &  & & \\
  \hline
\end{tabular}
\caption{Connections of leptonic four-fermion WET WCs to SMEFT WCs {and various classes, at the tree-level}.} \label{tab:wetlepSMEFT}
\end{center}
\end{table}

\begin{table}[tbp]
\begin{center}
\renewcommand{\arraystretch}{1.3}
\begin{tabular}{|c|c|c|c|c|c|}
\hline
WET & SMEFT & Class & WET & SMEFT & Class\\
\hline
$\wcL[]{u \gamma}{} $ & $\wc[]{u B}{}, \wc[]{uW}{}$& 7 &
$\wcL[]{d \gamma}{} $ &  $\wc[]{dB}{},\,   \wc[]{dW}{}  $ & 2,7\\
$\wcL[]{u G}{} $ & $\wc[]{uG}{} $&  7
& $\wcL[]{ dG }{} $ & $\wc[]{dG}{}$ &  2,7 \\
$\wcL[]{ e\gamma}{} $ & $\wc[]{eB}{}$, $\wc[]{eW}{}$ &  5,6,7
&   &   &  \\
  \hline
\end{tabular}
\caption{Connections of dipole WET WCs to SMEFT WCs and various classes, at the tree-level.} \label{tab:wetdipSMEFT}
\end{center}
\end{table}

Once a given set of relevant WCs is identified from the data one can find out, inspecting Tabs.~\ref{tab:cor-bosonic}-\ref{tab:cor-fermionic} and Tabs.~\ref{tab:mixing-bosonic-color}-\ref{tab:mixing-fermionic-color4} in conjunction with formulae for various NP scenarios in Sec.~\ref{sec:5}, which of the models has a chance to describe the data. In order to make this search more efficient it is useful to have also tables that list the particles which can generate a given WC at $\Lambda$, so that it is guaranteed that this operator will also contribute at the hadronic scale. However, one should be aware of the fact that due to the operator mixing a given coefficient can be generated also at the electroweak scale, even if it was vanishing at the NP scale. The charts presented in Part II and various formulae presented there demonstrate it in explicit terms. Yet, as demonstrated there, in most cases the WCs of such operators will be suppressed relatively to the leading ones that enter already at the tree-level. On the other hand one should not forget that in particular in the case of non-leptonic decays this suppression can be compensated and even overcompensated by the enhanced values of the corresponding hadronic matrix elements and also by large anomalous dimensions of certain operators, whose contributions can be significantly enhanced through RG evolution for a large NP scale $\Lambda$. This is in fact the case for left-right vector operators, as well as scalar and tensor operators.

The listings of particles which can generate a given WC at $\Lambda$ are given in Tabs.~\ref{tab:NP-bosonic-fermionic} and \ref{tab:NP-fermionic}. Here we have separated three cases 
\begin{align}
\wc[]{eH}{}:& \qquad  Z^\prime\,,  Z^\prime_1\,, W^\prime\,,  W^\prime_1\,, \phi_0\,, {S}\,, E\,, \Delta_1\,, \Delta_3\,, T_0\,,{T_e\,,}\\
\wc[]{uH}{}:& \qquad  Z^\prime\,, Z^\prime_1\,, W^\prime\,,  W^\prime_1\,, \phi_0\,, {S}\,, U\,, Q_V\,, Q_u\,, T_d\,, T_u\,,\\
\wc[]{dH}{}:& \qquad  Z^\prime\,, Z^\prime_1\,, W^\prime\,,  W^\prime_1\,, \phi_0\,, {S}\,, D\,, Q_V\,, Q_d\,, T_d\,, T_u\,,
\end{align}
which can be generated by numerous NP scenarios reviewed in Sec.~\ref{sec:5}.

Having Tabs.~\ref{tab:NP-bosonic-fermionic} and \ref{tab:NP-fermionic} at hand and inspecting the classes presented in Part II one could proceed as follows.

{\bf Step 1:}
For all classes of decays one can inspect for each observable which WCs are the most important for a given decay. These are in most cases the ones already present at the electroweak scale. Dependently on the decay considered various flavour indices will be involved. We just present the examples in Class 1, Class 2 and Class 5 in Tabs.~\ref{tab:WCs-class1}-\ref{tab:WCs-class5}. It is evident from these examples how to construct the corresponding tables for other classes.

{\bf Step 2:}
Once an anomaly in a given decay is found such tables give the first indication which WCs contributing to this decay could be  affected by NP.

{\bf Step 3:}
Inspecting then Tabs.~\ref{tab:NP-bosonic-fermionic} and \ref{tab:NP-fermionic} one can get hints which particles could be responsible for the observed anomaly.

{\bf Step 4:} This in turn would select NP models that could have a chance to explain the observed anomalies. For these models one could then perform a fully-fledged global fit analysis following the six steps of the top-down approach, described in Sec.~\ref{sec:GV}. This will finally allow to make predictions for a multitude of observables within this concrete model, including those that have been already measured and those for which only bounds exist. To this end various codes described in Sec.~\ref{SMEFTtools} are essential.

However, other avenues can be followed. In particular in the context of the top-down approach the DNA strategies of \cite{Buras:2013ooa} could select efficiently models on the basis of correlations and/or anti-correlations between different observables predicted in a given model. This can be done even at the qualitative level. For instance if the data implies enhancements of two observables compared to the SM predictions, but a given model implies the suppression of one of them in the presence of the enhancement of the other observable, then such a model is automatically ruled out and a global fit for it can be avoided.

In this context the $\vcb$-independent strategies developed in~\cite{Buras:2021nns,Buras:2022wpw,Buras:2022qip} could be used to obtain the first hints of the type of NP at work. The pattern of enhancements and suppressions of $\vcb$-independent ratios relative to their SM values and the correlations and/or anti-correlations between them in a given NP scenario, in the spirit of DNA strategies of \cite{Buras:2013ooa}, are powerful as recently stressed in \cite{Buras:2024per}. In particular suitable $\vcb$-independent ratios of the magnificent seven among the rare $K$ and $B$ decays
\be
\kpn,\quad \klpn,\quad K_S\to\mu^+\mu^-,\quad 
B^+\to K^+(K^*)\nu\bar\nu, \quad B_{s,d}\to\mu^+\mu^-,
\ee
being already precisely predicted within the SM, will be particularly helpful when their measurements will be improved in this decade.

A number of simplified models have been analyzed in \cite{Buras:2015yca} and also listed at the beginning of Section 16.1 in \cite{Buras:2020xsm}. In this context imposing flavour symmetries in the SMEFT, discussed in Sec.~\ref{Fsym}, could be helpful. 

\begin{table}[H]
\begin{center}
\renewcommand{\arraystretch}{1.3}
\begin{tabular}{|l|l|l|l|l|l|}
\hline
$C_i(\Lambda)$   &    New Particles  &    $C_i(\Lambda)$   &   New Particles  &    $C_i(\Lambda)$   &   New Particles           \\
\hline
\hline
$\wc[(1)]{H\ell}{}$      & ${ Z^\prime}$, $N$, $E$, $T_0$, $T_e$ & $\wc[]{eW}{}$    & $\Delta_1$, $T_e$  & {$\wc[]{eH}{}$} & $Z^\prime$, $Z^\prime_1$, $W^\prime$, $W^\prime_1$, $\phi_0$, $S$, \\ 
$\wc[(3)]{H\ell}{}$  & ${ W^\prime}$, $T_0$, $T_e$, $N$, $E$  & $\wc[]{eB}{}$    & $\Delta_1$, $E$  &  &  $E$, $\Delta_1$, $\Delta_3$, $T_0$, $T_e$  \\
$\wc[(1)]{Hq}{}$    & ${ Z^\prime}$, $D$, $U$, $T_u$, $T_d$ & $\wc[]{uG}{}$    & $U$, $Q_V$  & {$\wc[]{dH}{}$}  &  $Z^\prime$, $Z^\prime_1$, $W^\prime$, $W^\prime_1$, $\phi_0$, $S$, \\
$\wc[(3)]{Hq}{}$   & ${ W^\prime}$, $D$, $U$, $T_u$, $T_d$ &  $\wc[]{dG}{}$    & $D$, $Q_V$  &  & $D$, $Q_V$, $Q_d$, $T_d$, $T_u$  \\
$\wc[]{Hd}{}$    & ${ Z^\prime}$, $Q_V$, $Q_d$ & $\wc[]{dW}{}$    & $Q_V$, $T_d$  & {$\wc[]{uH}{}$} &  $Z^\prime$, $Z^\prime_1$, $W^\prime$, $W^\prime_1$, $\phi_0$, $S$, \\ 
$\wc[]{Hu}{}$    & ${ Z^\prime}$, $Q_V$, $Q_u$ &  $\wc[]{uW}{}$    & $Q_V$,  $T_u$  &  &   $U$, $Q_V$, $Q_u$, $T_d$, $T_u$
\\       
$\wc[]{Hud}{}$ &  $Z^\prime_1$, $Q_V$ &$\wc[]{uB}{}$    & $U$, $Q_V$  &  &   \\ 
$\wc[]{He}{}$  &  ${ Z^\prime}$, $\Delta_1$, $\Delta_3$ &
$\wc[]{dB}{}$    & $D$, $Q_V$  &  &   \\
\hline
\end{tabular}
\caption{New mediators generating a given Fermionic-Bosonic operator in the SMEFT.}
\label{tab:NP-bosonic-fermionic}
\end{center}
\end{table}

One should also observe that once a given model has been selected for a test, the values of the relevant $C_i(\Lambda)$ found using the technology of the previous section can then be used to find at least approximately the values of the couplings in a given model which could also be useful for a fully-fledged global fit analysis in this model. In this context the complete tree-level dictionary of this kind presented in \cite{deBlas:2017xtg} will certainly play an important role.

\begin{table}[H]
\begin{center}
\renewcommand{\arraystretch}{1.3}
\begin{tabular}{|l|l|l|l|}
\hline
$C_i(\Lambda)$   &  New Particles  &    $C_i(\Lambda)$   &   New Particles           \\
\hline
\hline
$\wc[]{H^4}{}$      &  $Z^\prime_1$, $W^\prime$, $W^\prime_1$, $\phi_0$ & $\wc[]{HB}{}$    & $\phi_0$    \\
$\wc[]{\nu\nu}{}$ & $N$, $T_0$ & $\wc[]{H\tilde B}{}$    & $\phi_0$    \\
$\wc[]{H}{}$      & $Z^\prime_1$, $W^\prime$, $W^\prime_1$, $\phi_0$, $S$  & $\wc[]{HW}{}$    & $\phi_0$    \\ 
$\wc[]{HD}{}$      & $Z^\prime$, $Z^\prime_1$, $W^\prime$, $W^\prime_1$ & $\wc[]{H\tilde W}{}$    & $\phi_0$    \\
$\wc[]{H\Box}{}$      & $Z^\prime$, $Z^\prime_1$, $W^\prime$, $W^\prime_1$, $\phi_0$ & $\wc[]{HG}{}$    & $\phi_0$   \\
& &  $\wc[]{H\tilde G}{}$    & $\phi_0$   \\
\hline
\end{tabular}
\caption{{New mediators generating the dim-4 Higgs self coupling $\wc[]{H^4}{}$, the dim-5 Weinberg WC $\wc[]{\nu\nu}{}$, or bosonic dim-6 operators in the SMEFT.}}
\label{tab:NP-bosonic-fermionic}
\end{center}
\end{table}

\begin{table}[H]
\begin{center}
\renewcommand{\arraystretch}{1.3}
\begin{tabular}{|l|l|l|l|l|l|}
\hline
$C_i(\Lambda)$   &    {New Particles}  & $C_i(\Lambda)$   &    {New Particles}     & $C_i(\Lambda)$   &    {New Particles}               \\
\hline
\hline
$\wc[]{\ell \ell}{}$  & ${ Z^\prime}$,  ${ W^\prime}$, $\phi_1$ &
$\wc[(1)]{qu}{}$    & ${ Z^\prime}$,  {$S$}, $\Phi$ &
$\wc[]{ee}{}$    & ${ Z^\prime}$,  $\phi_2$
\\
$\wc[(1)]{qq}{}$  & ${ Z^\prime}$,  ${ G^\prime}$ &
$\wc[(8)]{qu}{}$    & ${ G^\prime}$, {$S$},  $\Phi$ &
$\wc[]{uu}{}$    & ${ Z^\prime}$,  ${ G^\prime}$
\\
$\wc[(3)]{qq}{}$    & ${ W^\prime}$,  ${ G^\prime}$  &
$\wc[(1)]{qd}{}$    & ${ Z^\prime}$,  {$S$},  $\Phi$ &
$\wc[]{dd}{}$    & ${ Z^\prime}$,  ${ G^\prime}$\\
$\wc[(1)]{\ell q}{}$  & ${ Z^\prime}$,  $S_1$,  $S_3$,  $U_1$, $U_3$ &
$\wc[(8)]{qd}{}$    & ${ G^\prime}$, {$S$},  $\Phi$ &
$\wc[]{eu}{}$    & ${ Z^\prime}$\\
$\wc[(3)]{\ell q}{}$  & $W^\prime$, $S_1$,   $S_3$,  $U_1$,  $U_3$ &
$\wc[]{\ell e dq}{}$    & {$S$} &
$\wc[]{ed}{}$    & ${ Z^\prime}$ \\
$\wc[]{\ell e}{}$    & ${ Z^\prime}$, $S$ &
$\wc[(1)]{quqd}{}$    & {$S$} &
$\wc[(1)]{ud}{}$    & ${ Z^\prime}$,  $Z^\prime_1$, $G^\prime_1$ \\
$\wc[]{\ell u}{}$    & ${ Z^\prime}$ &
$\wc[(8)]{quqd}{}$    & $\Phi$ &
$\wc[(8)]{ud}{}$    & $Z^\prime_1$,  $G^\prime$, $G^\prime_1$\\
$\wc[]{\ell d}{}$    & $Z^\prime$,  $\tilde R_2$,  $V_2$ &
$\wc[(1)]{\ell e qu}{}$    & $S$,  & &
\\  
$\wc[]{qe}{}$  & ${ Z^\prime}$ & 
$\wc[(3)]{\ell e qu}{}$    &  & & \\ 
\hline
\end{tabular}
\caption{\small New mediators generating a given Four-Fermion operator in the SMEFT.}
\label{tab:NP-fermionic}
\end{center}
\end{table}

\begin{table}[H]
\begin{center}
\renewcommand{\arraystretch}{1.0}
\begin{tabular}{|l|l|}
\hline
Observable  &    Dominant WCs  \\
\hline
\hline
$\Delta M_s$, $S_{\psi\phi}$  & $\wc[(1)]{qq}{2323},\wc[(3)]{qq}{2323},\wc[(1)]{qd}{2323},\wc[(8)]{qd}{2323},\wc[]{dd}{2323}$ \\
$\Delta M_d$, $S_{\psi K_S}$  & $\wc[(1)]{qq}{1313},\wc[(3)]{qq}{1313},\wc[(1)]{qd}{1313},\wc[(8)]{qd}{1313},\wc[]{dd}{1313}$ \\
$\Delta M_K$, $\varepsilon_K$    & $\wc[(1)]{qq}{1212},\wc[(3)]{qq}{1212},\wc[(1)]{qd}{1212},\wc[(8)]{qd}{1212},\wc[]{dd}{1212}$ \\
$\Delta M_D$  &  $\wcup[(1)]{qq}{1212},\wcup[(3)]{qq}{1212},\wcup[(1)]{qu}{1212},\wcup[(8)]{qu}{1212},\wcup[]{uu}{1212}$ \\
\hline
\end{tabular}
\caption{\small Leading WCs for different observables in Class 1.}
\label{tab:WCs-class1}
\end{center}
\end{table}

\begin{table}[H]
\begin{center}
\renewcommand{\arraystretch}{1.0}
\begin{tabular}{|l|l|}
\hline
Observable  &    Dominant WCs    \\
\hline
\hline
$\kpn$,   $\klpn$   &$\wc[(1,3)]{l q}{pp12},\wc[]{l d}{pp12},\wc[(1,3)]{Hq}{12}, \wc[]{Hd}{12}$ \\
$B\to  K\nu\bar\nu$, $B\to  K^*\nu\bar\nu$  & $\wc[(1,3)]{l q}{pp23},\wc[]{l d}{pp23},\wc[(1,3)]{Hq}{23}, \wc[]{Hd}{23}$\\
$B\to K \ell^+\ell^-$, $B\to K^* \ell^+\ell^-$  &$\wc[(1,3)]{l q}{pp23},\wc[]{l d}{pp23},\wc[(1,3)]{Hq}{23}, \wc[]{Hd}{23}$, $\wc[]{ed}{pp23}$, $\wc[]{qe}{23pp}$ \\
$ K_{L}\to\mu^+\mu^-$,  $ K_{S}\to\mu^+\mu^-$     &$\wc[(1,3)]{l q}{2212},\wc[]{l d}{2212},\wc[(1,3)]{Hq}{12},\wc[]{Hd}{12}$, {$\wc[]{ed}{2212}$}, {$\wc[]{qe}{1222}$}\\
$B_{s}\to\mu^+\mu^- $ &$\wc[(1,3)]{l q}{2223},\wc[]{l d}{2223},\wc[(1,3)]{Hq}{23}, \wc[]{Hd}{23}$, {$\wc[]{ed}{2223}$}, {$\wc[]{qe}{2322}$}\\
$B_{d}\to\mu^+\mu^- $ &$\wc[(1,3)]{l q}{2213},\wc[]{l d}{2213}, \wc[(1,3)]{Hq}{13}, \wc[]{Hd}{13}$, {$\wc[]{ed}{2213}$}, {$\wc[]{qe}{1322}$}\\
$K_L\to \pi^0 \mu^+\mu^-$   &$\wc[(1,3)]{l q}{2212}, \wc[]{l d}{2212},\wc[(1,3)]{Hq}{12},\wc[]{Hd}{12}$, $\wc[]{ed}{2212}$, $\wc[]{qe}{1222}$\\
$K_L\to \pi^0 e^+e^-$ &$\wc[(1,3)]{l q}{1112}, \wc[]{l d}{1112}, \wc[(1,3)]{Hq}{12},\wc[]{Hd}{12}$, $\wc[]{ed}{1112}$, $\wc[]{qe}{1211}$ \\
\hline
\end{tabular}
\caption{\small Leading WCs for different observables in Class 2.}
\label{tab:WCs-class2}
\end{center}
\end{table}

\begin{table}[H]
\begin{center}
\renewcommand{\arraystretch}{1.0}
\begin{tabular}{|l|l|}
\hline
Observable  &    Dominant WCs    \\
\hline
\hline
$\mu^-\to e^-\gamma$  &$\wc[]{eW}{12}$, $\wc[]{eB}{12}$ \\
$\tau^-\to e^-\gamma$   & $\wc[]{eW}{13}$, $\wc[]{eB}{13}$,   \\
$\tau^-\to \mu^-\gamma$    &$\wc[]{eW}{23}$, $\wc[]{eB}{23}$ \\
$\mu^-\to e^-e^+e^-$  &$\wc[]{eW}{12}$, $\wc[]{eB}{12}$,
$\wc[(1)]{Hl}{12}$,$\wc[(3)]{Hl}{12}$, $\wc[]{He}{12}$, $\wc[]{ll}{1112}$,
$\wc[]{ee}{1112}$, $\wc[]{le}{1112}$, $\wc[]{le}{1211}$
\\
$\tau^-\to \mu^-\mu^+\mu^-$ &$\wc[]{eW}{23}$, $\wc[]{eB}{23}$,
$\wc[(1)]{Hl}{23}$,$\wc[(3)]{Hl}{23}$, $\wc[]{He}{23}$, $\wc[]{ll}{2223}$,
$\wc[]{ee}{2223}$, $\wc[]{le}{2223}$, $\wc[]{le}{2322}$
\\
$\tau^-\to e^-e^+e^-$& $\wc[]{eW}{13}$, $\wc[]{eB}{13}$,
$\wc[(1)]{Hl}{13}$,$\wc[(3)]{Hl}{13}$,
$\wc[]{He}{13}$, $\wc[]{ll}{1113}$,
$\wc[]{ee}{1113}$, $\wc[]{le}{1113}$,  $\wc[]{le}{1311}$
\\
$\tau^-\to \mu^-e^+e^-$ &$\wc[]{eW}{23}$, $\wc[]{eB}{23}$,
$\wc[(1)]{Hl}{23}$,$\wc[(3)]{Hl}{23}$,
$\wc[]{He}{23}$, $\wc[]{ll}{1123}$, {$\wc[]{ll}{2113}$,}
$\wc[]{ee}{1123}$, $\wc[]{le}{1123}$, {$\wc[]{le}{2113}$,}  $\wc[]{le}{2311}$
\\
$\tau^-\to e^-\mu^+\mu^-$&$\wc[]{eW}{13}$, $\wc[]{eB}{13}$,
$\wc[(1)]{Hl}{13}$,$\wc[(3)]{Hl}{13}$,
$\wc[]{He}{13}$, $\wc[]{ll}{2213}$, {$\wc[]{ll}{1223}$,}
$\wc[]{ee}{2213}$, $\wc[]{le}{2213}$, {$\wc[]{le}{1223}$,} $\wc[]{le}{1322}$
\\
\hline
\end{tabular}
\caption{\small Leading WCs for different observables in Class 5.}
\label{tab:WCs-class5}
\end{center}
\end{table}

\section{Summary and Outlook} \label{sec:7}

The SMEFT will undoubtedly play a very important role in the search for phenomena beyond the SM and is the framework in which  footprints of NP in the landscape between the electroweak scale and the lowest scale of some UV completion should be analyzed. Our review collects insights on the SMEFT and WET gained over the last decade by numerous authors with the main goal to exhibit various virtues of this approach in a user-friendly manner. While, eventually, due to the presence of very many operators one cannot avoid numerical codes, we think it is essential to get a clear picture of the correlations between various observables implied not only by the full unbroken SM gauge symmetry but in particular by the renormalization group evolution and the related operator mixing characteristic for a given NP scenario. These correlations not only depend on possible flavour symmetries but also on the nature of the particles involved, such as scalars, fermions and vector bosons.

We hope that the numerous tables and in particular the RG charts will facilitate decoding the hints of NP, hidden  in the enormous landscape above the electroweak scale. In order to analyze them it is not only crucial to use efficiently the technology described in our review but in particular to have a multitude of precise measurements of various observables and precise theory predictions not only within the SM but also within various UV completions that we collected in our review. We expect significant progress in this direction in the current and the following decade.

Most importantly we are looking forward to a broad spectrum of experiments in the rest of this decade and in the next decade which will give us a new insight into the physics of flavour and CP violation. In particular, we will witness new results on $B_s$, $B_d$ and $\Lambda_b$ decays at LHCb \cite{LHCb:2018roe} and Belle II \cite{Belle-II:2018jsg}, on $K$ decays from  NA62 and KOTO \cite{NA62KLEVER:2022nea,Goudzovski:2022scl,Aebischer:2025mwl}, on charm decays from BESIII \cite{BESIII:2020nme} and STCF \cite{Lyu:2021tlb}, searches for lepton flavor violation, like at Mu2e~\cite{Bernstein:2019fyh}, Mu3e \cite{Hesketh:2022wgw} and MEG II \cite{MEGII:2023ltw}, and also searches for electric dipole moments \cite{n2EDM:2021yah,Alarcon:2022ero,Athanasakis-Kaklamanakis:2025xcg}.
Finally, the opportunities in flavour physics at the FCC \cite{FCC:2025lpp}, HL-LHC and HE-LHC \cite{Cerri:2018ypt,CidVidal:2018eel} should be emphasized.

\vspace{0.4cm}
{\bf Acknowledgements} 

We would like to thank a number of our colleagues for informative discussions during the writing of our review. These are in particular Gerhard Buchalla, Emanuele Mereghetti and Peter Stangl. The work of J.A. is supported by the European Union's Horizon 2020 research and innovation program under the Marie Skłodowska-Curie grant agreement No. 101145975 - EFT-NLO. Financial support of A.J.B by the Excellence Cluster ORIGINS, funded by the Deutsche Forschungsgemeinschaft (DFG, German Research Foundation), Excellence Strategy, EXC-2094, 390783311 is acknowledged. The work of J.K. was supported by the US Department of Energy Office and by the Laboratory Directed Research and Development (LDRD) program of Los Alamos National Laboratory under project numbers 20230047DR, 20220706PRD1. Los Alamos National Laboratory is operated by Triad National Security, LLC, for the National Nuclear Security Administration of the U.S. Department of Energy (Contract No. 89233218CNA000001). J.K. acknowledges the Physical Research Laboratory, Ahmedabad, where this work was finalized.

\appendix
\section{The JMS Basis}
\label{app:jmsbasis}
In this appendix we report the operator basis of the WET in the JMS basis \cite{Jenkins:2017jig}. We will restrict the discussion to operators relevant for this review. For the complete basis, including bosonic and baryon number violating operators we refer to Appendix B of \cite{Jenkins:2017jig}. The EOMs of the WET including corrections due to higher dimensional operators and a discussion on the associated symmetry currents can be found in \cite{Helset:2018dht}.

\subsection{Dipole operators} \label{app:wetdipole}

The dim-5 WET dipole operators have the following general form
\be
\mathcal{O}_{\psi X} = \bar \psi_{Lp} \sigma^{\mu \nu} \psi_{Rr} X_{\mu \nu}\,,
\ee
with the fields $\psi= e, u, d$ and $X= F_{\mu \nu}$ for $e$ and $F_{\mu \nu}$ or $G_{\mu \nu}$ for $u$ and $d$. The subscripts $p,r$ refer to the flavour indices. In the quark case $\text{SU(3)}_C$ generators have to be added to the fermion current.

\subsection{Four-fermion operators}

In this subsection we collect the dimension-six four-fermion operators. The semileptonic operators are given in Tab.~\ref{tab:wetsemilep}, the four-quark operators in Tab.~\ref{tab:wetnonlep} and the four-lepton operators in Tab.~\ref{tab:wetlep}.

\begin{table}[H]
\begin{center}
\renewcommand{\arraystretch}{1.3}
\begin{tabular}{|cc|cc|}
\hline

$\op{\nu u}{V}{LL}$       & $(\bar \nu_{Lp} \gamma^\mu \nu_{Lr}) (\bar u_{Ls}  \gamma_\mu u_{Lt})$ &
$\op{\nu u}{V}{LR}$         & $(\bar \nu_{Lp} \gamma^\mu \nu_{Lr})(\bar u_{Rs}  \gamma_\mu u_{Rt})$  \\
$\op{\nu d}{V}{LL}$       & $(\bar \nu_{Lp} \gamma^\mu \nu_{Lr})(\bar d_{Ls} \gamma_\mu d_{Lt})$   &
$\op{\nu d}{V}{LR}$         & $(\bar \nu_{Lp} \gamma^\mu \nu_{Lr})(\bar d_{Rs} \gamma_\mu d_{Rt})$  \\
$\op{eu}{V}{LL}$      & $(\bar e_{Lp}  \gamma^\mu e_{Lr})(\bar u_{Ls} \gamma_\mu u_{Lt})$ &
$\op{eu}{V}{LR}$        & $(\bar e_{Lp}  \gamma^\mu e_{Lr})(\bar u_{Rs} \gamma_\mu u_{Rt}) $   \\
$\op{ed}{V}{LL}$       & $(\bar e_{Lp}  \gamma^\mu e_{Lr})(\bar d_{Ls} \gamma_\mu d_{Lt})$  &
$\op{ed}{V}{LR}$        & $(\bar e_{Lp}  \gamma^\mu e_{Lr})(\bar d_{Rs} \gamma_\mu d_{Rt}) $\\
$\op{\nu edu}{V}{LL}$      & $(\bar \nu_{Lp} \gamma^\mu e_{Lr}) (\bar d_{Ls} \gamma_\mu u_{Lt})  + \hc$ &
$\op{ue}{V}{LR}$        & $(\bar u_{Lp} \gamma^\mu u_{Lr})(\bar e_{Rs}  \gamma_\mu e_{Rt}) $   \\
$\op{eu}{V}{RR}$       & $(\bar e_{Rp}  \gamma^\mu e_{Rr})(\bar u_{Rs} \gamma_\mu u_{Rt})$  &
$\op{de}{V}{LR}$         & $(\bar d_{Lp} \gamma^\mu d_{Lr}) (\bar e_{Rs} \gamma_\mu e_{Rt}) $ \\
$\op{ed}{V}{RR}$     & $(\bar e_{Rp} \gamma^\mu e_{Rr})  (\bar d_{Rs} \gamma_\mu d_{Rt})$&
$\op{\nu edu}{V}{LR}$        & $(\bar \nu_{Lp} \gamma^\mu e_{Lr})(\bar d_{Rs} \gamma_\mu u_{Rt})  +\hc$   \\
$\op{eu}{S}{RR}$  & $(\bar e_{Lp}   e_{Rr}) (\bar u_{Ls} u_{Rt})$ &
$\op{eu}{S}{RL}$  & $(\bar e_{Lp} e_{Rr}) (\bar u_{Rs}  u_{Lt})$   \\
$\op{eu}{T}{RR}$ & $(\bar e_{Lp}   \sigma^{\mu \nu}   e_{Rr}) (\bar u_{Ls}  \sigma_{\mu \nu}  u_{Rt})$ &
$\op{ed}{S}{RL}$ & $(\bar e_{Lp} e_{Rr}) (\bar d_{Rs} d_{Lt}) $ \\
$\op{ed}{S}{RR}$  & $(\bar e_{Lp} e_{Rr})(\bar d_{Ls} d_{Rt})$  &
$\op{\nu edu}{S}{RL}$  & $(\bar \nu_{Lp} e_{Rr}) (\bar d_{Rs}  u_{Lt})$  \\
$\op{ed}{T}{RR}$ & $(\bar e_{Lp} \sigma^{\mu \nu} e_{Rr}) (\bar d_{Ls} \sigma_{\mu \nu} d_{Rt})$ &&  \\
$\op{\nu edu}{S}{RR}$ & $(\bar   \nu_{Lp} e_{Rr})  (\bar d_{Ls} u_{Rt} )$ && \\
$\op{\nu edu}{T}{RR}$ &  $(\bar  \nu_{Lp}  \sigma^{\mu \nu} e_{Rr} )  (\bar  d_{Ls}  \sigma_{\mu \nu} u_{Rt} )$ &&   \\
\hline
\end{tabular}
\caption{Semileptonic {WET operators} in the JMS basis.} \label{tab:wetsemilep}
\end{center}
\end{table}

\begin{table}[H]
\begin{center}
\renewcommand{\arraystretch}{1.3}
\begin{tabular}{|cc|cc|}
\hline

$\op{uu}{V}{LL}        $&$ (\bar u_{Lp} \gamma^\mu u_{Lr})(\bar u_{Ls} \gamma_\mu u_{Lt}) $&
$\op{uu}{V1}{LR}        $&$ (\bar u_{Lp} \gamma^\mu u_{Lr})(\bar u_{Rs} \gamma_\mu u_{Rt})$    \\
$\op{dd}{V}{LL}   $&$ (\bar d_{Lp} \gamma^\mu d_{Lr})(\bar d_{Ls} \gamma_\mu d_{Lt})$  &
$\op{uu}{V8}{LR}       $&$ (\bar u_{Lp} \gamma^\mu T^A u_{Lr})(\bar u_{Rs} \gamma_\mu T^A u_{Rt})$   \\
$\op{ud}{V1}{LL}     $&$ (\bar u_{Lp} \gamma^\mu u_{Lr}) (\bar d_{Ls} \gamma_\mu d_{Lt})$ &
$\op{ud}{V1}{LR}       $&$ (\bar u_{Lp} \gamma^\mu u_{Lr}) (\bar d_{Rs} \gamma_\mu d_{Rt})$ \\
$\op{ud}{V8}{LL}     $&$ (\bar u_{Lp} \gamma^\mu T^A u_{Lr}) (\bar d_{Ls} \gamma_\mu T^A d_{Lt})$&
$\op{ud}{V8}{LR}       $&$ (\bar u_{Lp} \gamma^\mu T^A u_{Lr})  (\bar d_{Rs} \gamma_\mu T^A d_{Rt})$    \\
$\op{uu}{V}{RR}      $&$ (\bar u_{Rp} \gamma^\mu u_{Rr})(\bar u_{Rs} \gamma_\mu u_{Rt})$ &
$\op{du}{V1}{LR}       $&$ (\bar d_{Lp} \gamma^\mu d_{Lr})(\bar u_{Rs} \gamma_\mu u_{Rt})$ \\
$\op{dd}{V}{RR}      $&$ (\bar d_{Rp} \gamma^\mu d_{Rr})(\bar d_{Rs} \gamma_\mu d_{Rt}) $ &
$\op{du}{V8}{LR}       $&$ (\bar d_{Lp} \gamma^\mu T^A d_{Lr})(\bar u_{Rs} \gamma_\mu T^A u_{Rt})$  \\
$\op{ud}{V1}{RR}       $&$ (\bar u_{Rp} \gamma^\mu u_{Rr}) (\bar d_{Rs} \gamma_\mu d_{Rt})$ &
$\op{dd}{V1}{LR}      $&$ (\bar d_{Lp} \gamma^\mu d_{Lr})(\bar d_{Rs} \gamma_\mu d_{Rt})$ \\
$\op{ud}{V8}{RR}    $&$ (\bar u_{Rp} \gamma^\mu T^A u_{Rr}) (\bar d_{Rs} \gamma_\mu T^A d_{Rt})$ &
$\op{dd}{V8}{LR}   $&$ (\bar d_{Lp} \gamma^\mu T^A d_{Lr})(\bar d_{Rs} \gamma_\mu T^A d_{Rt})$ \\
$\op{uu}{S1}{RR}  $&$ (\bar u_{Lp}   u_{Rr}) (\bar u_{Ls} u_{Rt})$   &
$\op{uddu}{V1}{LR}   $&$ (\bar u_{Lp} \gamma^\mu d_{Lr})(\bar d_{Rs} \gamma_\mu u_{Rt})  + \hc$  \\
$\op{uu}{S8}{RR}   $&$ (\bar u_{Lp}   T^A u_{Rr}) (\bar u_{Ls} T^A u_{Rt})$ &
$\op{uddu}{V8}{LR}      $&$ (\bar u_{Lp} \gamma^\mu T^A d_{Lr})(\bar d_{Rs} \gamma_\mu T^A  u_{Rt})  + \hc$   \\
$\op{ud}{S1}{RR}   $&$ (\bar u_{Lp} u_{Rr})  (\bar d_{Ls} d_{Rt}) $ &
$\op{ud}{S8}{RR}  $&$ (\bar u_{Lp} T^A u_{Rr})  (\bar d_{Ls} T^A d_{Rt})$   \\
$\op{dd}{S1}{RR}   $&$ (\bar d_{Lp} d_{Rr}) (\bar d_{Ls} d_{Rt})$&
$\op{dd}{S8}{RR}  $&$ (\bar d_{Lp} T^A d_{Rr}) (\bar d_{Ls} T^A d_{Rt})$ \\
$\op{uddu}{S1}{RR} $&$  (\bar u_{Lp} d_{Rr}) (\bar d_{Ls}  u_{Rt})$&
$\op{uddu}{S8}{RR}  $&$  (\bar u_{Lp} T^A d_{Rr}) (\bar d_{Ls}  T^A u_{Rt})$\\
  \hline
\end{tabular}
\caption{Non-leptonic WET operators in the JMS basis.} \label{tab:wetnonlep}
\end{center}
\end{table}

\begin{table}[H]
\begin{center}
\renewcommand{\arraystretch}{1.3}
\begin{tabular}{|cc|cc|}
\hline
 $\op{\nu\nu}{V}{LL}$& $(\bar \nu_{Lp}  \gamma^\mu \nu_{Lr} )(\bar \nu_{Ls} \gamma_\mu \nu_{Lt})$
& $\op{ee}{V}{RR}$ & $(\bar e_{Rp} \gamma^\mu e_{Rr})(\bar e_{Rs} \gamma_\mu e_{Rt})$   \\
$ \op{ee}{V}{LL}$& $(\bar e_{Lp}  \gamma^\mu e_{Lr})(\bar e_{Ls} \gamma_\mu e_{Lt})$
& $\op{\nu e}{V}{LR}$ & $(\bar \nu_{Lp} \gamma^\mu \nu_{Lr})(\bar e_{Rs}  \gamma_\mu e_{Rt})$  \\
  $\op{\nu e}{V}{LL}$& $(\bar \nu_{Lp} \gamma^\mu \nu_{Lr})(\bar e_{Ls}  \gamma_\mu e_{Lt})$
& $\op{ee}{V}{LR}$ &  $(\bar e_{Lp}  \gamma^\mu e_{Lr})(\bar e_{Rs} \gamma_\mu e_{Rt})$  \\
$\op{ee}{S}{RR}$ & $(\bar e_{Lp}   e_{Rr}) (\bar e_{Ls} e_{Rt})$  &  & \\
  \hline
\end{tabular}
\caption{Leptonic WET operators in the JMS basis.} \label{tab:wetlep}
\end{center}
\end{table}

\section{The Warsaw Basis}\label{Warsawbasis}
In this appendix we introduce the notation for the Warsaw basis \cite{Grzadkowski:2010es} up to mass-dimension six.

\boldmath
\subsection{Weinberg operator}
\unboldmath
The only gauge-invariant dimension-five operator involving the SM fields is the Weinberg operator, which was introduced for the first time by Steven Weinberg in \cite{Weinberg:1979sa}. It involves two Higgs fields and two lepton doublets and has the form
\begin{equation}
  \Op[]{\nu \nu} = (\tilde H^\dagger l_p)^T C ( \tilde H^\dagger l_r) \,,
\end{equation}
where $C$ denotes the charge-conjugation operator and $\tilde H$ is the conjugate Higgs-doublet. After EW symmetry breaking this operator contributes to the Majorana mass term of neutrinos.

\subsection{Dimension-six operators}
The dimension-six bosonic and semi-bosonic SMEFT operators are given in Tab.~\ref{tab:no4fermsmeft} and the four-fermi operators are collected in Tab.~\ref{tab:4fermsmeft}.

\begin{table}[H]
\centering
\renewcommand{\arraystretch}{1.0}
{\begin{tabular}{||c|c||c|c||c|c||}
\hline \hline
\multicolumn{2}{||c||}{$X^3$} &
\multicolumn{2}{|c||}{$\vp^6$~ and~ $\vp^4 D^2$} &
\multicolumn{2}{|c||}{$f^2\vp^3$}\\
\hline
${\cal O}_G$                & $f^{ABC} G_\mu^{A\nu} G_\nu^{B\rho} G_\rho^{C\mu} $ &
${\cal O}_\vp$       & $(\vp^\dag\vp)^3$ &
${\cal O}_{e\vp}$           & $(\vp^\dag \vp)(\bar l_p e_r \vp)$\\
${\cal O}_{\wt G}$          & $f^{ABC} \wt G_\mu^{A\nu} G_\nu^{B\rho} G_\rho^{C\mu} $ &
${\cal O}_{\vp\Box}$ & $(\vp^\dag \vp)\raisebox{-.5mm}{$\Box$}(\vp^\dag \vp)$ &
${\cal O}_{u\vp}$           & $(\vp^\dag \vp)(\bar q_p u_r \tvp)$\\
${\cal O}_W$                & $\eps^{IJK} W_\mu^{I\nu} W_\nu^{J\rho} W_\rho^{K\mu}$ &
${\cal O}_{\vp D}$   & $\left(\vp^\dag D^\mu\vp\right)^\star \left(\vp^\dag D_\mu\vp\right)$ &
${\cal O}_{d\vp}$           & $(\vp^\dag \vp)(\bar q_p d_r \vp)$\\
${\cal O}_{\wt W}$          & $\eps^{IJK} \wt W_\mu^{I\nu} W_\nu^{J\rho} W_\rho^{K\mu}$ &&&&\\
\hline \hline
\multicolumn{2}{||c||}{$X^2\vp^2$} &
\multicolumn{2}{|c||}{$f^2 X\vp$} &
\multicolumn{2}{|c||}{$f^2\vp^2 D$}\\
\hline
${\cal O}_{\vp G}$     & $\vp^\dag \vp\, G^A_{\mu\nu} G^{A\mu\nu}$ &
${\cal O}_{eW}$               & $(\bar \ell_p \sigma^{\mu\nu} e_r) \tau^I \vp W_{\mu\nu}^I$ &
${\cal O}_{\vp l}^{(1)}$      & $(\vpj)(\bar \ell_p \gamma^\mu \ell_r)$\\
${\cal O}_{\vp\wt G}$         & $\vp^\dag \vp\, \wt G^A_{\mu\nu} G^{A\mu\nu}$ &
${\cal O}_{eB}$        & $(\bar \ell_p \sigma^{\mu\nu} e_r) \vp B_{\mu\nu}$ &
${\cal O}_{\vp l}^{(3)}$      & $(\vpjt)(\bar \ell_p \tau^I \gamma^\mu \ell_r)$\\
${\cal O}_{\vp W}$     & $\vp^\dag \vp\, W^I_{\mu\nu} W^{I\mu\nu}$ &
${\cal O}_{uG}$        & $(\bar q_p \sigma^{\mu\nu} T^A u_r) \tvp\, G_{\mu\nu}^A$ &
${\cal O}_{\vp e}$            & $(\vpj)(\bar e_p \gamma^\mu e_r)$\\
${\cal O}_{\vp\wt W}$         & $\vp^\dag \vp\, \wt W^I_{\mu\nu} W^{I\mu\nu}$ &
${\cal O}_{uW}$               & $(\bar q_p \sigma^{\mu\nu} u_r) \tau^I \tvp\, W_{\mu\nu}^I$ &
${\cal O}_{\vp q}^{(1)}$      & $(\vpj)(\bar q_p \gamma^\mu q_r)$\\
${\cal O}_{\vp B}$     & $ \vp^\dag \vp\, B_{\mu\nu} B^{\mu\nu}$ &
${\cal O}_{uB}$        & $(\bar q_p \sigma^{\mu\nu} u_r) \tvp\, B_{\mu\nu}$&
${\cal O}_{\vp q}^{(3)}$      & $(\vpjt)(\bar q_p \tau^I \gamma^\mu q_r)$\\
${\cal O}_{\vp\wt B}$         & $\vp^\dag \vp\, \wt B_{\mu\nu} B^{\mu\nu}$ &
${\cal O}_{dG}$        & $(\bar q_p \sigma^{\mu\nu} T^A d_r) \vp\, G_{\mu\nu}^A$ &
${\cal O}_{\vp u}$            & $(\vpj)(\bar u_p \gamma^\mu u_r)$\\
${\cal O}_{\vp WB}$     & $ \vp^\dag \tau^I \vp\, W^I_{\mu\nu} B^{\mu\nu}$ &
${\cal O}_{dW}$               & $(\bar q_p \sigma^{\mu\nu} d_r) \tau^I \vp\, W_{\mu\nu}^I$ &
${\cal O}_{\vp d}$            & $(\vpj)(\bar d_p \gamma^\mu d_r)$\\
${\cal O}_{\vp\wt WB}$ & $\vp^\dag \tau^I \vp\, \wt W^I_{\mu\nu} B^{\mu\nu}$ &
${\cal O}_{dB}$        & $(\bar q_p \sigma^{\mu\nu} d_r) \vp\, B_{\mu\nu}$ &
${\cal O}_{\vp u d}$   & $i(\tvp^\dag D_\mu \vp)(\bar u_p \gamma^\mu d_r)$\\
\hline \hline
\end{tabular}}
\caption{Dimension-six bosonic and semi-bosonic SMEFT operators.\label{tab:no4fermsmeft}}
\end{table}

\begin{table}[H]
\begin{center}
\renewcommand{\arraystretch}{1.0}
{\begin{tabular}{||c|c||c|c||c|c||}
\hline\hline
\multicolumn{2}{||c||}{$(\bar LL)(\bar LL)$} &
\multicolumn{2}{|c||}{$(\bar RR)(\bar RR)$} &
\multicolumn{2}{|c||}{$(\bar LL)(\bar RR)$}\\
\hline
${\cal O}_{ll}$        & $(\bar \ell_p \gamma_\mu \ell_r)(\bar \ell_s \gamma^\mu \ell_t)$ &
${\cal O}_{ee}$               & $(\bar e_p \gamma_\mu e_r)(\bar e_s \gamma^\mu e_t)$ &
${\cal O}_{le}$               & $(\bar \ell_p \gamma_\mu \ell_r)(\bar e_s \gamma^\mu e_t)$ \\
${\cal O}_{qq}^{(1)}$  & $(\bar q_p \gamma_\mu q_r)(\bar q_s \gamma^\mu q_t)$ &
${\cal O}_{uu}$        & $(\bar u_p \gamma_\mu u_r)(\bar u_s \gamma^\mu u_t)$ &
${\cal O}_{lu}$               & $(\bar \ell_p \gamma_\mu \ell_r)(\bar u_s \gamma^\mu u_t)$ \\
${\cal O}_{qq}^{(3)}$  & $(\bar q_p \gamma_\mu \tau^I q_r)(\bar q_s \gamma^\mu \tau^I q_t)$ &
${\cal O}_{dd}$        & $(\bar d_p \gamma_\mu d_r)(\bar d_s \gamma^\mu d_t)$ &
${\cal O}_{ld}$               & $(\bar \ell_p \gamma_\mu \ell_r)(\bar d_s \gamma^\mu d_t)$ \\
${\cal O}_{lq}^{(1)}$                & $(\bar \ell_p \gamma_\mu \ell_r)(\bar q_s \gamma^\mu q_t)$ &
${\cal O}_{eu}$                      & $(\bar e_p \gamma_\mu e_r)(\bar u_s \gamma^\mu u_t)$ &
${\cal O}_{qe}$               & $(\bar q_p \gamma_\mu q_r)(\bar e_s \gamma^\mu e_t)$ \\
${\cal O}_{lq}^{(3)}$                & $(\bar \ell_p \gamma_\mu \tau^I \ell_r)(\bar q_s \gamma^\mu \tau^I q_t)$ &
${\cal O}_{ed}$                      & $(\bar e_p \gamma_\mu e_r)(\bar d_s\gamma^\mu d_t)$ &
${\cal O}_{qu}^{(1)}$         & $(\bar q_p \gamma_\mu q_r)(\bar u_s \gamma^\mu u_t)$ \\
&&
${\cal O}_{ud}^{(1)}$                & $(\bar u_p \gamma_\mu u_r)(\bar d_s \gamma^\mu d_t)$ &
${\cal O}_{qu}^{(8)}$         & $(\bar q_p \gamma_\mu T^A q_r)(\bar u_s \gamma^\mu T^A u_t)$ \\
&&
${\cal O}_{ud}^{(8)}$                & $(\bar u_p \gamma_\mu T^A u_r)(\bar d_s \gamma^\mu T^A d_t)$ &
${\cal O}_{qd}^{(1)}$ & $(\bar q_p \gamma_\mu q_r)(\bar d_s \gamma^\mu d_t)$ \\
&&&&
${\cal O}_{qd}^{(8)}$ & $(\bar q_p \gamma_\mu T^A q_r)(\bar d_s \gamma^\mu T^A d_t)$\\
\hline\hline
\multicolumn{2}{||c||}{$(\bar LR)(\bar RL)$ and $(\bar LR)(\bar LR)$} &
\multicolumn{4}{|c||}{$B$-violating}\\\hline
${\cal O}_{ledq}$ & $(\bar \ell_p^j e_r)(\bar d_s q_t^j)$ &
${\cal O}_{duq}$ & \multicolumn{3}{|c||}{$\eps^{\alpha\beta\gamma} \eps_{jk}
 \left[ (d^\alpha_p)^T C u^\beta_r \right]\left[(q^{\gamma j}_s)^T C \ell^k_t\right]$}\\
${\cal O}_{quqd}^{(1)}$ & $(\bar q_p^j u_r) \eps_{jk} (\bar q_s^k d_t)$ &
${\cal O}_{qqu}$ & \multicolumn{3}{|c||}{$\eps^{\alpha\beta\gamma} \eps_{jk}
  \left[ (q^{\alpha j}_p)^T C q^{\beta k}_r \right]\left[(u^\gamma_s)^T C e_t\right]$}\\
${\cal O}_{quqd}^{(8)}$ & $(\bar q_p^j T^A u_r) \eps_{jk} (\bar q_s^k T^A d_t)$ &
${\cal O}_{qqq}$ & \multicolumn{3}{|c||}{$\eps^{\alpha\beta\gamma} \eps_{jn} \eps_{km}
  \left[ (q^{\alpha j}_p)^T C q^{\beta k}_r \right]\left[(q^{\gamma m}_s)^T C l^n_t\right]$}\\
${\cal O}_{lequ}^{(1)}$ & $(\bar \ell_p^j e_r) \eps_{jk} (\bar q_s^k u_t)$ & ${\cal O}_{duu}$
     & \multicolumn{3}{|c||}{$\eps^{\alpha\beta\gamma}
  \left[ (d^\alpha_p)^T C u^\beta_r \right]\left[(u^\gamma_s)^T C e_t\right]$}\\
${\cal O}_{lequ}^{(3)}$ & $(\bar \ell_p^j \sigma_{\mu\nu} e_r) \eps_{jk} (\bar q_s^k \sigma^{\mu\nu} u_t)$ &
& \multicolumn{3}{|c||}{}\\
\hline\hline
\end{tabular}}
\caption{Four-fermion SMEFT operators. \label{tab:4fermsmeft}}
\end{center}
\end{table}

\section{Extensions and Alternative Formulations of the SMEFT and WET}

In this appendix we give a concise overview over studied extensions and alternative descriptions of the SMEFT and WET.

\subsection{Scalar extensions}

\subsubsection{ALP-SMEFT and ALP-WET}
There are several studies in the literature of the SMEFT and WET extended with axion-like particles (ALPs), which are motivated due to their ability to solve the strong CP problem. The full one-loop running up to dim-5 operators was performed for the first time in \cite{Chala:2020wvs}. The results were also obtained in a more general context in \cite{DasBakshi:2023lca}, without imposing a shift-symmetry. Furthermore, the RGEs were computed using amplitude methods in \cite{Bresciani:2024shu}. Higher-order loop corrections in both the matching and running were then considered in \cite{Bauer:2020jbp}. Baryon number violating nucleon decays involving ALPs were studied in \cite{Fan:2025xhi}. A phenomenological study of this theory including collider constraints was carried out in \cite{Bonilla:2021ufe,Biekotter:2025fll}. Finally, the impact of ALP running effects on the stability of the EW vacuum as well as on various SM couplings was recently studied in \cite{Galda:2025tqs}.

\subsection{$\phi$SMEFT}
Monojet and direct detection constraints of the SMEFT with an additional scalar singlet extension $\phi$ that acts as a dark matter candidate were studied in \cite{Roy:2025pht}.

\subsubsection{2HDM-SMEFT}
SMEFT-type effective field theories with two Higgs doublets can be found in \cite{Crivellin:2016ihg,Karmakar:2017yek,Anisha:2019nzx,Dermisek:2024ohe}. The complete list of Feynman rules for type I-III 2HDMs was recently worked out in \cite{Dermisek:2024btn}.

\subsubsection{Real triplet}
The operator $O_{HWB}$ in the SMEFT, as well as an extension of the SMEFT with a real scalar triplet in the context of Abelian-Non-Abelian Kinetic Mixing was studied in \cite{Tran:2024srm}.

\subsection{Dark SMEFT and WET}
The SMEFT and WET can be extended by adding additional spin 0, 1/2 and 1 dark matter (DM) particles. Such dark matter effective field theories were presented in \cite{Aebischer:2022wnl} under the assumption of singlet DM particles, and in \cite{Criado:2021trs}, assuming the DM particles to be $\text{SU(2)}_L$ multiplets. In \cite{Aebischer:2022wnl}, in addition to the classification of all gauge invariant interactions in the Lagrangian up to terms of dimension six, the tree-level matching conditions between the two theories at the electroweak scale were worked out. Generalizations up to dim-7 and dim-8 were presented in \cite{Liang:2023yta} and \cite{Song:2023jqm}, respectively. Finally, a study of top EFTs that goes beyond the SMEFT, containing light degrees of freedom is described in \cite{Lessa:2023tqc}.

\subsection{$\nu$SMEFT}

The $\nu$SMEFT is the SMEFT, augmented with right-handed neutrinos. It is a subset of the Dark SMEFT presented in \cite{Aebischer:2022wnl}, where only spin 1/2 singlets are added to the SM field content. The operator basis of the $\nu$SMEFT was derived in \cite{Liao:2016qyd} up to dim-7. The gauge-coupling dependence of the one-loop ADM is given in \cite{Datta:2020ocb}, and the Yukawa dependence for four-fermi operators was first derived in \cite{Datta:2021akg}. The complete RGEs of the $\nu$SMEFT were finally presented in \cite{Ardu:2024tzb}. The two-loop ADM in the $\nu$SMEFT up to dim-5 was computed very recently in \cite{Zhang:2025ywe}.

Possible UV completions for the $\nu$SMEFT are discussed for instance in \cite{Beltran:2023ymm} for dim-6 and dim-7 operators. Recently in \cite{Colangelo:2024sbf} the constraints on the WCs of the $\nu$SMEFT enlarged by a $\text{U(1)}^\prime$ symmetry were derived, in particular those from gauge anomaly cancellation. In \cite{Grojean:2024qdm} a classification of the Jarlskog-like flavor invariants entering CP-violating observables was performed. The authors find 459 invariants, out of which 208 are CP-even and 251 are CP-odd. The one-loop matching from a model with right-handed neutrinos, vector-like fermions and a new scalar onto the $\nu$SMEFT is given in \cite{Chala:2020vqp}, together with partial results for the one-loop ADM of the $\nu$WET. Furthermore, a geometric description of the theory, described by the geo$\nu$SMEFT is presented in \cite{Talbert:2022unj}. 

Concerning phenomenology, a recent analysis of possible constraints from FCC-ee can be found in \cite{Duarte:2025zrg,Bolton:2025tqw} and in \cite{Zapata:2023wsz} the sensitivity to collider processes like lepton number (and flavour) violating decays involving jets (j), represented by the processes $pe^-\to \mu^\pm +3j$ are discussed. The authors of \cite{Cirigliano:2021peb} investigate the letpon anomalous magnetic moment at the one- and two-loop level in the $\nu$SMEFT, and the observables $\mathcal{B}(B\to K^{(*)}\nu\nu)$ were analyzed recently in \cite{Rosauro-Alcaraz:2024mvx}. The sensitivity of COHERENT-like experiments to non-standard interactions within the $\nu$WET was studied in \cite{Breso-Pla:2025pds}. The implications from the $\nu$SMEFT on leptogensis are studied in \cite{Fuyuto:2025feh}. The sensitivity of $W$-boson measurements to low-mass right-handed neutrinos is discussed in \cite{Alonso:2025gzl}. Baryon number violating decays at dim-6 and dim-7 in the $\nu$SMEFT were studied in \cite{Li:2025slp}. Finally, sterile neutrino dark matter in the context of the $\nu$SMEFT was discussed in \cite{Fuyuto:2024oii}.

\subsection{The Higgs Effective Field Theory (HEFT)}

An effective Lagrangian for a spontaneously broken gauge theory with three eaten Goldstone bosons and one additional neutral scalar particle offers the most general low-energy description of the scalar particle detected in 2012 at the LHC. The only implicit assumption of this HEFT framework \cite{Feruglio:1992wf,Grinstein:2007iv} is that there are no other light states in the few hundred GeV range, that couple to SM particles. An important special case of the HEFT is the SMEFT, where the scalar fields transform linearly as a complex scalar doublet.

A very compact introduction to the phenomenological motivation for this framework can be found in \cite{Buchalla:2015wfa} and an introductory, review-style account is contained in Sec. II.2.4 of the CERN Yellow report \cite{LHCHiggsCrossSectionWorkingGroup:2016ypw}.

Over the years there have been a number of developments in the HEFT that, similar to chiral perturbation theory, studied $v^2/\Lambda^2$ and higher order contributions, deriving various power-counting formulas and presenting complete sets of operators at leading and next-to-leading order in the chiral expansion \cite{Buchalla:2012qq,Alonso:2012px,Buchalla:2013rka,Buchalla:2013eza,Gavela:2016bzc,Buchalla:2016sop}. The complete one-loop renormalization of the Higgs-electroweak chiral Lagrangian, which is equivalent to the HEFT Lagrangian, was derived in \cite{Buchalla:2017jlu,Alonso:2017tdy}.

A geometric formulation of HEFT was presented in \cite{Alonso:2015fsp,Alonso:2016btr,Alonso:2016oah}. This approach was adopted in \cite{Cohen:2020xca}, to identify UV theories that match solely onto the HEFT. It was concluded, that the SMEFT might often not be enough to capture all NP effects from UV theories.

The question then arises how to distinguish the SMEFT from the HEFT. Here we summarize a number of papers related to this issue.

An early analysis in which a fit to Higgs data in the HEFT was performed is given in \cite{Buchalla:2015qju}. The distinction between the SMEFT and HEFT was studied for Higgs self-interactions in \cite{Falkowski:2019tft}, for double and triple Higgs production via vector boson fusion in \cite{Anisha:2024ryj,Domenech:2025gmn} and gluon fusion using bootstrap methods in \cite{Grober:2025vse}, and for $2\to 4$ vector boson scattering in \cite{Mahmud:2025wye}. A possible falsification of the SMEFT via multi-Higgs measurements was studied in \cite{Gomez-Ambrosio:2022why}, and CP-violating Higgs couplings to gauge bosons in the HEFT/SMEFT and their effect on Higgs and di-Higgs production were studied in \cite{Bhardwaj:2024lyr}. Higgs production in the Regge limit was studied at the two-loop level in \cite{DelDuca:2025vux}. A distinction between the SMFET and HEFT using unitarity, analyticity and geometry is discussed in \cite{Cohen:2021ucp} in the context of unitarity violation. A critical examination of Di-Higgs production via gluon fusion can be found in \cite{Cadamuro:2025car}, and differences of the SMEFT and HEFT in predicting vector-pair production at colliders were outlined in \cite{Goldberg:2024eot}. Possible distinctions of the SMEFT and HEFT$\backslash $SMEFT from cosmology and LHC constraints were discussed in \cite{Alonso:2023jsi}. Differences in SMEFT and HEFT phenomenology concerning the muon $(g-2)$ were studied in \cite{Fortuna:2024rqp}. Mappings from the HEFT to the SMEFT and vice versa were studied in the bundle formalism in \cite{Alminawi:2023qtf}, and various aspects of the HEFT (and SMEFT) like the relation between perturbative unitarity and gauge-invariance are discussed in \cite{Liu:2023jbq}, using amplitude methods.

Furthermore, there are several analyses that go beyond the HEFT, which we will mention in the following: The matching of different UV models onto the HEFT was considered in \cite{Dawson:2023oce}, and the matching of the real Higgs triplet extension onto the HEFT is performed in \cite{Song:2025kjp}. NLO positivity bounds in the HEFT, resulting in the 15-dimensional allowed region called HEFT-hedron were studied in \cite{Chakraborty:2024ciu}. General scalar extensions in the HEFT were recently considered in \cite{Song:2024kos}. Finally, a novel framework called broken phase effective field theory (bEFT) was presented in \cite{Liao:2025bmn}, which allows to describe UV completions with Higgs mixing or new particles with a mass close to the EW scale, that goes beyond the SMEFT/HEFT applicability.

\subsection{GeoSMEFT}

The GeoSMEFT is the geometric interpretation of the SMEFT, of which a detailed review can be found for instance in \cite{Helset:2020yio}.\footnote{A general discussion on the geometry of EFTs can be found in \cite{Cohen:2024bml}.} The renormalization of the bosonic SMEFT sector up to dim-8 was performed in \cite{Helset:2022pde} using geometric arguments, and the inclusion of fermions in the geometric picture is discussed in \cite{Assi:2023zid}, together with the renormalization of SMEFT dim-8 operators. Furthermore, two-fermion operators were renormalized in \cite{Assi:2025fsm}, using supergeometry invariants. Dirac masses and mixings, as well as CP-violating phases in this framework were studied in \cite{Talbert:2021iqn}. Finally, general properties of scattering amplitudes for arbitrary scalar theories from a geometric viewpoint were studied in \cite{Cheung:2021yog}, and one-loop effects on the curvature of the field-space manifold were investigated for scalar theories in \cite{Aigner:2025xyt}.

\newpage

\section{\boldmath More on Matching SMEFT to WET}\label{app:WETmatching}

At the EW scale one has to go one step down in the EFT ladder by matching the SMEFT onto the WET. This is achieved by integrating out the heavy particles such as the $W$ and $Z$ bosons, the Higgs boson and the top-quark from the SMEFT spectrum. In addition to the tree-level matching, the one-loop matching effects are important for several reasons as discussed already in Sec.~\ref{sec:smeft-beyond-leading} and in other sections of our review:

\begin{enumerate}
\item Certain SMEFT operators can be mapped onto the WET only through one-loop matching and not via one-loop RG evolution and tree-level matching. 
\item For the SMEFT operators giving simultaneous contributions via one-loop matching and running, the inclusion of the former helps in eliminating the uncertainties due to the choice of matching scale $\muEW$. 
\item Inclusion of the one-loop matching is also necessary to cancel the renormalization scheme dependence present in the two-loop ADMs.
\end{enumerate}
Furthermore, depending upon the observables under consideration, the matching contributions can be numerically important for suitable values of $\Lambda$. The inclusion of the matching effects for specific processes was already studied in \cite{Aebischer:2015fzz} (for $\Delta F=1 $ processes) and \cite{Bobeth:2017xry} (for $\Delta F=2$ processes), thanks to recent results \cite{Dekens:2019ept} where the matching conditions have been extended to the full set of operators at the one-loop level.  

The issue of the renormalization scheme dependence and its removal through one-loop matching has already been discussed in general terms in Sec.~\ref{RSDEP} and cancellation of the scale dependence in Sec.~\ref{SCALEDEP}. Here we want to present two concrete examples to stress in particular the first two points above in explicit terms.

For the cases in which both running and matching effects are present, naively the former is expected to dominate because of log enhancement. In those cases the matching effects could be subdominant corrections to the overall effect. The relevance of these corrections must be assessed case by case because some observables can be sensitive to small effects. There are however cases of SMEFT operators which cannot contribute via running but only through matching. In such cases, the matching obviously becomes important.

As an example, we consider the contribution of the $\wc[]{Hud}{3i}$ WC to the WC of the dipole operator, $\wcL[]{d\gamma}{ii}$ in the WET. This can happen only through the finite part and not the operator mixing in the SMEFT. Specifically, we have the matching condition 
\be
\begin{aligned}
\wcL[]{d\gamma}{ii}(\muEW)  =g_1 c_w V^\dagger_{i3}   \wc[]{Hud}{3i}(\muEW) \sqrt{x_t} { F_1 (x_t) \over \pi^2},
\end{aligned}
\ee
where $x_t={m_t^2}/M_W^2$. Here the loop function is given by
\be
\begin{aligned}
F_1(x_t,\muEW) & = 
{  7x_t^2 + 7 x_t - 8   \over 192 (x_t-1)^2  }
- {3 x_t^2 - 2 x_t \over 32 (x_t-1)^3    } 
\ln x_t  \,.
\end{aligned}
\ee
Note that in the WET $\wc[]{Hud}{3i}$ matches onto VLR four fermion operators at tree-level and in this manner can additionally contribute to the EDMs using two-loop running \cite{Kumar:2024yuu}. But the above mentioned effect due to the finite parts is qualitatively and quantitatively different from the two-loop running effect. 

Concerning $\muEW$ dependence, its worth mentioning an example discussed in \cite{Bobeth:2017xry} for the $\Delta F=2$ operators. There is an operator mixing between $\wc[]{Hd}{ij}$ and $\wc[(1)]{qd}{ijij}$ which originates from the divergent parts of the matrix elements. So the combination of the 1stLL RG running and the tree-level matching of $\wc[(1)]{qd}{ijij}$ to the low energy $\Delta F=2$ operator  
\be 
{G_F^2 \over 4 \pi^2} M_W^2 \wc[ij]{LR,1}{}  (\lambda_t^{ij})^2 (\bar d_i \gamma_\mu P_L d_j ) 
(\bar d_i \gamma^\mu P_R d_j)\,,
\ee
{is given by}
\be
[\wc[ij]{LR,1}{}(\muEW)]_{\textrm{1st LL}} = \wc[]{Hd}{ij}(\Lambda) {v^2 \over \lambda_t^{ij}  }
x_t \ln {\Lambda \over \muEW}\,.
\ee
Adding the matching contribution resulting from the finite parts leads to \cite{ Bobeth:2017xry} 
\be \label{CLR1NLO}
[\wc[ij]{LR,1}{}(\muEW)]_{\textrm{NLO}} = \wc[]{Hd}{ij}(\Lambda) {v^2 \over \lambda_t^{ij}  }
x_t \left [ \ln {\Lambda \over \muEW}  + H_1(x_t, \muEW)   \right ]\,.
\ee
Here $H_1(x_t, \muEW)$ is the loop function given by
\be
H_1(x_t,\muEW) = \ln {\muEW \over M_W} - { x_t-7 \over 4 (x_t-1)  }
- {x_t^2-2x_t+4 \over 2 (x_t-1)^2} \ln x_t  \, .
\ee
The dependence on the matching scale $\muEW$ is then canceled by the terms proportional to $\ln {\Lambda / \muEW}$ (divergent part) and $\ln {\muEW / M_W}$ (finite part). Note that in \eqref{CLR1NLO} we have ignored self-running of $\wc[]{Hd}{}$ from $\Lambda$ to $\muEW$. Quantitatively, in this case the matching effects are of the order of $15-30\%$ for $x_t \simeq 4$ and  {$\Lambda=1-10$} TeV.

\section{$\rho$ and $\eta$ Parameters for Various Classes}\label{App:etas}
In this appendix, we present the values of the $\rho$ and $\eta$ parameters for various classes. These parameters, defined in Eq.~\eqref{ABKfinal}, are dimensionless in general. However, for certain cases, such as Class 2, we adopt the CWET basis for their definition, resulting in parameters with mass dimension $[\mathrm{TeV}^{2}]$. For simplicity, we truncate the $\rho$ and $\eta$ values below a class specific threshold. If needed, smaller values can be easily computed using publicly available tools such as {\tt wilson}. Its worth reminding that the $\eta$ factors do not contain log-enhancement factors. Further, these are computed using 1stLL solutions and by definition are independent of NP scale $\Lambda$.

\begin{table}\centering\begin{tabular}{|cc|cc|ccc|}\toprule $C^{\rm WET}_i(C^{\rm SMEFT}_ m)$ & size & $C^{\rm WET}_i(C^{\rm SMEFT}_ m)$ & size & $C^{\rm WET}_i(C^{\rm SMEFT}_ m)$ & size& \\ \hline
$\wcL[V,LL]{dd}{2323} [\wc[(1)]{qq}{2323}]$&$1$&
$\wcL[V,LL]{dd}{2323} [\wc[(3)]{qq}{2323}]$&$1$&
$\wcL[V,LL]{dd}{2323} (\wc[(3)]{qq}{2333})$&$-3\cdot 10^{-4}$&
\\ 
$\wcL[V,LL]{dd}{2323} (\wc[(1)]{qq}{2333})$&$-3\cdot 10^{-4}$&
$\wcL[V,LL]{dd}{2323} (\wc[(1)]{qu}{2333})$&$2\cdot 10^{-4}$&
$\wcL[V,LL]{dd}{2323} (\wc[(1)]{Hq}{23})$&$-2\cdot 10^{-4}$&
\\ 
$\wcL[V,LL]{dd}{2323} (\wc[(3)]{Hq}{23})$&$2\cdot 10^{-4}$&
$\wcL[V,LL]{dd}{2323} (\wc[(3)]{qq}{2223})$&$1\cdot 10^{-4}$&
$\wcL[V,LL]{dd}{2323} (\wc[(1)]{qq}{2223})$&$1\cdot 10^{-4}$&
\\ 
$\wcL[V,LL]{dd}{2323} (\wc[(8)]{qu}{2333})$&$7\cdot 10^{-5}$&
$\wcL[V,LL]{dd}{2323} (\wc[(3)]{qq}{1232})$&$-3\cdot 10^{-5}$&
$\wcL[V,LL]{dd}{2323} (\wc[(1)]{qq}{1232})$&$-3\cdot 10^{-5}$&
\\ 
$\wcL[V,LL]{dd}{2323} (\wc[(1)]{qu}{2323})$&$-2\cdot 10^{-5}$&
$\wcL[V,LL]{dd}{2323} (\wc[(8)]{qu}{2323})$&$-6\cdot 10^{-6}$&
$\wcL[V,RR]{dd}{2323} [\wc[]{dd}{2323}]$&$1$&
\\ 
$\wcL[V,RR]{dd}{2323} (\wc[]{dd}{2223})$&$1\cdot 10^{-5}$&
$\wcL[V,RR]{dd}{2323} (\wc[]{dd}{2333})$&$-1\cdot 10^{-5}$&
$\wcL[V1,LR]{dd}{2323} [\wc[(1)]{qd}{2323}]$&$1$&
\\ 
$\wcL[V1,LR]{dd}{2323} (\wc[(8)]{qd}{2323})$&$-0.02$&
$\wcL[V1,LR]{dd}{2323} (\wc[(1)]{qd}{3323})$&$-4\cdot 10^{-4}$&
$\wcL[V1,LR]{dd}{2323} (\wc[(1)]{ud}{3323})$&$2\cdot 10^{-4}$&
\\ 
$\wcL[V1,LR]{dd}{2323} (\wc[]{Hd}{23})$&$-2\cdot 10^{-4}$&
$\wcL[V1,LR]{dd}{2323} (\wc[(1)]{qd}{2223})$&$2\cdot 10^{-4}$&
$\wcL[V1,LR]{dd}{2323} (\wc[(1)]{qd}{1232})$&$-3\cdot 10^{-5}$&
\\ 
$\wcL[V1,LR]{dd}{2323} (\wc[(1)]{ud}{2323})$&$-2\cdot 10^{-5}$&
$\wcL[V1,LR]{dd}{2323} (\wc[(8)]{qd}{3323})$&$-1\cdot 10^{-5}$&
$\wcL[V1,LR]{dd}{2323} (\wc[(8)]{qd}{2223})$&$1\cdot 10^{-5}$&
\\ 
$\wcL[V1,LR]{dd}{2323} (\wc[(1)]{qd}{2322})$&$6\cdot 10^{-6}$&
$\wcL[V1,LR]{dd}{2323} (\wc[(1)]{qd}{2333})$&$-6\cdot 10^{-6}$&
$\wcL[V8,LR]{dd}{2323} [\wc[(8)]{qd}{2323}]$&$1$&
\\ 
$\wcL[V8,LR]{dd}{2323} (\wc[(1)]{qd}{2323})$&$-0.09$&
$\wcL[V8,LR]{dd}{2323} (\wc[(8)]{qd}{3323})$&$-4\cdot 10^{-4}$&
$\wcL[V8,LR]{dd}{2323} (\wc[(8)]{qd}{2223})$&$2\cdot 10^{-4}$&
\\ 
$\wcL[V8,LR]{dd}{2323} (\wc[(8)]{ud}{3323})$&$2\cdot 10^{-4}$&
$\wcL[V8,LR]{dd}{2323} (\wc[(1)]{qd}{2223})$&$5\cdot 10^{-5}$&
$\wcL[V8,LR]{dd}{2323} (\wc[(1)]{qd}{3323})$&$-5\cdot 10^{-5}$&
\\ 
$\wcL[V8,LR]{dd}{2323} (\wc[(8)]{qd}{1232})$&$-5\cdot 10^{-5}$&
$\wcL[V8,LR]{dd}{2323} (\wc[(8)]{ud}{2323})$&$-2\cdot 10^{-5}$&
$\wcL[V8,LR]{dd}{2323} (\wc[(1)]{qd}{1232})$&$-1\cdot 10^{-5}$&
\\ 
$\wcL[V8,LR]{dd}{2323} (\wc[(1)]{quqd}{3332})$&$-7\cdot 10^{-6}$&
$\wcL[V8,LR]{dd}{2323} (\wc[(8)]{qd}{2322})$&$6\cdot 10^{-6}$&
$\wcL[V8,LR]{dd}{2323} (\wc[(8)]{qd}{2333})$&$-5\cdot 10^{-6}$&
\\ 
\bottomrule \end{tabular} \caption{The $\rho$ and $\eta$-parameters for Class 1, relevant for $B_s$-mixing.} 
\label{tab:etaclass1Bs} \end{table}

\begin{table}\centering\begin{tabular}{|cc|cc|ccc|}\toprule $C^{\rm WET}_i(C^{\rm SMEFT}_ m)$ & size & $C^{\rm WET}_i(C^{\rm SMEFT}_ m)$ & size & $C^{\rm WET}_i(C^{\rm SMEFT}_ m)$ & size& \\ \hline
$\wcL[V,LL]{dd}{1212} [\wc[(1)]{qq}{1212}]$&$1$&
$\wcL[V,LL]{dd}{1212} [\wc[(3)]{qq}{1212}]$&$1$&
$\wcL[V,LL]{dd}{1212} (\wc[(1)]{qq}{1213})$&$-3\cdot 10^{-4}$&
\\ 
$\wcL[V,LL]{dd}{1212} (\wc[(3)]{qq}{1213})$&$-3\cdot 10^{-4}$&
$\wcL[V,LL]{dd}{1212} (\wc[(3)]{qq}{1232})$&$7\cdot 10^{-5}$&
$\wcL[V,LL]{dd}{1212} (\wc[(1)]{qq}{1232})$&$7\cdot 10^{-5}$&
\\ 
$\wcL[V,RR]{dd}{1212} [\wc[]{dd}{1212}]$&$1$&
$\wcL[V,RR]{dd}{1212} (\wc[]{dd}{1213})$&$-1\cdot 10^{-5}$&
$\wcL[V1,LR]{dd}{1212} [\wc[(1)]{qd}{1212}]$&$1$&
\\ 
$\wcL[V1,LR]{dd}{1212} (\wc[(8)]{qd}{1212})$&$-0.02$&
$\wcL[V1,LR]{dd}{1212} (\wc[(1)]{qd}{1312})$&$-4\cdot 10^{-4}$&
$\wcL[V1,LR]{dd}{1212} (\wc[(1)]{qd}{2321})$&$7\cdot 10^{-5}$&
\\ 
$\wcL[V1,LR]{dd}{1212} (\wc[(8)]{qd}{1312})$&$-1\cdot 10^{-5}$&
$\wcL[V1,LR]{dd}{1212} (\wc[(1)]{qd}{1222})$&$1\cdot 10^{-5}$&
$\wcL[V1,LR]{dd}{1212} (\wc[(1)]{qd}{1211})$&$-1\cdot 10^{-5}$&
\\ 
$\wcL[V1,LR]{dd}{1212} (\wc[(1)]{qd}{1213})$&$-1\cdot 10^{-5}$&
$\wcL[V1,LR]{dd}{1212} (\wc[(1)]{qd}{1112})$&$-9\cdot 10^{-6}$&
$\wcL[V1,LR]{dd}{1212} (\wc[(1)]{qd}{2212})$&$8\cdot 10^{-6}$&
\\ 
$\wcL[V8,LR]{dd}{1212} [\wc[(8)]{qd}{1212}]$&$1$&
$\wcL[V8,LR]{dd}{1212} (\wc[(1)]{qd}{1212})$&$-0.09$&
$\wcL[V8,LR]{dd}{1212} (\wc[(8)]{qd}{1312})$&$-4\cdot 10^{-4}$&
\\ 
$\wcL[V8,LR]{dd}{1212} (\wc[(8)]{qd}{2321})$&$9\cdot 10^{-5}$&
$\wcL[V8,LR]{dd}{1212} (\wc[(1)]{qd}{1312})$&$-5\cdot 10^{-5}$&
$\wcL[V8,LR]{dd}{1212} (\wc[(8)]{qd}{1222})$&$2\cdot 10^{-5}$&
\\ 
$\wcL[V8,LR]{dd}{1212} (\wc[(8)]{qd}{1211})$&$-2\cdot 10^{-5}$&
$\wcL[V8,LR]{dd}{1212} (\wc[(8)]{qd}{1112})$&$-2\cdot 10^{-5}$&
$\wcL[V8,LR]{dd}{1212} (\wc[(8)]{qd}{2212})$&$1\cdot 10^{-5}$&
\\ 
$\wcL[V8,LR]{dd}{1212} (\wc[(8)]{qd}{1213})$&$-1\cdot 10^{-5}$&
$\wcL[V8,LR]{dd}{1212} (\wc[(1)]{qd}{2321})$&$1\cdot 10^{-5}$&
&&\\
\bottomrule \end{tabular} \caption{The $\rho$ and $\eta$-parameters for Class 1, relevant for Kaon mixing.} \label{tab:etaclass1K} \end{table}

\begin{table}[H]
\centering\begin{tabular}{|cc|cc|ccc|}\toprule $C^{\rm CWET}_i(C^{\rm SMEFT}_ m)$ & size & $C^{\rm CWET}_i(C^{\rm SMEFT}_ m)$ & size & $C^{\rm CWET}_i(C^{\rm SMEFT}_ m)$ & size &\\ \hline

$C_{9bs}^{\mu\mu} [\wc[(1)]{lq}{2223}]$&$-589$&
$C_{9bs}^{\mu\mu} [\wc[(3)]{lq}{2223}]$&$-589$&
$C_{9bs}^{\mu\mu} [\wc[]{qe}{2322}]$&$-589$&
\\ 
$C_{9bs}^{\mu\mu} [\wc[(1)]{Hq}{23}]$&$46$&
$C_{9bs}^{\mu\mu} [\wc[(3)]{Hq}{23}]$&$46$&
$C_{9bs}^{\mu\mu} (\wc[(3)]{qq}{1123})$&$-3$&
\\ 
$C_{9bs}^{\mu\mu} (\wc[(3)]{qq}{2333})$&$-2$&
$C_{9bs}^{\mu\mu} (\wc[(1)]{qu}{2311})$&$2$&
$C_{9bs}^{\mu\mu} (\wc[(1)]{qu}{2322})$&$2$&
\\ 
$C_{9bs}^{\mu\mu} (\wc[(1)]{qq}{2333})$&$2$&
$C_{9bs}^{\mu\mu} (\wc[(3)]{qq}{2223})$&$-2$&
$C_{9bs}^{\mu\mu} (\wc[(1)]{qq}{1123})$&$1$&
\\ 
$C_{9bs}^{\mu\mu} (\wc[(1)]{lq}{3323})$&$-1$&
$C_{9bs}^{\mu\mu} (\wc[(1)]{lq}{1123})$&$-1$&
$C_{9bs}^{\mu\mu} (\wc[]{qe}{2311})$&$-1$&
\\ 
$C_{9bs}^{\mu\mu} (\wc[(1)]{qd}{2311})$&$-1$&
$C_{9bs}^{\mu\mu} (\wc[(1)]{qd}{2322})$&$-1$&
$C_{9bs}^{\mu\mu} (\wc[]{qe}{2333})$&$-1$&
\\ 
$C_{9bs}^{\mu\mu} (\wc[(1)]{qd}{2333})$&$-1$&
$C_{10bs}^{\mu\mu} [\wc[(1)]{Hq}{23}]$&$-589$&
$C_{10bs}^{\mu\mu} [\wc[(3)]{Hq}{23}]$&$-589$&
\\ 
$C_{10bs}^{\mu\mu} [\wc[(1)]{lq}{2223}]$&$589$&
$C_{10bs}^{\mu\mu} [\wc[(3)]{lq}{2223}]$&$589$&
$C_{10bs}^{\mu\mu} [\wc[]{qe}{2322}]$&$-589$&
\\ 
$C_{10bs}^{\mu\mu} (\wc[(1)]{qu}{2333})$&$17$&
$C_{10bs}^{\mu\mu} (\wc[(1)]{qq}{2333})$&$-17$&
$C_{10bs}^{\mu\mu} (\wc[(3)]{qq}{2333})$&$6$&
\\ 
$C_{10bs}^{\mu\mu} (\wc[(1)]{qq}{2323})$&$1$&
$C_{9bs}^{\prime \mu\mu} [\wc[]{ed}{2223}]$&$-589$&
$C_{9bs}^{\prime \mu\mu} [\wc[]{ld}{2223}]$&$-589$&
\\ 
$C_{9bs}^{\prime \mu\mu} [\wc[]{Hd}{23}]$&$46$&
$C_{9bs}^{\prime \mu\mu} (\wc[(1)]{qd}{3323})$&$2$&
$C_{9bs}^{\prime \mu\mu} (\wc[(1)]{ud}{1123})$&$2$&
\\ 
$C_{9bs}^{\prime \mu\mu} (\wc[(1)]{ud}{2223})$&$2$&
$C_{9bs}^{\prime \mu\mu} (\wc[]{dd}{2223})$&$-1$&
$C_{9bs}^{\prime \mu\mu} (\wc[]{dd}{2333})$&$-1$&
\\ 
$C_{9bs}^{\prime \mu\mu} (\wc[(1)]{qd}{2223})$&$1$&
$C_{9bs}^{\prime \mu\mu} (\wc[(1)]{qd}{1123})$&$1$&
$C_{9bs}^{\prime \mu\mu} (\wc[]{ld}{3323})$&$-1$&
\\ 
$C_{9bs}^{\prime \mu\mu} (\wc[]{ld}{1123})$&$-1$&
$C_{9bs}^{\prime \mu\mu} (\wc[]{ed}{1123})$&$-1$&
$C_{9bs}^{\prime \mu\mu} (\wc[]{dd}{1123})$&$-1$&
\\ 
$C_{9bs}^{\prime \mu\mu} (\wc[]{ed}{3323})$&$-1$&
$C_{10bs}^{\prime \mu\mu} [\wc[]{ed}{2223}]$&$-589$&
$C_{10bs}^{\prime \mu\mu} [\wc[]{ld}{2223}]$&$589$&
\\ 
$C_{10bs}^{\prime \mu\mu} [\wc[]{Hd}{23}]$&$-589$&
$C_{10bs}^{\prime \mu\mu} (\wc[(1)]{ud}{3323})$&$17$&
$C_{10bs}^{\prime \mu\mu} (\wc[(1)]{qd}{3323})$&$-17$&
\\ 
\bottomrule \end{tabular}
\caption{Part 1 of the $\rho$ and $\eta$-parameters (in units of TeV$^{2}$) for Class 2A, relevant for $b\to s \mu\mu$. All entries having absolute value below one have been suppressed.}  
\label{tab:2A1}
\end{table}

\begin{table}[H]
\centering
\begin{tabular}{|cc|cc|ccc|}\toprule $C^{\rm CWET}_i(C^{\rm SMEFT}_ m)$ & size & $C^{\rm CWET}_i(C^{\rm SMEFT}_ m)$ & size & $C^{\rm CWET}_i(C^{\rm SMEFT}_ m)$ & size & \\ \hline
$C_{Lbs}^{\nu_\mu\nu_\mu} [\wc[(1)]{Hq}{23}]$&$-589$&
$C_{Lbs}^{\nu_\mu\nu_\mu} [\wc[(3)]{Hq}{23}]$&$-589$&
$C_{Lbs}^{\nu_\mu\nu_\mu} [\wc[(1)]{lq}{2223}]$&$-589$&
\\ 
$C_{Lbs}^{\nu_\mu\nu_\mu} [\wc[(3)]{lq}{2223}]$&$589$&
$C_{Lbs}^{\nu_\mu\nu_\mu} (\wc[(1)]{qu}{2333})$&$17$&
$C_{Lbs}^{\nu_\mu\nu_\mu} (\wc[(1)]{qq}{2333})$&$-17$&
\\ 
$C_{Lbs}^{\nu_\mu\nu_\mu} (\wc[(3)]{qq}{2333})$&$6$&
$C_{Lbs}^{\nu_\mu\nu_\mu} (\wc[(1)]{qq}{2323})$&$1$&
$C_{Rbs}^{\nu_\mu\nu_\mu} [\wc[]{Hd}{23}]$&$-589$&
\\ 
$C_{Rbs}^{\nu_\mu\nu_\mu} [\wc[]{ld}{2223}]$&$-589$&
$C_{Rbs}^{\nu_\mu\nu_\mu} (\wc[(1)]{ud}{3323})$&$17$&
$C_{Rbs}^{\nu_\mu\nu_\mu} (\wc[(1)]{qd}{3323})$&$-17$&
\\ 
\bottomrule
\end{tabular}
\caption{Part 2 of the $\rho$ and $\eta$-parameters (in units of TeV$^{2}$) for Class 2A, relevant for $b\to s \nu_\mu \bar \nu_\mu$. All entries having absolute value below one have been suppressed.}  
\label{tab:2A2} 
\end{table}

\begin{table}[H]\centering\begin{tabular}{|cc|cc|ccc|}\toprule $C^{\rm WET}_i(C^{\rm SMEFT}_ m)$ & size & $C^{\rm WET}_i(C^{\rm SMEFT}_ m)$ & size & $C^{\rm WET}_i(C^{\rm SMEFT}_ m)$ & size &\\ \hline
$\wcL[V,LL]{ed}{2212} [\wc[(1)]{lq}{2212}]$&$1$&
$\wcL[V,LL]{ed}{2212} [\wc[(3)]{lq}{2212}]$&$1$&
$\wcL[V,LL]{ed}{2212} [\wc[(1)]{Hq}{12}]$&$-0.5$&
\\ 
$\wcL[V,LL]{ed}{2212} [\wc[(3)]{Hq}{12}]$&$-0.5$&
$\wcL[V,LL]{ed}{2212} (\wc[(3)]{qq}{1233})$&$0.02$&
$\wcL[V,LL]{ed}{2212} (\wc[(1)]{qq}{1233})$&$-0.02$&
\\ 
$\wcL[V,LL]{ed}{2212} (\wc[(1)]{qu}{1233})$&$0.01$&
$\wcL[V,LL]{ed}{2212} (\wc[(3)]{qq}{1332})$&$-0.01$&
$\wcL[V,LL]{ed}{2212} (\wc[(1)]{qu}{1211})$&$-2\cdot 10^{-3}$&
\\ 
$\wcL[V,LL]{ed}{2212} (\wc[(1)]{qu}{1222})$&$-2\cdot 10^{-3}$&
$\wcL[V,LL]{ed}{2212} (\wc[(3)]{qq}{1222})$&$2\cdot 10^{-3}$&
$\wcL[V,LL]{ed}{2212} (\wc[(3)]{qq}{1112})$&$2\cdot 10^{-3}$&
\\ 
$\wcL[V,RR]{ed}{2212} [\wc[]{ed}{2212}]$&$1$&
$\wcL[V,RR]{ed}{2212} [\wc[]{Hd}{12}]$&$0.5$&
$\wcL[V,RR]{ed}{2212} (\wc[(1)]{ud}{3312})$&$-0.01$&
\\ 
$\wcL[V,RR]{ed}{2212} (\wc[(1)]{qd}{3312})$&$0.01$&
$\wcL[V,RR]{ed}{2212} (\wc[(1)]{ud}{2212})$&$-2\cdot 10^{-3}$&
$\wcL[V,RR]{ed}{2212} (\wc[(1)]{ud}{1112})$&$-2\cdot 10^{-3}$&
\\ 
$\wcL[V,RR]{ed}{2212} (\wc[]{dd}{1222})$&$1\cdot 10^{-3}$&
$\wcL[V,RR]{ed}{2212} (\wc[]{dd}{1112})$&$1\cdot 10^{-3}$&
$\wcL[V,LR]{ed}{2212} [\wc[]{ld}{2212}]$&$1$&
\\ 
$\wcL[V,LR]{ed}{2212} [\wc[]{Hd}{12}]$&$-0.5$&
$\wcL[V,LR]{ed}{2212} (\wc[(1)]{qd}{3312})$&$-0.02$&
$\wcL[V,LR]{ed}{2212} (\wc[(1)]{ud}{3312})$&$0.01$&
\\ 
$\wcL[V,LR]{ed}{2212} (\wc[(1)]{ud}{1112})$&$-2\cdot 10^{-3}$&
$\wcL[V,LR]{ed}{2212} (\wc[(1)]{ud}{2212})$&$-2\cdot 10^{-3}$&
$\wcL[V,LR]{ed}{2212} (\wc[]{dd}{1112})$&$1\cdot 10^{-3}$&
\\ 
$\wcL[V,LR]{ed}{2212} (\wc[]{dd}{1222})$&$1\cdot 10^{-3}$&
$\wcL[V,LR]{de}{1222} [\wc[]{qe}{1222}]$&$1$&
$\wcL[V,LR]{de}{1222} [\wc[(1)]{Hq}{12}]$&$0.5$&
\\ 
$\wcL[V,LR]{de}{1222} [\wc[(3)]{Hq}{12}]$&$0.5$&
$\wcL[V,LR]{de}{1222} (\wc[(1)]{qu}{1233})$&$-0.01$&
$\wcL[V,LR]{de}{1222} (\wc[(1)]{qq}{1233})$&$0.01$&
\\ 
$\wcL[V,LR]{de}{1222} (\wc[(3)]{qq}{1233})$&$-0.01$&
$\wcL[V,LR]{de}{1222} (\wc[(3)]{qq}{1332})$&$8\cdot 10^{-3}$&
$\wcL[V,LR]{de}{1222} (\wc[(1)]{qu}{1222})$&$-2\cdot 10^{-3}$&
\\ 
$\wcL[V,LR]{de}{1222} (\wc[(1)]{qu}{1211})$&$-2\cdot 10^{-3}$&
$\wcL[V,LR]{de}{1222} (\wc[(3)]{qq}{1112})$&$2\cdot 10^{-3}$&
$\wcL[V,LR]{de}{1222} (\wc[(3)]{qq}{1222})$&$2\cdot 10^{-3}$&
\\ 
$\wcL[S,RL]{ed}{2212} [\wc[]{ledq}{2212}]$&$1$&
&&&&\\
\bottomrule \end{tabular} \caption{Part 1 of the $\rho$ and $\eta$-parameters for Class 2A and 2B for $s\to d \mu^+ \mu^-$.} \label{tab:class2-3} \end{table}

\begin{table}[H]\centering\begin{tabular}{|cc|cc|ccc|}\toprule $C^{\rm WET}_i(C^{\rm SMEFT}_ m)$ & size & $C^{\rm WET}_i(C^{\rm SMEFT}_ m)$ & size & $C^{\rm WET}_i(C^{\rm SMEFT}_ m)$ & size &\\ \hline
$\wcL[V,LL]{\nu d}{2212} [\wc[(1)]{Hq}{12}]$&$1$&
$\wcL[V,LL]{\nu d}{2212} [\wc[(3)]{Hq}{12}]$&$1$&
&&
\\ 
$\wcL[V,LL]{\nu d}{2212} [\wc[(1)]{lq}{2212}]$&$1$&
$\wcL[V,LL]{\nu d}{2212} [\wc[(3)]{lq}{2212}]$&$-1$&
$\wcL[V,LL]{\nu d}{2212} (\wc[(1)]{qu}{1233})$&$-0.03$&
\\ 
$\wcL[V,LL]{\nu d}{2212} (\wc[(3)]{qq}{1233})$&$-0.03$&
$\wcL[V,LL]{\nu d}{2212} (\wc[(1)]{qq}{1233})$&$0.03$&
$\wcL[V,LL]{\nu d}{2212} (\wc[(3)]{qq}{1332})$&$0.02$&
\\ 
$\wcL[V,LL]{\nu d}{2212} (\wc[(3)]{qq}{1223})$&$1\cdot 10^{-3}$&
$\wcL[V,LL]{\nu d}{2212} (\wc[(1)]{qq}{1232})$&$-1\cdot 10^{-3}$&
$\wcL[V,LL]{\nu d}{2212} (\wc[(1)]{qq}{1223})$&$-1\cdot 10^{-3}$&
\\ 
$\wcL[V,LR]{\nu d}{2212} [\wc[]{Hd}{12}]$&$1$&
$\wcL[V,LR]{\nu d}{2212} [\wc[]{ld}{2212}]$&$1$&
$\wcL[V,LR]{\nu d}{2212} (\wc[(1)]{ud}{3312})$&$-0.03$&
\\ 
$\wcL[V,LR]{\nu d}{2212} (\wc[(1)]{qd}{3312})$&$0.03$&
$\wcL[V,LR]{\nu d}{2212} (\wc[(1)]{qd}{2321})$&$-1\cdot 10^{-3}$&
$\wcL[V,LR]{\nu d}{2212} (\wc[(1)]{qd}{2312})$&$-1\cdot 10^{-3}$&
\\ 
\bottomrule \end{tabular} \caption{Part 2 of the $\rho$ and $\eta$-parameters for Class 2A and 2B for $s\to d \nu_\mu  \bar \nu_\mu$.} \label{tab:class2-4} \end{table}

\begin{table}[H]\centering\begin{tabular}{|cc|cc|ccc|}\toprule $C^{\rm WET}_i(C^{\rm SMEFT}_ m)$ & size & $C^{\rm WET}_i(C^{\rm SMEFT}_ m)$ & size & $C^{\rm WET}_i(C^{\rm SMEFT}_ m)$ & size &\\ \hline
$\wcL[]{d\gamma}{12} [\wc[]{dB}{12}]$&$153$&
$\wcL[]{d\gamma}{12} [\wc[]{dW}{12}]$&$-84$&
$\wcL[]{d\gamma}{12} (\wc[]{dW}{21})$&$-4$&
\\ 
$\wcL[]{d\gamma}{12} (\wc[]{dG}{12})$&$-1$&
$\wcL[]{d\gamma}{12} (\wc[(8)]{quqd}{1332})$&$-0.3$&
$\wcL[]{d\gamma}{12} (\wc[(1)]{quqd}{1332})$&$-0.2$&
\\ 
$\wcL[]{d\gamma}{12} (\wc[]{dW}{22})$&$0.1$&
$\wcL[]{d\gamma}{12} (\wc[]{dW}{11})$&$-0.03$&
$\wcL[]{d\gamma}{12} (\wc[]{dB}{21})$&$0.02$&
\\ 
$\wcL[]{d\gamma}{12} (\wc[(8)]{quqd}{1322})$&$0.01$&
$\wcL[]{d\gamma}{12} (\wc[]{dW}{32})$&$-0.01$&
$\wcL[]{d\gamma}{12} (\wc[(1)]{quqd}{1322})$&$8\cdot 10^{-3}$&
\\ 
$\wcL[]{d\gamma}{12} (\wc[(8)]{quqd}{1312})$&$-2\cdot 10^{-3}$&
$\wcL[]{d\gamma}{12} (\wc[(1)]{quqd}{1312})$&$-2\cdot 10^{-3}$&
$\wcL[]{d\gamma}{12} (\wc[]{dB}{32})$&$-2\cdot 10^{-3}$&
\\ 
$\wcL[]{d\gamma}{12} (\wc[]{dB}{13})$&$-1\cdot 10^{-3}$&
$\wcL[]{dG}{12} [\wc[]{dG}{12}]$&$174$&
$\wcL[]{dG}{12} (\wc[(1)]{quqd}{1332})$&$-1.0$&
\\ 
$\wcL[]{dG}{12} (\wc[]{dW}{21})$&$-0.6$&
$\wcL[]{dG}{12} (\wc[]{dB}{12})$&$-0.3$&
$\wcL[]{dG}{12} (\wc[]{dW}{12})$&$0.2$&
\\ 
$\wcL[]{dG}{12} (\wc[(8)]{quqd}{1332})$&$0.2$&
$\wcL[]{dG}{12} (\wc[(1)]{quqd}{1322})$&$0.04$&
$\wcL[]{dG}{12} (\wc[]{dW}{22})$&$0.01$&
\\ 
$\wcL[]{dG}{12} (\wc[(1)]{quqd}{1312})$&$-1\cdot 10^{-2}$&
$\wcL[]{dG}{12} (\wc[(8)]{quqd}{1322})$&$-7\cdot 10^{-3}$&
$\wcL[]{dG}{12} (\wc[]{dW}{11})$&$-5\cdot 10^{-3}$&
\\ 
$\wcL[]{dG}{12} (\wc[(1)]{quqd}{1222})$&$-4\cdot 10^{-3}$&
$\wcL[]{dG}{12} (\wc[]{dG}{13})$&$-2\cdot 10^{-3}$&
$\wcL[]{dG}{12} (\wc[(1)]{quqd}{2331})$&$2\cdot 10^{-3}$&
\\ 
$\wcL[]{dG}{12} (\wc[(8)]{quqd}{1312})$&$1\cdot 10^{-3}$&
&&&&\\
\bottomrule \end{tabular} \caption{The $\rho$ and $\eta$-parameters for Class 2C for $s\to d \gamma$.} \label{tab:class2-5} \end{table}

\begin{table}[H]\centering\begin{tabular}{|cc|cc|ccc|}\toprule $C^{\rm WET}_i(C^{\rm SMEFT}_ m)$ & size & $C^{\rm WET}_i(C^{\rm SMEFT}_ m)$ & size & $C^{\rm WET}_i(C^{\rm SMEFT}_ m)$ & size &\\ \hline
$\wcL[V,LL]{ee}{2223} [\wc[]{ll}{2223}]$&$1$&
$\wcL[V,LL]{ee}{2223} [\wc[(1)]{Hl}{23}]$&$-0.5$&
$\wcL[V,LL]{ee}{2223} [\wc[(3)]{Hl}{23}]$&$-0.5$&
\\ 
$\wcL[V,LL]{ee}{2223} (\wc[(3)]{lq}{2333})$&$0.02$&
$\wcL[V,LL]{ee}{2223} (\wc[(1)]{lq}{2333})$&$-0.02$&
$\wcL[V,LL]{ee}{2223} (\wc[]{lu}{2333})$&$0.01$&
\\ 
$\wcL[V,LL]{ee}{2223} (\wc[(3)]{lq}{2322})$&$2\cdot 10^{-3}$&
$\wcL[V,LL]{ee}{2223} (\wc[(3)]{lq}{2311})$&$2\cdot 10^{-3}$&
$\wcL[V,LL]{ee}{2223} (\wc[]{lu}{2311})$&$-2\cdot 10^{-3}$&
\\ 
$\wcL[V,LL]{ee}{2223} (\wc[]{lu}{2322})$&$-2\cdot 10^{-3}$&
$\wcL[V,LL]{ee}{2223} (\wc[]{ll}{2333})$&$2\cdot 10^{-3}$&
$\wcL[V,LL]{ee}{2223} (\wc[(1)]{lq}{2322})$&$-9\cdot 10^{-4}$&
\\ 
$\wcL[V,LL]{ee}{2223} (\wc[(1)]{lq}{2311})$&$-9\cdot 10^{-4}$&
$\wcL[V,LL]{ee}{2223} (\wc[]{ll}{1123})$&$9\cdot 10^{-4}$&
$\wcL[V,LL]{ee}{2223} (\wc[]{le}{2311})$&$9\cdot 10^{-4}$&
\\ 
$\wcL[V,LL]{ee}{2223} (\wc[]{ld}{2311})$&$9\cdot 10^{-4}$&
$\wcL[V,LL]{ee}{2223} (\wc[]{ld}{2322})$&$9\cdot 10^{-4}$&
$\wcL[V,LL]{ee}{2223} (\wc[]{le}{2322})$&$-9\cdot 10^{-4}$&
\\ 
$\wcL[V,LL]{ee}{2223} (\wc[]{le}{2333})$&$-9\cdot 10^{-4}$&
$\wcL[V,LL]{ee}{2223} (\wc[]{ld}{2333})$&$8\cdot 10^{-4}$&
$\wcL[V,LL]{ee}{2223} (\wc[]{ll}{1231})$&$8\cdot 10^{-4}$&
\\ 
$\wcL[V,LL]{ee}{2223} (\wc[(3)]{lq}{2323})$&$-6\cdot 10^{-4}$&
$\wcL[V,LL]{ee}{2223} (\wc[(1)]{lq}{2323})$&$6\cdot 10^{-4}$&
$\wcL[V,LL]{ee}{2223} (\wc[(1)]{lq}{2332})$&$6\cdot 10^{-4}$&
\\ 
$\wcL[V,LL]{ee}{2223} (\wc[(3)]{lq}{2332})$&$-6\cdot 10^{-4}$&
$\wcL[V,LL]{ee}{2223} (\wc[(3)]{lq}{2313})$&$1\cdot 10^{-4}$&
$\wcL[V,LL]{ee}{2223} (\wc[(1)]{lq}{2331})$&$-1\cdot 10^{-4}$&
\\ 
$\wcL[V,LL]{ee}{2223} (\wc[(1)]{lq}{2313})$&$-1\cdot 10^{-4}$&
$\wcL[V,LL]{ee}{2223} (\wc[(3)]{lq}{2331})$&$1\cdot 10^{-4}$&
$\wcL[V,LL]{ee}{2223} (\wc[(1)]{lq}{2321})$&$6\cdot 10^{-6}$&
\\ 
$\wcL[V,LL]{ee}{2223} (\wc[(1)]{lq}{2312})$&$6\cdot 10^{-6}$&
$\wcL[V,LL]{ee}{2223} (\wc[(3)]{lq}{2321})$&$-6\cdot 10^{-6}$&
$\wcL[V,LL]{ee}{2223} (\wc[(3)]{lq}{2312})$&$-6\cdot 10^{-6}$&
\\ 
$\wcL[V,RR]{ee}{2223} [\wc[]{ee}{2223}]$&$1$&
$\wcL[V,RR]{ee}{2223} [\wc[]{He}{23}]$&$0.5$&
$\wcL[V,RR]{ee}{2223} (\wc[]{eu}{2333})$&$-0.01$&
\\ 
$\wcL[V,RR]{ee}{2223} (\wc[]{qe}{3323})$&$0.01$&
$\wcL[V,RR]{ee}{2223} (\wc[]{ee}{2333})$&$2\cdot 10^{-3}$&
$\wcL[V,RR]{ee}{2223} (\wc[]{eu}{2322})$&$-2\cdot 10^{-3}$&
\\ 
$\wcL[V,RR]{ee}{2223} (\wc[]{eu}{2311})$&$-2\cdot 10^{-3}$&
$\wcL[V,RR]{ee}{2223} (\wc[]{ed}{2333})$&$9\cdot 10^{-4}$&
$\wcL[V,RR]{ee}{2223} (\wc[]{ed}{2322})$&$9\cdot 10^{-4}$&
\\ 
$\wcL[V,RR]{ee}{2223} (\wc[]{ed}{2311})$&$9\cdot 10^{-4}$&
$\wcL[V,RR]{ee}{2223} (\wc[]{ee}{1123})$&$9\cdot 10^{-4}$&
$\wcL[V,RR]{ee}{2223} (\wc[]{le}{1123})$&$9\cdot 10^{-4}$&
\\ 
$\wcL[V,RR]{ee}{2223} (\wc[]{le}{2223})$&$-9\cdot 10^{-4}$&
$\wcL[V,RR]{ee}{2223} (\wc[]{le}{3323})$&$-9\cdot 10^{-4}$&
$\wcL[V,RR]{ee}{2223} (\wc[]{qe}{1123})$&$-9\cdot 10^{-4}$&
\\ 
$\wcL[V,RR]{ee}{2223} (\wc[]{qe}{2223})$&$-8\cdot 10^{-4}$&
$\wcL[V,RR]{ee}{2223} (\wc[]{qe}{2332})$&$-6\cdot 10^{-4}$&
$\wcL[V,RR]{ee}{2223} (\wc[]{qe}{2323})$&$-6\cdot 10^{-4}$&
\\ 
$\wcL[V,RR]{ee}{2223} (\wc[]{qe}{1323})$&$1\cdot 10^{-4}$&
$\wcL[V,RR]{ee}{2223} (\wc[]{qe}{1332})$&$1\cdot 10^{-4}$&
$\wcL[V,RR]{ee}{2223} (\wc[]{qe}{1223})$&$-5\cdot 10^{-6}$&
\\ 
$\wcL[V,RR]{ee}{2223} (\wc[]{qe}{1232})$&$-5\cdot 10^{-6}$&
&&&&\\
\bottomrule 
\end{tabular} 
\caption{The $\rho$ and $\eta$-parameters for Class 5A (part 1), for the transition $\tau\to 3\mu$ and $\tau \to \mu \gamma$, respectively.} 
\label{tab:rhoeta5-1} \end{table}

\begin{table}[H]\centering\begin{tabular}{|cc|cc|ccc|}\toprule $C^{\rm WET}_i(C^{\rm SMEFT}_ m)$ & size & $C^{\rm WET}_i(C^{\rm SMEFT}_ m)$ & size & $C^{\rm WET}_i(C^{\rm SMEFT}_ m)$ & size &\\ \hline
$\wcL[V,LR]{ee}{2223} [\wc[]{le}{2223}]$&$1$&
$\wcL[V,LR]{ee}{2223} [\wc[]{He}{23}]$&$-0.5$&
&&\\ 
$\wcL[V,LR]{ee}{2223} (\wc[]{qe}{3323})$&$-0.02$&
$\wcL[V,LR]{ee}{2223} (\wc[]{eu}{2333})$&$0.01$&
$\wcL[V,LR]{ee}{2223} (\wc[]{eu}{2311})$&$-2\cdot 10^{-3}$&
\\ 
$\wcL[V,LR]{ee}{2223} (\wc[]{ee}{2223})$&$2\cdot 10^{-3}$&
$\wcL[V,LR]{ee}{2223} (\wc[]{eu}{2322})$&$-2\cdot 10^{-3}$&
$\wcL[V,LR]{ee}{2223} (\wc[]{ee}{2333})$&$2\cdot 10^{-3}$&
\\ 
$\wcL[V,LR]{ee}{2223} (\wc[]{qe}{2223})$&$-9\cdot 10^{-4}$&
$\wcL[V,LR]{ee}{2223} (\wc[]{qe}{1123})$&$-9\cdot 10^{-4}$&
$\wcL[V,LR]{ee}{2223} (\wc[]{le}{3323})$&$-9\cdot 10^{-4}$&
\\ 
$\wcL[V,LR]{ee}{2223} (\wc[]{le}{1123})$&$9\cdot 10^{-4}$&
$\wcL[V,LR]{ee}{2223} (\wc[]{ee}{1123})$&$9\cdot 10^{-4}$&
$\wcL[V,LR]{ee}{2223} (\wc[]{ed}{2311})$&$9\cdot 10^{-4}$&
\\ 
$\wcL[V,LR]{ee}{2223} (\wc[]{ed}{2322})$&$9\cdot 10^{-4}$&
$\wcL[V,LR]{ee}{2223} (\wc[]{ed}{2333})$&$8\cdot 10^{-4}$&
$\wcL[V,LR]{ee}{2223} (\wc[]{qe}{2332})$&$6\cdot 10^{-4}$&
\\ 
$\wcL[V,LR]{ee}{2223} (\wc[]{qe}{2323})$&$6\cdot 10^{-4}$&
$\wcL[V,LR]{ee}{2223} (\wc[]{qe}{1323})$&$-1\cdot 10^{-4}$&
$\wcL[V,LR]{ee}{2223} (\wc[]{qe}{1332})$&$-1\cdot 10^{-4}$&
\\ 
$\wcL[V,LR]{ee}{2223} (\wc[(1)]{lequ}{2333})$&$-1\cdot 10^{-5}$&
$\wcL[V,LR]{ee}{2223} (\wc[]{qe}{1223})$&$6\cdot 10^{-6}$&
$\wcL[V,LR]{ee}{2223} (\wc[]{qe}{1232})$&$6\cdot 10^{-6}$&
\\ 
$\wcL[V,LR]{ee}{2322} [\wc[]{le}{2322}]$&$1$&
$\wcL[V,LR]{ee}{2322} [\wc[(1)]{Hl}{23}]$&$0.5$&
$\wcL[V,LR]{ee}{2322} [\wc[(3)]{Hl}{23}]$&$0.5$&
\\ 
$\wcL[V,LR]{ee}{2322} (\wc[]{lu}{2333})$&$-0.01$&
$\wcL[V,LR]{ee}{2322} (\wc[(1)]{lq}{2333})$&$0.01$&
$\wcL[V,LR]{ee}{2322} (\wc[(3)]{lq}{2333})$&$-0.01$&
\\ 
$\wcL[V,LR]{ee}{2322} (\wc[(3)]{lq}{2311})$&$2\cdot 10^{-3}$&
$\wcL[V,LR]{ee}{2322} (\wc[(3)]{lq}{2322})$&$2\cdot 10^{-3}$&
$\wcL[V,LR]{ee}{2322} (\wc[]{lu}{2322})$&$-2\cdot 10^{-3}$&
\\ 
$\wcL[V,LR]{ee}{2322} (\wc[]{lu}{2311})$&$-2\cdot 10^{-3}$&
$\wcL[V,LR]{ee}{2322} (\wc[]{ll}{2223})$&$2\cdot 10^{-3}$&
$\wcL[V,LR]{ee}{2322} (\wc[]{ll}{2333})$&$2\cdot 10^{-3}$&
\\ 
$\wcL[V,LR]{ee}{2322} (\wc[]{ld}{2333})$&$9\cdot 10^{-4}$&
$\wcL[V,LR]{ee}{2322} (\wc[]{le}{2333})$&$-9\cdot 10^{-4}$&
$\wcL[V,LR]{ee}{2322} (\wc[]{ld}{2322})$&$9\cdot 10^{-4}$&
\\ 
$\wcL[V,LR]{ee}{2322} (\wc[]{ld}{2311})$&$9\cdot 10^{-4}$&
$\wcL[V,LR]{ee}{2322} (\wc[]{le}{2311})$&$9\cdot 10^{-4}$&
$\wcL[V,LR]{ee}{2322} (\wc[]{ll}{1123})$&$9\cdot 10^{-4}$&
\\ 
$\wcL[V,LR]{ee}{2322} (\wc[(1)]{lq}{2311})$&$-9\cdot 10^{-4}$&
$\wcL[V,LR]{ee}{2322} (\wc[(1)]{lq}{2322})$&$-8\cdot 10^{-4}$&
$\wcL[V,LR]{ee}{2322} (\wc[]{ll}{1231})$&$8\cdot 10^{-4}$&
\\ 
$\wcL[V,LR]{ee}{2322} (\wc[(3)]{lq}{2323})$&$6\cdot 10^{-4}$&
$\wcL[V,LR]{ee}{2322} (\wc[(1)]{lq}{2323})$&$-6\cdot 10^{-4}$&
$\wcL[V,LR]{ee}{2322} (\wc[(1)]{lq}{2332})$&$-6\cdot 10^{-4}$&
\\ 
$\wcL[V,LR]{ee}{2322} (\wc[(3)]{lq}{2332})$&$6\cdot 10^{-4}$&
$\wcL[V,LR]{ee}{2322} (\wc[(3)]{lq}{2313})$&$-1\cdot 10^{-4}$&
$\wcL[V,LR]{ee}{2322} (\wc[(1)]{lq}{2331})$&$1\cdot 10^{-4}$&
\\ 
$\wcL[V,LR]{ee}{2322} (\wc[(1)]{lq}{2313})$&$1\cdot 10^{-4}$&
$\wcL[V,LR]{ee}{2322} (\wc[(3)]{lq}{2331})$&$-1\cdot 10^{-4}$&
$\wcL[V,LR]{ee}{2322} (\wc[(1)]{lequ}{3233})$&$-1\cdot 10^{-5}$&
\\ 
$\wcL[V,LR]{ee}{2322} (\wc[(1)]{lq}{2321})$&$-5\cdot 10^{-6}$&
$\wcL[V,LR]{ee}{2322} (\wc[(1)]{lq}{2312})$&$-5\cdot 10^{-6}$&
$\wcL[V,LR]{ee}{2322} (\wc[(3)]{lq}{2321})$&$5\cdot 10^{-6}$&
\\ 
$\wcL[V,LR]{ee}{2322} (\wc[(3)]{lq}{2312})$&$5\cdot 10^{-6}$&
&&&&\\
\bottomrule \end{tabular} \caption{The $\rho$ and $\eta$-parameters for Class 5A (part 2), for the transition $\tau\to 3\mu$ and $\tau \to \mu \gamma$, respectively.} \label{tab:rhoeta5-2} \end{table}

\begin{table}[H]
\centering
\begin{tabular}{|cc|cc|ccc|}\toprule $C^{\rm WET}_i(C^{\rm SMEFT}_ m)$ & size & $C^{\rm WET}_i(C^{\rm SMEFT}_ m)$ & size & $C^{\rm WET}_i(C^{\rm SMEFT}_ m)$ & size &\\ \hline
$\wcL[]{e\gamma}{23} [\wc[]{eB}{23}]$&$153$&
$\wcL[]{e\gamma}{23} [\wc[]{eW}{23}]$&$-84$&
$\wcL[]{e\gamma}{23} (\wc[(3)]{lequ}{2333})$&$5$&
\\ 
$\wcL[]{e\gamma}{23} (\wc[(3)]{lequ}{2323})$&$-0.2$&
$\wcL[]{e\gamma}{23} (\wc[(3)]{lequ}{2313})$&$0.04$&
$\wcL[]{e\gamma}{23} (\wc[]{eW}{32})$&$-0.04$&
\\ 
$\wcL[]{e\gamma}{23} (\wc[(3)]{lequ}{2322})$&$0.02$&
$\wcL[]{e\gamma}{23} (\wc[(3)]{lequ}{2312})$&$-4\cdot 10^{-3}$&
$\wcL[]{e\gamma}{23} (\wc[(3)]{lequ}{2332})$&$8\cdot 10^{-4}$&
\\ 
$\wcL[]{e\gamma}{23} (\wc[(3)]{lequ}{2311})$&$3\cdot 10^{-5}$&
$\wcL[]{e\gamma}{23} (\wc[(3)]{lequ}{2321})$&$8\cdot 10^{-6}$&
$\wcL[]{e\gamma}{23} (\wc[]{eB}{32})$&$6\cdot 10^{-6}$&
\\ 
\bottomrule \end{tabular} \caption{The $\rho$ and $\eta$-parameters for Class 5C, for the transition $\tau\to \mu \gamma$.} \label{tab:rhoeta5-3} \end{table}

\begin{table}[H]\centering\begin{tabular}{|cc|cc|ccc|}\toprule $C^{\rm WET}_i(C^{\rm SMEFT}_ m)$ & size & $C^{\rm WET}_i(C^{\rm SMEFT}_ m)$ & size & $C^{\rm WET}_i(C^{\rm SMEFT}_ m)$ & size &\\ \hline
$\wcL[V,LL]{eu}{2311} [\wc[(1)]{lq}{2311}]$&$0.9$&
$\wcL[V,LL]{eu}{2311} [\wc[(3)]{lq}{2311}]$&$-0.9$&
$\wcL[V,LL]{eu}{2311} [\wc[(1)]{Hl}{23}]$&$0.7$&
\\ 
$\wcL[V,LL]{eu}{2311} [\wc[(3)]{Hl}{23}]$&$0.7$&
$\wcL[V,LL]{eu}{2311} [\wc[(1)]{lq}{2312}]$&$0.2$&
$\wcL[V,LL]{eu}{2311} [\wc[(1)]{lq}{2321}]$&$0.2$&
\\ 
$\wcL[V,LL]{eu}{2311} [\wc[(3)]{lq}{2312}]$&$-0.2$&
$\wcL[V,LL]{eu}{2311} [\wc[(3)]{lq}{2321}]$&$-0.2$&
$\wcL[V,LL]{eu}{2311} [\wc[(1)]{lq}{2322}]$&$0.05$&
\\ 
$\wcL[V,LL]{eu}{2311} [\wc[(3)]{lq}{2322}]$&$-0.05$&
$\wcL[V,LL]{eu}{2311} (\wc[(3)]{lq}{2333})$&$-0.02$&
$\wcL[V,LL]{eu}{2311} (\wc[(1)]{lq}{2333})$&$0.02$&
\\ 
$\wcL[V,LL]{eu}{2311} (\wc[]{lu}{2333})$&$-0.02$&
$\wcL[V,LL]{eu}{2311} (\wc[]{lu}{2311})$&$1\cdot 10^{-3}$&
$\wcL[V,LL]{eu}{2311} (\wc[]{lu}{2322})$&$1\cdot 10^{-3}$&
\\ 
$\wcL[V,LL]{eu}{2311} (\wc[]{ll}{2333})$&$-1\cdot 10^{-3}$&
$\wcL[V,LL]{eu}{2311} (\wc[]{ll}{2223})$&$-1\cdot 10^{-3}$&
$\wcL[V,LL]{eu}{2311} (\wc[(3)]{lq}{2323})$&$9\cdot 10^{-4}$&
\\ 
$\wcL[V,LL]{eu}{2311} (\wc[(3)]{lq}{2332})$&$9\cdot 10^{-4}$&
$\wcL[V,LL]{eu}{2311} (\wc[(1)]{lq}{2323})$&$-9\cdot 10^{-4}$&
$\wcL[V,LL]{eu}{2311} (\wc[(1)]{lq}{2332})$&$-9\cdot 10^{-4}$&
\\ 
$\wcL[V,LL]{eu}{2311} (\wc[]{ll}{1123})$&$-6\cdot 10^{-4}$&
$\wcL[V,LL]{eu}{2311} (\wc[]{le}{2311})$&$-6\cdot 10^{-4}$&
$\wcL[V,LL]{eu}{2311} (\wc[]{ld}{2311})$&$-6\cdot 10^{-4}$&
\\ 
$\wcL[V,LL]{eu}{2311} (\wc[]{ld}{2322})$&$-6\cdot 10^{-4}$&
$\wcL[V,LL]{eu}{2311} (\wc[]{le}{2322})$&$6\cdot 10^{-4}$&
$\wcL[V,LL]{eu}{2311} (\wc[]{le}{2333})$&$6\cdot 10^{-4}$&
\\ 
$\wcL[V,LL]{eu}{2311} (\wc[]{ld}{2333})$&$-6\cdot 10^{-4}$&
$\wcL[V,LL]{eu}{2311} (\wc[]{ll}{1231})$&$-5\cdot 10^{-4}$&
$\wcL[V,LL]{eu}{2311} (\wc[(3)]{lq}{2313})$&$3\cdot 10^{-4}$&
\\ 
$\wcL[V,LL]{eu}{2311} (\wc[(3)]{lq}{2331})$&$3\cdot 10^{-4}$&
$\wcL[V,LL]{eu}{2311} (\wc[(1)]{lq}{2331})$&$-3\cdot 10^{-4}$&
$\wcL[V,LL]{eu}{2311} (\wc[(1)]{lq}{2313})$&$-3\cdot 10^{-4}$&
\\ 
$\wcL[V,LR]{eu}{2311} [\wc[]{lu}{2311}]$&$1$&
$\wcL[V,LR]{eu}{2311} [\wc[(1)]{Hl}{23}]$&$-0.3$&
$\wcL[V,LR]{eu}{2311} [\wc[(3)]{Hl}{23}]$&$-0.3$&
\\ 
$\wcL[V,LR]{eu}{2311} (\wc[]{lu}{2333})$&$1\cdot 10^{-2}$&
$\wcL[V,LR]{eu}{2311} (\wc[(1)]{lq}{2333})$&$-8\cdot 10^{-3}$&
$\wcL[V,LR]{eu}{2311} (\wc[(3)]{lq}{2333})$&$7\cdot 10^{-3}$&
\\ 
$\wcL[V,LR]{eu}{2311} (\wc[(3)]{lq}{2311})$&$-2\cdot 10^{-3}$&
$\wcL[V,LR]{eu}{2311} (\wc[(3)]{lq}{2322})$&$-2\cdot 10^{-3}$&
$\wcL[V,LR]{eu}{2311} (\wc[]{lu}{2322})$&$1\cdot 10^{-3}$&
\\ 
$\wcL[V,LR]{eu}{2311} (\wc[]{ll}{2223})$&$-1\cdot 10^{-3}$&
$\wcL[V,LR]{eu}{2311} (\wc[]{ll}{2333})$&$-1\cdot 10^{-3}$&
$\wcL[V,LR]{eu}{2311} (\wc[]{ld}{2333})$&$-6\cdot 10^{-4}$&
\\ 
$\wcL[V,LR]{eu}{2311} (\wc[]{le}{2333})$&$6\cdot 10^{-4}$&
$\wcL[V,LR]{eu}{2311} (\wc[]{le}{2322})$&$6\cdot 10^{-4}$&
$\wcL[V,LR]{eu}{2311} (\wc[]{ld}{2322})$&$-6\cdot 10^{-4}$&
\\ 
$\wcL[V,LR]{eu}{2311} (\wc[]{ld}{2311})$&$-6\cdot 10^{-4}$&
$\wcL[V,LR]{eu}{2311} (\wc[]{le}{2311})$&$-6\cdot 10^{-4}$&
$\wcL[V,LR]{eu}{2311} (\wc[]{ll}{1123})$&$-6\cdot 10^{-4}$&
\\ 
$\wcL[V,LR]{eu}{2311} (\wc[(1)]{lq}{2311})$&$6\cdot 10^{-4}$&
$\wcL[V,LR]{eu}{2311} (\wc[(1)]{lq}{2322})$&$6\cdot 10^{-4}$&
$\wcL[V,LR]{eu}{2311} (\wc[]{ll}{1231})$&$-5\cdot 10^{-4}$&
\\ 
$\wcL[V,LR]{eu}{2311} (\wc[(3)]{lq}{2323})$&$-4\cdot 10^{-4}$&
$\wcL[V,LR]{eu}{2311} (\wc[(3)]{lq}{2332})$&$-4\cdot 10^{-4}$&
$\wcL[V,LR]{eu}{2311} (\wc[(1)]{lq}{2323})$&$4\cdot 10^{-4}$&
\\ 
$\wcL[V,LR]{eu}{2311} (\wc[(1)]{lq}{2332})$&$4\cdot 10^{-4}$&
$\wcL[V,LR]{eu}{2311} (\wc[(1)]{lq}{2331})$&$-7\cdot 10^{-5}$&
$\wcL[V,LR]{eu}{2311} (\wc[(1)]{lq}{2313})$&$-7\cdot 10^{-5}$&
\\ 
$\wcL[V,LR]{eu}{2311} (\wc[(3)]{lq}{2313})$&$7\cdot 10^{-5}$&
$\wcL[V,LR]{eu}{2311} (\wc[(3)]{lq}{2331})$&$7\cdot 10^{-5}$&
$\wcL[V,LR]{eu}{2311} (\wc[(3)]{lq}{2321})$&$-3\cdot 10^{-6}$&
\\ 
$\wcL[V,LR]{eu}{2311} (\wc[(3)]{lq}{2312})$&$-3\cdot 10^{-6}$&
$\wcL[V,LR]{eu}{2311} (\wc[(1)]{lq}{2321})$&$3\cdot 10^{-6}$&
$\wcL[V,LR]{eu}{2311} (\wc[(1)]{lq}{2312})$&$3\cdot 10^{-6}$&
\\ 
\bottomrule \end{tabular} \caption{The $\rho$ and $\eta$-parameters for Class 6A for $\tau \to \mu u \bar u$  type transitions, part 1.} \label{tab:6A-1} \end{table}
\begin{table}[H]\centering\begin{tabular}{|cc|cc|ccc|}\toprule $C^{\rm WET}_i(C^{\rm SMEFT}_ m)$ & size & $C^{\rm WET}_i(C^{\rm SMEFT}_ m)$ & size & $C^{\rm WET}_i(C^{\rm SMEFT}_ m)$ & size &\\ \hline
$\wcL[V,LR]{ue}{1123} [\wc[]{qe}{1123}]$&$0.9$&
$\wcL[V,LR]{ue}{1123} [\wc[]{He}{23}]$&$0.7$&
$\wcL[V,LR]{ue}{1123} [\wc[]{qe}{1223}]$&$0.2$&
\\ 
$\wcL[V,LR]{ue}{1123} [\wc[]{qe}{1232}]$&$0.2$&
$\wcL[V,LR]{ue}{1123} [\wc[]{qe}{2223}]$&$0.05$&
$\wcL[V,LR]{ue}{1123} (\wc[]{qe}{3323})$&$0.02$&
\\ 
$\wcL[V,LR]{ue}{1123} (\wc[]{eu}{2333})$&$-0.02$&
$\wcL[V,LR]{ue}{1123} (\wc[]{eu}{2311})$&$1\cdot 10^{-3}$&
$\wcL[V,LR]{ue}{1123} (\wc[]{ee}{2223})$&$-1\cdot 10^{-3}$&
\\ 
$\wcL[V,LR]{ue}{1123} (\wc[]{eu}{2322})$&$1\cdot 10^{-3}$&
$\wcL[V,LR]{ue}{1123} (\wc[]{ee}{2333})$&$-1\cdot 10^{-3}$&
$\wcL[V,LR]{ue}{1123} (\wc[]{qe}{2332})$&$-9\cdot 10^{-4}$&
\\ 
$\wcL[V,LR]{ue}{1123} (\wc[]{qe}{2323})$&$-9\cdot 10^{-4}$&
$\wcL[V,LR]{ue}{1123} (\wc[]{le}{3323})$&$6\cdot 10^{-4}$&
$\wcL[V,LR]{ue}{1123} (\wc[]{le}{2223})$&$6\cdot 10^{-4}$&
\\ 
$\wcL[V,LR]{ue}{1123} (\wc[]{le}{1123})$&$-6\cdot 10^{-4}$&
$\wcL[V,LR]{ue}{1123} (\wc[]{ee}{1123})$&$-6\cdot 10^{-4}$&
$\wcL[V,LR]{ue}{1123} (\wc[]{ed}{2311})$&$-6\cdot 10^{-4}$&
\\ 
$\wcL[V,LR]{ue}{1123} (\wc[]{ed}{2322})$&$-6\cdot 10^{-4}$&
$\wcL[V,LR]{ue}{1123} (\wc[]{ed}{2333})$&$-6\cdot 10^{-4}$&
$\wcL[V,LR]{ue}{1123} (\wc[]{qe}{1323})$&$-3\cdot 10^{-4}$&
\\ 
$\wcL[V,LR]{ue}{1123} (\wc[]{qe}{1332})$&$-3\cdot 10^{-4}$&
$\wcL[V,RR]{eu}{2311} [\wc[]{eu}{2311}]$&$1$&
$\wcL[V,RR]{eu}{2311} [\wc[]{He}{23}]$&$-0.3$&
\\ 
$\wcL[V,RR]{eu}{2311} (\wc[]{eu}{2333})$&$1\cdot 10^{-2}$&
$\wcL[V,RR]{eu}{2311} (\wc[]{qe}{3323})$&$-8\cdot 10^{-3}$&
$\wcL[V,RR]{eu}{2311} (\wc[]{ee}{2333})$&$-1\cdot 10^{-3}$&
\\ 
$\wcL[V,RR]{eu}{2311} (\wc[]{eu}{2322})$&$1\cdot 10^{-3}$&
$\wcL[V,RR]{eu}{2311} (\wc[]{ee}{2223})$&$-1\cdot 10^{-3}$&
$\wcL[V,RR]{eu}{2311} (\wc[]{ed}{2333})$&$-6\cdot 10^{-4}$&
\\ 
$\wcL[V,RR]{eu}{2311} (\wc[]{ed}{2322})$&$-6\cdot 10^{-4}$&
$\wcL[V,RR]{eu}{2311} (\wc[]{ed}{2311})$&$-6\cdot 10^{-4}$&
$\wcL[V,RR]{eu}{2311} (\wc[]{ee}{1123})$&$-6\cdot 10^{-4}$&
\\ 
$\wcL[V,RR]{eu}{2311} (\wc[]{le}{1123})$&$-6\cdot 10^{-4}$&
$\wcL[V,RR]{eu}{2311} (\wc[]{le}{2223})$&$6\cdot 10^{-4}$&
$\wcL[V,RR]{eu}{2311} (\wc[]{le}{3323})$&$6\cdot 10^{-4}$&
\\ 
$\wcL[V,RR]{eu}{2311} (\wc[]{qe}{1123})$&$6\cdot 10^{-4}$&
$\wcL[V,RR]{eu}{2311} (\wc[]{qe}{2223})$&$6\cdot 10^{-4}$&
$\wcL[V,RR]{eu}{2311} (\wc[]{qe}{2332})$&$4\cdot 10^{-4}$&
\\ 
$\wcL[V,RR]{eu}{2311} (\wc[]{qe}{2323})$&$4\cdot 10^{-4}$&
$\wcL[V,RR]{eu}{2311} (\wc[]{qe}{1323})$&$-7\cdot 10^{-5}$&
$\wcL[V,RR]{eu}{2311} (\wc[]{qe}{1332})$&$-7\cdot 10^{-5}$&
\\ 
$\wcL[V,RR]{eu}{2311} (\wc[]{qe}{1223})$&$3\cdot 10^{-6}$&
$\wcL[V,RR]{eu}{2311} (\wc[]{qe}{1232})$&$3\cdot 10^{-6}$&
$\wcL[T,RR]{eu}{2311} [\wc[(3)]{lequ}{2311}]$&$-1.0$&
\\ 
$\wcL[T,RR]{eu}{2311} [\wc[(3)]{lequ}{2321}]$&$-0.2$&
$\wcL[T,RR]{eu}{2311} (\wc[(1)]{lequ}{2311})$&$1\cdot 10^{-3}$&
$\wcL[T,RR]{eu}{2311} (\wc[(3)]{lequ}{2331})$&$4\cdot 10^{-4}$&
\\ 
$\wcL[T,RR]{eu}{2311} (\wc[(1)]{lequ}{2321})$&$3\cdot 10^{-4}$&
$\wcL[T,RR]{eu}{2311} (\wc[(1)]{lequ}{2331})$&$-2\cdot 10^{-6}$&
$\wcL[S,RR]{eu}{2311} [\wc[(1)]{lequ}{2311}]$&$-1.0$&
\\ 
$\wcL[S,RR]{eu}{2311} [\wc[(1)]{lequ}{2321}]$&$-0.2$&
$\wcL[S,RR]{eu}{2311} (\wc[(3)]{lequ}{2311})$&$-0.07$&
$\wcL[S,RR]{eu}{2311} (\wc[(3)]{lequ}{2321})$&$-0.02$&
\\ 
$\wcL[S,RR]{eu}{2311} (\wc[(1)]{lequ}{2331})$&$5\cdot 10^{-4}$&
$\wcL[S,RR]{eu}{2311} (\wc[(3)]{lequ}{2331})$&$-8\cdot 10^{-5}$&
$\wcL[T,RR]{eu}{3211} [\wc[(3)]{lequ}{3211}]$&$-1.0$&
\\ 
$\wcL[T,RR]{eu}{3211} [\wc[(3)]{lequ}{3221}]$&$-0.2$&
$\wcL[T,RR]{eu}{3211} (\wc[(1)]{lequ}{3211})$&$-1\cdot 10^{-3}$&
$\wcL[T,RR]{eu}{3211} (\wc[(3)]{lequ}{3231})$&$4\cdot 10^{-4}$&
\\ 
$\wcL[T,RR]{eu}{3211} (\wc[(1)]{lequ}{3221})$&$-3\cdot 10^{-4}$&
$\wcL[T,RR]{eu}{3211} (\wc[(1)]{lequ}{3231})$&$-2\cdot 10^{-6}$&
$\wcL[S,RR]{eu}{3211} [\wc[(1)]{lequ}{3211}]$&$-1.0$&
\\ 
$\wcL[S,RR]{eu}{3211} [\wc[(1)]{lequ}{3221}]$&$-0.2$&
$\wcL[S,RR]{eu}{3211} (\wc[(3)]{lequ}{3211})$&$-0.07$&
$\wcL[S,RR]{eu}{3211} (\wc[(3)]{lequ}{3221})$&$-0.02$&
\\ 
$\wcL[S,RR]{eu}{3211} (\wc[(1)]{lequ}{3231})$&$5\cdot 10^{-4}$&
$\wcL[S,RR]{eu}{3211} (\wc[(3)]{lequ}{3231})$&$-8\cdot 10^{-5}$&
&&\\
\bottomrule \end{tabular} \caption{The $\rho$ and $\eta$-parameters for Class 6A for the $\tau \to \mu u \bar u$  type transitions, part 2.} \label{tab:6A-2} \end{table}

\begin{table}[H]\centering\begin{tabular}{|cc|cc|ccc|}\toprule $C^{\rm WET}_i(C^{\rm SMEFT}_ m)$ & size & $C^{\rm WET}_i(C^{\rm SMEFT}_ m)$ & size & $C^{\rm WET}_i(C^{\rm SMEFT}_ m)$ & size &\\ \hline
$\wcL[V,LL]{ed}{2323} [\wc[(1)]{lq}{2323}]$&$1$&
$\wcL[V,LL]{ed}{2323} [\wc[(3)]{lq}{2323}]$&$1$&
$\wcL[V,LL]{ed}{2323} (\wc[(1)]{lq}{2333})$&$-4\cdot 10^{-4}$&
\\ 
$\wcL[V,LL]{ed}{2323} (\wc[(3)]{lq}{2333})$&$-4\cdot 10^{-4}$&
$\wcL[V,LL]{ed}{2323} (\wc[]{lu}{2333})$&$2\cdot 10^{-4}$&
$\wcL[V,LL]{ed}{2323} (\wc[(1)]{Hl}{23})$&$-2\cdot 10^{-4}$&
\\ 
$\wcL[V,LL]{ed}{2323} (\wc[(3)]{Hl}{23})$&$2\cdot 10^{-4}$&
$\wcL[V,LL]{ed}{2323} (\wc[(1)]{lq}{2322})$&$2\cdot 10^{-4}$&
$\wcL[V,LL]{ed}{2323} (\wc[(3)]{lq}{2322})$&$2\cdot 10^{-4}$&
\\ 
$\wcL[V,LL]{ed}{2323} (\wc[(3)]{lq}{2321})$&$-3\cdot 10^{-5}$&
$\wcL[V,LL]{ed}{2323} (\wc[(1)]{lq}{2321})$&$-3\cdot 10^{-5}$&
$\wcL[V,LL]{ed}{2323} (\wc[]{lu}{2323})$&$-2\cdot 10^{-5}$&
\\ 
$\wcL[V,LL]{ed}{2323} (\wc[(1)]{lq}{2313})$&$1\cdot 10^{-6}$&
$\wcL[V,LL]{ed}{2323} (\wc[(3)]{lq}{2313})$&$1\cdot 10^{-6}$&
$\wcL[V,LR]{ed}{2323} [\wc[]{ld}{2323}]$&$1$&
\\ 
$\wcL[V,LR]{ed}{2323} (\wc[]{ld}{2322})$&$1\cdot 10^{-5}$&
$\wcL[V,LR]{ed}{2323} (\wc[]{ld}{2333})$&$-1\cdot 10^{-5}$&
$\wcL[V,LR]{de}{2323} [\wc[]{qe}{2323}]$&$1$&
\\ 
$\wcL[V,LR]{de}{2323} (\wc[]{qe}{3323})$&$-4\cdot 10^{-4}$&
$\wcL[V,LR]{de}{2323} (\wc[]{eu}{2333})$&$2\cdot 10^{-4}$&
$\wcL[V,LR]{de}{2323} (\wc[]{He}{23})$&$2\cdot 10^{-4}$&
\\ 
$\wcL[V,LR]{de}{2323} (\wc[]{qe}{2223})$&$2\cdot 10^{-4}$&
$\wcL[V,LR]{de}{2323} (\wc[]{qe}{1232})$&$-3\cdot 10^{-5}$&
$\wcL[V,LR]{de}{2323} (\wc[(3)]{lequ}{2323})$&$-2\cdot 10^{-5}$&
\\ 
$\wcL[V,LR]{de}{2323} (\wc[]{eu}{2323})$&$-2\cdot 10^{-5}$&
$\wcL[V,LR]{de}{2323} (\wc[(3)]{lequ}{3233})$&$1\cdot 10^{-5}$&
$\wcL[V,LR]{de}{2323} (\wc[(1)]{lequ}{2323})$&$-2\cdot 10^{-6}$&
\\ 
$\wcL[V,LR]{de}{2323} (\wc[]{qe}{1323})$&$1\cdot 10^{-6}$&
$\wcL[V,LR]{de}{2323} (\wc[(3)]{lequ}{3232})$&$-1\cdot 10^{-6}$&
$\wcL[V,LR]{de}{2323} (\wc[(1)]{lequ}{3233})$&$-1\cdot 10^{-6}$&
\\ 
$\wcL[V,RR]{ed}{2323} [\wc[]{ed}{2323}]$&$1$&
$\wcL[V,RR]{ed}{2323} (\wc[]{ed}{2322})$&$1\cdot 10^{-5}$&
$\wcL[V,RR]{ed}{2323} (\wc[]{ed}{2333})$&$-1\cdot 10^{-5}$&
\\ 
$\wcL[S,RL]{ed}{2332} [\wc[]{ledq}{2332}]$&$1$&
$\wcL[S,RL]{ed}{2332} (\wc[]{ledq}{2333})$&$4\cdot 10^{-4}$&
$\wcL[S,RL]{ed}{2332} (\wc[]{ledq}{2322})$&$-1\cdot 10^{-5}$&
\\ 
$\wcL[S,RL]{ed}{2332} (\wc[(1)]{lequ}{2333})$&$7\cdot 10^{-6}$&
$\wcL[S,RL]{ed}{2332} (\wc[(3)]{lq}{2332})$&$6\cdot 10^{-6}$&
$\wcL[S,RL]{ed}{2332} (\wc[(1)]{lq}{2332})$&$2\cdot 10^{-6}$&
\\ 
$\wcL[S,RL]{ed}{2332} (\wc[]{ledq}{2331})$&$2\cdot 10^{-6}$&
$\wcL[S,RL]{ed}{3232} [\wc[]{ledq}{3232}]$&$1$&
$\wcL[S,RL]{ed}{3232} (\wc[]{ledq}{3233})$&$4\cdot 10^{-4}$&
\\ 
$\wcL[S,RL]{ed}{3232} (\wc[]{ledq}{3222})$&$-1\cdot 10^{-5}$&
$\wcL[S,RL]{ed}{3232} (\wc[(1)]{lequ}{3233})$&$7\cdot 10^{-6}$&
$\wcL[S,RL]{ed}{3232} (\wc[]{qe}{2323})$&$-2\cdot 10^{-6}$&
\\ 
$\wcL[S,RL]{ed}{3232} (\wc[]{ledq}{3231})$&$2\cdot 10^{-6}$&
&&&&\\
\bottomrule \end{tabular} \caption{The $\rho$ and $\eta$-parameters for Class 6B operators relevant for $b \to s \tau^+ \mu^-$ transitions. The $\rho$ and $\eta$-parameters for other similar transitions in this class will be provided in the supplemental material. Moreover, for brevity the parameters with values less than $10^{-6}$ are not shown.} \label{tab:6B} \end{table}

\begin{table}[H]\centering\begin{tabular}{|cc|cc|ccc|}\toprule $C^{\rm WET}_i(C^{\rm SMEFT}_ m)$ & size & $C^{\rm WET}_i(C^{\rm SMEFT}_ m)$ & size & $C^{\rm WET}_i(C^{\rm SMEFT}_ m)$ & size &\\ \hline
$\wcL[V,LL]{\nu edu}{1111} [\wc[(3)]{Hl}{11}]$&$-3$&
$\wcL[V,LL]{\nu edu}{1111} [\wc[(3)]{Hq}{11}]$&$-2$&
$\wcL[V,LL]{\nu edu}{1111} [\wc[(3)]{lq}{1111}]$&$2$&
\\ 
$\wcL[V,LL]{\nu edu}{1111} [\wc[(3)]{Hl}{22}]$&$-1$&
$\wcL[V,LL]{\nu edu}{1111} [\wc[]{ll}{1221}]$&$-1.0$&
$\wcL[V,LL]{\nu edu}{1111} [\wc[(3)]{Hq}{12}]$&$-0.4$&
\\ 
$\wcL[V,LL]{\nu edu}{1111} [\wc[(3)]{lq}{1112}]$&$0.4$&
$\wcL[V,LL]{\nu edu}{1111} (\wc[(3)]{lq}{2233})$&$-0.06$&
$\wcL[V,LL]{\nu edu}{1111} (\wc[(3)]{qq}{1133})$&$0.06$&
\\ 
$\wcL[V,LL]{\nu edu}{1111} (\wc[]{ll}{1122})$&$-0.02$&
$\wcL[V,LL]{\nu edu}{1111} (\wc[(1)]{lq}{1111})$&$0.02$&
$\wcL[V,LL]{\nu edu}{1111} (\wc[(3)]{qq}{1233})$&$0.01$&
\\ 
$\wcL[V,LL]{\nu edu}{1111} (\wc[(3)]{qq}{1331})$&$-9\cdot 10^{-3}$&
$\wcL[V,LL]{\nu edu}{1111} (\wc[(1)]{qq}{1331})$&$9\cdot 10^{-3}$&
$\wcL[V,LL]{\nu edu}{1111} (\wc[(3)]{qq}{1123})$&$-5\cdot 10^{-3}$&
\\ 
$\wcL[V,LL]{\nu edu}{1111} (\wc[(3)]{lq}{2223})$&$5\cdot 10^{-3}$&
$\wcL[V,LL]{\nu edu}{1111} (\wc[(1)]{lq}{1112})$&$3\cdot 10^{-3}$&
$\wcL[V,LL]{\nu edu}{1111} [\wc[(3)]{Hq}{13}]$&$-2\cdot 10^{-3}$&
\\ 
$\wcL[V,LL]{\nu edu}{1111} [\wc[(3)]{lq}{1113}]$&$2\cdot 10^{-3}$&
$\wcL[V,LL]{\nu edu}{1111} (\wc[(3)]{qq}{1332})$&$-2\cdot 10^{-3}$&
$\wcL[V,LL]{\nu edu}{1111} (\wc[(1)]{qq}{1332})$&$2\cdot 10^{-3}$&
\\ 
$\wcL[V,LL]{\nu edu}{1111} (\wc[(3)]{lq}{2213})$&$-9\cdot 10^{-4}$&
$\wcL[V,LL]{\nu edu}{1111} (\wc[(3)]{qq}{1231})$&$9\cdot 10^{-4}$&
$\wcL[V,LL]{\nu edu}{1111} (\wc[(3)]{Hl}{23})$&$-8\cdot 10^{-4}$&
\\ 
$\wcL[V,LL]{\nu edu}{1111} (\wc[(3)]{Hl}{12})$&$8\cdot 10^{-4}$&
$\wcL[V,LL]{\nu edu}{1111} (\wc[(3)]{qq}{1113})$&$8\cdot 10^{-4}$&
$\wcL[V,LL]{\nu edu}{1111} (\wc[(1)]{qq}{1231})$&$-8\cdot 10^{-4}$&
\\ 
$\wcL[V,LL]{\nu edu}{1111} (\wc[(3)]{qq}{1223})$&$-5\cdot 10^{-4}$&
$\wcL[V,LL]{\nu edu}{1111} (\wc[(3)]{qq}{1232})$&$-4\cdot 10^{-4}$&
$\wcL[V,LL]{\nu edu}{1111} (\wc[(1)]{qq}{1113})$&$2\cdot 10^{-4}$&
\\ 
$\wcL[V,LL]{\nu edu}{1111} (\wc[(1)]{Hq}{13})$&$1\cdot 10^{-4}$&
$\wcL[V,LR]{\nu edu}{1111} [\wc[]{Hud}{11}]$&$-1$&
$\wcL[V,LR]{\nu edu}{1111} (\wc[(8)]{ud}{1331})$&$-4\cdot 10^{-4}$&
\\ 
$\wcL[V,LR]{\nu edu}{1111} (\wc[(1)]{ud}{1331})$&$-3\cdot 10^{-4}$&
$\wcL[S,RL]{\nu edu}{1111} [\wc[]{ledq}{1111}]$&$1.0$&
$\wcL[S,RL]{\nu edu}{1111} [\wc[]{ledq}{1112}]$&$0.2$&
\\ 
$\wcL[S,RL]{\nu edu}{1111} [\wc[]{ledq}{1113}]$&$1\cdot 10^{-3}$&
$\wcL[S,RR]{\nu edu}{1111} [\wc[(1)]{lequ}{1111}]$&$1$&
$\wcL[S,RR]{\nu edu}{1111} (\wc[(3)]{lequ}{1111})$&$0.07$&
\\ 
$\wcL[T,RR]{\nu edu}{1111} [\wc[(3)]{lequ}{1111}]$&$1$&
$\wcL[T,RR]{\nu edu}{1111} (\wc[(1)]{lequ}{1111})$&$1\cdot 10^{-3}$&
&&   \\
\bottomrule \end{tabular} \caption{The $\rho$ and $\eta$-parameters for Class 8 for $d \to u e \bar \nu_e$ transitions governing processes such as nuclear beta or neutron decay.} \label{tab:8-1} \end{table}

\begin{table}[H]\centering\begin{tabular}{|cc|cc|ccc|}\toprule $C^{\rm WET}_i(C^{\rm SMEFT}_ m)$ & size & $C^{\rm WET}_i(C^{\rm SMEFT}_ m)$ & size & $C^{\rm WET}_i(C^{\rm SMEFT}_ m)$ & size &\\ \hline
$\wcL[V,LL]{\nu edu}{3332} [\wc[(3)]{Hq}{23}]$&$-2$&
$\wcL[V,LL]{\nu edu}{3332} [\wc[(3)]{lq}{3323}]$&$2$&
$\wcL[V,LL]{\nu edu}{3332} [\wc[(3)]{lq}{3313}]$&$-0.4$&
\\ 
$\wcL[V,LL]{\nu edu}{3332} [\wc[(3)]{Hq}{13}]$&$0.4$&
$\wcL[V,LL]{\nu edu}{3332} [\wc[(3)]{Hl}{33}]$&$-0.08$&
$\wcL[V,LL]{\nu edu}{3332} [\wc[(3)]{Hq}{33}]$&$-0.08$&
\\ 
$\wcL[V,LL]{\nu edu}{3332} [\wc[(3)]{lq}{3333}]$&$0.08$&
$\wcL[V,LL]{\nu edu}{3332} [\wc[(3)]{Hl}{11}]$&$-0.06$&
$\wcL[V,LL]{\nu edu}{3332} [\wc[(3)]{Hl}{22}]$&$-0.06$&
\\ 
$\wcL[V,LL]{\nu edu}{3332} (\wc[(3)]{qq}{2333})$&$0.05$&
$\wcL[V,LL]{\nu edu}{3332} [\wc[]{ll}{1221}]$&$-0.04$&
$\wcL[V,LL]{\nu edu}{3332} (\wc[(1)]{lq}{3323})$&$0.02$&
\\ 
$\wcL[V,LL]{\nu edu}{3332} (\wc[(1)]{Hq}{23})$&$0.01$&
$\wcL[V,LL]{\nu edu}{3332} (\wc[(3)]{qq}{1333})$&$-0.01$&
$\wcL[V,LL]{\nu edu}{3332} (\wc[(1)]{qq}{2333})$&$9\cdot 10^{-3}$&
\\ 
$\wcL[V,LL]{\nu edu}{3332} (\wc[(3)]{qq}{3333})$&$4\cdot 10^{-3}$&
$\wcL[V,LL]{\nu edu}{3332} (\wc[(3)]{qq}{2323})$&$-4\cdot 10^{-3}$&
$\wcL[V,LL]{\nu edu}{3332} (\wc[(1)]{lq}{3313})$&$-3\cdot 10^{-3}$&
\\ 
$\wcL[V,LL]{\nu edu}{3332} (\wc[(1)]{Hq}{13})$&$-3\cdot 10^{-3}$&
$\wcL[V,LL]{\nu edu}{3332} (\wc[(3)]{lq}{1133})$&$-2\cdot 10^{-3}$&
$\wcL[V,LL]{\nu edu}{3332} (\wc[(3)]{lq}{2233})$&$-2\cdot 10^{-3}$&
\\ 
$\wcL[V,LL]{\nu edu}{3332} (\wc[(3)]{qq}{2332})$&$-2\cdot 10^{-3}$&
$\wcL[V,LL]{\nu edu}{3332} (\wc[(1)]{qq}{1333})$&$-2\cdot 10^{-3}$&
$\wcL[V,LL]{\nu edu}{3332} (\wc[(3)]{qq}{1332})$&$1\cdot 10^{-3}$&
\\ 
$\wcL[V,LL]{\nu edu}{3332} (\wc[(3)]{qq}{1323})$&$8\cdot 10^{-4}$&
$\wcL[V,LL]{\nu edu}{3332} (\wc[(1)]{qq}{3333})$&$8\cdot 10^{-4}$&
$\wcL[V,LL]{\nu edu}{3332} (\wc[(1)]{qq}{2323})$&$-8\cdot 10^{-4}$&
\\ 
$\wcL[V,LL]{\nu edu}{3332} (\wc[(1)]{lq}{3333})$&$7\cdot 10^{-4}$&
$\wcL[V,LL]{\nu edu}{3332} (\wc[]{ll}{1122})$&$-7\cdot 10^{-4}$&
$\wcL[V,LL]{\nu edu}{3332} (\wc[(1)]{Hq}{33})$&$6\cdot 10^{-4}$&
\\ 
$\wcL[V,LL]{\nu edu}{3332} (\wc[(1)]{Hq}{22})$&$-6\cdot 10^{-4}$&
$\wcL[V,LL]{\nu edu}{3332} (\wc[(1)]{qq}{2233})$&$-4\cdot 10^{-4}$&
$\wcL[V,LL]{\nu edu}{3332} (\wc[(3)]{qq}{2233})$&$4\cdot 10^{-4}$&
\\ 
$\wcL[V,LL]{\nu edu}{3332} (\wc[(3)]{lq}{3322})$&$3\cdot 10^{-4}$&
$\wcL[V,LL]{\nu edu}{3332} (\wc[(3)]{lequ}{3323})$&$3\cdot 10^{-4}$&
$\wcL[V,LL]{\nu edu}{3332} (\wc[(1)]{Hq}{12})$&$3\cdot 10^{-4}$&
\\ 
$\wcL[V,LL]{\nu edu}{3332} (\wc[(3)]{lq}{2223})$&$2\cdot 10^{-4}$&
$\wcL[V,LL]{\nu edu}{3332} (\wc[(3)]{lq}{1123})$&$2\cdot 10^{-4}$&
$\wcL[V,LL]{\nu edu}{3332} (\wc[(3)]{qq}{1313})$&$-2\cdot 10^{-4}$&
\\ 
$\wcL[V,LL]{\nu edu}{3332} (\wc[(1)]{qq}{1233})$&$2\cdot 10^{-4}$&
$\wcL[V,LL]{\nu edu}{3332} (\wc[(1)]{qq}{1323})$&$2\cdot 10^{-4}$&
$\wcL[V,LL]{\nu edu}{3332} (\wc[(3)]{qq}{1233})$&$-2\cdot 10^{-4}$&
\\ 
$\wcL[V,LL]{\nu edu}{3332} (\wc[(3)]{lq}{3312})$&$-1\cdot 10^{-4}$&
$\wcL[V,LL]{\nu edu}{3332} (\wc[(3)]{qq}{1331})$&$-1\cdot 10^{-4}$&
$\wcL[V,LL]{\nu edu}{3332} (\wc[(3)]{Hq}{22})$&$-1\cdot 10^{-4}$&
\\ 
$\wcL[V,LR]{\nu edu}{3332} [\wc[]{Hud}{23}]$&$-1$&
$\wcL[V,LR]{\nu edu}{3332} (\wc[(8)]{ud}{2333})$&$-4\cdot 10^{-4}$&
$\wcL[V,LR]{\nu edu}{3332} (\wc[(1)]{ud}{2333})$&$-3\cdot 10^{-4}$&
\\ 
$\wcL[V,LR]{\nu edu}{3332} (\wc[]{Hu}{23})$&$-2\cdot 10^{-4}$&
$\wcL[S,RL]{\nu edu}{3332} [\wc[]{ledq}{3332}]$&$1.0$&
$\wcL[S,RL]{\nu edu}{3332} [\wc[]{ledq}{3331}]$&$-0.2$&
\\ 
$\wcL[S,RL]{\nu edu}{3332} [\wc[]{ledq}{3333}]$&$0.04$&
$\wcL[S,RL]{\nu edu}{3332} (\wc[(1)]{quqd}{3323})$&$3\cdot 10^{-4}$&
$\wcL[S,RR]{\nu edu}{3332} [\wc[(1)]{lequ}{3332}]$&$1$&
\\ 
$\wcL[S,RR]{\nu edu}{3332} (\wc[(3)]{lequ}{3332})$&$0.07$&
$\wcL[S,RR]{\nu edu}{3332} (\wc[(8)]{qu}{3323})$&$3\cdot 10^{-4}$&
$\wcL[S,RR]{\nu edu}{3332} (\wc[(1)]{qu}{3323})$&$2\cdot 10^{-4}$&
\\ 
$\wcL[S,RR]{\nu edu}{3332} (\wc[(1)]{lequ}{3322})$&$2\cdot 10^{-4}$&
$\wcL[S,RR]{\nu edu}{3332} (\wc[]{eu}{3323})$&$1\cdot 10^{-4}$&
$\wcL[S,RR]{\nu edu}{3332} (\wc[]{lu}{3323})$&$-1\cdot 10^{-4}$&
\\ 
$\wcL[T,RR]{\nu edu}{3332} [\wc[(3)]{lequ}{3332}]$&$1$&
$\wcL[T,RR]{\nu edu}{3332} (\wc[(1)]{lequ}{3332})$&$1\cdot 10^{-3}$&
$\wcL[T,RR]{\nu edu}{3332} (\wc[(3)]{lequ}{3322})$&$2\cdot 10^{-4}$&
\\ 
\bottomrule \end{tabular} \caption{The $\rho$ and $\eta$-parameters for Class 8 for $b \to c \tau \bar \nu$ type transitions.} \label{tab:8-2} \end{table}

\renewcommand{\refname}{R\lowercase{eferences}}
\addcontentsline{toc}{section}{References}

\bibliographystyle{JHEP}

\small

\bibliography{Bookallrefs}

\providecommand{\href}[2]{#2}\begingroup\raggedright\begin{thebibliography}{1000}

\bibitem{Buras:2015nta}
A.~Buras, {\it {Flavour Expedition to the Zeptouniverse}},  {\em PoS} {\bf FWNP} (2015) 003, [\href{http://arxiv.org/abs/1505.00618}{{\tt arXiv:1505.00618}}].

\bibitem{Glashow:1970gm}
S.~L. Glashow, J.~Iliopoulos, and L.~Maiani, {\it {Weak Interactions with Lepton-Hadron Symmetry}},  {\em Phys. Rev.} {\bf D2} (1970) 1285--1292.

\bibitem{Gaillard:1974hs}
M.~Gaillard and B.~W. Lee, {\it {Rare Decay Modes of the K-Mesons in Gauge Theories}},  {\em Phys.~Rev.} {\bf D10} (1974) 897.

\bibitem{Cabibbo:1963yz}
N.~Cabibbo, {\it {Unitary Symmetry and Leptonic Decays}},  {\em Phys. Rev. Lett.} {\bf 10} (1963) 531--533. [648(1963)].

\bibitem{Kobayashi:1973fv}
M.~Kobayashi and T.~Maskawa, {\it {CP Violation in the Renormalizable Theory of Weak Interaction}},  {\em Prog.~Theor.~Phys.} {\bf 49} (1973) 652--657.

\bibitem{Buchalla:1995vs}
G.~Buchalla, A.~J. Buras, and M.~E. Lautenbacher, {\it {Weak decays beyond leading logarithms}},  {\em Rev.~Mod.~Phys.} {\bf 68} (1996) 1125--1144, [\href{http://arxiv.org/abs/hep-ph/9512380}{{\tt hep-ph/9512380}}].

\bibitem{Buras:1998raa}
A.~J. Buras, {\it {Weak Hamiltonian, CP violation and rare decays}},  in {\em {Probing the standard model of particle interactions. Proceedings, Summer School in Theoretical Physics, NATO Advanced Study Institute, 68th session, Les Houches, France, July 28-September 5, 1997. Pt. 1, 2}}, pp.~281--539, 1998.
\newblock \href{http://arxiv.org/abs/hep-ph/9806471}{{\tt hep-ph/9806471}}.

\bibitem{Branco:1999fs}
G.~C. Branco, L.~Lavoura, and J.~P. Silva, {\it {CP Violation}},  {\em Int.Ser.Monogr.Phys.} {\bf 103} (1999) 1--536.

\bibitem{Bigi:2000yz}
I.~I. Bigi and A.~Sanda, {\it {CP violation}},  {\em Camb.Monogr.Part.Phys.Nucl.~Phys.Cosmol.} {\bf 9} (2000) 1--382.

\bibitem{Zupan:2019uoi}
J.~Zupan, {\it {Introduction to flavour physics}},  {\em CERN Yellow Rep. School Proc.} {\bf 6} (2019) 181--212, [\href{http://arxiv.org/abs/1903.05062}{{\tt arXiv:1903.05062}}].

\bibitem{Silvestrini:2019sey}
L.~Silvestrini, {\it {Effective Theories for Quark Flavour Physics}},  \href{http://arxiv.org/abs/1905.00798}{{\tt arXiv:1905.00798}}.

\bibitem{Buras:2020xsm}
A.~J. Buras, {\em {Gauge Theory of Weak Decays}}.
\newblock Cambridge University Press, 6, 2020.

\bibitem{Bigi:2021hxw}
I.~I. Bigi, G.~Ricciardi, and M.~Pallavicini, {\em {New Era for CP Asymmetries}}.
\newblock World Scientific, 7, 2021.

\bibitem{Artuso:2022ijh}
M.~Artuso, G.~Isidori, and S.~Stone, {\em {New Physics in b Decays}}.
\newblock World Scientific, 5, 2022.

\bibitem{Buras:2011we}
A.~J. Buras, {\it {Climbing NLO and NNLO Summits of Weak Decays: 1988-2023}},  {\em Phys. Rep.} {\bf 1025} (2023) 1--64, [\href{http://arxiv.org/abs/1102.5650}{{\tt arXiv:1102.5650}}].

\bibitem{Grossman:2023wrq}
Y.~Grossman and Y.~Nir, {\em {The Standard Model: From Fundamental Symmetries to Experimental Tests}}.
\newblock Princeton University Press, 10, 2023.

\bibitem{Altmannshofer:2024ykf}
W.~Altmannshofer, {\it {TASI 2022 lectures on flavor physics}},  {\em PoS} {\bf TASI2022} (2024) 001.

\bibitem{Isidori:2025iyu}
G.~Isidori, {\it {Flavour Physics and CP Violation}},  3, 2025.
\newblock \href{http://arxiv.org/abs/2503.14042}{{\tt arXiv:2503.14042}}.

\bibitem{Buras:2003td}
A.~J. Buras, {\it {Relations between $\Delta M_{s,d}$ and $B_{s,d} \to \mu^+ \mu^-$ in models with minimal flavour violation}},  {\em Phys.~Lett.} {\bf B566} (2003) 115--119, [\href{http://arxiv.org/abs/hep-ph/0303060}{{\tt hep-ph/0303060}}].

\bibitem{Buras:2021nns}
A.~J. Buras and E.~Venturini, {\it {Searching for New Physics in Rare $K$ and $B$ Decays without $|V_{cb}|$ and $|V_{ub}|$ Uncertainties}},  {\em Acta Phys. Polon. B} {\bf 53} (9, 2021) A1, [\href{http://arxiv.org/abs/2109.11032}{{\tt arXiv:2109.11032}}].

\bibitem{Buras:2022wpw}
A.~J. Buras and E.~Venturini, {\it {The exclusive vision of rare K and B decays and of the quark mixing in the standard model}},  {\em Eur. Phys. J. C} {\bf 82} (2022), no.~7 615, [\href{http://arxiv.org/abs/2203.11960}{{\tt arXiv:2203.11960}}].

\bibitem{Buras:2022qip}
A.~J. Buras, {\it {Standard Model predictions for rare K and B decays without new physics infection}},  {\em Eur. Phys. J. C} {\bf 83} (2023), no.~1 66, [\href{http://arxiv.org/abs/2209.03968}{{\tt arXiv:2209.03968}}].

\bibitem{Aebischer:2023mbz}
J.~Aebischer, A.~J. Buras, and J.~Kumar, {\it {Kaon physics without new physics in $ \varepsilon _K$}},  {\em Eur. Phys. J. C} {\bf 83} (2023), no.~5 368, [\href{http://arxiv.org/abs/2302.00013}{{\tt arXiv:2302.00013}}].

\bibitem{Buras:2024per}
A.~J. Buras, {\it {Hunting new animalcula with rare K and B decays}},  {\em EPJ Web Conf.} {\bf 314} (2024) 00002, [\href{http://arxiv.org/abs/2411.03440}{{\tt arXiv:2411.03440}}].

\bibitem{Crivellin:2025txc}
A.~Crivellin and B.~Mellado, {\it {Anomalies in Particle Physcis}},  {\em PoS} {\bf DIS2024} (2025) 007.

\bibitem{Brivio:2017vri}
I.~Brivio and M.~Trott, {\it {The Standard Model as an Effective Field Theory}},  {\em Phys. Rept.} {\bf 793} (2019) 1--98, [\href{http://arxiv.org/abs/1706.08945}{{\tt arXiv:1706.08945}}].

\bibitem{Isidori:2023pyp}
G.~Isidori, F.~Wilsch, and D.~Wyler, {\it {The standard model effective field theory at work}},  {\em Rev. Mod. Phys.} {\bf 96} (2024), no.~1 015006, [\href{http://arxiv.org/abs/2303.16922}{{\tt arXiv:2303.16922}}].

\bibitem{Buchmuller:1985jz}
W.~Buchmuller and D.~Wyler, {\it {Effective Lagrangian Analysis of New Interactions and Flavor Conservation}},  {\em Nucl.~Phys.} {\bf B268} (1986) 621--653.

\bibitem{Grzadkowski:2010es}
B.~Grzadkowski, M.~Iskrzynski, M.~Misiak, and J.~Rosiek, {\it {Dimension-Six Terms in the Standard Model Lagrangian}},  {\em JHEP} {\bf 1010} (2010) 085, [\href{http://arxiv.org/abs/1008.4884}{{\tt arXiv:1008.4884}}].

\bibitem{Jenkins:2013zja}
E.~E. Jenkins, A.~V. Manohar, and M.~Trott, {\it {Renormalization Group Evolution of the Standard Model Dimension Six Operators I: Formalism and lambda Dependence}},  {\em JHEP} {\bf 10} (2013) 087, [\href{http://arxiv.org/abs/1308.2627}{{\tt arXiv:1308.2627}}].

\bibitem{Jenkins:2013wua}
E.~E. Jenkins, A.~V. Manohar, and M.~Trott, {\it {Renormalization Group Evolution of the Standard Model Dimension Six Operators II: Yukawa Dependence}},  {\em JHEP} {\bf 01} (2014) 035, [\href{http://arxiv.org/abs/1310.4838}{{\tt arXiv:1310.4838}}].

\bibitem{Alonso:2013hga}
R.~Alonso, E.~E. Jenkins, A.~V. Manohar, and M.~Trott, {\it {Renormalization Group Evolution of the Standard Model Dimension Six Operators III: Gauge Coupling Dependence and Phenomenology}},  {\em JHEP} {\bf 04} (2014) 159, [\href{http://arxiv.org/abs/1312.2014}{{\tt arXiv:1312.2014}}].

\bibitem{Alonso:2014zka}
R.~Alonso, H.-M. Chang, E.~E. Jenkins, A.~V. Manohar, and B.~Shotwell, {\it {Renormalization group evolution of dimension-six baryon number violating operators}},  {\em Phys. Lett.} {\bf B734} (2014) 302--307, [\href{http://arxiv.org/abs/1405.0486}{{\tt arXiv:1405.0486}}].

\bibitem{Weinberg:1979sa}
S.~Weinberg, {\it {Baryon and Lepton Nonconserving Processes}},  {\em Phys. Rev. Lett.} {\bf 43} (1979) 1566--1570.

\bibitem{Wilczek:1979hc}
F.~Wilczek and A.~Zee, {\it {Operator Analysis of Nucleon Decay}},  {\em Phys. Rev. Lett.} {\bf 43} (1979) 1571--1573.

\bibitem{Abbott:1980zj}
L.~F. Abbott and M.~B. Wise, {\it {The Effective Hamiltonian for Nucleon Decay}},  {\em Phys. Rev.} {\bf D22} (1980) 2208.

\bibitem{Jenkins:2017jig}
E.~E. Jenkins, A.~V. Manohar, and P.~Stoffer, {\it {Low-Energy Effective Field Theory below the Electroweak Scale: Operators and Matching}},  {\em JHEP} {\bf 03} (2018) 016, [\href{http://arxiv.org/abs/1709.04486}{{\tt arXiv:1709.04486}}].

\bibitem{Appelquist:1974tg}
T.~Appelquist and J.~Carazzone, {\it {Infrared Singularities and Massive Fields}},  {\em Phys. Rev. D} {\bf 11} (1975) 2856.

\bibitem{DeAngelis:2023bmd}
S.~De~Angelis and G.~Durieux, {\it {EFT matching from analyticity and unitarity}},  {\em SciPost Phys.} {\bf 16} (2024) 071, [\href{http://arxiv.org/abs/2308.00035}{{\tt arXiv:2308.00035}}].

\bibitem{Chala:2024llp}
M.~Chala, J.~L{\'o}pez~Miras, J.~Santiago, and F.~Vilches, {\it {Efficient on-shell matching}},  {\em SciPost Phys.} {\bf 18} (2025), no.~6 185, [\href{http://arxiv.org/abs/2411.12798}{{\tt arXiv:2411.12798}}].

\bibitem{Dong:2025wvf}
Z.~Dong, C.~Li, T.~Ma, J.~Shu, and Z.~Zhou, {\it {An Efficient On-shell Framework for EFT Matching}},  \href{http://arxiv.org/abs/2507.17829}{{\tt arXiv:2507.17829}}.

\bibitem{Zhang:2016pja}
Z.~Zhang, {\it {Covariant diagrams for one-loop matching}},  {\em JHEP} {\bf 05} (2017) 152, [\href{http://arxiv.org/abs/1610.00710}{{\tt arXiv:1610.00710}}].

\bibitem{Wilson:1969zs}
K.~G. Wilson, {\it {Nonlagrangian models of current algebra}},  {\em Phys. Rev.} {\bf 179} (1969) 1499--1512.

\bibitem{Zimmermann:1972tv}
W.~Zimmermann, {\it {Normal products and the short distance expansion in the perturbation theory of renormalizable interactions}},  {\em Annals Phys.} {\bf 77} (1973) 570--601. [Lect. Notes Phys.558,278(2000)].

\bibitem{Helset:2018fgq}
A.~Helset, M.~Paraskevas, and M.~Trott, {\it {Gauge fixing the Standard Model Effective Field Theory}},  {\em Phys. Rev. Lett.} {\bf 120} (2018), no.~25 251801, [\href{http://arxiv.org/abs/1803.08001}{{\tt arXiv:1803.08001}}].

\bibitem{Misiak:2018gvl}
M.~Misiak, M.~Paraskevas, J.~Rosiek, K.~Suxho, and B.~Zglinicki, {\it {Effective Field Theories in R$_\xi$ gauges}},  {\em JHEP} {\bf 02} (2019) 051, [\href{http://arxiv.org/abs/1812.11513}{{\tt arXiv:1812.11513}}].

\bibitem{Henning:2017fpj}
B.~Henning, X.~Lu, T.~Melia, and H.~Murayama, {\it {Operator bases, $S$-matrices, and their partition functions}},  {\em JHEP} {\bf 10} (2017) 199, [\href{http://arxiv.org/abs/1706.08520}{{\tt arXiv:1706.08520}}].

\bibitem{Henning:2015alf}
B.~Henning, X.~Lu, T.~Melia, and H.~Murayama, {\it {2, 84, 30, 993, 560, 15456, 11962, 261485, ...: Higher dimension operators in the SM EFT}},  {\em JHEP} {\bf 08} (2017) 016, [\href{http://arxiv.org/abs/1512.03433}{{\tt arXiv:1512.03433}}]. [Erratum: JHEP 09, 019 (2019)].

\bibitem{Fonseca:2017lem}
R.~M. Fonseca, {\it {The Sym2Int program: going from symmetries to interactions}},  {\em J. Phys. Conf. Ser.} {\bf 873} (2017), no.~1 012045, [\href{http://arxiv.org/abs/1703.05221}{{\tt arXiv:1703.05221}}].

\bibitem{Criado:2019ugp}
J.~C. Criado, {\it {BasisGen: automatic generation of operator bases}},  {\em Eur. Phys. J. C} {\bf 79} (2019), no.~3 256, [\href{http://arxiv.org/abs/1901.03501}{{\tt arXiv:1901.03501}}].

\bibitem{Gripaios:2018zrz}
B.~Gripaios and D.~Sutherland, {\it {DEFT: A program for operators in EFT}},  {\em JHEP} {\bf 01} (2019) 128, [\href{http://arxiv.org/abs/1807.07546}{{\tt arXiv:1807.07546}}].

\bibitem{Li:2022tec}
H.-L. Li, Z.~Ren, M.-L. Xiao, J.-H. Yu, and Y.-H. Zheng, {\it {Operators for generic effective field theory at any dimension: on-shell amplitude basis construction}},  {\em JHEP} {\bf 04} (2022) 140, [\href{http://arxiv.org/abs/2201.04639}{{\tt arXiv:2201.04639}}].

\bibitem{Shadmi:2018xan}
Y.~Shadmi and Y.~Weiss, {\it {Effective Field Theory Amplitudes the On-Shell Way: Scalar and Vector Couplings to Gluons}},  {\em JHEP} {\bf 02} (2019) 165, [\href{http://arxiv.org/abs/1809.09644}{{\tt arXiv:1809.09644}}].

\bibitem{Ma:2019gtx}
T.~Ma, J.~Shu, and M.-L. Xiao, {\it {Standard model effective field theory from on-shell amplitudes*}},  {\em Chin. Phys. C} {\bf 47} (2023), no.~2 023105, [\href{http://arxiv.org/abs/1902.06752}{{\tt arXiv:1902.06752}}].

\bibitem{Durieux:2019siw}
G.~Durieux and C.~S. Machado, {\it {Enumerating higher-dimensional operators with on-shell amplitudes}},  {\em Phys. Rev. D} {\bf 101} (2020), no.~9 095021, [\href{http://arxiv.org/abs/1912.08827}{{\tt arXiv:1912.08827}}].

\bibitem{AccettulliHuber:2021uoa}
M.~Accettulli~Huber and S.~De~Angelis, {\it {Standard Model EFTs via on-shell methods}},  {\em JHEP} {\bf 11} (2021) 221, [\href{http://arxiv.org/abs/2108.03669}{{\tt arXiv:2108.03669}}].

\bibitem{Balkin:2021dko}
R.~Balkin, G.~Durieux, T.~Kitahara, Y.~Shadmi, and Y.~Weiss, {\it {On-shell Higgsing for EFTs}},  {\em JHEP} {\bf 03} (2022) 129, [\href{http://arxiv.org/abs/2112.09688}{{\tt arXiv:2112.09688}}].

\bibitem{Durieux:2019eor}
G.~Durieux, T.~Kitahara, Y.~Shadmi, and Y.~Weiss, {\it {The electroweak effective field theory from on-shell amplitudes}},  {\em JHEP} {\bf 01} (2020) 119, [\href{http://arxiv.org/abs/1909.10551}{{\tt arXiv:1909.10551}}].

\bibitem{Durieux:2020gip}
G.~Durieux, T.~Kitahara, C.~S. Machado, Y.~Shadmi, and Y.~Weiss, {\it {Constructing massive on-shell contact terms}},  {\em JHEP} {\bf 12} (2020) 175, [\href{http://arxiv.org/abs/2008.09652}{{\tt arXiv:2008.09652}}].

\bibitem{Dong:2021vxo}
Z.-Y. Dong, T.~Ma, and J.~Shu, {\it {Constructing on-shell operator basis for all masses and spins}},  {\em Phys. Rev. D} {\bf 107} (2023), no.~11 L111901, [\href{http://arxiv.org/abs/2103.15837}{{\tt arXiv:2103.15837}}].

\bibitem{Li:2020gnx}
H.-L. Li, Z.~Ren, J.~Shu, M.-L. Xiao, J.-H. Yu, and Y.-H. Zheng, {\it {Complete set of dimension-eight operators in the standard model effective field theory}},  {\em Phys. Rev. D} {\bf 104} (2021), no.~1 015026, [\href{http://arxiv.org/abs/2005.00008}{{\tt arXiv:2005.00008}}].

\bibitem{Li:2020xlh}
H.-L. Li, Z.~Ren, M.-L. Xiao, J.-H. Yu, and Y.-H. Zheng, {\it {Complete set of dimension-nine operators in the standard model effective field theory}},  {\em Phys. Rev. D} {\bf 104} (2021), no.~1 015025, [\href{http://arxiv.org/abs/2007.07899}{{\tt arXiv:2007.07899}}].

\bibitem{Li:2020zfq}
H.-L. Li, J.~Shu, M.-L. Xiao, and J.-H. Yu, {\it {Depicting the Landscape of Generic Effective Field Theories}},  \href{http://arxiv.org/abs/2012.11615}{{\tt arXiv:2012.11615}}.

\bibitem{Arzt:1994gp}
C.~Arzt, M.~B. Einhorn, and J.~Wudka, {\it {Patterns of deviation from the standard model}},  {\em Nucl. Phys. B} {\bf 433} (1995) 41--66, [\href{http://arxiv.org/abs/hep-ph/9405214}{{\tt hep-ph/9405214}}].

\bibitem{Einhorn:2013kja}
M.~B. Einhorn and J.~Wudka, {\it {The Bases of Effective Field Theories}},  {\em Nucl. Phys. B} {\bf 876} (2013) 556--574, [\href{http://arxiv.org/abs/1307.0478}{{\tt arXiv:1307.0478}}].

\bibitem{Aebischer:2017ugx}
J.~Aebischer et~al., {\it {WCxf: an exchange format for Wilson coefficients beyond the Standard Model}},  {\em Comput. Phys. Commun.} {\bf 232} (2018) 71--83, [\href{http://arxiv.org/abs/1712.05298}{{\tt arXiv:1712.05298}}].

\bibitem{Giudice:2007fh}
G.~F. Giudice, C.~Grojean, A.~Pomarol, and R.~Rattazzi, {\it {The Strongly-Interacting Light Higgs}},  {\em JHEP} {\bf 06} (2007) 045, [\href{http://arxiv.org/abs/hep-ph/0703164}{{\tt hep-ph/0703164}}].

\bibitem{Contino:2013kra}
R.~Contino, M.~Ghezzi, C.~Grojean, M.~Muhlleitner, and M.~Spira, {\it {Effective Lagrangian for a light Higgs-like scalar}},  {\em JHEP} {\bf 07} (2013) 035, [\href{http://arxiv.org/abs/1303.3876}{{\tt arXiv:1303.3876}}].

\bibitem{Aoude:2019tzn}
R.~Aoude and C.~S. Machado, {\it {The Rise of SMEFT On-shell Amplitudes}},  {\em JHEP} {\bf 12} (2019) 058, [\href{http://arxiv.org/abs/1905.11433}{{\tt arXiv:1905.11433}}].

\bibitem{Dedes:2017zog}
A.~Dedes, W.~Materkowska, M.~Paraskevas, J.~Rosiek, and K.~Suxho, {\it {Feynman rules for the Standard Model Effective Field Theory in $R_{\xi}$ -gauges}},  {\em JHEP} {\bf 06} (2017) 143, [\href{http://arxiv.org/abs/1704.03888}{{\tt arXiv:1704.03888}}].

\bibitem{Dedes:2023zws}
A.~Dedes, J.~Rosiek, M.~Ryczkowski, K.~Suxho, and L.~Trifyllis, {\it {SmeftFR v3 \textendash{} Feynman rules generator for the Standard Model Effective Field Theory}},  {\em Comput. Phys. Commun.} {\bf 294} (2024) 108943, [\href{http://arxiv.org/abs/2302.01353}{{\tt arXiv:2302.01353}}].

\bibitem{Corbett:2020bqv}
T.~Corbett, {\it {The Feynman rules for the SMEFT in the background field gauge}},  {\em JHEP} {\bf 03} (2021) 001, [\href{http://arxiv.org/abs/2010.15852}{{\tt arXiv:2010.15852}}].

\bibitem{Corbett:2019cwl}
T.~Corbett, A.~Helset, and M.~Trott, {\it {Ward Identities for the Standard Model Effective Field Theory}},  {\em Phys. Rev. D} {\bf 101} (2020), no.~1 013005, [\href{http://arxiv.org/abs/1909.08470}{{\tt arXiv:1909.08470}}].

\bibitem{Corbett:2020ymv}
T.~Corbett and M.~Trott, {\it {One loop verification of SMEFT Ward Identities}},  {\em SciPost Phys.} {\bf 10} (2021), no.~6 144, [\href{http://arxiv.org/abs/2010.08451}{{\tt arXiv:2010.08451}}].

\bibitem{Li:2025ikn}
W.-F. Li, J.~Chen, Q.-J. Wang, and Z.-H. Yu, {\it {A Numerical Study on Gauge Symmetry of Electroweak Amplitudes}},  \href{http://arxiv.org/abs/2507.17429}{{\tt arXiv:2507.17429}}.

\bibitem{Kribs:2020jgn}
G.~D. Kribs, X.~Lu, A.~Martin, and T.~Tong, {\it {Custodial symmetry violation in the SMEFT}},  {\em Phys. Rev. D} {\bf 104} (2021), no.~5 056006, [\href{http://arxiv.org/abs/2009.10725}{{\tt arXiv:2009.10725}}].

\bibitem{Dekens:2019ept}
W.~Dekens and P.~Stoffer, {\it {Low-energy effective field theory below the electroweak scale: matching at one loop}},  {\em JHEP} {\bf 10} (2019) 197, [\href{http://arxiv.org/abs/1908.05295}{{\tt arXiv:1908.05295}}]. [Erratum: JHEP 11, 148 (2022)].

\bibitem{Passarino:2019yjx}
G.~Passarino, {\it {XEFT, the challenging path up the hill: dim = 6 and dim = 8}},  \href{http://arxiv.org/abs/1901.04177}{{\tt arXiv:1901.04177}}.

\bibitem{Alioli:2020kez}
S.~Alioli, R.~Boughezal, E.~Mereghetti, and F.~Petriello, {\it {Novel angular dependence in Drell-Yan lepton production via dimension-8 operators}},  {\em Phys. Lett. B} {\bf 809} (2020) 135703, [\href{http://arxiv.org/abs/2003.11615}{{\tt arXiv:2003.11615}}].

\bibitem{Alioli:2022fng}
S.~Alioli et~al., {\it {Theoretical developments in the SMEFT at dimension-8 and beyond}},  in {\em {2022 Snowmass Summer Study}}, 3, 2022.
\newblock \href{http://arxiv.org/abs/2203.06771}{{\tt arXiv:2203.06771}}.

\bibitem{Assi:2024zap}
B.~Assi and A.~Martin, {\it {Energy-enhanced dimension eight SMEFT effects in VBF Higgs production}},  {\em JHEP} {\bf 02} (2025) 029, [\href{http://arxiv.org/abs/2410.21563}{{\tt arXiv:2410.21563}}].

\bibitem{Degrande:2013kka}
C.~Degrande, {\it {A basis of dimension-eight operators for anomalous neutral triple gauge boson interactions}},  {\em JHEP} {\bf 02} (2014) 101, [\href{http://arxiv.org/abs/1308.6323}{{\tt arXiv:1308.6323}}].

\bibitem{Ellis:2024omd}
J.~Ellis, H.-J. He, R.-Q. Xiao, S.-P. Zeng, and J.~Zheng, {\it {UV completion of neutral triple gauge couplings}},  {\em Phys. Rev. D} {\bf 111} (2025), no.~1 015007, [\href{http://arxiv.org/abs/2408.12508}{{\tt arXiv:2408.12508}}].

\bibitem{Corbett:2021eux}
T.~Corbett, A.~Helset, A.~Martin, and M.~Trott, {\it {EWPD in the SMEFT to dimension eight}},  {\em JHEP} {\bf 06} (2021) 076, [\href{http://arxiv.org/abs/2102.02819}{{\tt arXiv:2102.02819}}].

\bibitem{Adhikary:2025gdh}
N.~Adhikary, T.~Biswas, J.~Chakrabortty, C.~Englert, and M.~Spannowsky, {\it {Electroweak Scalar Effects Beyond Dimension-6 in SMEFT}},  \href{http://arxiv.org/abs/2501.12160}{{\tt arXiv:2501.12160}}.

\bibitem{Ardu:2021koz}
M.~Ardu and S.~Davidson, {\it {What is Leading Order for LFV in SMEFT?}},  {\em JHEP} {\bf 08} (2021) 002, [\href{http://arxiv.org/abs/2103.07212}{{\tt arXiv:2103.07212}}].

\bibitem{Burgess:2021ylu}
C.~P. Burgess, S.~Hamoudou, J.~Kumar, and D.~London, {\it {Beyond the standard model effective field theory with $B \rightarrow c \tau^- \overline{\nu}$}},  {\em Phys. Rev. D} {\bf 105} (2022), no.~7 073008, [\href{http://arxiv.org/abs/2111.07421}{{\tt arXiv:2111.07421}}].

\bibitem{Azatov:2016sqh}
A.~Azatov, R.~Contino, C.~S. Machado, and F.~Riva, {\it {Helicity selection rules and noninterference for BSM amplitudes}},  {\em Phys. Rev. D} {\bf 95} (2017), no.~6 065014, [\href{http://arxiv.org/abs/1607.05236}{{\tt arXiv:1607.05236}}].

\bibitem{Kobach:2016ami}
A.~Kobach, {\it {Baryon Number, Lepton Number, and Operator Dimension in the Standard Model}},  {\em Phys. Lett. B} {\bf 758} (2016) 455--457, [\href{http://arxiv.org/abs/1604.05726}{{\tt arXiv:1604.05726}}].

\bibitem{Helset:2019eyc}
A.~Helset and A.~Kobach, {\it {Baryon Number, Lepton Number, and Operator Dimension in the SMEFT with Flavor Symmetries}},  {\em Phys. Lett. B} {\bf 800} (2020) 135132, [\href{http://arxiv.org/abs/1909.05853}{{\tt arXiv:1909.05853}}].

\bibitem{Heeck:2025btc}
J.~Heeck and D.~Sokhashvili, {\it {Revisiting the connection of baryon number, lepton number, and operator dimension}},  {\em Phys. Lett. B} {\bf 868} (2025) 139791, [\href{http://arxiv.org/abs/2505.06172}{{\tt arXiv:2505.06172}}].

\bibitem{Lehman:2014jma}
L.~Lehman, {\it {Extending the Standard Model Effective Field Theory with the Complete Set of Dimension-7 Operators}},  {\em Phys. Rev. D} {\bf 90} (2014), no.~12 125023, [\href{http://arxiv.org/abs/1410.4193}{{\tt arXiv:1410.4193}}].

\bibitem{Liao:2016hru}
Y.~Liao and X.-D. Ma, {\it {Renormalization Group Evolution of Dimension-seven Baryon- and Lepton-number-violating Operators}},  {\em JHEP} {\bf 11} (2016) 043, [\href{http://arxiv.org/abs/1607.07309}{{\tt arXiv:1607.07309}}].

\bibitem{Lehman:2015coa}
L.~Lehman and A.~Martin, {\it {Low-derivative operators of the Standard Model effective field theory via Hilbert series methods}},  {\em JHEP} {\bf 02} (2016) 081, [\href{http://arxiv.org/abs/1510.00372}{{\tt arXiv:1510.00372}}].

\bibitem{Murphy:2020rsh}
C.~W. Murphy, {\it {Dimension-8 operators in the Standard Model Effective Field Theory}},  {\em JHEP} {\bf 10} (2020) 174, [\href{http://arxiv.org/abs/2005.00059}{{\tt arXiv:2005.00059}}].

\bibitem{Durieux:2024zrg}
G.~Durieux, G.~N. Remmen, N.~L. Rodd, O.~J.~P. \'Eboli, M.~C. Gonzalez-Garcia, D.~Kondo, H.~Murayama, and R.~Okabe, {\it {LHC EFT WG note: Basis for anomalous quartic gauge couplings}},  \href{http://arxiv.org/abs/2411.02483}{{\tt arXiv:2411.02483}}.

\bibitem{Corbett:2024yoy}
T.~Corbett, J.~Desai, O.~J.~P. Eboli, and M.~C. Gonzalez-Garcia, {\it {Dimension-eight operator basis for universal standard model effective field theory}},  {\em Phys. Rev. D} {\bf 110} (2024), no.~3 033003, [\href{http://arxiv.org/abs/2404.03720}{{\tt arXiv:2404.03720}}].

\bibitem{Liao:2020jmn}
Y.~Liao and X.-D. Ma, {\it {An explicit construction of the dimension-9 operator basis in the standard model effective field theory}},  {\em JHEP} {\bf 11} (2020) 152, [\href{http://arxiv.org/abs/2007.08125}{{\tt arXiv:2007.08125}}].

\bibitem{Harlander:2023psl}
R.~V. Harlander, T.~Kempkens, and M.~C. Schaaf, {\it {Standard model effective field theory up to mass dimension 12}},  {\em Phys. Rev. D} {\bf 108} (2023), no.~5 055020, [\href{http://arxiv.org/abs/2305.06832}{{\tt arXiv:2305.06832}}].

\bibitem{Gherardi:2020det}
V.~Gherardi, D.~Marzocca, and E.~Venturini, {\it {Matching scalar leptoquarks to the SMEFT at one loop}},  {\em JHEP} {\bf 07} (2020) 225, [\href{http://arxiv.org/abs/2003.12525}{{\tt arXiv:2003.12525}}]. [Erratum: JHEP 01, 006 (2021)].

\bibitem{Zhang:2023kvw}
D.~Zhang, {\it {Renormalization group equations for the SMEFT operators up to dimension seven}},  {\em JHEP} {\bf 10} (2023) 148, [\href{http://arxiv.org/abs/2306.03008}{{\tt arXiv:2306.03008}}].

\bibitem{Chala:2021cgt}
M.~Chala, A.~D\'\i{}az-Carmona, and G.~Guedes, {\it {A Green\textquoteright{}s basis for the bosonic SMEFT to dimension 8}},  {\em JHEP} {\bf 05} (2022) 138, [\href{http://arxiv.org/abs/2112.12724}{{\tt arXiv:2112.12724}}].

\bibitem{Ren:2022tvi}
Z.~Ren and J.-H. Yu, {\it {A complete set of the dimension-8 Green\textquoteright{}s basis operators in the Standard Model effective field theory}},  {\em JHEP} {\bf 02} (2024) 134, [\href{http://arxiv.org/abs/2211.01420}{{\tt arXiv:2211.01420}}].

\bibitem{Naterop:2023dek}
L.~Naterop and P.~Stoffer, {\it {Low-energy effective field theory below the electroweak scale: one-loop renormalization in the \textquoteright{}t Hooft-Veltman scheme}},  {\em JHEP} {\bf 02} (2024) 068, [\href{http://arxiv.org/abs/2310.13051}{{\tt arXiv:2310.13051}}].

\bibitem{Liao:2020zyx}
Y.~Liao, X.-D. Ma, and Q.-Y. Wang, {\it {Extending low energy effective field theory with a complete set of dimension-7 operators}},  {\em JHEP} {\bf 08} (2020) 162, [\href{http://arxiv.org/abs/2005.08013}{{\tt arXiv:2005.08013}}].

\bibitem{Murphy:2020cly}
C.~W. Murphy, {\it {Low-Energy Effective Field Theory below the Electroweak Scale: Dimension-8 Operators}},  {\em JHEP} {\bf 04} (2021) 101, [\href{http://arxiv.org/abs/2012.13291}{{\tt arXiv:2012.13291}}].

\bibitem{Li:2020tsi}
H.-L. Li, Z.~Ren, M.-L. Xiao, J.-H. Yu, and Y.-H. Zheng, {\it {Low energy effective field theory operator basis at d \ensuremath{\leq} 9}},  {\em JHEP} {\bf 06} (2021) 138, [\href{http://arxiv.org/abs/2012.09188}{{\tt arXiv:2012.09188}}].

\bibitem{Gerard:1982mm}
J.~M. Gerard, {\it {Fermion Mass Spectrum in $SU(2)_L\times U(1)$}},  {\em Z. Phys. C} {\bf 18} (1983) 145.

\bibitem{Aebischer:2015fzz}
J.~Aebischer, A.~Crivellin, M.~Fael, and C.~Greub, {\it {Matching of gauge invariant dimension-six operators for $b\to s$ and $b\to c$ transitions}},  {\em JHEP} {\bf 05} (2016) 037, [\href{http://arxiv.org/abs/1512.02830}{{\tt arXiv:1512.02830}}].

\bibitem{Aebischer:2016xmn}
J.~Aebischer, A.~Crivellin, M.~Fael, and C.~Greub, {\it {1-Loop Matching of gauge invariant dim-6 operators for B decays}},  {\em PoS} {\bf BEAUTY2016} (2016) 064, [\href{http://arxiv.org/abs/1606.02588}{{\tt arXiv:1606.02588}}].

\bibitem{LHCHiggsCrossSectionWorkingGroup:2016ypw}
{\bf LHC Higgs Cross Section Working Group} Collaboration, D.~de~Florian et~al., {\it {Handbook of LHC Higgs Cross Sections: 4. Deciphering the Nature of the Higgs Sector}},  \href{http://arxiv.org/abs/1610.07922}{{\tt arXiv:1610.07922}}.

\bibitem{Silvestrini:2018dos}
L.~Silvestrini and M.~Valli, {\it {Model-independent Bounds on the Standard Model Effective Theory from Flavour Physics}},  {\em Phys. Lett. B} {\bf 799} (2019) 135062, [\href{http://arxiv.org/abs/1812.10913}{{\tt arXiv:1812.10913}}].

\bibitem{Bordone:2017bld}
M.~Bordone, C.~Cornella, J.~Fuentes-Martin, and G.~Isidori, {\it {A three-site gauge model for flavor hierarchies and flavor anomalies}},  {\em Phys. Lett.} {\bf B779} (2018) 317--323, [\href{http://arxiv.org/abs/1712.01368}{{\tt arXiv:1712.01368}}].

\bibitem{Gherardi:2019zil}
V.~Gherardi, D.~Marzocca, M.~Nardecchia, and A.~Romanino, {\it {Rank-One Flavor Violation and B-meson anomalies}},  {\em JHEP} {\bf 10} (2019) 112, [\href{http://arxiv.org/abs/1903.10954}{{\tt arXiv:1903.10954}}].

\bibitem{Davighi:2023evx}
J.~Davighi and B.~A. Stefanek, {\it {Deconstructed hypercharge: a natural model of flavour}},  {\em JHEP} {\bf 11} (2023) 100, [\href{http://arxiv.org/abs/2305.16280}{{\tt arXiv:2305.16280}}].

\bibitem{Davighi:2022bqf}
J.~Davighi, G.~Isidori, and M.~Pesut, {\it {Electroweak-flavour and quark-lepton unification: a family non-universal path}},  {\em JHEP} {\bf 04} (2023) 030, [\href{http://arxiv.org/abs/2212.06163}{{\tt arXiv:2212.06163}}].

\bibitem{Davighi:2023iks}
J.~Davighi and G.~Isidori, {\it {Non-universal gauge interactions addressing the inescapable link between Higgs and flavour}},  {\em JHEP} {\bf 07} (2023) 147, [\href{http://arxiv.org/abs/2303.01520}{{\tt arXiv:2303.01520}}].

\bibitem{Allwicher:2023shc}
L.~Allwicher, C.~Cornella, G.~Isidori, and B.~A. Stefanek, {\it {New physics in the third generation. A comprehensive SMEFT analysis and future prospects}},  {\em JHEP} {\bf 03} (2024) 049, [\href{http://arxiv.org/abs/2311.00020}{{\tt arXiv:2311.00020}}].

\bibitem{Davighi:2023xqn}
J.~Davighi, A.~Gosnay, D.~J. Miller, and S.~Renner, {\it {Phenomenology of a Deconstructed Electroweak Force}},  {\em JHEP} {\bf 05} (2024) 085, [\href{http://arxiv.org/abs/2312.13346}{{\tt arXiv:2312.13346}}].

\bibitem{Marzocca:2024hua}
D.~Marzocca, M.~Nardecchia, A.~Stanzione, and C.~Toni, {\it {Implications of $B \rightarrow K \nu {\bar{\nu }}$ under rank-one flavor violation hypothesis}},  {\em Eur. Phys. J. C} {\bf 84} (2024), no.~11 1217, [\href{http://arxiv.org/abs/2404.06533}{{\tt arXiv:2404.06533}}].

\bibitem{Moreno-Sanchez:2025bzz}
A.~Moreno-S{\'a}nchez and A.~Palavri{\'c}, {\it {Leptonic flavor from a modular A4 symmetry: UV mediators and SMEFT realizations}},  {\em Phys. Rev. D} {\bf 112} (2025), no.~7 075002, [\href{http://arxiv.org/abs/2505.01535}{{\tt arXiv:2505.01535}}].

\bibitem{Giarnetti:2025idu}
A.~Giarnetti, S.~Marciano, D.~Meloni, and M.~Rettaroli, {\it {A Roadmap for neutrino charge assignments in $U(2)_F$ Flavor Models: Implications for LFV processes and leptonic anomalous magnetic moments}},  \href{http://arxiv.org/abs/2505.20281}{{\tt arXiv:2505.20281}}.

\bibitem{Banks:2025baf}
H.~Banks, G.~Crawford, M.~McCullough, and D.~Sutherland, {\it {Flavour, Accidentally}},  \href{http://arxiv.org/abs/2510.03403}{{\tt arXiv:2510.03403}}.

\bibitem{Isidori:2025rci}
G.~Isidori, P.~Paradisi, A.~Sainaghi, and N.~Selimovic, {\it {Anarchic neutrinos from flavor deconstruction: phenomenology of the lepton sector}},  \href{http://arxiv.org/abs/2510.23703}{{\tt arXiv:2510.23703}}.

\bibitem{Alonso:2014csa}
R.~Alonso, B.~Grinstein, and J.~Martin~Camalich, {\it {$SU(2)\times U(1)$ gauge invariance and the shape of new physics in rare $B$ decays}},  {\em Phys. Rev. Lett.} {\bf 113} (2014) 241802, [\href{http://arxiv.org/abs/1407.7044}{{\tt arXiv:1407.7044}}].

\bibitem{Feruglio:2017rjo}
F.~Feruglio, P.~Paradisi, and A.~Pattori, {\it {On the Importance of Electroweak Corrections for B Anomalies}},  {\em JHEP} {\bf 09} (2017) 061, [\href{http://arxiv.org/abs/1705.00929}{{\tt arXiv:1705.00929}}].

\bibitem{Buttazzo:2017ixm}
D.~Buttazzo, A.~Greljo, G.~Isidori, and D.~Marzocca, {\it {B-physics anomalies: a guide to combined explanations}},  {\em JHEP} {\bf 11} (2017) 044, [\href{http://arxiv.org/abs/1706.07808}{{\tt arXiv:1706.07808}}].

\bibitem{Kumar:2018kmr}
J.~Kumar, D.~London, and R.~Watanabe, {\it {Combined Explanations of the $b \to s \mu^+ \mu^-$ and $b \to c \tau^- {\bar\nu}$ Anomalies: a General Model Analysis}},  {\em Phys. Rev.} {\bf D99} (2019), no.~1 015007, [\href{http://arxiv.org/abs/1806.07403}{{\tt arXiv:1806.07403}}].

\bibitem{Aebischer:2018iyb}
J.~Aebischer, J.~Kumar, P.~Stangl, and D.~M. Straub, {\it {A Global Likelihood for Precision Constraints and Flavour Anomalies}},  {\em Eur. Phys. J. C} {\bf 79} (2019), no.~6 509, [\href{http://arxiv.org/abs/1810.07698}{{\tt arXiv:1810.07698}}].

\bibitem{Ciuchini:2019usw}
M.~Ciuchini, A.~M. Coutinho, M.~Fedele, E.~Franco, A.~Paul, L.~Silvestrini, and M.~Valli, {\it {New Physics in $b \to s \ell^+ \ell^-$ confronts new data on Lepton Universality}},  {\em Eur. Phys. J. C} {\bf 79} (2019), no.~8 719, [\href{http://arxiv.org/abs/1903.09632}{{\tt arXiv:1903.09632}}].

\bibitem{Aebischer:2019mlg}
J.~Aebischer, W.~Altmannshofer, D.~Guadagnoli, M.~Reboud, P.~Stangl, and D.~M. Straub, {\it {$B$-decay discrepancies after Moriond 2019}},  {\em Eur. Phys. J. C} {\bf 80} (2020), no.~3 252, [\href{http://arxiv.org/abs/1903.10434}{{\tt arXiv:1903.10434}}].

\bibitem{Aebischer:2018bkb}
J.~Aebischer, J.~Kumar, and D.~M. Straub, {\it {Wilson: a Python package for the running and matching of Wilson coefficients above and below the electroweak scale}},  {\em Eur. Phys. J.} {\bf C78} (2018), no.~12 1026, [\href{http://arxiv.org/abs/1804.05033}{{\tt arXiv:1804.05033}}].

\bibitem{Aebischer:2020lsx}
J.~Aebischer and J.~Kumar, {\it {Flavour violating effects of Yukawa running in SMEFT}},  {\em JHEP} {\bf 09} (2020) 187, [\href{http://arxiv.org/abs/2005.12283}{{\tt arXiv:2005.12283}}].

\bibitem{Datta:2025csr}
A.~Datta, J.-F. Fortin, J.~Kumar, D.~London, D.~Marfatia, and N.~Sanfa\c{c}on, {\it {Model dependence in SMEFT}},  \href{http://arxiv.org/abs/2502.04634}{{\tt arXiv:2502.04634}}.

\bibitem{Jenkins:2017dyc}
E.~E. Jenkins, A.~V. Manohar, and P.~Stoffer, {\it {Low-Energy Effective Field Theory below the Electroweak Scale: Anomalous Dimensions}},  {\em JHEP} {\bf 01} (2018) 084, [\href{http://arxiv.org/abs/1711.05270}{{\tt arXiv:1711.05270}}].

\bibitem{Manton:2024eli}
N.~S. Manton, {\it {A $ {\boldmath{CP}}^2 $ SMEFT}},  {\em JHEP} {\bf 04} (2025) 180, [\href{http://arxiv.org/abs/2411.09521}{{\tt arXiv:2411.09521}}].

\bibitem{Brivio:2017bnu}
I.~Brivio and M.~Trott, {\it {Scheming in the SMEFT... and a reparameterization invariance!}},  {\em JHEP} {\bf 07} (2017) 148, [\href{http://arxiv.org/abs/1701.06424}{{\tt arXiv:1701.06424}}]. [Addendum: JHEP 05, 136 (2018)].

\bibitem{Brivio:2017btx}
I.~Brivio, Y.~Jiang, and M.~Trott, {\it {The SMEFTsim package, theory and tools}},  {\em JHEP} {\bf 12} (2017) 070, [\href{http://arxiv.org/abs/1709.06492}{{\tt arXiv:1709.06492}}].

\bibitem{Falkowski:2014tna}
A.~Falkowski and F.~Riva, {\it {Model-independent precision constraints on dimension-6 operators}},  {\em JHEP} {\bf 02} (2015) 039, [\href{http://arxiv.org/abs/1411.0669}{{\tt arXiv:1411.0669}}].

\bibitem{Descotes-Genon:2018foz}
S.~Descotes-Genon, A.~Falkowski, M.~Fedele, M.~Gonz\'alez-Alonso, and J.~Virto, {\it {The CKM parameters in the SMEFT}},  {\em JHEP} {\bf 05} (2019) 172, [\href{http://arxiv.org/abs/1812.08163}{{\tt arXiv:1812.08163}}].

\bibitem{Wolfenstein:1983yz}
L.~Wolfenstein, {\it {Parametrization of the Kobayashi-Maskawa Matrix}},  {\em Phys.~Rev.~Lett.} {\bf 51} (1983) 1945.

\bibitem{Buras:1994ec}
A.~J. Buras, M.~E. Lautenbacher, and G.~Ostermaier, {\it {Waiting for the top quark mass, $K^+ \to \pi^+ \nu\bar\nu$, $B_s^0 - \bar B_s^0$ mixing and CP asymmetries in $B$ decays}},  {\em Phys.~Rev.} {\bf D50} (1994) 3433--3446, [\href{http://arxiv.org/abs/hep-ph/9403384}{{\tt hep-ph/9403384}}].

\bibitem{Trott:2023jrw}
M.~Trott, {\it {$\alpha_s$ as an input parameter in the SMEFT}},  \href{http://arxiv.org/abs/2306.14784}{{\tt arXiv:2306.14784}}.

\bibitem{FlavourLatticeAveragingGroup:2019iem}
{\bf Flavour Lattice Averaging Group} Collaboration, S.~Aoki et~al., {\it {FLAG Review 2019: Flavour Lattice Averaging Group (FLAG)}},  {\em Eur. Phys. J. C} {\bf 80} (2020), no.~2 113, [\href{http://arxiv.org/abs/1902.08191}{{\tt arXiv:1902.08191}}].

\bibitem{Feruglio:2016gvd}
F.~Feruglio, P.~Paradisi, and A.~Pattori, {\it {Revisiting Lepton Flavor Universality in B Decays}},  {\em Phys. Rev. Lett.} {\bf 118} (2017), no.~1 011801, [\href{http://arxiv.org/abs/1606.00524}{{\tt arXiv:1606.00524}}].

\bibitem{Bobeth:2016llm}
C.~Bobeth, A.~J. Buras, A.~Celis, and M.~Jung, {\it {Patterns of Flavour Violation in Models with Vector-Like Quarks}},  {\em JHEP} {\bf 04} (2017) 079, [\href{http://arxiv.org/abs/1609.04783}{{\tt arXiv:1609.04783}}].

\bibitem{Bobeth:2017xry}
C.~Bobeth, A.~J. Buras, A.~Celis, and M.~Jung, {\it {Yukawa enhancement of $Z$-mediated new physics in $\Delta S = 2$ and $\Delta B = 2$ processes}},  {\em JHEP} {\bf 07} (2017) 124, [\href{http://arxiv.org/abs/1703.04753}{{\tt arXiv:1703.04753}}].

\bibitem{Gonzalez-Alonso:2017iyc}
M.~González-Alonso, J.~Martin~Camalich, and K.~Mimouni, {\it {Renormalization-group evolution of new physics contributions to (semi)leptonic meson decays}},  {\em Phys. Lett.} {\bf B772} (2017) 777--785, [\href{http://arxiv.org/abs/1706.00410}{{\tt arXiv:1706.00410}}].

\bibitem{Buchalla:2022vjp}
G.~Buchalla, G.~Heinrich, C.~M\"uller-Salditt, and F.~Pandler, {\it {Loop counting matters in SMEFT}},  {\em SciPost Phys.} {\bf 15} (2023), no.~3 088, [\href{http://arxiv.org/abs/2204.11808}{{\tt arXiv:2204.11808}}].

\bibitem{Jenkins:2013fya}
E.~E. Jenkins, A.~V. Manohar, and M.~Trott, {\it {On Gauge Invariance and Minimal Coupling}},  {\em JHEP} {\bf 09} (2013) 063, [\href{http://arxiv.org/abs/1305.0017}{{\tt arXiv:1305.0017}}].

\bibitem{Jenkins:2013sda}
E.~E. Jenkins, A.~V. Manohar, and M.~Trott, {\it {Naive Dimensional Analysis Counting of Gauge Theory Amplitudes and Anomalous Dimensions}},  {\em Phys. Lett. B} {\bf 726} (2013) 697--702, [\href{http://arxiv.org/abs/1309.0819}{{\tt arXiv:1309.0819}}].

\bibitem{Celis:2017hod}
A.~Celis, J.~Fuentes-Martin, A.~Vicente, and J.~Virto, {\it {DsixTools: The Standard Model Effective Field Theory Toolkit}},  {\em Eur. Phys. J.} {\bf C77} (2017), no.~6 405, [\href{http://arxiv.org/abs/1704.04504}{{\tt arXiv:1704.04504}}].

\bibitem{Chankowski:1993tx}
P.~H. Chankowski and Z.~Pluciennik, {\it {Renormalization group equations for seesaw neutrino masses}},  {\em Phys. Lett. B} {\bf 316} (1993) 312--317, [\href{http://arxiv.org/abs/hep-ph/9306333}{{\tt hep-ph/9306333}}].

\bibitem{Babu:1993qv}
K.~S. Babu, C.~N. Leung, and J.~T. Pantaleone, {\it {Renormalization of the neutrino mass operator}},  {\em Phys. Lett. B} {\bf 319} (1993) 191--198, [\href{http://arxiv.org/abs/hep-ph/9309223}{{\tt hep-ph/9309223}}].

\bibitem{Antusch:2001ck}
S.~Antusch, M.~Drees, J.~Kersten, M.~Lindner, and M.~Ratz, {\it {Neutrino mass operator renormalization revisited}},  {\em Phys. Lett. B} {\bf 519} (2001) 238--242, [\href{http://arxiv.org/abs/hep-ph/0108005}{{\tt hep-ph/0108005}}].

\bibitem{Ibarra:2024tpt}
A.~Ibarra, N.~Leister, and D.~Zhang, {\it {Complete two-loop renormalization group equation of the Weinberg operator}},  {\em JHEP} {\bf 03} (2025) 214, [\href{http://arxiv.org/abs/2411.08011}{{\tt arXiv:2411.08011}}].

\bibitem{Davidson:2018zuo}
S.~Davidson, M.~Gorbahn, and M.~Leak, {\it {Majorana neutrino masses in the renormalization group equations for lepton flavor violation}},  {\em Phys. Rev. D} {\bf 98} (2018), no.~9 095014, [\href{http://arxiv.org/abs/1807.04283}{{\tt arXiv:1807.04283}}].

\bibitem{Naterop:2025nqv}
L.~Naterop and P.~Stoffer, {\it {Renormalization-group equations of the LEFT at two loops: dimension-five effects}},  {\em JHEP} {\bf 06} (2025) 007, [\href{http://arxiv.org/abs/2412.13251}{{\tt arXiv:2412.13251}}].

\bibitem{deBlas:2017xtg}
J.~de~Blas, J.~C. Criado, M.~Perez-Victoria, and J.~Santiago, {\it {Effective description of general extensions of the Standard Model: the complete tree-level dictionary}},  {\em JHEP} {\bf 03} (2018) 109, [\href{http://arxiv.org/abs/1711.10391}{{\tt arXiv:1711.10391}}].

\bibitem{Guedes:2023azv}
G.~Guedes, P.~Olgoso, and J.~Santiago, {\it {Towards the one loop IR/UV dictionary in the SMEFT: One loop generated operators from new scalars and fermions}},  {\em SciPost Phys.} {\bf 15} (2023), no.~4 143, [\href{http://arxiv.org/abs/2303.16965}{{\tt arXiv:2303.16965}}].

\bibitem{Guedes:2024vuf}
G.~Guedes and P.~Olgoso, {\it {From the EFT to the UV: the complete SMEFT one-loop dictionary}},  \href{http://arxiv.org/abs/2412.14253}{{\tt arXiv:2412.14253}}.

\bibitem{Buchalla:2019wsc}
G.~Buchalla, A.~Celis, C.~Krause, and J.-N. Toelstede, {\it {Master Formula for One-Loop Renormalization of Bosonic SMEFT Operators}},  \href{http://arxiv.org/abs/1904.07840}{{\tt arXiv:1904.07840}}.

\bibitem{Fonseca:2025zjb}
R.~M. Fonseca, P.~Olgoso, and J.~Santiago, {\it {Renormalization of general Effective Field Theories: Formalism and renormalization of bosonic operators}},  \href{http://arxiv.org/abs/2501.13185}{{\tt arXiv:2501.13185}}.

\bibitem{Misiak:2025xzq}
M.~Misiak and I.~Na{\l}{\k{e}}cz, {\it {One-loop renormalization group equations in generic effective field theories. Part I. Bosonic operators}},  {\em JHEP} {\bf 06} (2025) 210, [\href{http://arxiv.org/abs/2501.17134}{{\tt arXiv:2501.17134}}].

\bibitem{Aebischer:2025zxg}
J.~Aebischer, L.~C. Bresciani, and N.~Selimovic, {\it {Anomalous dimension of a general effective gauge theory. Part I. Bosonic sector}},  {\em JHEP} {\bf 08} (2025) 209, [\href{http://arxiv.org/abs/2502.14030}{{\tt arXiv:2502.14030}}].

\bibitem{Henriksson:2025hwi}
J.~Henriksson, F.~Herzog, S.~R. Kousvos, and J.~Roosmale~Nepveu, {\it {Multi-loop spectra in general scalar EFTs and CFTs}},  \href{http://arxiv.org/abs/2507.12518}{{\tt arXiv:2507.12518}}.

\bibitem{Baratella:2020lzz}
P.~Baratella, C.~Fernandez, and A.~Pomarol, {\it {Renormalization of Higher-Dimensional Operators from On-shell Amplitudes}},  {\em Nucl. Phys. B} {\bf 959} (2020) 115155, [\href{http://arxiv.org/abs/2005.07129}{{\tt arXiv:2005.07129}}].

\bibitem{Machado:2022ozb}
C.~S. Machado, S.~Renner, and D.~Sutherland, {\it {Building blocks of the flavourful SMEFT RG}},  {\em JHEP} {\bf 03} (2023) 226, [\href{http://arxiv.org/abs/2210.09316}{{\tt arXiv:2210.09316}}].

\bibitem{Aebischer:2022anv}
J.~Aebischer, A.~J. Buras, and J.~Kumar, {\it {NLO QCD renormalization group evolution for nonleptonic \ensuremath{\Delta}F=2 transitions in the SMEFT}},  {\em Phys. Rev. D} {\bf 106} (2022), no.~3 035003, [\href{http://arxiv.org/abs/2203.11224}{{\tt arXiv:2203.11224}}].

\bibitem{Aebischer:2020dsw}
J.~Aebischer, C.~Bobeth, A.~J. Buras, and J.~Kumar, {\it {SMEFT ATLAS of $\Delta$F = 2 transitions}},  {\em JHEP} {\bf 12} (2020) 187, [\href{http://arxiv.org/abs/2009.07276}{{\tt arXiv:2009.07276}}].

\bibitem{Banik:2023ogi}
S.~Banik and A.~Crivellin, {\it {Renormalization group evolution with scalar leptoquarks}},  {\em JHEP} {\bf 11} (2023) 121, [\href{http://arxiv.org/abs/2307.06800}{{\tt arXiv:2307.06800}}].

\bibitem{EliasMiro:2020tdv}
J.~Elias~Mir\'o, J.~Ingoldby, and M.~Riembau, {\it {EFT anomalous dimensions from the S-matrix}},  {\em JHEP} {\bf 09} (2020) 163, [\href{http://arxiv.org/abs/2005.06983}{{\tt arXiv:2005.06983}}].

\bibitem{Bern:2020ikv}
Z.~Bern, J.~Parra-Martinez, and E.~Sawyer, {\it {Structure of two-loop SMEFT anomalous dimensions via on-shell methods}},  {\em JHEP} {\bf 10} (2020) 211, [\href{http://arxiv.org/abs/2005.12917}{{\tt arXiv:2005.12917}}].

\bibitem{Baratella:2022nog}
P.~Baratella, S.~Maggio, M.~Stadlbauer, and T.~Theil, {\it {Two-loop infrared renormalization with on-shell methods}},  {\em Eur. Phys. J. C} {\bf 83} (2023), no.~8 751, [\href{http://arxiv.org/abs/2207.08831}{{\tt arXiv:2207.08831}}].

\bibitem{Baratella:2020dvw}
P.~Baratella, C.~Fernandez, B.~von Harling, and A.~Pomarol, {\it {Anomalous Dimensions of Effective Theories from Partial Waves}},  {\em JHEP} {\bf 03} (2021) 287, [\href{http://arxiv.org/abs/2010.13809}{{\tt arXiv:2010.13809}}].

\bibitem{Duhr:2025zqw}
C.~Duhr, A.~Vasquez, G.~Ventura, and E.~Vryonidou, {\it {Two-loop renormalisation of quark and gluon fields in the SMEFT}},  {\em JHEP} {\bf 07} (2025) 160, [\href{http://arxiv.org/abs/2503.01954}{{\tt arXiv:2503.01954}}].

\bibitem{DiNoi:2025arz}
S.~Di~Noi and R.~Gr{\"o}ber, {\it {Two loops, four tops and two {\ensuremath{\gamma}}5 schemes: A renormalization story}},  {\em Phys. Lett. B} {\bf 869} (2025) 139878, [\href{http://arxiv.org/abs/2507.10295}{{\tt arXiv:2507.10295}}].

\bibitem{DiNoi:2024ajj}
S.~Di~Noi, R.~Gr\"ober, and M.~K. Mandal, {\it {Two-loop running effects in Higgs physics in Standard Model Effective Field Theory}},  {\em JHEP} {\bf 12} (2025) 220, [\href{http://arxiv.org/abs/2408.03252}{{\tt arXiv:2408.03252}}].

\bibitem{DiNoi:2025tka}
S.~Di~Noi, B.~A. Erdelyi, and R.~Gr{\"o}ber, {\it {Complete two-loop Yukawa-induced running of the Higgs-gluon coupling in SMEFT}},  \href{http://arxiv.org/abs/2510.14680}{{\tt arXiv:2510.14680}}.

\bibitem{Banik:2025wpi}
S.~Banik, A.~Crivellin, L.~Naterop, and P.~Stoffer, {\it {Two-loop anomalous dimensions for baryon-number-violating operators in SMEFT}},  \href{http://arxiv.org/abs/2510.08682}{{\tt arXiv:2510.08682}}.

\bibitem{deVries:2019nsu}
J.~de~Vries, G.~Falcioni, F.~Herzog, and B.~Ruijl, {\it {Two- and three-loop anomalous dimensions of Weinberg\textquoteright{}s dimension-six CP-odd gluonic operator}},  {\em Phys. Rev. D} {\bf 102} (2020), no.~1 016010, [\href{http://arxiv.org/abs/1907.04923}{{\tt arXiv:1907.04923}}].

\bibitem{Panico:2018hal}
G.~Panico, A.~Pomarol, and M.~Riembau, {\it {EFT approach to the electron Electric Dipole Moment at the two-loop level}},  {\em JHEP} {\bf 04} (2019) 090, [\href{http://arxiv.org/abs/1810.09413}{{\tt arXiv:1810.09413}}].

\bibitem{Jenkins:2023rtg}
E.~E. Jenkins, A.~V. Manohar, L.~Naterop, and J.~Pag\`es, {\it {An algebraic formula for two loop renormalization of scalar quantum field theory}},  {\em JHEP} {\bf 12} (2023) 165, [\href{http://arxiv.org/abs/2308.06315}{{\tt arXiv:2308.06315}}].

\bibitem{Jenkins:2023bls}
E.~E. Jenkins, A.~V. Manohar, L.~Naterop, and J.~Pag\`es, {\it {Two loop renormalization of scalar theories using a geometric approach}},  {\em JHEP} {\bf 02} (2024) 131, [\href{http://arxiv.org/abs/2310.19883}{{\tt arXiv:2310.19883}}].

\bibitem{Born:2024mgz}
L.~Born, J.~Fuentes-Mart\'\i{}n, S.~Kvedarait\.{e}, and A.~E. Thomsen, {\it {Two-loop running in the bosonic SMEFT using functional methods}},  {\em JHEP} {\bf 05} (2025) 121, [\href{http://arxiv.org/abs/2410.07320}{{\tt arXiv:2410.07320}}].

\bibitem{Fuentes-Martin:2024agf}
J.~Fuentes-Mart{\'\i}n, A.~Moreno-S{\'a}nchez, A.~Palavri{\'c}, and A.~E. Thomsen, {\it {A guide to functional methods beyond one-loop order}},  {\em JHEP} {\bf 08} (2025) 099, [\href{http://arxiv.org/abs/2412.12270}{{\tt arXiv:2412.12270}}].

\bibitem{Fuentes-Martin:2025meq}
J.~Fuentes-Mart{\'\i}n, A.~Moreno-S{\'a}nchez, and A.~E. Thomsen, {\it {Symmetry Restoration in the SMEFT: Finite Counterterms for a Non-Anticommuting $\gamma_5$}},  \href{http://arxiv.org/abs/2507.19589}{{\tt arXiv:2507.19589}}.

\bibitem{Duhr:2025yor}
C.~Duhr, G.~Ventura, and E.~Vryonidou, {\it {Two-loop renormalisation of quark and gluon fields in the SMEFT in the on-shell scheme}},  \href{http://arxiv.org/abs/2508.04500}{{\tt arXiv:2508.04500}}.

\bibitem{Hurth:2019ula}
T.~Hurth, S.~Renner, and W.~Shepherd, {\it {Matching for FCNC effects in the flavour-symmetric SMEFT}},  {\em JHEP} {\bf 06} (2019) 029, [\href{http://arxiv.org/abs/1903.00500}{{\tt arXiv:1903.00500}}].

\bibitem{Endo:2018gdn}
M.~Endo, T.~Kitahara, and D.~Ueda, {\it {SMEFT top-quark effects on $\Delta F=2$ observables}},  {\em JHEP} {\bf 07} (2019) 182, [\href{http://arxiv.org/abs/1811.04961}{{\tt arXiv:1811.04961}}].

\bibitem{Grzadkowski:2008mf}
B.~Grzadkowski and M.~Misiak, {\it {Anomalous Wtb coupling effects in the weak radiative B-meson decay}},  {\em Phys.~Rev.} {\bf D78} (2008) 077501, [\href{http://arxiv.org/abs/0802.1413}{{\tt arXiv:0802.1413}}].

\bibitem{Haisch:2024wnw}
U.~Haisch and L.~Schnell, {\it {Precision tests of third-generation four-quark operators: one- and two-loop matching}},  {\em JHEP} {\bf 02} (2025) 038, [\href{http://arxiv.org/abs/2410.13304}{{\tt arXiv:2410.13304}}].

\bibitem{Aebischer:2017gaw}
J.~Aebischer, M.~Fael, C.~Greub, and J.~Virto, {\it {B physics Beyond the Standard Model at One Loop: Complete Renormalization Group Evolution below the Electroweak Scale}},  {\em JHEP} {\bf 09} (2017) 158, [\href{http://arxiv.org/abs/1704.06639}{{\tt arXiv:1704.06639}}].

\bibitem{Renner:2025cmd}
S.~Renner, B.~Smith, and D.~Sutherland, {\it {Diagonalising the LEFT}},  \href{http://arxiv.org/abs/2507.18689}{{\tt arXiv:2507.18689}}.

\bibitem{Buras:2000if}
A.~J. Buras, M.~Misiak, and J.~Urban, {\it {Two loop QCD anomalous dimensions of flavor changing four quark operators within and beyond the standard model}},  {\em Nucl.~Phys.} {\bf B586} (2000) 397--426, [\href{http://arxiv.org/abs/hep-ph/0005183}{{\tt hep-ph/0005183}}].

\bibitem{Aebischer:2021raf}
J.~Aebischer, C.~Bobeth, A.~J. Buras, J.~Kumar, and M.~Misiak, {\it {General non-leptonic $\Delta F = 1$ WET at the NLO in QCD}},  {\em JHEP} {\bf 11} (2021) 227, [\href{http://arxiv.org/abs/2107.10262}{{\tt arXiv:2107.10262}}].

\bibitem{Aebischer:2025hsx}
J.~Aebischer, P.~Morell, M.~Pesut, and J.~Virto, {\it {Two-Loop Anomalous Dimensions in the LEFT: Dimension-Six Four-Fermion Operators in NDR}},  \href{http://arxiv.org/abs/2501.08384}{{\tt arXiv:2501.08384}}.

\bibitem{Naterop:2025lzc}
L.~Naterop and P.~Stoffer, {\it {Renormalization-group equations of the LEFT at two loops: dimension-six baryon-number-violating operators}},  {\em JHEP} {\bf 07} (2025) 237, [\href{http://arxiv.org/abs/2505.03871}{{\tt arXiv:2505.03871}}].

\bibitem{Naterop:2025cwg}
L.~Naterop and P.~Stoffer, {\it {Renormalization-group equations of the LEFT at two loops: dimension-six operators}},  \href{http://arxiv.org/abs/2507.08926}{{\tt arXiv:2507.08926}}.

\bibitem{Liao:2019tep}
Y.~Liao and X.-D. Ma, {\it {Renormalization Group Evolution of Dimension-seven Operators in Standard Model Effective Field Theory and Relevant Phenomenology}},  {\em JHEP} {\bf 03} (2019) 179, [\href{http://arxiv.org/abs/1901.10302}{{\tt arXiv:1901.10302}}].

\bibitem{Zhang:2023ndw}
D.~Zhang, {\it {Revisiting renormalization group equations of the SMEFT dimension-seven operators}},  {\em JHEP} {\bf 02} (2024) 133, [\href{http://arxiv.org/abs/2310.11055}{{\tt arXiv:2310.11055}}].

\bibitem{Zhang:2024clp}
D.~Zhang, {\it {Renormalization Group Equations for the Dimension-7 SMEFT Operators}},  {\em PoS} {\bf ICHEP2024} (2025) 776, [\href{http://arxiv.org/abs/2409.02622}{{\tt arXiv:2409.02622}}].

\bibitem{Jiang:2020mhe}
M.~Jiang, T.~Ma, and J.~Shu, {\it {Renormalization Group Evolution from On-shell SMEFT}},  {\em JHEP} {\bf 01} (2021) 101, [\href{http://arxiv.org/abs/2005.10261}{{\tt arXiv:2005.10261}}].

\bibitem{DasBakshi:2022mwk}
S.~Das~Bakshi, M.~Chala, A.~D\'\i{}az-Carmona, and G.~Guedes, {\it {Towards the renormalisation of the Standard Model effective field theory to dimension eight: bosonic interactions II}},  {\em Eur. Phys. J. Plus} {\bf 137} (2022), no.~8 973, [\href{http://arxiv.org/abs/2205.03301}{{\tt arXiv:2205.03301}}].

\bibitem{Chala:2021pll}
M.~Chala, G.~Guedes, M.~Ramos, and J.~Santiago, {\it {Towards the renormalisation of the Standard Model effective field theory to dimension eight: Bosonic interactions I}},  {\em SciPost Phys.} {\bf 11} (2021) 065, [\href{http://arxiv.org/abs/2106.05291}{{\tt arXiv:2106.05291}}].

\bibitem{DasBakshi:2023htx}
S.~Das~Bakshi and A.~D\'\i{}az-Carmona, {\it {Renormalisation of SMEFT bosonic interactions up to dimension eight by LNV operators}},  {\em JHEP} {\bf 06} (2023) 123, [\href{http://arxiv.org/abs/2301.07151}{{\tt arXiv:2301.07151}}].

\bibitem{Bakshi:2024wzz}
S.~D. Bakshi, M.~Chala, A.~D\'\i{}az-Carmona, Z.~Ren, and F.~Vilches, {\it {Renormalization of the SMEFT to dimension eight: Fermionic interactions I}},  {\em JHEP} {\bf 12} (2025) 214, [\href{http://arxiv.org/abs/2409.15408}{{\tt arXiv:2409.15408}}].

\bibitem{Boughezal:2024zqa}
R.~Boughezal, Y.~Huang, and F.~Petriello, {\it {Renormalization-group running of dimension-8 four-fermion operators in the SMEFT}},  {\em Phys. Rev. D} {\bf 110} (2024), no.~11 116015, [\href{http://arxiv.org/abs/2408.15378}{{\tt arXiv:2408.15378}}].

\bibitem{DiNoi:2025uan}
S.~Di~Noi, R.~Gr{\"o}ber, and P.~Olgoso, {\it {Mapping between {\ensuremath{\gamma}}$_{5}$ schemes in the Standard Model Effective Field Theory}},  {\em JHEP} {\bf 09} (2025) 027, [\href{http://arxiv.org/abs/2504.00112}{{\tt arXiv:2504.00112}}].

\bibitem{Kley:2021yhn}
J.~Kley, T.~Theil, E.~Venturini, and A.~Weiler, {\it {Electric dipole moments at one-loop in the dimension-6 SMEFT}},  {\em Eur. Phys. J. C} {\bf 82} (2022), no.~10 926, [\href{http://arxiv.org/abs/2109.15085}{{\tt arXiv:2109.15085}}].

\bibitem{Kumar:2024yuu}
J.~Kumar and E.~Mereghetti, {\it {Electric dipole moments in 5+3 flavor weak effective theory}},  {\em JHEP} {\bf 09} (2024) 028, [\href{http://arxiv.org/abs/2404.00516}{{\tt arXiv:2404.00516}}].

\bibitem{Chetyrkin:1997gb}
K.~G. Chetyrkin, M.~Misiak, and M.~M\"unz, {\it {$|\Delta F| = 1$ nonleptonic effective Hamiltonian in a simpler scheme}},  {\em Nucl. Phys. B} {\bf 520} (1998) 279--297, [\href{http://arxiv.org/abs/hep-ph/9711280}{{\tt hep-ph/9711280}}].

\bibitem{Gorbahn:2004my}
M.~Gorbahn and U.~Haisch, {\it {Effective Hamiltonian for non-leptonic $|\Delta F| = 1$ decays at NNLO in QCD}},  {\em Nucl.~Phys.} {\bf B713} (2005) 291--332, [\href{http://arxiv.org/abs/hep-ph/0411071}{{\tt hep-ph/0411071}}].

\bibitem{Herrlich:1994kh}
S.~Herrlich and U.~Nierste, {\it {Evanescent operators, scheme dependences and double insertions}},  {\em Nucl. Phys.} {\bf B455} (1995) 39--58, [\href{http://arxiv.org/abs/hep-ph/9412375}{{\tt hep-ph/9412375}}].

\bibitem{Tracas:1982gp}
N.~Tracas and N.~Vlachos, {\it {Two Loop Calculations in {QCD} and the $\Delta I = 1/2$ Rule in Nonleptonic Weak Decays}},  {\em Phys. Lett.} {\bf B115} (1982) 419.

\bibitem{Buras:1989xd}
A.~J. Buras and P.~H. Weisz, {\it {QCD Nonleading Corrections to Weak Decays in Dimensional Regularization and 't Hooft-Veltman Schemes}},  {\em Nucl. Phys.} {\bf B333} (1990) 66--99.

\bibitem{Criado:2018sdb}
J.~C. Criado and M.~P\'erez-Victoria, {\it {Field redefinitions in effective theories at higher orders}},  {\em JHEP} {\bf 03} (2019) 038, [\href{http://arxiv.org/abs/1811.09413}{{\tt arXiv:1811.09413}}].

\bibitem{Cohen:2022uuw}
T.~Cohen, N.~Craig, X.~Lu, and D.~Sutherland, {\it {On-Shell Covariance of Quantum Field Theory Amplitudes}},  {\em Phys. Rev. Lett.} {\bf 130} (2023), no.~4 041603, [\href{http://arxiv.org/abs/2202.06965}{{\tt arXiv:2202.06965}}].

\bibitem{Cohen:2023ekv}
T.~Cohen, X.~Lu, and D.~Sutherland, {\it {On amplitudes and field redefinitions}},  {\em JHEP} {\bf 06} (2024) 149, [\href{http://arxiv.org/abs/2312.06748}{{\tt arXiv:2312.06748}}].

\bibitem{Manohar:2024xbh}
A.~V. Manohar, J.~Pag\`es, and J.~Roosmale~Nepveu, {\it {Field redefinitions and infinite field anomalous dimensions}},  {\em JHEP} {\bf 05} (2024) 018, [\href{http://arxiv.org/abs/2402.08715}{{\tt arXiv:2402.08715}}].

\bibitem{Cohen:2024fak}
T.~Cohen, M.~Forslund, and A.~Helset, {\it {Field redefinitions can be nonlocal}},  {\em JHEP} {\bf 10} (2025) 019, [\href{http://arxiv.org/abs/2412.12247}{{\tt arXiv:2412.12247}}].

\bibitem{Criado:2024mpx}
J.~C. Criado, J.~Jaeckel, and M.~Spannowsky, {\it {Field redefinitions in classical field theory with some quantum perspectives}},  {\em Phys. Rev. D} {\bf 111} (2025), no.~7 076019, [\href{http://arxiv.org/abs/2408.03369}{{\tt arXiv:2408.03369}}].

\bibitem{Arzt:1993gz}
C.~Arzt, {\it {Reduced effective Lagrangians}},  {\em Phys. Lett. B} {\bf 342} (1995) 189--195, [\href{http://arxiv.org/abs/hep-ph/9304230}{{\tt hep-ph/9304230}}].

\bibitem{Aebischer:2022tvz}
J.~Aebischer, A.~J. Buras, and J.~Kumar, {\it {Simple rules for evanescent operators in one-loop basis transformations}},  {\em Phys. Rev. D} {\bf 107} (2023), no.~7 075007, [\href{http://arxiv.org/abs/2202.01225}{{\tt arXiv:2202.01225}}].

\bibitem{Aebischer:2022aze}
J.~Aebischer and M.~Pesut, {\it {One-loop Fierz transformations}},  {\em JHEP} {\bf 10} (2022) 090, [\href{http://arxiv.org/abs/2208.10513}{{\tt arXiv:2208.10513}}].

\bibitem{Aebischer:2022rxf}
J.~Aebischer, M.~Pesut, and Z.~Polonsky, {\it {Dipole operators in Fierz identities}},  {\em Phys. Lett. B} {\bf 842} (2023) 137968, [\href{http://arxiv.org/abs/2211.01379}{{\tt arXiv:2211.01379}}].

\bibitem{Aebischer:2023djt}
J.~Aebischer, M.~Pesut, and Z.~Polonsky, {\it {Renormalization scheme factorization of one-loop Fierz identities}},  {\em JHEP} {\bf 01} (2024) 060, [\href{http://arxiv.org/abs/2306.16449}{{\tt arXiv:2306.16449}}].

\bibitem{Aebischer:2024xnf}
J.~Aebischer, M.~Pesut, and Z.~Polonsky, {\it {A simple dirac prescription for two-loop anomalous dimension matrices}},  {\em Eur. Phys. J. C} {\bf 84} (2024), no.~7 750, [\href{http://arxiv.org/abs/2401.16904}{{\tt arXiv:2401.16904}}].

\bibitem{Fuentes-Martin:2022vvu}
J.~Fuentes-Mart\'\i{}n, M.~K\"onig, J.~Pag\`es, A.~E. Thomsen, and F.~Wilsch, {\it {Evanescent operators in one-loop matching computations}},  {\em JHEP} {\bf 02} (2023) 031, [\href{http://arxiv.org/abs/2211.09144}{{\tt arXiv:2211.09144}}].

\bibitem{Falkowski:2015wza}
A.~Falkowski, B.~Fuks, K.~Mawatari, K.~Mimasu, F.~Riva, and V.~Sanz, {\it {Rosetta: an operator basis translator for Standard Model effective field theory}},  {\em Eur. Phys. J. C} {\bf 75} (2015), no.~12 583, [\href{http://arxiv.org/abs/1508.05895}{{\tt arXiv:1508.05895}}].

\bibitem{Altmannshofer:2024jyv}
W.~Altmannshofer and A.~Greljo, {\it {Recent Progress in Flavor Model Building}},  \href{http://arxiv.org/abs/2412.04549}{{\tt arXiv:2412.04549}}.

\bibitem{Altmannshofer:2025rxc}
W.~Altmannshofer and P.~Stangl, {\it {Flavour Physics Beyond the Standard Model}},  \href{http://arxiv.org/abs/2508.03950}{{\tt arXiv:2508.03950}}.

\bibitem{DAmbrosio:2002vsn}
G.~D'Ambrosio, G.~F. Giudice, G.~Isidori, and A.~Strumia, {\it {Minimal flavor violation: An Effective field theory approach}},  {\em Nucl. Phys. B} {\bf 645} (2002) 155--187, [\href{http://arxiv.org/abs/hep-ph/0207036}{{\tt hep-ph/0207036}}].

\bibitem{Cirigliano:2005ck}
V.~Cirigliano, B.~Grinstein, G.~Isidori, and M.~B. Wise, {\it {Minimal flavor violation in the lepton sector}},  {\em Nucl.~Phys.} {\bf B728} (2005) 121--134, [\href{http://arxiv.org/abs/hep-ph/0507001}{{\tt hep-ph/0507001}}].

\bibitem{Kagan:2009bn}
A.~L. Kagan, G.~Perez, T.~Volansky, and J.~Zupan, {\it {General Minimal Flavor Violation}},  {\em Phys.~Rev.} {\bf D80} (2009) 076002, [\href{http://arxiv.org/abs/0903.1794}{{\tt arXiv:0903.1794}}].

\bibitem{Feldmann:2006jk}
T.~Feldmann and T.~Mannel, {\it {Minimal Flavour Violation and Beyond}},  {\em JHEP} {\bf 0702} (2007) 067, [\href{http://arxiv.org/abs/hep-ph/0611095}{{\tt hep-ph/0611095}}].

\bibitem{Paradisi:2008qh}
P.~Paradisi, M.~Ratz, R.~Schieren, and C.~Simonetto, {\it {Running minimal flavor violation}},  {\em Phys.~Lett.} {\bf B668} (2008) 202--209, [\href{http://arxiv.org/abs/0805.3989}{{\tt arXiv:0805.3989}}].

\bibitem{Mercolli:2009ns}
L.~Mercolli and C.~Smith, {\it {EDM constraints on flavored CP-violating phases}},  {\em Nucl.~Phys.} {\bf B817} (2009) 1--24, [\href{http://arxiv.org/abs/0902.1949}{{\tt arXiv:0902.1949}}].

\bibitem{Feldmann:2009dc}
T.~Feldmann, M.~Jung, and T.~Mannel, {\it {Sequential Flavour Symmetry Breaking}},  {\em Phys.~Rev.} {\bf D80} (2009) 033003, [\href{http://arxiv.org/abs/0906.1523}{{\tt arXiv:0906.1523}}].

\bibitem{Paradisi:2009ey}
P.~Paradisi and D.~M. Straub, {\it {The SUSY CP Problem and the MFV Principle}},  {\em Phys.~Lett.} {\bf B684} (2010) 147--153, [\href{http://arxiv.org/abs/0906.4551}{{\tt arXiv:0906.4551}}].

\bibitem{Isidori:2010gz}
G.~Isidori, {\it {B Physics in the LHC Era}},  \href{http://arxiv.org/abs/1001.3431}{{\tt arXiv:1001.3431}}.

\bibitem{Nir:2007xn}
Y.~Nir, {\it {Probing new physics with flavor physics (and probing flavor physics with new physics)}},  in {\em {Prospects in Theoretical Physics (PiTP) summer program on The Standard Model and Beyond IAS, Princeton, NJ, June 16-27, 2007}}, 2007.
\newblock \href{http://arxiv.org/abs/0708.1872}{{\tt arXiv:0708.1872}}.

\bibitem{Hurth:2008jc}
T.~Hurth, G.~Isidori, J.~F. Kamenik, and F.~Mescia, {\it {Constraints on New Physics in MFV models: A Model-independent analysis of $\Delta F = $1 processes}},  {\em Nucl.~Phys.} {\bf B808} (2009) 326--346, [\href{http://arxiv.org/abs/0807.5039}{{\tt arXiv:0807.5039}}].

\bibitem{Isidori:2012ts}
G.~Isidori and D.~M. Straub, {\it {Minimal Flavour Violation and Beyond}},  {\em Eur.~Phys.~J.} {\bf C72} (2012) 2103, [\href{http://arxiv.org/abs/1202.0464}{{\tt arXiv:1202.0464}}].

\bibitem{Aoude:2020dwv}
R.~Aoude, T.~Hurth, S.~Renner, and W.~Shepherd, {\it {The impact of flavour data on global fits of the MFV SMEFT}},  {\em JHEP} {\bf 12} (2020) 113, [\href{http://arxiv.org/abs/2003.05432}{{\tt arXiv:2003.05432}}].

\bibitem{Faroughy:2020ina}
D.~A. Faroughy, G.~Isidori, F.~Wilsch, and K.~Yamamoto, {\it {Flavour symmetries in the SMEFT}},  {\em JHEP} {\bf 08} (2020) 166, [\href{http://arxiv.org/abs/2005.05366}{{\tt arXiv:2005.05366}}].

\bibitem{Bruggisser:2021duo}
S.~Bruggisser, R.~Sch\"afer, D.~van Dyk, and S.~Westhoff, {\it {The Flavor of UV Physics}},  {\em JHEP} {\bf 05} (2021) 257, [\href{http://arxiv.org/abs/2101.07273}{{\tt arXiv:2101.07273}}].

\bibitem{Bruggisser:2022rhb}
S.~Bruggisser, D.~van Dyk, and S.~Westhoff, {\it {Resolving the flavor structure in the MFV-SMEFT}},  {\em JHEP} {\bf 02} (2023) 225, [\href{http://arxiv.org/abs/2212.02532}{{\tt arXiv:2212.02532}}].

\bibitem{Greljo:2022cah}
A.~Greljo, A.~Palavri\'c, and A.~E. Thomsen, {\it {Adding Flavor to the SMEFT}},  {\em JHEP} {\bf 10} (2022) 010, [\href{http://arxiv.org/abs/2203.09561}{{\tt arXiv:2203.09561}}].

\bibitem{Bartocci:2023nvp}
R.~Bartocci, A.~Biek\"otter, and T.~Hurth, {\it {A global analysis of the SMEFT under the minimal MFV assumption}},  {\em JHEP} {\bf 05} (2024) 074, [\href{http://arxiv.org/abs/2311.04963}{{\tt arXiv:2311.04963}}].

\bibitem{Fajfer:2023gie}
S.~Fajfer, J.~F. Kamenik, N.~Ko\v{s}nik, A.~Smolkovi\v{c}, and M.~Tammaro, {\it {New Physics in CP violating and flavour changing quark dipole transitions}},  {\em JHEP} {\bf 10} (2023) 133, [\href{http://arxiv.org/abs/2306.16471}{{\tt arXiv:2306.16471}}].

\bibitem{Grunwald:2023nli}
C.~Grunwald, G.~Hiller, K.~Kr\"oninger, and L.~Nollen, {\it {More synergies from beauty, top, Z and Drell-Yan measurements in SMEFT}},  {\em JHEP} {\bf 11} (2023) 110, [\href{http://arxiv.org/abs/2304.12837}{{\tt arXiv:2304.12837}}].

\bibitem{Grunwald:2024yuq}
C.~Grunwald, G.~Hiller, K.~Kr\"oninger, and L.~Nollen, {\it {Predicting ${{B}}(B\to K^{(*)}\nu \bar \nu)$ within the MFV-SMEFT using $B$, Top, $Z$ and Drell-Yan data}},  {\em PoS} {\bf EPS-HEP2023} (2024) 298.

\bibitem{Bartocci:2024fmm}
R.~Bartocci, A.~Biek\"otter, and T.~Hurth, {\it {Renormalisation group evolution effects on global SMEFT analyses}},  {\em JHEP} {\bf 05} (2025) 203, [\href{http://arxiv.org/abs/2412.09674}{{\tt arXiv:2412.09674}}].

\bibitem{Sun:2025axx}
H.~Sun and J.-H. Yu, {\it {Flavor and CP symmetries in the standard model effective field theory}},  {\em Phys. Rev. D} {\bf 112} (2025), no.~7 075022, [\href{http://arxiv.org/abs/2502.03526}{{\tt arXiv:2502.03526}}].

\bibitem{Greljo:2023adz}
A.~Greljo and A.~Palavri\'c, {\it {Leading directions in the SMEFT}},  {\em JHEP} {\bf 09} (2023) 009, [\href{http://arxiv.org/abs/2305.08898}{{\tt arXiv:2305.08898}}].

\bibitem{Greljo:2023bdy}
A.~Greljo, A.~Palavri\'c, and A.~Smolkovi\v{c}, {\it {Leading directions in the SMEFT: Renormalization effects}},  {\em Phys. Rev. D} {\bf 109} (2024), no.~7 075033, [\href{http://arxiv.org/abs/2312.09179}{{\tt arXiv:2312.09179}}].

\bibitem{Grinstein:2024iyf}
B.~Grinstein, X.~Lu, C.~Mir{\'o}, and P.~Qu{\'\i}lez, {\it {Most general EFTs from spurion analysis Hilbert series and minimal lepton flavor violation}},  {\em JHEP} {\bf 07} (2025) 259, [\href{http://arxiv.org/abs/2412.16285}{{\tt arXiv:2412.16285}}].

\bibitem{Song:2025snz}
C.-Q. Song, H.~Sun, and J.-H. Yu, {\it {Systematic Spurion Matching between Low Energy EFT and Chiral Lagrangian}},  \href{http://arxiv.org/abs/2501.09787}{{\tt arXiv:2501.09787}}.

\bibitem{Barbieri:2011ci}
R.~Barbieri, G.~Isidori, J.~Jones-Perez, P.~Lodone, and D.~M. Straub, {\it {U(2) and Minimal Flavour Violation in Supersymmetry}},  {\em Eur.~Phys.~J.} {\bf C71} (2011) 1725, [\href{http://arxiv.org/abs/1105.2296}{{\tt arXiv:1105.2296}}].

\bibitem{Barbieri:2012uh}
R.~Barbieri, D.~Buttazzo, F.~Sala, and D.~M. Straub, {\it {Flavour physics from an approximate $U(2)^3$ symmetry}},  {\em JHEP} {\bf 1207} (2012) 181, [\href{http://arxiv.org/abs/1203.4218}{{\tt arXiv:1203.4218}}].

\bibitem{Smolkovic:2019jow}
A.~Smolkovi\v{c}, M.~Tammaro, and J.~Zupan, {\it {Anomaly free Froggatt-Nielsen models of flavor}},  {\em JHEP} {\bf 10} (2019) 188, [\href{http://arxiv.org/abs/1907.10063}{{\tt arXiv:1907.10063}}]. [Erratum: JHEP 02, 033 (2022)].

\bibitem{Bordone:2019uzc}
M.~Bordone, O.~Cat\`a, and T.~Feldmann, {\it {Effective Theory Approach to New Physics with Flavour: General Framework and a Leptoquark Example}},  {\em JHEP} {\bf 01} (2020) 067, [\href{http://arxiv.org/abs/1910.02641}{{\tt arXiv:1910.02641}}].

\bibitem{Antusch:2023shi}
S.~Antusch, A.~Greljo, B.~A. Stefanek, and A.~E. Thomsen, {\it {U(2) Is Right for Leptons and Left for Quarks}},  {\em Phys. Rev. Lett.} {\bf 132} (2024), no.~15 151802, [\href{http://arxiv.org/abs/2311.09288}{{\tt arXiv:2311.09288}}].

\bibitem{Capdevila:2024gki}
B.~Capdevila, A.~Crivellin, J.~M. Lizana, and S.~Pokorski, {\it {SU(2)$_{L}$ deconstruction and flavour (non)-universality}},  {\em JHEP} {\bf 08} (2024) 031, [\href{http://arxiv.org/abs/2401.00848}{{\tt arXiv:2401.00848}}].

\bibitem{Grinstein:2023njq}
B.~Grinstein, X.~Lu, L.~Merlo, and P.~Qu\'\i{}lez, {\it {Hilbert series for covariants and their applications to minimal flavor violation}},  {\em JHEP} {\bf 2024} (2024) 154, [\href{http://arxiv.org/abs/2312.13349}{{\tt arXiv:2312.13349}}].

\bibitem{Greljo:2023bix}
A.~Greljo and A.~E. Thomsen, {\it {Rising through the ranks: flavor hierarchies from a gauged SU(2) symmetry}},  {\em Eur. Phys. J. C} {\bf 84} (2024), no.~2 213, [\href{http://arxiv.org/abs/2309.11547}{{\tt arXiv:2309.11547}}].

\bibitem{Greljo:2024zrj}
A.~Greljo, A.~E. Thomsen, and H.~Tiblom, {\it {Flavor hierarchies from SU(2) flavor and quark-lepton unification}},  {\em JHEP} {\bf 08} (2024) 143, [\href{http://arxiv.org/abs/2406.02687}{{\tt arXiv:2406.02687}}].

\bibitem{Calibbi:2025fzi}
L.~Calibbi, C.~Hagedorn, M.~A. Schmidt, and J.~Vandeleur, {\it {Selection rules for charged lepton flavour violating processes from residual flavour groups}},  \href{http://arxiv.org/abs/2505.24350}{{\tt arXiv:2505.24350}}.

\bibitem{Glioti:2024hye}
A.~Glioti, R.~Rattazzi, L.~Ricci, and L.~Vecchi, {\it {Exploring the Flavor Symmetry Landscape}},  {\em SciPost Phys.} {\bf 18} (2025) 201, [\href{http://arxiv.org/abs/2402.09503}{{\tt arXiv:2402.09503}}].

\bibitem{Buras:2000dm}
A.~J. Buras, P.~Gambino, M.~Gorbahn, S.~J{\"a}ger, and L.~Silvestrini, {\it {Universal unitarity triangle and physics beyond the standard model}},  {\em Phys.~Lett.} {\bf B500} (2001) 161--167, [\href{http://arxiv.org/abs/hep-ph/0007085}{{\tt hep-ph/0007085}}].

\bibitem{Buras:2003jf}
A.~J. Buras, {\it {Minimal flavor violation}},  {\em Acta Phys. Polon.} {\bf B34} (2003) 5615--5668, [\href{http://arxiv.org/abs/hep-ph/0310208}{{\tt hep-ph/0310208}}].

\bibitem{Blanke:2006ig}
M.~Blanke, A.~J. Buras, D.~Guadagnoli, and C.~Tarantino, {\it {Minimal Flavour Violation Waiting for Precise Measurements of $\Delta M_s$, $S_{\psi \phi}$, $A^s_\text{SL}$, $|V_{ub}|$, $\gamma$ and $B^0_{s,d} \to \mu^+ \mu^-$}},  {\em JHEP} {\bf 10} (2006) 003, [\href{http://arxiv.org/abs/hep-ph/0604057}{{\tt hep-ph/0604057}}].

\bibitem{Buras:2012ts}
A.~J. Buras and J.~Girrbach, {\it {BSM models facing the recent LHCb data: A First look}},  {\em Acta Phys.Polon.} {\bf B43} (2012) 1427, [\href{http://arxiv.org/abs/1204.5064}{{\tt arXiv:1204.5064}}].

\bibitem{Blanke:2016bhf}
M.~Blanke and A.~J. Buras, {\it {Universal Unitarity Triangle 2016 and the tension between $\Delta M_{s,d}$ and $\varepsilon _K$ in CMFV models}},  {\em Eur. Phys. J.} {\bf C76} (2016), no.~4 197, [\href{http://arxiv.org/abs/1602.04020}{{\tt arXiv:1602.04020}}].

\bibitem{Inami:1980fz}
T.~Inami and C.~Lim, {\it {Effects of Superheavy Quarks and Leptons in Low-Energy Weak Processes $K_L\to\mu^+\mu^-$, $K^+\to\pi^+\nu\bar\nu$ and $K^0-\bar K^0$}},  {\em Prog.~Theor.~Phys.} {\bf 65} (1981) 297.

\bibitem{Buchalla:1990qz}
G.~Buchalla, A.~J. Buras, and M.~K. Harlander, {\it {Penguin box expansion: Flavor changing neutral current processes and a heavy top quark}},  {\em Nucl.~Phys.} {\bf B349} (1991) 1--47.

\bibitem{Feldmann:2008ja}
T.~Feldmann and T.~Mannel, {\it {Large Top Mass and Non-Linear Representation of Flavour Symmetry}},  {\em Phys. Rev. Lett.} {\bf 100} (2008) 171601, [\href{http://arxiv.org/abs/0801.1802}{{\tt arXiv:0801.1802}}].

\bibitem{D'Ambrosio:2002ex}
G.~D'Ambrosio, G.~F. Giudice, G.~Isidori, and A.~Strumia, {\it {Minimal flavour violation: An effective field theory approach}},  {\em Nucl.~Phys.} {\bf B645} (2002) 155--187, [\href{http://arxiv.org/abs/hep-ph/0207036}{{\tt hep-ph/0207036}}].

\bibitem{Aebischer:2023irs}
J.~Aebischer et~al., {\it {Computing Tools for Effective Field Theories}},  7, 2023.
\newblock \href{http://arxiv.org/abs/2307.08745}{{\tt arXiv:2307.08745}}.

\bibitem{Proceedings:2019rnh}
J.~Aebischer, M.~Fael, A.~Lenz, M.~Spannowsky, and J.~Virto, eds., {\em {Computing Tools for the SMEFT}}, 10, 2019.

\bibitem{Dawson:2022ewj}
S.~Dawson et~al., {\it {LHC EFT WG Note: Precision matching of microscopic physics to the Standard Model Effective Field Theory (SMEFT)}},  \href{http://arxiv.org/abs/2212.02905}{{\tt arXiv:2212.02905}}.

\bibitem{Belvedere:2024nzh}
A.~Belvedere et~al., {\it {LHC EFT WG Note: SMEFT predictions, event reweighting, and simulation}},  \href{http://arxiv.org/abs/2406.14620}{{\tt arXiv:2406.14620}}.

\bibitem{Fonseca:2020vke}
R.~M. Fonseca, {\it {GroupMath: A Mathematica package for group theory calculations}},  {\em Comput. Phys. Commun.} {\bf 267} (2021) 108085, [\href{http://arxiv.org/abs/2011.01764}{{\tt arXiv:2011.01764}}].

\bibitem{Feger:2019tvk}
R.~Feger, T.~W. Kephart, and R.~J. Saskowski, {\it {LieART 2.0 {\textendash} A Mathematica application for Lie Algebras and Representation Theory}},  {\em Comput. Phys. Commun.} {\bf 257} (2020) 107490, [\href{http://arxiv.org/abs/1912.10969}{{\tt arXiv:1912.10969}}].

\bibitem{Lehman:2015via}
L.~Lehman and A.~Martin, {\it {Hilbert Series for Constructing Lagrangians: expanding the phenomenologist's toolbox}},  {\em Phys. Rev. D} {\bf 91} (2015) 105014, [\href{http://arxiv.org/abs/1503.07537}{{\tt arXiv:1503.07537}}].

\bibitem{Henning:2015daa}
B.~Henning, X.~Lu, T.~Melia, and H.~Murayama, {\it {Hilbert series and operator bases with derivatives in effective field theories}},  {\em Commun. Math. Phys.} {\bf 347} (2016), no.~2 363--388, [\href{http://arxiv.org/abs/1507.07240}{{\tt arXiv:1507.07240}}].

\bibitem{Harlander:2023ozs}
R.~V. Harlander and M.~C. Schaaf, {\it {AutoEFT: Automated operator construction for effective field theories}},  {\em Comput. Phys. Commun.} {\bf 300} (2024) 109198, [\href{http://arxiv.org/abs/2309.15783}{{\tt arXiv:2309.15783}}].

\bibitem{Schaaf:2023mpw}
M.~C. Schaaf, {\it {AutoEFT: Constructing and exploring on-shell bases of effective field theories}},  in {\em {2023 European Physical Society Conference on High Energy Physics~}}, 10, 2023.
\newblock \href{http://arxiv.org/abs/2310.19606}{{\tt arXiv:2310.19606}}.

\bibitem{Criado:2017khh}
J.~C. Criado, {\it {MatchingTools: a Python library for symbolic effective field theory calculations}},  {\em Comput. Phys. Commun.} {\bf 227} (2018) 42--50, [\href{http://arxiv.org/abs/1710.06445}{{\tt arXiv:1710.06445}}].

\bibitem{LopezMiras:2025gar}
J.~L\'opez~Miras and F.~Vilches, {\it {Automation of a Matching On-Shell Calculator}},  \href{http://arxiv.org/abs/2505.21353}{{\tt arXiv:2505.21353}}.

\bibitem{Bakshi:2018ics}
S.~Das~Bakshi, J.~Chakrabortty, and S.~K. Patra, {\it {CoDEx: Wilson coefficient calculator connecting SMEFT to UV theory}},  {\em Eur. Phys. J.} {\bf C79} (2019), no.~1 21, [\href{http://arxiv.org/abs/1808.04403}{{\tt arXiv:1808.04403}}].

\bibitem{Henning:2014wua}
B.~Henning, X.~Lu, and H.~Murayama, {\it {How to use the Standard Model effective field theory}},  {\em JHEP} {\bf 01} (2016) 023, [\href{http://arxiv.org/abs/1412.1837}{{\tt arXiv:1412.1837}}].

\bibitem{Drozd:2015rsp}
A.~Drozd, J.~Ellis, J.~Quevillon, and T.~You, {\it {The Universal One-Loop Effective Action}},  {\em JHEP} {\bf 03} (2016) 180, [\href{http://arxiv.org/abs/1512.03003}{{\tt arXiv:1512.03003}}].

\bibitem{Fuentes-Martin:2016uol}
J.~Fuentes-Martin, J.~Portoles, and P.~Ruiz-Femenia, {\it {Integrating out heavy particles with functional methods: a simplified framework}},  {\em JHEP} {\bf 09} (2016) 156, [\href{http://arxiv.org/abs/1607.02142}{{\tt arXiv:1607.02142}}].

\bibitem{Ellis:2017jns}
S.~A.~R. Ellis, J.~Quevillon, T.~You, and Z.~Zhang, {\it {Extending the Universal One-Loop Effective Action: Heavy-Light Coefficients}},  {\em JHEP} {\bf 08} (2017) 054, [\href{http://arxiv.org/abs/1706.07765}{{\tt arXiv:1706.07765}}].

\bibitem{Banerjee:2023iiv}
U.~Banerjee, J.~Chakrabortty, S.~U. Rahaman, and K.~Ramkumar, {\it {One-loop effective action up to dimension eight: integrating out heavy scalar(s)}},  {\em Eur. Phys. J. Plus} {\bf 139} (2024), no.~2 159, [\href{http://arxiv.org/abs/2306.09103}{{\tt arXiv:2306.09103}}].

\bibitem{Banerjee:2022thk}
U.~Banerjee, J.~Chakrabortty, C.~Englert, S.~U. Rahaman, and M.~Spannowsky, {\it {Integrating out heavy scalars with modified equations of motion: Matching computation of dimension-eight SMEFT coefficients}},  {\em Phys. Rev. D} {\bf 107} (2023), no.~5 055007, [\href{http://arxiv.org/abs/2210.14761}{{\tt arXiv:2210.14761}}].

\bibitem{Chakrabortty:2023yke}
J.~Chakrabortty, S.~U. Rahaman, and K.~Ramkumar, {\it {One-loop effective action up to dimension eight: Integrating out heavy fermion(s)}},  {\em Nucl. Phys. B} {\bf 1000} (2024) 116488, [\href{http://arxiv.org/abs/2308.03849}{{\tt arXiv:2308.03849}}].

\bibitem{Fuentes-Martin:2022jrf}
J.~Fuentes-Mart\'\i{}n, M.~K\"onig, J.~Pag\`es, A.~E. Thomsen, and F.~Wilsch, {\it {A proof of concept for matchete: an automated tool for matching effective theories}},  {\em Eur. Phys. J. C} {\bf 83} (2023), no.~7 662, [\href{http://arxiv.org/abs/2212.04510}{{\tt arXiv:2212.04510}}].

\bibitem{Fuentes-Martin:2020udw}
J.~Fuentes-Martin, M.~K\"onig, J.~Pag\`es, A.~E. Thomsen, and F.~Wilsch, {\it {SuperTracer: A Calculator of Functional Supertraces for One-Loop EFT Matching}},  {\em JHEP} {\bf 04} (2021) 281, [\href{http://arxiv.org/abs/2012.08506}{{\tt arXiv:2012.08506}}].

\bibitem{Cohen:2020qvb}
T.~Cohen, X.~Lu, and Z.~Zhang, {\it {STrEAMlining EFT Matching}},  {\em SciPost Phys.} {\bf 10} (2021), no.~5 098, [\href{http://arxiv.org/abs/2012.07851}{{\tt arXiv:2012.07851}}].

\bibitem{Cohen:2020fcu}
T.~Cohen, X.~Lu, and Z.~Zhang, {\it {Functional Prescription for EFT Matching}},  {\em JHEP} {\bf 02} (2021) 228, [\href{http://arxiv.org/abs/2011.02484}{{\tt arXiv:2011.02484}}].

\bibitem{Carmona:2021xtq}
A.~Carmona, A.~Lazopoulos, P.~Olgoso, and J.~Santiago, {\it {Matchmakereft: automated tree-level and one-loop matching}},  {\em SciPost Phys.} {\bf 12} (2022), no.~6 198, [\href{http://arxiv.org/abs/2112.10787}{{\tt arXiv:2112.10787}}].

\bibitem{NOGUEIRA1993279}
P.~Nogueira, {\it Automatic feynman graph generation},  {\em Journal of Computational Physics} {\bf 105} (1993), no.~2 279--289.

\bibitem{Alloul:2013bka}
A.~Alloul, N.~D. Christensen, C.~Degrande, C.~Duhr, and B.~Fuks, {\it {FeynRules 2.0 - A complete toolbox for tree-level phenomenology}},  {\em Comput. Phys. Commun.} {\bf 185} (2014) 2250--2300, [\href{http://arxiv.org/abs/1310.1921}{{\tt arXiv:1310.1921}}].

\bibitem{terHoeve:2023pvs}
J.~ter Hoeve, G.~Magni, J.~Rojo, A.~N. Rossia, and E.~Vryonidou, {\it {The automation of SMEFT-assisted constraints on UV-complete models}},  {\em JHEP} {\bf 01} (2024) 179, [\href{http://arxiv.org/abs/2309.04523}{{\tt arXiv:2309.04523}}].

\bibitem{Hartland:2019bjb}
N.~P. Hartland, F.~Maltoni, E.~R. Nocera, J.~Rojo, E.~Slade, E.~Vryonidou, and C.~Zhang, {\it {A Monte Carlo global analysis of the Standard Model Effective Field Theory: the top quark sector}},  {\em JHEP} {\bf 04} (2019) 100, [\href{http://arxiv.org/abs/1901.05965}{{\tt arXiv:1901.05965}}].

\bibitem{Braun:2025afl}
J.~Braun, B.~Campillo~Aveleira, G.~Heinrich, M.~H{\"o}fer, S.~P. Jones, M.~Kerner, J.~Lang, and V.~Magerya, {\it {One-Loop Calculations in Effective Field Theories with GoSam-3.0}},  \href{http://arxiv.org/abs/2507.23549}{{\tt arXiv:2507.23549}}.

\bibitem{Fuentes-Martin:2020zaz}
J.~Fuentes-Martin, P.~Ruiz-Femenia, A.~Vicente, and J.~Virto, {\it {DsixTools 2.0: The Effective Field Theory Toolkit}},  {\em Eur. Phys. J. C} {\bf 81} (2021), no.~2 167, [\href{http://arxiv.org/abs/2010.16341}{{\tt arXiv:2010.16341}}].

\bibitem{Aebischer:2024csk}
J.~Aebischer, T.~Kapoor, and J.~Kumar, {\it {wilson: A package for renormalization group running in the SMEFT with Sterile Neutrinos}},  \href{http://arxiv.org/abs/2411.07220}{{\tt arXiv:2411.07220}}.

\bibitem{DiNoi:2022ejg}
S.~Di~Noi and L.~Silvestrini, {\it {RGESolver: a C++ library to perform renormalization group evolution in the Standard Model Effective Theory}},  {\em Eur. Phys. J. C} {\bf 83} (2023), no.~3 200, [\href{http://arxiv.org/abs/2210.06838}{{\tt arXiv:2210.06838}}].

\bibitem{Liao:2025qwp}
Y.~Liao, X.-D. Ma, H.-L. Wang, and X.~Zhao, {\it {RGE solver for the complete dim-7 SMEFT interactions and its application to $0\nu\beta\beta$ decay}},  \href{http://arxiv.org/abs/2505.06499}{{\tt arXiv:2505.06499}}.

\bibitem{Straub:2018kue}
D.~M. Straub, {\it {flavio: a Python package for flavour and precision phenomenology in the Standard Model and beyond}},  \href{http://arxiv.org/abs/1810.08132}{{\tt arXiv:1810.08132}}.

\bibitem{EOSAuthors:2021xpv}
{\bf EOS Authors} Collaboration, D.~van Dyk et~al., {\it {EOS: a software for flavor physics phenomenology}},  {\em Eur. Phys. J. C} {\bf 82} (2022), no.~6 569, [\href{http://arxiv.org/abs/2111.15428}{{\tt arXiv:2111.15428}}].

\bibitem{Fael:2024fkt}
M.~Fael, I.~S. Milutin, and K.~K. Vos, {\it {Kolya: An open-source package for inclusive semileptonic B decays}},  {\em SciPost Phys. Codeb.} {\bf 55} (2025) 1, [\href{http://arxiv.org/abs/2409.15007}{{\tt arXiv:2409.15007}}].

\bibitem{Chetyrkin:2000yt}
K.~G. Chetyrkin, J.~H. Kuhn, and M.~Steinhauser, {\it {RunDec: A Mathematica package for running and decoupling of the strong coupling and quark masses}},  {\em Comput. Phys. Commun.} {\bf 133} (2000) 43--65, [\href{http://arxiv.org/abs/hep-ph/0004189}{{\tt hep-ph/0004189}}].

\bibitem{Schmidt:2012az}
B.~Schmidt and M.~Steinhauser, {\it {CRunDec: a C++ package for running and decoupling of the strong coupling and quark masses}},  {\em Comput. Phys. Commun.} {\bf 183} (2012) 1845--1848, [\href{http://arxiv.org/abs/1201.6149}{{\tt arXiv:1201.6149}}].

\bibitem{Herren:2017osy}
F.~Herren and M.~Steinhauser, {\it {Version 3 of RunDec and CRunDec}},  {\em Comput. Phys. Commun.} {\bf 224} (2018) 333--345, [\href{http://arxiv.org/abs/1703.03751}{{\tt arXiv:1703.03751}}].

\bibitem{Allwicher:2022mcg}
L.~Allwicher, D.~A. Faroughy, F.~Jaffredo, O.~Sumensari, and F.~Wilsch, {\it {HighPT: A tool for high-pT Drell-Yan tails beyond the standard model}},  {\em Comput. Phys. Commun.} {\bf 289} (2023) 108749, [\href{http://arxiv.org/abs/2207.10756}{{\tt arXiv:2207.10756}}].

\bibitem{Allwicher:2022gkm}
L.~Allwicher, D.~A. Faroughy, F.~Jaffredo, O.~Sumensari, and F.~Wilsch, {\it {Drell-Yan tails beyond the Standard Model}},  {\em JHEP} {\bf 03} (2023) 064, [\href{http://arxiv.org/abs/2207.10714}{{\tt arXiv:2207.10714}}].

\bibitem{Giani:2023gfq}
T.~Giani, G.~Magni, and J.~Rojo, {\it {SMEFiT: a flexible toolbox for global interpretations of particle physics data with effective field theories}},  {\em Eur. Phys. J. C} {\bf 83} (2023), no.~5 393, [\href{http://arxiv.org/abs/2302.06660}{{\tt arXiv:2302.06660}}].

\bibitem{Celada:2024mcf}
E.~Celada, T.~Giani, J.~ter Hoeve, L.~Mantani, J.~Rojo, A.~N. Rossia, M.~O.~A. Thomas, and E.~Vryonidou, {\it {Mapping the SMEFT at high-energy colliders: from LEP and the (HL-)LHC to the FCC-ee}},  {\em JHEP} {\bf 09} (2024) 091, [\href{http://arxiv.org/abs/2404.12809}{{\tt arXiv:2404.12809}}].

\bibitem{DeBlas:2019ehy}
J.~De~Blas et~al., {\it {$\texttt{HEPfit}$: a code for the combination of indirect and direct constraints on high energy physics models}},  {\em Eur. Phys. J. C} {\bf 80} (2020), no.~5 456, [\href{http://arxiv.org/abs/1910.14012}{{\tt arXiv:1910.14012}}].

\bibitem{Costantini:2024xae}
{\bf PBSP} Collaboration, M.~N. Costantini, E.~Hammou, Z.~Kassabov, M.~Madigan, L.~Mantani, M.~Morales~Alvarado, J.~M. Moore, and M.~Ubiali, {\it {SIMUnet: an open-source tool for simultaneous global fits of EFT Wilson coefficients and PDFs}},  {\em Eur. Phys. J. C} {\bf 84} (2024), no.~8 805, [\href{http://arxiv.org/abs/2402.03308}{{\tt arXiv:2402.03308}}].

\bibitem{Krause:2025qnl}
C.~Krause, D.~Wang, and R.~Winterhalder, {\it {BitHEP -- The Limits of Low-Precision ML in HEP}},  \href{http://arxiv.org/abs/2504.03387}{{\tt arXiv:2504.03387}}.

\bibitem{Keaveney:2021dfa}
J.~Keaveney, {\it {Constraining the SMEFT with a differential cross section measurement of tWZ production at the HL-LHC}},  {\em Phys. Rev. D} {\bf 107} (2023), no.~3 036021, [\href{http://arxiv.org/abs/2107.01053}{{\tt arXiv:2107.01053}}].

\bibitem{Workgroup:2017myk}
{\bf The GAMBIT Flavour Workgroup} Collaboration, F.~U. Bernlochner et~al., {\it {FlavBit: A GAMBIT module for computing flavour observables and likelihoods}},  {\em Eur. Phys. J.} {\bf C77} (2017), no.~11 786, [\href{http://arxiv.org/abs/1705.07933}{{\tt arXiv:1705.07933}}].

\bibitem{GAMBIT:2017yxo}
{\bf GAMBIT} Collaboration, P.~Athron et~al., {\it {GAMBIT: The Global and Modular Beyond-the-Standard-Model Inference Tool}},  {\em Eur. Phys. J. C} {\bf 77} (2017), no.~11 784, [\href{http://arxiv.org/abs/1705.07908}{{\tt arXiv:1705.07908}}]. [Addendum: Eur.Phys.J.C 78, 98 (2018)].

\bibitem{Degrande:2020evl}
C.~Degrande, G.~Durieux, F.~Maltoni, K.~Mimasu, E.~Vryonidou, and C.~Zhang, {\it {Automated one-loop computations in the standard model effective field theory}},  {\em Phys. Rev. D} {\bf 103} (2021), no.~9 096024, [\href{http://arxiv.org/abs/2008.11743}{{\tt arXiv:2008.11743}}].

\bibitem{Alwall:2014hca}
J.~Alwall, R.~Frederix, S.~Frixione, V.~Hirschi, F.~Maltoni, O.~Mattelaer, H.~S. Shao, T.~Stelzer, P.~Torrielli, and M.~Zaro, {\it {The automated computation of tree-level and next-to-leading order differential cross sections, and their matching to parton shower simulations}},  {\em JHEP} {\bf 07} (2014) 079, [\href{http://arxiv.org/abs/1405.0301}{{\tt arXiv:1405.0301}}].

\bibitem{Hahn:2000kx}
T.~Hahn, {\it {Generating Feynman diagrams and amplitudes with FeynArts 3}},  {\em Comput. Phys. Commun.} {\bf 140} (2001) 418--431, [\href{http://arxiv.org/abs/hep-ph/0012260}{{\tt hep-ph/0012260}}].

\bibitem{Gleisberg:2008ta}
T.~Gleisberg, S.~Hoeche, F.~Krauss, M.~Schonherr, S.~Schumann, F.~Siegert, and J.~Winter, {\it {Event generation with SHERPA 1.1}},  {\em JHEP} {\bf 02} (2009) 007, [\href{http://arxiv.org/abs/0811.4622}{{\tt arXiv:0811.4622}}].

\bibitem{Aebischer:2019zoe}
J.~Aebischer, T.~Kuhr, and K.~Lieret, {\it {Clustering of $\bar B\to D^{(*)}\tau^-\bar\nu_\tau$ kinematic distributions with ClusterKinG}},  {\em JHEP} {\bf 04} (2020) 007, [\href{http://arxiv.org/abs/1909.11088}{{\tt arXiv:1909.11088}}]. [Erratum: JHEP 05, 147 (2021)].

\bibitem{Laa:2021dlg}
U.~Laa and G.~Valencia, {\it {Pandemonium: a clustering tool to partition parameter space\textemdash{}application to the B anomalies}},  {\em Eur. Phys. J. Plus} {\bf 137} (2022), no.~1 145, [\href{http://arxiv.org/abs/2103.07937}{{\tt arXiv:2103.07937}}].

\bibitem{Bechtle:2022tck}
P.~Bechtle, C.~Chall, M.~King, M.~Kraemer, P.~Maettig, and M.~St\"oltzner, {\it {Bottoms Up: Standard Model Effective Field Theory from a Model Perspective}},  \href{http://arxiv.org/abs/2201.08819}{{\tt arXiv:2201.08819}}.

\bibitem{Adams:2006sv}
A.~Adams, N.~Arkani-Hamed, S.~Dubovsky, A.~Nicolis, and R.~Rattazzi, {\it {Causality, analyticity and an IR obstruction to UV completion}},  {\em JHEP} {\bf 10} (2006) 014, [\href{http://arxiv.org/abs/hep-th/0602178}{{\tt hep-th/0602178}}].

\bibitem{Remmen:2019cyz}
G.~N. Remmen and N.~L. Rodd, {\it {Consistency of the Standard Model Effective Field Theory}},  {\em JHEP} {\bf 12} (2019) 032, [\href{http://arxiv.org/abs/1908.09845}{{\tt arXiv:1908.09845}}].

\bibitem{Remmen:2020vts}
G.~N. Remmen and N.~L. Rodd, {\it {Flavor Constraints from Unitarity and Analyticity}},  {\em Phys. Rev. Lett.} {\bf 125} (2020), no.~8 081601, [\href{http://arxiv.org/abs/2004.02885}{{\tt arXiv:2004.02885}}]. [Erratum: Phys.Rev.Lett. 127, 149901 (2021)].

\bibitem{Remmen:2020uze}
G.~N. Remmen and N.~L. Rodd, {\it {Signs, spin, SMEFT: Sum rules at dimension six}},  {\em Phys. Rev. D} {\bf 105} (2022), no.~3 036006, [\href{http://arxiv.org/abs/2010.04723}{{\tt arXiv:2010.04723}}].

\bibitem{Remmen:2022orj}
G.~N. Remmen and N.~L. Rodd, {\it {Spinning sum rules for the dimension-six SMEFT}},  {\em JHEP} {\bf 09} (2022) 030, [\href{http://arxiv.org/abs/2206.13524}{{\tt arXiv:2206.13524}}].

\bibitem{Cao:2024vfc}
Q.-H. Cao, Y.~Liu, and S.-R. Yuan, {\it {Unitarity bounds and basis transformations in SMEFT: An analysis of Warsaw and SILH bases}},  {\em Nucl. Phys. B} {\bf 1010} (2025) 116781.

\bibitem{Cohen:2021gdw}
T.~Cohen, J.~Doss, and X.~Lu, {\it {Unitarity bounds on effective field theories at the LHC}},  {\em JHEP} {\bf 04} (2022) 155, [\href{http://arxiv.org/abs/2111.09895}{{\tt arXiv:2111.09895}}].

\bibitem{Altmannshofer:2023bfk}
W.~Altmannshofer, S.~Gori, B.~V. Lehmann, and J.~Zuo, {\it {UV physics from IR features: New prospects from top flavor violation}},  {\em Phys. Rev. D} {\bf 107} (2023), no.~9 095025, [\href{http://arxiv.org/abs/2303.00781}{{\tt arXiv:2303.00781}}].

\bibitem{Altmannshofer:2025lun}
W.~Altmannshofer, Z.~Balme, C.~M. Donohue, S.~Gori, and S.~V. Mukundhan, {\it {Targets for flavor-violating top decays}},  {\em JHEP} {\bf 08} (2025) 191, [\href{http://arxiv.org/abs/2504.18664}{{\tt arXiv:2504.18664}}].

\bibitem{Ye:2024rzr}
Y.~Ye, B.~He, and J.~Gu, {\it {Positivity bounds in scalar Effective Field Theories at one-loop level}},  {\em JHEP} {\bf 12} (2024) 046, [\href{http://arxiv.org/abs/2408.10318}{{\tt arXiv:2408.10318}}].

\bibitem{Zhang:2018shp}
C.~Zhang and S.-Y. Zhou, {\it {Positivity bounds on vector boson scattering at the LHC}},  {\em Phys. Rev. D} {\bf 100} (2019), no.~9 095003, [\href{http://arxiv.org/abs/1808.00010}{{\tt arXiv:1808.00010}}].

\bibitem{Bi:2019phv}
Q.~Bi, C.~Zhang, and S.-Y. Zhou, {\it {Positivity constraints on aQGC: carving out the physical parameter space}},  {\em JHEP} {\bf 06} (2019) 137, [\href{http://arxiv.org/abs/1902.08977}{{\tt arXiv:1902.08977}}].

\bibitem{Yamashita:2020gtt}
K.~Yamashita, C.~Zhang, and S.-Y. Zhou, {\it {Elastic positivity vs extremal positivity bounds in SMEFT: a case study in transversal electroweak gauge-boson scatterings}},  {\em JHEP} {\bf 01} (2021) 095, [\href{http://arxiv.org/abs/2009.04490}{{\tt arXiv:2009.04490}}].

\bibitem{Remmen:2024hry}
G.~N. Remmen and N.~L. Rodd, {\it {Positively Identifying HEFT or SMEFT}},  \href{http://arxiv.org/abs/2412.07827}{{\tt arXiv:2412.07827}}.

\bibitem{Ghosh:2022qqq}
D.~Ghosh, R.~Sharma, and F.~Ullah, {\it {Amplitude\textquoteright{}s positivity vs. subluminality: causality and unitarity constraints on dimension 6 \& 8 gluonic operators in the SMEFT}},  {\em JHEP} {\bf 02} (2023) 199, [\href{http://arxiv.org/abs/2211.01322}{{\tt arXiv:2211.01322}}].

\bibitem{Chen:2023bhu}
Q.~Chen, K.~Mimasu, T.~A. Wu, G.-D. Zhang, and S.-Y. Zhou, {\it {Capping the positivity cone: dimension-8 Higgs operators in the SMEFT}},  {\em JHEP} {\bf 03} (2024) 180, [\href{http://arxiv.org/abs/2309.15922}{{\tt arXiv:2309.15922}}].

\bibitem{Zhang:2020jyn}
C.~Zhang and S.-Y. Zhou, {\it {Convex Geometry Perspective on the (Standard Model) Effective Field Theory Space}},  {\em Phys. Rev. Lett.} {\bf 125} (2020), no.~20 201601, [\href{http://arxiv.org/abs/2005.03047}{{\tt arXiv:2005.03047}}].

\bibitem{Yang:2023ncf}
C.~Yang, Z.~Ren, and J.-H. Yu, {\it {Positivity from J-Basis operators in the standard model effective Field Theory}},  {\em JHEP} {\bf 05} (2024) 221, [\href{http://arxiv.org/abs/2312.04663}{{\tt arXiv:2312.04663}}].

\bibitem{Liao:2025npz}
Y.-P. Liao, J.~Roosmale~Nepveu, and C.-H. Shen, {\it {Positivity in Perturbative Renormalization: an EFT $a$-theorem}},  \href{http://arxiv.org/abs/2505.02910}{{\tt arXiv:2505.02910}}.

\bibitem{Zhang:2021eeo}
C.~Zhang, {\it {SMEFTs living on the edge: determining the UV theories from positivity and extremality}},  {\em JHEP} {\bf 12} (2022) 096, [\href{http://arxiv.org/abs/2112.11665}{{\tt arXiv:2112.11665}}].

\bibitem{Chala:2023xjy}
M.~Chala and X.~Li, {\it {Positivity restrictions on the mixing of dimension-eight SMEFT operators}},  {\em Phys. Rev. D} {\bf 109} (2024), no.~6 065015, [\href{http://arxiv.org/abs/2309.16611}{{\tt arXiv:2309.16611}}].

\bibitem{Bonnefoy:2021tbt}
Q.~Bonnefoy, E.~Gendy, C.~Grojean, and J.~T. Ruderman, {\it {Beyond Jarlskog: 699 invariants for CP violation in SMEFT}},  {\em JHEP} {\bf 08} (2022) 032, [\href{http://arxiv.org/abs/2112.03889}{{\tt arXiv:2112.03889}}].

\bibitem{Jarlskog:1985ht}
C.~Jarlskog, {\it {Commutator of the Quark Mass Matrices in the Standard Electroweak Model and a Measure of Maximal CP Violation}},  {\em Phys. Rev. Lett.} {\bf 55} (1985) 1039.

\bibitem{Jarlskog:1985cw}
C.~Jarlskog, {\it {A Basis Independent Formulation of the Connection Between Quark Mass Matrices, CP Violation and Experiment}},  {\em Z. Phys.} {\bf C29} (1985) 491--497.

\bibitem{Darvishi:2023ckq}
N.~Darvishi, Y.~Wang, and J.-H. Yu, {\it {Automated ring-diagram framework for classifying CP invariants}},  {\em Phys. Rev. D} {\bf 108} (2023), no.~11 115030, [\href{http://arxiv.org/abs/2311.15422}{{\tt arXiv:2311.15422}}].

\bibitem{Darvishi:2024cwe}
N.~Darvishi, Y.~Wang, and J.-H. Yu, {\it {Identifying CP Basis Invariants in SMEFT}},  \href{http://arxiv.org/abs/2403.18732}{{\tt arXiv:2403.18732}}.

\bibitem{Wang:2021wdq}
Y.~Wang, B.~Yu, and S.~Zhou, {\it {Flavor invariants and renormalization-group equations in the leptonic sector with massive Majorana neutrinos}},  {\em JHEP} {\bf 09} (2021) 053, [\href{http://arxiv.org/abs/2107.06274}{{\tt arXiv:2107.06274}}].

\bibitem{Yu:2021cco}
B.~Yu and S.~Zhou, {\it {Hilbert series for leptonic flavor invariants in the minimal seesaw model}},  {\em JHEP} {\bf 10} (2021) 017, [\href{http://arxiv.org/abs/2107.11928}{{\tt arXiv:2107.11928}}].

\bibitem{Yu:2022ttm}
B.~Yu and S.~Zhou, {\it {CP violation and flavor invariants in the seesaw effective field theory}},  {\em JHEP} {\bf 08} (2022) 017, [\href{http://arxiv.org/abs/2203.10121}{{\tt arXiv:2203.10121}}].

\bibitem{Jenkins:2009dy}
E.~E. Jenkins and A.~V. Manohar, {\it {Algebraic Structure of Lepton and Quark Flavor Invariants and CP Violation}},  {\em JHEP} {\bf 10} (2009) 094, [\href{http://arxiv.org/abs/0907.4763}{{\tt arXiv:0907.4763}}].

\bibitem{Feldmann:2015nia}
T.~Feldmann, T.~Mannel, and S.~Schwertfeger, {\it {Renormalization Group Evolution of Flavour Invariants}},  {\em JHEP} {\bf 10} (2015) 007, [\href{http://arxiv.org/abs/1507.00328}{{\tt arXiv:1507.00328}}].

\bibitem{Camargo-Molina:2024sde}
E.~Camargo-Molina, R.~Enberg, and J.~L{\"o}fgren, {\it {A catalog of first-order electroweak phase transitions in the Standard Model Effective Field Theory}},  {\em JHEP} {\bf 08} (2025) 113, [\href{http://arxiv.org/abs/2410.23210}{{\tt arXiv:2410.23210}}].

\bibitem{Chala:2025xlk}
M.~Chala, M.~C. Fiore, and L.~Gil, {\it {Hot news on the phase-structure of the SMEFT}},  \href{http://arxiv.org/abs/2507.16905}{{\tt arXiv:2507.16905}}.

\bibitem{Chala:2025aiz}
M.~Chala and G.~Guedes, {\it {The high-temperature limit of the SM(EFT)}},  {\em JHEP} {\bf 07} (2025) 085, [\href{http://arxiv.org/abs/2503.20016}{{\tt arXiv:2503.20016}}].

\bibitem{Chala:2025oul}
M.~Chala, L.~Gil, and Z.~Ren, {\it {Phase transitions in dimensional reduction up to three loops*}},  {\em Chin. Phys.} {\bf 49} (2025), no.~12 123105, [\href{http://arxiv.org/abs/2505.14335}{{\tt arXiv:2505.14335}}].

\bibitem{Aebischer:2018csl}
J.~Aebischer, C.~Bobeth, A.~J. Buras, and D.~M. Straub, {\it {Anatomy of $\varepsilon '/\varepsilon $ beyond the standard model}},  {\em Eur. Phys. J.} {\bf C79} (2019), no.~3 219, [\href{http://arxiv.org/abs/1808.00466}{{\tt arXiv:1808.00466}}].

\bibitem{Aebischer:2020mkv}
J.~Aebischer, A.~J. Buras, and J.~Kumar, {\it {Another SMEFT story: $Z^\prime$ facing new results on $\epsilon^\prime/\epsilon$, $\Delta M_{K}$ and $K \to \pi \nu \overline{\nu} $}},  {\em JHEP} {\bf 12} (2020) 097, [\href{http://arxiv.org/abs/2006.01138}{{\tt arXiv:2006.01138}}].

\bibitem{Liao:2024xel}
Y.~Liao, X.-D. Ma, and H.-L. Wang, {\it {Probing dimension-8 SMEFT operators through neutral meson mixing}},  {\em JHEP} {\bf 03} (2025) 133, [\href{http://arxiv.org/abs/2409.10305}{{\tt arXiv:2409.10305}}].

\bibitem{Carvunis:2025vab}
A.~Carvunis, G.~Finauri, P.~Gambino, M.~Jung, and S.~M{\"a}chler, {\it {New Physics in inclusive semileptonic $B$ decays}},  \href{http://arxiv.org/abs/2507.22123}{{\tt arXiv:2507.22123}}.

\bibitem{Deppisch:2020oyx}
F.~F. Deppisch, K.~Fridell, and J.~Harz, {\it {Probing lepton number violating interactions in rare kaon decays}},  \href{http://arxiv.org/abs/2009.04494}{{\tt arXiv:2009.04494}}.

\bibitem{Karmakar:2023rdt}
S.~Karmakar, S.~Chattopadhyay, and A.~Dighe, {\it {Identifying physics beyond SMEFT in the angular distribution of \ensuremath{\Lambda}b\textrightarrow{}\ensuremath{\Lambda}c(\textrightarrow{}\ensuremath{\Lambda}\ensuremath{\pi})\ensuremath{\tau}\ensuremath{\nu}\textasciimacron{}\ensuremath{\tau} decay}},  {\em Phys. Rev. D} {\bf 110} (2024), no.~1 015010, [\href{http://arxiv.org/abs/2305.16007}{{\tt arXiv:2305.16007}}].

\bibitem{Bhattacharya:2023beo}
S.~Bhattacharya, S.~Jahedi, S.~Nandi, and A.~Sarkar, {\it {Probing flavor constrained SMEFT operators through tc production at the muon collider}},  {\em JHEP} {\bf 07} (2024) 061, [\href{http://arxiv.org/abs/2312.14872}{{\tt arXiv:2312.14872}}].

\bibitem{Alici:2024eez}
E.~Alici, {\it {Sensitivity analysis for the anomalous tq\ensuremath{\gamma} couplings via \ensuremath{\gamma}q\textrightarrow{}t\ensuremath{\gamma} subprocess in photon\textendash{}proton collisions at the FCC-\ensuremath{\mu}p}},  {\em Results Phys.} {\bf 70} (2025) 108167, [\href{http://arxiv.org/abs/2410.19329}{{\tt arXiv:2410.19329}}].

\bibitem{Kumar:2021yod}
J.~Kumar, {\it {Renormalization group improved implications of semileptonic operators in SMEFT}},  {\em JHEP} {\bf 01} (2022) 107, [\href{http://arxiv.org/abs/2107.13005}{{\tt arXiv:2107.13005}}].

\bibitem{Ali:2023kua}
M.~I. Ali, U.~Chattopadhyay, N.~Rajeev, and J.~Roy, {\it {SMEFT analysis of charged lepton flavor violating B-meson decays}},  {\em Phys. Rev. D} {\bf 109} (2024), no.~7 075028, [\href{http://arxiv.org/abs/2312.05071}{{\tt arXiv:2312.05071}}].

\bibitem{ThomasArun:2025rgx}
M.~Thomas~Arun, S.~M, and R.~Pal, {\it {RG evolution and effect of intermediate new physics on $\Delta B=2$ six-quark operators}},  {\em JHEP} {\bf 10} (2025) 032, [\href{http://arxiv.org/abs/2506.10105}{{\tt arXiv:2506.10105}}].

\bibitem{Chattopadhyay:2025air}
U.~Chattopadhyay, D.~Das, R.~Puri, and J.~Roy, {\it {Constraining lepton flavor violating $2q 2\ell$ operators from low-energy cLFV processes}},  \href{http://arxiv.org/abs/2507.13141}{{\tt arXiv:2507.13141}}.

\bibitem{Asadi:2025dii}
P.~Asadi, H.~Bagherian, K.~Fraser, S.~Homiller, and Q.~Lu, {\it {Lepton Flavor Violation: From Muon Decays to Muon Colliders}},  \href{http://arxiv.org/abs/2509.22771}{{\tt arXiv:2509.22771}}.

\bibitem{Crivellin:2013hpa}
A.~Crivellin, S.~Najjari, and J.~Rosiek, {\it {Lepton Flavor Violation in the Standard Model with general Dimension-Six Operators}},  {\em JHEP} {\bf 04} (2014) 167, [\href{http://arxiv.org/abs/1312.0634}{{\tt arXiv:1312.0634}}].

\bibitem{Ardu:2025awk}
M.~Ardu, S.~Davidson, and N.~Valori, {\it {Left-Handed Physics is not right for EDMs}},  \href{http://arxiv.org/abs/2507.19421}{{\tt arXiv:2507.19421}}.

\bibitem{Pruna:2014asa}
G.~M. Pruna and A.~Signer, {\it {The $\mu\to e\gamma$ decay in a systematic effective field theory approach with dimension 6 operators}},  {\em JHEP} {\bf 10} (2014) 014, [\href{http://arxiv.org/abs/1408.3565}{{\tt arXiv:1408.3565}}].

\bibitem{Ardu:2022pzk}
M.~Ardu, S.~Davidson, and M.~Gorbahn, {\it {Sensitivity of \ensuremath{\mu}\textrightarrow{}e processes to \ensuremath{\tau} flavor change}},  {\em Phys. Rev. D} {\bf 105} (2022), no.~9 096040, [\href{http://arxiv.org/abs/2202.09246}{{\tt arXiv:2202.09246}}].

\bibitem{Greljo:2025ljr}
A.~Greljo, A.~Palavri{\'c}, M.~Tunja, and J.~Zupan, {\it {Expanding the Landscape of Exotic Muon Decays}},  \href{http://arxiv.org/abs/2510.08674}{{\tt arXiv:2510.08674}}.

\bibitem{Pruna:2015jhf}
G.~M. Pruna and A.~Signer, {\it {Lepton-flavour violating decays in theories with dimension 6 operators}},  {\em EPJ Web Conf.} {\bf 118} (2016) 01031, [\href{http://arxiv.org/abs/1511.04421}{{\tt arXiv:1511.04421}}].

\bibitem{Ardu:2024bua}
M.~Ardu, S.~Davidson, and S.~Lavignac, {\it {Constraining new physics models from $\mu \rightarrow e $ observables in bottom-up EFT}},  {\em Eur. Phys. J. C} {\bf 84} (2024), no.~5 458, [\href{http://arxiv.org/abs/2401.06214}{{\tt arXiv:2401.06214}}].

\bibitem{Sarkar:2025bgo}
A.~Sarkar, {\it {Lepton flavor violating top quark FCNC at the $\mu$TRISTAN}},  \href{http://arxiv.org/abs/2506.18015}{{\tt arXiv:2506.18015}}.

\bibitem{Isidori:2021gqe}
G.~Isidori, J.~Pag\`es, and F.~Wilsch, {\it {Flavour alignment of New Physics in light of the (g \ensuremath{-} 2)$_{\mu}$ anomaly}},  {\em JHEP} {\bf 03} (2022) 011, [\href{http://arxiv.org/abs/2111.13724}{{\tt arXiv:2111.13724}}].

\bibitem{Calibbi:2021pyh}
L.~Calibbi, X.~Marcano, and J.~Roy, {\it {Z lepton flavour violation as a probe for new physics at future $e^+e^-$ colliders}},  {\em Eur. Phys. J. C} {\bf 81} (2021), no.~12 1054, [\href{http://arxiv.org/abs/2107.10273}{{\tt arXiv:2107.10273}}].

\bibitem{Altmannshofer:2023tsa}
W.~Altmannshofer, P.~Munbodh, and T.~Oh, {\it {Probing lepton flavor violation at Circular Electron-Positron Colliders}},  {\em JHEP} {\bf 08} (2023) 026, [\href{http://arxiv.org/abs/2305.03869}{{\tt arXiv:2305.03869}}].

\bibitem{Altmannshofer:2025nbp}
W.~Altmannshofer and P.~Munbodh, {\it {Probing lepton flavor violation at linear electron-positron colliders}},  {\em JHEP} {\bf 08} (2025) 147, [\href{http://arxiv.org/abs/2505.11653}{{\tt arXiv:2505.11653}}].

\bibitem{Jahedi:2024kvi}
S.~Jahedi and A.~Sarkar, {\it {Exploring optimal sensitivity of lepton flavor violating effective couplings at the $e^+e^-$ colliders}},  {\em Phys. Rev. D} {\bf 110} (2024), no.~9 095021, [\href{http://arxiv.org/abs/2408.00190}{{\tt arXiv:2408.00190}}].

\bibitem{Chala:2021juk}
M.~Chala and A.~Titov, {\it {Neutrino masses in the Standard Model effective field theory}},  {\em Phys. Rev. D} {\bf 104} (2021), no.~3 035002, [\href{http://arxiv.org/abs/2104.08248}{{\tt arXiv:2104.08248}}].

\bibitem{Falkowski:2019xoe}
A.~Falkowski, M.~Gonz\'alez-Alonso, and Z.~Tabrizi, {\it {Reactor neutrino oscillations as constraints on Effective Field Theory}},  {\em JHEP} {\bf 05} (2019) 173, [\href{http://arxiv.org/abs/1901.04553}{{\tt arXiv:1901.04553}}].

\bibitem{Du:2021rdg}
Y.~Du, H.-L. Li, J.~Tang, S.~Vihonen, and J.-H. Yu, {\it {Exploring SMEFT induced nonstandard interactions: From COHERENT to neutrino oscillations}},  {\em Phys. Rev. D} {\bf 105} (2022), no.~7 075022, [\href{http://arxiv.org/abs/2106.15800}{{\tt arXiv:2106.15800}}].

\bibitem{Du:2020dwr}
Y.~Du, H.-L. Li, J.~Tang, S.~Vihonen, and J.-H. Yu, {\it {Non-standard interactions in SMEFT confronted with terrestrial neutrino experiments}},  {\em JHEP} {\bf 03} (2021) 019, [\href{http://arxiv.org/abs/2011.14292}{{\tt arXiv:2011.14292}}].

\bibitem{Kopp:2025ffx}
J.~Kopp, Z.~Tabrizi, and S.~Urrea, {\it {Effective Field Theory in Long-Baseline Neutrino Oscillation Experiments}},  \href{http://arxiv.org/abs/2509.21537}{{\tt arXiv:2509.21537}}.

\bibitem{Graf:2025cfk}
L.~Gr\'af, C.~Hati, A.~Mart\'\i{}n-Gal\'an, and O.~Scholer, {\it {Importance of Loop Effects in Probing Lepton Number Violation}},  \href{http://arxiv.org/abs/2504.00081}{{\tt arXiv:2504.00081}}.

\bibitem{Donini:2025cuy}
A.~Donini, M.~Gonz{\'a}lez, M.~Hirsch, and N.~A. Neill, {\it {Leading large $N_c$ contributions to Lepton Number Violating Meson Decays}},  \href{http://arxiv.org/abs/2510.24835}{{\tt arXiv:2510.24835}}.

\bibitem{Fridell:2023rtr}
K.~Fridell, L.~Gr\'af, J.~Harz, and C.~Hati, {\it {Probing lepton number violation: a comprehensive survey of dimension-7 SMEFT}},  {\em JHEP} {\bf 05} (2024) 154, [\href{http://arxiv.org/abs/2306.08709}{{\tt arXiv:2306.08709}}].

\bibitem{Heeck:2024uiz}
J.~Heeck and M.~Sokhashvili, {\it {Lepton flavor violation by two units}},  {\em Phys. Lett. B} {\bf 852} (2024) 138621, [\href{http://arxiv.org/abs/2401.09580}{{\tt arXiv:2401.09580}}].

\bibitem{Bhattacharya:2025xwv}
S.~Bhattacharya, S.~Datta, and A.~Sarkar, {\it {Probing $\Delta L=2$ lepton number violating SMEFT operators at the same-sign muon collider}},  \href{http://arxiv.org/abs/2505.20936}{{\tt arXiv:2505.20936}}.

\bibitem{Heeck:2025jfs}
J.~Heeck, M.~Sokhashvili, and A.~Thapa, {\it {Lepton flavor violation by three units}},  {\em Phys. Rev. D} {\bf 112} (2025), no.~5 055045, [\href{http://arxiv.org/abs/2505.17178}{{\tt arXiv:2505.17178}}].

\bibitem{Aebischer:2021uvt}
J.~Aebischer, W.~Dekens, E.~E. Jenkins, A.~V. Manohar, D.~Sengupta, and P.~Stoffer, {\it {Effective field theory interpretation of lepton magnetic and electric dipole moments}},  {\em JHEP} {\bf 07} (2021) 107, [\href{http://arxiv.org/abs/2102.08954}{{\tt arXiv:2102.08954}}].

\bibitem{Fajfer:2021cxa}
S.~Fajfer, J.~F. Kamenik, and M.~Tammaro, {\it {Interplay of New Physics effects in $(g-2)_l$ and $h \to \ell^+\ell^-$ -Lessons from SMEFT}},  {\em JHEP} {\bf 06} (2021) 099, [\href{http://arxiv.org/abs/2103.10859}{{\tt arXiv:2103.10859}}].

\bibitem{Cirigliano:2019vfc}
V.~Cirigliano, A.~Crivellin, W.~Dekens, J.~de~Vries, M.~Hoferichter, and E.~Mereghetti, {\it {CP Violation in Higgs-Gauge Interactions: From Tabletop Experiments to the LHC}},  {\em Phys. Rev. Lett.} {\bf 123} (2019), no.~5 051801, [\href{http://arxiv.org/abs/1903.03625}{{\tt arXiv:1903.03625}}].

\bibitem{Unal:2023wct}
Y.~{\"U}nal, {\it {Electric dipole moments of charm baryons using dimension-six operators}},  {\em Phys. Rev. D} {\bf 108} (2023), no.~7 075023, [\href{http://arxiv.org/abs/2306.03639}{{\tt arXiv:2306.03639}}].

\bibitem{Bonnefoy:2024gca}
Q.~Bonnefoy, J.~Kley, D.~Liu, A.~N. Rossia, and C.-Y. Yao, {\it {Aligned yet large dipoles: a SMEFT study}},  {\em JHEP} {\bf 11} (2024) 046, [\href{http://arxiv.org/abs/2403.13065}{{\tt arXiv:2403.13065}}].

\bibitem{Bischer:2021jqn}
I.~Bischer, W.~Rodejohann, P.~S.~B. Dev, X.-J. Xu, and Y.~Zhang, {\it {Searching for new physics from SMEFT and leptoquarks at the P2 experiment}},  {\em Phys. Rev. D} {\bf 105} (2022), no.~9 095016, [\href{http://arxiv.org/abs/2112.12051}{{\tt arXiv:2112.12051}}].

\bibitem{Dawson:2021xei}
S.~Dawson, S.~Homiller, and M.~Sullivan, {\it {Impact of dimension-eight SMEFT contributions: A case study}},  {\em Phys. Rev. D} {\bf 104} (2021), no.~11 115013, [\href{http://arxiv.org/abs/2110.06929}{{\tt arXiv:2110.06929}}].

\bibitem{Guchait:2022ktz}
M.~Guchait and A.~Roy, {\it {Exploring SMEFT operators in the tHq production at the LHC}},  {\em JHEP} {\bf 10} (2023) 064, [\href{http://arxiv.org/abs/2210.05503}{{\tt arXiv:2210.05503}}].

\bibitem{Hays:2018zze}
C.~Hays, A.~Martin, V.~Sanz, and J.~Setford, {\it {On the impact of dimension-eight SMEFT operators on Higgs measurements}},  {\em JHEP} {\bf 02} (2019) 123, [\href{http://arxiv.org/abs/1808.00442}{{\tt arXiv:1808.00442}}].

\bibitem{Banerjee:2025dsh}
S.~Banerjee, R.~S. Gupta, S.~Jain, M.~Mangano, and E.~Venturini, {\it {An EFT study of the $pp \to \bar{t} t Z(ll) h(bb)$ process at the FCC-$\boldsymbol{hh}$}},  \href{http://arxiv.org/abs/2509.10449}{{\tt arXiv:2509.10449}}.

\bibitem{Subba:2024aut}
A.~Subba, R.~K. Singh, and R.~M. Godbole, {\it {Looking into the quantum entanglement in $H\to ZZ^\star$ at LHC within SMEFT framework}},  \href{http://arxiv.org/abs/2411.19171}{{\tt arXiv:2411.19171}}.

\bibitem{Biekotter:2020flu}
A.~Biek\"otter, R.~Gomez-Ambrosio, P.~Gregg, F.~Krauss, and M.~Sch\"onherr, {\it {Constraining SMEFT operators with associated $h\gamma$ production in weak boson fusion}},  {\em Phys. Lett. B} {\bf 814} (2021) 136079, [\href{http://arxiv.org/abs/2003.06379}{{\tt arXiv:2003.06379}}].

\bibitem{Araz:2020zyh}
J.~Y. Araz, S.~Banerjee, R.~S. Gupta, and M.~Spannowsky, {\it {Precision SMEFT bounds from the VBF Higgs at high transverse momentum}},  {\em JHEP} {\bf 04} (2021) 125, [\href{http://arxiv.org/abs/2011.03555}{{\tt arXiv:2011.03555}}].

\bibitem{Battaglia:2021nys}
M.~Battaglia, M.~Grazzini, M.~Spira, and M.~Wiesemann, {\it {Sensitivity to BSM effects in the Higgs p$_{T}$ spectrum within SMEFT}},  {\em JHEP} {\bf 11} (2021) 173, [\href{http://arxiv.org/abs/2109.02987}{{\tt arXiv:2109.02987}}].

\bibitem{Domenech:2022uud}
D.~Domenech, M.~J. Herrero, R.~A. Morales, and M.~Ramos, {\it {Double Higgs boson production at TeV $e^+e^-$ colliders with effective field theories: Sensitivity to BSM Higgs couplings}},  {\em Phys. Rev. D} {\bf 106} (2022), no.~11 115027, [\href{http://arxiv.org/abs/2208.05452}{{\tt arXiv:2208.05452}}].

\bibitem{Freitas:2019hbk}
F.~F. Freitas, C.~K. Khosa, and V.~Sanz, {\it {Exploring the standard model EFT in VH production with machine learning}},  {\em Phys. Rev. D} {\bf 100} (2019), no.~3 035040, [\href{http://arxiv.org/abs/1902.05803}{{\tt arXiv:1902.05803}}].

\bibitem{Silva:2025hzo}
M.~Silva, R.~Barru{\'e}, I.~Ochoa, and P.~Conde~Mu{\'\i}{\~n}o, {\it {Searching for HWW Anomalous Couplings with Simulation-Based Inference}},  \href{http://arxiv.org/abs/2509.03307}{{\tt arXiv:2509.03307}}.

\bibitem{Barger:2023wbg}
V.~Barger, K.~Hagiwara, and Y.-J. Zheng, {\it {CP-violating top-Higgs coupling in SMEFT}},  {\em Phys. Lett. B} {\bf 850} (2024) 138547, [\href{http://arxiv.org/abs/2310.10852}{{\tt arXiv:2310.10852}}].

\bibitem{Delgado:2023ynh}
R.~L. Delgado, R.~G\'omez-Ambrosio, J.~Mart\'\i{}nez-Mart\'\i{}n, A.~Salas-Bern\'ardez, and J.~J. Sanz-Cillero, {\it {Production of two, three, and four Higgs bosons: where SMEFT and HEFT depart}},  {\em JHEP} {\bf 03} (2024) 037, [\href{http://arxiv.org/abs/2311.04280}{{\tt arXiv:2311.04280}}].

\bibitem{Brivio:2019myy}
I.~Brivio, T.~Corbett, and M.~Trott, {\it {The Higgs width in the SMEFT}},  {\em JHEP} {\bf 10} (2019) 056, [\href{http://arxiv.org/abs/1906.06949}{{\tt arXiv:1906.06949}}].

\bibitem{Hays:2020scx}
C.~Hays, A.~Helset, A.~Martin, and M.~Trott, {\it {Exact SMEFT formulation and expansion to $\mathcal{O}(v^4/\Lambda^4)$}},  {\em JHEP} {\bf 11} (2020) 087, [\href{http://arxiv.org/abs/2007.00565}{{\tt arXiv:2007.00565}}].

\bibitem{Grojean:2024tcw}
C.~Grojean, G.~Guedes, J.~Roosmale~Nepveu, and G.~M. Salla, {\it {A log story short: running contributions to radiative Higgs decays in the SMEFT}},  {\em JHEP} {\bf 12} (2024) 065, [\href{http://arxiv.org/abs/2405.20371}{{\tt arXiv:2405.20371}}].

\bibitem{Bhattacharya:2014wla}
B.~Bhattacharya, A.~Datta, D.~London, and S.~Shivashankara, {\it {Simultaneous Explanation of the $R_K$ and $R(D^{(*)})$ Puzzles}},  {\em Phys. Lett.} {\bf B742} (2015) 370--374, [\href{http://arxiv.org/abs/1412.7164}{{\tt arXiv:1412.7164}}].

\bibitem{Bause:2020auq}
R.~Bause, H.~Gisbert, M.~Golz, and G.~Hiller, {\it {Lepton universality and lepton flavor conservation tests with dineutrino modes}},  {\em Eur. Phys. J. C} {\bf 82} (2022), no.~2 164, [\href{http://arxiv.org/abs/2007.05001}{{\tt arXiv:2007.05001}}].

\bibitem{Karmakar:2024gla}
S.~Karmakar, A.~Dighe, and R.~S. Gupta, {\it {SMEFT predictions for semileptonic processes}},  {\em Phys. Rev. D} {\bf 111} (2025), no.~5 055002, [\href{http://arxiv.org/abs/2404.10061}{{\tt arXiv:2404.10061}}].

\bibitem{Cirigliano:2017tqn}
V.~Cirigliano, A.~Crivellin, and M.~Hoferichter, {\it {No-go theorem for nonstandard explanations of the $\tau\to K_S\pi\nu_\tau$ CP asymmetry}},  {\em Phys. Rev. Lett.} {\bf 120} (2018), no.~14 141803, [\href{http://arxiv.org/abs/1712.06595}{{\tt arXiv:1712.06595}}].

\bibitem{Buras:2014fpa}
A.~J. Buras, J.~Girrbach-Noe, C.~Niehoff, and D.~M. Straub, {\it {$B\to K^{(*)}\nu\bar\nu$ decays in the Standard Model and beyond}},  {\em JHEP} {\bf 1502} (2015) 184, [\href{http://arxiv.org/abs/1409.4557}{{\tt arXiv:1409.4557}}].

\bibitem{Hamoudou:2022tdn}
S.~Hamoudou, J.~Kumar, and D.~London, {\it {Dimension-8 SMEFT matching conditions for the low-energy effective field theory}},  {\em JHEP} {\bf 03} (2023) 157, [\href{http://arxiv.org/abs/2207.08856}{{\tt arXiv:2207.08856}}].

\bibitem{Grinstein:2024jqt}
B.~Grinstein, X.~Lu, C.~Mir\'o, and P.~Qu\'\i{}lez, {\it {Accidental symmetries, Hilbert series, and friends}},  {\em JHEP} {\bf 03} (2025) 172, [\href{http://arxiv.org/abs/2412.05359}{{\tt arXiv:2412.05359}}].

\bibitem{deBoer:2025jhc}
T.~de~Boer, F.~Goertz, and A.~Incrocci, {\it {The goofy-symmetric Standard Model and the Hierarchy Problem}},  \href{http://arxiv.org/abs/2507.22111}{{\tt arXiv:2507.22111}}.

\bibitem{Trautner:2025prm}
A.~Trautner, {\it {Goofy transformations and the hierarchy problem}},  \href{http://arxiv.org/abs/2508.02646}{{\tt arXiv:2508.02646}}.

\bibitem{Bellazzini:2018paj}
B.~Bellazzini and F.~Riva, {\it {New phenomenological and theoretical perspective on anomalous ZZ and Z\ensuremath{\gamma} processes}},  {\em Phys. Rev. D} {\bf 98} (2018), no.~9 095021, [\href{http://arxiv.org/abs/1806.09640}{{\tt arXiv:1806.09640}}].

\bibitem{Degrande:2021zpv}
C.~Degrande and J.~Touch\`eque, {\it {A reduced basis for CP violation in SMEFT at colliders and its application to diboson production}},  {\em JHEP} {\bf 04} (2022) 032, [\href{http://arxiv.org/abs/2110.02993}{{\tt arXiv:2110.02993}}].

\bibitem{Degrande:2012wf}
C.~Degrande, N.~Greiner, W.~Kilian, O.~Mattelaer, H.~Mebane, T.~Stelzer, S.~Willenbrock, and C.~Zhang, {\it {Effective Field Theory: A Modern Approach to Anomalous Couplings}},  {\em Annals Phys.} {\bf 335} (2013) 21--32, [\href{http://arxiv.org/abs/1205.4231}{{\tt arXiv:1205.4231}}].

\bibitem{Falkowski:2016cxu}
A.~Falkowski, M.~Gonzalez-Alonso, A.~Greljo, D.~Marzocca, and M.~Son, {\it {Anomalous Triple Gauge Couplings in the Effective Field Theory Approach at the LHC}},  {\em JHEP} {\bf 02} (2017) 115, [\href{http://arxiv.org/abs/1609.06312}{{\tt arXiv:1609.06312}}].

\bibitem{ElFaham:2024uop}
H.~El~Faham, G.~Pelliccioli, and E.~Vryonidou, {\it {Triple-gauge couplings in LHC diboson production: a SMEFT view from every angle}},  {\em JHEP} {\bf 08} (2024) 087, [\href{http://arxiv.org/abs/2405.19083}{{\tt arXiv:2405.19083}}].

\bibitem{Aoude:2019cmc}
R.~Aoude and W.~Shepherd, {\it {Jet Substructure Measurements of Interference in Non-Interfering SMEFT Effects}},  {\em JHEP} {\bf 08} (2019) 009, [\href{http://arxiv.org/abs/1902.11262}{{\tt arXiv:1902.11262}}].

\bibitem{Celada:2024cxw}
E.~Celada, G.~Durieux, K.~Mimasu, and E.~Vryonidou, {\it {Triboson production in the SMEFT}},  {\em JHEP} {\bf 12} (2024) 055, [\href{http://arxiv.org/abs/2407.09600}{{\tt arXiv:2407.09600}}].

\bibitem{Chatterjee:2024pbp}
S.~Chatterjee, S.~S. Cruz, R.~Sch\"ofbeck, and D.~Schwarz, {\it {Rotation-equivariant graph neural network for learning hadronic SMEFT effects}},  {\em Phys. Rev. D} {\bf 109} (2024), no.~7 076012, [\href{http://arxiv.org/abs/2401.10323}{{\tt arXiv:2401.10323}}].

\bibitem{Martin:2023tvi}
A.~Martin, {\it {A case study of SMEFT $ \mathcal{O}\left(1/{\Lambda}^4\right) $ effects in diboson processes: pp \textrightarrow{} W$^{±}$(\ensuremath{\ell}$^{±}$\ensuremath{\nu})\ensuremath{\gamma}}},  {\em JHEP} {\bf 05} (2024) 223, [\href{http://arxiv.org/abs/2312.09867}{{\tt arXiv:2312.09867}}].

\bibitem{Degrande:2023iob}
C.~Degrande and H.-L. Li, {\it {Impact of dimension-8 SMEFT operators on diboson productions}},  {\em JHEP} {\bf 06} (2023) 149, [\href{http://arxiv.org/abs/2303.10493}{{\tt arXiv:2303.10493}}].

\bibitem{Ethier:2021ydt}
J.~J. Ethier, R.~Gomez-Ambrosio, G.~Magni, and J.~Rojo, {\it {SMEFT analysis of vector boson scattering and diboson data from the LHC Run II}},  {\em Eur. Phys. J. C} {\bf 81} (2021), no.~6 560, [\href{http://arxiv.org/abs/2101.03180}{{\tt arXiv:2101.03180}}].

\bibitem{Gauld:2024glt}
R.~Gauld, U.~Haisch, and J.~Weiss, {\it {A tale of $Z$+jet: SMEFT effects and the Lam-Tung relation}},  {\em SciPost Phys.} {\bf 18} (2025) 148, [\href{http://arxiv.org/abs/2412.13014}{{\tt arXiv:2412.13014}}].

\bibitem{Boughezal:2020klp}
R.~Boughezal, C.-Y. Chen, F.~Petriello, and D.~Wiegand, {\it {Four-lepton $Z$ boson decay constraints on the standard model EFT}},  {\em Phys. Rev. D} {\bf 103} (2021), no.~5 055015, [\href{http://arxiv.org/abs/2010.06685}{{\tt arXiv:2010.06685}}].

\bibitem{DiCanto:2025fpk}
A.~Di~Canto, T.~Hacheney, G.~Hiller, D.~S. Mitzel, S.~Monteil, L.~R{\"o}hrig, and D.~Suelmann, {\it {New opportunities for rare charm from $Z\to c\bar{c}$ decays}},  \href{http://arxiv.org/abs/2509.10447}{{\tt arXiv:2509.10447}}.

\bibitem{Bjorn:2016zlr}
M.~Bj\o{}rn and M.~Trott, {\it {Interpreting $W$ mass measurements in the SMEFT}},  {\em Phys. Lett. B} {\bf 762} (2016) 426--431, [\href{http://arxiv.org/abs/1606.06502}{{\tt arXiv:1606.06502}}].

\bibitem{Wells:2015uba}
J.~D. Wells and Z.~Zhang, {\it {Effective theories of universal theories}},  {\em JHEP} {\bf 01} (2016) 123, [\href{http://arxiv.org/abs/1510.08462}{{\tt arXiv:1510.08462}}].

\bibitem{Wells:2015cre}
J.~D. Wells and Z.~Zhang, {\it {Renormalization group evolution of the universal theories EFT}},  {\em JHEP} {\bf 06} (2016) 122, [\href{http://arxiv.org/abs/1512.03056}{{\tt arXiv:1512.03056}}].

\bibitem{Maltoni:2019aot}
F.~Maltoni, L.~Mantani, and K.~Mimasu, {\it {Top-quark electroweak interactions at high energy}},  {\em JHEP} {\bf 10} (2019) 004, [\href{http://arxiv.org/abs/1904.05637}{{\tt arXiv:1904.05637}}].

\bibitem{Barman:2022vjd}
R.~K. Barman and A.~Ismail, {\it {Constraining the top electroweak sector of the SMEFT through $Z$ associated top pair and single top production at the HL-LHC}},  \href{http://arxiv.org/abs/2205.07912}{{\tt arXiv:2205.07912}}.

\bibitem{Aoude:2022deh}
R.~Aoude, H.~El~Faham, F.~Maltoni, and E.~Vryonidou, {\it {Complete SMEFT predictions for four top quark production at hadron colliders}},  {\em JHEP} {\bf 10} (2022) 163, [\href{http://arxiv.org/abs/2208.04962}{{\tt arXiv:2208.04962}}].

\bibitem{Maltoni:2024csn}
F.~Maltoni, C.~Severi, S.~Tentori, and E.~Vryonidou, {\it {Quantum tops at circular lepton colliders}},  {\em JHEP} {\bf 09} (2024) 001, [\href{http://arxiv.org/abs/2404.08049}{{\tt arXiv:2404.08049}}].

\bibitem{Jahedi:2024wnw}
S.~Jahedi, J.~Lahiri, and A.~Subba, {\it {Optimal sensitivity of anomalous charged triple gauge couplings through W boson helicity at the e$^+$e$^-$ colliders}},  {\em JHEP} {\bf 07} (2025) 280, [\href{http://arxiv.org/abs/2411.13664}{{\tt arXiv:2411.13664}}].

\bibitem{Ellis:2025jgt}
J.~Ellis, H.-J. He, R.-Q. Xiao, and S.-P. Zeng, {\it {Probing Neutral Triple Gauge Couplings via $ZZ$ Production at $e^+e^-$ Colliders with Machine Learning}},  \href{http://arxiv.org/abs/2506.21433}{{\tt arXiv:2506.21433}}.

\bibitem{Ellis:2025ghl}
J.~Ellis, H.-J. He, and R.-Q. Xiao, {\it {Probing CP-violating neutral triple gauge couplings at electron-positron colliders}},  {\em Sci. China Phys. Mech. Astron.} {\bf 68} (2025), no.~12 121062, [\href{http://arxiv.org/abs/2504.13135}{{\tt arXiv:2504.13135}}].

\bibitem{Bhattacharya:2025jhs}
S.~Bhattacharya, A.~Subba, and A.~Sarkar, {\it {Optimal estimation of Higgs-Gauge Boson couplings at the future $e^+e^-$ colliders}},  \href{http://arxiv.org/abs/2508.08893}{{\tt arXiv:2508.08893}}.

\bibitem{Helset:2017mlf}
A.~Helset and M.~Trott, {\it {On interference and non-interference in the SMEFT}},  {\em JHEP} {\bf 04} (2018) 038, [\href{http://arxiv.org/abs/1711.07954}{{\tt arXiv:1711.07954}}].

\bibitem{Alte:2017pme}
S.~Alte, M.~K\"onig, and W.~Shepherd, {\it {Consistent Searches for SMEFT Effects in Non-Resonant Dijet Events}},  {\em JHEP} {\bf 01} (2018) 094, [\href{http://arxiv.org/abs/1711.07484}{{\tt arXiv:1711.07484}}].

\bibitem{Goldouzian:2020wdq}
R.~Goldouzian and M.~D. Hildreth, {\it {LHC dijet angular distributions as a probe for the dimension-six triple gluon vertex}},  {\em Phys. Lett. B} {\bf 811} (2020) 135889, [\href{http://arxiv.org/abs/2001.02736}{{\tt arXiv:2001.02736}}].

\bibitem{Keilmann:2019cbp}
E.~Keilmann and W.~Shepherd, {\it {Dijets at Tevatron Cannot Constrain SMEFT Four-Quark Operators}},  {\em JHEP} {\bf 09} (2019) 086, [\href{http://arxiv.org/abs/1907.13160}{{\tt arXiv:1907.13160}}].

\bibitem{Kim:2022amu}
T.~Kim and A.~Martin, {\it {Monolepton production in SMEFT to $ \mathcal{O} $(1/\ensuremath{\Lambda}$^{4}$) and beyond}},  {\em JHEP} {\bf 09} (2022) 124, [\href{http://arxiv.org/abs/2203.11976}{{\tt arXiv:2203.11976}}].

\bibitem{Alte:2018xgc}
S.~Alte, M.~K\"onig, and W.~Shepherd, {\it {Consistent Searches for SMEFT Effects in Non-Resonant Dilepton Events}},  {\em JHEP} {\bf 07} (2019) 144, [\href{http://arxiv.org/abs/1812.07575}{{\tt arXiv:1812.07575}}].

\bibitem{Horne:2020pot}
A.~Horne, J.~Pittman, M.~Snedeker, W.~Shepherd, and J.~W. Walker, {\it {Shift-Type SMEFT Effects in Dileptons at the LHC}},  {\em JHEP} {\bf 03} (2021) 118, [\href{http://arxiv.org/abs/2007.12698}{{\tt arXiv:2007.12698}}].

\bibitem{Boughezal:2020uwq}
R.~Boughezal, F.~Petriello, and D.~Wiegand, {\it {Removing flat directions in standard model EFT fits: How polarized electron-ion collider data can complement the LHC}},  {\em Phys. Rev. D} {\bf 101} (2020), no.~11 116002, [\href{http://arxiv.org/abs/2004.00748}{{\tt arXiv:2004.00748}}].

\bibitem{Grossi:2024tou}
S.~Grossi and R.~Torre, {\it {More variables or more bins? Impact on the EFT interpretation of Drell\textendash{}Yan measurements}},  {\em Eur. Phys. J. C} {\bf 84} (2024), no.~7 713, [\href{http://arxiv.org/abs/2404.10569}{{\tt arXiv:2404.10569}}].

\bibitem{Hiller:2025zov}
G.~Hiller, L.~Nollen, and D.~Wendler, {\it {Teaming up MET plus jet with Drell-Yan in the SMEFT}},  {\em PoS} {\bf DISCRETE2024} (2025) 002, [\href{http://arxiv.org/abs/2503.13638}{{\tt arXiv:2503.13638}}].

\bibitem{Berthier:2015oma}
L.~Berthier and M.~Trott, {\it {Towards consistent Electroweak Precision Data constraints in the SMEFT}},  {\em JHEP} {\bf 05} (2015) 024, [\href{http://arxiv.org/abs/1502.02570}{{\tt arXiv:1502.02570}}].

\bibitem{Berthier:2016tkq}
L.~Berthier, M.~Bj\o{}rn, and M.~Trott, {\it {Incorporating doubly resonant $W^\pm$ data in a global fit of SMEFT parameters to lift flat directions}},  {\em JHEP} {\bf 09} (2016) 157, [\href{http://arxiv.org/abs/1606.06693}{{\tt arXiv:1606.06693}}].

\bibitem{Liu:2024tcz}
D.~Liu, R.-Q. Xiao, S.~Li, J.~Ellis, H.-J. He, and R.~Yuan, {\it {Probing Neutral Triple Gauge Couplings via $\boldsymbol{Z\gamma\,(\ell^+\ell^-\gamma)}$ Production at $\boldsymbol{e^+e^-}$ Colliders}},  {\em Front. Phys. (Beijing)} {\bf 20} (2025), no.~1 015201, [\href{http://arxiv.org/abs/2404.15937}{{\tt arXiv:2404.15937}}].

\bibitem{Subba:2025pos}
A.~Subba and R.~K. Singh, {\it {Bounds on SMEFT affecting multi gauge and Higgs-gauge couplings using two and three body spin correlations in $e^-e^+\to 3l2j\slashed{E}$ process}},  \href{http://arxiv.org/abs/2509.01452}{{\tt arXiv:2509.01452}}.

\bibitem{Xie:2025izk}
W.~Xie and J.-C. Yang, {\it {Searching for the neutral triple gauge couplings in the process $\mu^+\mu^-\to \gamma \nu\bar \nu$ at muon colliders}},  \href{http://arxiv.org/abs/2507.08681}{{\tt arXiv:2507.08681}}.

\bibitem{Biondini:2025gpg}
S.~Biondini, L.~Tiberi, and O.~Panella, {\it {Connecting $t$-channel Dark Matter Models to the Standard Model Effective Field Theory}},  \href{http://arxiv.org/abs/2507.00925}{{\tt arXiv:2507.00925}}.

\bibitem{Olgoso:2025jot}
P.~Olgoso, P.~Paradisi, and N.~Selimovic, {\it {The Dark Side of a Tera-Z Factory}},  \href{http://arxiv.org/abs/2507.17803}{{\tt arXiv:2507.17803}}.

\bibitem{Hashino:2022ghd}
K.~Hashino and D.~Ueda, {\it {SMEFT effects on the gravitational wave spectrum from an electroweak phase transition}},  {\em Phys. Rev. D} {\bf 107} (2023), no.~9 095022, [\href{http://arxiv.org/abs/2210.11241}{{\tt arXiv:2210.11241}}].

\bibitem{Hashino:2025nku}
K.~Hashino and D.~Ueda, {\it {RGE effects on new physics searches via gravitational waves}},  {\em JHEP} {\bf 09} (2025) 094, [\href{http://arxiv.org/abs/2505.13074}{{\tt arXiv:2505.13074}}].

\bibitem{Banerjee:2024qiu}
U.~Banerjee, S.~Chakraborty, S.~Prakash, and S.~U. Rahaman, {\it {Feasibility of ultrarelativistic bubbles in SMEFT}},  {\em Phys. Rev. D} {\bf 110} (2024), no.~5 055002, [\href{http://arxiv.org/abs/2402.02914}{{\tt arXiv:2402.02914}}].

\bibitem{Wen:2023xxc}
X.-K. Wen, B.~Yan, Z.~Yu, and C.~P. Yuan, {\it {Single Transverse Spin Asymmetry as a New Probe of Standard-Model-Effective-Field-Theory Dipole Operators}},  {\em Phys. Rev. Lett.} {\bf 131} (2023), no.~24 241801, [\href{http://arxiv.org/abs/2307.05236}{{\tt arXiv:2307.05236}}].

\bibitem{Krauss:2016ely}
F.~Krauss, S.~Kuttimalai, and T.~Plehn, {\it {LHC multijet events as a probe for anomalous dimension-six gluon interactions}},  {\em Phys. Rev. D} {\bf 95} (2017), no.~3 035024, [\href{http://arxiv.org/abs/1611.00767}{{\tt arXiv:1611.00767}}].

\bibitem{Hirschi:2018etq}
V.~Hirschi, F.~Maltoni, I.~Tsinikos, and E.~Vryonidou, {\it {Constraining anomalous gluon self-interactions at the LHC: a reappraisal}},  {\em JHEP} {\bf 07} (2018) 093, [\href{http://arxiv.org/abs/1806.04696}{{\tt arXiv:1806.04696}}].

\bibitem{Degrande:2020tno}
C.~Degrande and M.~Maltoni, {\it {Reviving the interference: framework and proof-of-principle for the anomalous gluon self-interaction in the SMEFT}},  {\em Phys. Rev. D} {\bf 103} (2021), no.~9 095009, [\href{http://arxiv.org/abs/2012.06595}{{\tt arXiv:2012.06595}}].

\bibitem{Gargalionis:2024nij}
J.~Gargalionis, J.~Herrero-Garc\'\i{}a, and M.~A. Schmidt, {\it {Model-independent estimates for loop-induced baryon-number-violating nucleon decays}},  {\em JHEP} {\bf 06} (2024) 182, [\href{http://arxiv.org/abs/2401.04768}{{\tt arXiv:2401.04768}}].

\bibitem{Cata:2020crs}
O.~Cata, W.~Kilian, and N.~Kreher, {\it {Gauge anomalies in the Standard-Model Effective Field Theory}},  \href{http://arxiv.org/abs/2011.09976}{{\tt arXiv:2011.09976}}.

\bibitem{Beis:2025zzd}
D.~Beis and A.~Dedes, {\it {Chiral anomaly cancellation and neutral triple gauge boson vertices in the SM EFT}},  {\em Phys. Rev. D} {\bf 112} (2025), no.~5 056006, [\href{http://arxiv.org/abs/2503.18079}{{\tt arXiv:2503.18079}}].

\bibitem{Cohen:2023gap}
T.~Cohen, X.~Lu, and Z.~Zhang, {\it {Anomaly cancellation in effective field theories from the covariant derivative expansion}},  {\em Phys. Rev. D} {\bf 108} (2023), no.~5 056027, [\href{http://arxiv.org/abs/2301.00827}{{\tt arXiv:2301.00827}}].

\bibitem{Cohen:2023hmq}
T.~Cohen, X.~Lu, and Z.~Zhang, {\it {Anomalies from the covariant derivative expansion}},  {\em Phys. Rev. D} {\bf 107} (2023), no.~11 116015, [\href{http://arxiv.org/abs/2301.00821}{{\tt arXiv:2301.00821}}].

\bibitem{Biswas:2020abl}
A.~Biswas, A.~Kundu, and P.~Mondal, {\it {Hierarchy problem and dimension-six effective operators}},  {\em Phys. Rev. D} {\bf 102} (2020), no.~7 075022, [\href{http://arxiv.org/abs/2006.13513}{{\tt arXiv:2006.13513}}].

\bibitem{Endo:2020kie}
M.~Endo, S.~Mishima, and D.~Ueda, {\it {Revisiting electroweak radiative corrections to $b\to s\ell\ell$ in SMEFT}},  {\em JHEP} {\bf 05} (2021) 050, [\href{http://arxiv.org/abs/2012.06197}{{\tt arXiv:2012.06197}}].

\bibitem{EliasMiro:2021jgu}
J.~Elias~Miro, C.~Fernandez, M.~A. Gumus, and A.~Pomarol, {\it {Gearing up for the next generation of LFV experiments, via on-shell methods}},  {\em JHEP} {\bf 06} (2022) 126, [\href{http://arxiv.org/abs/2112.12131}{{\tt arXiv:2112.12131}}].

\bibitem{Haisch:2023upo}
U.~Haisch, L.~Schnell, and J.~Weiss, {\it {LHC tau-pair production constraints on $a_\tau$ and $d_\tau$}},  {\em SciPost Phys.} {\bf 16} (2024), no.~2 048, [\href{http://arxiv.org/abs/2307.14133}{{\tt arXiv:2307.14133}}].

\bibitem{Ardu:2025rqy}
M.~Ardu and N.~Valori, {\it {The equivalent Electric Dipole Moment in SMEFT}},  \href{http://arxiv.org/abs/2503.21920}{{\tt arXiv:2503.21920}}.

\bibitem{Dawid:2024wmp}
M.~Dawid, V.~Cirigliano, and W.~Dekens, {\it {One-loop analysis of \ensuremath{\beta} decays in SMEFT}},  {\em JHEP} {\bf 08} (2024) 175, [\href{http://arxiv.org/abs/2402.06723}{{\tt arXiv:2402.06723}}].

\bibitem{Endo:2019mxw}
M.~Endo and D.~Ueda, {\it {Nuclear EDM from SMEFT flavor-changing operator}},  {\em JHEP} {\bf 04} (2020) 053, [\href{http://arxiv.org/abs/1911.10805}{{\tt arXiv:1911.10805}}].

\bibitem{Grazzini:2018eyk}
M.~Grazzini, A.~Ilnicka, and M.~Spira, {\it {Higgs boson production at large transverse momentum within the SMEFT: analytical results}},  {\em Eur. Phys. J. C} {\bf 78} (2018), no.~10 808, [\href{http://arxiv.org/abs/1806.08832}{{\tt arXiv:1806.08832}}].

\bibitem{Maltoni:2016yxb}
F.~Maltoni, E.~Vryonidou, and C.~Zhang, {\it {Higgs production in association with a top-antitop pair in the Standard Model Effective Field Theory at NLO in QCD}},  {\em JHEP} {\bf 10} (2016) 123, [\href{http://arxiv.org/abs/1607.05330}{{\tt arXiv:1607.05330}}].

\bibitem{DiNoi:2023onw}
S.~Di~Noi and R.~Gr\"ober, {\it {Renormalisation group running effects in $pp\rightarrow t{\bar{t}}h$ in the Standard Model Effective Field Theory}},  {\em Eur. Phys. J. C} {\bf 84} (2024), no.~4 403, [\href{http://arxiv.org/abs/2312.11327}{{\tt arXiv:2312.11327}}].

\bibitem{Bhattacharya:2022kje}
S.~Bhattacharya, S.~Biswas, K.~Pal, and J.~Wudka, {\it {Associated production of Higgs and single top at the LHC in presence of the SMEFT operators}},  {\em JHEP} {\bf 08} (2023) 015, [\href{http://arxiv.org/abs/2211.05450}{{\tt arXiv:2211.05450}}].

\bibitem{Degrande:2018fog}
C.~Degrande, F.~Maltoni, K.~Mimasu, E.~Vryonidou, and C.~Zhang, {\it {Single-top associated production with a $Z$ or $H$ boson at the LHC: the SMEFT interpretation}},  {\em JHEP} {\bf 10} (2018) 005, [\href{http://arxiv.org/abs/1804.07773}{{\tt arXiv:1804.07773}}].

\bibitem{Degrande:2014tta}
C.~Degrande, F.~Maltoni, J.~Wang, and C.~Zhang, {\it {Automatic computations at next-to-leading order in QCD for top-quark flavor-changing neutral processes}},  {\em Phys. Rev. D} {\bf 91} (2015) 034024, [\href{http://arxiv.org/abs/1412.5594}{{\tt arXiv:1412.5594}}].

\bibitem{Alioli:2018ljm}
S.~Alioli, W.~Dekens, M.~Girard, and E.~Mereghetti, {\it {NLO QCD corrections to SM-EFT dilepton and electroweak Higgs boson production, matched to parton shower in POWHEG}},  {\em JHEP} {\bf 08} (2018) 205, [\href{http://arxiv.org/abs/1804.07407}{{\tt arXiv:1804.07407}}].

\bibitem{GomezAmbrosio:2022mpm}
R.~Gomez~Ambrosio, J.~ter Hoeve, M.~Madigan, J.~Rojo, and V.~Sanz, {\it {Unbinned multivariate observables for global SMEFT analyses from machine learning}},  {\em JHEP} {\bf 03} (2023) 033, [\href{http://arxiv.org/abs/2211.02058}{{\tt arXiv:2211.02058}}].

\bibitem{Gauld:2023gtb}
R.~Gauld, U.~Haisch, and L.~Schnell, {\it {SMEFT at NNLO+PS: Vh production}},  {\em JHEP} {\bf 01} (2024) 192, [\href{http://arxiv.org/abs/2311.06107}{{\tt arXiv:2311.06107}}].

\bibitem{Haisch:2022nwz}
U.~Haisch, D.~J. Scott, M.~Wiesemann, G.~Zanderighi, and S.~Zanoli, {\it {NNLO event generation for $ pp\to Zh\to {\mathrm{\ell}}^{+}{\mathrm{\ell}}^{-}b\overline{b} $ production in the SM effective field theory}},  {\em JHEP} {\bf 07} (2022) 054, [\href{http://arxiv.org/abs/2204.00663}{{\tt arXiv:2204.00663}}].

\bibitem{Degrande:2016dqg}
C.~Degrande, B.~Fuks, K.~Mawatari, K.~Mimasu, and V.~Sanz, {\it {Electroweak Higgs boson production in the standard model effective field theory beyond leading order in QCD}},  {\em Eur. Phys. J. C} {\bf 77} (2017), no.~4 262, [\href{http://arxiv.org/abs/1609.04833}{{\tt arXiv:1609.04833}}].

\bibitem{Rossia:2024rfo}
A.~N. Rossia and E.~Vryonidou, {\it {CP-odd effects at NLO in SMEFT WH and ZH production}},  {\em JHEP} {\bf 11} (2024) 142, [\href{http://arxiv.org/abs/2409.00168}{{\tt arXiv:2409.00168}}].

\bibitem{Bonetti:2025hnb}
M.~Bonetti, R.~V. Harlander, D.~Korneev, M.-M. Long, K.~Melnikov, R.~R{\"o}ntsch, and D.~M. Tagliabue, {\it {WH production at the LHC within SMEFT at next-to-next-to-leading order QCD}},  {\em Phys. Rev. D} {\bf 112} (2025), no.~3 034033, [\href{http://arxiv.org/abs/2502.12846}{{\tt arXiv:2502.12846}}].

\bibitem{Baglio:2020oqu}
J.~Baglio, S.~Dawson, S.~Homiller, S.~D. Lane, and I.~M. Lewis, {\it {Validity of standard model EFT studies of VH and VV production at NLO}},  {\em Phys. Rev. D} {\bf 101} (2020), no.~11 115004, [\href{http://arxiv.org/abs/2003.07862}{{\tt arXiv:2003.07862}}].

\bibitem{Bellan:2023efn}
R.~Bellan, S.~Bhattacharya, G.~Boldrini, F.~Cetorelli, P.~Govoni, A.~Massironi, A.~Mecca, C.~Tarricone, and A.~Vagnerini, {\it {A sensitivity study of triboson production processes to dimension-6 EFT operators at the LHC}},  {\em JHEP} {\bf 08} (2023) 158, [\href{http://arxiv.org/abs/2303.18215}{{\tt arXiv:2303.18215}}].

\bibitem{Banerjee:2019twi}
S.~Banerjee, R.~S. Gupta, J.~Y. Reiness, S.~Seth, and M.~Spannowsky, {\it {Towards the ultimate differential SMEFT analysis}},  {\em JHEP} {\bf 09} (2020) 170, [\href{http://arxiv.org/abs/1912.07628}{{\tt arXiv:1912.07628}}].

\bibitem{Goldouzian:2020ekx}
R.~Goldouzian, J.~H. Kim, K.~Lannon, A.~Martin, K.~Mohrman, and A.~Wightman, {\it {Matching in $ pp\to t\overline{t}W/Z/h+ $ jet SMEFT studies}},  {\em JHEP} {\bf 06} (2021) 151, [\href{http://arxiv.org/abs/2012.06872}{{\tt arXiv:2012.06872}}].

\bibitem{Asteriadis:2024xts}
K.~Asteriadis, S.~Dawson, P.~P. Giardino, and R.~Szafron, {\it {e$^{+}$e$^{-}$ \textrightarrow{} ZH process in the SMEFT beyond leading order}},  {\em JHEP} {\bf 02} (2025) 162, [\href{http://arxiv.org/abs/2409.11466}{{\tt arXiv:2409.11466}}].

\bibitem{Deutschmann:2017qum}
N.~Deutschmann, C.~Duhr, F.~Maltoni, and E.~Vryonidou, {\it {Gluon-fusion Higgs production in the Standard Model Effective Field Theory}},  {\em JHEP} {\bf 12} (2017) 063, [\href{http://arxiv.org/abs/1708.00460}{{\tt arXiv:1708.00460}}]. [Erratum: JHEP 02, 159 (2018)].

\bibitem{Asteriadis:2022ras}
K.~Asteriadis, S.~Dawson, and D.~Fontes, {\it {Double insertions of SMEFT operators in gluon fusion Higgs boson production}},  {\em Phys. Rev. D} {\bf 107} (2023), no.~5 055038, [\href{http://arxiv.org/abs/2212.03258}{{\tt arXiv:2212.03258}}].

\bibitem{Corbett:2021cil}
T.~Corbett, A.~Martin, and M.~Trott, {\it {Consistent higher order $ \sigma \left(\mathcal{GG}\to h\right) $, $ \Gamma \left(h\to \mathcal{GG}\right) $ and \ensuremath{\Gamma}(h \textrightarrow{} \ensuremath{\gamma}\ensuremath{\gamma}) in geoSMEFT}},  {\em JHEP} {\bf 12} (2021) 147, [\href{http://arxiv.org/abs/2107.07470}{{\tt arXiv:2107.07470}}].

\bibitem{Martin:2023fad}
A.~Martin and M.~Trott, {\it {More accurate $ \sigma \left(\mathcal{GG}\to h\right),\Gamma \left(h\to \mathcal{GG},\mathcal{AA},\overline{\Psi}\Psi \right) $ and Higgs width results via the geoSMEFT}},  {\em JHEP} {\bf 01} (2024) 170, [\href{http://arxiv.org/abs/2305.05879}{{\tt arXiv:2305.05879}}].

\bibitem{Haisch:2025vqj}
U.~Haisch and M.~Niggetiedt, {\it {Precision tests of third-generation four-quark operators: $gg \to h$ and $h \to \gamma\gamma$}},  \href{http://arxiv.org/abs/2507.20803}{{\tt arXiv:2507.20803}}.

\bibitem{Haisch:2025lvd}
U.~Haisch, {\it {Higgs production from anomalous gluon dynamics}},  {\em JHEP} {\bf 06} (2025) 004, [\href{http://arxiv.org/abs/2503.06249}{{\tt arXiv:2503.06249}}].

\bibitem{Heinrich:2022idm}
G.~Heinrich, J.~Lang, and L.~Scyboz, {\it {SMEFT predictions for gg \textrightarrow{} hh at full NLO QCD and truncation uncertainties}},  {\em JHEP} {\bf 08} (2022) 079, [\href{http://arxiv.org/abs/2204.13045}{{\tt arXiv:2204.13045}}].

\bibitem{Heinrich:2023rsd}
G.~Heinrich and J.~Lang, {\it {Combining chromomagnetic and four-fermion operators with leading SMEFT operators for gg \textrightarrow{} hh at NLO QCD}},  {\em JHEP} {\bf 05} (2024) 121, [\href{http://arxiv.org/abs/2311.15004}{{\tt arXiv:2311.15004}}].

\bibitem{DiNoi:2023ygk}
S.~Di~Noi, R.~Gr\"ober, G.~Heinrich, J.~Lang, and M.~Vitti, {\it {$\gamma_5$ schemes and the interplay of SMEFT operators in the Higgs-gluon coupling}},  {\em Phys. Rev. D} {\bf 109} (2024), no.~9 095024, [\href{http://arxiv.org/abs/2310.18221}{{\tt arXiv:2310.18221}}].

\bibitem{BessidskaiaBylund:2016jvp}
O.~Bessidskaia~Bylund, F.~Maltoni, I.~Tsinikos, E.~Vryonidou, and C.~Zhang, {\it {Probing top quark neutral couplings in the Standard Model Effective Field Theory at NLO in QCD}},  {\em JHEP} {\bf 05} (2016) 052, [\href{http://arxiv.org/abs/1601.08193}{{\tt arXiv:1601.08193}}].

\bibitem{Rossia:2023hen}
A.~Rossia, M.~Thomas, and E.~Vryonidou, {\it {Diboson production in the SMEFT from gluon fusion}},  {\em JHEP} {\bf 11} (2023) 132, [\href{http://arxiv.org/abs/2306.09963}{{\tt arXiv:2306.09963}}].

\bibitem{Thomas:2024dwd}
M.~O.~A. Thomas and E.~Vryonidou, {\it {CP violation in loop-induced diboson production}},  {\em JHEP} {\bf 03} (2025) 038, [\href{http://arxiv.org/abs/2411.00959}{{\tt arXiv:2411.00959}}].

\bibitem{Vryonidou:2018eyv}
E.~Vryonidou and C.~Zhang, {\it {Dimension-six electroweak top-loop effects in Higgs production and decay}},  {\em JHEP} {\bf 08} (2018) 036, [\href{http://arxiv.org/abs/1804.09766}{{\tt arXiv:1804.09766}}].

\bibitem{Alasfar:2022zyr}
L.~Alasfar, J.~de~Blas, and R.~Gr\"ober, {\it {Higgs probes of top quark contact interactions and their interplay with the Higgs self-coupling}},  {\em JHEP} {\bf 05} (2022) 111, [\href{http://arxiv.org/abs/2202.02333}{{\tt arXiv:2202.02333}}].

\bibitem{Gauld:2015lmb}
R.~Gauld, B.~D. Pecjak, and D.~J. Scott, {\it {One-loop corrections to $h\to b\bar b$ and $h\to \tau\bar \tau$ decays in the Standard Model Dimension-6 EFT: four-fermion operators and the large-$m_t$ limit}},  {\em JHEP} {\bf 05} (2016) 080, [\href{http://arxiv.org/abs/1512.02508}{{\tt arXiv:1512.02508}}].

\bibitem{Gauld:2016kuu}
R.~Gauld, B.~D. Pecjak, and D.~J. Scott, {\it {QCD radiative corrections for $h\to b\bar b$ in the Standard Model Dimension-6 EFT}},  {\em Phys. Rev. D} {\bf 94} (2016), no.~7 074045, [\href{http://arxiv.org/abs/1607.06354}{{\tt arXiv:1607.06354}}].

\bibitem{Cullen:2019nnr}
J.~M. Cullen, B.~D. Pecjak, and D.~J. Scott, {\it {NLO corrections to $h\to b\bar b$ decay in SMEFT}},  {\em JHEP} {\bf 08} (2019) 173, [\href{http://arxiv.org/abs/1904.06358}{{\tt arXiv:1904.06358}}].

\bibitem{Cullen:2020zof}
J.~M. Cullen and B.~D. Pecjak, {\it {Higgs decay to fermion pairs at NLO in SMEFT}},  {\em JHEP} {\bf 11} (2020) 079, [\href{http://arxiv.org/abs/2007.15238}{{\tt arXiv:2007.15238}}].

\bibitem{Banerjee:2020vtm}
S.~Banerjee, R.~S. Gupta, O.~Ochoa-Valeriano, M.~Spannowsky, and E.~Venturini, {\it {A fully differential SMEFT analysis of the golden channel using the method of moments}},  {\em JHEP} {\bf 06} (2021) 031, [\href{http://arxiv.org/abs/2012.11631}{{\tt arXiv:2012.11631}}].

\bibitem{Dawson:2024pft}
S.~Dawson, M.~Forslund, and P.~P. Giardino, {\it {NLO SMEFT electroweak corrections to Higgs boson decays to four leptons in the narrow width approximation}},  {\em Phys. Rev. D} {\bf 111} (2025), no.~1 015016, [\href{http://arxiv.org/abs/2411.08952}{{\tt arXiv:2411.08952}}].

\bibitem{Corbett:2021iob}
T.~Corbett and T.~Rasmussen, {\it {Higgs decays to two leptons and a photon beyond leading order in the SMEFT}},  {\em SciPost Phys.} {\bf 13} (2022) 112, [\href{http://arxiv.org/abs/2110.03694}{{\tt arXiv:2110.03694}}].

\bibitem{Manohar:2006gz}
A.~V. Manohar and M.~B. Wise, {\it {Modifications to the properties of the Higgs boson}},  {\em Phys. Lett. B} {\bf 636} (2006) 107--113, [\href{http://arxiv.org/abs/hep-ph/0601212}{{\tt hep-ph/0601212}}].

\bibitem{Grojean:2013kd}
C.~Grojean, E.~E. Jenkins, A.~V. Manohar, and M.~Trott, {\it {Renormalization Group Scaling of Higgs Operators and $\Gamma(h\to \gamma \gamma)$}},  {\em JHEP} {\bf 04} (2013) 016, [\href{http://arxiv.org/abs/1301.2588}{{\tt arXiv:1301.2588}}].

\bibitem{Ghezzi:2015vva}
M.~Ghezzi, R.~Gomez-Ambrosio, G.~Passarino, and S.~Uccirati, {\it {NLO Higgs effective field theory and \ensuremath{\kappa}-framework}},  {\em JHEP} {\bf 07} (2015) 175, [\href{http://arxiv.org/abs/1505.03706}{{\tt arXiv:1505.03706}}].

\bibitem{Hartmann:2015aia}
C.~Hartmann and M.~Trott, {\it {Higgs Decay to Two Photons at One Loop in the Standard Model Effective Field Theory}},  {\em Phys. Rev. Lett.} {\bf 115} (2015), no.~19 191801, [\href{http://arxiv.org/abs/1507.03568}{{\tt arXiv:1507.03568}}].

\bibitem{Hartmann:2015oia}
C.~Hartmann and M.~Trott, {\it {On one-loop corrections in the standard model effective field theory; the $\Gamma(h \rightarrow \gamma \, \gamma)$ case}},  {\em JHEP} {\bf 07} (2015) 151, [\href{http://arxiv.org/abs/1505.02646}{{\tt arXiv:1505.02646}}].

\bibitem{Dawson:2018liq}
S.~Dawson and P.~P. Giardino, {\it {Electroweak corrections to Higgs boson decays to $\gamma\gamma$ and $W^+W^-$ in standard model EFT}},  {\em Phys. Rev. D} {\bf 98} (2018), no.~9 095005, [\href{http://arxiv.org/abs/1807.11504}{{\tt arXiv:1807.11504}}].

\bibitem{Dedes:2018seb}
A.~Dedes, M.~Paraskevas, J.~Rosiek, K.~Suxho, and L.~Trifyllis, {\it {The decay $h\to \gamma\gamma$ in the Standard-Model Effective Field Theory}},  {\em JHEP} {\bf 08} (2018) 103, [\href{http://arxiv.org/abs/1805.00302}{{\tt arXiv:1805.00302}}].

\bibitem{Cirigliano:2016nyn}
V.~Cirigliano, W.~Dekens, J.~de~Vries, and E.~Mereghetti, {\it {Constraining the top-Higgs sector of the Standard Model Effective Field Theory}},  {\em Phys. Rev.} {\bf D94} (2016), no.~3 034031, [\href{http://arxiv.org/abs/1605.04311}{{\tt arXiv:1605.04311}}].

\bibitem{Dedes:2019bew}
A.~Dedes, K.~Suxho, and L.~Trifyllis, {\it {The decay $h\to Z \gamma$ in the Standard-Model Effective Field Theory}},  {\em JHEP} {\bf 06} (2019) 115, [\href{http://arxiv.org/abs/1903.12046}{{\tt arXiv:1903.12046}}].

\bibitem{Dawson:2018pyl}
S.~Dawson and P.~P. Giardino, {\it {Higgs decays to $ZZ$ and $Z\gamma$ in the standard model effective field theory: An NLO analysis}},  {\em Phys. Rev. D} {\bf 97} (2018), no.~9 093003, [\href{http://arxiv.org/abs/1801.01136}{{\tt arXiv:1801.01136}}].

\bibitem{Bishara:2022vsc}
F.~Bishara, P.~Englert, C.~Grojean, G.~Panico, and A.~N. Rossia, {\it {Revisiting Vh(\textrightarrow{}$ b\overline{b} $) at the LHC and FCC-hh}},  {\em JHEP} {\bf 06} (2023) 077, [\href{http://arxiv.org/abs/2208.11134}{{\tt arXiv:2208.11134}}].

\bibitem{Bellafronte:2025jbk}
L.~Bellafronte, S.~Dawson, C.~Del~Pio, M.~Forslund, and P.~P. Giardino, {\it {Complete NLO SMEFT Electroweak Corrections to Higgs Decays}},  \href{http://arxiv.org/abs/2508.14966}{{\tt arXiv:2508.14966}}.

\bibitem{Baglio:2019uty}
J.~Baglio, S.~Dawson, and S.~Homiller, {\it {QCD corrections in Standard Model EFT fits to $WZ$ and $WW$ production}},  {\em Phys. Rev. D} {\bf 100} (2019), no.~11 113010, [\href{http://arxiv.org/abs/1909.11576}{{\tt arXiv:1909.11576}}].

\bibitem{Degrande:2024bmd}
C.~Degrande and M.~Maltoni, {\it {EFT observable stability under NLO corrections through interference revival}},  {\em Phys. Lett. B} {\bf 856} (2024) 138970, [\href{http://arxiv.org/abs/2403.16894}{{\tt arXiv:2403.16894}}].

\bibitem{Haisch:2025jqr}
U.~Haisch, J.~Linder, G.~Pelliccioli, E.~Re, and G.~Zanderighi, {\it {Polarized-boson pairs at NLO in the SMEFT}},  \href{http://arxiv.org/abs/2507.21768}{{\tt arXiv:2507.21768}}.

\bibitem{Dawson:2021ofa}
S.~Dawson and P.~P. Giardino, {\it {New physics through Drell-Yan standard model EFT measurements at NLO}},  {\em Phys. Rev. D} {\bf 104} (2021), no.~7 073004, [\href{http://arxiv.org/abs/2105.05852}{{\tt arXiv:2105.05852}}].

\bibitem{Baglio:2017bfe}
J.~Baglio, S.~Dawson, and I.~M. Lewis, {\it {An NLO QCD effective field theory analysis of $W^+W^-$ production at the LHC including fermionic operators}},  {\em Phys. Rev. D} {\bf 96} (2017), no.~7 073003, [\href{http://arxiv.org/abs/1708.03332}{{\tt arXiv:1708.03332}}].

\bibitem{Boughezal:2021tih}
R.~Boughezal, E.~Mereghetti, and F.~Petriello, {\it {Dilepton production in the SMEFT at O(1/\ensuremath{\Lambda}4)}},  {\em Phys. Rev. D} {\bf 104} (2021), no.~9 095022, [\href{http://arxiv.org/abs/2106.05337}{{\tt arXiv:2106.05337}}].

\bibitem{Dawson:2018dxp}
S.~Dawson, P.~P. Giardino, and A.~Ismail, {\it {Standard model EFT and the Drell-Yan process at high energy}},  {\em Phys. Rev. D} {\bf 99} (2019), no.~3 035044, [\href{http://arxiv.org/abs/1811.12260}{{\tt arXiv:1811.12260}}].

\bibitem{Biekotter:2025nln}
A.~Biek{\"o}tter and B.~D. Pecjak, {\it {Analytic results for electroweak precision observables at NLO in SMEFT}},  {\em JHEP} {\bf 07} (2025) 134, [\href{http://arxiv.org/abs/2503.07724}{{\tt arXiv:2503.07724}}].

\bibitem{Dawson:2019clf}
S.~Dawson and P.~P. Giardino, {\it {Electroweak and QCD corrections to $Z$ and $W$ pole observables in the standard model EFT}},  {\em Phys. Rev. D} {\bf 101} (2020), no.~1 013001, [\href{http://arxiv.org/abs/1909.02000}{{\tt arXiv:1909.02000}}].

\bibitem{Dawson:2022bxd}
S.~Dawson and P.~P. Giardino, {\it {Flavorful electroweak precision observables in the Standard Model effective field theory}},  {\em Phys. Rev. D} {\bf 105} (2022), no.~7 073006, [\href{http://arxiv.org/abs/2201.09887}{{\tt arXiv:2201.09887}}].

\bibitem{Hartmann:2016pil}
C.~Hartmann, W.~Shepherd, and M.~Trott, {\it {The $Z$ decay width in the SMEFT: $y_t$ and $\lambda$ corrections at one loop}},  {\em JHEP} {\bf 03} (2017) 060, [\href{http://arxiv.org/abs/1611.09879}{{\tt arXiv:1611.09879}}].

\bibitem{Basan:2020btr}
A.~Basan, P.~Berta, L.~Masetti, E.~Vryonidou, and S.~Westhoff, {\it {Measuring the top energy asymmetry at the LHC: QCD and SMEFT interpretations}},  {\em JHEP} {\bf 03} (2020) 184, [\href{http://arxiv.org/abs/2001.07225}{{\tt arXiv:2001.07225}}].

\bibitem{ElFaham:2024egs}
H.~El~Faham, K.~Mimasu, D.~Pagani, C.~Severi, E.~Vryonidou, and M.~Zaro, {\it {Electroweak corrections in the SMEFT: four-fermion operators at high energies}},  {\em JHEP} {\bf 06} (2025) 241, [\href{http://arxiv.org/abs/2412.16076}{{\tt arXiv:2412.16076}}].

\bibitem{Faham:2021zet}
H.~E. Faham, F.~Maltoni, K.~Mimasu, and M.~Zaro, {\it {Single top production in association with a WZ pair at the LHC in the SMEFT}},  {\em JHEP} {\bf 01} (2022) 100, [\href{http://arxiv.org/abs/2111.03080}{{\tt arXiv:2111.03080}}].

\bibitem{Kidonakis:2023htm}
N.~Kidonakis and A.~Tonero, {\it {SMEFT chromomagnetic dipole operator contributions to $t{{\bar{t}}}$ production at approximate NNLO in QCD}},  {\em Eur. Phys. J. C} {\bf 84} (2024), no.~6 591, [\href{http://arxiv.org/abs/2309.16758}{{\tt arXiv:2309.16758}}].

\bibitem{Aoude:2022imd}
R.~Aoude, E.~Madge, F.~Maltoni, and L.~Mantani, {\it {Quantum SMEFT tomography: Top quark pair production at the LHC}},  {\em Phys. Rev. D} {\bf 106} (2022), no.~5 055007, [\href{http://arxiv.org/abs/2203.05619}{{\tt arXiv:2203.05619}}].

\bibitem{Zhang:2016omx}
C.~Zhang, {\it {Single Top Production at Next-to-Leading Order in the Standard Model Effective Field Theory}},  {\em Phys. Rev. Lett.} {\bf 116} (2016), no.~16 162002, [\href{http://arxiv.org/abs/1601.06163}{{\tt arXiv:1601.06163}}].

\bibitem{Boughezal:2019xpp}
R.~Boughezal, C.-Y. Chen, F.~Petriello, and D.~Wiegand, {\it {Top quark decay at next-to-leading order in the Standard Model Effective Field Theory}},  {\em Phys. Rev. D} {\bf 100} (2019), no.~5 056023, [\href{http://arxiv.org/abs/1907.00997}{{\tt arXiv:1907.00997}}].

\bibitem{Zhang:2014rja}
C.~Zhang, {\it {Effective field theory approach to top-quark decay at next-to-leading order in QCD}},  {\em Phys. Rev. D} {\bf 90} (2014), no.~1 014008, [\href{http://arxiv.org/abs/1404.1264}{{\tt arXiv:1404.1264}}].

\bibitem{Severi:2022qjy}
C.~Severi and E.~Vryonidou, {\it {Quantum entanglement and top spin correlations in SMEFT at higher orders}},  {\em JHEP} {\bf 01} (2023) 148, [\href{http://arxiv.org/abs/2210.09330}{{\tt arXiv:2210.09330}}].

\bibitem{Durieux:2014xla}
G.~Durieux, F.~Maltoni, and C.~Zhang, {\it {Global approach to top-quark flavor-changing interactions}},  {\em Phys. Rev. D} {\bf 91} (2015), no.~7 074017, [\href{http://arxiv.org/abs/1412.7166}{{\tt arXiv:1412.7166}}].

\bibitem{Hiller:2024vtr}
G.~Hiller and D.~Wendler, {\it {Missing energy plus jet in the SMEFT}},  {\em JHEP} {\bf 09} (2024) 009, [\href{http://arxiv.org/abs/2403.17063}{{\tt arXiv:2403.17063}}].

\bibitem{Trott:2021vqa}
M.~Trott, {\it {Methodology for theory uncertainties in the standard model effective field theory}},  {\em Phys. Rev. D} {\bf 104} (2021), no.~9 095023, [\href{http://arxiv.org/abs/2106.13794}{{\tt arXiv:2106.13794}}].

\bibitem{Iranipour:2022iak}
S.~Iranipour and M.~Ubiali, {\it {A new generation of simultaneous fits to LHC data using deep learning}},  {\em JHEP} {\bf 05} (2022) 032, [\href{http://arxiv.org/abs/2201.07240}{{\tt arXiv:2201.07240}}].

\bibitem{vanBeek:2019evb}
S.~van Beek, E.~R. Nocera, J.~Rojo, and E.~Slade, {\it {Constraining the SMEFT with Bayesian reweighting}},  {\em SciPost Phys.} {\bf 7} (2019), no.~5 070, [\href{http://arxiv.org/abs/1906.05296}{{\tt arXiv:1906.05296}}].

\bibitem{Hirsch:2025qya}
M.~Hirsch, L.~Mantani, and V.~Sanz, {\it {Can SMEFT discover New Physics?}},  \href{http://arxiv.org/abs/2507.11109}{{\tt arXiv:2507.11109}}.

\bibitem{Brivio:2022hrb}
I.~Brivio, S.~Bruggisser, N.~Elmer, E.~Geoffray, M.~Luchmann, and T.~Plehn, {\it {To profile or to marginalize - A SMEFT case study}},  {\em SciPost Phys.} {\bf 16} (2024), no.~1 035, [\href{http://arxiv.org/abs/2208.08454}{{\tt arXiv:2208.08454}}].

\bibitem{Heimel:2024drk}
T.~Heimel, T.~Plehn, and N.~Schmal, {\it {Profile Likelihoods on ML-Steroids}},  \href{http://arxiv.org/abs/2411.00942}{{\tt arXiv:2411.00942}}.

\bibitem{Dawson:2020oco}
S.~Dawson, S.~Homiller, and S.~D. Lane, {\it {Putting standard model EFT fits to work}},  {\em Phys. Rev. D} {\bf 102} (2020), no.~5 055012, [\href{http://arxiv.org/abs/2007.01296}{{\tt arXiv:2007.01296}}].

\bibitem{Chang:2025ohh}
S.~Chang, M.~A. Luty, T.~Ma, F.~Montagno, and A.~Wulzer, {\it {Quantifying EFT Uncertainties in LHC Searches}},  \href{http://arxiv.org/abs/2507.15954}{{\tt arXiv:2507.15954}}.

\bibitem{Camponovo:2022wwn}
F.~Camponovo and G.~Passarino, {\it {SMEFT deviations}},  {\em Eur. Phys. J. C} {\bf 83} (2023), no.~1 67, [\href{http://arxiv.org/abs/2211.12718}{{\tt arXiv:2211.12718}}].

\bibitem{Greljo:2023bab}
A.~Greljo, J.~Salko, A.~Smolkovi\v{c}, and P.~Stangl, {\it {SMEFT restrictions on exclusive b \textrightarrow{} u\ensuremath{\ell}\ensuremath{\nu} decays}},  {\em JHEP} {\bf 11} (2023) 023, [\href{http://arxiv.org/abs/2306.09401}{{\tt arXiv:2306.09401}}].

\bibitem{Jung:2018lfu}
M.~Jung and D.~M. Straub, {\it {Constraining new physics in $b\to c\ell\nu$ transitions}},  {\em JHEP} {\bf 01} (2019) 009, [\href{http://arxiv.org/abs/1801.01112}{{\tt arXiv:1801.01112}}].

\bibitem{Greljo:2022jac}
A.~Greljo, J.~Salko, A.~Smolkovi\v{c}, and P.~Stangl, {\it {Rare b decays meet high-mass Drell-Yan}},  {\em JHEP} {\bf 05} (2023) 087, [\href{http://arxiv.org/abs/2212.10497}{{\tt arXiv:2212.10497}}].

\bibitem{Panda:2024ygr}
D.~Panda, M.~K. Mohapatra, and R.~Mohanta, {\it {Exploring the lepton flavor violating decay modes $b \to s\mu^\pm\tau^\mp$ in SMEFT approach}},  {\em Nucl. Phys. B} {\bf 1008} (2024) 116720, [\href{http://arxiv.org/abs/2403.09393}{{\tt arXiv:2403.09393}}].

\bibitem{Karmakar:2024dml}
S.~Karmakar and A.~Dighe, {\it {Exploring observable effects of scalar operators beyond SMEFT in the angular distribution of $B\rightarrow K^{0*}\tau^+\tau^-$}},  {\em Phys. Rev. D} {\bf 110} (2024), no.~11 115041, [\href{http://arxiv.org/abs/2408.13069}{{\tt arXiv:2408.13069}}].

\bibitem{Chen:2024jlj}
F.-Z. Chen, Q.~Wen, and F.~Xu, {\it {Correlating $B\to K^{(\ast)} \nu\bar{\nu}$ and flavor anomalies in SMEFT}},  {\em Eur. Phys. J. C} {\bf 84} (2024) 1012, [\href{http://arxiv.org/abs/2401.11552}{{\tt arXiv:2401.11552}}].

\bibitem{Hou:2024vyw}
B.-F. Hou, X.-Q. Li, M.~Shen, Y.-D. Yang, and X.-B. Yuan, {\it {Deciphering the Belle II data on $ B\to K\nu \overline{\nu} $ decay in the (dark) SMEFT with minimal flavour violation}},  {\em JHEP} {\bf 06} (2024) 172, [\href{http://arxiv.org/abs/2402.19208}{{\tt arXiv:2402.19208}}].

\bibitem{Guadagnoli:2023ddc}
D.~Guadagnoli, C.~Normand, S.~Simula, and L.~Vittorio, {\it {Insights on the current semi-leptonic B-decay discrepancies \textemdash{} and how B$_{s}$\textrightarrow{} \ensuremath{\mu}$^{+}$\ensuremath{\mu}$^{-}$\ensuremath{\gamma} can help}},  {\em JHEP} {\bf 10} (2023) 102, [\href{http://arxiv.org/abs/2308.00034}{{\tt arXiv:2308.00034}}].

\bibitem{Mohapatra:2024knf}
M.~K. Mohapatra, D.~Panda, and R.~Mohanta, {\it {Imprints of new physics operators in the semileptonic $B\rightarrow a_1(1260)\ell^-\nu_{\ell}$ process in SMEFT approach}},  {\em Phys. Lett. B} {\bf 855} (2024) 138866, [\href{http://arxiv.org/abs/2402.18410}{{\tt arXiv:2402.18410}}].

\bibitem{Das:2023kch}
N.~Das and R.~Dutta, {\it {New physics analysis of $\Lambda_b\to (\Lambda^*(\to pK^-), \Lambda(\to p\pi))({\mu}^{+}\mu^{-},\,\nu\bar{\nu})$ baryonic decays under SMEFT framework}},  {\em Phys. Rev. D} {\bf 108} (2023), no.~9 095051, [\href{http://arxiv.org/abs/2307.03615}{{\tt arXiv:2307.03615}}].

\bibitem{Boora:2025odj}
P.~Boora, S.~Karmakar, D.~Kumar, and K.~Lalwani, {\it {A comprehensive study of $\Lambda_c^- \to \Lambda(\to p \pi) \mu^- \bar \nu_\mu$ incorporating SMEFT implications and right-handed neutrino}},  \href{http://arxiv.org/abs/2510.05050}{{\tt arXiv:2510.05050}}.

\bibitem{Fernandez-Martinez:2024bxg}
E.~Fern\'andez-Mart\'\i{}nez, X.~Marcano, and D.~Naredo-Tuero, {\it {Global lepton flavour violating constraints on new physics}},  {\em Eur. Phys. J. C} {\bf 84} (2024), no.~7 666, [\href{http://arxiv.org/abs/2403.09772}{{\tt arXiv:2403.09772}}].

\bibitem{Glioti:2025zpn}
A.~Glioti, D.~Marzocca, and A.~Wulzer, {\it {Flavor physics at high-energy muon colliders}},  \href{http://arxiv.org/abs/2509.08132}{{\tt arXiv:2509.08132}}.

\bibitem{Englert:2016aei}
C.~Englert, L.~Moore, K.~Nordstr\"om, and M.~Russell, {\it {Giving top quark effective operators a boost}},  {\em Phys. Lett. B} {\bf 763} (2016) 9--15, [\href{http://arxiv.org/abs/1607.04304}{{\tt arXiv:1607.04304}}].

\bibitem{Aguilar-Saavedra:2018ksv}
D.~Barducci et~al., {\it {Interpreting top-quark LHC measurements in the standard-model effective field theory}},  \href{http://arxiv.org/abs/1802.07237}{{\tt arXiv:1802.07237}}.

\bibitem{Egle:2025buk}
F.~Egle, C.~Englert, M.~M\"uhlleitner, and M.~Spannowsky, {\it {Distorting the Top Resonance with Effective Interactions}},  \href{http://arxiv.org/abs/2503.19841}{{\tt arXiv:2503.19841}}.

\bibitem{Aoude:2022aro}
R.~Aoude, F.~Maltoni, O.~Mattelaer, C.~Severi, and E.~Vryonidou, {\it {Renormalisation group effects on SMEFT interpretations of LHC data}},  {\em JHEP} {\bf 09} (2023) 191, [\href{http://arxiv.org/abs/2212.05067}{{\tt arXiv:2212.05067}}].

\bibitem{deBlas:2021wap}
J.~de~Blas, M.~Ciuchini, E.~Franco, A.~Goncalves, S.~Mishima, M.~Pierini, L.~Reina, and L.~Silvestrini, {\it {Global analysis of electroweak data in the Standard Model}},  {\em Phys. Rev. D} {\bf 106} (2022), no.~3 033003, [\href{http://arxiv.org/abs/2112.07274}{{\tt arXiv:2112.07274}}].

\bibitem{deBlas:2022ofj}
J.~de~Blas, Y.~Du, C.~Grojean, J.~Gu, V.~Miralles, M.~E. Peskin, J.~Tian, M.~Vos, and E.~Vryonidou, {\it {Global SMEFT Fits at Future Colliders}},  in {\em {Snowmass 2021}}, 6, 2022.
\newblock \href{http://arxiv.org/abs/2206.08326}{{\tt arXiv:2206.08326}}.

\bibitem{Celada:2024oax}
E.~Celada, {\it {Constraining the SMEFT at Present and Future Colliders}},  in {\em {17th International Workshop on Top Quark Physics}}, 12, 2024.
\newblock \href{http://arxiv.org/abs/2412.08311}{{\tt arXiv:2412.08311}}.

\bibitem{Allwicher:2024sso}
L.~Allwicher, M.~McCullough, and S.~Renner, {\it {New physics at Tera-Z: precision renormalised}},  {\em JHEP} {\bf 02} (2025) 164, [\href{http://arxiv.org/abs/2408.03992}{{\tt arXiv:2408.03992}}].

\bibitem{terHoeve:2025gey}
J.~ter Hoeve, L.~Mantani, J.~Rojo, A.~N. Rossia, and E.~Vryonidou, {\it {Connecting scales: RGE effects in the SMEFT at the LHC and future colliders}},  {\em JHEP} {\bf 06} (2025) 125, [\href{http://arxiv.org/abs/2502.20453}{{\tt arXiv:2502.20453}}].

\bibitem{terHoeve:2025zmp}
J.~ter Hoeve, {\it {Fingerprinting New Physics with Effective Field Theories}},  other thesis, 1, 2025.

\bibitem{Allanach:2025wfi}
B.~Allanach and E.~Loisa, {\it {Computation of FCC-ee Sensitivity to Heavy New Physics with Interactions of Any Flavor Structure}},  \href{http://arxiv.org/abs/2501.08321}{{\tt arXiv:2501.08321}}.

\bibitem{Chala:2025utt}
M.~Chala, J.~C. Criado, and M.~Spannowsky, {\it {The TeraZ Mirage: New Physics Lost in Blind Directions}},  \href{http://arxiv.org/abs/2504.16558}{{\tt arXiv:2504.16558}}.

\bibitem{Englert:2025onf}
C.~Englert, W.~Naskar, and M.~Spannowsky, {\it {Impact of new physics on momentum-dependent particle widths and propagators}},  {\em Phys. Rev. D} {\bf 111} (2025), no.~5 055017, [\href{http://arxiv.org/abs/2501.08407}{{\tt arXiv:2501.08407}}].

\bibitem{deBlas:2019rxi}
J.~de~Blas et~al., {\it {Higgs Boson Studies at Future Particle Colliders}},  {\em JHEP} {\bf 01} (2020) 139, [\href{http://arxiv.org/abs/1905.03764}{{\tt arXiv:1905.03764}}].

\bibitem{Blondel:2024mry}
A.~Blondel, C.~Grojean, P.~Janot, and G.~Wilkinson, {\it {Higgs Factory options for CERN: A comparative study}},  \href{http://arxiv.org/abs/2412.13130}{{\tt arXiv:2412.13130}}.

\bibitem{Assi:2025zmp}
B.~Assi and A.~Martin, {\it {Energy-enhanced expansion of the standard model effective field theory}},  {\em Phys. Rev. D} {\bf 112} (2025), no.~1 015024, [\href{http://arxiv.org/abs/2504.10617}{{\tt arXiv:2504.10617}}].

\bibitem{Breso-Pla:2023tnz}
V.~Bres\'o-Pla, A.~Falkowski, M.~Gonz\'alez-Alonso, and K.~Mons\'alvez-Pozo, {\it {EFT analysis of New Physics at COHERENT}},  {\em JHEP} {\bf 05} (2023) 074, [\href{http://arxiv.org/abs/2301.07036}{{\tt arXiv:2301.07036}}].

\bibitem{Coloma:2024ict}
P.~Coloma, E.~Fern\'andez-Mart\'\i{}nez, J.~L\'opez-Pav\'on, X.~Marcano, D.~Naredo-Tuero, and S.~Urrea, {\it {Improving the global SMEFT picture with bounds on neutrino NSI}},  {\em JHEP} {\bf 02} (2025) 137, [\href{http://arxiv.org/abs/2411.00090}{{\tt arXiv:2411.00090}}].

\bibitem{Beltran:2025ilg}
R.~Beltr{\'a}n, G.~Cottin, J.~G{\"u}nther, M.~Hirsch, A.~Titov, and Z.~S. Wang, {\it {Heavy neutral leptons and top quarks in effective field theory}},  {\em JHEP} {\bf 05} (2025) 238, [\href{http://arxiv.org/abs/2501.09065}{{\tt arXiv:2501.09065}}].

\bibitem{Buckley:2015lku}
A.~Buckley, C.~Englert, J.~Ferrando, D.~J. Miller, L.~Moore, M.~Russell, and C.~D. White, {\it {Constraining top quark effective theory in the LHC Run II era}},  {\em JHEP} {\bf 04} (2016) 015, [\href{http://arxiv.org/abs/1512.03360}{{\tt arXiv:1512.03360}}].

\bibitem{Buckley:2015nca}
A.~Buckley, C.~Englert, J.~Ferrando, D.~J. Miller, L.~Moore, M.~Russell, and C.~D. White, {\it {Global fit of top quark effective theory to data}},  {\em Phys. Rev. D} {\bf 92} (2015), no.~9 091501, [\href{http://arxiv.org/abs/1506.08845}{{\tt arXiv:1506.08845}}].

\bibitem{Bissmann:2019qcd}
S.~Bi\ss{}mann, J.~Erdmann, C.~Grunwald, G.~Hiller, and K.~Kr\"oninger, {\it {Correlating uncertainties in global analyses within SMEFT matters}},  {\em Phys. Rev. D} {\bf 102} (2020) 115019, [\href{http://arxiv.org/abs/1912.06090}{{\tt arXiv:1912.06090}}].

\bibitem{Brivio:2019ius}
I.~Brivio, S.~Bruggisser, F.~Maltoni, R.~Moutafis, T.~Plehn, E.~Vryonidou, S.~Westhoff, and C.~Zhang, {\it {O new physics, where art thou? A global search in the top sector}},  {\em JHEP} {\bf 02} (2020) 131, [\href{http://arxiv.org/abs/1910.03606}{{\tt arXiv:1910.03606}}].

\bibitem{Elmer:2023wtr}
N.~Elmer, M.~Madigan, T.~Plehn, and N.~Schmal, {\it {Staying on Top of SMEFT-Likelihood Analyses}},  {\em SciPost Phys.} {\bf 18} (2025) 108, [\href{http://arxiv.org/abs/2312.12502}{{\tt arXiv:2312.12502}}].

\bibitem{Kassabov:2023hbm}
Z.~Kassabov, M.~Madigan, L.~Mantani, J.~Moore, M.~Morales~Alvarado, J.~Rojo, and M.~Ubiali, {\it {The top quark legacy of the LHC Run II for PDF and SMEFT analyses}},  {\em JHEP} {\bf 05} (2023) 205, [\href{http://arxiv.org/abs/2303.06159}{{\tt arXiv:2303.06159}}].

\bibitem{Gao:2022srd}
J.~Gao, M.~Gao, T.~J. Hobbs, D.~Liu, and X.~Shen, {\it {Simultaneous CTEQ-TEA extraction of PDFs and SMEFT parameters from jet and $ t\overline{t} $ data}},  {\em JHEP} {\bf 05} (2023) 003, [\href{http://arxiv.org/abs/2211.01094}{{\tt arXiv:2211.01094}}].

\bibitem{DiNoi:2025uhu}
S.~Di~Noi, H.~El~Faham, R.~Gr{\"o}ber, M.~Vitti, and E.~Vryonidou, {\it {Constraining four-heavy-quark operators with top-quark, Higgs, and electroweak precision data}},  \href{http://arxiv.org/abs/2507.01137}{{\tt arXiv:2507.01137}}.

\bibitem{Bissmann:2020mfi}
S.~Bi\ss{}mann, C.~Grunwald, G.~Hiller, and K.~Kr\"oninger, {\it {Top and Beauty synergies in SMEFT-fits at present and future colliders}},  {\em JHEP} {\bf 06} (2021) 010, [\href{http://arxiv.org/abs/2012.10456}{{\tt arXiv:2012.10456}}].

\bibitem{Ellis:2020unq}
J.~Ellis, M.~Madigan, K.~Mimasu, V.~Sanz, and T.~You, {\it {Top, Higgs, Diboson and Electroweak Fit to the Standard Model Effective Field Theory}},  {\em JHEP} {\bf 04} (2021) 279, [\href{http://arxiv.org/abs/2012.02779}{{\tt arXiv:2012.02779}}].

\bibitem{Ethier:2021bye}
{\bf SMEFiT} Collaboration, J.~J. Ethier, G.~Magni, F.~Maltoni, L.~Mantani, E.~R. Nocera, J.~Rojo, E.~Slade, E.~Vryonidou, and C.~Zhang, {\it {Combined SMEFT interpretation of Higgs, diboson, and top quark data from the LHC}},  {\em JHEP} {\bf 11} (2021) 089, [\href{http://arxiv.org/abs/2105.00006}{{\tt arXiv:2105.00006}}].

\bibitem{Biekotter:2023mpd}
A.~Biek\"otter, J.~Fuentes-Mart\'\i{}n, A.~M. Galda, and M.~Neubert, {\it {A global analysis of axion-like particle interactions using SMEFT fits}},  {\em JHEP} {\bf 09} (2023) 120, [\href{http://arxiv.org/abs/2307.10372}{{\tt arXiv:2307.10372}}].

\bibitem{Garosi:2023yxg}
F.~Garosi, D.~Marzocca, A.~R. S\'anchez, and A.~Stanzione, {\it {Indirect constraints on top quark operators from a global SMEFT analysis}},  {\em JHEP} {\bf 12} (2023) 129, [\href{http://arxiv.org/abs/2310.00047}{{\tt arXiv:2310.00047}}].

\bibitem{deBlas:2025xhe}
J.~de~Blas, A.~Goncalves, V.~Miralles, L.~Reina, L.~Silvestrini, and M.~Valli, {\it {Constraining new physics effective interactions via a global fit of electroweak, Drell-Yan, Higgs, top, and flavour observables}},  \href{http://arxiv.org/abs/2507.06191}{{\tt arXiv:2507.06191}}.

\bibitem{Maltoni:2024dpn}
F.~Maltoni, G.~Ventura, and E.~Vryonidou, {\it {Impact of SMEFT renormalisation group running on Higgs production at the LHC}},  {\em JHEP} {\bf 12} (2024) 183, [\href{http://arxiv.org/abs/2406.06670}{{\tt arXiv:2406.06670}}].

\bibitem{Gisbert:2024sjw}
H.~Gisbert, A.~Rodr{\'\i}guez-S{\'a}nchez, and L.~Vale~Silva, {\it {Constraints on baryon-number-violating top-quark operators in standard model effective field theory}},  {\em Phys. Rev. D} {\bf 112} (2025), no.~1 015026, [\href{http://arxiv.org/abs/2409.00218}{{\tt arXiv:2409.00218}}].

\bibitem{Hiller:2025hpf}
G.~Hiller, L.~Nollen, and D.~Wendler, {\it {Total Drell{\textendash}Yan in the flavorful SMEFT}},  {\em Eur. Phys. J. C} {\bf 85} (2025), no.~6 657, [\href{http://arxiv.org/abs/2502.12250}{{\tt arXiv:2502.12250}}].

\bibitem{Corbett:2025oqk}
T.~Corbett, J.~Desai, O.~J.~P. Eboli, M.~C. Gonzalez-Garcia, M.~Martines, and P.~Reimitz, {\it {Drell-Yan production in universal theories beyond dimension-six SMEFT}},  {\em Phys. Rev. D} {\bf 112} (2025), no.~1 013009, [\href{http://arxiv.org/abs/2503.19962}{{\tt arXiv:2503.19962}}].

\bibitem{Kala:2025srq}
S.~Kala, L.~Kolay, L.~Mukherjee, and S.~Nandi, {\it {Constraining anomalous $W tb$ and related SMEFT couplings using low-energy and electroweak precision observables}},  \href{http://arxiv.org/abs/2505.07926}{{\tt arXiv:2505.07926}}.

\bibitem{Gomez-Ambrosio:2018pnl}
R.~Gomez-Ambrosio, {\it {Studies of Dimension-Six EFT effects in Vector Boson Scattering}},  {\em Eur. Phys. J. C} {\bf 79} (2019), no.~5 389, [\href{http://arxiv.org/abs/1809.04189}{{\tt arXiv:1809.04189}}].

\bibitem{Mildner:2024wbl}
H.~Mildner, {\it {An EWPD SMEFT likelihood for the LHC {\textemdash} and how to improve it with measurements of W and Z boson properties}},  {\em JHEP} {\bf 07} (2025) 089, [\href{http://arxiv.org/abs/2412.07651}{{\tt arXiv:2412.07651}}].

\bibitem{Schofbeck:2024zjo}
R.~Sch\"ofbeck, {\it {Refinable modeling for unbinned SMEFT analyses}},  {\em Mach. Learn. Sci. Tech.} {\bf 6} (2025), no.~1 015007, [\href{http://arxiv.org/abs/2406.19076}{{\tt arXiv:2406.19076}}].

\bibitem{Efrati:2015eaa}
A.~Efrati, A.~Falkowski, and Y.~Soreq, {\it {Electroweak constraints on flavorful effective theories}},  {\em JHEP} {\bf 07} (2015) 018, [\href{http://arxiv.org/abs/1503.07872}{{\tt arXiv:1503.07872}}].

\bibitem{Bellafronte:2023amz}
L.~Bellafronte, S.~Dawson, and P.~P. Giardino, {\it {The importance of flavor in SMEFT Electroweak Precision Fits}},  {\em JHEP} {\bf 05} (2023) 208, [\href{http://arxiv.org/abs/2304.00029}{{\tt arXiv:2304.00029}}].

\bibitem{ThomasArun:2023wbd}
M.~Thomas~Arun, K.~Deka, and T.~Srivastava, {\it {Constraining SMEFT BSM scenarios with EWPO and $\Delta _{\textrm{CKM}}$}},  {\em Pramana} {\bf 99} (2025), no.~4 145, [\href{http://arxiv.org/abs/2301.09273}{{\tt arXiv:2301.09273}}].

\bibitem{Fan:2022yly}
J.~Fan, L.~Li, T.~Liu, and K.-F. Lyu, {\it {W-boson mass, electroweak precision tests, and SMEFT}},  {\em Phys. Rev. D} {\bf 106} (2022), no.~7 073010, [\href{http://arxiv.org/abs/2204.04805}{{\tt arXiv:2204.04805}}].

\bibitem{Corbett:2023qtg}
T.~Corbett, J.~Desai, O.~J.~P. \'Eboli, M.~C. Gonzalez-Garcia, M.~Martines, and P.~Reimitz, {\it {Impact of dimension-eight SMEFT operators in the electroweak precision observables and triple gauge couplings analysis in universal SMEFT}},  {\em Phys. Rev. D} {\bf 107} (2023), no.~11 115013, [\href{http://arxiv.org/abs/2304.03305}{{\tt arXiv:2304.03305}}].

\bibitem{Elias-Miro:2013eta}
J.~Elias-Mir\'o, C.~Grojean, R.~S. Gupta, and D.~Marzocca, {\it {Scaling and tuning of EW and Higgs observables}},  {\em JHEP} {\bf 05} (2014) 019, [\href{http://arxiv.org/abs/1312.2928}{{\tt arXiv:1312.2928}}].

\bibitem{Ahmed:2024hpg}
A.~Ahmed, Z.~Chacko, I.~Flood, C.~Kilic, and S.~Najjari, {\it {General form of effective operators from hidden sectors}},  {\em JHEP} {\bf 05} (2025) 167, [\href{http://arxiv.org/abs/2412.15067}{{\tt arXiv:2412.15067}}].

\bibitem{Falkowski:2019hvp}
A.~Falkowski and D.~Straub, {\it {Flavourful SMEFT likelihood for Higgs and electroweak data}},  {\em JHEP} {\bf 04} (2020) 066, [\href{http://arxiv.org/abs/1911.07866}{{\tt arXiv:1911.07866}}].

\bibitem{Bagnaschi:2022whn}
E.~Bagnaschi, J.~Ellis, M.~Madigan, K.~Mimasu, V.~Sanz, and T.~You, {\it {SMEFT analysis of m$_{W}$}},  {\em JHEP} {\bf 08} (2022) 308, [\href{http://arxiv.org/abs/2204.05260}{{\tt arXiv:2204.05260}}].

\bibitem{Ellis:2018gqa}
J.~Ellis, C.~W. Murphy, V.~Sanz, and T.~You, {\it {Updated Global SMEFT Fit to Higgs, Diboson and Electroweak Data}},  {\em JHEP} {\bf 06} (2018) 146, [\href{http://arxiv.org/abs/1803.03252}{{\tt arXiv:1803.03252}}].

\bibitem{Cirigliano:2023nol}
V.~Cirigliano, W.~Dekens, J.~de~Vries, E.~Mereghetti, and T.~Tong, {\it {Anomalies in global SMEFT analyses. A case study of first-row CKM unitarity}},  {\em JHEP} {\bf 03} (2024) 033, [\href{http://arxiv.org/abs/2311.00021}{{\tt arXiv:2311.00021}}].

\bibitem{Allwicher:2024mzw}
L.~Allwicher, D.~A. Faroughy, M.~Martines, O.~Sumensari, and F.~Wilsch, {\it {On the EFT validity for Drell\textendash{}Yan tails at the LHC}},  {\em Eur. Phys. J. C} {\bf 85} (2025), no.~4 463, [\href{http://arxiv.org/abs/2412.14162}{{\tt arXiv:2412.14162}}].

\bibitem{Grojean:2018dqj}
C.~Grojean, M.~Montull, and M.~Riembau, {\it {Diboson at the LHC vs LEP}},  {\em JHEP} {\bf 03} (2019) 020, [\href{http://arxiv.org/abs/1810.05149}{{\tt arXiv:1810.05149}}].

\bibitem{Gillies:2024mqp}
D.~Gillies, A.~Banfi, A.~Martin, and M.~A. Lim, {\it {Dimension-8 operators in $W^+W^-$ production via gluon fusion}},  {\em JHEP} {\bf 06} (2025) 111, [\href{http://arxiv.org/abs/2412.16020}{{\tt arXiv:2412.16020}}].

\bibitem{Bhattacharya:2024sxl}
S.~Bhattacharya, A.~Sarkar, and S.~Biswas, {\it {Higgs couplings in SMEFT via Zh production at the HL-LHC}},  \href{http://arxiv.org/abs/2403.03001}{{\tt arXiv:2403.03001}}.

\bibitem{Bernlochner:2018opw}
F.~U. Bernlochner, C.~Englert, C.~Hays, K.~Lohwasser, H.~Mildner, A.~Pilkington, D.~D. Price, and M.~Spannowsky, {\it {Angles on CP-violation in Higgs boson interactions}},  {\em Phys. Lett. B} {\bf 790} (2019) 372--379, [\href{http://arxiv.org/abs/1808.06577}{{\tt arXiv:1808.06577}}].

\bibitem{Murphy:2017omb}
C.~W. Murphy, {\it {Statistical approach to Higgs boson couplings in the standard model effective field theory}},  {\em Phys. Rev. D} {\bf 97} (2018), no.~1 015007, [\href{http://arxiv.org/abs/1710.02008}{{\tt arXiv:1710.02008}}].

\bibitem{ATLAS:2022xyx}
{\bf ATLAS} Collaboration, ATLAS, {\it {Combined effective field theory interpretation of Higgs boson and weak boson production and decay with ATLAS data and electroweak precision observables}}, .

\bibitem{Falkowski:2015jaa}
A.~Falkowski, M.~Gonzalez-Alonso, A.~Greljo, and D.~Marzocca, {\it {Global constraints on anomalous triple gauge couplings in effective field theory approach}},  {\em Phys. Rev. Lett.} {\bf 116} (2016), no.~1 011801, [\href{http://arxiv.org/abs/1508.00581}{{\tt arXiv:1508.00581}}].

\bibitem{terHoeve:2025yup}
J.~ter Hoeve, L.~Mantani, J.~Rojo, A.~N. Rossia, and E.~Vryonidou, {\it {Higgs trilinear coupling in the standard model effective field theory at the high luminosity LHC and the FCC-ee}},  {\em Phys. Rev. D} {\bf 112} (2025), no.~1 013008, [\href{http://arxiv.org/abs/2504.05974}{{\tt arXiv:2504.05974}}].

\bibitem{Maura:2025rcv}
V.~Maura, B.~A. Stefanek, and T.~You, {\it {The Higgs Self-Coupling at FCC-ee}},  \href{http://arxiv.org/abs/2503.13719}{{\tt arXiv:2503.13719}}.

\bibitem{Allwicher:2025bub}
L.~Allwicher, G.~Isidori, and M.~Pesut, {\it {Flavored circular collider: cornering New Physics at FCC-ee via flavor-changing processes}},  {\em Eur. Phys. J. C} {\bf 85} (2025), no.~6 631, [\href{http://arxiv.org/abs/2503.17019}{{\tt arXiv:2503.17019}}].

\bibitem{Maura:2024zxz}
V.~Maura, B.~A. Stefanek, and T.~You, {\it {Accuracy complements energy: electroweak precision tests at Tera-Z}},  {\em JHEP} {\bf 10} (2025) 022, [\href{http://arxiv.org/abs/2412.14241}{{\tt arXiv:2412.14241}}].

\bibitem{Greljo:2025ggc}
A.~Greljo, B.~A. Stefanek, and A.~Valenti, {\it {Cornering Natural SUSY at a Tera-$Z$ Factory}},  \href{http://arxiv.org/abs/2507.03073}{{\tt arXiv:2507.03073}}.

\bibitem{Kosnik:2025srw}
N.~Ko{\v{s}}nik, A.~Palavri{\'c}, and A.~Smolkovi{\v{c}}, {\it {UV origins of CP-violating leptonic Yukawa couplings}},  \href{http://arxiv.org/abs/2509.04325}{{\tt arXiv:2509.04325}}.

\bibitem{Breso-Pla:2021qoe}
V.~Bres\'o-Pla, A.~Falkowski, and M.~Gonz\'alez-Alonso, {\it {A$_{FB}$ in the SMEFT: precision Z physics at the LHC}},  {\em JHEP} {\bf 08} (2021) 021, [\href{http://arxiv.org/abs/2103.12074}{{\tt arXiv:2103.12074}}].

\bibitem{Boughezal:2023nhe}
R.~Boughezal, Y.~Huang, and F.~Petriello, {\it {Impact of high invariant-mass Drell-Yan forward-backward asymmetry measurements on SMEFT fits}},  {\em Phys. Rev. D} {\bf 108} (2023), no.~7 076008, [\href{http://arxiv.org/abs/2303.08257}{{\tt arXiv:2303.08257}}].

\bibitem{Bissolotti:2023vdw}
C.~Bissolotti, R.~Boughezal, and K.~Simsek, {\it {SMEFT probes in future precision DIS experiments}},  {\em Phys. Rev. D} {\bf 108} (2023), no.~7 075007, [\href{http://arxiv.org/abs/2306.05564}{{\tt arXiv:2306.05564}}].

\bibitem{Boughezal:2022pmb}
R.~Boughezal, A.~Emmert, T.~Kutz, S.~Mantry, M.~Nycz, F.~Petriello, K.~\c{S}im\c{s}ek, D.~Wiegand, and X.~Zheng, {\it {Neutral-current electroweak physics and SMEFT studies at the EIC}},  {\em Phys. Rev. D} {\bf 106} (2022), no.~1 016006, [\href{http://arxiv.org/abs/2204.07557}{{\tt arXiv:2204.07557}}].

\bibitem{Greljo:2021kvv}
A.~Greljo, S.~Iranipour, Z.~Kassabov, M.~Madigan, J.~Moore, J.~Rojo, M.~Ubiali, and C.~Voisey, {\it {Parton distributions in the SMEFT from high-energy Drell-Yan tails}},  {\em JHEP} {\bf 07} (2021) 122, [\href{http://arxiv.org/abs/2104.02723}{{\tt arXiv:2104.02723}}].

\bibitem{Boughezal:2022nof}
R.~Boughezal, Y.~Huang, and F.~Petriello, {\it {Exploring the SMEFT at dimension eight with Drell-Yan transverse momentum measurements}},  {\em Phys. Rev. D} {\bf 106} (2022), no.~3 036020, [\href{http://arxiv.org/abs/2207.01703}{{\tt arXiv:2207.01703}}].

\bibitem{Abdolmaleki:2023jvw}
H.~Abdolmaleki et~al., {\it {Exploring SMEFT couplings using the forward\textendash{}backward asymmetry in neutral current Drell\textendash{}Yan production at the LHC}},  {\em Eur. Phys. J. C} {\bf 84} (2024), no.~12 1277, [\href{http://arxiv.org/abs/2310.19638}{{\tt arXiv:2310.19638}}].

\bibitem{Corbett:2024evt}
T.~Corbett, {\it {Top-down and bottom-up: Studying the SMEFT beyond leading order in $1/\Lambda^2$}},  {\em SciPost Phys. Core} {\bf 7} (2024) 053, [\href{http://arxiv.org/abs/2405.04570}{{\tt arXiv:2405.04570}}].

\bibitem{Boughezal:2021kla}
R.~Boughezal, F.~Petriello, and D.~Wiegand, {\it {Disentangling Standard Model EFT operators with future low-energy parity-violating electron scattering experiments}},  {\em Phys. Rev. D} {\bf 104} (2021), no.~1 016005, [\href{http://arxiv.org/abs/2104.03979}{{\tt arXiv:2104.03979}}].

\bibitem{Alioli:2017ces}
S.~Alioli, V.~Cirigliano, W.~Dekens, J.~de~Vries, and E.~Mereghetti, {\it {Right-handed charged currents in the era of the Large Hadron Collider}},  {\em JHEP} {\bf 05} (2017) 086, [\href{http://arxiv.org/abs/1703.04751}{{\tt arXiv:1703.04751}}].

\bibitem{Falkowski:2017pss}
A.~Falkowski, M.~Gonz\'alez-Alonso, and K.~Mimouni, {\it {Compilation of low-energy constraints on 4-fermion operators in the SMEFT}},  {\em JHEP} {\bf 08} (2017) 123, [\href{http://arxiv.org/abs/1706.03783}{{\tt arXiv:1706.03783}}].

\bibitem{Englert:2024nlj}
C.~Englert, C.~Mayer, W.~Naskar, and S.~Renner, {\it {Doubling down on down-type diquarks}},  {\em JHEP} {\bf 03} (2025) 011, [\href{http://arxiv.org/abs/2410.00952}{{\tt arXiv:2410.00952}}].

\bibitem{Greljo:2024ytg}
A.~Greljo, H.~Tiblom, and A.~Valenti, {\it {New physics through flavor tagging at FCC-ee}},  \href{http://arxiv.org/abs/2411.02485}{{\tt arXiv:2411.02485}}.

\bibitem{Mantani:2025bqu}
L.~Mantani and V.~Sanz, {\it {Probing the flavour-blind SMEFT: EFT validity and the interplay of energy scales}},  {\em JHEP} {\bf 06} (2025) 147, [\href{http://arxiv.org/abs/2503.02935}{{\tt arXiv:2503.02935}}].

\bibitem{Roy:2024avj}
A.~Roy and G.~Valencia, {\it {High-p$_{T}$ LHC constraints on SMEFT operators affecting rare kaon and hyperon decays}},  {\em JHEP} {\bf 05} (2025) 088, [\href{http://arxiv.org/abs/2410.05859}{{\tt arXiv:2410.05859}}].

\bibitem{Panda:2024oam}
D.~Panda, M.~K. Mohapatra, and R.~Mohanta, {\it {Analysis of $b \to c \ell \nu $ baryonic decay modes in SMEFT approach}},  \href{http://arxiv.org/abs/2411.19044}{{\tt arXiv:2411.19044}}.

\bibitem{Ali:2025xkw}
M.~I. Ali, U.~Chattopadhyay, D.~K. Ghosh, and N.~Rajeev, {\it {Constraints on lepton flavor universal and non-universal New Physics in $b\, \to\, s\, \ell^+ \ell^-$ decays: a global SMEFT survey}},  \href{http://arxiv.org/abs/2502.20145}{{\tt arXiv:2502.20145}}.

\bibitem{DiVita:2017vrr}
S.~Di~Vita, G.~Durieux, C.~Grojean, J.~Gu, Z.~Liu, G.~Panico, M.~Riembau, and T.~Vantalon, {\it {A global view on the Higgs self-coupling at lepton colliders}},  {\em JHEP} {\bf 02} (2018) 178, [\href{http://arxiv.org/abs/1711.03978}{{\tt arXiv:1711.03978}}].

\bibitem{Chala:2018ari}
M.~Chala, C.~Krause, and G.~Nardini, {\it {Signals of the electroweak phase transition at colliders and gravitational wave observatories}},  {\em JHEP} {\bf 07} (2018) 062, [\href{http://arxiv.org/abs/1802.02168}{{\tt arXiv:1802.02168}}].

\bibitem{Henning:2018kys}
B.~Henning, D.~Lombardo, M.~Riembau, and F.~Riva, {\it {Measuring Higgs Couplings without Higgs Bosons}},  {\em Phys. Rev. Lett.} {\bf 123} (2019), no.~18 181801, [\href{http://arxiv.org/abs/1812.09299}{{\tt arXiv:1812.09299}}].

\bibitem{Bobeth:2017ecx}
C.~Bobeth and A.~J. Buras, {\it {Leptoquarks meet $\varepsilon'/\varepsilon$ and rare Kaon processes}},  {\em JHEP} {\bf 02} (2018) 101, [\href{http://arxiv.org/abs/1712.01295}{{\tt arXiv:1712.01295}}].

\bibitem{Aebischer:2019blw}
J.~Aebischer, A.~J. Buras, M.~Cerd{\'a}-Sevilla, and F.~De~Fazio, {\it {Quark-lepton connections in Z' mediated FCNC processes: gauge anomaly cancellations at work}},  {\em JHEP} {\bf 02} (2020) 183, [\href{http://arxiv.org/abs/1912.09308}{{\tt arXiv:1912.09308}}].

\bibitem{Colangelo:2025nbd}
P.~Colangelo, F.~De~Fazio, and D.~Milillo, {\it {Correlating lepton flavour violating $b \to s$ and leptonic decay modes in a minimal abelian extension of the Standard Model}},  \href{http://arxiv.org/abs/2506.02552}{{\tt arXiv:2506.02552}}.

\bibitem{Buras:2018gto}
A.~J. Buras and M.~Jung, {\it {Analytic inclusion of the scale dependence of the anomalous dimension matrix in Standard Model Effective Theory}},  {\em JHEP} {\bf 06} (2018) 067, [\href{http://arxiv.org/abs/1804.05852}{{\tt arXiv:1804.05852}}].

\bibitem{Altmannshofer:2008dz}
W.~Altmannshofer, P.~Ball, A.~Bharucha, A.~J. Buras, D.~M. Straub, et~al., {\it {Symmetries and Asymmetries of $B \to K^{*} \mu^{+} \mu^{-}$ Decays in the Standard Model and Beyond}},  {\em JHEP} {\bf 0901} (2009) 019, [\href{http://arxiv.org/abs/0811.1214}{{\tt arXiv:0811.1214}}].

\bibitem{Altmannshofer:2009ma}
W.~Altmannshofer, A.~J. Buras, D.~M. Straub, and M.~Wick, {\it {New strategies for New Physics search in $B \to K^{*} \nu \bar{\nu}$, $B \to K \nu \bar{\nu}$ and $B \to X_{s} \nu \bar{\nu}$ decays}},  {\em JHEP} {\bf 04} (2009) 022, [\href{http://arxiv.org/abs/0902.0160}{{\tt arXiv:0902.0160}}].

\bibitem{Zyla:2020zbs}
{\bf Particle Data Group} Collaboration, P.~A. Zyla et~al., {\it {Review of Particle Physics}},  {\em PTEP} {\bf 2020} (2020), no.~8 083C01.

\bibitem{Aoki:2019cca}
{\bf Flavour Lattice Averaging Group} Collaboration, S.~Aoki et~al., {\it {FLAG Review 2019: Flavour Lattice Averaging Group (FLAG)}},  {\em Eur. Phys. J. C} {\bf 80} (2020), no.~2 113, [\href{http://arxiv.org/abs/1902.08191}{{\tt arXiv:1902.08191}}].

\bibitem{Aoki:2021kgd}
Y.~Aoki et~al., {\it {FLAG Review 2021}},  \href{http://arxiv.org/abs/2111.09849}{{\tt arXiv:2111.09849}}.

\bibitem{Dowdall:2019bea}
R.~J. Dowdall, C.~T.~H. Davies, R.~R. Horgan, G.~P. Lepage, C.~J. Monahan, J.~Shigemitsu, and M.~Wingate, {\it {Neutral $B$-meson mixing from full lattice QCD at the physical point}},  {\em Phys. Rev. D} {\bf 100} (2019), no.~9 094508, [\href{http://arxiv.org/abs/1907.01025}{{\tt arXiv:1907.01025}}].

\bibitem{Brod:2021hsj}
J.~Brod, M.~Gorbahn, and E.~Stamou, {\it {Updated Standard Model Prediction for $K \to \pi \nu \bar{\nu}$ and $\epsilon_K$}},  {\em PoS} {\bf BEAUTY2020} (2021) 056, [\href{http://arxiv.org/abs/2105.02868}{{\tt arXiv:2105.02868}}].

\bibitem{Brod:2019rzc}
J.~Brod, M.~Gorbahn, and E.~Stamou, {\it {Standard-Model Prediction of $\epsilon_K$ with Manifest Quark-Mixing Unitarity}},  {\em Phys. Rev. Lett.} {\bf 125} (2020), no.~17 171803, [\href{http://arxiv.org/abs/1911.06822}{{\tt arXiv:1911.06822}}].

\bibitem{Buras:2010pza}
A.~J. Buras, D.~Guadagnoli, and G.~Isidori, {\it {On $\epsilon_K$ beyond lowest order in the Operator Product Expansion}},  {\em Phys.~Lett.} {\bf B688} (2010) 309--313, [\href{http://arxiv.org/abs/1002.3612}{{\tt arXiv:1002.3612}}].

\bibitem{Buras:1990fn}
A.~J. Buras, M.~Jamin, and P.~H. Weisz, {\it {Leading and next-to-leading QCD corrections to $\varepsilon$ parameter and $B^0-\bar{B}^0$ mixing in the presence of a heavy top quark}},  {\em Nucl.~Phys.} {\bf B347} (1990) 491--536.

\bibitem{Urban:1997gw}
J.~Urban, F.~Krauss, U.~Jentschura, and G.~Soff, {\it {Next-to-leading order QCD corrections for the $B^0 - \bar B^0$ mixing with an extended Higgs sector}},  {\em Nucl.~Phys.} {\bf B523} (1998) 40--58, [\href{http://arxiv.org/abs/hep-ph/9710245}{{\tt hep-ph/9710245}}].

\bibitem{Amhis:2016xyh}
{\bf Heavy Flavor Averaging Group (HFAG)} Collaboration, Y.~Amhis et~al., {\it {Averages of $b$-hadron, $c$-hadron, and $\tau$-lepton properties as of summer 2016}},  \href{http://arxiv.org/abs/1612.07233}{{\tt arXiv:1612.07233}}. Updates on \url{https://urldefense.com/v3/__http://www.slac.stanford.edu/xorg/hfag*7D*7Bhttp:/*www.slac.stanford.edu/xorg/hfag__;JSUv!!Mih3wA!WzHhMs_LrEEC6iOCPmeHL0di2Fuq19ujdL9vDWRi-hsLF8Wz-c-B5FHxeTo320oJWwbLq1dv$ }.

\bibitem{Boyle:2024gge}
{\bf RBC, UKQCD} Collaboration, P.~A. Boyle, F.~Erben, J.~M. Flynn, N.~Garron, J.~Kettle, R.~Mukherjee, and J.~T. Tsang, {\it {Kaon mixing beyond the standard model with physical masses}},  {\em Phys. Rev. D} {\bf 110} (2024), no.~3 034501, [\href{http://arxiv.org/abs/2404.02297}{{\tt arXiv:2404.02297}}].

\bibitem{Gorbahn:2024qpe}
M.~Gorbahn, S.~J{\"a}ger, and S.~Kvedarait{\.{e}}, {\it {RI-(S)MOM to $ \overline{\textrm{MS}} $ conversion for $B_{K}$ at two-loop order}},  {\em JHEP} {\bf 09} (2025) 011, [\href{http://arxiv.org/abs/2411.19861}{{\tt arXiv:2411.19861}}].

\bibitem{Buras:2014maa}
A.~J. Buras, J.-M. G{\'e}rard, and W.~A. Bardeen, {\it {Large $N$ Approach to Kaon Decays and Mixing 28 Years Later: $\Delta I = 1/2$ Rule, $\hat B_K$ and $\Delta M_K$}},  {\em Eur.~Phys.~J.} {\bf C74} (2014), no.~5 2871, [\href{http://arxiv.org/abs/1401.1385}{{\tt arXiv:1401.1385}}].

\bibitem{Ciuchini:1997bw}
M.~Ciuchini, E.~Franco, V.~Lubicz, G.~Martinelli, I.~Scimemi, et~al., {\it {Next-to-leading order QCD corrections to $\Delta F = 2$ effective Hamiltonians}},  {\em Nucl.~Phys.} {\bf B523} (1998) 501--525, [\href{http://arxiv.org/abs/hep-ph/9711402}{{\tt hep-ph/9711402}}].

\bibitem{Buras:2018lgu}
A.~J. Buras and J.-M. G\'erard, {\it {Dual QCD Insight into BSM Hadronic Matrix Elements for $K^0-\bar K^0$ Mixing from Lattice QCD}},  {\em Acta Phys. Polon. B} {\bf 50} (2019) 121, [\href{http://arxiv.org/abs/1804.02401}{{\tt arXiv:1804.02401}}].

\bibitem{Aebischer:2018rrz}
J.~Aebischer, A.~J. Buras, and J.-M. G{\'e}rard, {\it {BSM hadronic matrix elements for $\epsilon'/\epsilon$ and $K\to\pi\pi$ decays in the Dual QCD approach}},  {\em JHEP} {\bf 02} (2019) 021, [\href{http://arxiv.org/abs/1807.01709}{{\tt arXiv:1807.01709}}].

\bibitem{Buras:2001ra}
A.~J. Buras, S.~J{\"a}ger, and J.~Urban, {\it {Master formulae for $\Delta F=2$ NLO QCD factors in the standard model and beyond}},  {\em Nucl.~Phys.} {\bf B605} (2001) 600--624, [\href{http://arxiv.org/abs/hep-ph/0102316}{{\tt hep-ph/0102316}}].

\bibitem{Endo:2016tnu}
M.~Endo, T.~Kitahara, S.~Mishima, and K.~Yamamoto, {\it {Revisiting Kaon Physics in General $Z$ Scenario}},  {\em Phys. Lett.} {\bf B771} (2017) 37--44, [\href{http://arxiv.org/abs/1612.08839}{{\tt arXiv:1612.08839}}].

\bibitem{Buras:2014sba}
A.~J. Buras, F.~De~Fazio, and J.~Girrbach, {\it {$\Delta I=1/2$ rule, $\varepsilon '/\varepsilon $ and $K\rightarrow \pi \nu \bar{\nu }$ in $Z' (Z)$ and $G' $ models with FCNC quark couplings}},  {\em Eur.~Phys.~J.} {\bf C74} (2014) 2950, [\href{http://arxiv.org/abs/1404.3824}{{\tt arXiv:1404.3824}}].

\bibitem{Buras:2014zga}
A.~J. Buras, D.~Buttazzo, J.~Girrbach-Noe, and R.~Knegjens, {\it {Can we reach the Zeptouniverse with rare $K$ and $B_{s,d}$ decays?}},  {\em JHEP} {\bf 1411} (2014) 121, [\href{http://arxiv.org/abs/1408.0728}{{\tt arXiv:1408.0728}}].

\bibitem{Crivellin:2015era}
A.~Crivellin, L.~Hofer, J.~Matias, U.~Nierste, S.~Pokorski, and J.~Rosiek, {\it {Lepton-flavour violating $B$ decays in generic $Z'$ models}},  {\em Phys. Rev.} {\bf D92} (2015), no.~5 054013, [\href{http://arxiv.org/abs/1504.07928}{{\tt arXiv:1504.07928}}].

\bibitem{Buras:2024mnq}
A.~J. Buras and P.~Stangl, {\it {On the interplay of constraints from $B_{s},$D,~ and K meson mixing in $Z^\prime $ models with implications for $b\to s \nu {\bar{\nu }}$ transitions}},  {\em Eur. Phys. J. C} {\bf 85} (2025), no.~5 519, [\href{http://arxiv.org/abs/2412.14254}{{\tt arXiv:2412.14254}}].

\bibitem{Gambino:2010jz}
P.~Gambino and J.~F. Kamenik, {\it {Lepton energy moments in semileptonic charm decays}},  {\em Nucl. Phys. B} {\bf 840} (2010) 424--437, [\href{http://arxiv.org/abs/1004.0114}{{\tt arXiv:1004.0114}}].

\bibitem{deBoer:2016dcg}
S.~de~Boer, B.~M\"uller, and D.~Seidel, {\it {Higher-order Wilson coefficients for $c \to u$ transitions in the standard model}},  {\em JHEP} {\bf 08} (2016) 091, [\href{http://arxiv.org/abs/1606.05521}{{\tt arXiv:1606.05521}}].

\bibitem{DeBoer:2018pdx}
S.~De~Boer and G.~Hiller, {\it {Null tests from angular distributions in $D \to P_1 P_2 l^+l^-$, $l=e,\mu$ decays on and off peak}},  {\em Phys. Rev. D} {\bf 98} (2018), no.~3 035041, [\href{http://arxiv.org/abs/1805.08516}{{\tt arXiv:1805.08516}}].

\bibitem{Fael:2019umf}
M.~Fael, T.~Mannel, and K.~K. Vos, {\it {The Heavy Quark Expansion for Inclusive Semileptonic Charm Decays Revisited}},  {\em JHEP} {\bf 12} (2019) 067, [\href{http://arxiv.org/abs/1910.05234}{{\tt arXiv:1910.05234}}].

\bibitem{Golz:2022alh}
M.~Golz, G.~Hiller, and T.~Magorsch, {\it {Pinning down $|\Delta c|=|\Delta u|=1$ couplings with rare charm baryon decays}},  {\em Eur. Phys. J. C} {\bf 82} (2022), no.~4 357, [\href{http://arxiv.org/abs/2202.02331}{{\tt arXiv:2202.02331}}].

\bibitem{Gisbert:2024kob}
H.~Gisbert, G.~Hiller, and D.~Suelmann, {\it {Effective field theory analysis of rare $|\Delta c|=|\Delta u|=1$ charm decays}},  {\em JHEP} {\bf 12} (2024) 102, [\href{http://arxiv.org/abs/2410.00115}{{\tt arXiv:2410.00115}}].

\bibitem{Isidori:2010kg}
G.~Isidori, Y.~Nir, and G.~Perez, {\it {Flavor Physics Constraints for Physics Beyond the Standard Model}},  {\em Ann.Rev.Nucl.Part.Sci.} {\bf 60} (2010) 355, [\href{http://arxiv.org/abs/1002.0900}{{\tt arXiv:1002.0900}}].

\bibitem{London:2021lfn}
D.~London and J.~Matias, {\it {$B$ Flavour Anomalies: 2021 Theoretical Status Report}},  {\em Ann. Rev. Nucl. Part. Sci.} {\bf 72} (2022) 37--68, [\href{http://arxiv.org/abs/2110.13270}{{\tt arXiv:2110.13270}}].

\bibitem{Capdevila:2023yhq}
B.~Capdevila, A.~Crivellin, and J.~Matias, {\it {Review of semileptonic B anomalies}},  {\em Eur. Phys. J. ST} {\bf 1} (2023) 20, [\href{http://arxiv.org/abs/2309.01311}{{\tt arXiv:2309.01311}}].

\bibitem{Bause:2021cna}
R.~Bause, H.~Gisbert, M.~Golz, and G.~Hiller, {\it {Interplay of dineutrino modes with semileptonic rare B-decays}},  {\em JHEP} {\bf 12} (2021) 061, [\href{http://arxiv.org/abs/2109.01675}{{\tt arXiv:2109.01675}}].

\bibitem{He:2021yoz}
X.~G. He and G.~Valencia, {\it {$R^\nu(K^{(*)})$ and non-standard neutrino interactions}},  {\em Phys. Lett. B} {\bf 821} (2021) 136607, [\href{http://arxiv.org/abs/2108.05033}{{\tt arXiv:2108.05033}}].

\bibitem{Bause:2022rrs}
R.~Bause, H.~Gisbert, M.~Golz, and G.~Hiller, {\it {Model-independent analysis of $b \rightarrow d$ processes}},  {\em Eur. Phys. J. C} {\bf 83} (2023), no.~5 419, [\href{http://arxiv.org/abs/2209.04457}{{\tt arXiv:2209.04457}}].

\bibitem{Becirevic:2023aov}
D.~Be\v{c}irevi\'c, G.~Piazza, and O.~Sumensari, {\it {Revisiting $B\rightarrow K^{(*)} \nu {\bar{\nu }}$ decays in the Standard Model and beyond}},  {\em Eur. Phys. J. C} {\bf 83} (2023), no.~3 252, [\href{http://arxiv.org/abs/2301.06990}{{\tt arXiv:2301.06990}}].

\bibitem{Bause:2023mfe}
R.~Bause, H.~Gisbert, and G.~Hiller, {\it {Implications of an enhanced $B\to K \nu\bar\nu$ branching ratio}},  {\em Phys. Rev. D} {\bf 109} (2024), no.~1 015006, [\href{http://arxiv.org/abs/2309.00075}{{\tt arXiv:2309.00075}}].

\bibitem{Allwicher:2023xba}
L.~Allwicher, D.~Becirevic, G.~Piazza, S.~Rosauro-Alcaraz, and O.~Sumensari, {\it {Understanding the first measurement of $\mathcal{B}(B\to K \nu \bar{\nu})$}},  {\em Phys. Lett. B} {\bf 848} (2024) 138411, [\href{http://arxiv.org/abs/2309.02246}{{\tt arXiv:2309.02246}}].

\bibitem{Wang:2023trd}
Z.~S. Wang, H.~K. Dreiner, and J.~Y. G\"unther, {\it {The decay $B\rightarrow K+\nu +\bar{\nu }$ at Belle II and a massless bino in R-parity-violating supersymmetry}},  {\em Eur. Phys. J. C} {\bf 85} (2025), no.~1 66, [\href{http://arxiv.org/abs/2309.03727}{{\tt arXiv:2309.03727}}].

\bibitem{Altmannshofer:2023hkn}
W.~Altmannshofer, A.~Crivellin, H.~Haigh, G.~Inguglia, and J.~Martin~Camalich, {\it {Light new physics in $B\to K^{(*)}\nu\bar\nu$?}},  {\em Phys. Rev. D} {\bf 109} (2024), no.~7 075008, [\href{http://arxiv.org/abs/2311.14629}{{\tt arXiv:2311.14629}}].

\bibitem{Gabrielli:2024wys}
E.~Gabrielli, L.~Marzola, K.~M\"u\"ursepp, and M.~Raidal, {\it {Explaining the $B^+\rightarrow K^+ \nu \bar{\nu }$ excess via a massless dark photon}},  {\em Eur. Phys. J. C} {\bf 84} (2024), no.~5 460, [\href{http://arxiv.org/abs/2402.05901}{{\tt arXiv:2402.05901}}].

\bibitem{He:2024iju}
X.-G. He, X.-D. Ma, M.~A. Schmidt, G.~Valencia, and R.~R. Volkas, {\it {Scalar dark matter explanation of the excess in the Belle II B$^{+}$\textrightarrow{} K$^{+}$+ invisible measurement}},  {\em JHEP} {\bf 07} (2024) 168, [\href{http://arxiv.org/abs/2403.12485}{{\tt arXiv:2403.12485}}].

\bibitem{Bolton:2024egx}
P.~D. Bolton, S.~Fajfer, J.~F. Kamenik, and M.~Novoa-Brunet, {\it {Signatures of light new particles in B\textrightarrow{}K(*)Emiss}},  {\em Phys. Rev. D} {\bf 110} (2024), no.~5 055001, [\href{http://arxiv.org/abs/2403.13887}{{\tt arXiv:2403.13887}}].

\bibitem{Buras:2024ewl}
A.~J. Buras, J.~Harz, and M.~A. Mojahed, {\it {Disentangling new physics in $ K\to \pi \nu \overline{\nu} $ and $ B\to K\left({K}^{\ast}\right)\nu \overline{\nu} $ observables}},  {\em JHEP} {\bf 10} (2024) 087, [\href{http://arxiv.org/abs/2405.06742}{{\tt arXiv:2405.06742}}].

\bibitem{Altmannshofer:2024kxb}
W.~Altmannshofer and S.~Roy, {\it {Joint explanation of the B\textrightarrow{}\ensuremath{\pi}K puzzle and the B\textrightarrow{}K\ensuremath{\nu}\ensuremath{\nu}\textasciimacron{} excess}},  {\em Phys. Rev. D} {\bf 111} (2025), no.~7 075029, [\href{http://arxiv.org/abs/2411.06592}{{\tt arXiv:2411.06592}}].

\bibitem{Hati:2024ppg}
C.~Hati, J.~Leite, N.~Nath, and J.~W.~F. Valle, {\it {QCD axion, color-mediated neutrino masses, and $B^+\rightarrow K^+ +\text{Emiss}$ anomaly}},  {\em Phys. Rev. D} {\bf 111} (2025), no.~1 015038, [\href{http://arxiv.org/abs/2408.00060}{{\tt arXiv:2408.00060}}].

\bibitem{Bolton:2025fsq}
P.~D. Bolton, S.~Fajfer, J.~F. Kamenik, and M.~Novoa-Brunet, {\it {Impact of new invisible particles on $B\to K^{(*)} E_{\rm miss}$ observables}},  {\em Phys. Rev. D} {\bf 112} (2025), no.~3 035010, [\href{http://arxiv.org/abs/2503.19025}{{\tt arXiv:2503.19025}}].

\bibitem{He:2025zfy}
X.-G. He, X.-D. Ma, J.~Tandean, and G.~Valencia, {\it {Light dark-matter window constrained by $K^+\to\pi^+$$+$$\not{\!\!E}$}},  {\em Phys. Rev. D} {\bf 112} (2025), no.~5 055025, [\href{http://arxiv.org/abs/2505.02031}{{\tt arXiv:2505.02031}}].

\bibitem{Crivellin:2025qsq}
A.~Crivellin, S.~Iguro, and T.~Kitahara, {\it {Correlating the $B$ anomalies to $K\to \pi\nu\bar\nu$ and $B\to K\nu\bar\nu$ via leptoquarks}},  \href{http://arxiv.org/abs/2505.05552}{{\tt arXiv:2505.05552}}.

\bibitem{Celis:2017doq}
A.~Celis, J.~Fuentes-Martin, A.~Vicente, and J.~Virto, {\it {Gauge-invariant implications of the LHCb measurements on lepton-flavor nonuniversality}},  {\em Phys. Rev.} {\bf D96} (2017), no.~3 035026, [\href{http://arxiv.org/abs/1704.05672}{{\tt arXiv:1704.05672}}].

\bibitem{Datta:2019zca}
A.~Datta, J.~Kumar, and D.~London, {\it {The $B$ anomalies and new physics in $b \to s e^+ e^-$}},  {\em Phys. Lett. B} {\bf 797} (2019) 134858, [\href{http://arxiv.org/abs/1903.10086}{{\tt arXiv:1903.10086}}].

\bibitem{Datta:2024zrl}
A.~Datta, J.~Kumar, S.~Kumbhakar, and D.~London, {\it {Uniting low-energy semileptonic and hadronic anomalies within SMEFT}},  {\em JHEP} {\bf 12} (2024) 175, [\href{http://arxiv.org/abs/2408.03380}{{\tt arXiv:2408.03380}}].

\bibitem{Pich:2013lsa}
A.~Pich, {\it {Precision Tau Physics}},  {\em Prog. Part. Nucl. Phys.} {\bf 75} (2014) 41--85, [\href{http://arxiv.org/abs/1310.7922}{{\tt arXiv:1310.7922}}].

\bibitem{Fajfer:2012jt}
S.~Fajfer, J.~F. Kamenik, I.~Nisandzic, and J.~Zupan, {\it {Implications of Lepton Flavor Universality Violations in B Decays}},  {\em Phys.~Rev.~Lett.} {\bf 109} (2012) 161801, [\href{http://arxiv.org/abs/1206.1872}{{\tt arXiv:1206.1872}}].

\bibitem{Glashow:2014iga}
S.~L. Glashow, D.~Guadagnoli, and K.~Lane, {\it {Lepton Flavor Violation in $B$ Decays?}},  {\em Phys. Rev. Lett.} {\bf 114} (2015) 091801, [\href{http://arxiv.org/abs/1411.0565}{{\tt arXiv:1411.0565}}].

\bibitem{Alonso:2015sja}
R.~Alonso, B.~Grinstein, and J.~Martin~Camalich, {\it {Lepton universality violation and lepton flavor conservation in $B$-meson decays}},  {\em JHEP} {\bf 10} (2015) 184, [\href{http://arxiv.org/abs/1505.05164}{{\tt arXiv:1505.05164}}].

\bibitem{Calibbi:2015kma}
L.~Calibbi, A.~Crivellin, and T.~Ota, {\it {Effective Field Theory Approach to $b\to s\ell\ell^{(\prime)}$, $B\to K^{(*)}\nu\bar\nu$ and $B\to D^{(*)}\tau\nu$ with Third Generation Couplings}},  {\em Phys. Rev. Lett.} {\bf 115} (2015) 181801, [\href{http://arxiv.org/abs/1506.02661}{{\tt arXiv:1506.02661}}].

\bibitem{Buras:2014yna}
A.~J. Buras, F.~De~Fazio, and J.~Girrbach-Noe, {\it {Z-Z' mixing and Z-mediated FCNCs in $SU(3)_C \times SU(3)_L \times U(1)_X$ Models}},  {\em JHEP} {\bf 1408} (2014) 039, [\href{http://arxiv.org/abs/1405.3850}{{\tt arXiv:1405.3850}}].

\bibitem{Buras:2001af}
A.~J. Buras and R.~Fleischer, {\it {Bounds on the unitarity triangle, $\sin2\beta$ and $K \to\pi \nu\bar\nu$ decays in models with minimal flavor violation}},  {\em Phys.~Rev.} {\bf D64} (2001) 115010, [\href{http://arxiv.org/abs/hep-ph/0104238}{{\tt hep-ph/0104238}}].

\bibitem{Buras:2015yca}
A.~J. Buras, D.~Buttazzo, and R.~Knegjens, {\it {$K\to\pi\nu\bar\nu$ and $\epsilon'/\epsilon$ in Simplified New Physics Models}},  {\em JHEP} {\bf 11} (2015) 166, [\href{http://arxiv.org/abs/1507.08672}{{\tt arXiv:1507.08672}}].

\bibitem{Dreiner:2023cms}
Z.~S. Wang, H.~K. Dreiner, and J.~Y. G\"unther, {\it {The decay $B\rightarrow K^+\nu \bar{\nu }$ at Belle II and a massless bino in R-parity-violating supersymmetry}},  {\em Eur. Phys. J. C} {\bf 85} (2025), no.~1 66, [\href{http://arxiv.org/abs/2309.03727}{{\tt arXiv:2309.03727}}].

\bibitem{He:2023bnk}
X.-G. He, X.-D. Ma, and G.~Valencia, {\it {Revisiting models that enhance $B^+\to K^+\nu\bar\nu$ in light of the new Belle II measurement}},  {\em Phys. Rev. D} {\bf 109} (2024), no.~7 075019, [\href{http://arxiv.org/abs/2309.12741}{{\tt arXiv:2309.12741}}].

\bibitem{Chen:2025npb}
C.-H. Chen, C.-W. Chiang, and L.~M.~G. de~la Vega, {\it {Leptoquark-mediated Dirac neutrino mass and its impact on $ B\to K\nu \overline{\nu} $ and $ K\to \pi \nu \overline{\nu} $ decays}},  {\em JHEP} {\bf 09} (2025) 055, [\href{http://arxiv.org/abs/2503.22431}{{\tt arXiv:2503.22431}}].

\bibitem{Belle-II:2023esi}
{\bf Belle-II} Collaboration, I.~Adachi et~al., {\it {Evidence for $B^+\to K^+\nu\bar\nu$ decays}},  {\em Phys. Rev. D} {\bf 109} (2024), no.~11 112006, [\href{http://arxiv.org/abs/2311.14647}{{\tt arXiv:2311.14647}}].

\bibitem{Morell:2024aml}
P.~Morell and J.~Virto, {\it {On the two-loop penguin contributions to the Anomalous Dimensions of four-quark operators}},  {\em JHEP} {\bf 04} (2024) 105, [\href{http://arxiv.org/abs/2402.00249}{{\tt arXiv:2402.00249}}].

\bibitem{Aebischer:2021hws}
J.~Aebischer, C.~Bobeth, A.~J. Buras, and J.~Kumar, {\it {BSM master formula for $\epe$ in the WET basis at NLO in QCD}},  {\em JHEP} {\bf 12} (2021) 043, [\href{http://arxiv.org/abs/2107.12391}{{\tt arXiv:2107.12391}}].

\bibitem{Buras:2022cyc}
A.~J. Buras, {\it {$\varepsilon'/\varepsilon$ in the Standard Model and Beyond: 2021}},  in {\em {11th International Workshop on the CKM Unitarity Triangle}}, 3, 2022.
\newblock \href{http://arxiv.org/abs/2203.12632}{{\tt arXiv:2203.12632}}.

\bibitem{Buras:2023qaf}
A.~J. Buras, {\it {Kaon Theory: 50 Years Later}},  7, 2023.
\newblock \href{http://arxiv.org/abs/2307.15737}{{\tt arXiv:2307.15737}}.

\bibitem{RBC:2020kdj}
{\bf RBC, UKQCD} Collaboration, R.~Abbott et~al., {\it {Direct CP violation and the $\Delta I=1/2$ rule in $K\to\pi\pi$ decay from the standard model}},  {\em Phys. Rev. D} {\bf 102} (2020), no.~5 054509, [\href{http://arxiv.org/abs/2004.09440}{{\tt arXiv:2004.09440}}].

\bibitem{Cirigliano:2019ani}
V.~Cirigliano, H.~Gisbert, A.~Pich, and A.~Rodr\'\i{}guez-S\'anchez, {\it {Theoretical status of $\varepsilon'/\varepsilon$}},  {\em J. Phys. Conf. Ser.} {\bf 1526} (2020) 012011, [\href{http://arxiv.org/abs/1912.04736}{{\tt arXiv:1912.04736}}].

\bibitem{Batley:2002gn}
{\bf NA48} Collaboration, J.~Batley et~al., {\it {A Precision measurement of direct CP violation in the decay of neutral kaons into two pions}},  {\em Phys.~Lett.} {\bf B544} (2002) 97--112, [\href{http://arxiv.org/abs/hep-ex/0208009}{{\tt hep-ex/0208009}}].

\bibitem{AlaviHarati:2002ye}
{\bf KTeV} Collaboration, A.~Alavi-Harati et~al., {\it {Measurements of direct CP violation, CPT symmetry, and other parameters in the neutral kaon system}},  {\em Phys.~Rev.} {\bf D67} (2003) 012005, [\href{http://arxiv.org/abs/hep-ex/0208007}{{\tt hep-ex/0208007}}].

\bibitem{Abouzaid:2010ny}
{\bf KTeV} Collaboration, E.~Abouzaid et~al., {\it {Precise Measurements of Direct CP Violation, CPT Symmetry, and Other Parameters in the Neutral Kaon System}},  {\em Phys. Rev.} {\bf D83} (2011) 092001, [\href{http://arxiv.org/abs/1011.0127}{{\tt arXiv:1011.0127}}].

\bibitem{Bardeen:1986vz}
W.~A. Bardeen, A.~J. Buras, and J.-M. G\'erard, {\it {A Consistent Analysis of the $\Delta I = 1/2$ Rule for K Decays}},  {\em Phys.~Lett.} {\bf B192} (1987) 138.

\bibitem{Buras:2015xba}
A.~J. Buras and J.-M. G\'erard, {\it {Upper Bounds on $\varepsilon'/\varepsilon$ Parameters $B_6^{(1/2)}$ and $B_8^{(3/2)}$ from Large N QCD and other News}},  {\em JHEP} {\bf 12} (2015) 008, [\href{http://arxiv.org/abs/1507.06326}{{\tt arXiv:1507.06326}}].

\bibitem{Buras:2016fys}
A.~J. Buras and J.-M. G\'erard, {\it {Final state interactions in $K\rightarrow \pi \pi $ decays: $\Delta I=1/2$ rule vs. $\varepsilon '/\varepsilon $}},  {\em Eur. Phys. J.} {\bf C77} (2017), no.~1 10, [\href{http://arxiv.org/abs/1603.05686}{{\tt arXiv:1603.05686}}].

\bibitem{Buras:2020wyv}
A.~J. Buras, {\it {The $\epsilon'/\epsilon$-Story: 1976-2021}},  {\em Acta Phys. Polon. B} {\bf 52} (2021), no.~1 7--41, [\href{http://arxiv.org/abs/2101.00020}{{\tt arXiv:2101.00020}}].

\bibitem{Constantinou:2017sgv}
{\bf ETM} Collaboration, M.~Constantinou, M.~Costa, R.~Frezzotti, V.~Lubicz, G.~Martinelli, D.~Meloni, H.~Panagopoulos, and S.~Simula, {\it {$K \to \pi$ matrix elements of the chromomagnetic operator on the lattice}},  {\em Phys. Rev.} {\bf D97} (2018), no.~7 074501, [\href{http://arxiv.org/abs/1712.09824}{{\tt arXiv:1712.09824}}].

\bibitem{Buras:2018evv}
A.~J. Buras and J.-M. Gérard, {\it {$K\to\pi\pi$ and $K-\pi$ Matrix Elements of the Chromomagnetic Operators from Dual QCD}},  {\em JHEP} {\bf 07} (2018) 126, [\href{http://arxiv.org/abs/1803.08052}{{\tt arXiv:1803.08052}}].

\bibitem{Buras:2018wmb}
A.~J. Buras, {\it {The Return of Kaon Flavour Physics}},  {\em Acta Phys. Polon.} {\bf B49} (2018) 1043, [\href{http://arxiv.org/abs/1805.11096}{{\tt arXiv:1805.11096}}].

\bibitem{Aebischer:2018quc}
J.~Aebischer, C.~Bobeth, A.~J. Buras, J.-M. G{\'e}rard, and D.~M. Straub, {\it {Master formula for $\varepsilon'/\varepsilon$ beyond the Standard Model}},  {\em Phys. Lett.} {\bf B792} (2019) 465--469, [\href{http://arxiv.org/abs/1807.02520}{{\tt arXiv:1807.02520}}].

\bibitem{GellMann:1955jx}
M.~Gell-Mann and A.~Pais, {\it {Behavior of neutral particles under charge conjugation}},  {\em Phys.~Rev.} {\bf 97} (1955) 1387--1389.

\bibitem{GellMann:1957wh}
M.~Gell-Mann and A.~Rosenfeld, {\it {Hyperons and heavy mesons (systematics and decay)}},  {\em Ann.Rev.Nucl.Part.Sci.} {\bf 7} (1957) 407--478.

\bibitem{ALEPH:2005ab}
{\bf ALEPH, DELPHI, L3, OPAL, SLD, LEP Electroweak Working Group, SLD Electroweak Group, SLD Heavy Flavour Group} Collaboration, S.~Schael et~al., {\it {Precision electroweak measurements on the $Z$ resonance}},  {\em Phys. Rept.} {\bf 427} (2006) 257--454, [\href{http://arxiv.org/abs/hep-ex/0509008}{{\tt hep-ex/0509008}}].

\bibitem{ALEPH:2013dgf}
{\bf ALEPH, DELPHI, L3, OPAL, LEP Electroweak} Collaboration, S.~Schael et~al., {\it {Electroweak Measurements in Electron-Positron Collisions at W-Boson-Pair Energies at LEP}},  {\em Phys. Rept.} {\bf 532} (2013) 119--244, [\href{http://arxiv.org/abs/1302.3415}{{\tt arXiv:1302.3415}}].

\bibitem{ALEPH:2006bhb}
{\bf ALEPH, DELPHI, L3, OPAL, LEP Electroweak Working Group} Collaboration, J.~Alcaraz et~al., {\it {A Combination of preliminary electroweak measurements and constraints on the standard model}},  \href{http://arxiv.org/abs/hep-ex/0612034}{{\tt hep-ex/0612034}}.

\bibitem{Davidson:2022jai}
S.~Davidson, B.~Echenard, R.~H. Bernstein, J.~Heeck, and D.~G. Hitlin, {\it {Charged Lepton Flavor Violation}},  \href{http://arxiv.org/abs/2209.00142}{{\tt arXiv:2209.00142}}.

\bibitem{Crivellin:2017rmk}
A.~Crivellin, S.~Davidson, G.~M. Pruna, and A.~Signer, {\it {Renormalisation-group improved analysis of $\mu\to e$ processes in a systematic effective-field-theory approach}},  {\em JHEP} {\bf 05} (2017) 117, [\href{http://arxiv.org/abs/1702.03020}{{\tt arXiv:1702.03020}}].

\bibitem{Davidson:2020ord}
S.~Davidson, Y.~Kuno, Y.~Uesaka, and M.~Yamanaka, {\it {Probing $\mu e \gamma \gamma$ contact interactions with $\mu \to e$ conversion}},  {\em Phys. Rev. D} {\bf 102} (2020), no.~11 115043, [\href{http://arxiv.org/abs/2007.09612}{{\tt arXiv:2007.09612}}].

\bibitem{Davidson:2020hkf}
S.~Davidson, {\it {Completeness and complementarity for $\mu \to e\gamma \mu \to e \bar e e$ and $\mu A \to eA$}},  {\em JHEP} {\bf 02} (2021) 172, [\href{http://arxiv.org/abs/2010.00317}{{\tt arXiv:2010.00317}}].

\bibitem{Cirigliano:2022ekw}
V.~Cirigliano, K.~Fuyuto, M.~J. Ramsey-Musolf, and E.~Rule, {\it {Next-to-leading order scalar contributions to \ensuremath{\mu}\textrightarrow{}e conversion}},  {\em Phys. Rev. C} {\bf 105} (2022), no.~5 055504, [\href{http://arxiv.org/abs/2203.09547}{{\tt arXiv:2203.09547}}].

\bibitem{Haxton:2022piv}
W.~C. Haxton, E.~Rule, K.~McElvain, and M.~J. Ramsey-Musolf, {\it {Nuclear-level effective theory of \ensuremath{\mu}\textrightarrow{}e conversion: Formalism and applications}},  {\em Phys. Rev. C} {\bf 107} (2023), no.~3 035504, [\href{http://arxiv.org/abs/2208.07945}{{\tt arXiv:2208.07945}}].

\bibitem{Ardu:2023yyw}
M.~Ardu, S.~Davidson, and S.~Lavignac, {\it {Distinguishing models with \ensuremath{\mu} \textrightarrow{} e observables}},  {\em JHEP} {\bf 11} (2023) 101, [\href{http://arxiv.org/abs/2308.16897}{{\tt arXiv:2308.16897}}].

\bibitem{Haxton:2024lyc}
W.~Haxton, K.~McElvain, T.~Menzo, E.~Rule, and J.~Zupan, {\it {Effective theory tower for $\mu\rightarrow e$ conversion}},  {\em JHEP} {\bf 11} (2024) 076, [\href{http://arxiv.org/abs/2406.13818}{{\tt arXiv:2406.13818}}].

\bibitem{Delzanno:2024ooj}
F.~Delzanno, K.~Fuyuto, S.~Gonz{\`a}lez-Sol{\'\i}s, and E.~Mereghetti, {\it {Global analysis of $\mu \to e$ interactions in the SMEFT}},  {\em JHEP} {\bf 07} (2025) 283, [\href{http://arxiv.org/abs/2411.13497}{{\tt arXiv:2411.13497}}].

\bibitem{Buras:2021btx}
A.~J. Buras, A.~Crivellin, F.~Kirk, C.~A. Manzari, and M.~Montull, {\it {Global analysis of leptophilic Z' bosons}},  {\em JHEP} {\bf 06} (2021) 068, [\href{http://arxiv.org/abs/2104.07680}{{\tt arXiv:2104.07680}}].

\bibitem{Vives:2025clr}
O.~Vives and N.~Valori, {\it {Beyond the Standard Model contributions to dipole moments}},  \href{http://arxiv.org/abs/2505.06345}{{\tt arXiv:2505.06345}}.

\bibitem{Engel:2013lsa}
J.~Engel, M.~J. Ramsey-Musolf, and U.~van Kolck, {\it {Electric Dipole Moments of Nucleons, Nuclei, and Atoms: The Standard Model and Beyond}},  {\em Prog.~Part.~Nucl.~Phys.} {\bf 71} (2013) 21--74, [\href{http://arxiv.org/abs/1303.2371}{{\tt arXiv:1303.2371}}].

\bibitem{Allwicher:2021jkr}
L.~Allwicher, L.~Di~Luzio, M.~Fedele, F.~Mescia, and M.~Nardecchia, {\it {What is the scale of new physics behind the muon g-2?}},  {\em Phys. Rev. D} {\bf 104} (2021), no.~5 055035, [\href{http://arxiv.org/abs/2105.13981}{{\tt arXiv:2105.13981}}].

\bibitem{Cirigliano:2021peb}
V.~Cirigliano, W.~Dekens, J.~de~Vries, K.~Fuyuto, E.~Mereghetti, and R.~Ruiz, {\it {Leptonic anomalous magnetic moments in \ensuremath{\nu} SMEFT}},  {\em JHEP} {\bf 08} (2021) 103, [\href{http://arxiv.org/abs/2105.11462}{{\tt arXiv:2105.11462}}].

\bibitem{Aliberti:2025beg}
R.~Aliberti et~al., {\it {The anomalous magnetic moment of the muon in the Standard Model: an update}},  \href{http://arxiv.org/abs/2505.21476}{{\tt arXiv:2505.21476}}.

\bibitem{Athron:2021iuf}
P.~Athron, C.~Bal\'azs, D.~H.~J. Jacob, W.~Kotlarski, D.~St\"ockinger, and H.~St\"ockinger-Kim, {\it {New physics explanations of $a_{\mu}$ in light of the FNAL muon $g-2$ measurement}},  {\em JHEP} {\bf 09} (2021) 080, [\href{http://arxiv.org/abs/2104.03691}{{\tt arXiv:2104.03691}}].

\bibitem{Misiak:1994zw}
M.~Misiak and M.~M\"unz, {\it {Two loop mixing of dimension five flavor changing operators}},  {\em Phys. Lett.} {\bf B344} (1995) 308--318, [\href{http://arxiv.org/abs/hep-ph/9409454}{{\tt hep-ph/9409454}}].

\bibitem{Gorbahn:2005sa}
M.~Gorbahn, U.~Haisch, and M.~Misiak, {\it {Three-loop mixing of dipole operators}},  {\em Phys. Rev. Lett.} {\bf 95} (2005) 102004, [\href{http://arxiv.org/abs/hep-ph/0504194}{{\tt hep-ph/0504194}}].

\bibitem{Brod:2018lbf}
J.~Brod and D.~Skodras, {\it {Electric dipole moment constraints on CP-violating light-quark Yukawas}},  {\em JHEP} {\bf 01} (2019) 233, [\href{http://arxiv.org/abs/1811.05480}{{\tt arXiv:1811.05480}}].

\bibitem{Brod:2018pli}
J.~Brod and E.~Stamou, {\it {Electric dipole moment constraints on CP-violating heavy-quark Yukawas at next-to-leading order}},  {\em JHEP} {\bf 07} (2021) 080, [\href{http://arxiv.org/abs/1810.12303}{{\tt arXiv:1810.12303}}].

\bibitem{Brod:2022bww}
J.~Brod, J.~M. Cornell, D.~Skodras, and E.~Stamou, {\it {Global constraints on Yukawa operators in the standard model effective theory}},  {\em JHEP} {\bf 08} (2022) 294, [\href{http://arxiv.org/abs/2203.03736}{{\tt arXiv:2203.03736}}].

\bibitem{Brod:2023wsh}
J.~Brod, Z.~Polonsky, and E.~Stamou, {\it {A precise electron EDM constraint on CP-odd heavy-quark Yukawas}},  {\em JHEP} {\bf 06} (2024) 091, [\href{http://arxiv.org/abs/2306.12478}{{\tt arXiv:2306.12478}}].

\bibitem{Alves:2025owr}
G.~H.~S. Alves and C.~C. Nishi, {\it {Effective description of Nelson-Barr models and the theta parameter}},  {\em JHEP} {\bf 09} (2025) 162, [\href{http://arxiv.org/abs/2506.03257}{{\tt arXiv:2506.03257}}].

\bibitem{Biancofiore:2013ki}
P.~Biancofiore, P.~Colangelo, and F.~De~Fazio, {\it {On the anomalous enhancement observed in $B \to D^{(*)}\tau{\bar \nu}_\tau$ decays}},  {\em Phys. Rev. D} {\bf 87} (2013), no.~7 074010, [\href{http://arxiv.org/abs/1302.1042}{{\tt arXiv:1302.1042}}].

\bibitem{Sakaki:2014sea}
Y.~Sakaki, M.~Tanaka, A.~Tayduganov, and R.~Watanabe, {\it {Probing New Physics with $q^2$ distributions in $\bar{B} \to D^{(*)} \tau \bar\nu$}},  {\em Phys. Rev. D} {\bf 91} (2015), no.~11 114028, [\href{http://arxiv.org/abs/1412.3761}{{\tt arXiv:1412.3761}}].

\bibitem{Duraisamy:2014sna}
M.~Duraisamy, P.~Sharma, and A.~Datta, {\it {Azimuthal $B \to D^{*} \tau^{-} \bar{\nu_\tau}$ angular distribution with tensor operators}},  {\em Phys. Rev. D} {\bf 90} (2014), no.~7 074013, [\href{http://arxiv.org/abs/1405.3719}{{\tt arXiv:1405.3719}}].

\bibitem{Freytsis:2015qca}
M.~Freytsis, Z.~Ligeti, and J.~T. Ruderman, {\it {Flavor models for $\bar{B} \to D^{(*)} \tau \bar{\nu}$}},  {\em Phys. Rev.} {\bf D92} (2015), no.~5 054018, [\href{http://arxiv.org/abs/1506.08896}{{\tt arXiv:1506.08896}}].

\bibitem{Becirevic:2016hea}
D.~Becirevic, S.~Fajfer, I.~Nisandzic, and A.~Tayduganov, {\it {Angular distributions of $\bar B \to D^{(\ast)}\ell\bar \nu_\ell$ decays and search of New Physics}},  {\em Nucl. Phys. B} {\bf 946} (2019) 114707, [\href{http://arxiv.org/abs/1602.03030}{{\tt arXiv:1602.03030}}].

\bibitem{Alonso:2016gym}
R.~Alonso, A.~Kobach, and J.~Martin~Camalich, {\it {New physics in the kinematic distributions of $\bar B\to D^{(*)}\tau^-(\to\ell^-\bar\nu_\ell\nu_\tau)\bar\nu_\tau$}},  {\em Phys. Rev. D} {\bf 94} (2016), no.~9 094021, [\href{http://arxiv.org/abs/1602.07671}{{\tt arXiv:1602.07671}}].

\bibitem{Colangelo:2016ymy}
P.~Colangelo and F.~De~Fazio, {\it {Tension in the inclusive versus exclusive determinations of $|V_{cb}|$: a possible role of new physics}},  {\em Phys. Rev.} {\bf D95} (2017), no.~1 011701, [\href{http://arxiv.org/abs/1611.07387}{{\tt arXiv:1611.07387}}].

\bibitem{Li:2016vvp}
X.-Q. Li, Y.-D. Yang, and X.~Zhang, {\it {Revisiting the one leptoquark solution to the $R(D)$ and $R(D^*)$ anomalies and its phenomenological implications}},  {\em JHEP} {\bf 08} (2016) 054, [\href{http://arxiv.org/abs/1605.09308}{{\tt arXiv:1605.09308}}].

\bibitem{Bardhan:2016uhr}
D.~Bardhan, P.~Byakti, and D.~Ghosh, {\it {A closer look at the R$_{D}$ and R$_{D^*}$ anomalies}},  {\em JHEP} {\bf 01} (2017) 125, [\href{http://arxiv.org/abs/1610.03038}{{\tt arXiv:1610.03038}}].

\bibitem{Bhattacharya:2016zcw}
S.~Bhattacharya, S.~Nandi, and S.~K. Patra, {\it {Looking for possible new physics in $B\to D^{(\ast)}\tau\nu_{\tau}$ in light of recent data}},  {\em Phys. Rev. D} {\bf 95} (2017), no.~7 075012, [\href{http://arxiv.org/abs/1611.04605}{{\tt arXiv:1611.04605}}].

\bibitem{Ivanov:2017mrj}
M.~A. Ivanov, J.~G. Körner, and C.-T. Tran, {\it {Probing new physics in $\bar{B}^0 \to D^{(\ast)} \tau^- \bar\nu_{\tau}$ using the longitudinal, transverse, and normal polarization components of the tau lepton}},  {\em Phys. Rev.} {\bf D95} (2017), no.~3 036021, [\href{http://arxiv.org/abs/1701.02937}{{\tt arXiv:1701.02937}}].

\bibitem{Chen:2017hir}
C.-H. Chen, T.~Nomura, and H.~Okada, {\it {Excesses of muon $g-2$, $R_{D^{(\ast)}}$, and $R_K$ in a leptoquark model}},  {\em Phys. Lett. B} {\bf 774} (2017) 456--464, [\href{http://arxiv.org/abs/1703.03251}{{\tt arXiv:1703.03251}}].

\bibitem{Feruglio:2018fxo}
F.~Feruglio, P.~Paradisi, and O.~Sumensari, {\it {Implications of scalar and tensor explanations of $R_{D^{(\ast)}}$}},  {\em JHEP} {\bf 11} (2018) 191, [\href{http://arxiv.org/abs/1806.10155}{{\tt arXiv:1806.10155}}].

\bibitem{Becirevic:2018afm}
D.~Becirevic, I.~Dorsner, S.~Fajfer, N.~Kosnik, D.~A. Faroughy, and O.~Sumensari, {\it {Scalar leptoquarks from grand unified theories to accommodate the $B$-physics anomalies}},  {\em Phys. Rev.} {\bf D98} (2018), no.~5 055003, [\href{http://arxiv.org/abs/1806.05689}{{\tt arXiv:1806.05689}}].

\bibitem{CMS:2022dwd}
{\bf CMS} Collaboration, A.~Tumasyan et~al., {\it {A portrait of the Higgs boson by the CMS experiment ten years after the discovery.}},  {\em Nature} {\bf 607} (2022), no.~7917 60--68, [\href{http://arxiv.org/abs/2207.00043}{{\tt arXiv:2207.00043}}]. [Erratum: Nature 623, (2023)].

\bibitem{ATLAS:2022vkf}
{\bf ATLAS} Collaboration, G.~Aad et~al., {\it {A detailed map of Higgs boson interactions by the ATLAS experiment ten years after the discovery}},  {\em Nature} {\bf 607} (2022), no.~7917 52--59, [\href{http://arxiv.org/abs/2207.00092}{{\tt arXiv:2207.00092}}]. [Erratum: Nature 612, E24 (2022)].

\bibitem{CMS:2025qmm}
{\bf CMS} Collaboration, V.~Chekhovsky et~al., {\it {Search for the associated production of a Higgs boson with a charm quark in the diphoton decay channel in pp collisions at $\sqrt{s}$ = 13 TeV}},  \href{http://arxiv.org/abs/2503.08797}{{\tt arXiv:2503.08797}}.

\bibitem{ATLAS:2024yzu}
{\bf ATLAS} Collaboration, G.~Aad et~al., {\it {Measurements of WH and ZH production with Higgs boson decays into bottom quarks and direct constraints on the charm Yukawa coupling in 13 TeV pp collisions with the ATLAS detector}},  {\em JHEP} {\bf 04} (2025) 075, [\href{http://arxiv.org/abs/2410.19611}{{\tt arXiv:2410.19611}}].

\bibitem{CMS:2025xkn}
{\bf CMS} Collaboration, V.~Chekhovsky et~al., {\it {Search for $\gamma$H production and constraints on the Yukawa couplings of light quarks to the Higgs boson}},  \href{http://arxiv.org/abs/2502.05665}{{\tt arXiv:2502.05665}}.

\bibitem{Cepeda:2019klc}
M.~Cepeda et~al., {\it {Report from Working Group 2}: {Higgs Physics at the HL-LHC and HE-LHC}},  {\em CERN Yellow Rep. Monogr.} {\bf 7} (2019) 221--584, [\href{http://arxiv.org/abs/1902.00134}{{\tt arXiv:1902.00134}}].

\bibitem{Mlynarikova:2023bvx}
{\bf ATLAS, CMS} Collaboration, M.~Mlynarikova, {\it {Higgs Physics at HL-LHC}},  in {\em {30th International Workshop on Deep-Inelastic Scattering and Related Subjects}}, 7, 2023.
\newblock \href{http://arxiv.org/abs/2307.07772}{{\tt arXiv:2307.07772}}.

\bibitem{LHCHiggsCrossSectionWorkingGroup:2012nn}
{\bf LHC Higgs Cross Section Working Group} Collaboration, A.~David, A.~Denner, M.~Duehrssen, M.~Grazzini, C.~Grojean, G.~Passarino, M.~Schumacher, M.~Spira, G.~Weiglein, and M.~Zanetti, {\it {LHC HXSWG interim recommendations to explore the coupling structure of a Higgs-like particle}},  \href{http://arxiv.org/abs/1209.0040}{{\tt arXiv:1209.0040}}.

\bibitem{Dittmaier:2012vm}
S.~Dittmaier et~al., {\it {Handbook of LHC Higgs Cross Sections: 2. Differential Distributions}},  \href{http://arxiv.org/abs/1201.3084}{{\tt arXiv:1201.3084}}.

\bibitem{Azatov:2022kbs}
A.~Azatov et~al., {\it {Off-shell Higgs Interpretations Task Force: Models and Effective Field Theories Subgroup Report}},  \href{http://arxiv.org/abs/2203.02418}{{\tt arXiv:2203.02418}}.

\bibitem{Greljo:2017vvb}
A.~Greljo and D.~Marzocca, {\it {High-$p_T$ dilepton tails and flavor physics}},  {\em Eur. Phys. J. C} {\bf 77} (2017), no.~8 548, [\href{http://arxiv.org/abs/1704.09015}{{\tt arXiv:1704.09015}}].

\bibitem{Falkowski:2015krw}
A.~Falkowski and K.~Mimouni, {\it {Model independent constraints on four-lepton operators}},  {\em JHEP} {\bf 02} (2016) 086, [\href{http://arxiv.org/abs/1511.07434}{{\tt arXiv:1511.07434}}].

\bibitem{Faroughy:2016osc}
D.~A. Faroughy, A.~Greljo, and J.~F. Kamenik, {\it {Confronting lepton flavor universality violation in B decays with high-$p_T$ tau lepton searches at LHC}},  {\em Phys. Lett.} {\bf B764} (2017) 126--134, [\href{http://arxiv.org/abs/1609.07138}{{\tt arXiv:1609.07138}}].

\bibitem{Aebischer:2022oqe}
J.~Aebischer, G.~Isidori, M.~Pesut, B.~A. Stefanek, and F.~Wilsch, {\it {Confronting the vector leptoquark hypothesis with new low- and high-energy data}},  {\em Eur. Phys. J. C} {\bf 83} (2023), no.~2 153, [\href{http://arxiv.org/abs/2210.13422}{{\tt arXiv:2210.13422}}].

\bibitem{Banta:2021dek}
I.~Banta, T.~Cohen, N.~Craig, X.~Lu, and D.~Sutherland, {\it {Non-decoupling new particles}},  {\em JHEP} {\bf 02} (2022) 029, [\href{http://arxiv.org/abs/2110.02967}{{\tt arXiv:2110.02967}}].

\bibitem{Koren:2025utp}
S.~Koren and A.~Martin, {\it {Phenomenology of Fractionally Charged Particles: Two Reps Are Better Than One}},  \href{http://arxiv.org/abs/2507.16900}{{\tt arXiv:2507.16900}}.

\bibitem{Crawford:2024nun}
G.~Crawford and D.~Sutherland, {\it {Scalars with non-decoupling phenomenology at future colliders}},  {\em JHEP} {\bf 04} (2025) 197, [\href{http://arxiv.org/abs/2409.18177}{{\tt arXiv:2409.18177}}].

\bibitem{delAguila:2000rc}
F.~del Aguila, M.~Perez-Victoria, and J.~Santiago, {\it {Observable contributions of new exotic quarks to quark mixing}},  {\em JHEP} {\bf 0009} (2000) 011, [\href{http://arxiv.org/abs/hep-ph/0007316}{{\tt hep-ph/0007316}}].

\bibitem{delAguila:2008pw}
F.~del Aguila, J.~de~Blas, and M.~Perez-Victoria, {\it {Effects of new leptons in Electroweak Precision Data}},  {\em Phys. Rev.} {\bf D78} (2008) 013010, [\href{http://arxiv.org/abs/0803.4008}{{\tt arXiv:0803.4008}}].

\bibitem{delAguila:2010mx}
F.~del Aguila, J.~de~Blas, and M.~Perez-Victoria, {\it {Electroweak Limits on General New Vector Bosons}},  {\em JHEP} {\bf 09} (2010) 033, [\href{http://arxiv.org/abs/1005.3998}{{\tt arXiv:1005.3998}}].

\bibitem{deBlas:2014mba}
J.~de~Blas, M.~Chala, M.~Perez-Victoria, and J.~Santiago, {\it {Observable Effects of General New Scalar Particles}},  {\em JHEP} {\bf 04} (2015) 078, [\href{http://arxiv.org/abs/1412.8480}{{\tt arXiv:1412.8480}}].

\bibitem{Cepedello:2024ogz}
R.~Cepedello, F.~Esser, M.~Hirsch, and V.~Sanz, {\it {Fermionic UV models for neutral triple gauge boson vertices}},  {\em JHEP} {\bf 07} (2024) 275, [\href{http://arxiv.org/abs/2402.04306}{{\tt arXiv:2402.04306}}].

\bibitem{Dawson:2024ozw}
S.~Dawson, M.~Forslund, and M.~Schnubel, {\it {SMEFT matching to Z' models at dimension eight}},  {\em Phys. Rev. D} {\bf 110} (2024), no.~1 015002, [\href{http://arxiv.org/abs/2404.01375}{{\tt arXiv:2404.01375}}].

\bibitem{Buras:2012fs}
A.~J. Buras and J.~Girrbach, {\it {Complete NLO QCD Corrections for Tree Level $\Delta F = 2$ FCNC Processes}},  {\em JHEP} {\bf 1203} (2012) 052, [\href{http://arxiv.org/abs/1201.1302}{{\tt arXiv:1201.1302}}].

\bibitem{Buras:2012gm}
A.~J. Buras and J.~Girrbach, {\it {Completing NLO QCD Corrections for Tree Level Non-Leptonic $\Delta F = 1$ Decays Beyond the Standard Model}},  {\em JHEP} {\bf 02} (2012) 143, [\href{http://arxiv.org/abs/1201.2563}{{\tt arXiv:1201.2563}}].

\bibitem{Brivio:2021alv}
I.~Brivio, S.~Bruggisser, E.~Geoffray, W.~Killian, M.~Kr\"amer, M.~Luchmann, T.~Plehn, and B.~Summ, {\it {From models to SMEFT and back?}},  {\em SciPost Phys.} {\bf 12} (2022), no.~1 036, [\href{http://arxiv.org/abs/2108.01094}{{\tt arXiv:2108.01094}}].

\bibitem{Henning:2014gca}
B.~Henning, X.~Lu, and H.~Murayama, {\it {What do precision Higgs measurements buy us?}},  \href{http://arxiv.org/abs/1404.1058}{{\tt arXiv:1404.1058}}.

\bibitem{Ellis:2023zim}
J.~Ellis, K.~Mimasu, and F.~Zampedri, {\it {Dimension-8 SMEFT analysis of minimal scalar field extensions of the Standard Model}},  {\em JHEP} {\bf 10} (2023) 051, [\href{http://arxiv.org/abs/2304.06663}{{\tt arXiv:2304.06663}}].

\bibitem{Dawson:2022cmu}
S.~Dawson, D.~Fontes, S.~Homiller, and M.~Sullivan, {\it {Role of dimension-eight operators in an EFT for the 2HDM}},  {\em Phys. Rev. D} {\bf 106} (2022), no.~5 055012, [\href{http://arxiv.org/abs/2205.01561}{{\tt arXiv:2205.01561}}].

\bibitem{Dawson:2023ebe}
S.~Dawson, D.~Fontes, C.~Quezada-Calonge, and J.~J. Sanz-Cillero, {\it {Matching the 2HDM to the HEFT and the SMEFT: Decoupling and perturbativity}},  {\em Phys. Rev. D} {\bf 108} (2023), no.~5 055034, [\href{http://arxiv.org/abs/2305.07689}{{\tt arXiv:2305.07689}}].

\bibitem{Haisch:2020ahr}
U.~Haisch, M.~Ruhdorfer, E.~Salvioni, E.~Venturini, and A.~Weiler, {\it {Singlet night in Feynman-ville: one-loop matching of a real scalar}},  {\em JHEP} {\bf 04} (2020) 164, [\href{http://arxiv.org/abs/2003.05936}{{\tt arXiv:2003.05936}}]. [Erratum: JHEP 07, 066 (2020)].

\bibitem{Jiang:2018pbd}
M.~Jiang, N.~Craig, Y.-Y. Li, and D.~Sutherland, {\it {Complete one-loop matching for a singlet scalar in the Standard Model EFT}},  {\em JHEP} {\bf 02} (2019) 031, [\href{http://arxiv.org/abs/1811.08878}{{\tt arXiv:1811.08878}}]. [Erratum: JHEP 01, 135 (2021)].

\bibitem{DasBakshi:2024krs}
S.~Das~Bakshi, S.~Dawson, D.~Fontes, and S.~Homiller, {\it {Relevance of one-loop SMEFT matching in the 2HDM}},  {\em Phys. Rev. D} {\bf 109} (2024), no.~7 075022, [\href{http://arxiv.org/abs/2401.12279}{{\tt arXiv:2401.12279}}].

\bibitem{Dekens:2018bci}
W.~Dekens, J.~de~Vries, M.~Jung, and K.~K. Vos, {\it {The phenomenology of electric dipole moments in models of scalar leptoquarks}},  {\em JHEP} {\bf 01} (2019) 069, [\href{http://arxiv.org/abs/1809.09114}{{\tt arXiv:1809.09114}}].

\bibitem{Jiang:2016czg}
Y.~Jiang and M.~Trott, {\it {On the non-minimal character of the SMEFT}},  {\em Phys. Lett. B} {\bf 770} (2017) 108--116, [\href{http://arxiv.org/abs/1612.02040}{{\tt arXiv:1612.02040}}].

\bibitem{Banta:2023prj}
I.~Banta, T.~Cohen, N.~Craig, X.~Lu, and D.~Sutherland, {\it {Effective field theory of the two Higgs doublet model}},  {\em JHEP} {\bf 06} (2023) 150, [\href{http://arxiv.org/abs/2304.09884}{{\tt arXiv:2304.09884}}].

\bibitem{Davila:2025goc}
J.~M. D{\'a}vila, A.~Karan, E.~Passemar, A.~Pich, and L.~Vale~Silva, {\it {The Electric Dipole Moment of the electron in the decoupling limit of the aligned two-Higgs doublet model}},  {\em JHEP} {\bf 10} (2025) 053, [\href{http://arxiv.org/abs/2504.16700}{{\tt arXiv:2504.16700}}].

\bibitem{Kraml:2025fpv}
S.~Kraml, A.~Lessa, S.~Prakash, and F.~Wilsch, {\it {SUSY meets SMEFT: Complete one-loop matching of the general MSSM}},  \href{http://arxiv.org/abs/2506.05201}{{\tt arXiv:2506.05201}}.

\bibitem{Li:2023cwy}
X.-X. Li, Z.~Ren, and J.-H. Yub, {\it {Complete tree-level dictionary between simplified BSM models and SMEFT d\ensuremath{\leq}7 operators}},  {\em Phys. Rev. D} {\bf 109} (2024), no.~9 095041, [\href{http://arxiv.org/abs/2307.10380}{{\tt arXiv:2307.10380}}].

\bibitem{Li:2022abx}
H.-L. Li, Y.-H. Ni, M.-L. Xiao, and J.-H. Yu, {\it {The bottom-up EFT: complete UV resonances of the SMEFT operators}},  {\em JHEP} {\bf 11} (2022) 170, [\href{http://arxiv.org/abs/2204.03660}{{\tt arXiv:2204.03660}}].

\bibitem{Li:2023pfw}
H.-L. Li, Y.-H. Ni, M.-L. Xiao, and J.-H. Yu, {\it {Complete UV resonances of the dimension-8 SMEFT operators}},  {\em JHEP} {\bf 05} (2024) 238, [\href{http://arxiv.org/abs/2309.15933}{{\tt arXiv:2309.15933}}].

\bibitem{Zhang:2021jdf}
D.~Zhang and S.~Zhou, {\it {Complete one-loop matching of the type-I seesaw model onto the Standard Model effective field theory}},  {\em JHEP} {\bf 09} (2021) 163, [\href{http://arxiv.org/abs/2107.12133}{{\tt arXiv:2107.12133}}].

\bibitem{Li:2022ipc}
X.~Li, D.~Zhang, and S.~Zhou, {\it {One-loop matching of the type-II seesaw model onto the Standard Model effective field theory}},  {\em JHEP} {\bf 04} (2022) 038, [\href{http://arxiv.org/abs/2201.05082}{{\tt arXiv:2201.05082}}].

\bibitem{Li:2023ohq}
X.~Li and S.~Zhou, {\it {One-loop matching of the type-III seesaw model onto the Standard Model Effective Field Theory}},  {\em JHEP} {\bf 05} (2024) 169, [\href{http://arxiv.org/abs/2309.14702}{{\tt arXiv:2309.14702}}].

\bibitem{Banerjee:2023xak}
U.~Banerjee, J.~Chakrabortty, S.~U. Rahaman, and K.~Ramkumar, {\it {One-loop effective action up to any mass-dimension for non-degenerate scalars and fermions including light\textendash{}heavy mixing}},  {\em Eur. Phys. J. Plus} {\bf 139} (2024), no.~2 169, [\href{http://arxiv.org/abs/2311.12757}{{\tt arXiv:2311.12757}}].

\bibitem{Loisa:2024xuk}
E.~Loisa and J.~Talbert, {\it {Froggatt-Nielsen meets the SMEFT}},  {\em JHEP} {\bf 10} (2024) 017, [\href{http://arxiv.org/abs/2402.16940}{{\tt arXiv:2402.16940}}].

\bibitem{DasBakshi:2021xbl}
S.~Das~Bakshi, J.~Chakrabortty, S.~Prakash, S.~U. Rahaman, and M.~Spannowsky, {\it {EFT diagrammatica: UV roots of the CP-conserving SMEFT}},  {\em JHEP} {\bf 06} (2021) 033, [\href{http://arxiv.org/abs/2103.11593}{{\tt arXiv:2103.11593}}].

\bibitem{Naskar:2022rpg}
W.~Naskar, S.~Prakash, and S.~U. Rahaman, {\it {EFT Diagrammatica. Part II. Tracing the UV origin of bosonic D6 CPV and D8 SMEFT operators}},  {\em JHEP} {\bf 08} (2022) 190, [\href{http://arxiv.org/abs/2205.00910}{{\tt arXiv:2205.00910}}].

\bibitem{Cepedello:2022pyx}
R.~Cepedello, F.~Esser, M.~Hirsch, and V.~Sanz, {\it {Mapping the SMEFT to discoverable models}},  {\em JHEP} {\bf 09} (2022) 229, [\href{http://arxiv.org/abs/2207.13714}{{\tt arXiv:2207.13714}}].

\bibitem{Chakrabortty:2020mbc}
J.~Chakrabortty, S.~Prakash, S.~U. Rahaman, and M.~Spannowsky, {\it {Uncovering the root of LEFT in SMEFT}},  {\em EPL} {\bf 136} (2021), no.~1 11002, [\href{http://arxiv.org/abs/2011.00859}{{\tt arXiv:2011.00859}}].

\bibitem{Li:2024nuo}
H.-L. Li and L.-X. Xu, {\it {The Standard Model Gauge Group, SMEFT, and Generalized Symmetries}},  \href{http://arxiv.org/abs/2404.04229}{{\tt arXiv:2404.04229}}.

\bibitem{Gargalionis:2024jaw}
J.~Gargalionis, J.~Quevillon, P.~N.~H. Vuong, and T.~You, {\it {Linear Standard Model extensions in the SMEFT at one loop and Tera-Z}},  {\em JHEP} {\bf 07} (2025) 136, [\href{http://arxiv.org/abs/2412.01759}{{\tt arXiv:2412.01759}}].

\bibitem{Adhikary:2025pbb}
N.~Adhikary, J.~Das, and D.~Dey, {\it {Two-loop dimension Six Effective Action: Integrating Out Heavy Scalar}},  \href{http://arxiv.org/abs/2501.01313}{{\tt arXiv:2501.01313}}.

\bibitem{Palavric:2024gvu}
A.~Palavri\'c, {\it {Discrete leptonic flavor symmetries: UV mediators and phenomenology}},  {\em Phys. Rev. D} {\bf 110} (2024), no.~11 115025, [\href{http://arxiv.org/abs/2408.16044}{{\tt arXiv:2408.16044}}].

\bibitem{Fridell:2024pmw}
K.~Fridell, L.~Gr{\'a}f, J.~Harz, and C.~Hati, {\it {Radiative neutrino masses from dim-7 SMEFT: a simplified multi-scale approach}},  {\em JHEP} {\bf 09} (2025) 050, [\href{http://arxiv.org/abs/2412.14268}{{\tt arXiv:2412.14268}}].

\bibitem{Buras:2012dp}
A.~J. Buras, F.~De~Fazio, J.~Girrbach, and M.~V. Carlucci, {\it {The Anatomy of Quark Flavour Observables in 331 Models in the Flavour Precision Era}},  {\em JHEP} {\bf 1302} (2013) 023, [\href{http://arxiv.org/abs/1211.1237}{{\tt arXiv:1211.1237}}].

\bibitem{Buras:2013dea}
A.~J. Buras, F.~De~Fazio, and J.~Girrbach, {\it {331 models facing new $b \to s\mu^+ \mu^-$ data}},  {\em JHEP} {\bf 1402} (2014) 112, [\href{http://arxiv.org/abs/1311.6729}{{\tt arXiv:1311.6729}}].

\bibitem{Buras:2015kwd}
A.~J. Buras and F.~De~Fazio, {\it {$\varepsilon'/\varepsilon$ in 331 Models}},  {\em JHEP} {\bf 03} (2016) 010, [\href{http://arxiv.org/abs/1512.02869}{{\tt arXiv:1512.02869}}].

\bibitem{Buras:2016dxz}
A.~J. Buras and F.~De~Fazio, {\it {331 Models Facing the Tensions in $\Delta F=2$ Processes with the Impact on $\varepsilon^\prime/\varepsilon$, $B_s\to\mu^+\mu^-$ and $B\to K^*\mu^+\mu^-$}},  {\em JHEP} {\bf 08} (2016) 115, [\href{http://arxiv.org/abs/1604.02344}{{\tt arXiv:1604.02344}}].

\bibitem{Buras:2021rdg}
A.~J. Buras, P.~Colangelo, F.~De~Fazio, and F.~Loparco, {\it {The charm of 331}},  {\em JHEP} {\bf 10} (2021) 021, [\href{http://arxiv.org/abs/2107.10866}{{\tt arXiv:2107.10866}}].

\bibitem{Buras:2023ldz}
A.~J. Buras and F.~De~Fazio, {\it {331 model predictions for rare B and K decays, and \ensuremath{\Delta}F = 2 processes: an update}},  {\em JHEP} {\bf 03} (2023) 219, [\href{http://arxiv.org/abs/2301.02649}{{\tt arXiv:2301.02649}}].

\bibitem{Escalona:2025rxu}
P.~Escalona, J.~P. Pinheiro, A.~Doff, and C.~A. de~S.~Pires, {\it {Meson mixing bounds on Z$^\prime$ mass in the alignment limit: establishing the phenomenological viability of the 331 model}},  {\em JHEP} {\bf 07} (2025) 105, [\href{http://arxiv.org/abs/2503.14653}{{\tt arXiv:2503.14653}}].

\bibitem{Buras:2012jb}
A.~J. Buras, F.~De~Fazio, and J.~Girrbach, {\it {The Anatomy of Z' and Z with Flavour Changing Neutral Currents in the Flavour Precision Era}},  {\em JHEP} {\bf 1302} (2013) 116, [\href{http://arxiv.org/abs/1211.1896}{{\tt arXiv:1211.1896}}].

\bibitem{Buras:2013qja}
A.~J. Buras and J.~Girrbach, {\it {Left-handed Z' and Z FCNC quark couplings facing new $b \to s \mu^+ \mu^-$ data}},  {\em JHEP} {\bf 1312} (2013) 009, [\href{http://arxiv.org/abs/1309.2466}{{\tt arXiv:1309.2466}}].

\bibitem{Aebischer:2022vky}
J.~Aebischer, A.~J. Buras, and J.~Kumar, {\it {On the Importance of Rare Kaon Decays: A Snowmass 2021 White Paper}},  in {\em {2022 Snowmass Summer Study}}, 3, 2022.
\newblock \href{http://arxiv.org/abs/2203.09524}{{\tt arXiv:2203.09524}}.

\bibitem{Colangelo:2024sbf}
P.~Colangelo, F.~De~Fazio, F.~Loparco, and N.~Losacco, {\it {Constraining \ensuremath{\nu}SMEFT coefficients: The case of the extra U(1)'}},  {\em Phys. Rev. D} {\bf 110} (2024), no.~3 035007, [\href{http://arxiv.org/abs/2406.07059}{{\tt arXiv:2406.07059}}].

\bibitem{Ishiwata:2015cga}
K.~Ishiwata, Z.~Ligeti, and M.~B. Wise, {\it {New Vector-Like Fermions and Flavor Physics}},  {\em JHEP} {\bf 10} (2015) 027, [\href{http://arxiv.org/abs/1506.03484}{{\tt arXiv:1506.03484}}].

\bibitem{Nir:1990yq}
Y.~Nir and D.~J. Silverman, {\it {$Z$ Mediated Flavor Changing Neutral Currents and Their Implications for {CP} Asymmetries in $B^0$ Decays}},  {\em Phys. Rev.} {\bf D42} (1990) 1477--1484.

\bibitem{Branco:1992wr}
G.~C. Branco, T.~Morozumi, P.~A. Parada, and M.~N. Rebelo, {\it {CP asymmetries in $B^0$ decays in the presence of flavor changing neutral currents}},  {\em Phys. Rev.} {\bf D48} (1993) 1167--1175.

\bibitem{Barenboim:2001fd}
G.~Barenboim, F.~J. Botella, and O.~Vives, {\it {Constraining models with vector - like fermions from FCNC in $K$ and $B$ physics}},  {\em Nucl. Phys.} {\bf B613} (2001) 285--305, [\href{http://arxiv.org/abs/hep-ph/0105306}{{\tt hep-ph/0105306}}].

\bibitem{Buras:2009ka}
A.~J. Buras, B.~Duling, and S.~Gori, {\it {The Impact of Kaluza-Klein Fermions on Standard Model Fermion Couplings in a RS Model with Custodial Protection}},  {\em JHEP} {\bf 0909} (2009) 076, [\href{http://arxiv.org/abs/0905.2318}{{\tt arXiv:0905.2318}}].

\bibitem{Botella:2012ju}
F.~Botella, G.~Branco, and M.~Nebot, {\it {The Hunt for New Physics in the Flavour Sector with up vector-like quarks}},  {\em JHEP} {\bf 1212} (2012) 040, [\href{http://arxiv.org/abs/1207.4440}{{\tt arXiv:1207.4440}}].

\bibitem{Fajfer:2013wca}
S.~Fajfer, A.~Greljo, J.~F. Kamenik, and I.~Mustac, {\it {Light Higgs and Vector-like Quarks without Prejudice}},  {\em JHEP} {\bf 07} (2013) 155, [\href{http://arxiv.org/abs/1304.4219}{{\tt arXiv:1304.4219}}].

\bibitem{Buras:2013td}
A.~J. Buras, J.~Girrbach, and R.~Ziegler, {\it {Particle-Antiparticle Mixing, CP Violation and Rare K and B Decays in a Minimal Theory of Fermion Masses}},  {\em JHEP} {\bf 1304} (2013) 168, [\href{http://arxiv.org/abs/1301.5498}{{\tt arXiv:1301.5498}}].

\bibitem{Altmannshofer:2014cfa}
W.~Altmannshofer, S.~Gori, M.~Pospelov, and I.~Yavin, {\it {Quark flavor transitions in $L_\mu-L_\tau$ models}},  {\em Phys. Rev.} {\bf D89} (2014) 095033, [\href{http://arxiv.org/abs/1403.1269}{{\tt arXiv:1403.1269}}].

\bibitem{Alok:2015iha}
A.~K. Alok, S.~Banerjee, D.~Kumar, S.~U. Sankar, and D.~London, {\it {New-physics signals of a model with a vector-singlet up-type quark}},  {\em Phys. Rev.} {\bf D92} (2015) 013002, [\href{http://arxiv.org/abs/1504.00517}{{\tt arXiv:1504.00517}}].

\bibitem{Arnan:2016cpy}
P.~Arnan, L.~Hofer, F.~Mescia, and A.~Crivellin, {\it {Loop effects of heavy new scalars and fermions in $b\to s\mu^+\mu^-$}},  {\em JHEP} {\bf 04} (2017) 043, [\href{http://arxiv.org/abs/1608.07832}{{\tt arXiv:1608.07832}}].

\bibitem{Cepedello:2024qmq}
R.~Cepedello, F.~Esser, M.~Hirsch, and V.~Sanz, {\it {Faking ZZZ vertices at the LHC}},  {\em JHEP} {\bf 12} (2024) 098, [\href{http://arxiv.org/abs/2409.06776}{{\tt arXiv:2409.06776}}].

\bibitem{Erdelyi:2024sls}
B.~A. Erdelyi, R.~Gr\"ober, and N.~Selimovic, {\it {How large can the light quark Yukawa couplings be?}},  {\em JHEP} {\bf 05} (2025) 189, [\href{http://arxiv.org/abs/2410.08272}{{\tt arXiv:2410.08272}}].

\bibitem{Crivellin:2022ctt}
A.~Crivellin, {\it {Explaining the Cabibbo Angle Anomaly}},  7, 2022.
\newblock \href{http://arxiv.org/abs/2207.02507}{{\tt arXiv:2207.02507}}.

\bibitem{Belfatto:2019swo}
B.~Belfatto, R.~Beradze, and Z.~Berezhiani, {\it {The CKM unitarity problem: A trace of new physics at the TeV scale?}},  {\em Eur. Phys. J. C} {\bf 80} (2020), no.~2 149, [\href{http://arxiv.org/abs/1906.02714}{{\tt arXiv:1906.02714}}].

\bibitem{Belfatto:2021jhf}
B.~Belfatto and Z.~Berezhiani, {\it {Are the CKM anomalies induced by vector-like quarks? Limits from flavor changing and Standard Model precision tests}},  {\em JHEP} {\bf 10} (2021) 079, [\href{http://arxiv.org/abs/2103.05549}{{\tt arXiv:2103.05549}}].

\bibitem{Botella:2021uxz}
F.~J. Botella, G.~C. Branco, M.~N. Rebelo, J.~I. Silva-Marcos, and J.~F. Bastos, {\it {Decays of the heavy top and new insights on $\epsilon _K$ in a one-VLQ minimal solution to the CKM unitarity problem}},  {\em Eur. Phys. J. C} {\bf 82} (2022), no.~4 360, [\href{http://arxiv.org/abs/2111.15401}{{\tt arXiv:2111.15401}}].

\bibitem{Crivellin:2022rhw}
A.~Crivellin, M.~Kirk, T.~Kitahara, and F.~Mescia, {\it {Global fit of modified quark couplings to EW gauge bosons and vector-like quarks in light of the Cabibbo angle anomaly}},  {\em JHEP} {\bf 03} (2023) 234, [\href{http://arxiv.org/abs/2212.06862}{{\tt arXiv:2212.06862}}].

\bibitem{Alves:2023ufm}
J.~a.~M. Alves, G.~C. Branco, A.~L. Cherchiglia, C.~C. Nishi, J.~T. Penedo, P.~M.~F. Pereira, M.~N. Rebelo, and J.~I. Silva-Marcos, {\it {Vector-like singlet quarks: A roadmap}},  {\em Phys. Rept.} {\bf 1057} (2024) 1--69, [\href{http://arxiv.org/abs/2304.10561}{{\tt arXiv:2304.10561}}].

\bibitem{Dermisek:2013gta}
R.~Dermisek and A.~Raval, {\it {Explanation of the Muon g-2 Anomaly with Vectorlike Leptons and its Implications for Higgs Decays}},  {\em Phys. Rev.} {\bf D88} (2013) 013017, [\href{http://arxiv.org/abs/1305.3522}{{\tt arXiv:1305.3522}}].

\bibitem{Sierra:2015fma}
D.~Aristizabal~Sierra, F.~Staub, and A.~Vicente, {\it {Shedding light on the $b\to s$ anomalies with a dark sector}},  {\em Phys. Rev.} {\bf D92} (2015), no.~1 015001, [\href{http://arxiv.org/abs/1503.06077}{{\tt arXiv:1503.06077}}].

\bibitem{Belanger:2015nma}
G.~Bélanger, C.~Delaunay, and S.~Westhoff, {\it {A Dark Matter Relic From Muon Anomalies}},  {\em Phys. Rev.} {\bf D92} (2015) 055021, [\href{http://arxiv.org/abs/1507.06660}{{\tt arXiv:1507.06660}}].

\bibitem{Altmannshofer:2016oaq}
W.~Altmannshofer, M.~Carena, and A.~Crivellin, {\it {$L_\mu - L_\tau$ theory of Higgs flavor violation and $(g-2)_\mu$}},  {\em Phys. Rev.} {\bf D94} (2016), no.~9 095026, [\href{http://arxiv.org/abs/1604.08221}{{\tt arXiv:1604.08221}}].

\bibitem{Kowalska:2017iqv}
K.~Kowalska and E.~M. Sessolo, {\it {Expectations for the muon g-2 in simplified models with dark matter}},  {\em JHEP} {\bf 09} (2017) 112, [\href{http://arxiv.org/abs/1707.00753}{{\tt arXiv:1707.00753}}].

\bibitem{Darme:2018hqg}
L.~Darmé, K.~Kowalska, L.~Roszkowski, and E.~M. Sessolo, {\it {Flavor anomalies and dark matter in SUSY with an extra U(1)}},  {\em JHEP} {\bf 10} (2018) 052, [\href{http://arxiv.org/abs/1806.06036}{{\tt arXiv:1806.06036}}].

\bibitem{Crivellin:2020ebi}
A.~Crivellin, F.~Kirk, C.~A. Manzari, and M.~Montull, {\it {Global Electroweak Fit and Vector-Like Leptons in Light of the Cabibbo Angle Anomaly}},  {\em JHEP} {\bf 12} (2020) 166, [\href{http://arxiv.org/abs/2008.01113}{{\tt arXiv:2008.01113}}].

\bibitem{Falkowski:2013jya}
A.~Falkowski, D.~M. Straub, and A.~Vicente, {\it {Vector-like leptons: Higgs decays and collider phenomenology}},  {\em JHEP} {\bf 05} (2014) 092, [\href{http://arxiv.org/abs/1312.5329}{{\tt arXiv:1312.5329}}].

\bibitem{Bissmann:2020lge}
S.~Bi\ss{}mann, G.~Hiller, C.~Hormigos-Feliu, and D.~F. Litim, {\it {Multi-lepton signatures of vector-like leptons with flavor}},  {\em Eur. Phys. J. C} {\bf 81} (2021), no.~2 101, [\href{http://arxiv.org/abs/2011.12964}{{\tt arXiv:2011.12964}}].

\bibitem{Erdelyi:2025axy}
B.~A. Erdelyi, R.~Gr{\"o}ber, and N.~Selimovic, {\it {Probing new physics with the electron Yukawa coupling}},  {\em JHEP} {\bf 05} (2025) 135, [\href{http://arxiv.org/abs/2501.07628}{{\tt arXiv:2501.07628}}].

\bibitem{Kawamura:2019rth}
J.~Kawamura, S.~Raby, and A.~Trautner, {\it {Complete vectorlike fourth family and new $U(1)^\prime$ for muon anomalies}},  {\em Phys. Rev.} {\bf D100} (2019), no.~5 055030, [\href{http://arxiv.org/abs/1906.11297}{{\tt arXiv:1906.11297}}].

\bibitem{Buchmuller:1986zs}
W.~Buchm{\"u}ller, R.~R{\"u}ckl, and D.~Wyler, {\it {Leptoquarks in Lepton - Quark Collisions}},  {\em Phys. Lett.} {\bf B191} (1987) 442--448. [Erratum: Phys. Lett.B448,320(1999)].

\bibitem{Davies:1990sc}
A.~J. Davies and X.-G. He, {\it {Tree Level Scalar Fermion Interactions Consistent With the Symmetries of the Standard Model}},  {\em Phys. Rev.} {\bf D43} (1991) 225--235.

\bibitem{Davidson:1993qk}
S.~Davidson, D.~C. Bailey, and B.~A. Campbell, {\it {Model independent constraints on leptoquarks from rare processes}},  {\em Z. Phys.} {\bf C61} (1994) 613--644, [\href{http://arxiv.org/abs/hep-ph/9309310}{{\tt hep-ph/9309310}}].

\bibitem{Dorsner:2016wpm}
I.~Dorsner, S.~Fajfer, A.~Greljo, J.~F. Kamenik, and N.~Kosnik, {\it {Physics of leptoquarks in precision experiments and at particle colliders}},  {\em Phys. Rept.} {\bf 641} (2016) 1--68, [\href{http://arxiv.org/abs/1603.04993}{{\tt arXiv:1603.04993}}].

\bibitem{Gherardi:2020qhc}
V.~Gherardi, D.~Marzocca, and E.~Venturini, {\it {Low-energy phenomenology of scalar leptoquarks at one-loop accuracy}},  {\em JHEP} {\bf 01} (2021) 138, [\href{http://arxiv.org/abs/2008.09548}{{\tt arXiv:2008.09548}}].

\bibitem{Marzocca:2021miv}
D.~Marzocca, S.~Trifinopoulos, and E.~Venturini, {\it {From B-meson anomalies to Kaon physics with scalar leptoquarks}},  {\em Eur. Phys. J. C} {\bf 82} (2022), no.~4 320, [\href{http://arxiv.org/abs/2106.15630}{{\tt arXiv:2106.15630}}].

\bibitem{Cornella:2021sby}
C.~Cornella, D.~A. Faroughy, J.~Fuentes-Martin, G.~Isidori, and M.~Neubert, {\it {Reading the footprints of the B-meson flavor anomalies}},  {\em JHEP} {\bf 08} (2021) 050, [\href{http://arxiv.org/abs/2103.16558}{{\tt arXiv:2103.16558}}].

\bibitem{Fleischer:2023zeo}
R.~Fleischer, E.~Malami, A.~Rehult, and K.~K. Vos, {\it {New perspectives for testing electron-muon universality}},  {\em JHEP} {\bf 06} (2023) 033, [\href{http://arxiv.org/abs/2303.08764}{{\tt arXiv:2303.08764}}].

\bibitem{Buras:2013ooa}
A.~J. Buras and J.~Girrbach, {\it {Towards the Identification of New Physics through Quark Flavour Violating Processes}},  {\em Rept.~Prog.~Phys.} {\bf 77} (2014) 086201, [\href{http://arxiv.org/abs/1306.3775}{{\tt arXiv:1306.3775}}].

\bibitem{LHCb:2018roe}
{\bf LHCb} Collaboration, R.~Aaij et~al., {\it {Physics case for an LHCb Upgrade II - Opportunities in flavour physics, and beyond, in the HL-LHC era}},  \href{http://arxiv.org/abs/1808.08865}{{\tt arXiv:1808.08865}}.

\bibitem{Belle-II:2018jsg}
{\bf Belle-II} Collaboration, W.~Altmannshofer et~al., {\it {The Belle II Physics Book}},  {\em PTEP} {\bf 2019} (2019), no.~12 123C01, [\href{http://arxiv.org/abs/1808.10567}{{\tt arXiv:1808.10567}}]. [Erratum: PTEP 2020, 029201 (2020)].

\bibitem{NA62KLEVER:2022nea}
{\bf NA62/KLEVER, US Kaon Interest Group, KOTO, LHCb} Collaboration, {\it {Searches for new physics with high-intensity kaon beams}},  in {\em {Snowmass 2021}}, 4, 2022.
\newblock \href{http://arxiv.org/abs/2204.13394}{{\tt arXiv:2204.13394}}.

\bibitem{Goudzovski:2022scl}
E.~Goudzovski et~al., {\it {Weak Decays of Strange and Light Quarks}},  \href{http://arxiv.org/abs/2209.07156}{{\tt arXiv:2209.07156}}.

\bibitem{Aebischer:2025mwl}
J.~Aebischer et~al., {\it {Kaon Physics: A Cornerstone for Future Discoveries}},  \href{http://arxiv.org/abs/2503.22256}{{\tt arXiv:2503.22256}}.

\bibitem{BESIII:2020nme}
{\bf BESIII} Collaboration, M.~Ablikim et~al., {\it {Future Physics Programme of BESIII}},  {\em Chin. Phys. C} {\bf 44} (2020), no.~4 040001, [\href{http://arxiv.org/abs/1912.05983}{{\tt arXiv:1912.05983}}].

\bibitem{Lyu:2021tlb}
{\bf STCF Working Group} Collaboration, X.-R. Lyu, {\it {Physics Program of the Super Tau-Charm Factory}},  {\em PoS} {\bf BEAUTY2020} (2021) 060.

\bibitem{Bernstein:2019fyh}
{\bf Mu2e} Collaboration, R.~H. Bernstein, {\it {The Mu2e Experiment}},  {\em Front. in Phys.} {\bf 7} (2019) 1, [\href{http://arxiv.org/abs/1901.11099}{{\tt arXiv:1901.11099}}].

\bibitem{Hesketh:2022wgw}
{\bf Mu3e} Collaboration, G.~Hesketh, S.~Hughes, A.-K. Perrevoort, and N.~Rompotis, {\it {The Mu3e Experiment}},  in {\em {Snowmass 2021}}, 4, 2022.
\newblock \href{http://arxiv.org/abs/2204.00001}{{\tt arXiv:2204.00001}}.

\bibitem{MEGII:2023ltw}
{\bf MEG II} Collaboration, K.~Afanaciev et~al., {\it {A search for $\mu^+\to e^+\gamma$ with the first dataset of the MEG II experiment}},  \href{http://arxiv.org/abs/2310.12614}{{\tt arXiv:2310.12614}}.

\bibitem{n2EDM:2021yah}
{\bf n2EDM} Collaboration, N.~J. Ayres et~al., {\it {The design of the n2EDM experiment: nEDM Collaboration}},  {\em Eur. Phys. J. C} {\bf 81} (2021), no.~6 512, [\href{http://arxiv.org/abs/2101.08730}{{\tt arXiv:2101.08730}}].

\bibitem{Alarcon:2022ero}
R.~Alarcon et~al., {\it {Electric dipole moments and the search for new physics}},  in {\em {Snowmass 2021}}, 3, 2022.
\newblock \href{http://arxiv.org/abs/2203.08103}{{\tt arXiv:2203.08103}}.

\bibitem{Athanasakis-Kaklamanakis:2025xcg}
M.~Athanasakis-Kaklamanakis et~al., {\it {Community input to the European Strategy on particle physics: Searches for Permanent Electric Dipole Moments}},  \href{http://arxiv.org/abs/2505.22281}{{\tt arXiv:2505.22281}}.

\bibitem{FCC:2025lpp}
{\bf FCC} Collaboration, M.~Benedikt et~al., {\it {Future Circular Collider Feasibility Study Report: Volume 1, Physics, Experiments, Detectors}},  \href{http://arxiv.org/abs/2505.00272}{{\tt arXiv:2505.00272}}.

\bibitem{Cerri:2018ypt}
A.~Cerri et~al., {\it {Report from Working Group 4}: {Opportunities in Flavour Physics at the HL-LHC and HE-LHC}},  {\em CERN Yellow Rep. Monogr.} {\bf 7} (2019) 867--1158, [\href{http://arxiv.org/abs/1812.07638}{{\tt arXiv:1812.07638}}].

\bibitem{CidVidal:2018eel}
X.~Cid~Vidal et~al., {\it {Beyond the Standard Model Physics at the HL-LHC and HE-LHC}},  \href{http://arxiv.org/abs/1812.07831}{{\tt arXiv:1812.07831}}.

\bibitem{Helset:2018dht}
A.~Helset and M.~Trott, {\it {Equations of motion, symmetry currents and EFT below the electroweak scale}},  {\em Phys. Lett. B} {\bf 795} (2019) 606--619, [\href{http://arxiv.org/abs/1812.02991}{{\tt arXiv:1812.02991}}].

\bibitem{Chala:2020wvs}
M.~Chala, G.~Guedes, M.~Ramos, and J.~Santiago, {\it {Running in the ALPs}},  {\em Eur. Phys. J. C} {\bf 81} (2021), no.~2 181, [\href{http://arxiv.org/abs/2012.09017}{{\tt arXiv:2012.09017}}].

\bibitem{DasBakshi:2023lca}
S.~Das~Bakshi, J.~Machado-Rodr\'\i{}guez, and M.~Ramos, {\it {Running beyond ALPs: shift-breaking and CP-violating effects}},  {\em JHEP} {\bf 11} (2023) 133, [\href{http://arxiv.org/abs/2306.08036}{{\tt arXiv:2306.08036}}].

\bibitem{Bresciani:2024shu}
L.~C. Bresciani, G.~Brunello, G.~Levati, P.~Mastrolia, and P.~Paradisi, {\it {Renormalization of effective field theories via on-shell methods: the case of axion-like particles}},  \href{http://arxiv.org/abs/2412.04160}{{\tt arXiv:2412.04160}}.

\bibitem{Bauer:2020jbp}
M.~Bauer, M.~Neubert, S.~Renner, M.~Schnubel, and A.~Thamm, {\it {The Low-Energy Effective Theory of Axions and ALPs}},  {\em JHEP} {\bf 04} (2021) 063, [\href{http://arxiv.org/abs/2012.12272}{{\tt arXiv:2012.12272}}].

\bibitem{Fan:2025xhi}
W.-Q. Fan, Y.~Liao, X.-D. Ma, and H.-L. Wang, {\it {Comprehensive investigation on baryon number violating nucleon decays involving an axion-like particle}},  \href{http://arxiv.org/abs/2507.11844}{{\tt arXiv:2507.11844}}.

\bibitem{Bonilla:2021ufe}
J.~Bonilla, I.~Brivio, M.~B. Gavela, and V.~Sanz, {\it {One-loop corrections to ALP couplings}},  {\em JHEP} {\bf 11} (2021) 168, [\href{http://arxiv.org/abs/2107.11392}{{\tt arXiv:2107.11392}}].

\bibitem{Biekotter:2025fll}
A.~Biek{\"o}tter and K.~Mimasu, {\em {Axions and Axion-like particles: collider searches}}.
\newblock 8, 2025.
\newblock \href{http://arxiv.org/abs/2508.19358}{{\tt arXiv:2508.19358}}.

\bibitem{Galda:2025tqs}
A.~M. Galda and M.~Neubert, {\it {Saving or Destroying the Universe with Axion-Like Particles}},  \href{http://arxiv.org/abs/2506.06426}{{\tt arXiv:2506.06426}}.

\bibitem{Roy:2025pht}
A.~Roy, M.~A. Schmidt, and G.~Valencia, {\it {Monojet and direct detection constraints on real scalar dark matter: EFT and a simple UV completion}},  \href{http://arxiv.org/abs/2509.14869}{{\tt arXiv:2509.14869}}.

\bibitem{Crivellin:2016ihg}
A.~Crivellin, M.~Ghezzi, and M.~Procura, {\it {Effective Field Theory with Two Higgs Doublets}},  {\em JHEP} {\bf 09} (2016) 160, [\href{http://arxiv.org/abs/1608.00975}{{\tt arXiv:1608.00975}}].

\bibitem{Karmakar:2017yek}
S.~Karmakar and S.~Rakshit, {\it {Higher dimensional operators in 2HDM}},  {\em JHEP} {\bf 10} (2017) 048, [\href{http://arxiv.org/abs/1707.00716}{{\tt arXiv:1707.00716}}].

\bibitem{Anisha:2019nzx}
Anisha, S.~Das~Bakshi, J.~Chakrabortty, and S.~Prakash, {\it {Hilbert Series and Plethystics: Paving the path towards 2HDM- and MLRSM-EFT}},  {\em JHEP} {\bf 09} (2019) 035, [\href{http://arxiv.org/abs/1905.11047}{{\tt arXiv:1905.11047}}].

\bibitem{Dermisek:2024ohe}
R.~Dermisek and K.~Hermanek, {\it {Two-Higgs-doublet model effective field theory}},  {\em Phys. Rev. D} {\bf 110} (2024), no.~3 035026, [\href{http://arxiv.org/abs/2405.20511}{{\tt arXiv:2405.20511}}].

\bibitem{Dermisek:2024btn}
R.~Dermisek and K.~Hermanek, {\it {Feynman Rules in the Two-Higgs Doublet Model Effective Field Theory}},  \href{http://arxiv.org/abs/2411.07337}{{\tt arXiv:2411.07337}}.

\bibitem{Tran:2024srm}
V.~Q. Tran and T.-C. Yuan, {\it {Exploring Abelian\textendash{}non-Abelian kinetic mixing in SMEFT and beyond}},  {\em Phys. Rev. D} {\bf 111} (2025), no.~1 013001, [\href{http://arxiv.org/abs/2408.11626}{{\tt arXiv:2408.11626}}].

\bibitem{Aebischer:2022wnl}
J.~Aebischer, W.~Altmannshofer, E.~E. Jenkins, and A.~V. Manohar, {\it {Dark matter effective field theory and an application to vector dark matter}},  {\em JHEP} {\bf 06} (2022) 086, [\href{http://arxiv.org/abs/2202.06968}{{\tt arXiv:2202.06968}}].

\bibitem{Criado:2021trs}
J.~C. Criado, A.~Djouadi, M.~Perez-Victoria, and J.~Santiago, {\it {A complete effective field theory for dark matter}},  {\em JHEP} {\bf 07} (2021) 081, [\href{http://arxiv.org/abs/2104.14443}{{\tt arXiv:2104.14443}}].

\bibitem{Liang:2023yta}
J.-H. Liang, Y.~Liao, X.-D. Ma, and H.-L. Wang, {\it {Dark sector effective field theory}},  {\em JHEP} {\bf 12} (2023) 172, [\href{http://arxiv.org/abs/2309.12166}{{\tt arXiv:2309.12166}}].

\bibitem{Song:2023jqm}
H.~Song, H.~Sun, and J.-H. Yu, {\it {Complete EFT operator bases for dark matter and weakly-interacting light particle}},  {\em JHEP} {\bf 05} (2024) 103, [\href{http://arxiv.org/abs/2306.05999}{{\tt arXiv:2306.05999}}].

\bibitem{Lessa:2023tqc}
A.~Lessa and V.~Sanz, {\it {Going beyond Top EFT}},  {\em JHEP} {\bf 04} (2024) 107, [\href{http://arxiv.org/abs/2312.00670}{{\tt arXiv:2312.00670}}].

\bibitem{Liao:2016qyd}
Y.~Liao and X.-D. Ma, {\it {Operators up to Dimension Seven in Standard Model Effective Field Theory Extended with Sterile Neutrinos}},  {\em Phys. Rev. D} {\bf 96} (2017), no.~1 015012, [\href{http://arxiv.org/abs/1612.04527}{{\tt arXiv:1612.04527}}].

\bibitem{Datta:2020ocb}
A.~Datta, J.~Kumar, H.~Liu, and D.~Marfatia, {\it {Anomalous dimensions from gauge couplings in SMEFT with right-handed neutrinos}},  {\em JHEP} {\bf 02} (2021) 015, [\href{http://arxiv.org/abs/2010.12109}{{\tt arXiv:2010.12109}}].

\bibitem{Datta:2021akg}
A.~Datta, J.~Kumar, H.~Liu, and D.~Marfatia, {\it {Anomalous dimensions from Yukawa couplings in SMNEFT: four-fermion operators}},  {\em JHEP} {\bf 05} (2021) 037, [\href{http://arxiv.org/abs/2103.04441}{{\tt arXiv:2103.04441}}].

\bibitem{Ardu:2024tzb}
M.~Ardu and X.~Marcano, {\it {Completing the one-loop \ensuremath{\nu}SMEFT renormalization group evolution}},  {\em JHEP} {\bf 10} (2024) 212, [\href{http://arxiv.org/abs/2407.16751}{{\tt arXiv:2407.16751}}].

\bibitem{Zhang:2025ywe}
D.~Zhang, {\it {Two-loop renormalization group equations in the \ensuremath{\nu}SMEFT}},  {\em JHEP} {\bf 06} (2025) 106, [\href{http://arxiv.org/abs/2504.00792}{{\tt arXiv:2504.00792}}].

\bibitem{Beltran:2023ymm}
R.~Beltr\'an, R.~Cepedello, and M.~Hirsch, {\it {Tree-level UV completions for $N_R$SMEFT $d=6$ and $d=7$ operators}},  {\em JHEP} {\bf 08} (2023) 166, [\href{http://arxiv.org/abs/2306.12578}{{\tt arXiv:2306.12578}}].

\bibitem{Grojean:2024qdm}
C.~Grojean, J.~Kley, D.~Leflot, and C.-Y. Yao, {\it {The flavor invariants of the \ensuremath{\nu}SM}},  {\em JHEP} {\bf 12} (2024) 069, [\href{http://arxiv.org/abs/2406.00094}{{\tt arXiv:2406.00094}}].

\bibitem{Chala:2020vqp}
M.~Chala and A.~Titov, {\it {One-loop matching in the SMEFT extended with a sterile neutrino}},  {\em JHEP} {\bf 05} (2020) 139, [\href{http://arxiv.org/abs/2001.07732}{{\tt arXiv:2001.07732}}].

\bibitem{Talbert:2022unj}
J.~Talbert, {\it {The geometric \ensuremath{\nu}SMEFT: operators and connections}},  {\em JHEP} {\bf 01} (2023) 069, [\href{http://arxiv.org/abs/2208.11139}{{\tt arXiv:2208.11139}}].

\bibitem{Duarte:2025zrg}
L.~Duarte, D.~C. Maisian, and T.~Urruzola, {\it {Future collider sensitivities to {\ensuremath{\nu}}SMEFT interactions}},  {\em J. Phys. G} {\bf 52} (2025), no.~6 065001, [\href{http://arxiv.org/abs/2501.07618}{{\tt arXiv:2501.07618}}].

\bibitem{Bolton:2025tqw}
P.~D. Bolton, F.~F. Deppisch, S.~Kulkarni, C.~Majumdar, and W.~Pei, {\it {Constraining the SMEFT Extended with Sterile Neutrinos at FCC-ee}},  \href{http://arxiv.org/abs/2502.06972}{{\tt arXiv:2502.06972}}.

\bibitem{Zapata:2023wsz}
G.~Zapata, T.~Urruzola, O.~A. Sampayo, and L.~Duarte, {\it {Sensitivity prospects for lepton-trijet signals in the $\nu $SMEFT at the LHeC}},  {\em Eur. Phys. J. C} {\bf 84} (2024), no.~3 326, [\href{http://arxiv.org/abs/2305.16991}{{\tt arXiv:2305.16991}}].

\bibitem{Rosauro-Alcaraz:2024mvx}
S.~Rosauro-Alcaraz and L.~P.~S. Leal, {\it {Disentangling left and right-handed neutrino effects in $B\rightarrow K^{(*)}\nu \nu $}},  {\em Eur. Phys. J. C} {\bf 84} (2024), no.~8 795, [\href{http://arxiv.org/abs/2404.17440}{{\tt arXiv:2404.17440}}].

\bibitem{Breso-Pla:2025pds}
V.~Bres\'o-Pla, S.~Cruz-Alzaga, M.~Gonz\'alez-Alonso, and S.~Prakash, {\it {EFT analysis of New Physics at COHERENT with Dirac neutrinos}},  \href{http://arxiv.org/abs/2505.01275}{{\tt arXiv:2505.01275}}.

\bibitem{Fuyuto:2025feh}
K.~Fuyuto, J.~Harz, and S.~Weber, {\it {Impact of $\nu$SMEFT operators on low-scale leptogenesis}},  \href{http://arxiv.org/abs/2510.24843}{{\tt arXiv:2510.24843}}.

\bibitem{Alonso:2025gzl}
R.~Alonso, M.~Spannowsky, S.~Bates, C.~Hays, and C.~Pollard, {\it {Sensitivity of $W$-boson measurements to low-mass right-handed neutrinos}},  \href{http://arxiv.org/abs/2508.08903}{{\tt arXiv:2508.08903}}.

\bibitem{Li:2025slp}
T.~Li, M.~A. Schmidt, and C.-Y. Yao, {\it {Baryon-number-violating nucleon decays in sterile neutrino effective field theories}},  {\em JHEP} {\bf 06} (2025) 077, [\href{http://arxiv.org/abs/2502.14303}{{\tt arXiv:2502.14303}}].

\bibitem{Fuyuto:2024oii}
K.~Fuyuto, J.~Kumar, E.~Mereghetti, S.~Sandner, and C.~Sun, {\it {Sterile neutrino dark matter within the \ensuremath{\nu}SMEFT}},  {\em JHEP} {\bf 09} (2024) 042, [\href{http://arxiv.org/abs/2405.00119}{{\tt arXiv:2405.00119}}].

\bibitem{Feruglio:1992wf}
F.~Feruglio, {\it {The Chiral approach to the electroweak interactions}},  {\em Int. J. Mod. Phys. A} {\bf 8} (1993) 4937--4972, [\href{http://arxiv.org/abs/hep-ph/9301281}{{\tt hep-ph/9301281}}].

\bibitem{Grinstein:2007iv}
B.~Grinstein and M.~Trott, {\it {A Higgs-Higgs bound state due to new physics at a TeV}},  {\em Phys. Rev. D} {\bf 76} (2007) 073002, [\href{http://arxiv.org/abs/0704.1505}{{\tt arXiv:0704.1505}}].

\bibitem{Buchalla:2015wfa}
G.~Buchalla, O.~Cata, A.~Celis, and C.~Krause, {\it {Note on Anomalous Higgs-Boson Couplings in Effective Field Theory}},  {\em Phys. Lett. B} {\bf 750} (2015) 298--301, [\href{http://arxiv.org/abs/1504.01707}{{\tt arXiv:1504.01707}}].

\bibitem{Buchalla:2012qq}
G.~Buchalla and O.~Cata, {\it {Effective Theory of a Dynamically Broken Electroweak Standard Model at NLO}},  {\em JHEP} {\bf 07} (2012) 101, [\href{http://arxiv.org/abs/1203.6510}{{\tt arXiv:1203.6510}}].

\bibitem{Alonso:2012px}
R.~Alonso, M.~B. Gavela, L.~Merlo, S.~Rigolin, and J.~Yepes, {\it {The Effective Chiral Lagrangian for a Light Dynamical ''Higgs Particle''}},  {\em Phys. Lett. B} {\bf 722} (2013) 330--335, [\href{http://arxiv.org/abs/1212.3305}{{\tt arXiv:1212.3305}}]. [Erratum: Phys.Lett.B 726, 926 (2013)].

\bibitem{Buchalla:2013rka}
G.~Buchalla, O.~Cat\`a, and C.~Krause, {\it {Complete Electroweak Chiral Lagrangian with a Light Higgs at NLO}},  {\em Nucl. Phys. B} {\bf 880} (2014) 552--573, [\href{http://arxiv.org/abs/1307.5017}{{\tt arXiv:1307.5017}}]. [Erratum: Nucl.Phys.B 913, 475--478 (2016)].

\bibitem{Buchalla:2013eza}
G.~Buchalla, O.~Cat\'a, and C.~Krause, {\it {On the Power Counting in Effective Field Theories}},  {\em Phys. Lett. B} {\bf 731} (2014) 80--86, [\href{http://arxiv.org/abs/1312.5624}{{\tt arXiv:1312.5624}}].

\bibitem{Gavela:2016bzc}
B.~M. Gavela, E.~E. Jenkins, A.~V. Manohar, and L.~Merlo, {\it {Analysis of General Power Counting Rules in Effective Field Theory}},  {\em Eur. Phys. J. C} {\bf 76} (2016), no.~9 485, [\href{http://arxiv.org/abs/1601.07551}{{\tt arXiv:1601.07551}}].

\bibitem{Buchalla:2016sop}
G.~Buchalla, O.~Cata, A.~Celis, and C.~Krause, {\it {Comment on ''Analysis of General Power Counting Rules in Effective Field Theory''}},  \href{http://arxiv.org/abs/1603.03062}{{\tt arXiv:1603.03062}}.

\bibitem{Buchalla:2017jlu}
G.~Buchalla, O.~Cata, A.~Celis, M.~Knecht, and C.~Krause, {\it {Complete One-Loop Renormalization of the Higgs-Electroweak Chiral Lagrangian}},  {\em Nucl. Phys. B} {\bf 928} (2018) 93--106, [\href{http://arxiv.org/abs/1710.06412}{{\tt arXiv:1710.06412}}].

\bibitem{Alonso:2017tdy}
R.~Alonso, K.~Kanshin, and S.~Saa, {\it {Renormalization group evolution of Higgs effective field theory}},  {\em Phys. Rev. D} {\bf 97} (2018), no.~3 035010, [\href{http://arxiv.org/abs/1710.06848}{{\tt arXiv:1710.06848}}].

\bibitem{Alonso:2015fsp}
R.~Alonso, E.~E. Jenkins, and A.~V. Manohar, {\it {A Geometric Formulation of Higgs Effective Field Theory: Measuring the Curvature of Scalar Field Space}},  {\em Phys. Lett. B} {\bf 754} (2016) 335--342, [\href{http://arxiv.org/abs/1511.00724}{{\tt arXiv:1511.00724}}].

\bibitem{Alonso:2016btr}
R.~Alonso, E.~E. Jenkins, and A.~V. Manohar, {\it {Sigma Models with Negative Curvature}},  {\em Phys. Lett. B} {\bf 756} (2016) 358--364, [\href{http://arxiv.org/abs/1602.00706}{{\tt arXiv:1602.00706}}].

\bibitem{Alonso:2016oah}
R.~Alonso, E.~E. Jenkins, and A.~V. Manohar, {\it {Geometry of the Scalar Sector}},  {\em JHEP} {\bf 08} (2016) 101, [\href{http://arxiv.org/abs/1605.03602}{{\tt arXiv:1605.03602}}].

\bibitem{Cohen:2020xca}
T.~Cohen, N.~Craig, X.~Lu, and D.~Sutherland, {\it {Is SMEFT Enough?}},  {\em JHEP} {\bf 03} (2021) 237, [\href{http://arxiv.org/abs/2008.08597}{{\tt arXiv:2008.08597}}].

\bibitem{Buchalla:2015qju}
G.~Buchalla, O.~Cata, A.~Celis, and C.~Krause, {\it {Fitting Higgs Data with Nonlinear Effective Theory}},  {\em Eur. Phys. J. C} {\bf 76} (2016), no.~5 233, [\href{http://arxiv.org/abs/1511.00988}{{\tt arXiv:1511.00988}}].

\bibitem{Falkowski:2019tft}
A.~Falkowski and R.~Rattazzi, {\it {Which EFT}},  {\em JHEP} {\bf 10} (2019) 255, [\href{http://arxiv.org/abs/1902.05936}{{\tt arXiv:1902.05936}}].

\bibitem{Anisha:2024ryj}
Anisha, D.~Domenech, C.~Englert, M.~J. Herrero, and R.~A. Morales, {\it {HEFT's appraisal of triple (versus double) Higgs weak boson fusion}},  {\em Phys. Rev. D} {\bf 111} (2025), no.~5 055004, [\href{http://arxiv.org/abs/2407.20706}{{\tt arXiv:2407.20706}}].

\bibitem{Domenech:2025gmn}
D.~Domenech, M.~Herrero, R.~A. Morales, and A.~Salas-Bern{\'a}rdez, {\it {Matching HEFT and SMEFT in double and triple Higgs production from weak boson fusion}},  \href{http://arxiv.org/abs/2506.21716}{{\tt arXiv:2506.21716}}.

\bibitem{Grober:2025vse}
R.~Gr{\"o}ber, A.~N. Rossia, and M.~Ryczkowski, {\it {Multi-Higgs Amplitudes Bootstrapped: Dissecting SMEFT and HEFT}},  \href{http://arxiv.org/abs/2509.02680}{{\tt arXiv:2509.02680}}.

\bibitem{Mahmud:2025wye}
S.~Mahmud and K.~Tobioka, {\it {High energy vector boson scattering in four-body final states to probe Higgs cubic, quartic, and HEFT interactions}},  {\em JHEP} {\bf 06} (2025) 153, [\href{http://arxiv.org/abs/2501.16439}{{\tt arXiv:2501.16439}}].

\bibitem{Gomez-Ambrosio:2022why}
R.~G\'omez-Ambrosio, F.~J. Llanes-Estrada, A.~Salas-Bern\'ardez, and J.~J. Sanz-Cillero, {\it {SMEFT is falsifiable through multi-Higgs measurements (even in the absence of new light particles)}},  {\em Commun. Theor. Phys.} {\bf 75} (2023), no.~9 095202, [\href{http://arxiv.org/abs/2207.09848}{{\tt arXiv:2207.09848}}].

\bibitem{Bhardwaj:2024lyr}
A.~Bhardwaj, C.~Englert, D.~Gon\c{c}alves, and A.~Navarro, {\it {Non-linear gauge-Higgs CP violation}},  {\em Phys. Rev. D} {\bf 110} (2024), no.~11 115011, [\href{http://arxiv.org/abs/2407.14608}{{\tt arXiv:2407.14608}}].

\bibitem{DelDuca:2025vux}
V.~Del~Duca and G.~Falcioni, {\it {The two-loop Higgs impact factor}},  {\em JHEP} {\bf 07} (2025) 018, [\href{http://arxiv.org/abs/2504.06184}{{\tt arXiv:2504.06184}}].

\bibitem{Cohen:2021ucp}
T.~Cohen, N.~Craig, X.~Lu, and D.~Sutherland, {\it {Unitarity violation and the geometry of Higgs EFTs}},  {\em JHEP} {\bf 12} (2021) 003, [\href{http://arxiv.org/abs/2108.03240}{{\tt arXiv:2108.03240}}].

\bibitem{Cadamuro:2025car}
L.~Cadamuro, T.~Ingebretsen~Carlson, and J.~Sj\"olin, {\it {Di-Higgs and Effective Field Theory: Signal Reweighting Beyond $m_{hh}$}},  \href{http://arxiv.org/abs/2502.20976}{{\tt arXiv:2502.20976}}.

\bibitem{Goldberg:2024eot}
J.~M. Goldberg, H.~Liu, and Y.~Shadmi, {\it {Dimension-8 SMEFT contact-terms for vector-pair production via on-shell Higgsing}},  {\em JHEP} {\bf 12} (2024) 057, [\href{http://arxiv.org/abs/2407.07945}{{\tt arXiv:2407.07945}}].

\bibitem{Alonso:2023jsi}
R.~Alonso, J.~C. Criado, R.~Houtz, and M.~West, {\it {Walls, bubbles and doom \textemdash{} the cosmology of HEFT}},  {\em JHEP} {\bf 05} (2024) 049, [\href{http://arxiv.org/abs/2312.00881}{{\tt arXiv:2312.00881}}].

\bibitem{Fortuna:2024rqp}
F.~Fortuna, J.~M. M\'arquez, and P.~Roig, {\it {HEFT approach to investigate the muon g-2 anomaly at a muon collider}},  {\em Phys. Rev. D} {\bf 111} (2025), no.~7 075012, [\href{http://arxiv.org/abs/2408.16954}{{\tt arXiv:2408.16954}}].

\bibitem{Alminawi:2023qtf}
M.~Alminawi, I.~Brivio, and J.~Davighi, {\it {Jet Bundle Geometry of Scalar Field Theories}},  {\em J. Phys. A} {\bf 57} (2024) 435401, [\href{http://arxiv.org/abs/2308.00017}{{\tt arXiv:2308.00017}}].

\bibitem{Liu:2023jbq}
H.~Liu, T.~Ma, Y.~Shadmi, and M.~Waterbury, {\it {An EFT hunter\textquoteright{}s guide to two-to-two scattering: HEFT and SMEFT on-shell amplitudes}},  {\em JHEP} {\bf 05} (2023) 241, [\href{http://arxiv.org/abs/2301.11349}{{\tt arXiv:2301.11349}}].

\bibitem{Dawson:2023oce}
S.~Dawson, D.~Fontes, C.~Quezada-Calonge, and J.~J. Sanz-Cillero, {\it {Is the HEFT matching unique?}},  {\em Phys. Rev. D} {\bf 109} (2024), no.~5 055037, [\href{http://arxiv.org/abs/2311.16897}{{\tt arXiv:2311.16897}}].

\bibitem{Song:2025kjp}
H.~Song and X.~Wan, {\it {Matching the real Higgs triplet extension of Standard Model to HEFT}},  {\em JHEP} {\bf 06} (2025) 249, [\href{http://arxiv.org/abs/2503.00707}{{\tt arXiv:2503.00707}}].

\bibitem{Chakraborty:2024ciu}
D.~Chakraborty, S.~Chattopadhyay, and R.~S. Gupta, {\it {Towards the HEFT-hedron: the complete set of positivity constraints at NLO}},  \href{http://arxiv.org/abs/2412.14155}{{\tt arXiv:2412.14155}}.

\bibitem{Song:2024kos}
H.~Song and X.~Wan, {\it {A non-linear representation of general scalar extensions of the Standard Model for HEFT matching}},  {\em JHEP} {\bf 06} (2025) 021, [\href{http://arxiv.org/abs/2412.00355}{{\tt arXiv:2412.00355}}].

\bibitem{Liao:2025bmn}
Y.~Liao, X.-D. Ma, and Y.~Uchida, {\it {Effective field theory for a type II seesaw model: Symmetric phase vs broken phase}},  {\em Phys. Rev. D} {\bf 112} (2025), no.~3 035005, [\href{http://arxiv.org/abs/2504.02580}{{\tt arXiv:2504.02580}}].

\bibitem{Helset:2020yio}
A.~Helset, A.~Martin, and M.~Trott, {\it {The Geometric Standard Model Effective Field Theory}},  {\em JHEP} {\bf 03} (2020) 163, [\href{http://arxiv.org/abs/2001.01453}{{\tt arXiv:2001.01453}}].

\bibitem{Cohen:2024bml}
T.~Cohen, X.~Lu, and Z.~Zhang, {\it {What is the geometry of effective field theories?}},  {\em Phys. Rev. D} {\bf 111} (2025), no.~8 085012, [\href{http://arxiv.org/abs/2410.21378}{{\tt arXiv:2410.21378}}].

\bibitem{Helset:2022pde}
A.~Helset, E.~E. Jenkins, and A.~V. Manohar, {\it {Renormalization of the Standard Model Effective Field Theory from geometry}},  {\em JHEP} {\bf 02} (2023) 063, [\href{http://arxiv.org/abs/2212.03253}{{\tt arXiv:2212.03253}}].

\bibitem{Assi:2023zid}
B.~Assi, A.~Helset, A.~V. Manohar, J.~Pag\`es, and C.-H. Shen, {\it {Fermion geometry and the renormalization of the Standard Model Effective Field Theory}},  {\em JHEP} {\bf 11} (2023) 201, [\href{http://arxiv.org/abs/2307.03187}{{\tt arXiv:2307.03187}}].

\bibitem{Assi:2025fsm}
B.~Assi, A.~Helset, J.~Pag\`es, and C.-H. Shen, {\it {Renormalizing Two-Fermion Operators in the SMEFT via Supergeometry}},  \href{http://arxiv.org/abs/2504.18537}{{\tt arXiv:2504.18537}}.

\bibitem{Talbert:2021iqn}
J.~Talbert and M.~Trott, {\it {Dirac masses and mixings in the (geo)SM(EFT) and beyond}},  {\em JHEP} {\bf 11} (2021) 009, [\href{http://arxiv.org/abs/2107.03951}{{\tt arXiv:2107.03951}}].

\bibitem{Cheung:2021yog}
C.~Cheung, A.~Helset, and J.~Parra-Martinez, {\it {Geometric soft theorems}},  {\em JHEP} {\bf 04} (2022) 011, [\href{http://arxiv.org/abs/2111.03045}{{\tt arXiv:2111.03045}}].

\bibitem{Aigner:2025xyt}
P.~Aigner, L.~Bellafronte, E.~Gendy, D.~Haslehner, and A.~Weiler, {\it {Renormalising the field-space geometry}},  {\em JHEP} {\bf 07} (2025) 167, [\href{http://arxiv.org/abs/2503.09785}{{\tt arXiv:2503.09785}}].

\end{thebibliography}\endgroup
\end{document}